\documentclass[a4paper,11pt]{article}
\pdfoutput=1 
\usepackage{jheppub} 
\usepackage{tikzit}
\usepackage{array}
\newcolumntype{P}[1]{>{\centering\arraybackslash}p{#1}}

\usepackage[bb=boondox]{mathalfa}
\usepackage{tikz}
\usepackage{url}
\usepackage{caption}
\usepackage{csquotes}
\usepackage{braket}
\usepackage{makecell}
\usepackage{array}
\usepackage[shortlabels]{enumitem}
\usepackage{xcolor}
\usepackage{physics}
\usepackage{comment}
\usepackage{mathtools,tikz-category}
\usepackage{subcaption}
\usepackage{bbm} 
\usepackage{dsfont}
\usepackage{mathrsfs}
\usepackage{float}
\usepackage{quiver}
\usepackage{amsfonts}
\usepackage{booktabs}
\usetikzlibrary{braids}

\usepackage{xcolor}
\usepackage{physics}
\usepackage{tikz-cd} 
\usepackage{float}
\usepackage{tikz} 
\usepackage{adjustbox}
\usepackage{comment}
\usepackage{ytableau}
\usepackage{graphicx,psfrag,color}
\usepackage{pstricks,pst-node,pst-text,pst-3d}
\usetikzlibrary{arrows}
\newcommand{\midarrow}{\tikz \draw[-triangle 90] (0,0) -- +(.1,0);}
\usepackage{fixme}
\fxsetup{status=draft} 
\usetikzlibrary{patterns}
\usepackage{xtab}
\newcommand{\wt}[1]{\widetilde{#1}}
\newcommand{\q}{\mathfrak{q}}
\newcommand{\im}{\mathrm i}
\newcommand{\uR}{\tau}
\newcommand{\R}{\mathrm R}

\newcommand{\Y}{\mathtt{Y}}
\newcommand{\y}{\mathtt{y}}
\newcommand{\Rn}{\mathfrak R}
\newcommand{\bchi}{{\mathcal{X}}}
\newcommand{\lnorm}[1]{\left\lVert#1\right\rVert}
\newcommand{\tu}{\tilde{u}}
\newcommand{\Ll}{R}
\newcommand{\gf}{\theta}
\newcommand{\vc}{v_{I}}
\newcommand{\vd}{v_{I}}
\newcommand{\vbu}{v_{B}}
\newcommand{\uuR}{\tilde{\tau}}
\newcommand{\lattD}{ \widehat{{\cal D}}^{ ( \rm latt)}}
\newcommand{\lattDS}{ \widehat{{\cal D}}^{( \rm latt),0}}

\usetikzlibrary{arrows}
\usepackage{array}
\usepackage{subcaption}
\usepackage{wrapfig}
\usepackage{tabularx}
\usepackage{ltablex}
\usetikzlibrary{shapes.multipart}

\renewcommand{\arraystretch}{1.1}

\makeatletter
\let\over\@@over
\makeatother

\title{\boldmath Integrability and lattice discretizations of all Topological Defect Lines in minimal CFTs\\
}

\author[a]{Madhav Sinha,}
\author[b]{Thiago Silva Tavares,}
\author[a]{Ananda Roy,}
\author[c,d]{and Hubert Saleur }

\affiliation[a]{ Department of Physics and Astronomy, Rutgers University, 126
Frelinghuysen Rd., Piscataway NJ 08854, USA}
\affiliation[b]{Institute of Physics, University of S\~{a}o Paulo, Rua do Mat\~{a}o, S\~{a}o Paulo, SP 05508-090, Brazil}
\affiliation[c]{Institut de physique théorique, CEA, CNRS, Université Paris-Saclay, France}
\affiliation[d]{Physics Department, University of Southern California, Los Angeles, USA}

\emailAdd{ms3066@physics.rutgers.edu, tavares@df.ufscar.br,  ananda.roy@physics.rutgers.edu, hubert.saleur@ipht.fr}

\abstract{We discuss in this paper the lattice discretizations of all topological defect lines (TDLs) for diagonal, minimal  CFTs, using integrable  restricted solid-on-solid (RSOS) models. For these CFTs, the TDLs can be labeled by the Kac labels. In the case of $(1,s)$ TDLs, lines that are exactly topological on the lattice can be obtained using the centralizer of the underlying Temperley-Lieb algebra, all the other lines become topological in the continuum limit only. Our general  construction relies on  insertions of rows/columns of faces with modified spectral parameters, and can therefore be studied using integrability techniques. We determine the regions of spectral parameters realizing the different $(r,s)$ TDLs, and in particular calculate analytically all the associated eigenvalues (and degeneracy factors). We also show  how fusion of TDLs can be obtained from fusion hierarchies in the algebraic approach to the Bethe-ansatz. All our results are checked numerically in detail for several minimal CFTs.}
\begin{document} 
\maketitle
\newpage
\section{Introduction}\label{sec:intro}

Topological defects and the associated (possibly non-invertible) symmetries play an increasingly important role in our understanding of quantum field theories, and of their possible phases. In view of potential applications to experiments and (quantum) simulations, it is important to understand how these defects, naturally formulated in the continuum, can emerge from discrete (lattice) regularizations. This is not a simple question. In this paper, we will restrict to the simplest case of 2D CFTs. 

For these  CFTs, some progress on the question was realized early for the Ising model, or for diagonal  minimal models in the case of the $(1,s)$ defects \cite{Oshikawa1997,Grimm2001,Aasen:2020jwb} (see below for the meaning of these labels). However there is, up to now, for general families of CFTs admitting  well known lattice regularizations (such as the $G\times G/G$ coset minimal models), no strategy to obtain {\sl all} the associated defects. This is particularly vexing since there are issues  (in particular, related with entanglement cuts in the presence of defects~\cite{Roy2021a, Roy2024}) which remain controversial at the field theory level, and could likely be solved by carrying out numerical simulations. The present paper aims at identifying all $(r,s)$ defects in diagonal Virasoro minimal models. We will also suggest a simple strategy to extend our results to other coset diagonal minimal models.  

Before launching into detail and a review of the relevant literature, we emphasize that a crucial question in this area has been whether one can build defects that are topological on the lattice already (see below for a more thorough definition of this concept), or whether one must resign to having topological invariance emerge in the continuum limit only. We will argue  that for the diagonal Virasoro minimal models, only one family of defects ($(1,s)$ in our case, but see remark below) is in the former category, while all the others ($(r,s)$, $r\geq 2$ ) are in the latter.  

  To start, we remind the reader that, in the continuum 2D CFT,  \textit{topological defect lines} (TDLs) (denoted generically by ${\cal D}$ below) are topological if all physical observables remain invariant under their deformation as long as these lines do not pass over each other or cross local operator insertions.   
 While studying the CFT on a cylinder or torus , there are two non-contractible loops of interest, one along the (imaginary\footnote{We will usually refrain from specifying that we are working in imaginary time from now on, and simply refer to this direction as time.})  time direction and the other along the space direction. TDLs along the space direction of the cylinder give rise to \textit{line operators}, which are operators acting  on the  usual Hilbert space\footnote{These must not be confused with what are usually called defect operators, viz. the operators living at the end of defect lines. These are the fields encoded in the defect Hilbert space below.}.  These operators (denoted generically by $\widehat{\cal D}$ below) must commute with the left and right   Virasoro  algebras \cite{Petkova:2000ip,Bachas:2004sy, Kormos:2009sk}. In contrast, TDLs extending  along the time direction of the cylinder give rise to a \textit{defect Hilbert space}, and a \textit{defect Hamiltonian} $H_{{\cal D}}$. 

The lattice counterparts of the continuum objects are defined analogously. In the direct channel, this leads to a modified set of interactions in an otherwise homogeneous lattice model. Examples include the duality-twisted Ising chain~\cite{Oshikawa1997, Grimm2001} and its generalizations~\cite{Sinha:2023hum}. In the crossed channel, the defect lines are built out of suitable lattice operators. Explicit examples of these have been worked out in several cases~\cite{Aasen:2016dop, Aasen:2020jwb, Belletete:2018eua, Sinha:2023hum,Bhardwaj:2024kvy,Vanhove:2021nav}. In the lattice framework, for this work, a defect is considered to be topological if the TDLs can be deformed in space and time directions, {\it both} of which are discretized. This is in contrast to recent proposals to consider a defect topological if the modified set of interactions in $H_{{\cal D}}$ can be moved using a {\it local}, unitary translation operator. While this definition is reasonable when one keeps space discrete but time to be continuous, we will follow the more conservative definition which is closer to the continuum definition described above. Indeed, the correct object describing discrete time evolution is not the Hamiltonian but the transfer matrix \footnote{We will also sometimes use the name defect transfer matrix.}, which expands (as a function of the spectral parameter) onto an infinite series of ``higher''  defect Hamiltonians, of which the first one is the usual Hamiltonian. As a result, for a TDL to be topological on the lattice, we will require that there exists a local, unitary translation operator for all these defect Hamiltonians. An equivalent, explicitly two-dimensional condition for defects to be topological on the lattice can be written  in terms of Boltzmann weights for defect faces, as discussed in Section \ref{sec:Top-defects}.

In what follows, we use lattice topological defect line ($l$TDL) to denote a realization of a defect that is already topological on the lattice and discretization of topological defect line (dTDL) to denote a realization of a defect that is only topological in the continuum limit. For simplicity, we may often denote the lattice realization or discretization of the defect and its continuum limit by the same notation (and same for Hamiltonians). When ambiguity might arise, we will use a more explicit notation for lattice quantities - such as $\lattD$.

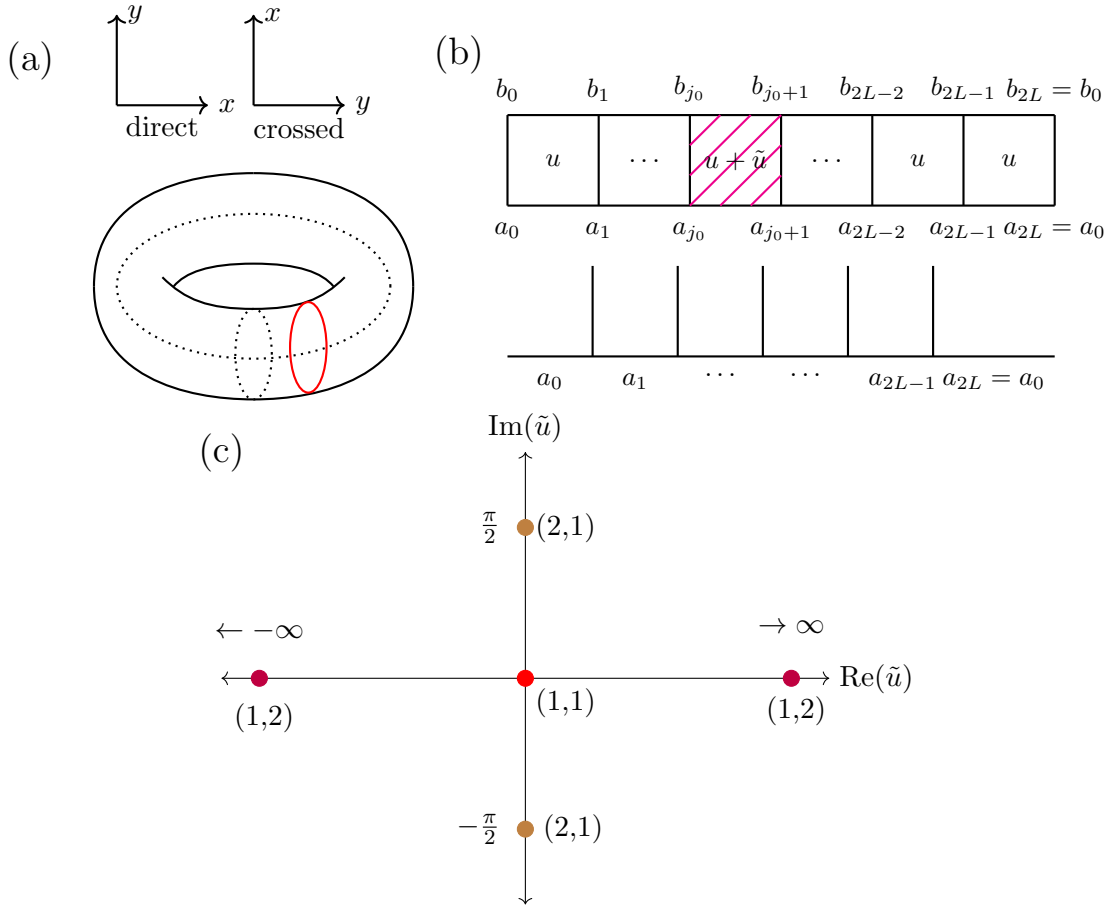
\begin{figure}[h]
\centering
\begin{tikzpicture}[thick, scale=0.6]
\node[anchor=center] at (9.1,3) {\Large{(a)}};
\begin{scope}[shift = {(14,-2)}]
\draw (-3.5,0) .. controls (-3.5,2) and (-1.5,2.5) .. (0,2.5);
\draw[xscale=-1] (-3.5,0) .. controls (-3.5,2) and (-1.5,2.5) .. (0,2.5);
\draw[rotate=180] (-3.5,0) .. controls (-3.5,2) and (-1.5,2.5) .. (0,2.5);
\draw[yscale=-1] (-3.5,0) .. controls (-3.5,2) and (-1.5,2.5) .. (0,2.5);

\draw (-2,.2) .. controls (-1.5,-0.3) and (-1,-0.5) .. (0,-.5) .. controls (1,-0.5) and (1.5,-0.3) .. (2,0.2);

\draw (-1.75,0) .. controls (-1.5,0.3) and (-1,0.5) .. (0,.5) .. controls (1,0.5) and (1.5,0.3) .. (1.75,0);
\draw[black, dotted] (0,-1.5) ellipse (0.4 cm and 1.0 cm);
\draw[red, thick] (1.2,-1.35) ellipse (0.4 cm and 1.0 cm);
\draw[black, dotted] (0.0,0) ellipse (3 cm and 1.6 cm);

\draw[<-] (0,6) -- (0,4) ;
\draw[->] (0,4) -- (2,4) ;

\draw[<-] (-3,6) -- (-3,4) ;
\draw[->] (-3,4) -- (-1,4) ;
\node[anchor=west] at (0,6) {$x$};
\node[anchor=west] at (2,4) {$y$};
\node[anchor=west] at (-3,6) {$y$};
\node[anchor=west] at (-1,4) {$x$};
\node[anchor = north] at (-2, 4) {direct} ; 
\node[anchor = north] at (1, 3.95) {crossed} ; 
\end{scope}

\end{tikzpicture}
            \begin{tikzpicture}[thick, scale=0.8, font = \small]
            \node[anchor=center] at (-0.8,2.5) {\Large{(b)}};
            \draw[black, thick] (0,0) -- (9,0);
            \draw[black, thick] (0,1.5) -- (9,1.5);
            \draw[black, thick] (0,0) -- (0,1.5);
            \draw[black, thick] (1.5,0) -- (1.5,1.5);
            \draw[black, thick] (3,0) -- (3,1.5);
            \draw[black, thick] (4.5,0) -- (4.5,1.5);
            \draw[black, thick] (6,0) -- (6,1.5);
            \draw[black, thick] (7.5,0) -- (7.5,1.5);
            \draw[black, thick] (9,0) -- (9,1.5);
            \draw[magenta] (3,0) -- (4.5,1.5) ;
            \draw[magenta] (3.5,0) -- (4.5,1) ;
            \draw[magenta] (4,0) -- (4.5,0.5) ;
            \draw[magenta] (3,0.5) -- (4,1.5) ;
            \draw[magenta] (3,1) -- (3.5,1.5) ;
            \node[below] at (0,-0.1) {{$a_0$}};
            \node[below] at (1.5,-0.1) {{$a_1$}};
            \node[below] at (3,-0.1) {{$a_{j_0}$}};
            \node[below] at (4.5,-0.1) {{$a_{j_0+1}$}};
            \node[below] at (6,-0.1) {{$a_{2L-2}$}};
            \node[below] at (7.5,-0.1) {{$a_{2L-1}$}};
            \node[below] at (9,-0.1) {{$a_{2L} = a_0$}};
            \node[above] at (0,1.55) {{$b_0$}};
            \node[above] at (1.5,1.55) {{$b_1$}};
            \node[above] at (3,1.55) {{$b_{j_0}$}};
            \node[above] at (4.5,1.55) {{$b_{j_0+1}$}};
            \node[above] at (6,1.55) {{$b_{2L-2}$}};
            \node[above] at (7.5,1.55) {{$b_{2L-1}$}};
            \node[above] at (9,1.55) {{$b_{2L} = b_0$}};
            \node[] at (0.75,0.75) {{$u$}};
            \node[] at (2.25,0.75) {{$\ldots$}};
            \node[] at (3.75,0.75) {{$u + \tu$}};
            \node[] at (5.25,0.75) {{$\ldots$}};
            \node[] at (6.75,0.75) {{$u$}};
            \node[] at (8.25,0.75) {{$u$}};

            \draw[black, thick] (0,-2.5) -- (9,-2.5);
            \draw[black, thick] (1.4,-2.5) -- (1.4,-1) ;
            \draw[black, thick] (2.8,-2.5) -- (2.8,-1) ;
            \draw[black, thick] (4.2,-2.5) -- (4.2,-1) ;
            \draw[black, thick] (5.6,-2.5) -- (5.6,-1) ;
            \draw[black, thick] (7,-2.5) -- (7,-1) ;
            
            \node[below ] at (0.7,-2.6) {{$a_0$}};
            \node[below  ] at (2.1,-2.6) {{$a_1$}};
            \node[below  ] at (3.5,-2.6) {{$\ldots$}};
            \node[below  ] at (4.9,-2.6) {{$\ldots$}};

            \node[below  ] at (6.5,-2.6) {$a_{2L-1}$};
            \node[below  ] at (8,-2.6) {{$a_{2L} = a_0$} };
           
    \end{tikzpicture}
\begin{tikzpicture}
    \node[anchor=center] at (20,3) {\Large{(c)}};

    \draw [<->] (20,0) -- (28,0);
    \draw [<->] (24,3) -- (24,-3);
    \node[anchor=west] at (28,0) {$\Re(\tu)$};
    \node[anchor=south] at (24,3) {$\Im(\tu)$};

    \filldraw [purple] (20.5,0) circle (3pt)node[anchor=north] {};
    \node[anchor=north] at (20.5,-0.2){{ (1,2) }} ; 
    \node[anchor= north] at (27.5,0.9){ $\rightarrow \infty$ } ;
    \node[anchor= north] at (20.5,0.9){ $ \leftarrow  -\infty  $}  ;
    \node[anchor = east] at (23.8,2) {{$\frac{\pi}{2}$}} ; 
    \node[anchor = east] at (23.8,-2) {{$-\frac{\pi}{2}$}} ; 

    \filldraw [purple] (27.5,0) circle (3pt)node[anchor=north] {};
    \node[anchor=north] at (27.5,-0.1){{ (1,2) }} ; 
    \filldraw [brown] (24,2) circle (3pt)node[anchor=east] {};
    \node[anchor=west] at (24,2){ {(2,1) }} ; 
    \filldraw [brown] (24,-2) circle (3pt)node[anchor=east] {};
    \node[anchor=west] at (24.1,-2){{(2,1) }} ; 
    \filldraw [red] (24,0) circle (3pt)node[anchor=east] {};
    \node[anchor= north west] at (24,0){{(1,1)}} ;

\end{tikzpicture}
\caption{\label{fig_1} (a) Schematic of a TDL (in red)  for a two-dimensional CFT. In the direct channel, the TDL gives rise to a defect Hamiltonian $H_{{\cal D}}$. In the crossed channel, it acts instead as an operator $\widehat{{\cal D}}$ ($x$ is the space direction along the chain, while $y$ is the imaginary time direction). Note the same interpretation carries over to more general defect lines such as the perturbed TDLs.~(b) Schematic of the construction of the dTDL (or $l$TDL)  of the restricted solid-on-solid~(RSOS) models with inhomogeneity parameter~$\tu$ and the equivalent anyonic chain. ~(c)  The different topological fixed points as a function of the real and imaginary parts of the inhomogeneity parameter.}
\end{figure}

All possible lattice topological defects have been built for the Ising model \cite{Aasen2016}. $l$TDLs of type $(1,s)$  have also been built~\cite{Aasen2020} for the RSOS models~\cite{Andrews:1984af} realizing the minimal diagonal CFTs with central charges $c=1-6/p(p+1)$. For the three-state Potts model, some $l$TDLs as well as dTDLs for all the remaining  defects of the continuum limit have been identified \cite{Sinha:2023hum}.  
The main purpose of the present paper is to build on  earlier works \cite{Chui_2003,Belletete2018,Belletete2020,Sinha:2023hum, tavares2024} to establish the discretization of  all topological defects  (and especially those of type $(r,1)$) in the diagonal RSOS models using  analytical and numerical techniques. In fact, all our proposed realizations (of $l$TDLs or dTDLs) are integrable. This is intriguing, as some of  the approaches to build  topological defects on the lattice have been quite different. In \cite{Aasen2016,Aasen2020} for instance, equations expressing the invariance of the partition function under deformations of the TDL were written down and solved explicitly, without ever appealing to the Yang-Baxter equation. In contrast, in \cite{Belletete2018,Belletete2020,Sinha:2023hum} - and in the present paper as well -  the interactions across the TDL are  obtained by using a specialization of the Yang-Baxter equation. This raises the question of what integrability has to do with lattice topological invariance. Here, we show that while it is perfectly possible to build lattice topological defects for non-integrable models, the basic moves underlying topological invariance can be interpreted in terms of the Yang-Baxter equation with special (infinite) values of the spectral parameters (see also Fig.~\ref{fig_1}).

We exclusively consider in this paper the minimal models ${\cal M}(p+1,p)$ with central charge $c=1-6/p(p+1)$. The associated conformal weights take values in the Kac table 
\begin{equation}\label{eq:conf-dim-min}
    h_{r,s} = \frac{\left((p+1)r - ps\right)^2 - 1}{4p (p+1)} \, , \quad {\rm where } \, \,  1 \leq r \leq p-1 \, , \, 1 \leq s \leq p \, ,  
\end{equation}
and are associated with chiral primary fields $\phi_{(r,s)}$.  We restrict moreover to  A-type theories, with diagonal partition functions 
\begin{equation}
Z_{A_p}={1\over 2}\sum_{r=1}^{p-1}\sum_{s=1}^p |\chi_{(r,s)}|^2\label{CFTZfct}
\end{equation}
where the symmetry 
 $h_{(r,s)} = h_{(p  - r, p +1 - s)}$ is used to guarantee that all non-chiral primaries $\phi_{(r,s)}$ come with degeneracy one. The number of independent such  primaries   is $\frac{(p-1)(p-2)}{2}$.

 To each primary is associated a topological defect called a Verlinde line\footnote{Since we restrict to diagonal theories, the question of left and right moving defects does not arise. In non-diagonal theories, the formalism can be extended to describe defect of different chiralities \cite{Belletete2020,Sinha:2023hum}.}
 ${\cal D}_{(r,s)}$ (sometimes we just label it by $(r,s)$).  In the crossed channel, this TDL corresponds to an operator, $\widehat{{\cal D}}_{(r,s)}$, which acts on the Hilbert space in the following way 
\begin{equation}\label{eq:verlinde-line-op}
    \widehat{{\cal D}}_{(r,s)} \ket{\phi_{(r',s')}} = \frac{S_{(r,s); {(r',s')}}}{S_{(1,1) ; (r',s')}}\ket{\phi_{(r',s')}} \, , 
\end{equation}
where $S$ is the S-matrix of the CFT, which in this case is given by  
\begin{eqnarray}
S_{(r,s); (r', s')} &=& 2 \sqrt{\frac{2}{p(p+1)}} (-1)^{1 + sr' + rs'} \sin \left( \pi \frac{(p+1)}{p} r r' \right) \sin \left( \pi \frac{p}{(p+1)} s s' \right)\nonumber\\
&=& 2\sqrt{\frac{2}{p(p+1)}} (-1)^{(r+s)(r'+s') } \sin \left( \pi \frac{rr'}{p} \right) \sin \left( \pi \frac{ss'}{p+1} \right)
    \, .
\end{eqnarray}
Note that since $\widehat{{\cal D}}$ commutes with the left and right Virasoro algebra $\hbox{Vir}\otimes \overline{\hbox{Vir}}$, the identity (\ref{eq:verlinde-line-op}) holds for all the fields in the corresponding modules 
$V_{(\rho,\sigma)}\otimes \overline{V}_{(\rho,\sigma)}$ with character $|\chi_{(\rho,\sigma)}|^2$. 

The ``degeneracy factors'' associated with these defects are given by 
\begin{equation}
g_{{\cal D}_{(r,s)}}={S_{(r,s);(1,1)}\over S_{(1,1);(1,1)}}={\sin{\pi r\over p}\sin{\pi s\over p+1}\over \sin{\pi\over p}\sin{\pi\over p+1}}\label{deg-factors}
\end{equation}
In the direct channel with a TDL ${\cal D}_{(r,s)}$ running parallel to the (imaginary) time axis inserted, the Hilbert space changes to ${\cal H}_{ {\cal D}_{(r,s)}}$. Via modular $S$ transformation, we can find the partition function of this theory to be \cite{Petkova:2000ip}
\begin{equation}\label{eq:twist-part-func}
    Z_{{\cal D}_{(r,s)}}(\tau, \bar{\tau}) = \sum_{(i,j), (l,m)}  N_{(r,s) (i,j)}^{(l,m)} \chi_{(i,j)} \bar{\chi}_{(l,m)}(\bar{\tau}) \, , 
\end{equation}
where the $N_{(r,s) (i,j)}^{(l,m)}$  are the fusion coefficients of the theory, given in \cite{DiFrancesco:1997nk}. Using the partition function given above, one can determine the spectrum of ${\cal H}_{ {\cal D}_{(r,s)}   }$, see \cite{Chang:2018iay} for more details.

Note that this whole paper is devoted to studying lattice regularizations of the theories (\ref{CFTZfct}) based on the six-vertex model, or equivalently the Temperley-Lieb algebra. It is for these models that the $(1,s)$ defects are topological on the lattice, and all others in the continuum limit only. It so happens that another regularization of the same CFTs is known based instead on the 19-vertex model and the dilute Temperley-Lieb algebra. For these models, it is the $(r,1)$ defects that are topological on the lattice, while all others are topological in the continuum only. For more details and references, see the related discussion in \cite{tavares2024}.

The manuscript is organized as follows.  In section \ref{sec:transfmatham} we recall aspects of  the basic integrability formalism to be used in the rest of the paper. In section \ref{sec:Top-defects} we discuss the concept of lattice topological defect $l$TDLs, analyze the allowed moves of topological defect lines and their relationship with the Yang-Baxter equation. In section \ref{sec:def-ham} we discuss realizations of basic $(1,2)$ and $(2,1)$ defect transfer matrices and defect Hamiltonians. The same discussion but for line operators, i.e. in the crossed channel is carried out in section \ref{sec:line-op}. In section \ref{sec:fused-Tmat-Th}, the whole construction is generalized to ``higher defects'' i.e. defects obtained by ``fusing'' faces. Section \ref{BASection} addresses the problem from the Bethe-ansatz point of view, both in the direct and the crossed channels. To a large extent, this section establishes analytically most of the claims of this paper. We do find it useful however - in particular, with a view on carrying out the same program for other, non-integrable models - to revisit our claims numerically in section \ref{sec:examples}. There, we examine in detail direct and crossed-channels again, with a particular emphasis on corrections to scaling. We also address numerically the issue of fusion, to be re-visited using the Bethe-ansatz in section \ref{continuum}. 

Many technical aspects are relegated to the appendices. Appendix \ref{sec:Tmatrix-appendix} proves a useful formula expressing the row to row transfer matrix in terms of the Temperley-Lieb algebra. Appendix \ref{sec:Yeigenvalues} discusses further the eigenvalues of the $Y$ operator in modules of the affine Temperley-Lieb algebra. Appendix \ref{sec:anyon-chain-RSOS} discusses the relationship between the $Y$ operator and early works on anyonic chains. Appendix \ref{AnotherFusion} discusses our strategy to obtain the fused weights, while appendix \ref{sec:fus:KP} compares our approach with the literature. Appendix \ref{StripsTMH} addresses two technical issues: the dependency of the defect identifications on the bulk spectral parameter $v_B$, and the emergence of unitary CFT results in case of non-hermitian lattice  Hamiltonians. Appendix \ref{sec:Yop-shao-seib} compares our defect construction with the one defined in \cite{Seiberg:2023cdc}, and later studied in \cite{Seiberg:2024gek, Zhang:2020pco}. Appendix \ref{sec:equiv_def_com_F_app} compares our results with those of the pioneering papers \cite{Aasen2016,Aasen2020}. In Appendix \ref{AppendixCorrespondence}, we go into the technicality of why we scale the energy and momentum eigenvalues, obtained via defect Hamiltonian and defect shift operator, for (1,s) defect with respect to  $L - s + 1$. While doing so, we show how to construct $l$TDLs using defect shift operator. 

A last couple of words before embarking on this long paper. 
First, the results presented here should have logically appeared  before \cite{tavares2024}. The latter reference deals with the fact that, for defects $(r,s)$ with $r>1$ - that is, whenever a finite spectral parameter is involved - the proposed model experiences in fact an RG flow between two defect fixed points, which is similar in some sense to the flow in the anisotropic Kondo model. This flow can be studied by scaling both the size of the system and the spectral parameters in the proper way
\cite{tavares2024}. We are not concerned by this aspect here, as we keep the spectral parameter fixed, and simply worry about identifying the corresponding defect in the scaling limit (or infrared limit from the point of view of defect flows) - viz, large lattices and low-energy. Also, some of the results presented here appeared in schematic  form in \cite{Belletete2020}, which can be considered as an (unpublished) precursor to the present work\footnote{Some misprints in \cite{Belletete2020} are corrected here.}. Finally, some of the results are also mentioned  - without in depth analysis - in the pioneering paper  \cite{Chui_2003}. Other early references (for $(1,s)$ defects include \cite{FINCH2014299,PhysRevB.94.085138}, and of course \cite{Aasen2016,Aasen2020}. 
For the convenience of the reader, a list of notations used in this paper is provided below.

\bigskip
 { 

 \centering
    \xentrystretch{0.3} 
    \bottomcaption{List of symbols and conventions}\label{tab:conventions}
\tablefirsthead{
\hline 
\multicolumn{1}{|c|}{\textbf{Notation}} &
\multicolumn{1}{c|}{\textbf{Description}}& \multicolumn{1}{c|}{\textbf{Reference}} 
\\ 
\hline 
}
\tablehead{\multicolumn{3}{c}%
{{\captionsize\bfseries \tablename\ \thetable{} --
continued from previous page}} 
\\
\hline 
\multicolumn{1}{|c|}{\textbf{Notation}} &
\multicolumn{1}{c|}{\textbf{Description}}& \multicolumn{1}{c|}{\textbf{Reference}} 
\\ 
\hline  
}
\tabletail{\hline \multicolumn{3}{|r|}{{Continued on next page}} \\ \hline}
\tablelasttail{\hline}
    \begin{mpxtabular}{|c|c|c|}
                     $\lattD  $ & \makecell{ Lattice realization of the\\ continuum line operator $\widehat{\cal{D}}$} & Sec. \ref{sec:intro} \\
    \hline
        $l$TDL&\makecell{Lattice Topological Defect Line}& Sec. \ref{sec:intro}\\
    \hline

    dTDL&\makecell{Discretization of Topological Defect Line}& Sec. \ref{sec:intro} \\
    \hline
    ${\cal M}(p+1,p)$&\makecell{A-type Minimal model CFT with central charge\\   $c = 1 - 6/(p(p+1))$}& Sec. \ref{sec:intro} \\
        \hline
        $S_{(r,s);(r',s')}$& \makecell{S-matrix element for minimal model CFT,\\ where $(r,s)$ and  $(r',s')$ are primary field labels} & Sec. \ref{sec:intro}\\
    \hline
     $\chi_{(r,s)}$& \makecell{ Virasoro (Kac's module) character}& Sec. \ref{sec:intro} \\
\hline
$g_{{\cal D}_{(r,s)}}$ & \makecell{Defect $g$-function }& Sec. \ref{sec:intro} \\
\hline
 \makecell{ $W \left( \, \cdot \,\lvert \,\cdot \,\right)$, $W_i \left( \, \cdot \,\lvert \,\cdot \,\right) $  \\ $\widetilde{W}_i \left( \, \cdot \,\lvert \,\cdot \,\right)$ } & \makecell{Boltzmann weights differing by a scale} &  \makecell{  Eq. \eqref{eq:standard_face_weight}, \eqref{eq:face_weight_scaled},\\ \eqref{eq:gauged_face_weight} }\\
             \hline
    \makecell{$a,~b,~c,~d,$\\$e,~f,~x$} & \makecell{Symbols to denote\\ ``canonical'' heights} & throughout this work 
    \\    
             \hline
         
             $\psi^{(1)}$ & \makecell{Perron-Frobenius eigenvector for A-type \\ Dynkin diagram}& Eq. \eqref{dynkeigvec} \\

         \hline
         
             $\theta_{a}$ & \makecell{ $= \sin \gamma a/ \sin \gamma$, Height dependent \\  factor appearing in Boltzmann weight } 
             & Eq. \eqref{eq:standard_face_weight} \\
         \hline
         
             $S_{a}$ & \makecell{ Gauge factor for height $a$ } 
             & Eq. \eqref{eq:standard_face_weight} \\
        
         \hline

             $e_i, q, \gamma$ & \makecell{Temperley-Lieb generators \\ and associated parameters} & Eq. \eqref{eq:TL_gen} \\
         \hline
             $g_i$ & Braid generators  & Eq. \eqref{eq:braid_op} \\
         \hline
             $\uR$ & Shift operator & Sec. \ref{sec:transfmatham} \\
             \hline
           
                 $u,\vbu $ & Spectral parameters : $u = {\rm i}  \, \vbu + \gamma/2$ 
                 & Sec. \ref{sec:transfmatham} ,  \ref{subsec:direct-bethe} \\
         \hline
         $2 L ~\&  \, ~2 \R  $ & \makecell{System-sizes in direct and\\  cross channels respectively}  
         &~\\
         \hline

        $\tilde{u},v_I $ & Defect spectral parameters : $\tilde{u} = {\rm i}  \, v_I$& Sec. \ref{subsec:imp-ham} , \ref{subsec:direct-bethe} \\
         \hline
        $T(\{u\})$ & \makecell{ Transfer matrix with the\\ set of spectral parameters  $\{ u\}$} & Eq. \eqref{eq:tmat_TLgen_KP} 
        \\ 
        \hline
    \makecell{$R_j(u),~\tilde{R}_j(u),$\\$~\Rn_j(u) $} & \makecell{ $R$ and normalized $R$ operators}& Eq. \eqref{eq:R&Rtop},  \eqref{unin}\\ 
         \hline

         $T\left(\{u,u+\tilde{u}\}_k \right)$ & \makecell{Defect transfer matrix with spectral parameter \\ $u$ at every site except $k$, where it is $u + \tilde{u}$} & Fig. \ref{fig:transfer-matrix-def-ham} \\
    \hline
\makecell{$q=e^{i\pi/p+1}$, \\ $\tilde{q}=e^{i\pi/p}$ } &q-numbers&   Eq. (\ref{eq:TL_gen}) and (\ref{Xqtilde}))
    \\
    \hline
    $\mathfrak{q} =e^{-2\pi \R/L}$ & Modular parameter  & \makecell{ Sec \ref{specialsubsec} }
\\
        \hline
        
         $H_n$ & \makecell{Homogeneous integrable Hamiltonians, $H_1 \equiv H$ } & Eq. \eqref{eq:tmat_transpose}, \eqref{eq:no-def-1-Ham} \\

        \hline
        
         $H^{k,k+1}(\tilde{u})$ & \makecell{Spectral parameter dependent defect Hamiltonian \\ with defects between sites $k\, \& \, k+1 $} & Eq. \eqref{eq:def_Ham_two_parts} \\
         \hline
             $f(u)$, $\bar{f}(u)$ & \makecell{ Spectral parameter dependent functions \\ which appear in defect Hamiltonians } & Eq. \eqref{eq:f-def-ham}, \eqref{eq:barf-def} \\
             \hline

$T^k(\tilde{u})$ & \makecell{ Local translation operator for\\ defect Hamiltonian $H^{k,k+1}(\tilde{u})$} & Sec. \ref{subsec:defect-Ham-direct-channel} \\
\hline
         $H_\mathcal{D}$ {$(H_\mathcal{D}^k)$ } & \makecell{Lattice Hamiltonian realizing defect\\ Hilbert space $\mathcal{H}_{\mathcal{D}}$ on the lattice} & Sec. \ref{subsec:defect-Ham-direct-channel}  (Appendix \ref{AppendixCorrespondence})\\
         \hline
         $T_\mathcal{D}$ {($T_\mathcal{D}^k$)} & \makecell{Local translation operator which \\ shifts the site of defect in $H_{\mathcal{D}}  $ ($ H^k_{\mathcal{D}}) $} & Sec. \ref{subsec:defect-Ham-direct-channel} (Appendix \ref{AppendixCorrespondence})
         \\

         \hline         
  \makecell{$ ^{(1J)}W \left( \, \cdot \,\lvert \,\cdot \,\right)  \, ,  $  \\  $^{(J1)}W \left( \, \cdot \,\lvert \,\cdot \,\right)$ }& \makecell{Fused Boltzmann weights in \\ horizontal and vertical directions} & Eq. \eqref{eq:Wt-Th-H}, \eqref{eq:Wt-Th-V} \\
         \hline

             $T^{(J)}  \, \&  \, T^{(J)}_{[k]}  $ & \makecell{Fused transfer matrix in the vertical direction \\ and its shifted counterpart }& Eq. \eqref{eq:fusion-T-M-def} \\
         \hline
         $P^{(J)}_k$  &\makecell{ Jones-Wenzl projector acting on $J$ strands \\($J+1$ sites)  ranging from $k$ to $k+J-1$. } & Appendix \ref{projectorsdef}
         \\ 
         \hline

    $\mathcal{R}$&\makecell{Reflection operator}& Eq. \eqref{def:heightreflection}\\
    \hline
        $Y_{\frac{k}{2}}$ &\makecell{Hoop operators}& Eq. (\ref{eq:prop-fac-high-fus}) \\
    \hline
 $\Lambda\left(\{\vbu,\vbu+\vd\} \right)$&  \makecell{ Eigenvalue of \\ $(-1)^L \, T\left(\{ \frac{\gamma}{2} + {\rm i} \vbu  , \frac{\gamma}{2} + {\rm i} (\vbu + \vd ) \}_k \right)$ } & Eq. \eqref{eq:eig-def-tmat-lamb}
\\ \hline
$\eta$, $l$ & \makecell{Twisting parameter in Bethe ansatz  \\ and discrete integer labeling its possible values   }& Eqs. \eqref{lambdapart} and \eqref{exponentsimp}
\\
    \hline
 $\Lambda^{(J)}(\vc)$  &  Eigenvalue of ${(-1)}^{ \R }~T^{(J)}\left({\gamma\over 2}+i\vc \right)$  & Eq. (\ref{signL}) 
 \\
 \hline 
 $q(\vc)$ and $\Phi(\vc)$ &  \makecell{$q$-function and vacuum function \\ in Bethe ansatz }& Eq. \eqref{eq:bethe-ansatz-qfunc}
\\
\hline 
$e_0^{(J)},\tilde{e}_0^{(J)}$ &  \makecell{Non-universal contribution to the eigenvalue \\ in Bethe ansatz
} & Eq. \eqref{bulk} , \eqref{eq:corr-norme}
\\
\hline
   $[x]_{e^{i\alpha}}$ &Quantum number ${\sin x\alpha\over \sin\alpha}$&  Eqs.  (\ref{1sentropies}) and (\ref{expectedvalues}) )
\\
    \hline
    $\y$, $\Y$&\makecell{Y system functions}& Eq. (\ref{yfunctions}) 
    \\
    \hline
$\lattDS$ & $\lattD$ without shift operator and phase factors.  & Eq. (\ref{eq:lattDSdef})
 \\
    \hline

   $\Pi$ &\makecell{$\pm\lim_{\R \to \infty} {\cal R}$}& Eq. (\ref{contReflection})
 \\
    \hline
   $\ell_2$ &\makecell{The Hilbert space of sequences $(\alpha_n)$ \\ such that $\sum_n |\alpha_n|^2<\infty$}&  \makecell{ Eqs. (\ref{normNFR}) and (\ref{norml2co})\\(induced matrix norm)}
    \\
    \hline
    $M_{\alpha \beta}$ & Double column operator & Eq. \eqref{Mdefined} and Fig. \ref{doublecolumnn}
\\

    \hline
    $\wt{F}$, $\hat{F}$ & $F$ - symbol for ${\rm su}(2)_{p+1}$ and ${\cal A}_{p}$ & Eq. \eqref{eq:F_symb_su2} $\&$ \eqref{eq:F_symb_A} 
\\
    \hline
    $\wt{Y}_{\frac{k}{2}}$ & \makecell{Topological symmetry operator defined \\ using $F$-symbol}
    & Eq. \eqref{eq:top-sym-op-su2}
\\
 
 \hline

     $v^{\pm}$ &\makecell{Symbols for non-canonical heights}& Appendix \ref{AnotherFusion}
    \\
   \hline

    \end{mpxtabular}

}

\newpage

\section{RSOS Model : General Features}\label{sec:transfmatham}

The RSOS models are labeled by Dynkin diagrams of ADE type (the same diagram  that labels their modular invariant partition function in the continuum limit \cite{Cappelli:1986hf}). In the following we will restrict  ourselves to the study of  $A$ type (although our construction carries over immediately to the $D$ and $E$ types - see e.g. \cite{Sinha:2023hum} for the analysis of the $D_4$ RSOS, {\it i.e.,} the three-state Potts model). 

\subsection{Transfer matrix} 
In the two-dimensional statistical mechanical formulation, the A$_p$ RSOS model is described by a square lattice, with \enquote{height} variables taking values   $1, 2,  \ldots,  p$ on each vertex. The heights are constrained by the rule that two nearest neighboring heights are also neighbors on the Dynkin diagram. 
The Boltzmann  weights are assigned to  faces as shown \footnote{Note that we used slightly different conventions in \cite{tavares2024} with $u={\gamma\over 2}+iv$.} in Fig. \ref{fig:face-RSOS} and Eq. (\ref{eq:standard_face_weight}):
\begin{figure}[H]
\begin{adjustbox}{max totalsize={.4\textwidth}{.4\textheight},center}
\begin{tikzpicture}[thick, scale=0.8]
    \draw[black, thick] (0,0) -- (3,0);
    \draw[black, thick] (3,0) -- (3,3);
    \draw[black, thick] (0,0) -- (0,3);
    \draw[black, thick] (0,3) -- (3,3);
     \node[] at (1.5,1.5) {\Large{$u$}};
     \node[] at (0,-0.3) {\Large{$a$}};
     \node[] at (3,-0.3) {\Large{$b$}};
     \node[] at (3,3.25) {\Large{$c$}};
     \node[] at (0,3.35) {\Large{$d$}};
    \draw[black, thick] (4,1.4) -- (4.5,1.4);
    \draw[black, thick] (4,1.6) -- (4.5,1.6);
    \draw[red,thick]  (0,0.3) arc (90:0:0.3);
    \node[] at (7,1.5) {$
     W  \! \!  \left(\begin{array}{ll}
d & c\\
a& b
\end{array} \Bigg| \ u \right)$} ;
\end{tikzpicture}
\end{adjustbox}
    \caption{Face of RSOS model and the weight attached to it.}
    \label{fig:face-RSOS}
\end{figure}

\begin{equation}\label{eq:standard_face_weight}
 \begin{split}
 W  \! \!  \left(\begin{array}{ll}
d & c \\
a & b
\end{array} \Bigg| \ u\right) & = \delta_{a , c} \frac{\sin (\gamma - u) }{ \sin \gamma} + 
\delta_{b,d} \sqrt{\frac{\gf_a \gf_c}{\gf_b \gf_d}} ~ \frac{S_a}{S_c} ~\frac{\sin u}{\sin \gamma} \, , \\ 
\text{where } \quad  & \gamma = \frac{\pi}{p + 1} \, ,  \quad \gf_t = \frac{\sin \left( \gamma t\right)}{  \sin \gamma}\, .
 \end{split}
\end{equation}
Here, $S_a$ are ``gauge'' factors (they disappear in the calculation of partition or correlation functions  on a periodic lattice, but will be useful later on when we discuss fusion). When we use the symbol $W$ for weights, these gauge factors are set to  1.

The weights in Eq. \eqref{eq:standard_face_weight} satisfy three important conditions. The first  is known as Yang-Baxter equation:
     \begin{equation}\label{eq:YBRSOS}
     \begin{split}
\sum_g W  \! \!  \left(\begin{array}{ll}
f & g \\
a & b
\end{array} \Bigg| \ u - v \right)  W  \! \!  \left(\begin{array}{ll}
f & e \\
g& d
\end{array} \Bigg| \ v\right) W  \! \!  \left(\begin{array}{ll}
g & d \\
b& c
\end{array} \Bigg| \ u \right) =  \\
          \sum_g    W \! \! \left(\begin{array}{ll}
f & e \\
a & g
\end{array} \Bigg| \ u \right)  W  \! \! \left(\begin{array}{ll}
a & g \\
b & c
\end{array} \Bigg| \ v\right) W  \! \!  \left(\begin{array}{ll}
e & d \\
g& c
\end{array} \Bigg| \ u -v \right) \, , 
\end{split}
     \end{equation}
the second is the Unitarity constraint:
    \begin{equation}\label{eq:YB-Eq-face}
           \sum_{e}  W  \! \!  \left(\begin{array}{ll}
d & c \\
e & b
\end{array} \Bigg| \ u\right)  W  \! \! \left(\begin{array}{ll}
d & e \\
a & b
\end{array} \Bigg| \ -u\right) = \frac{\sin(\gamma - u)\sin(\gamma + u)}{\sin^2\gamma} \delta_{a,c} \, .
     \end{equation}
In the following, it will be useful to represent diagrammatically eqs. (\ref{eq:YBRSOS}) and (\ref{eq:YB-Eq-face}) as follows: 
\begin{figure}[H]
        \begin{adjustbox}{max totalsize={.75\textwidth}{.7\textheight},center}
    \centering
    \begin{tikzpicture}[scale = 0.6]
        \draw[black, thick] (-10,0) -- (-6,0);
        \draw[black, thick] (-8,4) -- (-6,0);
        \draw[black, thick] (-8,-4) -- (-6,0);
        \draw[black, thick] (-10,0) -- (-12,4);
        \draw[black, thick] (-10,0) -- (-12,-4);
        \draw[black, thick] (-12,4) -- (-14,0);
        \draw[black, thick] (-12,-4) -- (-14,0);
        
        \draw[black, thick] (-12,4) -- (-8,4);
        \draw[black, thick] (-12,-4) -- (-8,-4);

        \node[] at (-14.5,0) {\Large{$a$}};
        \node[] at (-10.5,0) {\Large{$g$}};
        \node[] at (-0.5,0) {\Large{$a$}};
        \node[] at (3.8,0.3) {\Large{$g$}};
        \node[] at (8.5,0) {\Large{$d$}};
        \node[] at (-5.5,0) {\Large{$d$}};

        \node[] at (-12,0) {\Large{$u-v$}};
        \node[] at (6,0) {\Large{$u-v$}};
        \node[] at (-9,2) {\Large{$v$}};
        \node[] at (-9,-2) {\Large{$u$}};
        \node[] at (3,2) {\Large{$u$}};
        \node[] at (3,-2) {\Large{$v$}};
        \node[] at (-16.4,0) {\Large{$\sum\limits_{g}$}};
        \node[] at (-3.4,0) {\Large{$= \sum\limits_{g}$}};
        \node[] at (-12.5,4) {\Large{$f$}};
        \node[] at (-7.5,4) {\Large{$e$}};
        \node[] at (1.5,4) {\Large{$f$}};
        \node[] at (6.5,4) {\Large{$e$}};
        \node[] at (-12.5,-4) {\Large{$b$}};
        \node[] at (-7.5,-4) {\Large{$c$}};
        \node[] at (1.5,-4) {\Large{$b$}};
        \node[] at (6.5,-4) {\Large{$c$}};

        \draw[red,thick]  (0.3,0) arc (0:60:0.3);
        \draw[red,thick]  (-9.7,0) arc (0:120:0.3);
        \draw[red,thick]  (4.2,-0.24) arc (-60:68:0.3);
        \draw[red,thick]  (-11.7,-4) arc (0:60:0.3);
        \draw[red,thick]  (2.3,-4) arc (0:120:0.3);
        \draw[red,thick]  (-13.8,-0.25) arc (-60:60:0.3);

        \draw[black, thick] (0,0) -- (2,4);
        \draw[black, thick] (0,0) -- (2,-4);
        \draw[black, thick] (4,0) -- (6,4);
        \draw[black, thick] (8,0) -- (6,4);
        \draw[black, thick] (8,0) -- (6,-4);
        \draw[black, thick] (4,0) -- (6,-4);
        \draw[black, thick] (0,0) -- (4,0);model
        \draw[black, thick] (2,4) -- (6,4);
        \draw[black, thick] (2,-4) -- (6,-4);
        
    \end{tikzpicture}
\end{adjustbox}
    \caption{Yang-Baxter equation for face model.}
    \label{fig:YB-fig}
\end{figure}
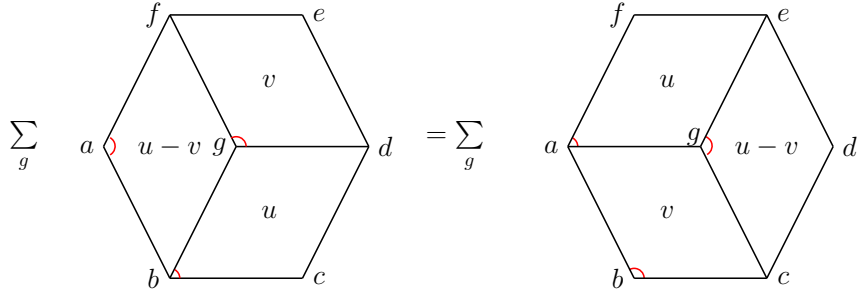
\begin{figure}[H]
\begin{adjustbox}{max totalsize={.75\textwidth}{.75\textheight},center}
    
    \centering
    \begin{tikzpicture}
        \node at (-4.2,-2) {\Large{$\sum\limits_{e}$}};
        \draw[black, thick] (0,0) -- (3,-2);
        \draw[black, thick] (0,0) -- (-3,-2);
        \draw[black, thick] (3,-2) -- (-3,-2);
        \draw[black, thick] (0,-4) -- (3,-2);
        \draw[black, thick] (0,-4) -- (-3,-2);
        \draw[red,thick]  (0.2,-2) arc (0:180:0.2);
        \draw[red,thick]  (0.19,-3.9) arc (60:120:0.3);
        \node[] at (0,-4.2) {\Large{$a$}};
        \node[] at (0,0.2) {\Large{$c$}};
        \node[] at (0,-2.3) {\Large{$e$}};
         \node[] at (3.3, -2) {\Large{$b$}};
        \node[] at (-3.3,-2) {\Large{$d$}};  
        \node[] at (0,-1) {\Large{$u$}};  
        \node[] at (0,-3) {\Large{$-u$}};  
        \node[] at (6,-2) {\Large{$ = \frac{\sin (\gamma + u) \sin(\gamma-u)}{\sin^2 \gamma} \delta_{a,c}$}};
        \end{tikzpicture}
\end{adjustbox}

    \caption{Unitarity for face model.}
    \label{fig:unitarity-fig}
\end{figure}
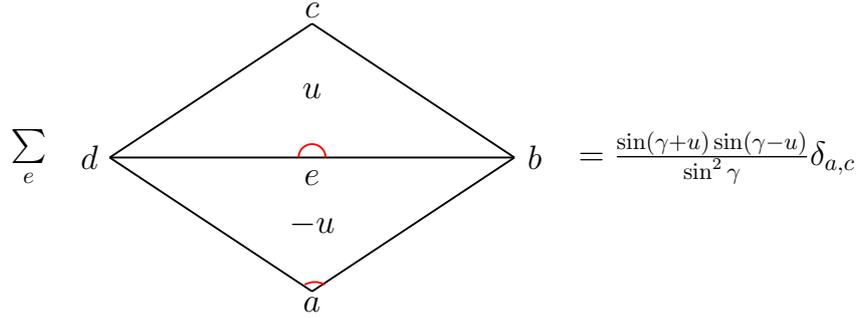

  The last property is the crossing relation. Its importance will become apparent when we discuss the topological nature of defects, and relate our work with \cite{Aasen:2020jwb}.  With the  special gauge choice\footnote{As we shall see below, this choice $S_a=S_c=1$   can be used  to produce hermitian hamiltonians in our problem. The same is not true for other gauge choices. See \autoref{subsec:imp-ham} for more details.} $S_a=S_c=1$ in Eq. \eqref{eq:standard_face_weight}, it reads:
\begin{equation}\label{eq:crossing_sym}    W  \! \!  \left(\begin{array}{ll}
d & c \\
a & b
\end{array} \Bigg| \ u\right)=\sqrt{\frac{\gf_a \gf_c}{\gf_b \gf_d}} \ W  \! \!  \left(\begin{array}{ll}
c & b \\
d & a
\end{array} \Bigg| \ \gamma - u \right) \, .
\end{equation}
It is easy to see if $W$ satisfies Yang-Baxter equation and Unitarity, then so will a weight of the form in Eq. \eqref{eq:standard_face_weight}, where $S_a \neq 1$, but the crossing relation is only satisfied when we choose the gauge factors to be 1.

Using the weights in Eq. \eqref{eq:standard_face_weight}, the transfer matrix is shown in Fig. \ref{fig:gen-transfer-matrix},
\begin{figure}[H]
        \begin{adjustbox}{max totalsize={1\textwidth}{1\textheight},center}
            \begin{tikzpicture}[thick, scale=0.75]
            \node[] at (-2.5,1.5) { \Large{ $ \langle  b  |  T(\{u\}) |  a 
            \rangle =  $ }};
            \draw[black, thick] (0,0) -- (18,0);
            \draw[black, thick] (0,3) -- (18,3);
            \draw[black, thick] (0,0) -- (0,3);
            \draw[black, thick] (3,0) -- (3,3);
            \draw[black, thick] (6,0) -- (6,3);
            \draw[black, thick] (9,0) -- (9,3);
            \draw[black, thick] (12,0) -- (12,3);
            \draw[black, thick] (15,0) -- (15,3);
            \draw[black, thick] (18,0) -- (18,3);
            \node[] at (0,-0.3) {\Large{$a_0$}};
            \node[] at (3,-0.3) {\Large{$a_1$}};
            \node[] at (6,-0.3) {\Large{$a_2$}};
            \node[] at (9,-0.3) {\Large{$a_3$}};
            \node[] at (10.5,-0.3) {\Large{$\ldots$}};
            \node[] at (12,-0.3) {\Large{$a_{2L-2}$}};
            \node[] at (15,-0.3) {\Large{$a_{2L-1}$}};
            \node[] at (18,-0.3) {\Large{$a_{2L} = a_0$}};
            \node[] at (0,3.4) {\Large{$b_0$}};
            \node[] at (3,3.4) {\Large{$b_1$}};
            \node[] at (6,3.4) {\Large{$b_2$}};
            \node[] at (9,3.4) {\Large{$b_3$}};
            \node[] at (12,3.4) {\Large{$b_{2L-2}$}};
            \node[] at (10.5,3.4) {\Large{$\ldots$}};
            \node[] at (15,3.4) {\Large{$b_{2L-1}$}};
            \node[] at (18,3.4) {\Large{$b_{2L} = b_0$}};
            \node[] at (1.5,1.5) {\Large{$u_0$}};
            \node[] at (4.5,1.5) {\Large{$u_1$}};
            \node[] at (7.5,1.5) {\Large{$u_2$}};
            \node[] at (10.5,1.5) {\Large{$\ldots$}};
            \node[] at (13.5,1.5) {\Large{$u_{2L-2}$}};
            \node[] at (16.5,1.5) {\Large{$u_{2L-1}$}};
    \end{tikzpicture}
\end{adjustbox}
    \caption{Inhomogeneous transfer matrix with spectral parameters $u_0, u_1, \ldots, u_{2L-1}$ for RSOS model with periodic boundary condition. }
    \label{fig:gen-transfer-matrix}
\end{figure}

\noindent where we have allowed the spectral parameter to be different in each column of the lattice. We shall only consider the case of  a lattice system with periodic boundary condition (PBC) in this paper: to impose PBC we must have even number of heights in a row. 
 $a_0, a_1, \ldots, a_{2L} = a_0$. Note that the choice in Fig. \ref{fig:gen-transfer-matrix} corresponds to propagation in the vertical time direction: the corresponding point of view is referred to as \textit{direct channel}.

To obtain a bulk CFT without defect, we need to consider the homogeneous case where all values of the spectral parameter are identical $u_0=\ldots=u_{2L-1}=u$. The corresponding transfer matrices $T(u)$ form a commuting family: 
\begin{equation}
    \left[T(u),T(v)\right] = 0 \,, 
\end{equation}
as the local face weights satisfy the Yang-Baxter equation and Unitarity, see  \cite{mccoy_2010} for a proof. Using the weight in Eq. \eqref{eq:standard_face_weight}, one can see that $T(u)$ for $u=0$  acts like the shift (translation) operator by one step to the right  - which we will denote $\uR$. Similarly, $T(0)^{-1}=\uR^{-1}$ is the shift operator in the opposite direction. 

For the quantum RSOS chain, define the Hilbert space, $\mathcal{H}_{2L}$ spanned by allowed configurations on a row, such as $\ket{a_0, a_1, \ldots, a_{2L - 1}}$ in Fig. \ref{fig:gen-transfer-matrix}. On this Hilbert space, we now define special local operators denoted by  $e_i$. To write their action, it is convenient to introduce  the adjacency matrix of the Dynkin diagram. In the case of A$_p$, this  matrix  - denoted by $G$ -  is a $p \times p$ matrix, with  $G_{ij}$  the number of links connecting nodes $i$ and $j$. We define the $2L$ operators $e_1, \ldots, e_{2L} $ 
\begin{equation}\label{eq:ei-operator}
   \bra{....,b_{i-1},b_{i},b_{i+1},....} e_i \ket{....,a_{i-1}, a_{i}, a_{i+1}....} = \left( \prod_{j \neq i}\delta_{a_j, b_j} \right)\frac{\left(\psi^{(1)}_{a_i}\psi^{(1)}_{b_i}\right)^{\frac{1}{2}}}{\psi^{(1)}_{a_{i-1}}}\delta_{a_{i-1}, a_{i+1}} \, ,
\end{equation}
where $a_{2L} \equiv a_0$ and $\psi^{(1)}$ is the eigenvector corresponding to the largest eigenvalue (Perron-Frobenius eigenvector) for the adjacency matrix of the Dynkin diagram.  For the A-type Dynkin diagrams that we study here, the eigenvector is given by \cite{Warnaar:1993ka}
\begin{equation}\label{dynkeigvec}
    \psi^{(1)} = \left(\sin(\frac{\pi}{p+1}), \, \sin(\frac{2\pi}{p+1}),  \ldots , \, \sin(\frac{p \pi}{p+1})   \right) \, , 
\end{equation}
which reduces Equation \eqref{eq:ei-operator} to the same form as in \cite{Feiguin:2006ydp}.
It can be shown \cite{Saleur:1990uz} that the $2L $ generators satisfy
\begin{eqnarray}\label{eq:TL_gen}
 e_j^2 &=& (q + q^{-1}) \, e_j \,,  \nonumber \\[0.2cm]
e_ie_{i\pm1}e_i &=& e_i \, , \\[0.2cm]
e_i e_j &=& e_j e_i \hspace{1cm} \text{ if }|i-j|\geq2 \nonumber \, ,
\end{eqnarray}
where $e_{2L+k} \equiv e_{k}$ and $q = {\rm e}^{{\rm i}\gamma}$. The (infinite dimensional) associative $\mathbb{C}$ algebra generated by these local operators satisfying  the properties listed above is called the Periodic Temperley-Lieb~(~TL$_{2L}(q)$~) algebra. 
 On the Hilbert space $\mathcal{H}_{2L}$, we can also define the shift operator $\uR$, which acts as 
\begin{equation}\label{eq:shift-op}
 \uR\ket{a_0,a_1,a_2, ......,a_{2L-1}} = \ket{a_{2L-1},a_0,a_1, ......,a_{2L-2}} \, .  
\end{equation}
 The following relations are then satisfied by $\uR$ and  the $e_i$'s
\begin{subequations}\label{eq:aTL-2}
\begin{equation}\label{eq:shift-ei-rel}
    \uR e_i = e_{i+1} \uR \, ,
\end{equation}
\begin{equation}\label{eq:prod_of_ei_rel}
\uR^2 e_{2L-1} = e_1  e_2.....e_{2L-1} \ .
\end{equation}
\end{subequations}
$\uR$, $\uR^{-1}$, and the periodic TL generators ($e_i$), which satisfy the relations in eqs. \eqref{eq:TL_gen} and \eqref{eq:aTL-2}, generate the affine TL algebra - aTL$_{2L}(q)$. aTL$_{2L}(q)$ is again an infinite dimensional associative $\mathbb{C}$ algebra which contains TL$_{2L}(q)$ as a subalgebra (see \cite{Belletete:2018eua} for more details).  Using the TL generators, we can also define the braid operators : 
    \begin{equation}\label{eq:braid_op}
  \quad       g_i^{\pm 1} = (-q)^{\pm 1/2} \mathbb{1} + (-q)^{\mp 1/2} e_i \, , 
    \end{equation}
 which satisfy the usual braid group relations
\begin{equation} 
\begin{split}
    g_{i} g_{i \pm 1} g_{i} & =  g_{i \pm 1} g_{i} g_{i \pm 1} \, ,  \\ 
g_i g_j & = g_j g_i \hspace{1cm} \text{ if }|i-j|\geq2  \, .
\end{split}
\end{equation}
Going back to the general case, the transfer matrix $T \left(  \{ u \} \right)$ can be expressed in terms of  generators in the affine Temperley-Lieb (aTL) algebra:
\begin{equation}\label{eq:tmat_TLgen_KP}
  T \left(  \{ u \} \right)   =  \frac{\sin u_0 }{\sin^{2L} \gamma } \left(\prod_{j = 1}^{2L-1}\tilde{R}_j(u_j) \right)  \uR^{-1} + \frac{\sin \left( \gamma - u_0\right)}{\sin^{2L} \gamma} \uR \prod_{j = 1}^{2L-1}R_{2L-j}(u_{2L-j})  = T_A (\{ u\}) + T_B (\{ u\})   \, ,  
 \end{equation}
where  
 \begin{equation}\label{eq:R&Rtop}
    R_j(u_j) = \sin(\gamma - u_j) \, \mathbb{1} + \sin(u_j) \, e_j , \ \tilde{R}_j(u_j) =   \sin(\gamma-u_j) \,  e_j + \sin(u_j) \, \mathbb{1} .  
\end{equation}
Here, the ``$R-$matrix'' is an object that  appears (although in a different - spin instead of heights - representation) in the symmetric 6 vertex model  solution of the  Yang-Baxter equation  \cite{ Baxter2013, mccoy_2010} 
\begin{equation}\label{eq:Yang-Baxter}
   R_j(u) R_{j+1}(u+v)R_j(v) =    R_{j+1}(v) R_{j}(u+v)R_{j+1}(u)   \, .
\end{equation}
To avoid ambiguities, let us specify how we interpret from now on products such as (\ref{eq:tmat_TLgen_KP}):  we have a  chain, where the first site is labeled by $0$ and the last site is labeled by $2L - 1$. Also note $2L + k \equiv k$ and 
\begin{equation}
\begin{split}
    \prod_{j = 1}^{2L-1} R_{2L-j}(u_{2L-j}) & \equiv R_{2L-1}(u_{2L-1}) \, R_{{2L-2}}(u_{2L-2}) \, \dots \, R_{1}(u_1) \, , \\\
 \prod_{j = 1}^{2L-1}\tilde{R}_j(u_j) & \equiv \tilde{R}_{1}(u_1) \, \tilde{R}_{2}(u_2) \, \dots \,  \tilde{R}_{2L-1}(u_{2L-1}) \, .
\end{split}
\end{equation}

We present the proof of the expression in Eq. \eqref{eq:tmat_TLgen_KP} in Appendix \ref{sec:Tmatrix-appendix}. Using Eq. \eqref{eq:tmat_TLgen_KP}, it can be seen that 
\begin{equation}\label{eq:tmat_transpose}
\begin{split}
  & \quad \quad \quad  T(\{ \gamma - u \}) = T(\{u\})^{T}  \, ,  \\ 
 \text{where} \quad \{ \gamma - u \} = & (\gamma - u_0, \gamma - u_1 , \ldots, \gamma - u_{2L-1}) \text{ if } \{u\} = (u_0, u_1, \ldots u_{2L - 1} ),
\end{split}
\end{equation}  
as $\uR^{T} = \uR^{-1}$ and $e_i^{T} = e_i$.  Here, $^T$ denotes the transpose action on the standard basis : $\ket{a_0, a_1, \ldots, a_{2L-1}}$. We also note that $e_{i}^{T} = e_i$ also implies $\left(g_i^{\pm 1}\right)^T = g_{i}^{\pm 1} $.\footnote{Throughout the paper whenever we talk about the transpose of an operator $A$, we mean the operator whose matrix representation is $A^{T}$ in the standard basis.}

\subsection{Hamiltonian}

 We now move on to consider the strongly anisotropic limit where the  2D statistical mechanics model becomes  a 1+1D quantum model \cite{RevModPhys.51.659}. 
Using the transfer matrix (\ref{fig:gen-transfer-matrix}) - first in the homogeneous case - we can define the following operators 
\begin{equation}\label{eq:tmat_ham}
H_n =  - \frac{\partial^n}{\partial u ^{n}} \log T(u) \bigg|_{u = 0} \, , 
\end{equation}
which commute with each other,  and with the homogeneous transfer matrix \cite{Saleur:1990uz}. Amongst all these Hamiltonians, the one which is linear in terms of TL generators is $H_1 \equiv H $, which obeys 
\begin{equation}
    H = - T^{-1}(0)\dot{T}(0) \, .
\end{equation}
Substituting Eq.  \eqref{eq:tmat_TLgen_KP} into this, we get  
\begin{equation}\label{eq:Ham_ei_rel}
\begin{split}
 H & =   - \uR^{-1}\frac{1}{\sin \gamma}\left( \prod_{i = 1}^{2L -1 } e_i\right) \uR^{-1} + \cot \gamma - \sum_{i = 1}^{2L - 1}\left( - \cot \gamma + \frac{1}{\sin \gamma} e_{2L - i}\right) \, ,  \\ 
  & = - \uR^{-2} \frac{1}{\sin \gamma} \prod_{i = 2}^{2L  } e_i  + \cot \gamma  - \sum_{i = 1}^{2L - 1}\left( - \cot \gamma + \frac{1}{\sin \gamma} e_{2L - i}\right)\, . 
\end{split}
\end{equation}
Note, from  Eq. \eqref{eq:prod_of_ei_rel} we have 
\begin{equation}
\begin{split}
    e_{2L - 1} &= \uR^{-2} e_1 e_2 \ldots e_{2L - 1} \, , \\
  \implies  \uR e_{2L-1} \uR^{-1} &= \uR^{-2} \uR e_1 e_2 \ldots e_{2L - 1} \uR^{-1} \, ,  \\ 
  \implies e_0  & = \uR^{-2} e_2 e_3 \ldots e_{2L} \, . 
\end{split}
\end{equation}
Substituting  into  \eqref{eq:Ham_ei_rel} we finally obtain
\begin{equation}\label{eq:no-def-1-Ham}
    H = - \sum_{i = 0}^{2L-1} \left( - \cot \gamma + \frac{1}{\sin \gamma} e_i \right) \, . 
\end{equation}
Another derivation of this Hamiltonian using face weights can be found in \cite{Frahm:2021ugl}. 

Note that the above Hamiltonian for the A$_p$ RSOS model coincides  -  up to a rescaling and a shift, with the Hamiltonian of the Anyonic chain built using the spin-$\frac{1}{2} \, \mathcal{A}_p $ fusion category \cite{Ardonne_2011}: we will show this in \autoref{sec:anyon-chain-RSOS}. In the scaling limit, it can also be shown that these A$_p$ RSOS models/anyonic chains give rise to  the A-type unitary Minimal CFT $\mathcal{M}(p+1,p)$ \cite{Feiguin:2006ydp,Gils_2009, Zini:2017bzi}.

\subsection{Special aspects of the continuum limit}\label{specialsubsec}

Clearly the row to row transfer matrix maps states where even sites carry even heights to states where they carry odd heights (we refer to these sometimes as  even and odd  sectors respectively). This means that we cannot restrict consistently our description of the model with only one type of states. $T(u)$ can however  be written as a block off-diagonal matrix, and its eigenvalues all come in pairs of  equal absolute value and opposite signs. We have in particular, after subtracting the free-energy per site as usual, and in the  regime of interest  $0<u<\gamma$, that 
\begin{equation}
\hbox{Tr}~\left[T(u)\right]^{2\R}\mapsto \sum_{r=1}^{p-1}\sum_{s=1}^p |\chi_{(r,s)}|^2.\label{lattZ}
\end{equation}
Here and elsewhere in the paper we reserve the symbol $\mapsto$ for the scaling limit, which might have slightly different but well known meanings depending on the question being discussed. In the present case  for instance, subtraction of the free-energy per site is implicit, while we must take the limit $2L,2\R\to\infty$ and the modular parameter $\q$ depends on the ratio $\R/L$ and the spectral parameter $u$ \cite{Koo_1994}, e.g. $\mathfrak{q}=e^{-2\pi \R/L}$ at the isotropic point $u={\gamma\over 2}$. Note that the partition function (\ref{lattZ}) is equal to {\sl twice} the conformal partition function (\ref{CFTZfct}): each pair of eigenvalues with equal absolute values  and opposite signs gives rise to twice the contribution of the corresponding  conformal state in (\ref{lattZ}). Furthermore, the existence of the pairs of eigenvalues guarantees that the partition function of the model with an odd length in the time direction vanishes exactly, as it should from the rules of the A$_p$ models:
\begin{equation}
\hbox{Tr}~\left[T(u)\right]^{2\R+1}=0 \, .
\end{equation}
Associated with the sign of the eigenvalues of $T(u)$ is another fact:   positive eigenvalues correspond to states with lattice momentum ${2\pi\over 2L}(h-\bar{h})$ , while those  with negative eigenvalues have a lattice momentum with a finite part: $\pi +{2\pi\over 2L}(h-\bar{h})$ 
(and of course, the same holds in the crossed channel with $\R$ and $L$ exchanged). This can be proven using the Bethe-ansatz: it is expected  as the isotropic point $u={\gamma\over 2}$ since then we can exchange behavior under time and space translations by one site, and follows in the whole domain $0<u<\gamma$ by analyticity. We note that the emergence of finite parts of lattice momentum  may have  deep origins in the context of  anomalies - see \cite{ChengSeiberg}. 

So when we identify  states in the CFT with  states in the lattice model (more precisely, a set of eigenstates of the transfer matrix or Hamiltonian, singled out as $L$ increases so they give rise to the proper continuum limit as usual - this is well known and discussed in detail recently in \cite{Zini:2017bzi,glhjs20}), we have to be careful that in this problem, we will have at our disposal two lattice states corresponding to every state in the module $V_{(r,s)}\otimes \overline{V}_{(r,s)}$, with opposite signs of eigenvalues, and finite parts of the lattice momentum equal to $0$ or $\pi$. The finite part of the momentum can be obtained by simply acting with the translation operator on these states, since, as $L\to\infty$, the conformal part of the lattice momentum vanishes.

The issue becomes important when one delves into the precise correspondence between the lattice model and the CFT. This is because the action of our defect lines will involve acting on the periodic Hilbert space of the RSOS model with a product of transfer matrices at specific values of the spectral parameter. Whenever this product is odd (like for the ${\cal D}_{(12)}$ and ${\cal D}_{(21)}$ lines), these lines operators, like the bulk transfer matrix, will have pairs of eigenvalues with equal absolute values and opposite signs. On the other hand, the eigenvalues of the defect operators in the CFT are uniquely determined by the conformal weights of the  states (see Eq. (\ref{eq:verlinde-line-op})): in order to reproduce these from the lattice, we will have to know how to handle the sign of the action of the lattice realizations such as $\lattD_{(12)}$ and $\lattD_{(21)}$. This is discussed in detail below.

\section{Lattice topological defects (LTDs)}\label{sec:Top-defects}
Recall that topological defects in the CFT  framework can be defined from a Euclidean point of view as  defect lines that can be arbitrarily deformed without changing the partition function - or, more generally, correlation functions of observables  provided the line does not cross over their insertion points. An analogous definition in the setting of a lattice discretization will involve typically an interface between two parts of the bulk model which are glued via a special \textit{seam} (examples of seams are the rows and columns of the previous section). This seam is the lattice discretization of a TDL, and thus, to have a $l$TDL, we should demand that it can be deformed  without, once again, modifying the partition function or correlation functions with insertions untouched by the seam. 
Of course we now need the potential deformations of the seam to be compatible with the underlying lattice: this restricts the possible types of moves, and sometimes makes definitions  a bit cumbersome.

In this section, we will explore local conditions that make the construction of a lattice TDL possible. 
As mentioned in the introduction, the concept of $l$TDL can be extended to  the quantum lattice model. In this case, the condition of  being topological can be reformulated  in terms of the existence of translation operators for a family of higher defect Hamiltonians, as we will see later. 

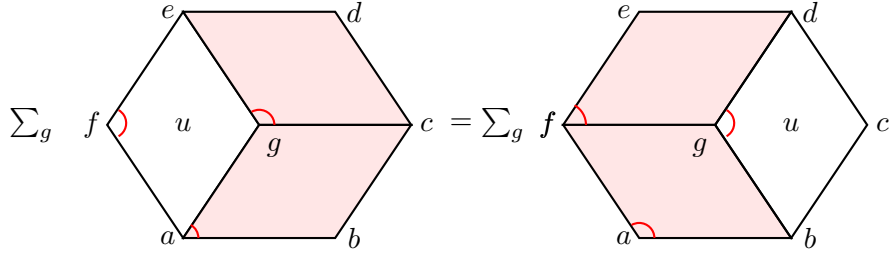
\begin{figure}[H]
    \centering
\begin{tikzpicture}[scale=1.0 ]

    \fill[red!10] (6,0) -- (8,0) -- (9,1.5) -- (7,1.5) -- cycle;
\draw[black, thick] (6,0) -- (8,0) -- (9,1.5) -- (7,1.5) -- cycle;
    \fill[red!10] (6,0) -- (8,0) -- (9,-1.5) -- (7,-1.5) -- cycle;
\draw[black, thick] (6,0) -- (8,0) -- (9,-1.5) -- (7,-1.5) -- cycle;
\draw[black, thick] (8,0) -- (9,1.5) -- (10,0) -- (9,-1.5) -- cycle;

\node at (5.8,0) {$f$} ; 
\node at (6.8,1.5) {$e$} ; 
\node at (9.25,1.5) {$d$} ; 
\node at (6.8,-1.5) {$a$} ; 
\node at (9.25,-1.5) {$b$} ; 
\node at (10.2,0) {$c$} ; 
\node at (7.8,-0.3) {$g$} ; 
\node at (5.8,0) {$f$} ; 
\draw[red, thick] (6.3,0) arc (0 : 60 : 0.3) ;
\draw[red, thick] (7.2,-1.5) arc (0 : 120 : 0.2) ;
\draw[red, thick] (8.15,-0.15) arc (-60 : 60 : 0.2) ;
\node at (9,0) {$u$} ; 
\node at (1,0) {$u$} ; 

\draw[black, thick] (0,0) -- (1,1.5) -- (2,0) -- (1,-1.5) -- cycle;
\fill[red!10] (3,1.5) -- (1,1.5) -- (2,0) -- (4,0) -- cycle;
\fill[red!10] (3,-1.5) -- (1,-1.5) -- (2,0) -- (4,0) -- cycle;
\draw[black, thick] (3,1.5) -- (1,1.5) -- (2,0) -- (4,0) -- cycle;
\draw[black, thick] (3,-1.5) -- (1,-1.5) -- (2,0) -- (4,0) -- cycle;

\node at (-.2,0) {$f$} ; 
\node at (0.8,1.5) {$e$} ; 
\node at (3.25,1.5) {$d$} ; 
\node at (0.8,-1.5) {$a$} ; 
\node at (3.25,-1.5) {$b$} ; 
\node at (4.2,0) {$c$} ; 
\node at (2.2,-0.3) {$g$} ; 
\node at (5,0) {   $= \sum_{g}$} ;
\node at (-1,0) {   $\sum_{g}$} ;

\draw[red, thick] (0.15,-0.15) arc (-60 : 60 : 0.2) ;
\draw[red, thick] (2.2,0) arc (0 : 120 : 0.2) ;
\draw[red, thick] (1.2,-1.5) arc (0 : 60 : 0.2) ;

\end{tikzpicture}
    \caption{The first condition for defect face to be topologically invariant. Note, how this is similar to Yang-Baxter for face models.}
    \label{fig:first_cond_top_inv}
\end{figure}

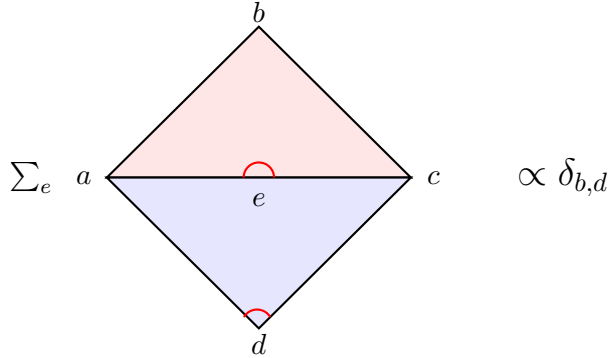
\begin{figure}[H]
    \centering
\begin{tikzpicture}
    \fill[red!10] (0,0) -- (4,0) -- (2,2) -- cycle;
\draw[black, thick] (0,0) -- (4,0) -- (2,2) -- cycle;

\fill[blue!10] (0,0) -- (4,0) -- (2,-2) -- cycle;
\draw[black, thick] (0,0) -- (4,0) -- (2,-2) -- cycle;

\node at (-0.3,0) {$a$};
\node at (4.3,0) {$c$};
\node at (2,2.2) {$b$};
\node at (2,-2.2) {$d$};
\node at (2,-0.3) {$e$};

\node at (-1,0) {$\sum_{e}$};

\node at (6,0) {\Large $\propto \delta_{b,d}$};

\draw[red, thick]  (2.2,0) arc (0 : 180 : 0.2)  ; 
\draw[red, thick]  (2.15,-1.85) arc (30 : 150 : 0.2)  ; 

\end{tikzpicture}   
\caption{Second condition for topological invariance. Notice how it resembles unitarity for usual face weights.}
    \label{fig:top-def-rel-2}
\end{figure}

We call a defect face - represented  in red below -   topological if it satisfies the two conditions given in  figures \ref{fig:first_cond_top_inv} and \ref{fig:top-def-rel-2}   (where note that the red faces do not carry  a spectral parameter). 

In \autoref{sec:equiv_def_com_F_app}, we show how these  conditions are equivalent to the defect commutation relations given in \cite{Aasen:2020jwb}. 

\begin{figure}[H]
    \centering
\begin{tikzpicture}
    \fill[blue!10] (0,0) -- (4,0) -- (2,2) -- cycle;
\draw[black, thick] (0,0) -- (4,0) -- (2,2) -- cycle;

\fill[red!10] (0,0) -- (4,0) -- (2,-2) -- cycle;
\draw[black, thick] (0,0) -- (4,0) -- (2,-2) -- cycle;

\node at (-0.3,0) {$a$};
\node at (4.3,0) {$c$};
\node at (2,2.2) {$b$};
\node at (2,-2.2) {$d$};
\node at (2,-0.3) {$e$};

\node at (-1,0) {$\sum_{e}$};

\node at (6,0) {\Large $\propto \delta_{b,d}$};

\draw[red, thick]  (2.2,0) arc (0 : 180 : 0.2)  ; 
\draw[red, thick]  (2.15,-1.85) arc (30 : 150 : 0.2)  ; 

\end{tikzpicture}   
\caption{The condition here is the inverse of the unitarity condition.}
    \label{fig:top-def-rel-2-inv}
\end{figure}
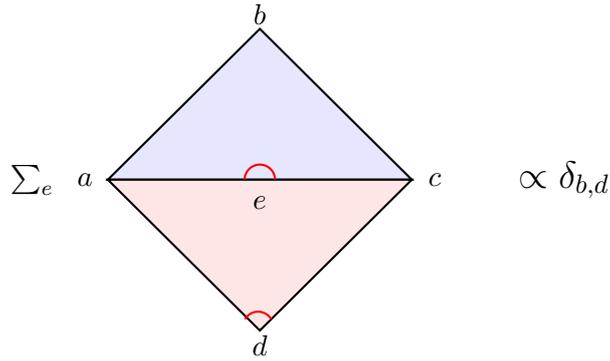

Note that  we assumed  the weights are non-degenerate. By this we mean more precisely that, 
defining $W_R$ and $W_B$ to be the weights corresponding to the red and the blue faces, figure \ref{fig:top-def-rel-2} reads
\begin{equation}
\begin{split}
 &  \sum_e  W_R  \! \!  \left(\begin{array}{ll}
a & b \\
e & c
\end{array} \right)     W_B  \! \!  \left(\begin{array}{ll}
a & e \\
d & c
\end{array} \right) = \delta_{b,d} \\
& \implies \sum_e \wt{M}_{d,e}   M_{e,b}  =\delta_{b,d}
\end{split}
\end{equation}
Assuming that $M$ is invertible, we choose  $\wt{M} = M^{-1} $, and  write 
\begin{equation}
\begin{split}
 & \sum_{e}  M_{d,e} \wt{M}_{e, b}  = \delta_{d,b} \\
 & \implies \sum_e  W_B \! \!  \left(\begin{array}{ll}
a & b \\
e & c
\end{array} \right)     W_R  \! \! \left(\begin{array}{ll}
a & e \\
d & c
\end{array} \right) = \delta_{b,d} 
\end{split}
\end{equation}
which leads to figure \ref{fig:top-def-rel-2-inv}.

Note that if we specialize the red and blue faces to be ordinary faces with spectral parameter set to $\rm{i} \infty$ and $-\rm{i} \infty$,  the conditions in figures \ref{fig:first_cond_top_inv} and \ref{fig:top-def-rel-2} are automatically satisfied due to the Yang-Baxter equation and Unitarity. This is an important case that we will discuss in detail below.

In the next subsection, we will show that if the above conditions are satisfied,  we can use the topological faces to build lattice topological defect lines.

\subsection{Moving defects via braids}
We now discuss how to deform seams made out of topological faces.  To start, we need  braid operators, as defined  in  figure \ref{fig:thread_faceweight} below.
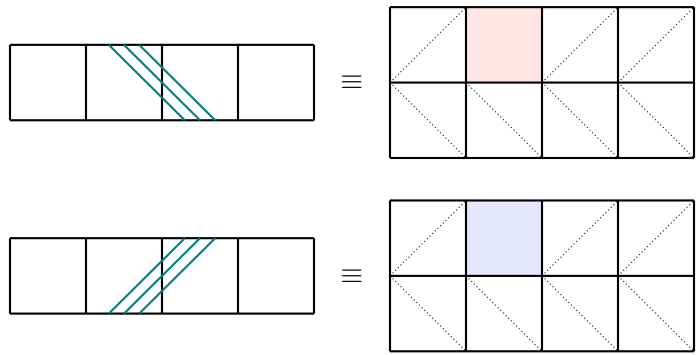
\begin{figure}[H]
    \centering
    \begin{tikzpicture}
        \fill[red!10] (6,1) -- (6,2) -- (7,2) -- (7,1) -- cycle;

        \draw[black,thick] (0,0.5) -- (4,0.5);
        \draw[black,thick] (0,1.5) -- (4,1.5);
        \draw[black,thick] (0,0.5) -- (0,1.5);
        \draw[black,thick] (1,0.5) -- (1,1.5);
        \draw[black,thick] (2,0.5) -- (2,1.5);
        \draw[black,thick] (3,0.5) -- (3,1.5);
        \draw[black,thick] (4,0.5) -- (4,1.5);
        
        \draw[teal,thick] (1.3,1.5) -- (2.3,0.5);
        \draw[teal,thick] (1.5,1.5) -- (2.5,0.5);
        \draw[teal,thick] (1.7,1.5) -- (2.7,0.5);

         \node[] at (4.5,1) {$\equiv$};

        \draw[black,thick] (5,0) -- (9,0);
        \draw[black,thick] (5,1) -- (9,1);
        \draw[black,thick] (5,2) -- (9,2);
        \draw[black,thick] (5,0) -- (5,2);
        \draw[black,thick] (6,0) -- (6,2);
        \draw[black,thick] (7,0) -- (7,2);
        \draw[black,thick] (8,0) -- (8,2);
        \draw[black,thick] (9,0) -- (9,2);
        \draw[black, densely dotted] (5,1) -- (6,0) ; 
        \draw[black, densely dotted] (6,1) -- (7,0) ; 
        \draw[black, densely dotted] (7,1) -- (8,0) ; 
        \draw[black, densely dotted] (8,1) -- (9,0) ; 
        
        \draw[black, densely dotted] (5,1) -- (6,2) ; 
        \draw[black, densely dotted] (7,1) -- (8,2) ; 
        \draw[black, densely dotted] (8,1) -- (9,2) ; 
    \end{tikzpicture}
    
    \begin{tikzpicture}
        \fill[blue!10] (6,1) -- (6,2) -- (7,2) -- (7,1) -- cycle;

        \draw[white, thick] (0,0) -- (0,2.5);
        \draw[black,thick] (0,0.5) -- (4,0.5);
        \draw[black,thick] (0,1.5) -- (4,1.5);
        \draw[black,thick] (0,0.5) -- (0,1.5);
        \draw[black,thick] (1,0.5) -- (1,1.5);
        \draw[black,thick] (2,0.5) -- (2,1.5);
        \draw[black,thick] (3,0.5) -- (3,1.5);
        \draw[black,thick] (4,0.5) -- (4,1.5);
        
        \draw[teal,thick] (1.3,0.5) -- (2.3,1.5);
        \draw[teal,thick] (1.5,0.5) -- (2.5,1.5);
        \draw[teal,thick] (1.7,0.5) -- (2.7,1.5);

         \node[] at (4.5,1) {$\equiv$};
    
        \draw[black,thick] (5,0) -- (9,0);
        \draw[black,thick] (5,1) -- (9,1);
        \draw[black,thick] (5,2) -- (9,2);
        \draw[black,thick] (5,0) -- (5,2);
        \draw[black,thick] (6,0) -- (6,2);
        \draw[black,thick] (7,0) -- (7,2);
        \draw[black,thick] (8,0) -- (8,2);
        \draw[black,thick] (9,0) -- (9,2);
        \draw[black, densely dotted] (5,1) -- (6,0) ; 
        \draw[black, densely dotted] (6,1) -- (7,0) ; 
        \draw[black, densely dotted] (7,1) -- (8,0) ; 
        \draw[black, densely dotted] (8,1) -- (9,0) ; 
        
        \draw[black, densely dotted] (5,1) -- (6,2) ; 
        \draw[black, densely dotted] (7,1) -- (8,2) ; 
        \draw[black, densely dotted] (8,1) -- (9,2) ; 

    \end{tikzpicture}
        \caption{We define the above rows to be the braid operators $g_2$, $g_2^{-1}$.}            \label{fig:thread_faceweight}

\end{figure}

On this figure, the  dotted lines represent delta functions of heights -  the  two heights at their extremities are forced to be equal (with a  Boltzmann weight $W=1$ for the  corresponding face).

If one considers the special case where these red and blue faces are ordinary  faces with spectral parameter i$\infty$ and $-$i$\infty$ respectively, then the operators defined above are actually proportional to the braid operators, which were defined in \eqref{eq:braid_op}.

In place of any row we can insert two braid operators, as they are inverses of each other, due to unitarity and its inverse in figures \ref{fig:top-def-rel-2} and \ref{fig:top-def-rel-2-inv}. 
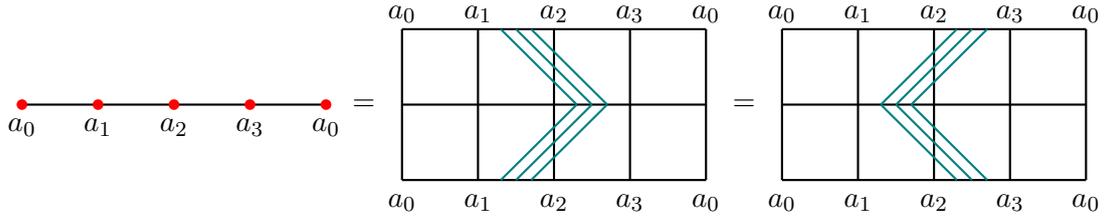
\begin{figure}[H]
    \centering
    \begin{tikzpicture}
        \draw[black, thick] (0,1) -- (4,1) ;
        \fill[red] (0,1)  circle (2pt);  
        \fill[red] (1,1)  circle (2pt);  
        \fill[red] (2,1)  circle (2pt);  
        \fill[red] (3,1)  circle (2pt);  
        \fill[red] (4,1)  circle (2pt);  
        \node[] at (0,0.7) {$a_0$} ;
        \node[] at (1,0.7) {$a_1$} ;
        \node[] at (2,0.7) {$a_2$} ;
        \node[] at (3,0.7) {$a_3$} ;
        \node[] at (4,0.7) {$a_0$} ;
        
        \node[] at (5,-0.3) {$a_0$} ;
        \node[] at (6,-0.3) {$a_1$} ;
        \node[] at (7,-0.3) {$a_2$} ;
        \node[] at (8,-0.3) {$a_3$} ;
        \node[] at (9,-0.3) {$a_0$} ;

        \node[] at (10,-0.3) {$a_0$} ;
        \node[] at (11,-0.3) {$a_1$} ;
        \node[] at (12,-0.3) {$a_2$} ;
        \node[] at (13,-0.3) {$a_3$} ;
        \node[] at (14,-0.3) {$a_0$} ;

        \node[] at (5,2.2) {$a_0$} ;
        \node[] at (6,2.2) {$a_1$} ;
        \node[] at (7,2.2) {$a_2$} ;
        \node[] at (8,2.2) {$a_3$} ;
        \node[] at (9,2.2) {$a_0$} ;

        \node[] at (10,2.2) {$a_0$} ;
        \node[] at (11,2.2) {$a_1$} ;
        \node[] at (12,2.2) {$a_2$} ;
        \node[] at (13,2.2) {$a_3$} ;
        \node[] at (14,2.2) {$a_0$} ;

        \node[] at (4.5,1) {$=$} ;
        \node[] at (9.5,1) {$=$} ;

        \draw[black, thick] (5,0) -- (9,0) ; 
        \draw[black, thick] (5,1) -- (9,1) ; 
        \draw[black, thick] (5,2) -- (9,2) ; 
        
        \draw[black, thick] (10,0) -- (14,0) ; 
        \draw[black, thick] (10,1) -- (14,1) ; 
        \draw[black, thick] (10,2) -- (14,2) ; 

        \draw[black, thick] (5,0) -- (5,2) ;
        \draw[black, thick] (6,0) -- (6,2) ;
        \draw[black, thick] (7,0) -- (7,2) ;
        \draw[black, thick] (8,0) -- (8,2) ;
        \draw[black, thick] (9,0) -- (9,2) ;

        \draw[black, thick] (10,0) -- (10,2) ;
        \draw[black, thick] (11,0) -- (11,2) ;
        \draw[black, thick] (12,0) -- (12,2) ;
        \draw[black, thick] (13,0) -- (13,2) ;
        \draw[black, thick] (14,0) -- (14,2) ;

        \draw[teal, thick] (6.5,0) -- (7.5,1);
        \draw[teal, thick] (6.5,2) -- (7.5,1);
        \draw[teal, thick] (6.3,0) -- (7.3,1);
        \draw[teal, thick] (6.3,2) -- (7.3,1);
        \draw[teal, thick] (6.7,0) -- (7.7,1);
        \draw[teal, thick] (6.7,2) -- (7.7,1);

        \draw[teal, thick] (11.5,1) -- (12.5,0);
        \draw[teal, thick] (11.5,1) -- (12.5,2);
        \draw[teal, thick] (11.3,1) -- (12.3,0);
        \draw[teal, thick] (11.3,1) -- (12.3,2);
        \draw[teal, thick] (11.7,1) -- (12.7,0);
        \draw[teal, thick] (11.7,1) -- (12.7,2);
        
    \end{tikzpicture}
    \caption{Inserting braids in the lattice}
    \label{fig:Briad_RSOS-red}
\end{figure}

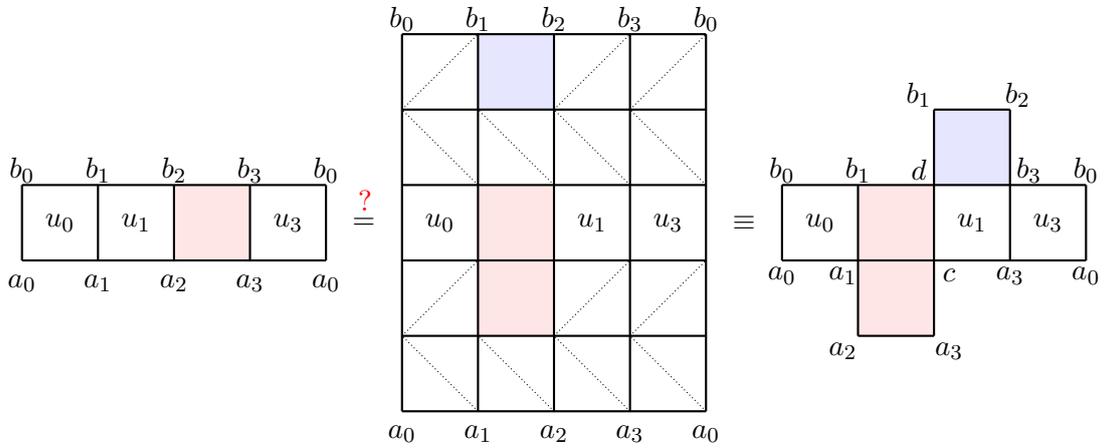
\begin{figure}[H]
 \begin{adjustbox}{max totalsize={1.2\textwidth}{1.2\textheight},center}
 \begin{tikzpicture}
\fill[red!10] (2,0) -- (2,1) -- (3,1) -- (3,0) -- cycle;
\fill[red!10] (6,0) -- (6,1) -- (7,1) -- (7,0) -- cycle;
\fill[red!10] (6,-1) -- (6,0) -- (7,0) -- (7,-1) -- cycle;
\fill[blue!10] (6,2) -- (6,3) -- (7,3) -- (7,2) -- cycle;
\fill[red!10] (11,0) -- (11,1) -- (12,1) -- (12,0) -- cycle;
\fill[red!10] (11,-1) -- (11,0) -- (12,0) -- (12,-1) -- cycle;

\fill[blue!10] (12,1) -- (12,2) -- (13,2) -- (13,1) -- cycle;

     \draw[black, thick] (0,0) -- (4,0) ; 
     \draw[black, thick] (0,1) -- (4,1) ; 
     \draw[black, thick] (0,0) -- (0,1) ; 
     \draw[black, thick] (1,0) -- (1,1) ; 
     \draw[black, thick] (2,0) -- (2,1) ; 
     \draw[black, thick] (3,0) -- (3,1) ; 
     \draw[black, thick] (4,0) -- (4,1) ; 
     \node[] at (0,-0.3) {$a_0$} ;
     \node[] at (1,-0.3) {$a_1$} ;
     \node[] at (2,-0.3) {$a_2$} ;
     \node[] at (3,-0.3) {$a_3$} ;
     \node[] at (4,-0.3) {$a_0$} ;
     \node[] at (0,1.2) {$b_0$} ;
     \node[] at (1,1.2) {$b_1$} ;
     \node[] at (2,1.2) {$b_2$} ;
     \node[] at (3,1.2) {$b_3$} ;
     \node[] at (4,1.2) {$b_0$} ;
     \node[] at (0.5,0.5) {$u_0$} ;
     \node[] at (1.5,0.5) {$u_1$} ;
     \node[] at (3.5,0.5) {$u_3$} ;
     \node[] at (4.5, 0.5) {$=$} ;
    \node[] at (4.45, 0.8) {{ \color{red} ?}} ;
     
     \draw[black, thick] (5,0) -- (9,0) ; 
     \draw[black, thick] (5,1) -- (9,1) ; 
     \draw[black, thick] (5,2) -- (9,2) ; 
     \draw[black, thick] (5,3) -- (9,3) ; 
     \draw[black, thick] (5,-1) -- (9,-1) ; 
     \draw[black, thick] (5,-2) -- (9,-2) ; 
     \draw[black, thick] (5,-2) -- (5,3) ; 
     \draw[black, thick] (6,-2) -- (6,3) ; 
     \draw[black, thick] (7,-2) -- (7,3) ; 
     \draw[black, thick] (8,-2) -- (8,3) ; 
     \draw[black, thick] (9,-2) -- (9,3) ; 
     
     \draw[black, densely dotted] (6,1)--(5,2);
     \draw[black, densely dotted] (7,1)--(6,2);
     \draw[black, densely dotted] (8,1)--(7,2);
     \draw[black, densely dotted] (9,1)--(8,2);
     \draw[black, densely dotted] (6,-2)--(5,-1);
     \draw[black, densely dotted] (7,-2)--(6,-1);
     \draw[black, densely dotted] (8,-2)--(7,-1);
     \draw[black, densely dotted] (9,-2)--(8,-1);

     \draw[black, densely dotted] (5,2)--(6,3);
     \draw[black, densely dotted] (7,2)--(8,3);
     \draw[black, densely dotted] (8,2)--(9,3);

     \draw[black, densely dotted] (5,-1)--(6,0);
     \draw[black, densely dotted] (7,-1)--(8,0);
     \draw[black, densely dotted] (8,-1)--(9,-0);

     \node[] at (5,-2.3) {$a_0$} ;
     \node[] at (6,-2.3) {$a_1$} ;
     \node[] at (7,-2.3) {$a_2$} ;
     \node[] at (8,-2.3) {$a_3$} ;
     \node[] at (9,-2.3) {$a_0$} ;
     \node[] at (5,3.2) {$b_0$} ;
     \node[] at (6,3.2) {$b_1$} ;
     \node[] at (7,3.2) {$b_2$} ;
     \node[] at (8,3.2) {$b_3$} ;
     \node[] at (9,3.2) {$b_0$} ;
     \node[] at (5.5,0.5) {$u_0$} ;

     \node[] at (7.5,0.5) {$u_1$} ;
     \node[] at (8.5,0.5) {$u_3$} ;
      
     \node[] at (9.5,0.5) {$\equiv$} ;
     
     \draw[black, thick] (10,0) -- (14,0) ; 
     \draw[black, thick] (10,1) -- (14,1) ; 
     \draw[black, thick] (10,0) -- (10,1) ; 
     \draw[black, thick] (11,0) -- (11,1) ; 
     \draw[black, thick] (12,0) -- (12,1) ; 
     \draw[black, thick] (13,0) -- (13,1) ; 
     \draw[black, thick] (14,0) -- (14,1) ; 
     \node[] at (10,-0.2) {$a_0$} ;
     \node[] at (10.8,-0.2) {$a_1$} ;
     \node[] at (10.8,-1.2) {$a_2$} ;
     \node[] at (12.2,-0.2) {$c$} ;
     \node[] at (12.2,-1.2) {$a_3$} ;

     \node[] at (13,-0.2) {$a_3$} ;
     \node[] at (14,-0.2) {$a_0$} ;
     \node[] at (10,1.2) {$b_0$} ;
     \node[] at (11,1.2) {$b_1$} ;
     \node[] at (11.8,1.2) {$d$} ;
     \node[] at (11.8,2.2) {$b_1$} ;
     \node[] at (13.1,2.2) {$b_2$} ;
     \node[] at (13.25,1.2) {$b_3$} ;
     \node[] at (14,1.2) {$b_0$} ;
     \node[] at (10.5,0.5) {$u_0$} ;
     \node[] at (12.5,0.5) {$u_1$} ;
     \node[] at (13.5,0.5) {$u_3$} ;
     \draw[black, thick] (12,1) -- (12,2);
     \draw[black, thick] (13,1) -- (13,2);
     \draw[black, thick] (12,2) -- (13,2);

     \draw[black, thick] (11,0) -- (11,-1) ;
     \draw[black, thick] (12,0) -- (12,-1) ;
     \draw[black, thick] (11,-1) -- (12,-1) ;
 \end{tikzpicture}
 \end{adjustbox}
    \caption{Here, we wish to prove the first equality. $c$ and $d$ are summed over in the figure on extreme right.}
    \label{fig:movig_face_red-tile}
\end{figure}

\begin{figure}[H]
     \begin{adjustbox}{max totalsize={1.2\textwidth}{1.2\textheight},center}
     \begin{tikzpicture}
     \fill[red!10] (11,0) -- (11,1) -- (12,1) -- (12,0) -- cycle;
     \fill[red!10] (11,-1) -- (11,0) -- (12,0) -- (12,-1) -- cycle;

     \fill[red!10] (22,0) -- (22,1) -- (23,1) -- (23,0) -- cycle;
     \fill[red!10] (17,0) -- (17,1) -- (18,1) -- (18,0) -- cycle;
     \fill[red!10] (17,1) -- (17,2) -- (18,2) -- (18,1) -- cycle;
     
     \fill[blue!10] (18,2) -- (18,3) -- (19,3) -- (19,2) -- cycle;
     
     \fill[blue!10] (12,1) -- (13,1) -- (13,2) -- (12,2) -- cycle;
     \draw[black, thick] (10,0) -- (11,0) ; 
     \draw[black, thick] (13,0) -- (14,0) ; 

     \draw[black, thick] (10,1) -- (11,1) ; 
     \draw[black, thick] (13,1) -- (14,1) ; 

     \draw[black, thick] (10,0) -- (10,1) ; 
     \draw[black, thick] (11,0) -- (11,1) ; 
     \draw[black, thick] (12,0) -- (12,1) ; 
     \draw[black, thick] (13,0) -- (13,1) ; 
     \draw[black, thick] (14,0) -- (14,1) ; 
     \node[] at (10,-0.2) {$a_0$} ;
     \node[] at (10.8,-0.2) {$a_1$} ;
     \node[] at (10.8,-1.2) {$a_2$} ;
     \node[] at (12.2,-0.2) {$c$} ;
     \node[] at (12.2,-1.2) {$a_3$} ;

     \node[] at (13,-0.2) {$a_3$} ;
     \node[] at (14,-0.2) {$a_0$} ;
     \node[] at (10,1.2) {$b_0$} ;
     \node[] at (11,1.2) {$b_1$} ;
     \node[] at (11.8,1.2) {$d$} ;
     \node[] at (11.8,2.2) {$b_1$} ;
     \node[] at (13.1,2.2) {$b_2$} ;
     \node[] at (13.25,1.2) {$b_3$} ;
     \node[] at (14,1.2) {$b_0$} ;
     \node[] at (10.5,0.5) {$u_0$} ;
     \node[] at (12.5,0.5) {$u_1$} ;
     \node[] at (13.5,0.5) {$u_3$} ;
     \draw[black, thick] (12,1) -- (12,2);
     \draw[black, thick] (13,1) -- (13,2);
     \draw[black, thick] (12,2) -- (13,2);
     \draw[black, thick] (11,1) -- (13,1) ; 
     \draw[black, thick] (11,0) -- (13,0) ; 
     \draw[black, thick] (11,0) -- (11,-1) ;
     \draw[black, thick] (12,0) -- (12,-1) ;
     \draw[black, thick] (11,-1) -- (12,-1) ;
    \node[] at (14.5,0.5) {$=$} ;

     \draw[black, thick] (15,0) -- (16,0) ; 
     \draw[black, thick] (18,0) -- (19,0) ; 

     \draw[black, thick] (15,1) -- (16,1) ; 
     \draw[black, thick] (16,1) -- (18,1) ; 
     \draw[black, thick] (16,0) -- (18,0) ; 

     \draw[black, thick] (18,1) -- (19,1) ; 
     \draw[black, thick] (17,2) -- (19,2) ; 
     \draw[black, thick] (15,0) -- (15,1) ; 
     \draw[black, thick] (16,0) -- (16,1) ; 
     \draw[black, thick] (17,0) -- (17,2) ; 
     \draw[black, thick] (18,0) -- (18,3) ; 
     \draw[black, thick] (19,0) -- (19,1) ;
     \draw[black, thick] (19,2) -- (19,3) ;
     \draw[black, thick] (18,3) -- (19,3) ;

     \node[] at (15,-0.2) {$a_0$} ;
     \node[] at (16,-0.2) {$a_1$} ;
     \node[] at (17,-0.2) {$a_2$} ;
     \node[] at (18,-0.2) {$a_3$} ;
     \node[] at (19,-0.2) {$a_0$} ;
     \node[] at (15,1.2) {$b_0$} ;
     \node[] at (16,1.2) {$b_1$} ;
     \node[] at (16.8,1.2) {$c$} ;
     \node[] at (16.8,2.2) {$b_1$} ;
     \node[] at (17.8,2.2) {$d$} ;
     \node[] at (17.8,3.2) {$b_1$} ;
     \node[] at (19.2,3.2) {$b_2$} ;
     \node[] at (19.2,2.2) {$b_3$} ;

     \node[] at (18.2,1.2) {$b_3$} ;
     \node[] at (19,1.2) {$b_0$} ;
     \node[] at (15.5,0.5) {$u_0$} ;
     
     \node[] at (16.5,0.5) {$u_1$} ;
     \node[] at (18.5,0.5) {$u_3$} ;

    \node[] at (19.5,0.5) {$=$} ;

     \draw[black, thick] (20,0) -- (21,0) ; 
     \draw[black, thick] (21,0) -- (23,0) ; 
     \draw[black, thick] (23,0) -- (24,0) ; 

     \draw[black, thick] (20,1) -- (21,1) ; 
     \draw[black, thick] (23,1) -- (24,1) ; 
     \draw[black, thick] (21,1) -- (23,1) ; 

     \draw[black, thick] (20,0) -- (20,1) ; 
     \draw[black, thick] (21,0) -- (21,1) ; 
     \draw[black, thick] (22,0) -- (22,1) ; 
     \draw[black, thick] (23,0) -- (23,1) ; 
     \draw[black, thick] (24,0) -- (24,1) ; 
     \node[] at (20,-0.3) {$a_0$} ;
     \node[] at (21,-0.3) {$a_1$} ;
     \node[] at (22,-0.3) {$a_2$} ;
     \node[] at (23,-0.3) {$a_3$} ;
     \node[] at (24,-0.3) {$a_0$} ;
     \node[] at (20,1.2) {$b_0$} ;
     \node[] at (21,1.2) {$b_1$} ;
     \node[] at (22,1.2) {$b_2$} ;
     \node[] at (23,1.2) {$b_3$} ;
     \node[] at (24,1.2) {$b_0$} ;
     \node[] at (20.5,0.5) {$u_0$} ;
     \node[] at (21.5,0.5) {$u_1$} ;
     \node[] at (23.5,0.5) {$u_3$} ;
     
     \end{tikzpicture}
    \end{adjustbox}
    \caption{The first equality follows from Yang-Baxter and the second from Unitarity. Again, $c$ and $d$ are summed over.}
    \label{fig:YBE-unitarity-simplification-red}
\end{figure}
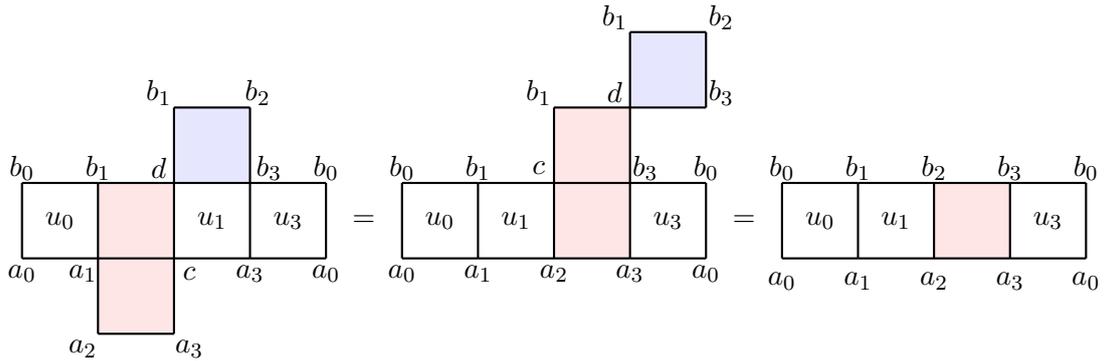

We will now move the defect face using the braid operator. To start, we   prove the first equality in figure \ref{fig:movig_face_red-tile}. This is accomplished in figure \ref{fig:YBE-unitarity-simplification-red}, by using identities in figures \ref{fig:first_cond_top_inv} and \ref{fig:top-def-rel-2}.

Note that the first relation in figure \ref{fig:movig_face_red-tile} can also be depicted in a more condensed form as shown  below in figure \ref{fig:condensing_braids-red}. 
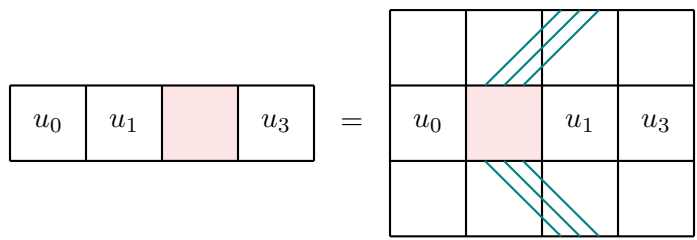
\begin{figure}[H]
\begin{adjustbox}{max totalsize={1.2\textwidth}{1.2\textheight},center}
\begin{tikzpicture}
        \fill[red!10] (7,1) -- (8,1) -- (8,2) -- (7,2) -- cycle;
     \fill[red!10] (11,1) -- (12,1) -- (12,2) -- (11,2) -- cycle;

        \draw[black, thick ] (5,1) -- (9,1) ;
        \draw[black, thick ] (5,2) -- (9,2) ;
        \draw[black, thick ] (5,1) -- (5,2) ;
        \draw[black, thick ] (6,1) -- (6,2) ;
        \draw[black, thick ] (7,1) -- (7,2) ;
        \draw[black, thick ] (8,1) -- (8,2) ;
        \draw[black, thick ] (9,1) -- (9,2) ;
        \node[] at  (9.5,1.5) {$=$} ; 
        \draw[black, thick] (10,0) -- (14,0) ; 
        \draw[black, thick] (10,1) -- (14,1) ; 
        \draw[black, thick] (10,2) -- (14,2) ; 
        \draw[black, thick] (10,3) -- (14,3) ;
        \draw[black, thick] (10,0) -- (10,3) ;
        \draw[black, thick] (11,0) -- (11,3) ;
        \draw[black, thick] (12,0) -- (12,3) ;
        \draw[black, thick] (13,0) -- (13,3) ;
        \draw[black, thick] (14,0) -- (14,3) ;
    
        \draw[teal, thick] (11.5,1) -- (12.5,0);
        \draw[teal, thick] (11.5,2) -- (12.5,3);
        \draw[teal, thick] (11.25,1) -- (12.25,0);
        \draw[teal, thick] (11.25,2) -- (12.25,3);
        \draw[teal, thick] (11.75,1) -- (12.75,0);
        \draw[teal, thick] (11.75,2) -- (12.75,3);

        \node[] at  (5.5,1.5) {$u_0$} ; 
        \node[] at  (6.5,1.5) {$u_1$} ; 
        \node[] at  (8.5,1.5) {$u_3$} ; 
        \node[] at  (10.5,1.5) {$u_0$} ; 
        \node[] at  (12.5,1.5) {$u_1$} ; 
        \node[] at  (13.5,1.5) {$u_3$} ; 
        
\end{tikzpicture}
    \end{adjustbox}
    \caption{Condensing the first equality in figure \ref{fig:movig_face_red-tile} by using relations in figure \ref{fig:Briad_RSOS-red}.} 
    \label{fig:condensing_braids-red}
\end{figure}
After these preliminaries, let us consider a configuration where we have a vertical seam (a column) of topological faces and use the relation in figure \ref{fig:condensing_braids-red} to  move them around as illustrated below in figure \ref{fig:moving_red-tile}.
\begin{figure}[H]
    \centering
     \begin{adjustbox}{max totalsize={1.0\textwidth}{1.0\textheight},center}
     \begin{tikzpicture}
     \fill[red!10] (2,0) -- (3,0) -- (3,4) -- (2,4) -- cycle ;
     \fill[red!10] (7,-1) -- (8,-1) -- (8,1) -- (7,1) -- cycle ;
     \fill[red!10] (7,4) -- (8,4) -- (8,5) -- (7,5) -- cycle ;
     \fill[red!10] (12,4) -- (13,4) -- (13,5) -- (12,5) -- cycle ;
     \fill[red!10] (11,2) -- (12,2) -- (12,3) -- (11,3) -- cycle ;
     \fill[red!10] (11,-1) -- (12,-1) -- (12,0) -- (11,0) -- cycle ;
     \fill[red!10] (12,-2) -- (13,-2) -- (13,-3) -- (12,-3) -- cycle ;

     \fill[red!10] (6,2) -- (7,2) -- (7,3) -- (6,3) -- cycle ;

         \draw[black, thick] (0,0) -- (4,0) ;
         \draw[black, thick] (0,1) -- (4,1) ;
         \draw[black, thick] (0,2) -- (4,2) ;
         \draw[black, thick] (0,3) -- (4,3) ;
         \draw[black, thick] (0,4) -- (4,4) ;
         \draw[black, thick] (0,0) -- (0,4) ;
         \draw[black, thick] (1,0) -- (1, 4) ;
         \draw[black, thick] (2,0) -- (2,4) ;
         \draw[black, thick] (3,0) -- (3,4) ;
         \draw[black, thick] (4,0) -- (4,4) ;

         \node[] at (1.5,0.5) {$u_3$} ;
         \node[] at (1.5,1.5) {$u_2$} ;
         \node[] at (1.5,2.5) {$u_1$} ;
         \node[] at (1.5,3.5) {$u_0$} ;

         \node[] at (4.5,2) {$=$} ;

        \draw[black, thick] (5,-1) -- (9,-1) ;
        \draw[black, thick] (5,0) -- (9,0) ; 
        \draw[black, thick] (5,1) -- (9,1) ; 
        \draw[black, thick] (5,2) -- (9,2) ; 
        \draw[black, thick] (5,3) -- (9,3) ; 
        \draw[black, thick] (5,4) -- (9,4) ;
        \draw[black, thick] (5,5) -- (9,5) ;
        \draw[black, thick] (5,-1) -- (5,5) ;
        \draw[black, thick] (6,-1) -- (6,5) ;
        \draw[black, thick] (7,-1) -- (7,5) ;
        \draw[black, thick] (8,-1) -- (8,5) ;
        \draw[black, thick] (9,-1) -- (9,5) ;

        \draw[teal, thick] (6.5,2) -- (7.5,1);
        \draw[teal, thick] (6.5,3) -- (7.5,4);
        \draw[teal, thick] (6.25,2) -- (7.25,1);
        \draw[teal, thick] (6.25,3) -- (7.25,4);
        \draw[teal, thick] (6.75,2) -- (7.75,1);
        \draw[teal, thick] (6.75,4) -- (7.75,4);
        \draw[teal, thick] (6.75,3) -- (7.75,4);

        \node[] at  (7.5,2.5) {$u_1$} ; 
        \node[] at  (6.5,0.5) {$u_2$} ; 
        \node[] at  (6.5,-0.5) {$u_3$} ; 
        \node[] at  (6.5,4.5) {$u_0$} ; 

        \node[] at (9.5,2) {$=$} ;
        \draw[black, thick] (10,-3) -- (14,-3) ;
        \draw[black, thick] (10,-2) -- (14,-2) ;
        \draw[black, thick] (10,-1) -- (14,-1) ;
        \draw[black, thick] (10,0) -- (14,0) ; 
        \draw[black, thick] (10,1) -- (14,1) ; 
        \draw[black, thick] (10,2) -- (14,2) ; 
        \draw[black, thick] (10,3) -- (14,3) ; 
        \draw[black, thick] (10,4) -- (14,4) ;
        \draw[black, thick] (10,5) -- (14,5) ;
        \draw[black, thick] (10,-3) -- (10,5) ;
        \draw[black, thick] (11,-3) -- (11,5) ;
        \draw[black, thick] (12,-3) -- (12,5) ;
        \draw[black, thick] (13,-3) -- (13,5) ;
        \draw[black, thick] (14,-3) -- (14,5) ;

        \draw[teal, thick] (11.5,2) -- (12.5,1);
        \draw[teal, thick] (11.5,-1) -- (12.5,-2);
        
        \draw[teal, thick] (11.5,3) -- (12.5,4);
        \draw[teal, thick] (11.5,0) -- (12.5,1);
        \draw[teal, thick] (11.25,2) -- (12.25,1);
        \draw[teal, thick] (11.25,-1) -- (12.25,-2);

        \draw[teal, thick] (11.25,3) -- (12.25,4);
        \draw[teal, thick] (11.25,0) -- (12.25,1);
        \draw[teal, thick] (11.75,2) -- (12.75,1);
        \draw[teal, thick] (11.75,-1) -- (12.75,-2);
        \draw[teal, thick] (11.75,3) -- (12.75,4);
        \draw[teal, thick] (11.75,0) -- (12.75,1);

        \node[] at  (12.5,2.5) {$u_1$} ; 
        \node[] at  (12.5,-0.5) {$u_2$} ; 
        \node[] at  (11.5,-2.5) {$u_3$} ; 
        \node[] at  (11.5,4.5) {$u_0$} ; 

     \end{tikzpicture}
          \end{adjustbox}\end{figure}
          
\begin{figure}[H]

     \begin{adjustbox}{max totalsize={1.2\textwidth}{1.2\textheight},center}
     \begin{tikzpicture}
      \fill[red!10] (17,0) -- (18,0) -- (18,-1) -- (17,-1) -- cycle ;
      \fill[red!10] (17,5) -- (18,5) -- (18,4) -- (17,4) -- cycle ;

           \fill[red!10] (16,1) -- (17,1) -- (17,3) -- (16,3) -- cycle ;
           \fill[red!10] (21,0) -- (22,0) -- (22,4) -- (21,4) -- cycle ;
 
    
     \node[] at (14.5,2) {$=$} ;
     
        \draw[black, thick] (15,-1) -- (19,-1) ;
        \draw[black, thick] (15,0) -- (19,0) ; 
        \draw[black, thick] (15,1) -- (19,1) ; 
        \draw[black, thick] (15,2) -- (19,2) ; 
        \draw[black, thick] (15,3) -- (19,3) ; 
        \draw[black, thick] (15,4) -- (19,4) ;
        \draw[black, thick] (15,5) -- (19,5) ;
        \draw[black, thick] (15,-1) -- (15,5) ;
        \draw[black, thick] (16,-1) -- (16,5) ;
        \draw[black, thick] (17,-1) -- (17,5) ;
        \draw[black, thick] (18,-1) -- (18,5) ;
        \draw[black, thick] (19,-1) -- (19,5) ;

        \draw[teal, thick] (16.5,3) -- (17.5,4);

        \draw[teal, thick] (16.75,1) -- (17.75,0);
        \draw[teal, thick] (16.25,1) -- (17.25,0);
        \draw[teal, thick] (16.5,1) -- (17.5,0);

        \draw[teal, thick] (16.25,3) -- (17.25,4);
        \draw[teal, thick] (16.75,3) -- (17.75,4);

        \node[] at  (17.5,2.5) {$u_1$} ; 
        \node[] at  (17.5,1.5) {$u_2$} ; 
        \node[] at  (16.5,-0.5) {$u_3$} ; 
        \node[] at  (16.5,4.5) {$u_0$} ; 

         \node[] at (19.5,2) {$=$} ;
        \draw[black, thick] (20,-1) -- (24,-1) ;
        \draw[black, thick] (20,0) -- (24,0) ; 
        \draw[black, thick] (20,1) -- (24,1) ; 
        \draw[black, thick] (20,2) -- (24,2) ; 
        \draw[black, thick] (20,3) -- (24,3) ; 
        \draw[black, thick] (20,4) -- (24,4) ;
        \draw[black, thick] (20,5) -- (24,5) ;
        \draw[black, thick] (20,-1) -- (20,5) ;
        \draw[black, thick] (21,-1) -- (21,5) ;
        \draw[black, thick] (22,-1) -- (22,5) ;
        \draw[black, thick] (23,-1) -- (23,5) ;
        \draw[black, thick] (24,-1) -- (24,5) ;

        \draw[teal, thick] (21.5,4) -- (22.5,5);

        \draw[teal, thick] (21.75,0) -- (22.75,-1);
        \draw[teal, thick] (21.25,0) -- (22.25,-1);
        \draw[teal, thick] (21.5,0) -- (22.5,-1);

        \draw[teal, thick] (21.25,4) -- (22.25,5);
        \draw[teal, thick] (21.75,4) -- (22.75,5);

        \node[] at  (22.5,2.5) {$u_1$} ; 
        \node[] at  (22.5,1.5) {$u_2$} ; 
        \node[] at  (22.5,0.5) {$u_3$} ; 
        \node[] at  (22.5,3.5) {$u_0$} ;

\end{tikzpicture}
     \end{adjustbox}
    \caption{Moving a column of topological defect}
    \label{fig:moving_red-tile}
\end{figure}
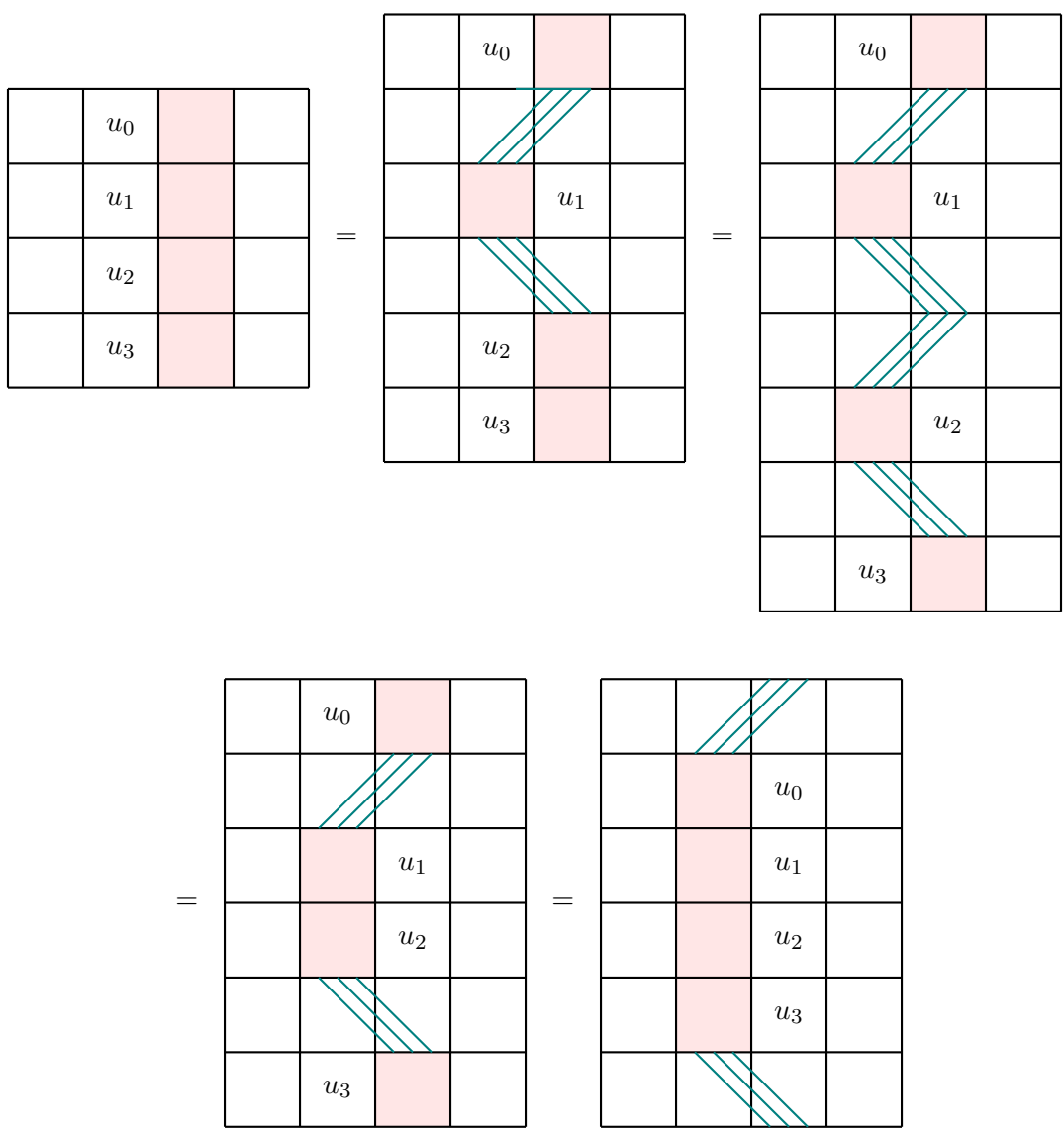
\begin{figure}[H]
     \begin{adjustbox}{max totalsize={1.0\textwidth}{1.0\textheight},center}
     \begin{tikzpicture}
           \fill[red!10] (2,0) -- (3,0) -- (3,4) -- (2,4) -- cycle ;
           \fill[red!10] (6,0) -- (7,0) -- (7,4) -- (6,4) -- cycle ;

         \draw[black, thick] (0,0) -- (4,0) ;
         \draw[black, thick] (0,1) -- (4,1) ;
         \draw[black, thick] (0,2) -- (4,2) ;
         \draw[black, thick] (0,3) -- (4,3) ;
         \draw[black, thick] (0,4) -- (4,4) ;

         \draw[black, thick] (0,0) -- (0,4) ;
         \draw[black, thick] (1,0) -- (1, 4) ;
         \draw[black, thick] (2,0) -- (2,4) ;
         \draw[black, thick] (3,0) -- (3,4) ;
         \draw[black, thick] (4,0) -- (4,4) ;

         \node[] at (1.5,0.5) {$u_3$} ;
         \node[] at (1.5,1.5) {$u_2$} ;
         \node[] at (1.5,2.5) {$u_1$} ;
         \node[] at (1.5,3.5) {$u_0$} ;

         
         \draw[black, thick] (5,0) -- (9,0) ;
         \draw[black, thick] (5,1) -- (9,1) ;
         \draw[black, thick] (5,2) -- (9,2) ;
         \draw[black, thick] (5,3) -- (9,3) ;
         \draw[black, thick] (5,4) -- (9,4) ;

         \draw[black, thick] (5,0) -- (5,4) ;
         \draw[black, thick] (6,0) -- (6, 4) ;
         \draw[black, thick] (7,0) -- (7,4) ;
         \draw[black, thick] (8,0) -- (8,4) ;
         \draw[black, thick] (9,0) -- (9,4) ;

         \node[] at (7.5,0.5) {$u_3$} ;
         \node[] at (7.5,1.5) {$u_2$} ;
         \node[] at (7.5,2.5) {$u_1$} ;
         \node[] at (7.5,3.5) {$u_0$} ;

     \end{tikzpicture}

    \end{adjustbox}

    \caption{The partition function of the above two configurations are the same. }
    \label{fig:TDL-partition}
\end{figure}
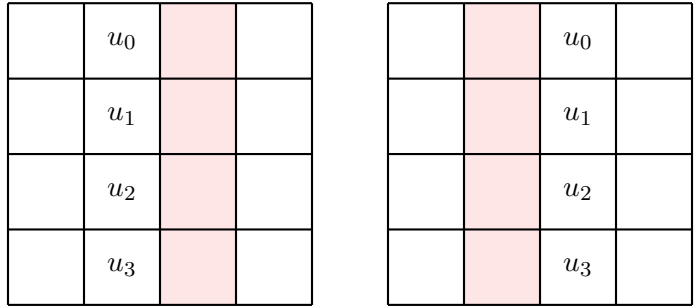

We can take the trace of all the diagrams in figure \ref{fig:moving_red-tile}: since   they are all equal as operators, the traces must be equal as well. When we take the trace of the last configuration in figure \ref{fig:moving_red-tile}, we can moreover get rid of the braids, since they are inverses of each other. Hence,  we see  that we can swap the column composed of topological  faces with its neighbors wihout changing the partition function (this is  illustrated further  in figure \ref{fig:TDL-partition}). 

Of course the partition function is invariant under more moves where the TDL is deformed into a zig-zag shape: in this case however, the interaction at the kinks has to be specified - see for instance \ref{fig:moving_red-tile}. These subtleties are absent in the continuum.

Note that this argument does not require the rest of the system to be homogeneous - all non topological faces in the lattice can have arbitrary spectral parameter, or even involve a  different kind of interaction altogether (corresponding in particular to the partition function in the presence of operator insertions - that is, correlation functions)  as long as they remain untouched by the  seam  and its deformations. This indeed is what is reasonably to require from a lattice TDL.

\subsection{A more abstract formulation}\label{subsec:loops}

The steps necessary to move the lattice TDL - and prove lattice topological invariance - become almost obvious if we stick to the case when the red and blue faces are usual faces with spectral parameters i$\infty$ and $-$i$\infty$ and change our formulation slightly. First, the conditions in  figures \ref{fig:first_cond_top_inv}, \ref{fig:top-def-rel-2}, and \ref{fig:top-def-rel-2-inv} trivially hold. 
 It is then convenient to interpret   $R$ matrices as acting on strings propagating on the dual lattice. The different faces we have encountered to far can then be  represented as in figure 
 \ref{fig:conv_loop} below:
\begin{figure}[H]
    \centering
    \begin{tikzpicture}[scale = 0.85]
        \draw[black,thick] (0,0) -- (1,0) ;
        \draw[black,thick] (0.5,0.5) -- (0.5,-0.5) ;
        \filldraw[black] (0.5,0) circle (2pt) node[anchor=north west]{$u$};
        \draw[black,thick] (2,0) -- (3,0) ;
        \draw[black,thick] (2.5,0.5) -- (2.5,0.05) ;
        \draw[black,thick] (2.5,-0.5) -- (2.5,-0.05);
         \node[] at (3.5,0) {$;$} ;
          \node[] at (1.5,0) {$;$} ;
           \node[] at (5.5,0) {$;$} ;
            \node[] at (7.5,0) {$;$} ;
        \filldraw[black] (2.5,0) circle (0pt) node[anchor=north west]{$-$i$\infty$};
        \draw[black,thick] (4,0) -- (4.45,0) ;
        \draw[black,thick] (4.55,0) -- (5,0) ;
        \draw[black,thick] (4.5,0.5) -- (4.5,-0.5) ;
        \filldraw[black] (4.5,0) circle (0pt) node[anchor=north west]{i$\infty$};
        \draw[black,thick]  (8.5,0.5) .. controls (8.5,0.025) and (8.525,0) .. (9,0);
        \draw[black,thick]  (8,0) .. controls (8.475,0) and (8.5,-0.025) .. (8.5,-0.5);
        \filldraw[black] (6.5,0) circle (0pt) node[anchor=north west]{$0$};

        \draw[black,thick]  (6.5,0.5) .. controls (6.5,0.025) and (6.475,0) .. (6,0);
        \draw[black,thick]  (6.5,-0.50) .. controls (6.5,-0.025) and (6.525,0) .. (7,0);

    \end{tikzpicture}
    \caption{The conventions for vertices with different spectral parameter, the last two figures correspond to shift operator and inverse shift operator respectively.}
    \label{fig:conv_loop}
\end{figure}
Second,  because  the expression of the Boltzmann weights  in terms of  Temperley-Lieb generators, the first  three faces can also be decomposed as a sum of two simpler diagrams as shown on the next figure:
\begin{figure}[H]
    \centering
    \begin{tikzpicture}[scale = 1]
   \draw[black,thick] (3,0) -- (4,0) ;
        \draw[black,thick] (3.5,0.5) -- (3.5,-0.5) ;
        \filldraw[black] (3.5,0) circle (2pt) node[anchor=north west]{$u$};
        \draw[black,thick]  (12.5,0.5) .. controls (12.5,0.025) and (12.525,0) .. (13,0);
        \draw[black,thick]  (12,0) .. controls (12.475,0) and (12.5,-0.025) .. (12.5,-0.5);
        \draw[black,thick]  (8.5,0.5) .. controls (8.5,0.025) and (8.475,0) .. (8,0);
        \draw[black,thick]  (8.5,-0.50) .. controls (8.5,-0.025) and (8.525,0) .. (9,0);
        \node[] at (5,-0) {\large{$=$}};
        \node[] at (6.5,-0) {\Large{$\frac{\sin (\gamma - u)}{\sin \gamma}$}};
        \node[] at (10.5,-0) {\Large{$ + \, \, \frac{\sin u }{\sin \gamma}$}};

         \end{tikzpicture}
         \caption{The first figure itself is a shorthand for the right hand side (if necessary, the  $g$ factors can be implicitly included in the diagrams)}
\end{figure}
This holds in particular for the special cases of ($u=\pm {\rm i}\infty$). This graphical representation is more than a bookkeeping device: the (Reidemeister) moves that are topologically allowed for the lines do in fact leave the partition function invariant. This is ultimately because the partition function (or correlation functions) can be expressed as traces over the aTL algebra and as such are independent of the particular degrees of freedom used in the representation. As a result, the lines which are involved in the $\tilde{u}=\pm i\infty$ vertices can be moved over the dotted vertices.

Summarizing the movement of defect faces in the last subsection as 
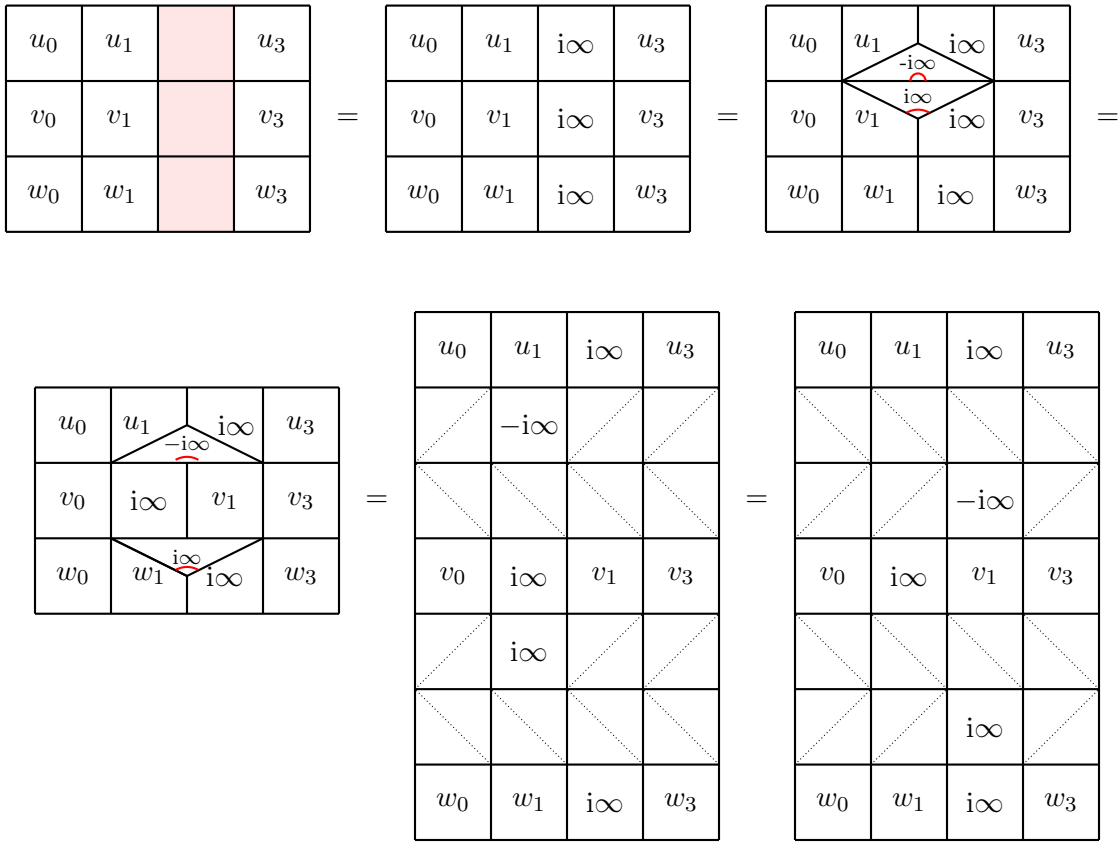
\begin{figure}[H]
     \begin{adjustbox}{max totalsize={1.0\textwidth}{1.0\textheight},center}
     \begin{tikzpicture}
        \fill[red!10] (2,0) -- (3,0) -- (3,3) -- (2,3) -- cycle ;

         \draw[black, thick] (0,0) -- (4,0) ;
         \draw[black, thick] (0,1) -- (4,1) ;
         \draw[black, thick] (0,2) -- (4,2) ;
         \draw[black, thick] (0,3) -- (4,3) ;

         \draw[black, thick] (0,0) -- (0,3) ;
         \draw[black, thick] (1,0) -- (1, 3) ;
         \draw[black, thick] (2,0) -- (2,3) ;
         \draw[black, thick] (3,0) -- (3,3) ;
         \draw[black, thick] (4,0) -- (4,3) ;
        
         \node[] at (0.5,0.5) {$w_0$} ;
         \node[] at (1.5,0.5) {$w_1$} ;
         \node[] at (3.5,0.5) {$w_3$} ;

         \node[] at (0.5,1.5) {$v_0$} ;
         \node[] at (1.5,1.5) {$v_1$} ;
         \node[] at (3.5,1.5) {$v_3$} ;

         \node[] at (0.5,2.5) {$u_0$} ;
         \node[] at (1.5,2.5) {$u_1$} ;
         \node[] at (3.5,2.5) {$u_3$} ;

         \node[] at (4.5,1.5) {$=$} ;
         
         \draw[black, thick] (5,0) -- (9,0) ;
         \draw[black, thick] (5,1) -- (9,1) ;
         \draw[black, thick] (5,2) -- (9,2) ;
         \draw[black, thick] (5,3) -- (9,3) ;

         \draw[black, thick] (5,0) -- (5,3) ;
         \draw[black, thick] (6,0) -- (6, 3) ;
         \draw[black, thick] (7,0) -- (7,3) ;
         \draw[black, thick] (8,0) -- (8,3) ;
         \draw[black, thick] (9,0) -- (9,3) ;

         \node[] at (5.5,0.5) {$w_0$} ;
         \node[] at (7.5,0.5) {i$\infty$} ;
         \node[] at (6.5,0.5) {$w_1$} ;
         \node[] at (8.5,0.5) {$w_3$} ;

         \node[] at (5.5,1.5) {$v_0$} ;
         \node[] at (6.5,1.5) {$v_1$} ;
         \node[] at (7.5,1.5) {i$\infty$} ;
         \node[] at (8.5,1.5) {$v_3$} ;

         \node[] at (5.5,2.5) {$u_0$} ;
         \node[] at (6.5,2.5) {$u_1$} ;        
         \node[] at (7.5,2.5) {i$\infty$} ;
         \node[] at (8.5,2.5) {$u_3$} ;

         \node[] at (9.5,1.5) {$=$} ;
         \draw[black, thick] (10,0) -- (14,0) ;
         \draw[black, thick] (10,1) -- (14,1) ;
         \draw[black, thick] (10,2) -- (14,2) ;
         \draw[black, thick] (10,3) -- (14,3) ;

         \draw[black, thick] (10,0) -- (10,3) ;
         \draw[black, thick] (11,0) -- (11, 3) ;
         \draw[black, thick] (11,2) -- (12, 2.5) ;
         \draw[black, thick] (11,2) -- (12, 1.5) ;
         \draw[black, thick] (13,2) -- (12, 2.5) ;
         \draw[black, thick] (13,2) -- (12, 1.5) ;         
         \draw[black, thick] (12,3) -- (12,2.5) ;
         \draw[black, thick] (12,0) -- (12,1.5) ;
         \draw[black, thick] (13,0) -- (13,3) ;
         \draw[black, thick] (14,0) -- (14,3) ;

         \node[] at (10.5,0.5) {$w_0$} ;
         \node[] at (12.5,0.5) {i$\infty$} ;
         \node[] at (11.5,0.5) {$w_1$} ;
         \node[] at (13.5,0.5) {$w_3$} ;

         \node[] at (10.5,1.5) {$v_0$} ;
         \node[] at (11.35,1.5) {$v_1$} ;
         \node[] at (12.65,1.5) {i$\infty$} ;
         \node[] at (13.5,1.5) {$v_3$} ;

         \node[] at (10.5,2.5) {$u_0$} ;
         \node[] at (11.35,2.5) {$u_1$} ;        
         \node[] at (12.65,2.5) {i$\infty$} ;
         \node[] at (13.5,2.5) {$u_3$} ;
         \node[] at (12,2.25) {\scriptsize{-i$\infty$}} ;
         \node[] at (12,1.8) {\scriptsize{i$\infty$}} ;

         \node[] at (14.5,1.5) {$=$} ;
        \draw[red,thick]  (12.1,2) arc (0:180:0.1);
         \draw[red,thick]  (12.15,1.58) arc (60:120:0.3);

     \end{tikzpicture}

    \end{adjustbox}
        
    \begin{adjustbox}{max totalsize={1.2\textwidth}{1.2\textheight},center}
        \begin{tikzpicture}
        \draw[black, thick] (0,0) -- (4,0) ;
         \draw[black, thick] (0,1) -- (4,1) ;
         \draw[black, thick] (0,2) -- (4,2) ;
         \draw[black, thick] (0,3) -- (4,3) ;

         \draw[black, thick] (0,0) -- (0,3) ;
         \draw[black, thick] (1,0) -- (1, 3) ;
         \draw[black, thick] (2,0) -- (2,0.5) ;
         \draw[black, thick] (2,1) -- (2,2) ;
         \draw[black, thick] (2,2.5) -- (2,3) ;

        \draw[black, thick] (1,1) -- (2,0.5) ;
        \draw[black, thick] (3,1) -- (2,0.5) ;
        \draw[black, thick] (2,2.5) -- (3,2) ;

        \draw[black, thick] (1,2) -- (2,2.5) ;
        \draw[black, thick] (1,1) -- (2,0.5) ;
         
         \draw[black, thick] (3,0) -- (3,3) ;
         \draw[black, thick] (4,0) -- (4,3) ;

         \node[] at (0.5,0.5) {$w_0$} ;
         \node[] at (1.5,0.5) {$w_1$} ;
         \node[] at (2.5,0.5) {i$\infty$} ;
         \node[] at (3.5,0.5) {$w_3$} ;

         \node[] at (0.5,1.5) {$v_0$} ;
         \node[] at (1.5,1.5) {i$\infty$} ;
         \node[] at (2.5,1.5) {$v_1$} ;
         \node[] at (3.5,1.5) {$v_3$} ;

         \node[] at (0.5,2.5) {$u_0$} ;
         \node[] at (2.65,2.5) {i$\infty$} ;
         \node[] at (1.35,2.5) {$u_1$} ;
         \node[] at (3.5,2.5) {$u_3$} ;

        \node[] at (2,0.75) {\scriptsize{i$\infty$}} ;
        \node[] at (2,2.25) {\scriptsize{$-$i$\infty$}} ;

         \draw[red,thick]  (2.15,2.04) arc (60:120:0.3);
         \draw[red,thick]  (2.15,0.58) arc (60:120:0.3);

         \node[] at (4.5,1.5) {$=$} ;
    \draw[white, thick] (0,4) -- (0,5) ; 
     \draw[black, thick] (5,0) -- (9,0) ; 
     \draw[black, thick] (5,1) -- (9,1) ; 
     \draw[black, thick] (5,2) -- (9,2) ; 
     \draw[black, thick] (5,3) -- (9,3) ; 
     \draw[black, thick] (5,4) -- (9,4) ;  
     \draw[black, thick] (5,-1) -- (9,-1) ; 
     \draw[black, thick] (5,-2) -- (9,-2) ; 
     \draw[black, thick] (5,-3) -- (9,-3) ; 
     \draw[black, thick] (5,-3) -- (5,4) ; 
     \draw[black, thick] (6,-3) -- (6,4) ; 
     \draw[black, thick] (7,-3) -- (7,4) ; 
     \draw[black, thick] (8,-3) -- (8,4) ; 
     \draw[black, thick] (9,-3) -- (9,4) ; 
     
     \draw[black, densely dotted] (6,1)--(5,2);
     \draw[black, densely dotted] (7,1)--(6,2);
     \draw[black, densely dotted] (8,1)--(7,2);
     \draw[black, densely dotted] (9,1)--(8,2);
     \draw[black, densely dotted] (6,-2)--(5,-1);
     \draw[black, densely dotted] (7,-2)--(6,-1);
     \draw[black, densely dotted] (8,-2)--(7,-1);
     \draw[black, densely dotted] (9,-2)--(8,-1);

     \draw[black, densely dotted] (5,2)--(6,3);
     \draw[black, densely dotted] (7,2)--(8,3);
     \draw[black, densely dotted] (8,2)--(9,3);

     \draw[black, densely dotted] (5,-1)--(6,0);
     \draw[black, densely dotted] (7,-1)--(8,0);
     \draw[black, densely dotted] (8,-1)--(9,-0);

     \node[] at (6.5,2.5) {$-$i$\infty$} ;
     \node[] at (6.5,-0.5) {i$\infty$} ;

     \node[] at (5.5,3.5) {$u_0$} ;
     \node[] at (6.5,3.5) {$u_1$} ;
     \node[] at (7.5,3.5) {i$\infty$} ;
     \node[] at (8.5,3.5) {$u_3$} ;
     
     \node[] at (5.5,0.5) {$v_0$} ;
     \node[] at (6.5,0.5) {i$\infty$} ;
     \node[] at (7.5,0.5) {$v_1$} ;
     \node[] at (8.5,0.5) {$v_3$} ;

     \node[] at (5.5,-2.5) {$w_0$} ;
     \node[] at (7.5,-2.5) {i$\infty$} ;
     \node[] at (6.5,-2.5) {$w_1$} ;
     \node[] at (8.5,-2.5) {$w_3$} ;

         \node[] at (9.5,1.5) {$=$} ;

     \draw[black, thick] (10,0) -- (14,0) ; 
     \draw[black, thick] (10,1) -- (14,1) ; 
     \draw[black, thick] (10,2) -- (14,2) ; 
     \draw[black, thick] (10,3) -- (14,3) ; 
     \draw[black, thick] (10,4) -- (14,4) ;  
     \draw[black, thick] (10,-1) -- (14,-1) ; 
     \draw[black, thick] (10,-2) -- (14,-2) ; 
     \draw[black, thick] (10,-3) -- (14,-3) ; 
     \draw[black, thick] (10,-3) -- (10,4) ; 
     \draw[black, thick] (11,-3) -- (11,4) ; 
     \draw[black, thick] (12,-3) -- (12,4) ; 
     \draw[black, thick] (13,-3) -- (13,4) ; 
     \draw[black, thick] (14,-3) -- (14,4) ; 
     
     \draw[black, densely dotted] (11,2)--(10,3);
     \draw[black, densely dotted] (12,2)--(11,3);
     \draw[black, densely dotted] (13,2)--(12,3);
     \draw[black, densely dotted] (14,2)--(13,3);
     \draw[black, densely dotted] (11,-1)--(10,0);
     \draw[black, densely dotted] (12,-1)--(11,0);
     \draw[black, densely dotted] (13,-1)--(12,0);
     \draw[black, densely dotted] (14,-1)--(13,0);

     \draw[black, densely dotted] (10,1)--(11,2);
     \draw[black, densely dotted] (11,1)--(12,2);
     \draw[black, densely dotted] (13,1)--(14,2);

     \draw[black, densely dotted] (10,-2)--(11,-1);
     \draw[black, densely dotted] (11,-2)--(12,-1);
     \draw[black, densely dotted] (13,-2)--(14,-1);

     \node[] at (12.5,1.5) {$-$i$\infty$} ;
     \node[] at (12.5,-1.5) {i$\infty$} ;

     \node[] at (10.5,3.5) {$u_0$} ;
     \node[] at (11.5,3.5) {$u_1$} ;
     \node[] at (12.5,3.5) {i$\infty$} ;
     \node[] at (13.5,3.5) {$u_3$} ;
     
     \node[] at (10.5,0.5) {$v_0$} ;
     \node[] at (11.5,0.5) {i$\infty$} ;
     \node[] at (12.5,0.5) {$v_1$} ;
     \node[] at (13.5,0.5) {$v_3$} ;

     \node[] at (10.5,-2.5) {$w_0$} ;
     \node[] at (12.5,-2.5) {i$\infty$} ;
     \node[] at (11.5,-2.5) {$w_1$} ;
     \node[] at (13.5,-2.5) {$w_3$} ;

        \end{tikzpicture}
    \end{adjustbox}
    
    \caption{Moving around defect face}
    \label{fig:def_line}
\end{figure}
We see that the above figure can be re-interpreted as shown in fig. \ref{fig:loop-equivalence}:
\begin{figure}[H]
    \centering
    \begin{tikzpicture}
        \draw[black,thick] (0,0) -- (1,0) ;
        \draw[black,thick] (0.5,0.5) -- (0.5,-0.5) ;
        \filldraw[black] (0.5,0) circle (2pt) node[anchor=north west]{$w_0$};
        \draw[black,thick] (1,0) -- (2,0) ;
        \draw[blue,thick] (1.5,0.5) -- (1.5,-0.5) ;
        \filldraw[black] (1.5,0) circle (2pt) node[anchor=north west]{$w_1$};
        \draw[black,thick] (2,0) -- (2.45,0) ;
        \draw[black,thick] (2.55,0) -- (3,0) ;
        \draw[red,thick] (2.5,0.5) -- (2.5,-0.5) ;
        \draw[black,thick] (3,0) -- (4,0) ;
        \draw[black,thick] (3.5,0.5) -- (3.5,-0.5) ;
        \filldraw[black] (3.5,0) circle (2pt) node[anchor=north west]{$w_3$};

        \draw[black,thick]  (0.5,-0.5) .. controls (0.5,-1.025) and (0.525,-1) .. (1,-1);
        \draw[black,thick]  (0,-1) .. controls (0.475,-1) and (0.5,-1.025) .. (0.5,-1.5);
        \draw[black,thick]  (0.5,-0.5) .. controls (0.5,-1.025) and (0.525,-1) .. (1,-1);
        \draw[black,thick]  (0,-1) .. controls (0.475,-1) and (0.5,-1.025) .. (0.5,-1.5);
        
        \draw[blue,thick]  (1.5,-0.5) .. controls (1.5,-1.025) and (1.525,-1) .. (2,-1);
        \draw[black,thick]  (1,-1) .. controls (1.475,-1) and (1.5,-1.025) .. (1.5,-1.5);
        \draw[red,thick]  (2.5,-0.5) .. controls (2.5,-1.025) and (2.525,-1) .. (3,-1);
        \draw[blue,thick]  (2,-1) .. controls (2.475,-1) and (2.5,-1.025) .. (2.5,-1.5);
        \draw[black,thick]  (3.5,-0.5) .. controls (3.5,-1.025) and (3.525,-1) .. (4,-1);
        \draw[red,thick]  (3,-1) .. controls (3.475,-1) and (3.5,-1.025) .. (3.5,-1.5);

        \draw[black,thick]  (0.5,-1.5) .. controls (0.5,-1.975) and (0.475,-2) .. (0,-2);
        \draw[black,thick]  (0.5,-2.50) .. controls (0.5,-2.025) and (0.525,-2) .. (1,-2);

        \draw[black,thick]  (1.5,-1.5) .. controls (1.5,-1.975) and (1.475,-2) .. (1,-2);
        \draw[red,thick]  (1.5,-2.50) .. controls (1.5,-2.025) and (1.525,-2) .. (2,-2);

        \draw[black,thick]  (1.5,-1.5) .. controls (1.5,-1.975) and (1.475,-2) .. (1,-2);

        \draw[red,thick]  (3.5,-1.5) .. controls (3.5,-1.975) and (3.475,-2) .. (3,-2);
        \draw[black,thick]  (3.5,-2.50) .. controls (3.5,-2.025) and (3.525,-2) .. (4,-2);
        \draw[blue,thick] (2.5,-1.5) -- (2.5,-1.95) ;
        \draw[blue,thick] (2.5,-2.5) -- (2.5,-2.05) ;
        \draw[red,thick] (2,-2) -- (3,-2) ;

        \draw[black,thick] (0,-3) -- (1,-3) ;
        \draw[black,thick] (0.5,-2.5) -- (0.5,-3.5) ;
        
        \filldraw[black] (0.5,-3) circle (2pt) node[anchor=north west]{$v_0$};
        
        \draw[black,thick] (2,-3) -- (3,-3) ;
        \draw[blue,thick] (2.5,-2.5) -- (2.5,-3.5) ;
        \filldraw[black] (2.5,-3) circle (2pt) node[anchor=north west]{$v_1$};
        \draw[black,thick] (1,-3) -- (1.45,-3) ;
        \draw[black,thick] (1.55,-3) -- (2,-3) ;
        \draw[black,thick] (3,-3) -- (4,-3) ;
        \draw[black,thick] (3.5,-3.5) -- (3.5,-2.5) ;
        \filldraw[black] (3.5,-3) circle (2pt) node[anchor=north west]{$v_3$};
         \draw[red,thick] (1.5,-2.5) -- (1.5,-3.5) ;

        \draw[black,thick]  (0.5,-3.5) .. controls (0.5,-3.975) and (0.475,-4) .. (0,-4);
        \draw[blue,thick]  (0.5,-4.50) .. controls (0.5,-4.025) and (0.525,-4) .. (1,-4);

        \draw[black,thick]  (1.5,-1.5) .. controls (1.5,-1.975) and (1.475,-2) .. (1,-2);
        \draw[black,thick]  (2.5,-4.50) .. controls (2.5,-4.025) and (2.525,-4) .. (3,-4);

        \draw[blue,thick]  (2.5,-3.5) .. controls (2.5,-3.975) and (2.475,-4) .. (2,-4);

        \draw[black,thick]  (3.5,-3.5) .. controls (3.5,-3.975) and (3.475,-4) .. (3,-4);
        \draw[black,thick]  (3.5,-4.50) .. controls (3.5,-4.025) and (3.525,-4) .. (4,-4);
        \draw[red,thick] (1.5,-3.5) -- (1.5,-4.5) ;
        \draw[blue,thick] (1,-4) -- (1.45,-4) ;
        \draw[blue,thick] (1.55,-4) -- (2,-4) ;

        \draw[blue,thick]  (0.5,-4.5) .. controls (0.5,-5.025) and (0.525,-5) .. (1,-5);
        \draw[black,thick]  (0,-5) .. controls (0.475,-5) and (0.5,-5.025) .. (0.5,-5.5);
        
        \draw[red,thick]  (1.5,-4.5) .. controls (1.5,-5.025) and (1.525,-5) .. (2,-5);
        \draw[blue,thick]  (1,-5) .. controls (1.475,-5) and (1.5,-5.025) .. (1.5,-5.5);

        \draw[black,thick]  (2.5,-4.5) .. controls (2.5,-5.025) and (2.525,-5) .. (3,-5);
        \draw[red,thick]  (2,-5) .. controls (2.475,-5) and (2.5,-5.025) .. (2.5,-5.5);
        
        \draw[black,thick]  (3.5,-4.5) .. controls (3.5,-5.025) and (3.525,-5) .. (4,-5);
        \draw[black,thick]  (3,-5) .. controls (3.475,-5) and (3.5,-5.025) .. (3.5,-5.5);

        \draw[black,thick] (0,-6) -- (1,-6) ;
        \draw[black,thick] (0.5,-5.5) -- (0.5,-6.5) ;
        \filldraw[black] (0.5,-6) circle (2pt) node[anchor=north west]{$u_0$};
        \draw[black,thick] (1,-6) -- (2,-6) ;
        \draw[blue,thick] (1.5,-5.5) -- (1.5,-6.5) ;
        \filldraw[black] (1.5,-6) circle (2pt) node[anchor=north west]{$u_1$};
        \draw[black,thick] (2,-6) -- (2.45,-6) ;
        \draw[black,thick] (2.55,-6) -- (3,-6) ;
        \draw[red,thick] (2.5,-5.5) -- (2.5,-6.5) ;
        \draw[black,thick] (3,-6) -- (4,-6) ;
        \draw[black,thick] (3.5,-5.5) -- (3.5,-6.5) ;
        \filldraw[black] (3.5,-6) circle (2pt) node[anchor=north west]{$u_3$};

        \draw[black,thick] (6,0) -- (7,0) ;
        \draw[black,thick] (6.5,0.5) -- (6.5,-0.5) ;
        \filldraw[black] (6.5,0) circle (2pt) node[anchor=north west]{$w_0$};
        \draw[black,thick] (7,0) -- (8,0) ;
        \draw[blue,thick] (7.5,0.5) -- (7.5,-0.5) ;
        \filldraw[black] (7.5,0) circle (2pt) node[anchor=north west]{$w_1$};
        \draw[black,thick] (8,0) -- (8.45,0) ;
        \draw[black,thick] (8.55,0) -- (9,0) ;
        \draw[red,thick] (8.5,0.5) -- (8.5,-0.5) ;
        \draw[black,thick] (9,0) -- (10,0) ;
        \draw[black,thick] (9.5,0.5) -- (9.5,-0.5) ;
        \filldraw[black] (9.5,0) circle (2pt) node[anchor=north west]{$w_3$};

        \draw[black,thick]  (6.5,-0.5) .. controls (6.5,-1.025) and (6.525,-1) .. (7,-1);
        \draw[black,thick]  (6,-1) .. controls (6.475,-1) and (6.5,-1.025) .. (6.5,-1.5);
        \draw[black,thick]  (6.5,-0.5) .. controls (6.5,-1.025) and (6.525,-1) .. (7,-1);
        \draw[black,thick]  (6,-1) .. controls (6.475,-1) and (6.5,-1.025) .. (6.5,-1.5);
        
        \draw[blue,thick]  (7.5,-0.5) .. controls (7.5,-1.025) and (7.525,-1) .. (8,-1);
        \draw[black,thick]  (7,-1) .. controls (7.475,-1) and (7.5,-1.025) .. (7.5,-1.5);
        \draw[red,thick]  (8.5,-0.5) .. controls (8.5,-1.025) and (8.525,-1) .. (9,-1);
        \draw[blue,thick]  (8,-1) .. controls (8.475,-1) and (8.5,-1.025) .. (8.5,-1.5);
        \draw[black,thick]  (9.5,-0.5) .. controls (9.5,-1.025) and (9.525,-1) .. (10,-1);
        \draw[red,thick]  (9,-1) .. controls (9.475,-1) and (9.5,-1.025) .. (9.5,-1.5);

       \draw[black,thick]  (6.5,-1.5) .. controls (6.5,-1.975) and (6.475,-2) .. (6,-2);
        \draw[black,thick]  (6.5,-2.50) .. controls (6.5,-2.025) and (6.525,-2) .. (7,-2);

        \draw[black,thick]  (7.5,-1.5) .. controls (7.5,-1.975) and (7.475,-2) .. (7,-2);
        \draw[blue,thick]  (7.5,-2.50) .. controls (7.5,-2.025) and (7.525,-2) .. (8,-2);

        \draw[red,thick]  (9.5,-1.5) .. controls (9.5,-1.975) and (9.475,-2) .. (9,-2);
        \draw[black,thick]  (9.5,-2.50) .. controls (9.5,-2.025) and (9.525,-2) .. (10,-2);
        \draw[blue,thick]  (8.5,-1.5) .. controls (8.5,-1.975) and (8.475,-2) .. (8,-2);
        \draw[red,thick]  (8.5,-2.50) .. controls (8.5,-2.025) and (8.525,-2) .. (9,-2);

        \draw[black,thick] (6,-3) -- (7,-3) ;
        \draw[black,thick] (6.5,-3.5) -- (6.5,-2.5) ;
        \filldraw[black] (6.5,-3) circle (2pt) node[anchor=north west]{$v_0$};
        \draw[black,thick] (7,-3) -- (8,-3) ;
        \draw[blue,thick] (7.5,-2.5) -- (7.5,-3.5) ;
        \filldraw[black] (7.5,-3) circle (2pt) node[anchor=north west]{$v_1$};
        \draw[black,thick] (8,-3) -- (8.45,-3) ;
        \draw[black,thick] (8.55,-3) -- (9,-3) ;
        \draw[red,thick] (8.5,-2.5) -- (8.5,-3.5) ;
        \draw[black,thick] (9,-3) -- (10,-3) ;
        \draw[black,thick] (9.5,-2.5) -- (9.5,-3.5) ;
        \filldraw[black] (9.5,-3) circle (2pt) node[anchor=north west]{$v_3$};

       \draw[black,thick]  (6.5,-3.5) .. controls (6.5,-3.975) and (6.475,-4) .. (6,-4);
        \draw[blue,thick]  (6.5,-4.50) .. controls (6.5,-4.025) and (6.525,-4) .. (7,-4);

        \draw[blue,thick]  (7.5,-3.5) .. controls (7.5,-3.975) and (7.475,-4) .. (7,-4);
        \draw[red,thick]  (7.5,-4.50) .. controls (7.5,-4.025) and (7.525,-4) .. (8,-4);

        \draw[black,thick]  (9.5,-3.5) .. controls (9.5,-3.975) and (9.475,-4) .. (9,-4);
        \draw[black,thick]  (9.5,-4.50) .. controls (9.5,-4.025) and (9.525,-4) .. (10,-4);
        \draw[red,thick]  (8.5,-3.5) .. controls (8.5,-3.975) and (8.475,-4) .. (8,-4);
        \draw[black,thick]  (8.5,-4.50) .. controls (8.5,-4.025) and (8.525,-4) .. (9,-4);

        \draw[blue,thick]  (6.5,-4.5) .. controls (6.5,-5.025) and (6.525,-5) .. (7,-5);
        \draw[black,thick]  (6,-5) .. controls (6.475,-5) and (6.5,-5.025) .. (6.5,-5.5);
        \draw[blue,thick]  (6.5,-4.5) .. controls (6.5,-5.025) and (6.525,-5) .. (7,-5);
        \draw[black,thick]  (6,-1) .. controls (6.475,-1) and (6.5,-1.025) .. (6.5,-1.5);
        
        \draw[red,thick]  (7.5,-4.5) .. controls (7.5,-5.025) and (7.525,-5) .. (8,-5);
        \draw[blue,thick]  (7,-5) .. controls (7.475,-5) and (7.5,-5.025) .. (7.5,-5.5);
        \draw[black,thick]  (8.5,-4.5) .. controls (8.5,-5.025) and (8.525,-5) .. (9,-5);
        \draw[red,thick]  (8,-5) .. controls (8.475,-5) and (8.5,-5.025) .. (8.5,-5.5);
        \draw[black,thick]  (9.5,-4.5) .. controls (9.5,-5.025) and (9.525,-5) .. (10,-5);
        \draw[black,thick]  (9,-5) .. controls (9.475,-5) and (9.5,-5.025) .. (9.5,-5.5);
        
        \draw[black,thick] (6,-6) -- (7,-6) ;
        \draw[black,thick] (6.5,-5.5) -- (6.5,-6.5) ;
        \filldraw[black] (6.5,-6) circle (2pt) node[anchor=north west]{$u_0$};
        \draw[black,thick] (7,-6) -- (8,-6) ;
        \draw[blue,thick] (7.5,-5.5) -- (7.5,-6.5) ;
        \filldraw[black] (7.5,-6) circle (2pt) node[anchor=north west]{$u_1$};
        \draw[black,thick] (8,-6) -- (8.45,-6) ;
        \draw[black,thick] (8.55,-6) -- (9,-6) ;
        \draw[red,thick] (8.5,-5.5) -- (8.5,-6.5) ;
        \draw[black,thick] (9,-6) -- (10,-6) ;
        \draw[black,thick] (09.5,-5.5) -- (9.5,-6.5) ;
        \filldraw[black] (9.5,-6) circle (2pt) node[anchor=north west]{$u_3$};

    \draw[black, thick] (4.75,-2.8) -- (5.25,-2.8) ; 
    \draw[black, thick] (4.75,-3.2) -- (5.25,-3.2) ; 
    \end{tikzpicture}
    \caption{The above two configurations have the same weight, the one on the  LHS is the sixth diagram in figure \ref{fig:def_line} and the one on the  RHS is the second diagram in the same figure.}
    \label{fig:loop-equivalence}
\end{figure}
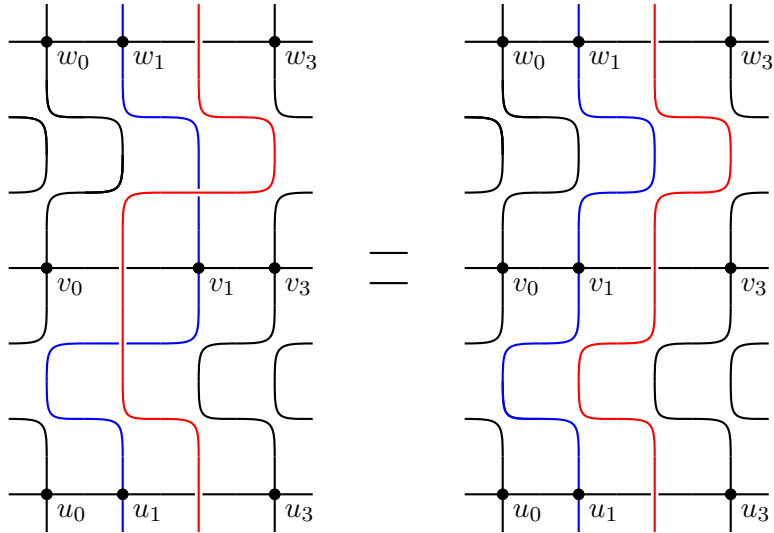
The basic move underlying the equivalence shown above is the Reidemeister move where one takes the red string towards  the right, and above the blue string.

A similar transformation could be carried out if the red string was going below the blue string, but the strings introduced in the middle, such as in figure \ref{fig:def_line}, have to be changed. 

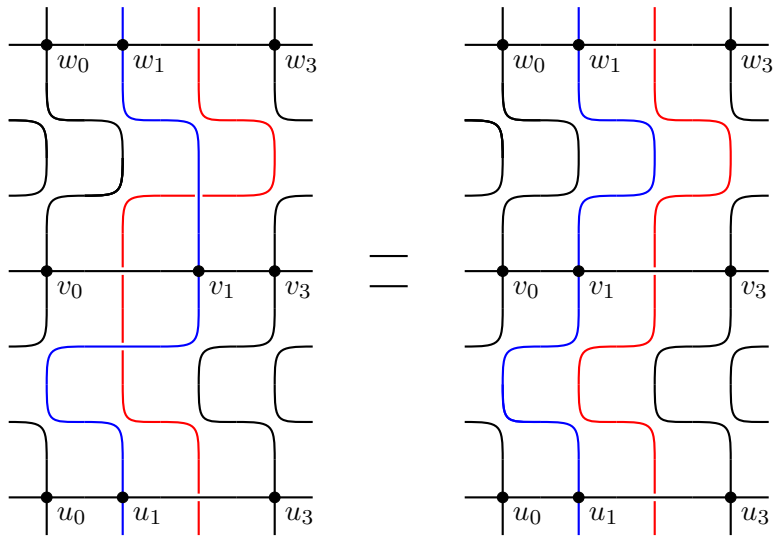
\begin{figure}[H]
    \centering
    \begin{tikzpicture}[scale = 1]
        \draw[black,thick] (0,0) -- (1,0) ;
        \draw[black,thick] (0.5,0.5) -- (0.5,-0.5) ;
        \filldraw[black] (0.5,0) circle (2pt) node[anchor=north west]{$w_0$};
        \draw[black,thick] (1,0) -- (2,0) ;
        \draw[blue,thick] (1.5,0.5) -- (1.5,-0.5) ;
        \filldraw[black] (1.5,0) circle (2pt) node[anchor=north west]{$w_1$};
        \draw[black,thick] (2,0) -- (3,0) ;
        \draw[red,thick] (2.5,0.5) -- (2.5,0.05) ;
        \draw[red,thick] (2.5,-0.05) -- (2.5,-0.5) ;
        \draw[black,thick] (3,0) -- (4,0) ;
        \draw[black,thick] (3.5,0.5) -- (3.5,-0.5) ;
        \filldraw[black] (3.5,0) circle (2pt) node[anchor=north west]{$w_3$};

        \draw[black,thick]  (0.5,-0.5) .. controls (0.5,-1.025) and (0.525,-1) .. (1,-1);
        \draw[black,thick]  (0,-1) .. controls (0.475,-1) and (0.5,-1.025) .. (0.5,-1.5);
        \draw[black,thick]  (0.5,-0.5) .. controls (0.5,-1.025) and (0.525,-1) .. (1,-1);
        \draw[black,thick]  (0,-1) .. controls (0.475,-1) and (0.5,-1.025) .. (0.5,-1.5);
        
        \draw[blue,thick]  (1.5,-0.5) .. controls (1.5,-1.025) and (1.525,-1) .. (2,-1);
        \draw[black,thick]  (1,-1) .. controls (1.475,-1) and (1.5,-1.025) .. (1.5,-1.5);
        \draw[red,thick]  (2.5,-0.5) .. controls (2.5,-1.025) and (2.525,-1) .. (3,-1);
        \draw[blue,thick]  (2,-1) .. controls (2.475,-1) and (2.5,-1.025) .. (2.5,-1.5);
        \draw[black,thick]  (3.5,-0.5) .. controls (3.5,-1.025) and (3.525,-1) .. (4,-1);
        \draw[red,thick]  (3,-1) .. controls (3.475,-1) and (3.5,-1.025) .. (3.5,-1.5);

        \draw[black,thick]  (0.5,-1.5) .. controls (0.5,-1.975) and (0.475,-2) .. (0,-2);
        \draw[black,thick]  (0.5,-2.50) .. controls (0.5,-2.025) and (0.525,-2) .. (1,-2);

        \draw[black,thick]  (1.5,-1.5) .. controls (1.5,-1.975) and (1.475,-2) .. (1,-2);
        \draw[red,thick]  (1.5,-2.50) .. controls (1.5,-2.025) and (1.525,-2) .. (2,-2);

        \draw[black,thick]  (1.5,-1.5) .. controls (1.5,-1.975) and (1.475,-2) .. (1,-2);

        \draw[red,thick]  (3.5,-1.5) .. controls (3.5,-1.975) and (3.475,-2) .. (3,-2);
        \draw[black,thick]  (3.5,-2.50) .. controls (3.5,-2.025) and (3.525,-2) .. (4,-2);
        \draw[blue,thick] (2.5,-1.5) -- (2.5,-2.5) ;
        \draw[red,thick] (2,-2) -- (2.45,-2) ;
        \draw[red,thick] (3,-2) -- (2.55,-2) ;

         \draw[black,thick] (0,-3) -- (1,-3) ;
        \draw[black,thick] (0.5,-2.5) -- (0.5,-3.5) ;
        \filldraw[black] (0.5,-3) circle (2pt) node[anchor=north west]{$v_0$};
        
        \draw[black,thick] (2,-3) -- (3,-3) ;
        \draw[blue,thick] (2.5,-2.5) -- (2.5,-3.5) ;
        \filldraw[black] (2.5,-3) circle (2pt) node[anchor=north west]{$v_1$};
        \draw[black,thick] (1,-3) -- (2,-3) ;
        \draw[black,thick] (3,-3) -- (4,-3) ;
        \draw[black,thick] (3.5,-3.5) -- (3.5,-2.5) ;
        \filldraw[black] (3.5,-3) circle (2pt) node[anchor=north west]{$v_3$};
         \draw[red,thick] (1.5,-2.5) -- (1.5,-2.95) ;
         \draw[red,thick] (1.5,-3.05) -- (1.5,-3.5) ;

        \draw[black,thick]  (0.5,-3.5) .. controls (0.5,-3.975) and (0.475,-4) .. (0,-4);
        \draw[blue,thick]  (0.5,-4.50) .. controls (0.5,-4.025) and (0.525,-4) .. (1,-4);

        \draw[black,thick]  (1.5,-1.5) .. controls (1.5,-1.975) and (1.475,-2) .. (1,-2);
        \draw[black,thick]  (2.5,-4.50) .. controls (2.5,-4.025) and (2.525,-4) .. (3,-4);

        \draw[blue,thick]  (2.5,-3.5) .. controls (2.5,-3.975) and (2.475,-4) .. (2,-4);

        \draw[black,thick]  (3.5,-3.5) .. controls (3.5,-3.975) and (3.475,-4) .. (3,-4);
        \draw[black,thick]  (3.5,-4.50) .. controls (3.5,-4.025) and (3.525,-4) .. (4,-4);
        \draw[red,thick] (1.5,-3.5) -- (1.5,-3.95) ;
        \draw[red,thick] (1.5,-4.05) -- (1.5,-4.5) ;
        \draw[blue,thick] (1,-4) -- (2,-4) ;

        \draw[blue,thick]  (0.5,-4.5) .. controls (0.5,-5.025) and (0.525,-5) .. (1,-5);
        \draw[black,thick]  (0,-5) .. controls (0.475,-5) and (0.5,-5.025) .. (0.5,-5.5);
        
        \draw[red,thick]  (1.5,-4.5) .. controls (1.5,-5.025) and (1.525,-5) .. (2,-5);
        \draw[blue,thick]  (1,-5) .. controls (1.475,-5) and (1.5,-5.025) .. (1.5,-5.5);

        \draw[black,thick]  (2.5,-4.5) .. controls (2.5,-5.025) and (2.525,-5) .. (3,-5);
        \draw[red,thick]  (2,-5) .. controls (2.475,-5) and (2.5,-5.025) .. (2.5,-5.5);
        
        \draw[black,thick]  (3.5,-4.5) .. controls (3.5,-5.025) and (3.525,-5) .. (4,-5);
        \draw[black,thick]  (3,-5) .. controls (3.475,-5) and (3.5,-5.025) .. (3.5,-5.5);

        \draw[black,thick] (0,-6) -- (1,-6) ;
        \draw[black,thick] (0.5,-5.5) -- (0.5,-6.5) ;
        \filldraw[black] (0.5,-6) circle (2pt) node[anchor=north west]{$u_0$};
        \draw[black,thick] (1,-6) -- (2,-6) ;
        \draw[blue,thick] (1.5,-5.5) -- (1.5,-6.5) ;
        \filldraw[black] (1.5,-6) circle (2pt) node[anchor=north west]{$u_1$};
        \draw[black,thick] (2,-6) -- (3,-6) ;
        \draw[red,thick] (2.5,-5.5) -- (2.5,-5.95) ;
        \draw[red,thick] (2.5,-6.05) -- (2.5,-6.5) ;
        \draw[black,thick] (3,-6) -- (4,-6) ;
        \draw[black,thick] (3.5,-5.5) -- (3.5,-6.5) ;
        \filldraw[black] (3.5,-6) circle (2pt) node[anchor=north west]{$u_3$};

        \draw[black,thick] (6,0) -- (7,0) ;
        \draw[black,thick] (6.5,0.5) -- (6.5,-0.5) ;
        \filldraw[black] (6.5,0) circle (2pt) node[anchor=north west]{$w_0$};
        \draw[black,thick] (7,0) -- (8,0) ;
        \draw[blue,thick] (7.5,0.5) -- (7.5,-0.5) ;
        \filldraw[black] (7.5,0) circle (2pt) node[anchor=north west]{$w_1$};
        \draw[black,thick] (8,0) -- (9,0) ;
        \draw[red,thick] (8.5,0.5) -- (8.5,0.05) ;
        \draw[red,thick] (8.5,-0.5) -- (8.5,-0.05) ;
        \draw[black,thick] (9,0) -- (10,0) ;
        \draw[black,thick] (9.5,0.5) -- (9.5,-0.5) ;
        \filldraw[black] (9.5,0) circle (2pt) node[anchor=north west]{$w_3$};

        \draw[black,thick]  (6.5,-0.5) .. controls (6.5,-1.025) and (6.525,-1) .. (7,-1);
        \draw[black,thick]  (6,-1) .. controls (6.475,-1) and (6.5,-1.025) .. (6.5,-1.5);
        \draw[black,thick]  (6.5,-0.5) .. controls (6.5,-1.025) and (6.525,-1) .. (7,-1);
        \draw[black,thick]  (6,-1) .. controls (6.475,-1) and (6.5,-1.025) .. (6.5,-1.5);
        
        \draw[blue,thick]  (7.5,-0.5) .. controls (7.5,-1.025) and (7.525,-1) .. (8,-1);
        \draw[black,thick]  (7,-1) .. controls (7.475,-1) and (7.5,-1.025) .. (7.5,-1.5);
        \draw[red,thick]  (8.5,-0.5) .. controls (8.5,-1.025) and (8.525,-1) .. (9,-1);
        \draw[blue,thick]  (8,-1) .. controls (8.475,-1) and (8.5,-1.025) .. (8.5,-1.5);
        \draw[black,thick]  (9.5,-0.5) .. controls (9.5,-1.025) and (9.525,-1) .. (10,-1);
        \draw[red,thick]  (9,-1) .. controls (9.475,-1) and (9.5,-1.025) .. (9.5,-1.5);

       \draw[black,thick]  (6.5,-1.5) .. controls (6.5,-1.975) and (6.475,-2) .. (6,-2);
        \draw[black,thick]  (6.5,-2.50) .. controls (6.5,-2.025) and (6.525,-2) .. (7,-2);

        \draw[black,thick]  (7.5,-1.5) .. controls (7.5,-1.975) and (7.475,-2) .. (7,-2);
        \draw[blue,thick]  (7.5,-2.50) .. controls (7.5,-2.025) and (7.525,-2) .. (8,-2);

        \draw[red,thick]  (9.5,-1.5) .. controls (9.5,-1.975) and (9.475,-2) .. (9,-2);
        \draw[black,thick]  (9.5,-2.50) .. controls (9.5,-2.025) and (9.525,-2) .. (10,-2);
        \draw[blue,thick]  (8.5,-1.5) .. controls (8.5,-1.975) and (8.475,-2) .. (8,-2);
        \draw[red,thick]  (8.5,-2.50) .. controls (8.5,-2.025) and (8.525,-2) .. (9,-2);

        \draw[black,thick] (6,-3) -- (7,-3) ;
        \draw[black,thick] (6.5,-3.5) -- (6.5,-2.5) ;
        \filldraw[black] (6.5,-3) circle (2pt) node[anchor=north west]{$v_0$};
        \draw[black,thick] (7,-3) -- (8,-3) ;
        \draw[blue,thick] (7.5,-2.5) -- (7.5,-3.5) ;
        \filldraw[black] (7.5,-3) circle (2pt) node[anchor=north west]{$v_1$};
        \draw[black,thick] (8,-3) -- (9,-3) ;
        \draw[red,thick] (8.5,-2.5) -- (8.5,-2.95) ;
        \draw[red,thick] (8.5,-3.5) -- (8.5,-3.05) ;
        \draw[black,thick] (9,-3) -- (10,-3) ;
        \draw[black,thick] (9.5,-2.5) -- (9.5,-3.5) ;
        \filldraw[black] (9.5,-3) circle (2pt) node[anchor=north west]{$v_3$};

       \draw[black,thick]  (6.5,-3.5) .. controls (6.5,-3.975) and (6.475,-4) .. (6,-4);
        \draw[blue,thick]  (6.5,-4.50) .. controls (6.5,-4.025) and (6.525,-4) .. (7,-4);

        \draw[blue,thick]  (7.5,-3.5) .. controls (7.5,-3.975) and (7.475,-4) .. (7,-4);
        \draw[red,thick]  (7.5,-4.50) .. controls (7.5,-4.025) and (7.525,-4) .. (8,-4);

        \draw[black,thick]  (9.5,-3.5) .. controls (9.5,-3.975) and (9.475,-4) .. (9,-4);
        \draw[black,thick]  (9.5,-4.50) .. controls (9.5,-4.025) and (9.525,-4) .. (10,-4);
        \draw[red,thick]  (8.5,-3.5) .. controls (8.5,-3.975) and (8.475,-4) .. (8,-4);
        \draw[black,thick]  (8.5,-4.50) .. controls (8.5,-4.025) and (8.525,-4) .. (9,-4);

        \draw[blue,thick]  (6.5,-4.5) .. controls (6.5,-5.025) and (6.525,-5) .. (7,-5);
        \draw[black,thick]  (6,-5) .. controls (6.475,-5) and (6.5,-5.025) .. (6.5,-5.5);
        \draw[blue,thick]  (6.5,-4.5) .. controls (6.5,-5.025) and (6.525,-5) .. (7,-5);
        \draw[black,thick]  (6,-1) .. controls (6.475,-1) and (6.5,-1.025) .. (6.5,-1.5);
        
        \draw[red,thick]  (7.5,-4.5) .. controls (7.5,-5.025) and (7.525,-5) .. (8,-5);
        \draw[blue,thick]  (7,-5) .. controls (7.475,-5) and (7.5,-5.025) .. (7.5,-5.5);
        \draw[black,thick]  (8.5,-4.5) .. controls (8.5,-5.025) and (8.525,-5) .. (9,-5);
        \draw[red,thick]  (8,-5) .. controls (8.475,-5) and (8.5,-5.025) .. (8.5,-5.5);
        \draw[black,thick]  (9.5,-4.5) .. controls (9.5,-5.025) and (9.525,-5) .. (10,-5);
        \draw[black,thick]  (9,-5) .. controls (9.475,-5) and (9.5,-5.025) .. (9.5,-5.5);
        
        \draw[black,thick] (6,-6) -- (7,-6) ;
        \draw[black,thick] (6.5,-5.5) -- (6.5,-6.5) ;
        \filldraw[black] (6.5,-6) circle (2pt) node[anchor=north west]{$u_0$};
        \draw[black,thick] (7,-6) -- (8,-6) ;
        \draw[blue,thick] (7.5,-5.5) -- (7.5,-6.5) ;
        \filldraw[black] (7.5,-6) circle (2pt) node[anchor=north west]{$u_1$};
        \draw[black,thick] (8,-6) -- (9,-6) ;
        \draw[red,thick] (8.5,-5.5) -- (8.5,-5.95) ;
        \draw[red,thick] (8.5,-6.05) -- (8.5,-6.5) ;
        \draw[black,thick] (9,-6) -- (10,-6) ;
        \draw[black,thick] (09.5,-5.5) -- (9.5,-6.5) ;
        \filldraw[black] (9.5,-6) circle (2pt) node[anchor=north west]{$u_3$};

    \draw[black, thick] (4.75,-2.8) -- (5.25,-2.8) ; 
    \draw[black, thick] (4.75,-3.2) -- (5.25,-3.2) ; 
    \end{tikzpicture}
    \caption{Here the red line runs below instead of above the blue line, as in figure \ref{fig:loop-equivalence}.}
    \label{fig:loop-equivalence-2}
\end{figure}
In this case the Reidemeister move corresponds to  moving the red string towards the  right, and below the blue string. 

We note that this re-interpretation matches the construction of topological defects in loop models as discussed in \cite{JSloopTopo}. An important point is that, in this formulation, the topological defect line can be moved without changing the Boltzmann weight of the configuration. This is different (and of course simpler) from what happens in the usual spins/heights formulations, where invariance is only obtained after summing over all spins/heights configurations.

\section{RSOS model with  column impurity and defect Hamiltonian in the direct channel}\label{sec:def-ham}

In  section \ref{sec:transfmatham}, we  only considered transfer matrices for a model  with the same spectral parameter on each face, and, correspondingly, a Hamiltonian with no defect. We now move on to the introduction of defects, realized via heterogeneities in the spectral parameters.

\subsection{Impurity column and defect Hamiltonian}\label{subsec:imp-ham}

Various cases with such heterogeneities have been considered in the past. We focus here on the case where all but 
one value of the spectral parameters are equal - that is,  there is a face in the transfer matrix and thus, a seam in the in the 2D lattice model, with a different value. By convention, we choose  the face between heights with indices $k$ and $k+1$ with $k>0$ (the value of $k$ is irrelevant since we use periodic boundary conditions), for which  we take $u + \tilde{u}$ instead of $u$. Let us denote this set of spectral parameter by $\{ u , u + \tilde{u}\}_k$. The corresponding transfer matrix is represented in figure \ref{fig:transfer-matrix-def-ham}.
\begin{figure}[H]
        \begin{adjustbox}{max totalsize={1\textwidth}{1\textheight},center}
            \begin{tikzpicture}[thick, scale=1]
            \node[] at (-3.6,1.5) { \LARGE{ $ \langle  b  |  T(\{u, u + \tilde{u}\}_k) |  a 
            \rangle  = $ }};
            \draw[black, thick] (0,0) -- (18,0);
            \draw[black, thick] (0,3) -- (18,3);
            \draw[black, thick] (0,0) -- (0,3);
            \draw[black, thick] (3,0) -- (3,3);
            \draw[black, thick] (6,0) -- (6,3);
            \draw[black, thick] (9,0) -- (9,3);
            \draw[black, thick] (12,0) -- (12,3);
            \draw[black, thick] (15,0) -- (15,3);
            \draw[black, thick] (18,0) -- (18,3);
            \node[] at (0,-0.3) {\Large{$a_0$}};
            \node[] at (3,-0.3) {\Large{$a_1$}};
            \node[] at (6,-0.3) {\Large{$a_k$}};
            \node[] at (9,-0.3) {\Large{$a_{k+1}$}};
            \node[] at (10.5,-0.3) {\Large{$\ldots$}};
            \node[] at (12,-0.3) {\Large{$a_{2L-2}$}};
            \node[] at (15,-0.3) {\Large{$a_{2L-1}$}};
            \node[] at (18,-0.3) {\Large{$a_{2L} = a_0$}};
            \node[] at (0,3.4) {\Large{$b_0$}};
            \node[] at (3,3.4) {\Large{$b_1$}};
            \node[] at (6,3.4) {\Large{$b_k$}};
            \node[] at (9,3.4) {\Large{$b_{k+1}$}};
            \node[] at (12,3.4) {\Large{$b_{2L-2}$}};
            \node[] at (10.5,3.4) {\Large{$\ldots$}};
            \node[] at (15,3.4) {\Large{$b_{2L-1}$}};
            \node[] at (18,3.4) {\Large{$b_{2L} = b_0$}};
            \node[] at (1.5,1.5) {\Large{$u$}};
            \node[] at (4.5,1.5) {\Large{$\ldots$}};
            \node[] at (7.5,1.5) {\Large{$u+\tilde{u}$}};
            \node[] at (10.5,1.5) {\Large{$\ldots$}};
            \node[] at (13.5,1.5) {\Large{$u$}};
            \node[] at (16.5,1.5) {\Large{$u$}};
    \end{tikzpicture}
\end{adjustbox}
    \caption{Transfer matrix with spectral parameter $\{u, u + \tilde{u} \}_k$ with PBC. }
    \label{fig:transfer-matrix-def-ham}
\end{figure}
We still have a commutation property:
\begin{equation}\label{eq:tmat-imp-com}
\left[T( \{u,u+\tilde{u}\}_k),T(\{v,v+\tilde{u}\}_k)\right]=0 \, .
\end{equation}
For other properties of this system see next section.

We also define  the defect Hamiltonian  as 
\begin{equation}\label{eq:def_Ham_two_parts}
\begin{split}
   H^{k, k+1}(\tilde{u}) & = - T^{-1}(\{ 0, \tilde{u}\}_k)  \frac{\partial}{\partial u} T(\{ u, u + \tilde{u}\}_k) \bigg|_{u = 0} \, . 
\end{split}
\end{equation}
We use Eq. \eqref{eq:tmat_TLgen_KP} to express $H^{k,k+1}$ in terms of generators of affine TL algebra. For the first part of Eq.\eqref{eq:def_Ham_two_parts}, it is not hard to see that 
\begin{equation}
\begin{split}
       & T(\{ 0, \tilde{u} \}_k) = \uR \left( \frac{\sin (\gamma - \tilde{u})}{\sin \gamma} + \frac{\sin \tilde{u}}{\sin \gamma} e_k \right) =  \frac{1}{\sin \gamma}\uR \,  R_{k}(\tilde{u})\\
   \implies  &  T^{-1}(\{ 0, \tilde{u} \}_k) = \frac{\sin \gamma}{\sin (\gamma - \tilde{u})}\left( \mathbb{1} - \frac{\sin \tilde{u}}{\sin (\gamma + \tilde{u})} e_k\right) \uR^{-1} \, .
\end{split}
\end{equation}
As for  the second part of Eq. \eqref{eq:def_Ham_two_parts} we have 
\begin{equation}
\frac{\partial}{\partial u} T(\{ u, u + \tilde{u}\}_k) \bigg|_{u = 0}   = \frac{\partial}{\partial u} T_A(\{ u, u + \tilde{u}\}_k) \bigg|_{u = 0}  + \frac{\partial}{\partial u} T_B(\{ u, u + \tilde{u}\}_k) \bigg|_{u = 0}  \, , 
\end{equation}
see Eq. \eqref{eq:tmat_TLgen_KP} for the definitions of $T_A$ and $T_B$. For the first term above we have  
\begin{equation}
    \begin{split}
        T^{-1}(\{ 0, \tilde{u} \}_k ) \frac{\partial}{\partial u} T_A(\{ u, u + \tilde{u}\}_k) \bigg|_{u = 0}  = \frac{1}{\sin \gamma} e_0 \, ,
    \end{split}
\end{equation}
while for the second one
\begin{equation}
\begin{split}
        T^{-1}(\{ 0, \tilde{u} \}_k ) \frac{\partial}{\partial u} T_B(\{ u, u + \tilde{u}\}_k) \bigg|_{u = 0}  & =  (- \cot \gamma) +\sum_{_{\substack{i=1 \\ i \neq k, k + 1}}}^{2L - 1} \left( - \cot \gamma +  \frac{1}{ \sin \gamma} e_i\right)   \\
 +    \left( -\cot \gamma + \frac{1}{\sin \gamma} R_{k}(\tilde{u})^{-1}e_{k + 1} R_{k}(\tilde{u}) \right) &+ \left( - \cot(\gamma - \tilde{u} ) + \frac{\sin \gamma}{\sin (\gamma + \tilde{u} ) \sin (\gamma - \tilde{u}) }e_k \right) \, . 
\end{split}
\end{equation}
Hence, the complete Hamiltonian can be written as 
\begin{equation}\label{eq:def_ham}
\begin{split}
    H^{k,k+1}(\tilde{u}) = & - \sum_{_{\substack{i=0 \\ i \neq k, k + 1}}}^{2L - 1} \left( - \cot \gamma +  \frac{1}{ \sin \gamma} e_i\right)  -   \left( -\cot \gamma + \frac{1}{\sin \gamma} R_{k}(\tilde{u})^{-1}e_{k + 1} R_{k}(\tilde{u}) \right)  \\
& -  \left( - \cot(\gamma - \tilde{u} ) + \frac{\sin \gamma}{\sin (\gamma + \tilde{u} ) \sin (\gamma - \tilde{u}) }e_k \right) \, . 
 \end{split}
\end{equation}
This form can be further simplified using using  identities in Eq.  \eqref{eq:TL_gen} to get 
\begin{equation}\label{eq:def_Ham_simp}
      H^{k,k+1} (\tilde{u}) =  H  +\frac{1}{\sin \gamma} f(\tilde{u}) \,  e_k e_{k+1}  + \frac{1}{\sin \gamma} f(- \tilde{u}) \,  e_{k + 1} e_k +\hbox{constant}\, ,
\end{equation}
where the constant  ($\cot(\gamma - \tilde{u}) - \cot \gamma $) is irrelevant and will be discarded from now on, and   
\begin{equation}\label{eq:f-def-ham}
    f(\tilde{u}) = \frac{\sin (\tilde{u})}{\sin (\gamma + \tilde{u})}  \, .
\end{equation}
Note, as TL generators are Hermitian, the Hamiltonian in Eq. \eqref{eq:def_Ham_simp} is Hermitian if and only if 
\begin{equation}\label{eq:cond-Ham-Herm}
   \left( \frac{\sin \tilde{u} }{\sin (\gamma - \tilde{u})} \right)^{*} = - \frac{\sin \tilde{u}}{\sin(\gamma + \tilde{u})}  \, . 
\end{equation}
The above equation is satisfied when  $\Re \tu$ is multiple of ${\pi\over 2}$, while the imaginary part can be  arbitrary (so in particular it holds when $\tilde{u}$ is purely imaginary, or $\tilde{u}$ is a multiple of $\frac{\pi}{2}$).

It is of course possible to define systems with more complex patterns of heterogeneities: for instance by modifying two rows etc. We will get back to this issue later.

\subsection{(Topological) Defect Hamiltonian in the direct channel}\label{subsec:defect-Ham-direct-channel}

We now move on to explain how the introduction of heterogeneities will allow us to build topological defects lines. These will be of two types: either topological on the lattice (what we call $l$TDLs)  or topological in the continuum limit (what we call dTDLs). To proceed, it is easier to keep considering the Hamiltonian point of view. 

We now claim that when  the spectral parameter  $\tilde{u}$ is purely complex or it is a multiple of $\frac{\pi}{2}$ (that is, as mentioned earlier, exactly when the defect Hamiltonian is hermitian) then there exists a local unitary operator which shifts the location of the defect link in the defect Hamiltonian. Such defects are called (lattice)  topological defects in \cite{Seifnashri:2023dpa},\cite{ Seiberg:2024gek} - they are expected to give rise to a conformal topological defect in the continuum limit, but not necessarily on the lattice,  as discussed in the introduction - see below as well.

The following (spectral parameter dependent) operator  
\begin{equation}\label{eq:transl_op}
    U^{k}(\tilde{u}) = \frac{\sin \gamma}{\sin(\gamma - \tilde{u}) }T(\{0, \tilde{u}\}_k) = \uR  \,  \left( \mathbb{1} + \frac{\sin \tilde{u} }{\sin(\gamma - \tilde{u})}e_{k} \right) \,    , 
\end{equation}
commutes with the (spectral parameter dependent) Hamiltonian  in Eq. \eqref{eq:def_ham} due to the relation in Eq. \eqref{eq:tmat-imp-com}.  We will call this operator  the \textit{defect shift operator}: it reduces  when  $\tu$ to 0, to the usual shift operator $\tau$. Note that  $U^{k}(\tilde{u})$ is an invertible operator 
\begin{equation}
    U^{k}(\tilde{u})^{-1} =  \left( \mathbb{1} - \frac{\sin(\tilde{u})}{\sin(\gamma + \tilde{u})} e_{k}  \right) \uR^{-1} \, . 
\end{equation}
We also define the (invertible) operator 
\begin{equation}\label{TvsU}
    T^{k}(\tilde{u}) = \uR^{-1}  U^{k}(\tilde{u}) = \left( \mathbb{1} + \frac{\sin \tilde{u} }{\sin(\gamma - \tilde{u})}e_{k} \right) \, . 
\end{equation}
As $U^{k}(\tilde{u})$ commutes with $H^{(k,k+1)}(\tilde{u})$, we have 
\begin{equation}\label{eq:action-shift}
\begin{split}
       & U^{k}(\tilde{u}) \,  H^{(k,k+1)}(\tilde{u}) \,  U^{k}(\tilde{u})^{-1} = H^{(k,k+1)}(\tilde{u}) \, , \\ 
\implies & T^{k}(\tilde{u}) \,  H^{(k,k+1)}(\tilde{u}) \,  T^{k}(\tilde{u})^{-1} = \uR^{-1} \, H^{(k,k+1)}(\tilde{u}) \, \uR \, , \\
\implies & T^{k}(\tilde{u}) \,  H^{(k,k+1)}(\tilde{u}) \,  T^{k}(\tilde{u})^{-1} =  H^{(k-1,k)}(\tilde{u}) \, ,
\end{split}
\end{equation}
where we have used  \eqref{eq:shift-ei-rel} in the last line. Hence, the operator $T^{k}(\tilde{u})$ can be interpreted as  a local translation operator. Note, this operator is not unique, for instance $ \alpha T^{k}(\tilde{u})$ is also a local translation operator, when $\alpha$ is a non-zero constant. 
The local translation operator is unitary if and only if 
\begin{equation}\label{eq:cond-trnsop-unit}
   \left( \frac{\sin \tilde{u}}{\sin (\gamma - \tilde{u})} \right)^{\star} =  - \frac{\sin \tilde{u}}{\sin (\gamma + \tilde{u})} \, , 
\end{equation}
as $e_i$ operators are Hermitian. Remarkably, this equation is the same as   Eq. \eqref{eq:cond-Ham-Herm}, hence  {\sl the Hamiltonian is Hermitian if and only if there exists a local unitary translation operator}. Note that below we will sometimes find it convenient to explore values of the spectral parameter outside of these special lines, as they will still give physical results in the scaling limit.
\bigskip 
Let us consider three distinct cases now\footnote{In notations of \cite{tavares2024} we have $\tilde{u}=iv_I$.}: 
\begin{enumerate}[(a)]
    \item $\tilde{u} = 0$ : In this case we get 
    \begin{equation}
      H^{(k,k+1)}(0) = H  \, , 
    \end{equation}
i.e. we recover the periodic RSOS Hamiltonian. Note, to connect with earlier works, such as \cite{Koo_1994}, we work with a slightly different normalization. We define 
\begin{equation}
     H_{{\cal D}_{(1,1)}} \equiv \frac{\gamma}{\pi} H = - \frac{\gamma  }{\pi \sin \gamma} \sum_{i = 0 }^{2L-1} e_i
\end{equation}

For this case, $U^{k}(0) =\uR$, hence 
\begin{equation}
T_{{\cal D}_{(1,1)}} = \mathbb{1}    \, ,
\end{equation}
 the identity operator.
 \item $\tilde{u} \to \pm {\rm i} \infty$ : In the two limits, we get the following two duality defect Hamiltonians respectively
 \begin{equation}\label{eq:H12}
H_{{\cal D}_{(1,2)}} = -\frac{\gamma}{\pi\sin\gamma}\sum_{i = 1}^{2L}e_i + \frac{\gamma}{\pi\sin\gamma}\left(qe_ke_{k+1} + q^{-1}e_{k+1}e_k\right) \, ,  
\end{equation}
\begin{align}
\label{eq:anti-chiral-(1,2)}
H_{{\cal D}_{\overline{(1,2)}}} &= -\frac{\gamma}{\pi\sin\gamma}\sum_{i = 1}^{2L}e_i + \frac{\gamma}{\pi\sin\gamma}\left(q^{-1}e_k e_{k+1} + qe_{k+1}e_k\right) \, , 
\end{align}
where again we have scaled the Hamiltonian in Eq. \eqref{eq:def_Ham_simp} by a factor of $ \gamma/\pi$. In the limit $\tilde{u} \to \pm {\rm i} \infty$,  $\sin \tilde{u} / \sin (\gamma - \tilde{u}) \to (-q)^{\mp 1}$, where $q = {\rm e}^{i \gamma}$. Hence, using Eq. \eqref{eq:transl_op}, it can be shown that 
\begin{equation}
            T_{{\cal D}_{(1,2)}} = (-q)^{- \frac{1}{2}} g_k \, ,
\end{equation}
    \begin{equation}
       T_{{\cal D}_{\overline{(1,2)}}} = (-q)^{ \frac{1}{2}} g_k^{- 1 } \, ,
    \end{equation}
    where $g_k^{\pm 1}$ are braid operators, which can be written in terms of TL generators cf Eq. \eqref{eq:braid_op}. 
    Hence, $g_k$ and $g_k^{-1}$ are also local translation operator for the defect Hamiltonian in Eq. \eqref{eq:H12} and \eqref{eq:anti-chiral-(1,2)} respectively.
\item $\tilde{u} = \pm \,   \pi / 2$ : In this case, again after normalizing properly, we get 
\begin{equation}\label{eq:H21Hbar21}
H_{{\cal D}_{(2,1)}} = H_{{\cal D}_{\overline{(2,1)}}} =  -\frac{\gamma}{\pi\sin\gamma}\sum_{i = 1}^{2L}e_i + \frac{\gamma}{\pi\sin\gamma\cos\gamma}\left(e_k e_{k+1} + e_{k+1}e_k\right) \, . 
\end{equation}
Using Eq. \eqref{eq:transl_op}, it is easy to show that 
\begin{equation}
    T_{{\cal D}_{(2,1)}} = T_{{\cal D}_{\overline{(2,1)}}} = \mathbb{1} - \frac{ e_k }{ \cos \gamma} \,  ,
\end{equation}
is the local translation operator.
\end{enumerate}

\bigskip

To conclude, we have seen that, when the real part of the defect spectral parameter $\tu$ is an integer multiple of $\frac{\pi}{2}$,  there exists a local unitary operator which shifts the location of the defect link in the defect Hamiltonian. Such defects are called topological defects in \cite{Seifnashri:2023dpa},\cite{ Seiberg:2024gek} - and indeed, this is all one can reasonably ask in a  point of view where time is already continuous. As discussed in the introduction however, since we are interested in lattice systems discretized in space {\sl and} time, we cannot consider this a sufficient condition for the defect to be a lattice topological defect. 

On the other hand, we can always expand face weights such as the one in figure \ref{fig:first_cond_top_inv} in powers of the spectral parameter. Such expansion gives rise, when considering transfer matrices with a row of defects, to an infinite family of higher Hamiltonians, each with their own defect. It follows that the condition for having a lattice topological defect is the existence of a local unitary able to  translate defects in {\sl all} these  higher Hamiltonians.

This is obviously a stronger condition than the existence of a local unitary for the usual Hamiltonian only: 
while the former allows the transmission of a face weight with varying parameter $u$ uniform over the bulk, the latter only requires the derivative at the initial point to be ``transmitted'',  as illustrated on figure  \ref{YBatInitial} below. 
\begin{figure}[H]
    \centering
\begin{tikzpicture}[scale=1.0 ]

\draw[black, thick] (6,0) -- (8,0) -- (9,1.5) -- (7,1.5) -- cycle;
\draw[black, thick] (6,0) -- (8,0) -- (9,-1.5) -- (7,-1.5) -- cycle;
\draw[black, thick] (8,0) -- (9,1.5) -- (10,0) -- (9,-1.5) -- cycle;

\node at (5.9,0) {$f$} ; 
\node at (6.8,1.5) {$e$} ; 
\node at (9.25,1.5) {$d$} ; 
\node at (6.8,-1.5) {$a$} ; 
\node at (9.25,-1.5) {$b$} ; 
\node at (10.2,0) {$c$} ; 
\node at (7.8,-0.3) {$g$} ; 
\draw[red, thick] (6.3,0) arc (0 : 60 : 0.3) ;
\draw[red, thick] (7.2,-1.5) arc (0 : 120 : 0.2) ;
\draw[red, thick] (8.15,-0.15) arc (-60 : 60 : 0.2) ;
\node at (9,0) {$u + \tu $} ; 
\node at (1,0) {$u + \tu$} ; 
\node at (2.5,0.75) {$-\tu $} ; 
\node at (2.5,-0.75) {$u $} ; 
\node at (7.5,0.75) {$u $} ; 
\node at (7.5,-0.75) {$-\tu $} ;

\draw[black, thick] (0,0) -- (1,1.5) -- (2,0) -- (1,-1.5) -- cycle;
\draw[black, thick] (3,1.5) -- (1,1.5) -- (2,0) -- (4,0) -- cycle;
\draw[black, thick] (3,-1.5) -- (1,-1.5) -- (2,0) -- (4,0) -- cycle;

\node at (-.2,0) {$f$} ; 
\node at (0.8,1.5) {$e$} ; 
\node at (3.25,1.5) {$d$} ; 
\node at (0.8,-1.5) {$a$} ; 
\node at (3.25,-1.5) {$b$} ; 
\node at (4.1,0) {$c$} ; 
\node at (2.2,-0.3) {$g$} ; 
\node at (5,0) {   $=\partial_u \sum_{g}$} ;
\node at (-1,0) {   $\partial_u\sum_{g}$} ;

\draw[red, thick] (0.15,-0.15) arc (-60 : 60 : 0.2) ;
\draw[red, thick] (2.2,0) arc (0 : 120 : 0.2) ;
\draw[red, thick] (1.2,-1.5) arc (0 : 60 : 0.2) ;

\end{tikzpicture}
    \caption{Yang-Baxter identity used to shift the defect position in the first charge, resp. to Eq. (\ref{eq:def_Ham_two_parts}). The derivatives above are evaluated at $u = 0$.}
\label{YBatInitial}
\end{figure}

Conversely, one can then argue that the existence of the local unitary for all higher Hamiltonians can, by resummation, be interpreted precisely as the condition in figure \ref{fig:first_cond_top_inv}. Hence we see that lattice topological invariance, from our point of view, is equivalent to a degenerate form of the Yang-Baxter equation where only one face has a spectral parameter dependency  (actually, combined with unitarity as in figure \ref{fig:top-def-rel-2}) .

This conclusion may seem at odds with ref. 
\cite{Aasen:2020jwb} which claims results unrelated with integrability - and, in particular, topological defect lines (see the crossed channel in the section below) that commute with  generic,  non-integrable and non-homogeneous, Hamiltonians.  The point is that the degenerate form of the Yang-Baxter equation in figure \ref{fig:first_cond_top_inv} in fact allows deformations of the $l$TDL through a system with arbitrary spectral parameters for each row, corresponding indeed to much more generic Hamiltonians. This can be formulated in a  more elegant way \cite{Belletete:2018eua} by recognizing that 
figure \ref{fig:first_cond_top_inv} leads to topological defect operators as centralizers of the algebra generated by local transfer matrices as the spectral parameter is varied - in other words, the centralizer of the (affine) Temperley-Lieb algebra. It is ultimately the nature of this centralizer that makes the construction of $l$TDL possible for the $(1,s)$ defects, and only for these.

\section{RSOS model with row impurity and line operator in crossed channel}\label{sec:line-op}

\subsection{Direct and crossed channel}

So far, we have seen the lattice realizations of topological defect lines in two-dimensional statistical mechanical models in section \ref{sec:Top-defects}, and in quantum impurity Hamiltonians in section \ref{sec:def-ham}. We now go back to considering lattice topological defect lines, but this time from the point of view of line  operators $\widehat{\cal D}$.  

 In doing so, one has to be very careful with comparing the natural geometry in the (conformal) field theory description with the one provided by the square lattice. So far, we considered a system of length $2L$ (for obvious heights-parity reasons in the RSOS construction), and the defect line was running parallel to the vertical axis.  For general values of the spectral parameter however, the geometry of the lattice gets distorted in the continuum limit \cite{Koo_1994} , introducing technical complications which are not essential to the study of defects. In what follows, we will therefore restrict to the isotropic version of the Euclidean case ($u= {\gamma\over 2}$) or the Hamiltonian case ( i.e. evolution by e$^{- \beta H}$, where $H$ is the Hamiltonian in Eq.  \eqref{eq:def_Ham_two_parts} when we study the one-impurity case for instance). 
 In both these cases, the horizontal axis is indeed ``space", and propagation occurs  vertically on the lattice as well as in the continuum, meaning   the vertical axis can be considered as the ``imaginary time" axis indeed. The two perspectives for the direct channel, the Euclidean and the Hamiltonian case, lead to the same results in the continuum limit, i.e. the state with largest eigenvalue of the transfer matrix at isotropic point ($u =$ $ \gamma \over 2$) corresponds to the ground-state of the Hamiltonian and so on. For more details, see Appendix \ref{app:def_Ham-tmat}.

The line operator then corresponds to inserting a {\sl row} of defect tiles (instead of a column). In the isotropic case where the transfer matrix is  $T(\{{\gamma\over 2}, {\gamma\over 2}+\tilde{u}\})$ we can immediately use crossing (see Eq. (\ref{eq:crossing_sym})) to see that 
 \begin{equation}
\lattD \propto \uR^{-1} T\left({\gamma\over 2}-\tu\right)\label{Dguess}
 \end{equation}
 Here and everywhere below, the $\propto$ symbol indicates the   normalization required to handle correctly the bulk term as discussed later in this section. Note also that we added the superscript latt to indicate that these may be topological defects only in the continuum limit. Finally, as discussed in section \ref{specialsubsec}, we also included a $\uR^{-1}$ factor to account correctly for  the issue  of finite values of the lattice momentum. $\tau$ commutes with the transfer matrix, and simply multiplies the result by  a $\pm$ sign in the scaling limit\footnote{The normalization of the bulk term may also include a phase depending on $\tu$ - this is discussed below.}. 
 
 We will in what follows denote the length in this crossed channel by $2\R$. Of course there is no difference at all between the systems of length $2L$ and $2\R$ (except for the shift in spectral parameters): we use the notational distinction mostly as a bookkeeping device to make discussions clearer.

Note that, since $T$ exchanges sectors with even heights on even sites and with even heights on odd sites, and so does $\tau$, the product (\ref{Dguess}) leaves these sectors invariant.

Eq. (\ref{Dguess})  is in fact the relevant formula for the Hamiltonian limit as well: this follows from the fact  that propagation occurs vertically in both cases, and that the eigenstates of the (isotropic) transfer matrix and Hamiltonians are the same.

\bigskip

That the  line operator is essentially  the action of a  now homogeneous transfer matrix $T( \tilde{u})$  (not to be confused with the transfer matrix $T(u)$  of the system without defect: although formally similar, the two objects play different roles because of the domains of values of spectral parameters $u,\tu$) has interesting consequences. In particular, since,   due to integrability, we have $[T(u_1),T(u_2)]=0$, and since $T(u)$ is the Euclidean time evolution operator (which gives rise to the Hamiltonian in the limit $u\to 0$), this guarantees that the line operators  will commute with $L_0+\overline{L}_0$ in the continuum limit.  Similarly, since the shift operator is also obtained from the transfer matrix by specializing the spectral parameter $T(0)$ = $\uR$, it also commutes with the line operator, and thus, with  the momentum operator, (the log of shift operator). Hence, we are guaranteed that the  lattice line operator will commute with $L_0$ and $\overline{L}_0$ in the continuum limit. Of course, to be topological, the line operator should then commute in fact  with the whole set of Virasoro generators $\hbox{Vir}\otimes \overline{\hbox{Vir}}$. Moreover, this line operator should also have eigenvalues on the Virasoro modules  determined from general conformal invariance principles as in Eq. 
(\ref{eq:verlinde-line-op}). While these properties follow logically from the discussion in the two previous sections, we will also investigate them directly below, both from the point of view of the Bethe-ansatz and from a numerical point of view  in \autoref{sec:examples} (see also \cite{Sinha:2023hum}).

 In addition, one of the most interesting properties of  line operators in the CFT is that they satisfy a non-trivial  fusion algebra. We will also discuss below to what extent this algebra is reproduced by lattice quantities -  an  a priori more challenging problem,  since the continuum limit of products of operators does not, in general, have to be simply related to the product of their continuum limits \cite{Koo_1994}. 
\bigskip 

A trivial example of our construction is  $\widehat{\cal{D}}_{(1,1)}$, or the identity line operator, whose lattice realization is simply the identity operator in the RSOS model.  However, like defect Hamiltonians are obtained using Hamiltonians at a particular value of the inhomogeneity parameter, we wish to similarly express this line operator as transfer matrix at a specific spectral parameter. Recall that  in the direct channel we obtained the (1,1) defect Hamiltonian when $\tu = 0$, hence 
 $\uR^{-1}T\left({\gamma\over 2}\right)$ should behave like the lattice realization of $\widehat{\cal{D}}_{(1,1)}$, i.e. $\lattD_{(1,1)}$.
 
 Note the role of the $\uR^{-1}$ factor: without it, $T\left({\gamma\over 2}\right)$ would act as the identity on all states in the scaling limit up to a sign. We will discuss this point in greater detail in \autoref{subsec:tau_op_sign}.   
 
 \bigskip

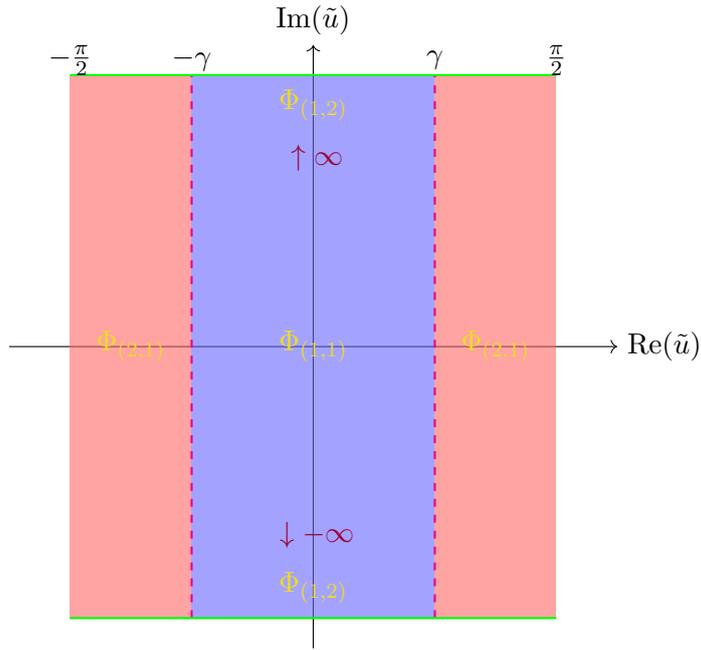
\begin{figure}[h]
    \centering
\begin{tikzpicture}
  \draw[->] (-4, 0) -- (4, 0) node[right] {\text{Re}$(\tu)$};
  \draw[->] (0, -4) -- (0, 4) node[above] {\text{Im}$(\tu)$};

  \fill[red!60, opacity=0.6] (1.6, -3.6) rectangle (3.2,3.6 ); 
  \fill[red!60, opacity=0.6] (-3.2,-3.6 ) rectangle (-1.6,3.6 ); 
  \fill[blue!60, opacity=0.6] (-1.6,-3.6 ) rectangle (1.6,3.6 ); 

  \draw[dashed, thick, magenta] (1.6,-3.6 ) -- (1.6, 3.6);
  \draw[dashed, thick, magenta] (-1.6,-3.6 ) -- (-1.6, 3.6);

  \draw[green, thick] (-3.2,-3.6 ) -- (3.2, -3.6);
  \draw[green, thick] (-3.2,3.6 ) -- ( 3.2,3.6);

  \node at (1.6,3.8 ) {$\gamma$};
  \node at (-1.6,3.8 ) {$-\gamma$};
  \node at (3.2,3.8 ) {$\frac{\pi}{2}$};
  \node at (-3.2,3.8 ) {$-\frac{\pi}{2}$};

  \node[yellow!90!black, font=\bfseries] at (0.,0 ) {$\Phi_{(1,1)}$};
  \node[yellow!90!black, font=\bfseries] at (0,-3.2 ) {$\Phi_{(1,2)}$};
  \node[yellow!90!black, font=\bfseries] at (0,3.2 ) {$\Phi_{(1,2)}$};
  \node[yellow!90!black, font=\bfseries] at (2.4,0 ) {$\Phi_{(2,1)}$};
  \node[yellow!90!black, font=\bfseries] at (-2.4,0 ) {$\Phi_{(2,1)}$};

  \node[align=center, purple!80!black] at (0.05,-2.5 ) {$\downarrow -\infty$};
  \node[align=center, purple!80!black] at (0.05,2.5 ) {$\uparrow \infty$};
\end{tikzpicture}
    \caption{In this figure, we show what defect line operator is obtained when we take the continuum limit of 
    $\uR^{-1}T( \frac{\gamma}{2}{\pm}\tu)$. }
    \label{fig:tu-cont-limit-def}
\end{figure}

 It turns out,  as we will see later for all the other defects - that  the same line operator is obtained in the continuum limit for a range of values of the spectral parameter.  In \autoref{fig:tu-cont-limit-def}, we indicate the continuum limit of the  defect line operator  $
 \uR^{-1}T\left({\gamma\over 2}-\tu\right)$ as a function of $\tu$. We will derive the result in this figure using Bethe ansatz in \autoref{BASection}.  
Recall now Eq. \eqref{eq:tmat_transpose}, where we showed that $T^{T}(\gamma - u) = T(u)$, hence $T^{T}\left(\frac{\gamma}{2} - u\right) = T\left(\frac{\gamma}{2} + u\right)$. As we expect line operators to be diagonal in the continuum limit, see Eq. \eqref{eq:verlinde-line-op}, therefore figure \ref{fig:tu-cont-limit-def} is symmetric about the y-axis.
In particular,  $\uR^{-1}T\left({\gamma\over 2}{\pm}\tu\right)$  can be used to obtain $\lattD_{(1,1)} $ 
 in the following range 
\begin{equation}
   \lvert \Re (\tu) \rvert < \gamma \, , \quad {\rm and} \quad \lvert \Im(\tu)\rvert < \infty \, ,
\end{equation}
 which includes the point $\tu = {\gamma\over 2}$ as well as the point $\tu=0$. This point is of course also a very natural choice.

We shall now discuss $\lattD_{(1,2)}$  and $\lattD_{(2,1)}$.
Using $\lattD_{(1,2)} $ and the fusion rules for defect operators of type $\widehat{\cal{D}}_{(1,s)}$, we can then construct $\lattD_{(1,s)}$. Similarly, $\widehat{{\cal D}}_{(r,1)}^{(\rm latt)}$ can be constructed by using $\widehat{{\cal D}}_{(2,1)}^{(\rm latt)}$ and the fusion rules. Now, using  $\widehat{{\cal D}}_{(r,1)}^{(\rm latt)}$ and $\widehat{{\cal D}}_{(1,s)}^{(\rm latt)}$ we can construct the lattice realization of any general line operator of the unitary, diagonal minimal model CFT. These fusions are discussed in detail in the following sections.

\subsection{(1,2) line operator}\label{sec:Y-12-op}
The lattice realization of the (1,2) line operator was already obtained  in terms of $F$-symbol for the $\mathcal{A}_p$ fusion category in the works of \cite{Feiguin:2006ydp, Buican:2017rxc, Aasen:2020jwb}, among others. Here, following \cite{Belletete:2018eua}, we approach this operator from the point of view of the center of the affine Temperley-Lieb algebra.

To start, let us see what happens to the  transfer matrix with a  homogeneous spectral parameter in the limit   $ u =\pm {\rm i} \infty $. The foregoing expressions lead to divergences, but we can always rescale the weights in Eq. \eqref{eq:standard_face_weight}, and thus  define

\begin{equation}\label{eq:face_weight_scaled}
\begin{split}
       W_1  \! \!  \left(\begin{array}{ll}
d & c \\
a & b
\end{array} \Bigg| \ u\right)= (-q)^{\frac{1}{2}} \left( \delta_{a , c} + \delta_{b , d}  \sqrt{\frac{\gf_a \gf_c}{\gf_b \gf_d}} \frac{\sin u }{\sin(\gamma - u)}  \right) \, ,  
\\
   W_2  \! \! \left(\begin{array}{ll}
d & c \\
a & b
\end{array} \Bigg| \ u\right)= (-q)^{-\frac{1}{2}} \left( \delta_{a , c}+ \delta_{b , d}  \sqrt{\frac{\gf_a \gf_c}{\gf_b \gf_d}} \frac{\sin u }{\sin(\gamma - u)}   \right) \, . 
\end{split}
\end{equation}
The transfer matrix defined using these weights, $T_1$ and $T_2$, differ from $T$ by a multiplicative factor. To be more precise 
\begin{equation}\label{eq:phase-renorm-q-t}
T_1(\{u \} ) =  \prod_{i= 0}^{2\R - 1} \left( (-q)^{\frac{1}{2}} \frac{\sin \gamma}{\sin(\gamma  - u_i)  }  \right) \  T(\{u \}) \, , \quad T_2(\{u \} ) =  \prod_{i= 0}^{2\R - 1} \left( (-q)^{-\frac{1}{2}} \frac{\sin \gamma }{\sin(\gamma- u_i)  } \right) \  T(\{u \}) \,  .    
\end{equation}
For the special spectral parameter $ \pm { \rm i } \infty $, using Eq. \eqref{eq:tmat_TLgen_KP}, the transfer matrices can then be shown to be 
\begin{subequations}\label{eq:spec_para_inf_Y}
    \begin{equation}
    T_1(\text{i} \infty) =  Y \, , \quad  T_1( - \text{i} \infty) = (-q)^{2\R} \, \overline{Y}  \, , 
\end{equation}
    \begin{equation}
   T_2( {\rm i} \infty ) =  (-q)^{-2\R} Y \, , \quad T_2( - {\rm i} \infty ) = \overline{Y}  \, ,
\end{equation}
\end{subequations}
where $Y$ and $\overline{Y}$ can be written in terms of affine TL generators as 
\begin{subequations}\label{Yoperator}
    \begin{equation}
        Y = (-q)^{-\frac{1}{2}} \,   \, g^{-1}_{1} \ldots g_{2\R - 1}^{-1}  \uR^{-1} 
        +  (-q)^{\frac{1}{2}} \, \uR \, g_{2\R -1}  \ldots g_1    \,  ,
    \end{equation}
    \begin{equation}
        \overline{Y} =  (-q)^{-\frac{1}{2}} \,  \uR  \, g^{-1}_{2\R - 1} \ldots g_{1}^{-1} +  (-q)^{\frac{1}{2}} \, g_1 \ldots g_{2\R -1}  \, \uR^{-1}    \,  ,
    \end{equation}
\end{subequations}    
where $g_i^{\pm 1}$ are braid operators of Eq. \eqref{eq:braid_op}. We had noted earlier that $\left(g_i^{\pm 1}\right)^{T} = g_i^{\pm 1}$ and $\uR^{T} = \uR^{-1} $. It is easy to show that $Y^{T} = \overline{Y}$.

Now it turns out that $Y$ and $\overline{Y}$ lie in the center of $aTL_{N}(q)$ with $N=2\R$, in fact they generate  (for $q$ generic) the center as shown in \cite{Belletete:2018eua}. Since on the other hand it is well known that $\hbox{Vir}\otimes \overline{\hbox{Vir}}$ can be approximated on the lattice by elements of $aTL_N(q)$ \cite{Koo_1994}, this guarantees that $Y$ and $\overline{Y}$ will give rise to topological line operators in the continuum limit. In fact, since elementary moves of the lattice TDL can be realized entirely in terms of actions with elements of $aTL_N$, it turns out that $Y$ and $\overline{Y}$ are topological on the lattice as well - a point we discussed in detail in \autoref{sec:Top-defects}. 

It follows that we can write 
\begin{equation}
    \widehat{{\cal D}}_{(1,2)}^{(\rm latt)} = \widehat{{\cal D}}_{\overline{(1,2)}}^{(\rm latt)} = \uR^{-1}Y = \uR^{-1}\overline{Y} \, ,
\end{equation}
Note that in these formulas, the parity of $\R$ doesn't play any role, thanks to the normalization we have used. Such simplification will not always be possible for other defects.   For A - type RSOS models, we show moreover in \autoref{sec:anyon-chain-RSOS} that $Y = \overline{Y}$. This matches  what is expected in the continuum CFTs for diagonal minimal models \cite{Belletete2020}. More generally, we shall see that  the lattice operators $\widehat{\cal{D}}_{(r,s)}^{(\rm latt)}$ acts in the same way as $\widehat{\cal{D}}_{\overline{(r,s)}}^{(\rm latt)}$. The same is not true for the three-state Potts model for instance: in that case  $\overline{Y}$ and $Y$ realize instead the  two different line operators, $\widehat{N}$ and $\widehat{N'}$, of the continuum CFT \cite{Sinha:2023hum}. 
 
 For the models considered in this paper however, this also means that we can consider as defect operators $T({\gamma\over 2}\pm \tu)$: we shall often  choose below the most convenient sign for our discussions.

In Appendix \ref{sec:Yop-F-symb} we will also show that $Y$ is unitarily equivalent to the lattice realization of $(1,2)$ operator in \cite{Aasen:2020jwb} up to a sign. We also remark here that $Y$ and $\overline{Y}$, are not only realized as transfer matrix at spectral parameter ${\rm i} \infty$ and  $-{\rm i} \infty$, but, thanks to the magic of integrability and analyticity properties (see below), one can add any finite real part to the defect parameter. Whenever 
\begin{equation}\label{eq:region-(1,2)-realization}
   \lvert \Re(\tilde{u}) \rvert < \infty \, , \quad  {\rm and} \quad \lvert\Im(\tilde{u})\rvert \to \infty \, , 
\end{equation}
 $\uR^{-1}T(\frac{\gamma}{2} + \tilde{u})$ can be used to realize the $(1,2)$ line operator \footnote{ The first condition $\lvert \Re(\tilde{u}) \rvert < \infty$ can  in fact  be dropped, since  the complex plane is compactified when the  Boltzmann weights are periodic functions as in our case where   $u\equiv u~\text{mod}~\pi$.}.

\subsection{(2,1) line operator and normalization issues}\label{sec:21-discretization}

Since we identified the defect Hamiltonian with the value $\tilde{u}=\pm {\pi\over 2}$, following our general logic, we see that the lattice realization of the $(2,1)$ line operator should be  
\begin{equation}\label{eq:realization-(2,1)-op}
 \widehat{{\cal D}}_{(2,1)}^{(\rm latt)} =  \widehat{{\cal D}}_{\overline{(2,1)}}^{(\rm latt)} \propto 
 \uR^{-1}T\left({\gamma\over 2}\pm\frac{\pi}{2}\right)     \, . 
\end{equation} 
Recall that all the objects we are interested in are periodic under shifts of the spectral parameter by $\pi$. 

At this point we must finally face the crucial question of the normalization of the discretization of the topological defect line. In CFT, this normalization follows automatically from the spectrum of the  defect Hamiltonian and a modular transformation of the corresponding partition function. Our construction on the lattice, which relies on comparing direct and crossed channel, is not as powerful, and provides the identification of $\widehat{{\cal D}}^{(\rm latt)}$ only up to a global normalization. This corresponds to a general  ambiguity   in the definition of the local Boltzmann weights, which can always be multiplied by a common factor.  Such a factor affects the spectrum of $H_{{\cal D}}$ by a global shift, and the one of  $\widehat{{\cal D}}^{(\rm latt)}$ by a multiplicative factor. More precisely, we expect the spectrum of  $\widehat{{\cal D}}^{(\rm latt)}$ in the  scaling limit to differ from the conformal results  by a factor $\exp[ 2\R e_0(\tilde{u})]$, where $e_0(\tilde{u})$ is some function, and recall $2\R$ is the length in this channel. The correct normalization would correspond to having  $\langle 0|\widehat{{\cal D}}|0\rangle=g_{\cal D}$ in the limit $\R\to\infty$ where $g_{\cal D}$ is the defect degeneracy (see below). In general, this quantity can be extracted from the lattice by calculating what amounts physically to a thermodynamic entropy. Setting $\omega=\langle 0|\widehat{{\cal D}}^{(\rm latt)}|0\rangle$, it is clear that   
\begin{equation}\label{entropydef}
s=\ln \omega-\R{\partial\ln\omega\over \partial \R} \, , 
\end{equation}
 is now independent of the  $e_0(\tilde{u})$ factor. If the regularization of the topological defect has been correctly identified, we should then have $s=\ln g_{\cal D}$ in the scaling limit. Note that in the  Bethe-ansatz calculations that follow, and as long as we are interested in properties at the critical point (contrast with \cite{tavares2024}) the $e_0$ term can be  determined analytically - see Eq. (\ref{bulk}) below. Note also that the issue does not arise for the $(1,s)$ defects - their exact normalization is a bonus of the construction in \cite{Belletete2018}\footnote{ The normalization for $(1,s)$ defects is determined  from the fact that  the lattice realization satisfy the fusion relations, such as $ \widehat{{\cal D}}_{(1,2)} \times  \widehat{{\cal D}}_{(1,2)} =  \widehat{{\cal D}}_{(1,1)} +  \widehat{{\cal D}}_{(1,3)}$ exactly on the lattice.}.

The normalization issue being settled, things are not particularly  nice for the line operator (\ref{eq:realization-(2,1)-op}) unfortunately. Nothing special happens in finite size, although of course the lattice operator at least commutes with the transfer matrix (Hamiltonian) and the momentum operator. But it does not commute with lattice approximations of the other Virasoro generators \cite{Koo_1994}. Similarly, the lattice TDL cannot be deformed without affecting the partition function\footnote{Of course, thanks to the usual Yang-Baxter and the unitarity equations it is possible to devise a series of deformations of the inserted line, without affecting the partition function. However, the nature of such deformations are quite different in that they include spectral parameters that are neither present in the original line nor do they correspond to their inverse.  } (see \autoref{sec:Top-defects}). It turns out however that things change in the  continuum limit, a point we explore numerically in great detail in sections below.

 We note now that, like in the case of $\widehat{{\cal D}}_{(1,1)}$, because of the regions, the line operator $ \widehat{{\cal D}}_{(2,1)}$ is not realized (via $T(\frac{\gamma}{2} + \tu)$) only at a single point in the continuum limit, but in a whole region where the spectral parameter $\tilde{u}$ obeys the following condition \cite{tavares2024}
\begin{equation}\label{eq:region-(2,1)-realization}
    \gamma < | \Re (\tilde{u}) |\leq  \frac{\pi}{2} \quad \text{and} \quad | \Im (\tilde{u})| < \infty \, .
\end{equation}
(see also Fig. \ref{fig:tu-cont-limit-def}). Recall that $\tilde{u}$, like all spectral parameters in this paper, is defined modulo $\pi$. 
This region contains the simple values $\pm {\pi\over 2}$: this is because, setting ${\gamma\over 2}+\tilde{u}={\pi\over 2}$ we have $\Re (\tilde{u})<\gamma$ provided $\gamma<{\pi\over 3}$, which is the case for $\gamma={\pi\over p+1}$, $p\geq 3$. Hence we can use the following discretization too
\begin{equation}\label{eq:otherrealization-(2,1)-op}
 \widehat{{\cal D}}_{(2,1)}^{(\rm latt)} =  \widehat{{\cal D}}_{\overline{(2,1)}}^{(\rm latt)} \propto  \uR^{-1} T\left(\pm \frac{\pi}{2}\right)     \, . 
\end{equation} 
We emphasize that, in finite size, these operators do not coincide with those  in (\ref{eq:realization-(2,1)-op}) - but they do  so in the continuum limit (in any case recall it does not seem to be possible to realize topological invariance on the lattice).

We note here that  in  the A$_3$ RSOS (which realizes the Ising CFT) it can be shown that $(-1)^\R\uR^{-1} T(\pi/2)  $ acts like the spin flip operator of Transverse Field Ising (TFI) model, see section \ref{sec:Ising} for details (the bulk normalization is not an issue in this case, apart from the $(-1)^\R$ factor, discussed below). The spin flip operator in TFI model is known to be the lattice realization of $\widehat{\cal{D}}_{(2,1)} = \widehat{\cal{D}}_{(1,3)}$ \cite{Aasen2016} in Ising CFT.
Note it is not hard to check from Eq. \eqref{eq:tmat_TLgen_KP} that $T(\tilde{u} + \pi) = T(\tilde{u})$, so one could also use $(-1)^\R\uR^{-1}T(-\pi/2) $ as $\widehat{{\cal D}}_{(2,1)}^{(\rm latt)}$, which is what we used in Potts model to realize the Fibonacci operator $\widehat{W}$ in Potts CFT \cite{Sinha:2023hum}, as the TDL $W$ in Potts CFT behaves like $\widehat{{\cal D}}_{(2,1)}$ in $\mathcal{M}(6,5)$\footnote{The role of the parity of $\R$ was not discussed in \cite{Sinha:2023hum}, where we implicitly corrected the signs of the expectation value of the line operators by comparing them with those for the ground state.}.

\section{Composition of topological defects}\label{sec:fused-Tmat-Th}

\begin{figure}[hb]
  \begin{center}
  \includegraphics[width=0.24 \linewidth]{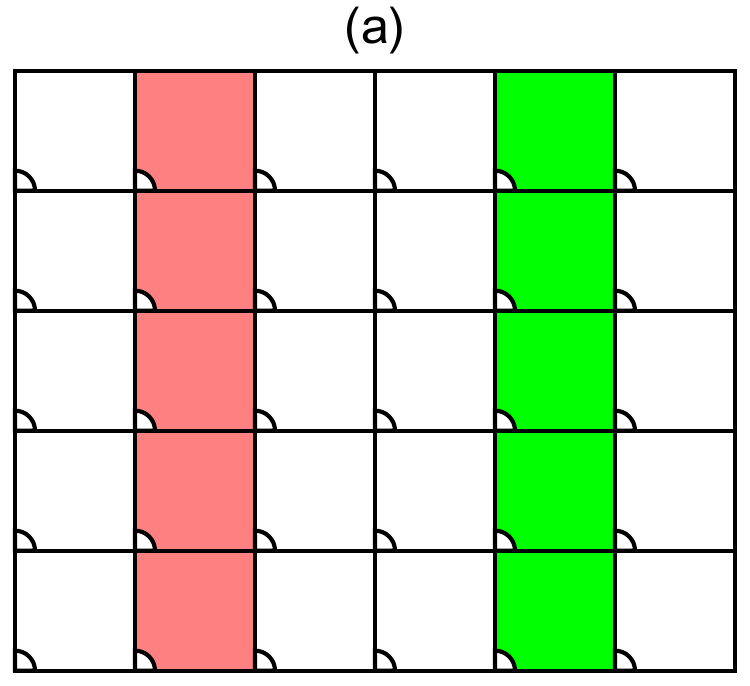}
  \includegraphics[width=0.24 \linewidth]{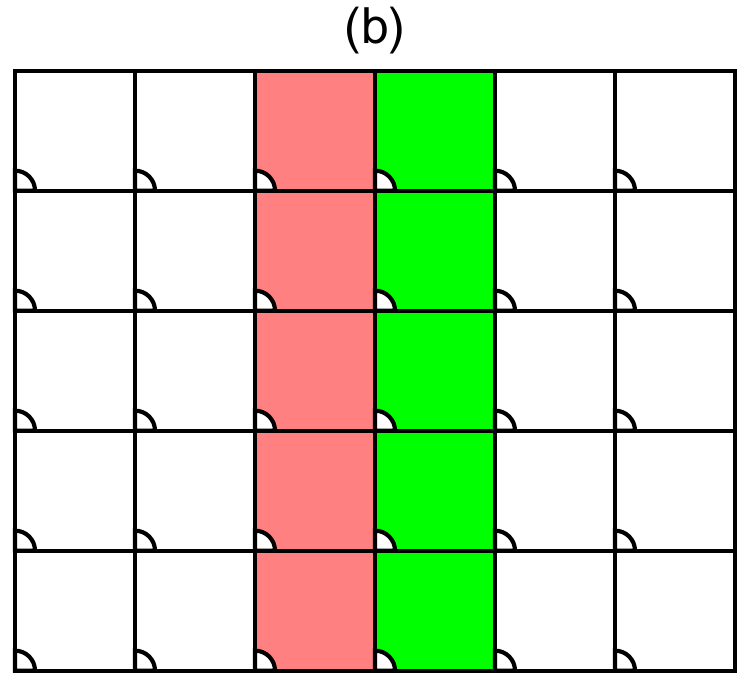}
  \includegraphics[width=0.24 \linewidth]{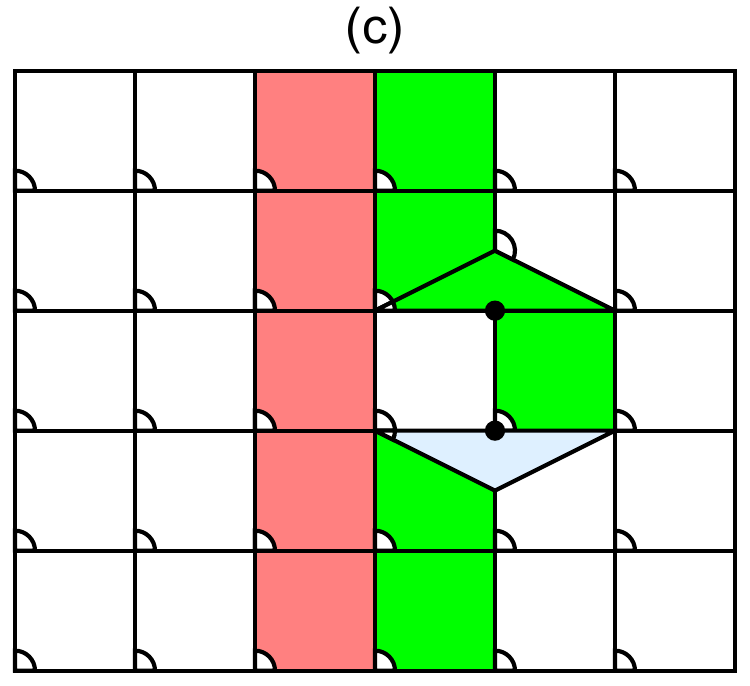}
  \includegraphics[width=0.24 \linewidth]{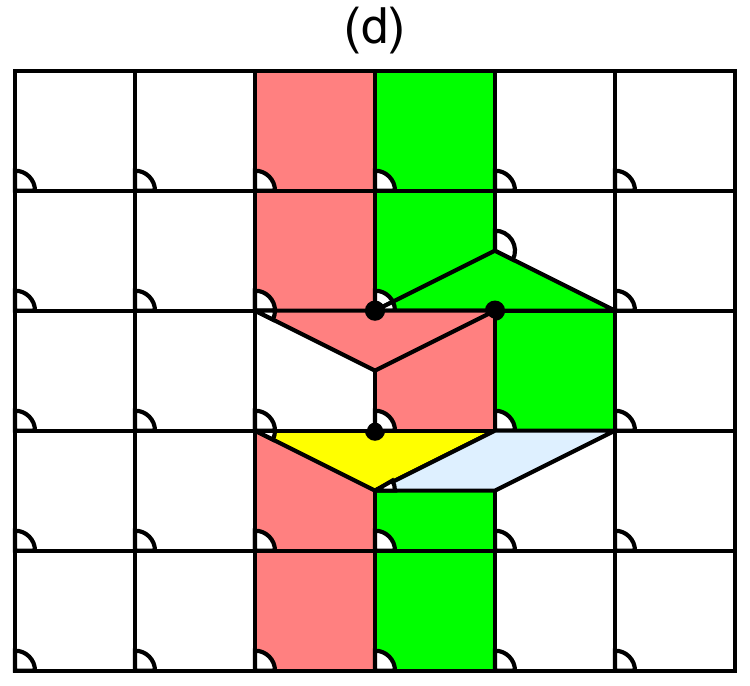}
\caption{(a) Two loop lines along the direct channel. (b) The defects are brought together. (c-d) Local condition for topological defects imply that an associative composition exists.  }
\label{Comp}
\end{center}
\end{figure}

An important consequence of our construction  of lattice topological defects  is that one can compose them to build ``higher defects''. This is simply due to  the Yang-Baxter equation, and our general definition of defect lines (topological or not) in this context as columns with a modified value of the spectral parameter. In figure \ref{Comp}, we illustrate how two such lines  can be moved next to each other  (in, say,  an otherwise homogeneous system) - the resulting   object  now being considered   as a new unique seam. Of course, if the lines are already topological on the lattice (what we called a $l$TDL), by using again the Yang-Baxter and unitarity equations, this seam can be deformed just like each of its individual components, leading in practice to a new $l$TDL. If the line is only a discretization of a TDL (what we called a dTDL), the same argument  will turn out to apply in the continuum limit as we discuss below. \footnote{We recall that one has  in general to be careful when exchanging products and continuum limits \cite{Koo_1994}.}

We note that, since the product of matrices is associative and the identity can be considered as a trivial defect, the foregoing construction results in what is called technically a monoid 
\footnote{Recall that in algebra, a monoid is a set equipped with an associative binary operation and an identity element.} of lattice topological defects for  the model.
In what follows, we shall clarify other aspects of this algebraic structure using the RSOS  models as a prototype. For example, it is desirable to understand whether it is possible to  decompose the  defect resulting from an operation such as the one in figure \ref{Comp} in terms of what should be called ``irreducible elements'', and whether one can define ``generators'' for the whole set of topological defects. The importance of these questions becomes evident when one considers the crossed channel, where the topological defects may be regarded as generalized symmetries, which can be used to resolve the operator content of the homogeneous theory itself.    

In order to proceed, we need more solutions to the relations expressed in figures \ref{fig:first_cond_top_inv} and \ref{fig:top-def-rel-2}. For this, we shall  explore the Yang-Baxter equation for higher representations.

\subsection{Higher spin Yang-Baxter equation: Fusion in the direct channel}

The idea underlying the construction is to introduce columns with different faces, corresponding technically (in the Yang-Baxter framework) to spins $J>1$. There is a large literature on the topic, see for instance \cite{DATE1987231, Klumper:1992vt}. Derivations are however not always explicit, which can create difficulties in particular when considering generalizations to RSOS models other than A type. For this reason we discuss what seems to be a new way to obtain the corresponding ``fused'' Boltzmann weights in appendix  \ref{AnotherFusion}, and give only the main results in this section. Our approach closely follows \cite{Gohmann_2010}, however  adapted to the face models. While it is possible instead to rely on the representation theory of the quantum deformed algebra \cite{KRS81, KauffmanLins+1994+13+21, DATE1987231}, the approach we use shows  how one may proceed without such  knowledge. This opens the way for further developments where no group-theoretic classification of solutions to the Yang-Baxter equation is known \cite{MARTINS2013243}.      

In a nutshell, the  fusion procedure consists of first assembling $J$ faces of spin $1$ with spectral parameters for neighboring faces  differing by $\gamma$ when moving from left to right (or right to left). This special value makes some faces ``singular'' - that is, they behave as projectors -  and allows one to restrict to linear combinations of internal heights.

The Boltzmann weights of fused faces we shall work with are given by the formulae

\begin{equation}\label{eq:Wt-Th-H}
    ^{(1J)}W  \! \!  \left(\begin{array}{ll}
d & c \\
a & b
\end{array} \Bigg| \ u\right)
 = 
(-1)^{\left(-\frac{(1+J)}{2}+\frac{(d-b+c-a) (a-c)}{4}\right)}\sqrt{\frac{\gf_{\frac{c+a-1-J}{2}} \gf_{\frac{c+a+1+J}{2}}}{\gf_b \gf_d}} \frac{\sin \left(u + \frac{(b d-a c-1) \gamma}{2}\right)}{\sin \gamma},  \nonumber
\end{equation}
for $b-a=c-d$ \, ,
\begin{equation}
    ^{(1J)}W  \! \! \left(\begin{array}{ll}
d & c \\
a & b
\end{array} \Bigg| \ u\right)
=-(-1)^{J \left( \frac{(a+c-b-d)}{4}+ \frac{1}{2}\right)}\sqrt{\frac{\gf_{\frac{c-a+1+J}{2}} \gf_{\frac{a-c+1+J}{2}}}{\gf_b \gf_d}} \frac{\sin\left( u+ \frac{(a c-b d-1) \gamma}{2}\right)}{\sin \gamma} \, , \label{IRFweightsSd0}
\end{equation}otherwise (recall factors $\gf_t$ are defined at the beginning) \footnote{Results in the literature may differ from these by some gauge factors.}. These  expressions, when $J = 1$, reduce to Eq. \eqref{eq:standard_face_weight} for the choice of  gauge factor there $S_a = 1$.
They must be supplemented by adjacency rules for the heights, which are conveniently expressed using new adjacency matrices. To each defect of spin $J$, one can associate the fused adjacency matrix $G^{(J)}$ obtained recursively from 
\begin{equation}
{ G} \cdot { G}^{(J-1)} = { G}^{(J)}+{ G}^{(J-2)}, 
\end{equation}
with the initial condition $G^{(1)}=G$, which, recall, is the adjacency matrix for an  A-type Dynkin diagram. As usual, these matrices encode incidence rules - rows and columns of these matrices take values that label the heights, and neighboring heights along a column (resp. a row)  in (\ref{IRFweightsSd0})  must correspond to a non-zero matrix element in the adjacency matrix $G^{(1)}$ (resp. $G^{(J)}$) (for the A$_p$ models we consider here, elements of $G^{(J)}$ are  always 0 or 1 so the question of multiplicities does not arise). For example, for the $A_4$ model we have 
\begin{equation}
\underbrace{\left(
\begin{array}{cccc}
 1 & 0 & 0 & 0 \\
 0 & 1 & 0 & 0 \\
 0 & 0 & 1 & 0 \\
 0 & 0 & 0 & 1 \\
\end{array}
\right)}_{G^{(0)}},~~
\underbrace{\left(
\begin{array}{cccc}
 0 & 1 & 0 & 0 \\
 1 & 0 & 1 & 0 \\
 0 & 1 & 0 & 1 \\
 0 & 0 & 1 & 0 \\
\end{array}
\right)}_{G^{(1)}},~~
\underbrace{\left(
\begin{array}{cccc}
 0 & 0 & 1 & 0 \\
 0 & 1 & 0 & 1 \\
 1 & 0 & 1 & 0 \\
 0 & 1 & 0 & 0 \\
\end{array}
\right)}_{G^{(2)}},~~
\underbrace{\left(
\begin{array}{cccc}
 0 & 0 & 0 & 1 \\
 0 & 0 & 1 & 0 \\
 0 & 1 & 0 & 0 \\
 1 & 0 & 0 & 0 \\
\end{array} 
\right)}_{G^{(3)}} \, .
\end{equation}

When  $J=p-1$, the weights $~^{(1J)}W$ in Eq. (\ref{IRFweightsSd0})  become proportional to the height-reflection (in the horizontal direction), i.e. it connects reflected heights: $a \leftrightarrow p+1-a$. In consequence, no new defect is obtained at $J=p$, since
\begin{equation}
{ G} \cdot { G}^{(p-1)} = {G}^{(p-2)} \, . 
\end{equation}
In the direct channel, the introduction of a fused face generally modifies the local Hilbert space. Later, we will show that it cannot affect the bulk contribution to the free-energy. It may, however, recombine different sectors of left/right chiral parts of the continuum limit theory.

\subsection{Higher spin Yang-Baxter equation: Fusion in the crossed channel}

Fusion can as well  be implemented vertically (i.e. along a column), instead of horizontally along a row. The corresponding Boltzmann weights are then given by 
\begin{equation}\label{eq:Wt-Th-V}
    ^{(J1)}W  \! \! \left(\begin{array}{ll}
d & c \\
a & b
\end{array} \Bigg| \ u\right) = (-1)^{\left(\frac{(1+J)}{2} + \frac{(b-d+c-a) (a-c)}{4}\right)}\sqrt{\frac{\gf_{\frac{c+a-1-J}{2}} \gf_{\frac{c+a+1+J}{2}}}{\gf_b \gf_d}} \frac{\sin(u+ \frac{(b d-a c-1) \gamma}{2})}{\sin \gamma},  \nonumber
\end{equation}
for $c-b=d-a$,
\begin{equation}
    ^{(J1)}W  \! \!  \left(\begin{array}{ll}
d & c \\
a & b
\end{array} \Bigg| \ u\right) = 
(-1)^{J \left(\frac{(a+c-b-d)}{4} - \frac{1}{2}\right)}\sqrt{\frac{\gf_{\frac{c-a+1+J}{2}} \gf_{\frac{a-c+1+J}{2}}}{\gf_b \gf_d}} \frac{\sin( u+ \frac{(a c-b d-1) \gamma}{2})}{\sin \gamma}, \label{IRFweightsS}
\end{equation}
otherwise. It is easy to see that when $J = 1$, the weight in the equation above is the same as the weight in Eq. \eqref{eq:Wt-Th-H}. Now neighboring heights along a row (resp. a column) must correspond to a non-zero matrix element in the adjacency matrix $G^{(1)}$ (resp. $G^{(J)}$). In \autoref{sec:fus:KP}, we discuss how to compare the Boltzmann weight in this paper with that of \cite{Klumper:1992vt}. Obviously these weights will allow us to build  lattice defect line operators $\widehat{{\cal D}}^{(\rm latt)}$ in the crossed channel. 

Now, calculations in the crossed channel can be formulated in a powerful  algebraic way - rendering  the detailed expression of the weights unnecessary - since the underlying operation of building fused faces by concatenating faces with different spectral parameters can now be interpreted as the multiplication of transfer matrices. It turns out that the following relations (fusion hierarchy) hold
\cite{Bazhanov:1989yk}%
\begin{equation}\label{eq:fusion-Th}
 T^{(1)}_{\left[ \frac{J}{2} \right]} \,  T^{(J)}_{\left[-\frac{1}{2}\right]} = T^{(0)}_{\left[ \frac{J +1}{2}\right]} \, T^{(J-1)}_{\left[ -1 \right]} + T^{(0)}_{\left[ \frac{J -1}{2}\right]} \,  T^{(J+1)}_{\left[0 \right]}  \, , 
\end{equation}
where we have defined 
\begin{equation}\label{eq:fusion-T-M-def}
\begin{split}
T^{(J)}_{[k]} := T^{(J)}(u +  k \gamma )  \, , \quad T^{(0)}_{[0]} &:= \left(\frac{\sin \left(u - \frac{\gamma}{2} \right)}{\sin \gamma}\right)^{2 \R} \, \mathbb{1} \, , \quad T^{(-1)}_{[0]} := 0\, , \quad
T^{(p)}_{[0]} := 0  \, , 
\\ 
& \left(T^{(J)}(u)\right)_{\mathbf{a}}^{\mathbf{b}} := \prod_{i=1}^{2 \R} ~^{(J1)}
W \! \! \left(\begin{array}{ll}
b_{i} & b_{i+1} \\  
a_{i} & a_{i+1} 
\end{array} \Bigg| \ u\right)
 \, .
\end{split}
\end{equation}
The latter object is of course the transfer matrix  propagating states over  one (periodic)  row of fused faces. In this hierarchy,  $J=0$ corresponds to a transfer matrix proportional to identity and $J=1$ gives the fundamental transfer matrix. We will later see that $J=p-1$ is proportional to the height reflection operator of Eq. \eqref{def:heightreflection}.

We notice that we are free to perform gauge transformations of the type 
\begin{equation}
~^{(J1)}W  \! \!  \left(\begin{array}{ll}
d & c \\
a & b
\end{array} \Bigg| \ u\right) \rightarrow ~^{(J1)}W  \! \!  \left(\begin{array}{ll}
d & c \\
a & b
\end{array} \Bigg| \ u\right)\frac{\kappa(a, d)}{\kappa(b,c)} \, ,
\end{equation}
with an arbitrary  function $\kappa(a,d)$  of the vertical edge heights. This does not change the transfer matrix $T^{(J)}$ because 
 of the horizontal periodic boundary conditions.  It is also
possible to include such a transformation for the horizontal edges, say with another function $\kappa_2(a, b)$, which will cancel out of the vertical periodic boundary
conditions. This corresponds now  to a similarity transformation between different  transfer matrices.

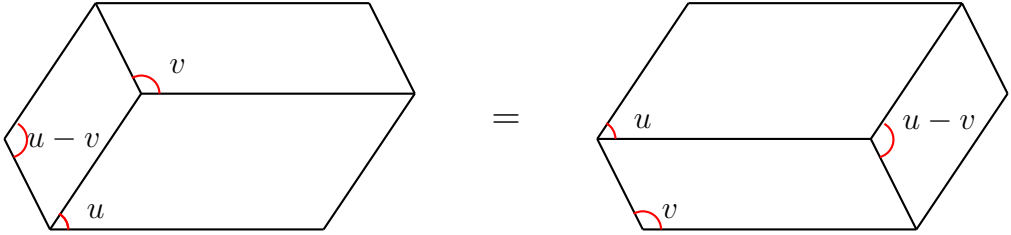
\begin{figure}[H]
    \centering
\begin{tikzpicture}[scale=1.2]
    \draw[black, thick] (0,0) -- (3,0);
    \draw[black, thick] (1,1.5) -- (4,1.5);
    \draw[black, thick] (0,0) -- (1,1.5);
    \draw[black, thick] (3,0) -- (4,1.5);
    \draw[black, thick] (0.5,2.5) -- (3.5,2.5);
    \draw[black, thick] (3.5,2.5) -- (4,1.5);
    \draw[black, thick] (0.5,2.5) -- (1,1.5);
   \draw[black, thick] (0,0) -- (-0.5,1);
   \draw[black, thick] (0.5,2.5) -- (-0.5,1);

     \node[] at (5,1.2) {\Large{$=$}};

    \draw[black, thick] (6,1) -- (9,1);
    \draw[black, thick] (7,2.5) -- (10,2.5);
    \draw[black, thick] (7,2.5) -- (6,1);
    \draw[black, thick] (9,1) -- (10,2.5);
    \draw[black, thick] (6.5,0) -- (9.5,0);
    \draw[black, thick] (6,1) -- (6.5,0);
    \draw[black, thick] (9,1) -- (9.5,0);
    \draw[black, thick] (10,2.5) -- (10.5,1.5);
    \draw[black, thick] (10.5,1.5) -- (9.5,0);
      \node[] at (0.5,0.2) {\large{$u$}};
      \node[] at (1.4,1.8) {\large{$v$}};
      \node[] at (0.15,1) {\large{$u -  v$}};

      \node[] at (6.8,0.2) {\large{$v$}};
      \node[] at (6.5,1.2) {\large{$u$}};
      \node[] at (9.75,1.2) {\large{$u -  v$}};

    \draw[red,thick]  (0.2,0) arc (0:60:0.2);
    \draw[red,thick]  (1.2,1.5) arc (0:120:0.2);
    \draw[red,thick]  (-.4,0.8) arc (-75:60:0.2);

    \draw[red,thick]  (6.2,1) arc (0:60:0.2);    
    \draw[red,thick]  (6.7,0) arc (0:120:0.2);
    \draw[red,thick]  (0.2,0) arc (0:60:0.2);
    \draw[red,thick]  (9.1,0.8) arc (-75:60:0.2);

\end{tikzpicture}    
\caption{Yang-Baxter equation for fused faces.}
  \label{YFIRFfusg}
\end{figure}

Generally, one can of course perform fusion in both directions. To illustrate what happens,  we represent in figure \ref{YFIRFfusg} the spin of each space by the length of the face edges, which come in three different sizes. 
One may turn the weights around to write the  Yang-Baxter equation satisfied by the fused weights in the operatorial form
\begin{equation}
 R_{i+1}^{(q s)}(v) R_i^{( r s)}(u) R_{i+1}^{( r q)}(u-v) =  R_{i}^{( r q)}(u-v) R_{i+1}^{( r s)}(u) R_{i}^{(q s)}(v), \label{HigherspinYBEIRF}
 \end{equation}
and 
\begin{equation}\label{eq:higher-R-def}
\bra{b} R_i^{(q s)}(u) \ket{a}
 = \prod_{\stackrel{j=1}{j \neq i }} \delta_{a_j,b_j}  ~^{(qs)}W  \! \! 
\left(\begin{array}{ll}
a_{i-1} & b_i \\
a_i & a_{i+1}
\end{array} \Bigg| \ u\right)
\, ,
\end{equation}
where we refrain from giving explicit expressions of the weights  $^{(qs)}W$.  It can be checked that if we set $q = s = 1$, then we recover Eq. \eqref{eq:R&Rtop}.

Note that we  are mainly concerned in this paper with topological defects in a bulk model defined via the fundamental weights (diagonal minimal models) and thus will mostly need fused weights either in the vertical(direct channel) or in the horizontal (crossed channel) directions. However, weights obtained by fusion in both directions are also important in analyzing situations where two defect lines intersect. For example, one might be interested in studying the associated generalized symmetries in an integrable impurity model. In this case, a defect operator in the crossed channel acts on a defect Hilbert space leading, in general, to arbitrary fused faces at the intersection of the lines.

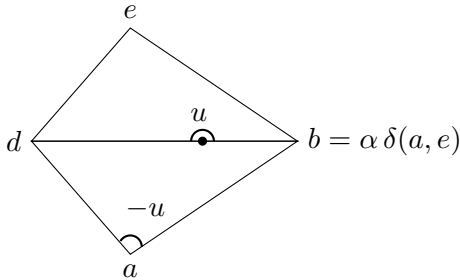
\begin{figure}[H]
    \centering
\begin{tikzpicture}
        \coordinate (A) at (-1.5,0);
    \coordinate (B) at (2,0);
    \coordinate (C) at (-0.2,1.5);
        \coordinate (D) at (-0.2,-1.5);

    \draw (A) -- (B) -- (C) -- cycle;
    \draw (A) -- (B) -- (D) -- cycle;


    \node[left] at (A) {\( d \)};
    \node[right] at (B) {\( b = \alpha \, \delta(a, e) \)};
    \node[below] at (D) {\( a \)};
    \node[above] at (C) {\( e \)};
    
    \node at (0.7,0.35) {\( u \)};
    \node at (0,-0.9) {\( -u \)};
    
    \filldraw[black] (0.75,0) circle (1.5pt);

\draw[black, thick] (0.9,0) arc (0 : 180 : 0.15) ;
\draw[black, thick] (-0.05,-1.4) arc (0 : 150 : 0.15) ;

\end{tikzpicture}  
\caption{Unitarity condition for fused faces. We draw the edges with different lengths  to represent two different higher spin representations. The two largest edges
  correspond to one spin and the two smallest ones to the other.}
\label{Unitfused}
\end{figure}

We now note that the weights obey a  generalized unitarity relation as represented in figure \ref{Unitfused}. 

{
The unitarity relation  can be written in terms of the weights as follows
\begin{equation} \label{eq:unifused}
 \sum_{c}  \,  ^{(1 J)}W  \! \!  \left(\begin{array}{ll} d & c \\ a & b \end{array} \, \Bigg| \, u  \right) \,    ^{(J1)} W  \! \!  \left(\begin{array}{ll} d & e \\ c & b \end{array} \, \Bigg| \, -u  \right) \propto \delta_{a,e} \, .
\end{equation}
}
Finally, we notice that  one can relate the $^{(1 J)}W$ and $^{(J 1)}W$  in the direct and crossed channels  simply by taking into account the crossing relation of
the fundamental weights (\ref{eq:crossing_sym}) which naturally extends to the fused faces, in view of the fusion process itself:
\begin{equation}
~^{(1 J)}W \! \! \left(\begin{array}{ll} d & c \\ a & b \end{array} \, \Bigg| \, u  \right) = \sqrt{\frac{\gf_a \gf_c}{\gf_b \gf_d}} ~^{(J 1)}W  \! \! 
\left(\begin{array}{ll} a & d \\ b & c \end{array} \, \Bigg| \, \gamma-u  \right). \label{crossinghigh}
\end{equation} 
Further, the following can be seen 
\begin{equation}
   \left( R_{i}^{(1J)}(u) R_{i}^{(J1)}(-u) \right) \propto \mathbb{1} \, , 
\end{equation}
using Eq. \eqref{eq:higher-R-def} and \eqref{eq:unifused}.

\subsection{Defect lines based on higher faces: a preview}

In the remainder of this paper, we will show that the topological defect lines can be realized, at least in the continuum limit, by introducing modified rows of columns based on the  local Boltzmann weights defined above.
 
Let us start by stating the results. We first discuss the case of defects of type $(1,s)$. We have discussed the most fundamental among them in \autoref{sec:Y-12-op}, where we have shown that this defect can be obtained using the transfer matrix at spectral parameter $\pm {\rm i} \infty$. More generally, the $(1,s)$ defect can be obtained at the same spectral parameter, $\tilde{u} = {\rm i} \infty$, but using fused transfer matrices, i.e.
\begin{equation}\label{eq:1s-discretization}
       \widehat{{\cal D}}_{(1,s)}^{(\rm latt)} = \widehat{{\cal D}}_{\overline{(1,s)}}^{(\rm latt)}
        \propto~\uR^{1-s}\lim_{\tilde{u} \to {\rm i } \infty} T^{(s-1)}( \tilde{u})  \, , 
\end{equation}
  where $1 \leq s \leq p $. Note, the claim above for $s = 1$ is obvious from Eq. \eqref{eq:fusion-T-M-def} and for $s = 2$ was discussed in \autoref{sec:Y-12-op}. For higher $s$, we show that the fused transfer matrix indeed are the lattice realization of $\widehat{\cal{D}}_{(1,s)}$ in \autoref{sec:fusion-Tmatrices-(1s)}. The proportionality factor is easy to figure out, and is given in Eq. \eqref{eq:prop-fac-high-fus}. Note that all operators where the power of $\uR$ would be replaced by one if $s$ is even and $0$ if $s$ is odd would give rise to the same defect in the continuum limit. In particular,  for $s$ odd, there is in fact no need to insert the $\uR$ term since eigenvalues of the lattice defect operator then have the correct sign for all states (and the defect operator then maps odd/even sectors to themselves). Again, these operators lie in the center of $aTL_{N}(q)$ and are topological on the lattice. 

We now turn to defects of type $(r,1)$, which, exactly like for the fundamental case, are obtained at ${\gamma\over 2}+\tilde{u} = \frac{\pi}{2}$ using fused transfer matrices, i.e. 
\begin{equation}\label{eq:r1-discretization}
    \widehat{{\cal D}}_{(r,1)}^{(\rm latt)} = \widehat{{\cal D}}_{\overline{(r,1)}}^{(\rm latt)} \propto  \uR^{1-r} T^{(r-1)}\left(\pm\frac{\pi}{2} \right) \, , 
\end{equation}
where $1 \leq r \leq p-2$ and the same remark about the power of $r$ as for the defects in Eq. (\ref{eq:1s-discretization}) hold as well. While the case $r = 1$ has been discussed in \autoref{sec:21-discretization}, we discuss the fused transfer matrix case in section \ref{BASection}. Note, these lattice line operators become topological only in the continuum limit, unlike the operators in Eq. \eqref{eq:1s-discretization}. Therefore, the line operators in Eq. \eqref{eq:1s-discretization} are $l$TDLs and the ones in Eq. \eqref{eq:r1-discretization} are dTDLs.

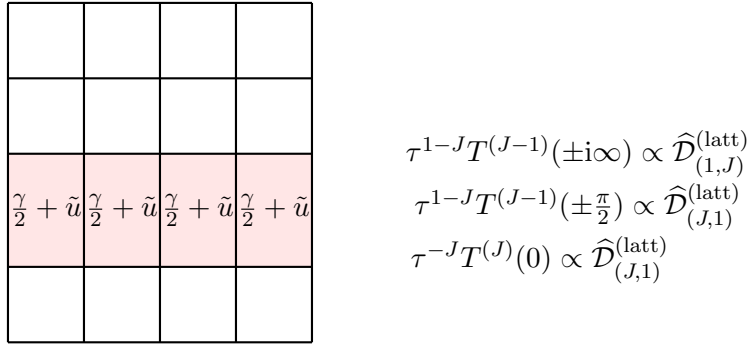
\begin{figure}[H]
     \begin{adjustbox}{max totalsize={1\textwidth}{1.\textheight},center}
     \begin{tikzpicture}
           \fill[red!10] (0,2.5) -- (4,2.5) -- (4,1.0) -- (0,1.0) -- cycle ;

         \draw[black, thick] (0,0) -- (4,0) ;
         \draw[black, thick] (0,1) -- (4,1) ;
         \draw[black, thick] (0,2.5) -- (4,2.5) ;
         \draw[black, thick] (0,3.5) -- (4,3.5) ;
         \draw[black, thick] (0,4.5) -- (4,4.5) ;

         \draw[black, thick] (0,0) -- (0,4.5) ;
         \draw[black, thick] (1,0) -- (1, 4.5) ;
         \draw[black, thick] (2,0) -- (2,4.5) ;
         \draw[black, thick] (3,0) -- (3,4.5) ;
         \draw[black, thick] (4,0) -- (4,4.5) ;

        \node[] at (0.5,1.75) {${\footnotesize  \gamma\over 2}+\tilde{u}$} ;
         \node[] at (1.5,1.75) {${\gamma\over 2}+\tilde{u}$} ;
         \node[] at (2.5,1.75) {${\gamma\over 2}+\tilde{u}$} ;
        \node[] at (3.5,1.75) {${\gamma\over 2}+\tilde{u}$} ;
       \node[] at (7.5,2.5) {$\uR^{1-J} T^{(J-1)}(\pm {\rm i } \infty) \propto \widehat{{\cal D}}_{(1,J)}^{(\rm latt)} $} ;
          \node[] at (7.5,1.8) {$\uR^{1-J} T^{(J-1)}(\pm \frac{\pi}{2}) \propto \widehat{{\cal D}}_{(J,1)}^{(\rm latt)} $} ;
             \node[] at (7,1.1) {$ \uR^{-J}T^{(J)}(0) \propto \widehat{{\cal D}}_{(J,1)}^{(\rm latt)} $} ;
 
     \end{tikzpicture}

    \end{adjustbox}

    \caption{A defect transfer matrix, $T^{(J)}({\gamma\over 2}+\tilde{u})$ (here $J=2$), constructed using weight $^{(1J)}W({\gamma\over 2}+\tu)$, in the crossed channel in an otherwise homogeneous system.
 }
    \label{fig:crossed-channel }
\end{figure}

In practice there are various ways to manufacture a system with  a row or column of modified faces for a given spin $J$. One is to simply ``add" the row or column to the original model, hence changing the size of the system (and potentially creating problems with the implementation of periodic boundary conditions). Another is to keep the initial system with fundamental faces (spin one) only,   adjust the spectral parameters to manufacture the  array necessary for fusion (see the appendix \ref{AnotherFusion}), and finally implement a projection. We will discuss in detail how these approaches compare on some simple examples below. For a more formal point of view, see \cite{Belletete:2018eua}.

Another important point to mention is that the defect identifications - like the simple examples before - hold not only for special values of the spectral parameters, but in fact for whole regions in the complex plane. The reason for this will become more apparent when we discuss the Non-linear Integral Equation (NLIE) corresponding to these defects later in section \ref{BASection}. The most important example of this phenomenon occurs for defects of type   $(r,1)$. If one sets $\tilde{u}=0$, then 
\begin{equation}\label{eq:r2-discretization}
    \widehat{{\cal D}}_{(r,1)}^{(\rm latt)} = \widehat{{\cal D}}_{\overline{(r,1)}}^{(\rm latt)} \propto  \uR^{-r} T^{(r)}\left(0 \right) \, , 
\end{equation}
notice the difference in the spin indices as compared to Eq. (\ref{eq:r1-discretization}).  
We also note here that  transfer matrices at generic spectral parameter are non-hermitian, and may  have   complex eigenvalues.  
In the (finite) complex plane, they become hermitian along the lines $\Re (u) = \gamma/2 ~\mbox{mod}~\pi/2$. For generic values, however, in the large system size limit, we observe using the  Bethe-ansatz that the complex part of eigenvalues of $T^{(J)}(u)$ go to 0 - as expected if they  flow to the defect operators of the CFT,  which of course only have real eigenvalues. 

\subsection{Higher defect Hamiltonian}\label{subsec:two-imp-ham}
Using the fused faces, one can define the higher defect Hamiltonian, as in the figure below.

\begin{figure}[H]
    \centering
\begin{tikzpicture}

\draw[black, thick] (0,0) -- (1,1.5) -- (2,0) -- (1,-1.5) -- cycle;
\fill[yellow!100] (3,1.5) -- (1,1.5) -- (2,0) -- (4,0) -- cycle;
\fill[red!50] (3,-1.5) -- (1,-1.5) -- (2,0) -- (4,0) -- cycle;
\draw[black, thick] (3,1.5) -- (1,1.5) -- (2,0) -- (4,0) -- cycle;
\draw[black, thick] (3,-1.5) -- (1,-1.5) -- (2,0) -- (4,0) -- cycle;

\node at (-1,0) {$H^{\rm imp} = \partial_u$} ; 

\draw[black, thick] (2.2,0) arc (0 : 120 : 0.2) ;
\draw[black, thick] (2.8,-1.5) arc (180 : 60 : 0.2) ;

\draw[black, thick] (1.1,-1.35) arc (60 : 120 : 0.2) ;

\draw[black, densely dotted] (1,-1.5) -- (1,-2.) ;
\draw[black, densely dotted] (1,2) -- (1,1.5) ;
\draw[black, densely dotted] (0,2) -- (0,-2.) ;
\draw[black, densely dotted] (4,2) -- (4,-2.) ;
\draw[black, densely dotted] (3,-1.5) -- (3,-2.) ;
\draw[black, densely dotted] (3,2) -- (3,1.5) ;

\node at (0,-2.3) {$k-1$} ; 
\node at (1,-2.3) {$k$} ; 
\node at (3,-2.3) {$k+1$} ; 
\node at (4,-2.3) {$k+2$} ; 
\node at (5.4,-2.3) {$u \to 0$} ; 

\node at (1,0) {$u$} ; 
\node at (1,0) {$u$} ; 
\node at (2.5,0.75) {$u + \tilde{u}_k$} ; 
\node at (2.5,-0.75) {$-\tilde{u}_k$} ; 


\end{tikzpicture}
    \caption{Face representation of the impurity contribution to the Hamiltonian. White face stands for the homogeneous model, carrying spin $(11)$. Yellow face carries spin $(1J)$, while the pink face carries spin $(J1)$.}
    \label{fig:impurity}
\end{figure}
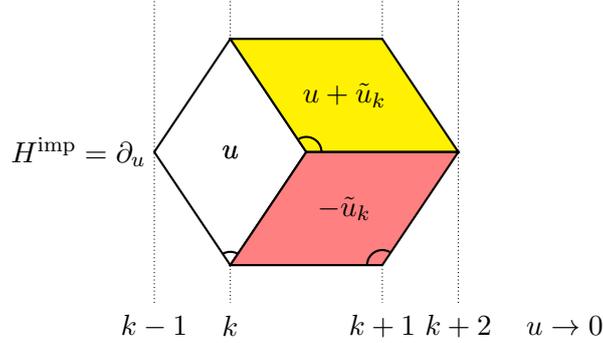
In this section, we will now derive the two impurity Hamiltonian, in the same way as the the one impurity defect Hamiltonian was derived in Section \ref{subsec:imp-ham}. We have
\begin{equation}\label{eq:def-two-imp-ham}
    H^{k,k+1,k+2}\left(\tu_{k},\tu_{k+1} \right) = - T^{-1}\left(\{0,\tu_k,\tu_{k+1} \}_{k,k+1} \right) \frac{\partial}{\partial u}T\left(\{ u, u + \tu_k, u + \tu_{k+1} \}_{k,k+1} \right)\bigg|_{u = 0} \, ,
\end{equation}
where $ \langle  b  |  T(\{u, u + \tu_k, u + \tu_{k+1}\}_{k,k+1}) |  a \rangle$ is given by 
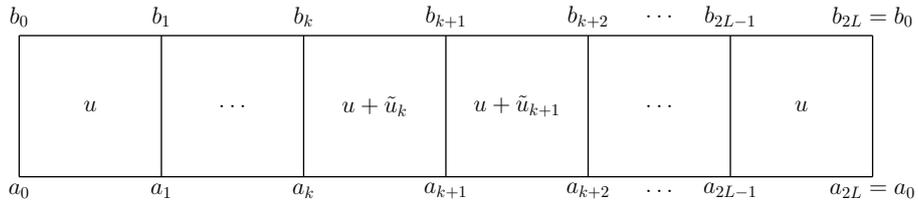
\begin{figure}[H]
        \begin{adjustbox}{max totalsize={0.8\textwidth}{0.8\textheight},center}
            \begin{tikzpicture}[thick, scale=1]
            \draw[black, thick] (0,0) -- (18,0);
            \draw[black, thick] (0,3) -- (18,3);
            \draw[black, thick] (0,0) -- (0,3);
            \draw[black, thick] (3,0) -- (3,3);
            \draw[black, thick] (6,0) -- (6,3);
            \draw[black, thick] (9,0) -- (9,3);
            \draw[black, thick] (12,0) -- (12,3);
            \draw[black, thick] (15,0) -- (15,3);
            \draw[black, thick] (18,0) -- (18,3);
            \node[] at (0,-0.3) {\Large{$a_0$}};
            \node[] at (3,-0.3) {\Large{$a_1$}};
            \node[] at (6,-0.3) {\Large{$a_k$}};
            \node[] at (9,-0.3) {\Large{$a_{k+1}$}};
            \node[] at (13.5,-0.3) {\Large{$\ldots$}};
            \node[] at (12,-0.3) {\Large{$a_{k+2}$}};
            \node[] at (15,-0.3) {\Large{$a_{2L-1}$}};
            \node[] at (18,-0.3) {\Large{$a_{2L} = a_0$}};
            \node[] at (0,3.4) {\Large{$b_0$}};
            \node[] at (3,3.4) {\Large{$b_1$}};
            \node[] at (6,3.4) {\Large{$b_k$}};
            \node[] at (9,3.4) {\Large{$b_{k+1}$}};
            \node[] at (12,3.4) {\Large{$b_{k+2}$}};
            \node[] at (13.5,3.4) {\Large{$\ldots$}};
            \node[] at (15,3.4) {\Large{$b_{2L-1}$}};
            \node[] at (18,3.4) {\Large{$b_{2L} = b_0$}};
            \node[] at (1.5,1.5) {\Large{$u$}};
            \node[] at (4.5,1.5) {\Large{$\ldots$}};
            \node[] at (7.5,1.5) {\Large{$u+\tu_k$}};
            \node[] at (10.5,1.5) {\Large{$u+\tu_{k+1}$}};
            \node[] at (13.5,1.5) {\Large{$\ldots$}};

            \node[] at (16.5,1.5) {\Large{$u$}};
    \end{tikzpicture}
\end{adjustbox}
    \caption{Transfer matrix with spectral parameter $\{u, u + \tu_k,u + \tu_{k+1} \}_{k,k+1}$ with PBC. }
    \label{fig:transfer-matrix-two-imp def-ham}
\end{figure}

Using Eq. \eqref{eq:tmat_TLgen_KP}, one can show that
\begin{equation}\label{eq:two-imp-mom-op}
 T\left(\{0,\tu_{k},\tu_{k+1} \}_{k,k+1} \right) = \frac{1}{\sin^2 \gamma} \uR \,  R_{k+1}\left(\tu_{k+1} \right) R_{k}\left(\tu_{k} \right) \, , 
\end{equation}
which implies 
\begin{equation}
\begin{split}
& T^{-1}\left(\{0,\tu_{k},\tu_{k+1} \}_{k,k+1} \right) = \sin^2{\gamma} \,  R_{k}^{-1}\left(\tu_{k} \right)   R^{-1}_{k+1}\left(\tu_{k+1} \right) \uR^{-1} \\
& = \frac{\sin^2\gamma}{\sin(\gamma - \tu_k)\sin(\gamma - \tu_{k+1})} \left( \mathbb{1} - \frac{\sin \tu_k}{\sin (\gamma + \tu_k)} e_{k} \right) \left( \mathbb{1} - \frac{\sin \tu_{k+1}}{\sin (\gamma + \tu_{k+1})}e_{k+1} \right) \uR^{-1} \, . 
\end{split}
\end{equation}
Similar to section \ref{subsec:imp-ham}, using Eq. \eqref{eq:tmat_TLgen_KP} we break the transfer matrix into two parts $T_A$ and $T_B$. We first calculate 
\begin{equation}\label{eq:two-imp-Ham-a}
    T^{-1}\left(\{0,\tu_{k},\tu_{k+1} \}_{k,k+1} \right)\frac{\partial}{\partial u}T_A\left(\{ u, u + \tu_k, u + \tu_{k+1} \}_{k,k+1} \right)\bigg|_{u = 0}  = \frac{1}{\sin \gamma} e_0 \, . 
\end{equation}
Then, finally we see that 
\begin{equation}\label{eq:two-imp-Ham-b}
\begin{split}    
    &T^{-1}\left(\{0,\tu_{k},\tu_{k+1} \}_{k,k+1} \right)\frac{\partial}{\partial u}T_B\left(\{ u, u + \tu_k, u + \tu_{k+1} \}_{k,k+1} \right)\bigg|_{u = 0}  = - \cot \gamma \mathbb{1}   \, + \\
    & \sum_{_{\substack{j=1 \\ j \neq k, k + 1, k+2}}}^{2L - 1} \left( - \cot \gamma \mathbb{1}  + \frac{e_j}{\sin \gamma } \right) + \left( - \cot \gamma \mathbb{1}  \, + \frac{1}{\sin \gamma}R_k^{-1}(\tu_k)R_{k+1}^{-1}(\tu_{k+1}) e_{k+2} R_{k+1}(\tu_{k+1})R_k(\tu_k) \right) \\ 
    & + \left( - \cot(\gamma - \tu_{k+1}) \mathbb{1}  +\bar{f}(\tu_{k+1}) \,  R_k^{-1} (\tu_{k}) e_{k+1} R_k (\tu_{k}) \right) + \left( - \cot(\gamma - \tu_{k}) \mathbb{1} +  \bar{f}(\tu_k) \,  e_k\right) \, , 
\end{split}
\end{equation}
where
\begin{equation}\label{eq:barf-def}
    \bar{f}(u) = \frac{\sin \gamma }{\sin (\gamma - u) \sin (\gamma + u)} \, . 
\end{equation}
Combining the two equations Eq. \eqref{eq:two-imp-Ham-a} and \eqref{eq:two-imp-Ham-b}, we get that 
\begin{equation}\label{eq:two-imp-Ham-n-simp}
\begin{split}
&    H^{k,k+1,k+2}\left(\tu_{k},\tu_{k+1} \right)  = - \sum_{_{\substack{j=0 \\ j \neq k, k + 1, k+2}}}^{2L - 1} \left( - \cot \gamma \mathbb{1}  + \frac{e_j}{\sin \gamma } \right)  - \left( - \cot(\gamma - \tu_{k}) \mathbb{1} +\bar{f}(\tu_{k}) \, e_k\right) \\
  & - \left( - \cot(\gamma - \tu_{k+1}) \mathbb{1}  +\bar{f}(\tu_{k+1}) \,  R_k^{-1} (\tu_{k}) e_{k+1} R_k (\tu_{k}) \right) \\ 
&  - \left( - \cot \gamma \mathbb{1}  \, + \frac{1}{\sin \gamma}R_k^{-1}(\tu_k)R_{k+1}^{-1}(\tu_{k+1}) e_{k+2} R_{k+1}(\tu_{k+1})R_k(\tu_k) \right) \, .
\end{split}
\end{equation}
One can simplify this Hamiltonian further 
\begin{equation}\label{eq:two-imp-ham-two-spec}
\begin{split}
    & H^{k,k+1,k+2}\left(\tu_{k},\tu_{k+1} \right)  =   \left( \cot (\gamma - \tu_{k}) \mathbb{1}+ \cot (\gamma - \tu_{k+1})\mathbb{1} - 2 \cot \gamma \right)\mathbb{1} - \sum_{j = 0}^{2L-1} \left( - \cot \gamma \mathbb{1} + \frac{e_j}{\sin \gamma} \right) \\
    & + \frac{1}{\sin \gamma}\left( f(\tu_{k}) \,  e_ke_{k+1}  + f(-\tu_k) \,  e_{k+1}e_k +f(\tu_{k+1}) \, e_{k+1}e_{k+2}  + f(-\tu_{k+1}) \,  e_{k+2}e_{k+1} \right) \\
    & + 2 \cot \gamma f(\tu_{k})f(\tu_{k+1}) f(-\tu_{k})f(-\tu_{k+1}) \,  e_k e_{k+2} \\
    & - \frac{1}{\sin \gamma} \left( 
   f(\tu_k)f(\tu_{k+1}) \,  e_k e_{k+1}e_{k+2} +  f(-\tu_k)f(-\tu_{k+1}) \, e_{k+2}e_{k+1}e_k  \right. \\
& \quad \quad  \quad  \left. +  f(-\tu_k)f(\tu_{k+1}) \,  e_{k+1}e_{k+2}e_{k}  +  f(\tu_k)f(-\tu_{k+1}) \, e_{k}e_{k+2}e_{k+1}  \right)    \, .
\end{split}
\end{equation}
 One can further show that the operator in Eq. \eqref{eq:two-imp-mom-op} commutes with the Hamiltonian above, and can be used to calculate the momentum eigenvalue of eigenstates of the Hamiltonian.

 Similar to the works of \cite{Kulish:1981bi}, we obtain higher defect Hamiltonian by setting the spectral parameters to be $u \pm \frac{\gamma}{2}$ and then applying a projector. Hence, we define the Hamiltonians 
\begin{equation}\label{eq:two-imp-ham-no-JW}
    H^{k,k+1,k+2}(\tu) = H^{k,k+1,k+2}\left(\tu + \frac{\gamma}{2},\tu - \frac{\gamma}{2} \right) \, , 
\end{equation}
\begin{equation}\label{eq:two-imp-ham-JW}
    H_{\rm JW}^{k,k+1,k+2}(\tu) = P^{(1)}_{k+1}H^{k,k+1,k+2}\left(\tu + \frac{\gamma}{2},\tu - \frac{\gamma}{2} \right)  P^{(1)}_{k+1} \, , 
\end{equation}
where 
\begin{equation}
    P^{(1)}_{j} = \mathbb{1} - \frac{1}{q+q^{-1}}e_j =  \frac{1}{\sin \left( 2 \gamma \right) } \, R_{j}(-\gamma)  \, .
\end{equation}
The above operator is a projector, i.e. $\left(P_j^{(1)}\right)^2 = P_j^{(1)}$.  The spectral parameters in the Hamiltonian of Eq. \eqref{eq:two-imp-ham-no-JW} are also carefully chosen, as they ensure that the Hamiltonian acts faithfully on the projected subspace corresponding to $P_{k+1}$, i.e. 
\begin{equation}
\begin{split}
   &   P^{(1)}_{k+1}H^{k,k+1,k+2}\left(\tu + \frac{\gamma}{2},\tu - \frac{\gamma}{2} \right)  P^{(1)}_{k+1}  = H^{k,k+1,k+2}\left(\tu + \frac{\gamma}{2},\tu - \frac{\gamma}{2} \right)  P^{(1)}_{k+1}  \\ & = P^{(1)}_{k+1}H^{k,k+1,k+2}\left(\tu + \frac{\gamma}{2},\tu - \frac{\gamma}{2} \right)  \, .
\end{split}
\end{equation}
The above relation is a consequence of the Yang-Baxter equation in \eqref{eq:Yang-Baxter}. We will analyze the role of the projection operator and the  two Hamiltonians in section \ref{sec:examples} via examples.

Recall that in section \ref{subsec:defect-Ham-direct-channel} we saw that by setting the spectral parameter to  $\pm { i} \infty$ or $\pm \frac{\pi}{2}$, we obtained the $(1,2)$ and $(2,1)$ defects. When we set both the impurity parameters $\tu_k$ and $\tu_{k+1}$ in \eqref{eq:two-imp-ham-two-spec}, to $\pm { i} \infty$ or $\pm \frac{\pi}{2}$, we are basically fusing two $(1,2)$ or $(2,1)$ defects with themselves, which from fusion of primary fields in continuum, we know should lead to $(1,1) \oplus (1,3)$ and $(1,1) \oplus (3,1)$ defects respectively. 

Now, for the case of $\tu_{k},\tu_{k+1} = {\rm i} \infty$, the Hamiltonian in Eq. \eqref{eq:two-imp-Ham-n-simp} can be shown to be 
\begin{equation}
    H^{k,k+1,k+2}_{{\cal D}_{(1,2)},{\cal D}_{(1,2)}} =  - \frac{\gamma}{\pi}\sum_{_{\substack{j=0 \\ j \neq k, k + 1, k+2}}}^{2L - 1} \left( - \cot \gamma \mathbb{1}  + \frac{e_j}{\sin \gamma } \right) - \frac{\gamma}{ \pi\sin \gamma} g_k^{-1}g_{k+1}^{-1}e_{k+2}g_{k+1}g_k \, , 
    \end{equation}
where we have ignored constant shifts. Further, it can be shown than that 
\begin{equation}
T_{{\cal D}_{(2,1)}, {\cal D}_{(2,1)} } =  g_{k+1} g_k    \, , 
\end{equation}
is the local translation operator for the above Hamiltonian.
\par
One can also show that the defect Hamiltonian with two impurities, i.e. 
\begin{equation}
\begin{split}
    &H^{k,k+1, l, l+1}(\tu_k, \tu_l) :=    - \sum_{_{\substack{i=0 \\ i \neq k, k + 1, l , l + 1}}}^{2L - 1} \left( - \cot \gamma +  \frac{1}{ \sin \gamma} e_i\right)  -    \left( -\cot \gamma + \frac{1}{\sin \gamma} R_{k}(\tilde{u}_k)^{-1}e_{k + 1} R_{k}(\tilde{u}_k) \right)  \\
&- \left( -\cot \gamma + \frac{1}{\sin \gamma} R_{l}(\tilde{u}_l)^{-1}e_{l + 1} R_{l}(\tilde{u}_l) \right) - \left( - \cot(\gamma - \tilde{u}_k ) + \frac{\sin \gamma}{\sin (\gamma + \tilde{u}_k ) \sin (\gamma - \tilde{u}_k) }e_k \right) \,  \\ 
& - \left( - \cot(\gamma - \tilde{u}_l ) + \frac{\sin \gamma}{\sin (\gamma + \tilde{u}_l ) \sin (\gamma - \tilde{u}_l) }e_l \right) \, ,  
\end{split}
\end{equation}
where $l>k+1$, is unitarily equivalent to $H^{k,k+1,k+2}(\tu_k, \tu_l)$. To see this one has to use translation operator of Eq. \eqref{TvsU}, to move the defect at site $l$ close to the defect at site $k$. 

Further, we had discussed in section \ref{sec:line-op}, that line operators can be realized as a transfer matrix not just for  one  value of the spectral parameter, but in a region. The same holds for  defect Hamiltonians (or defect transfer matrices) at the isotropic point - see a more thorough discussion of this point in section \ref{BASection}. Note now that $\frac{\pi}{2} \pm \frac{\gamma}{2}$ lies in the region given in Eq. \eqref{eq:region-(2,1)-realization} and $\pm {\rm i} \infty \pm \frac{\gamma}{2}$ lies in the region given in Eq. \eqref{eq:region-(1,2)-realization}. Hence, if we set $\tu$ to i$\infty$ or $\frac{\pi}{2}$ in Eq. \eqref{eq:two-imp-ham-no-JW}, we are again fusing two $(1,2)$ and $(2,1)$ defects respectively.

\subsection{Fusion and Topological Defect Lines: the \texorpdfstring{$(1,s)$}{Lg} case.}\label{sec:fusion-Tmatrices-(1s)}

While we will justify below our claim that impurity rows/columns built out of the fused Boltzmann weights provide lattice regularization of topological defect lines of higher types (e.g., $s>2,r>2$), the idea underlying the claim is simply that fusion on the lattice should be related with fusion in the continuum. Obviously this cannot work without {\sl some} caveats -  
for instance, our proposal for the fundamental defects $(1,2)$ and $(2,1)$ is based on the same fundamental faces with different defect parameters, but it is known that fusion in the continuum leads to  different fusion algebras for either defect types, i.e. $(r,1)$ and $(s,1)$. 

Fusion of $(r,1)$ lines will occupy us in sections below. Fusion of $(1,s)$ lines in contrast is in fact quite straightforward, since our construction leads to $l$TDL - i.e. objects which are already topological on the lattice. If that is the case, it is only reasonable to expect that their lattice fusion should also match the continuum fusion, something we now address.

In this subsection we shall discuss how fused transfer matrices can be used to realize the  $(1,s)$ type of defect line operators on the lattice by  using the fusion hierarchy  \eqref{eq:fusion-Th}. 

We first note the special case 
\begin{equation}\label{eq:spec_case_fusion}
T^{(p-1)}_{\left[\frac{p-2}{2}\right]} = (-1)^{p \R} \, T^{(0)}_{\left[ p-\frac{1}{2} \right]} \mathcal{R} \, , 
\end{equation}
where $\mathcal{R}$ is the height reflection operator\footnote{In \cite{Bazhanov:1989yk,Klumper:1992vt}, the height reflection operator is denoted by $Y$}, defined by
\begin{equation}\label{def:heightreflection}
\begin{split}
    \bra{b} \mathcal{R} \ket{a} & := \prod_{j = 0}^{2\R-1} \delta_{a_j, \,p + 1 - b_j} \, ,  \\
    \text{ i.e.  }  \mathcal{R} \ket{a_0, a_1, \ldots, a_{2\R - 1}} & =  \ket{p+1 - a_0, p+1 - a_1, \ldots , p+1 - a_{2\R - 1}  } \, .
\end{split} 
\end{equation}
The relation in Eq. \eqref{eq:spec_case_fusion} is a consequence of the definition in Eq. \eqref{eq:Wt-Th-V}. Using Eq.\eqref{eq:phase-renorm-q-t} and \eqref{eq:spec_para_inf_Y}, we obtain  we have the following relation
\begin{equation}\label{eq:T-Y-div-fac}
   \lim_{\tilde{u} \to {\rm i} \infty} \left( (-q)^{\frac{1}{2}} \frac{\sin \gamma}{\sin(\gamma - \tilde{u})}\right)^{2\R} T \left(\tilde{u} \right)=  Y := Y_{\frac{1}{2}} \, ,
\end{equation}
where a renormalization factor has been introduced to obtain $Y$.  Similarly, for higher fusion matrices, we can define the following operators
\begin{equation}\label{eq:prop-fac-high-fus}
      Y_{\frac{k}{2}} :=  \lim_{\tilde{u} \to {\rm i} \infty} \left( (-q)^{\frac{k}{2}} \frac{\sin\gamma }{\sin \left( \frac{k+1}{2}\gamma - \tilde{u} \right)}\right)^{2\R}  T^{(k)}_{[0]} \, , 
\end{equation}
where $k \leq p-1 $ , and 
\begin{equation}\label{eq:Y-rel-reflec}
Y_{\frac{p-1}{2}} = \mathcal{R} \, ,
\end{equation}
 which can be seen by substituting Eq. \eqref{eq:spec_case_fusion} into Eq. \eqref{eq:prop-fac-high-fus}. These operators satisfy the following relations
\begin{equation}\label{eq:fusion-Y-TL}
    \begin{split}
    Y_{1} &= Y_{\frac{1}{2}}^{2} - 1  \, , \\
    Y_{\frac{3}{2}} &= Y_{\frac{1}{2}}^{3} - 2Y_{\frac{1}{2}}  \, , \\
    Y_{2} &= Y_{\frac{1}{2}}^{4} - 3Y_{\frac{1}{2}}^2 + \mathbb{1}  \, , \\
    \ldots & = \ldots  \, , 
\end{split}
\end{equation}
The above relations can be proven using the fusion Eq. \eqref{eq:fusion-Th}. To see this,  let us first set $J = 1$ and take $\tu \to {\rm i} \infty$
\begin{equation}\label{eq:fus-1}
   \lim_{\tilde{u} \to {\rm i} \infty}  T\left(\tilde{u} + \frac{\gamma}{2}\right)\,  T\left(\tilde{u} -  \frac{\gamma}{2}\right) = \lim_{\tilde{u} \to {\rm i} \infty} \left(\frac{\sin \left(\tilde{u} + \frac{\gamma}{2} \right) \sin \left(\tilde{u} - \frac{3\gamma}{2} \right)}{ \sin^2 \gamma} \right)^{2\R} + \lim_{\tilde{u} \to {\rm i} \infty}\left(\frac{ \sin \left(\tilde{u} - \frac{\gamma}{2} \right)}{ \sin \gamma} \right)^{2\R} T^{(2)}_{[0]} \, .
\end{equation}
If we multiply the above equation by $\left((-q) \frac{\sin^2 \gamma}{\sin(\gamma/2 - \tilde{u}) \sin(3\gamma/2 - \tilde{u} )}\right)^{2\R}$, we get
\begin{equation}
  Y_{\frac{1}{2}}^2 = \mathbb{1} +    \lim_{\tilde{u} \to {\rm i } \infty}\left( (-q) \frac{\sin \gamma}{\sin \left( \frac{3 \gamma}{2} - \tilde{u}\right)} \right)^{2\R} T^{(2)}_{[0]} = \mathbb{1} + Y_{1} \, .
\end{equation}
Similarly, the second equation in Eq. \eqref{eq:fusion-Y-TL} can be obtained by setting $J = 2$ and $\tilde{u} \to {\rm i }\infty$
\begin{equation}\label{eq:fus-2}
\begin{split}    
    & \lim_{\tilde{u} \to {\rm i } \infty} T^{(1)}_{[1]} \, T^{(2)}_{[-\frac{1}{2}]} = \lim_{\tilde{u} \to {\rm i } \infty} \left( \frac{\sin (\tilde{u} + \gamma)}{\sin \gamma} \right)^{2\R} T^{(1)}_{[-1]} + \lim_{\tilde{u} \to {\rm i } \infty} \left(\frac{\sin \tilde{u}}{\sin \gamma}\right)^{2\R} T^{(3)}_{[0]}  \, ,\\
    \implies & Y Y_{1} = Y +  \lim_{\tilde{u} \to {\rm i } \infty} \left( (-q)^{\frac{3}{2}} \frac{\sin \gamma}{\sin(2 \gamma - \tilde{u})} \right)^{2\R} T^{(3)}_{[0]} = Y + Y_{\frac{3}{2}} \, ,
\end{split}
\end{equation}
where to get the second equation above we have multiplied both sides by $\left((-q)^{\frac{3}{2}}\frac{\sin^2 \gamma}{\sin(-\tilde{u}) \sin(2 \gamma - \tilde{u})}\right)^{2\R}$. 
As a final example, let us do the fusion for $J = 3$ 
\begin{equation}\label{eq:fus-3}
\begin{split}
&        \lim_{\tilde{u} \to {\rm i } \infty}  T^{(1)}_{[\frac{3}{2}]} \, T ^{(3)}_{[-\frac{1}{2}]}  = \lim_{\tilde{u} \to {\rm i } \infty} \left( \frac{\sin \left( \tilde{u} + \frac{3 \gamma}{2}\right)}{\sin \gamma}\right)^{2\R} T^{(2)}_{[-1]} + \lim_{\tilde{u} \to {\rm i } \infty}\left( \frac{\sin \left( \tilde{u} + \frac{ \gamma}{2}\right)}{\sin \gamma}\right)^{2\R}T^{(4)}_{[0]}  \, , \\
\implies & Y \, Y_{\frac{3}{2}} = Y_{1} + \lim_{\tilde{u} \to {\rm i } \infty}\left( (-q)^2 \frac{\sin \gamma}{\sin \left( \frac{5 \gamma}{2} - \tilde{u} \right)} \right)^{2\R} T^{(4)}_{[0]} = Y_1 + Y_{2} \, ,
\end{split}
\end{equation}
where we multiply both sides by $\left((-q)^2 \frac{\sin^2 \gamma}{\sin (- \gamma/2 - \tilde{u} )\sin (5 \gamma/2  - \tilde{u} ) }\right)^{2\R}$ to get the second equation from first.

Let us first analyze the case of A$_3$ RSOS, using equations \eqref{eq:fus-1} and \eqref{eq:fus-2} we get 
\begin{equation}\label{eq:fusion-TFI}
    \begin{split}
       Y^2 &= \mathbb{1} + \mathcal{R}  \, ,\\
       Y \mathcal{R}  &= Y  \, ,
    \end{split}
\end{equation}
where we have used $T^{(3)} = 0 $ for A$_3$ RSOS - Eq. \eqref{eq:fusion-T-M-def}. Similarly, we can study the case when $p = 4$, then using \eqref{eq:fus-1}, \eqref{eq:fus-2}, and \eqref{eq:fus-3}, we observe 
\begin{equation}\label{eq:fusion-TCI}
    \begin{split}
        Y^2 & = \mathbb{1} + Y_{1} \, , \\
        Y \, Y_{1} & = Y + \mathcal{R} \, ,  \\
        Y \mathcal{R} & = Y_1 \, . 
    \end{split}
\end{equation}
 The relations in Eq. \eqref{eq:fusion-Y-TL} were also observed in \cite{Belletete:2017gwt}, where $Y_{k}$'s were called higher spin topological defects. Like $Y_k$, we can define $\overline{Y}_{k}$ using the transfer matrix at spectral parameter $- {\rm i} \infty$
\begin{equation}
          \overline{Y}_{\frac{k}{2}}  =  \lim_{\tilde{u} \to \, - {\rm i} \infty} \left( (-q)^{-\frac{k}{2}} \frac{\sin\gamma }{\sin \left(\frac{k+1}{2} \gamma - \tilde{u}\right)}\right)^{2\R}  T^{(k)}_{[0]} \, 
 \,  . 
\end{equation}
As mentioned before, it turns out that $Y_{\frac{1}{2}} = \overline{Y}_{\frac{1}{2}}$ for A-type RSOS model - an identity  we show in \autoref{sec:anyon-chain-RSOS}. From this it follows  that $Y_k = \overline{Y}_{k}$, since both of these families have the same fusion rules.

While the relations in Eq. (\ref{eq:fusion-Y-TL}) are very satisfactory, they do not translate immediately to corresponding relations for $l$TDL in finite size. This is because of the issue of the sign of eigenvalues (and finite part of the lattice momentum) discussed earlier, in particular in section \ref{specialsubsec}.

In \autoref{sec:Y-12-op} we saw that $\uR^{-1}Y$ is the lattice realization of the $(1,2)$ line operator. This also agrees with our earlier discussion that $\uR^{-1}T(\frac{\gamma}{2} + \tu)$ can be used to realize the (1,2) line operator, as $\tu = -\frac{\gamma}{2} + {\rm i } \infty$ satisfies the conditions in Eq. \eqref{eq:region-(1,2)-realization}. The fusion relations for the $Y_k$ operators then become the fusion rules of the $(1,s)$ line operators  since parity of the number of $\uR$ insertions is the same in the left and right hand sides of  Eq. \eqref{eq:region-(1,2)-realization}, and since the extra factors of $\uR^{-2}$ all go to one in the scaling limit. This confirms that we can identify $Y_k$ as the lattice realization of $(1,2k+1)$ Verlinde line - $\widehat{\cal{D}}_{(1,2k+1)}$
\begin{equation}\label{eq:higher-k-Y-(1k)}
    \widehat{{\cal D}}_{(1,k + 1)}^{(\rm latt)} = \widehat{{\cal D}}_{\overline{(1,k + 1)}}^{(\rm latt)} = \uR^{-k}Y_{\frac{k}{2}} = \uR^{-k}\overline{Y}_{\frac{k}{2}} \, .
\end{equation}  
Note that in these equations it is $\uR$ and not $\uuR := (-1)^{\R} \tau$ that appears: the $(-1)^\R$ has been absorbed in the definition of $Y$, which is adopted here  for historical reasons. Fusion of these operators is almost exact on the lattice. Indeed, although relations (\ref{eq:fusion-Y-TL}) definitely hold in finite size, the same is not true when we insert the relevant powers of $\uuR$. Of course, parities of these powers on the left and right hand sides match, so the correct fusion relations hold in the scaling limit (this is discussed in more detail in section \ref{continuum} below). 

We can in fact render  this fusion exact on the lattice  as well with some small modifications. We can for instance decide to focus only on states with vanishing finite value of the lattice momentum, as we did in our earlier paper  \cite{Sinha:2023hum}. Then, for those states, the prescrption becomes Eq. (\ref{eq:higher-k-Y-(1k)}) without any $\uR$ factor and we truly have an exact realization of the topological defects even in finite size. Note also that the factor of $\uR$ is not necessary when $k$ is even, as then its contribution is always equal to one in the scaling limit. However, if we wish to write a general prescription valid for all defects and all states in the lattice model, we are forced to use Eq. (\ref{eq:higher-k-Y-(1k)}), with the associates slight unpleasantness.

Note that the fusion in Eq. \eqref{eq:fusion-TFI} and \eqref{eq:fusion-TCI}, and in general for any RSOS model, always terminates with the height reflection operator, $\mathcal{R}$. For $\mathcal{M}(p + 1, p)$  CFT, whose lattice realizations are A$_p$ RSOS models, the reflection operators are the lattice discretization of $(1,p)$ Verlinde line. Using the definition in \eqref{def:heightreflection}, it is not hard to show that 
\begin{equation}\label{invertibledef}
    \mathcal{R}^{2} = \mathbb{1} \ ,
\end{equation}
which is also satisfied by $\widehat{\cal{D}}_{(1,p)}$, an invertible topological defect line. $\mathcal{R} = Y_{\frac{p-1}{2}}$ is the lattice lattice realization of this invertible TDL. Note that we also have 
\begin{equation}
\widehat{\cal D}^{(\rm latt)}_{(1,p)}=\tau^{-(p-1)} {\cal R} \ .
\end{equation}

\section{Bethe-ansatz and Defects} \label{BASection}

The integrability of the  lattice regularizations of defect lines gives access to powerful tools to study the scaling limit. Eventually, this will allow comparison with results from conformal field theory, be it the spectrum of the theory in the presence of TDLs, or the fusion of these TDLs. 

In view of the current  state of the quantum integrability toolbox, many approaches are possible at this stage, from numerical solutions of the bare Bethe-ansatz equations \cite{PhysRevLett.58.771} to considerations involving the relationship with ordinary differential equations (the ODE-QISM correspondence, see e.g. \cite{GEHRMANN2024116624} and references therein). In this paper, our main emphasis will be on using  Bethe ansatz calculations in the form of Non-linear Integral equations (NLIE). Those we shall need are either in the literature \cite{Klumper:1992vt, DestriDeVega1992}, or can easily be derived using known techniques (see below), so we will be brief about this. 

The philosophy behind the NLIE approach is that one can solve functional relations for the eigenvalues or a set of auxiliary functions by exploiting analyticity properties implied by the Bethe ansatz or the fusion relations. It so happens that the analyticity of some functions severely constrains the possible solutions to the functional relations, from which one may determine what one may call elementary analytical factors \cite{Tavares:2023nma}. The crucial advantage of such an approach is that the system-size enters the equations as a parameter, in contrast with ``ordinary'' direct Bethe ansatz calculations where it defines e.g.  the number of Bethe roots for the  charge sector under study. The drawback of the approach is that  it rapidly becomes impossible to analyze all analytical properties for the desired eigenvalues, e.g. those corresponding to the low-lying excitations of the Hamiltonian or sub-leading eigenvalues of the transfer matrix. In this respect, as  will become evident later, the crossed channel is the easiest to study.  Since we are only interested here in  rational theories, all we have to do to identify the defects is to consider the (finite) set of eigenvalues  of the $\widehat{\cal{D}}$ operators. In contrast, in the direct channel, the proper identification of the defect Hamiltonian $H_{{\cal D}}$ (and associated defect Hilbert space) requires, in principle, the study of leading exponents and infinite towers of the Virasoro algebra - a more daunting task in practice. In view of this, we shall only  briefly discuss direct channel results here, and focus mostly on the crossed channel.

\subsection{The Direct Channel}\label{subsec:direct-bethe}

As mentioned earlier, we shall mostly use the technique of Non-Linear Integral Equations (NLIE) which is summarized below when we analyze the cross-channel. We do however start here with the direct channel in order to match the general logic of this paper. 

We start by considering the regime in the``vicinity" of the isotropic transfer matrix, but with a defect, that is $T(\{{\gamma\over2},{\gamma\over 2}+\tilde{u}\}
_k)$. We discuss the Hamiltonian limit at the end of this subsection.
It is thus convenient to perform a redefinition of the spectral parameters by setting \footnote{We emphasize  that $v_I$ here is the same as in \cite{tavares2024}.} 
\begin{equation}\label{eq:u-to-v-conv}
u= {\rm i} \vbu +\frac{\gamma}{2}, ~~~\tilde{u}= {\rm i} \vd \, . 
\end{equation}
so the isotropic point per se corresponds to $\vbu=0$. In what follows we use the following notation -  
\begin{equation}\label{eq:eig-def-tmat-lamb}
\Lambda\left(\{\vbu,\vbu+\vd\} \right) =\text{eigenvalue of}\, {(-1)}^L~T\left(\{u,u+\tilde{u}\}_k \right) \, . 
\end{equation}

In terms of Bethe ansatz, the  eigenvalue expression for a single impurity of spin $J$ is given by
\begin{equation}\label{eq:bethe_eig_timp}
\Lambda(\{\vbu,\vd\})= \lambda_1(\{\vbu,\vd\})+\lambda_2(\{\vbu,\vd\}),
\end{equation}
\begin{equation}
\lambda_{j}(\{\vbu,\vd\})= {\rm e}^{ \im \eta (3-2 j)}\Phi_{\left[{1\over 2}-(j-1)\right]}\left(\vbu \right) \phi_{\left[J({1\over 2}-(j-1))\right]}(\vbu+\vd) \frac{q_{[2 j-3]}(\vbu)}{q(\vbu)},
\label{lambdapart}
\end{equation}
where
\begin{equation}
q(v)= \prod_{i=1}^{2L \over 2} \sinh (v-v_i), ~~ \Phi(v)= {\left(\sinh v\over \sin \gamma\right)}^{2L-J},~~ \phi(v)= {\left(\sinh v\over \sin \gamma\right)}.    %
\end{equation}
The Bethe roots $v_n$ are fixed through the Bethe equations
\begin{equation}
{\rm e}^{2 \im \eta} \frac{\Phi_{[{1\over 2}]}(v_n) \phi_{[\frac{J}{2}]}(v_n+\vd) q_{[-1]}(v_n)}{\Phi_{[-{1 \over 2}]}(v_n) \phi_{[-\frac{J}{2}]}(v_n+\vd) q_{[1]}(v_n)}=-1 \, ,\qquad n=1,\ldots,{2L \over 2}.
\end{equation}
For the A$_p$ models, the parameter $\eta$ runs over the exponents of the corresponding diagram 
\begin{equation}
\gamma=\frac{\pi}{p+1}, \qquad \eta = \gamma l,~~\qquad l=1,~2,\ldots,p \,.\label{exponentsimp}
\end{equation}
Compared to the Bethe ansatz of the six-vertex model (resp. the XXZ chain), the RSOS expression corresponds to zero flux of arrows (resp. vanishing magnetization) and several values of twist parameter $\eta$, corresponding to what we call  sectors. Note that  periodic boundary conditions $(\eta=0)$ for the underlying six-vertex model (resp. XXZ spin chain) are not encountered \cite{Andrews:1984af,Bazhanov:1989yk}. 

In the direct channel, the form of the NLIE’s generically depends both on the excitation and the type of impurity. We shall content ourselves with an impurity based on the fundamental representation ($J=1$) with varying defect parameter. As we have already seen, this is sufficient to cover both generators of the defect fusion algebra, namely the defects of types $(1,2)$ and $(2,1)$. Moreover, we will focus on ground-state results, which are sufficient to distinguish the two  cases, and will give us further confirmation of the lattice/discrete regularizations identification of defect lines.

\subsubsection{(1,1) defect}

One may write the eigenvalues  of the transfer matrix  in terms of the Y-system (defined in what follows) as:
\begin{multline}
\log \frac{\Lambda(\{\vbu,\vbu+\vd\})}{\Lambda_{\infty}}= (2L-1) e_0^{(1)}(\vbu)+e_0^{(1)}(\vbu+\vd)+  \sum_{\Theta \in 1}\log\left[\frac{{\rm e}^{(p+1) (\vbu-\Theta)} -1}{{\rm e}^{(p+1) (\vbu-\Theta)} +1 }\right]         \\ +\frac{1}{2 \pi {\rm i}}\int_{-\infty}^{\infty} \log\left[\frac{{\rm e}^{(p+1) (\vbu-s)} -{\rm i}}{{\rm e}^{(p+1) (\vbu-s)} +{\rm i} }\right] {\log}' \Y^{(1)}(s) {{\rm d}}s, \label{EigImpurity}
\end{multline}
where the  $e_0$ functions 
\begin{equation}
e_0^{({J})}(v)=
-\int {\rm e}^{\im k v} \left[\frac{  \cosh \left( \frac{k \pi}{2(p+1)} \left(p-J \right)\right) -\cosh {\pi k \over 2 (p+1)}}{2 k \sinh\frac{k \pi}{2} \cosh \frac{k \pi }{2 (p+1)} }\right]{\rm d}k+\log \frac{\cosh(v)}{\sin \gamma}, \label{bulk}
\end{equation}
give the extensive as well as the non-extensive non-universal contributions to the eigenvalue. The constant $\Lambda_{\infty}$ is the asymptotic limit $\lim_{v \to \infty} \Lambda(\{v,v+\vd\})/{(\cosh^{2L-1}(v) \cosh(v+\vd))}$. 
One can use this limit value to obtain the twist sector of Bethe Ansatz expression (\ref{eq:bethe_eig_timp}-\ref{exponentsimp}), so it can be used to define it.
For the models of interest, it takes values  
\begin{equation} \label{quantum2}
\Lambda_\infty= [2]_{{q}^{l}}=2\cos{l\pi\over p+1}~,
\end{equation}
with $l=1,~2,\ldots, p$. We also have introduced the short-hand notation $\Theta \in J$ to refer to the set of all zeros (generically denoted by $\Theta$) of the fused eigenvalue of spin $J$ such that $|\Im \Theta|< \gamma/2$. Alternatively, one may say they are zeros to the equations $\Y^{(J)}\left(\Theta \pm i \gamma/2\right)=0$. Finally,  we also used the short-hand notation  ${\log'} \Y^{(J)}(v) = {(\Y^{(J)}(v))}^{-1}~{\Y^{(J)}}'(v)$. 

The $\Y^{(J)}$ functions themselves  are solutions to the following set of NLIE’s:
\begin{multline}
    \log \frac{\y^{(J)}(v)}{\y^{(J)}(\infty)} =\delta_{J1} \left( (2L-1) \log \left[\frac{{\rm e}^{(p+1)v}-1}{{\rm e}^{(p+1)v}+1}\right] +\log \left[\frac{{\rm e}^{(p+1)(v+\vd)}-1}{{\rm e}^{(p+1)(v+\vd)}+1}\right] \right) + \\ \sum_{M} G_{JM}^{[p-2]} \left[ \sum_{\Theta \in M}\log\left[\frac{{\rm e}^{(p+1) (v-\Theta)} -1}{{\rm e}^{(p+1) (v-\Theta)} +1 }\right]          + \frac{1}{2 \pi i}\int_{-\infty}^{\infty} \log\left[\frac{{\rm e}^{(p+1) (v-s)} -i}{{\rm e}^{(p+1) (v-s)} +i }\right] {\log'} \Y^{(M)}(s) {\rm d}s\right], \ \label{NLIEDirect}
\end{multline}
where $G^{[p-2]}$ is the adjacency matrix of $A_{p-2}$ diagram\footnote{We shall denote $G^{[p]}$ the adjacency matrix of $A_p$ diagram. Notice the slightly different notation with regards to $G^{(J)}$. The latter refers to the fused adjacency matrix when no misinterpretation about the fundamental one $(J=1)$ is possible.} and the lower-case function is
\begin{equation}
\y^{(J)}(v)=\Y^{(J)}(v)-1 \, .
\end{equation}
Again, the asymptotic limit  $\y^{(J)}(\infty)$ defines the twist sector. For example, for the ground-state we have $\y^{(J)}(\infty)= \frac{\sin (J-1) \gamma ~\sin(J+1)\gamma}{\sin^2 \gamma}$ and no zero $\Theta$ appear in the summations above. More generally, for the  different twist sectors labeled by $l=1,\ldots,p$ we have $\y^{(J)}(\infty)= \frac{\sin l (J-1) \gamma ~\sin l(J+1)\gamma}{\sin^2 l \gamma}$.

It is possible to extend expression (\ref{EigImpurity}) from $\vbu,~\vbu+\vd$ on the real line to a region in the complex plane, say $|\Im v_B|\leq\gamma/2$, thanks to the Cauchy theorem which allows the deformation of the contours of  integration to parallel lines, as long they do not sweep other zeros of the eigenvalue function. In this case, the convolution terms can also be shifted into the complex plane.

The finite-size corrections to the eigenvalue, especially the corrections of order $O(L^{-1})$, require studying the scaling behavior. For example, defining $\y^{(J)}_{\pm }(v)=\y^{(J)}( v \pm {\log 2 L \over p+1})$ produces  equations that are independent of the system-size  in the large $L$ limit, and correspond to the left and right moving sectors of the CFT:
\begin{multline}
    \log \frac{\y_{\pm}^{(J)}(v)}{\y^{(J)}(\pm\infty)} =-2\delta_{J1} {\rm e}^{-(p+1)|v|} +  \sum_{M} G_{JM}^{[p-2]} \Bigg[ \sum_{\Theta_{\pm} \in M}\log\pm\left[\frac{{\rm e}^{(p+1) (v-\Theta_{\pm})} -1}{{\rm e}^{(p+1) (v-\Theta_{\pm})} +1 } \right] \\         + \frac{1}{2 \pi i}\int_{-\infty}^{\infty} \log\pm \left[\frac{{\rm e}^{(p+1) (v-s)} -i}{{\rm e}^{(p+1) (v-s)} +i }\right] {\log}' \Y_{\pm}^{(M)}(s) {\rm d}s\Bigg], \ \label{scaledNLIE}
\end{multline}
where $\Theta_{\pm}=\Theta\mp {\log 2 L \over p+1}$ refers only to a subset of zeros for which $\Theta_{\pm}$ results in a finite number as $L\to\infty$.  We shall refer to (\ref{scaledNLIE}) as the scaled version of the NLIE.  One can see that different behaviors exist when studying excitations. If $\Theta$ does not scale appropriately with the system-size, either it does not affect $O(L^{-1})$ corrections, or it does only modify twist sectors for right/left movers. To see this, one may separate the different contributions in (\ref{EigImpurity}). We have
\begin{multline}
\log \Lambda(\{\vbu,\vbu+\vd\}) \approx\\ (2L-1) e_0^{(1)}(\vbu)+e_0^{(1)}(\vbu+\vd)+  \sum_{\Theta \in ~1, ~ ns<}\log\left[\frac{{\rm e}^{(p+1) (\vbu-\Theta)} -1}{{\rm e}^{(p+1) (\vbu-\Theta)} +1 }\right]+ \log {(-1)^{\nu_+}} {\mbox{sgn}~ \Lambda_{\infty}}        \\ +{\log \Y_+^{(1)}(-\infty) \over 2}+ \frac{2\pi }{2 L} {\rm e}^{(p+1) \vbu} \left[\frac{1}{2 \pi^2 }{\rm }\int_{-\infty}^{\infty} {\rm e}^{-(p+1)s} \log' \Y_+^{(1)}(s) {\rm d}s-\frac{1}{\pi}\sum_{\Theta_+\in ~1}{\rm e}^{-(p+1) \Theta_+}\right]\\-\frac{2\pi}{2L} {\rm e}^{-(p+1) \vbu}\left[ \frac{1}{2 \pi^2 }{\rm }\int_{-\infty}^{\infty} {\rm e}^{(p+1)s} \log' \Y_-^{(1)}(s) {\rm d}s 
+\frac{1}{\pi}\sum_{\Theta_-\in ~1}{\rm e}^{(p+1) \Theta_-} \right],\label{EigImpurityScaled}
\end{multline}
where $\nu_+$ are the number of positive zeros which render  $\Theta_+=\Theta-{\log 2 L\over p+1}$ either finite or $+\infty$ as $L\to\infty$. By means of the notation $\Theta \in 1,ns<$ we refer to ``slow'' non scaling zeros, e.g. 
  the positive zeros  such that $\Theta-{\log 2 L\over p+1} \to -\infty$ as $L\to\infty$. In the present case, the only slow non scaling  zeros  are such that $|\lim_{L \to \infty} \Theta| < \infty$, so no subleading corrections of order $L^{-\alpha}$ with $0<\alpha<1$ exist. Therefore we may identify
 \begin{equation}
 {c\over 24}-h_{\pm}-N_{\pm}= \pm\frac{1}{2 \pi^2 }{\rm }\int_{-\infty}^{\infty} {\rm e}^{\mp(p+1)s} \log' \Y_{\pm}^{(1)}(s) {\rm d}s 
-\frac{1}{\pi}\sum_{\Theta_{\pm}\in ~1}{\rm e}^{\mp(p+1) \Theta_{\pm}}. \label{Eexponents}
 \end{equation} 

 The critical exponents depend only on the solutions of the scaled version of the NLIE's.  Despite our slightly different presentation,  formula (\ref{Eexponents}) is essentially equivalent to results in \cite{Bazhanov:1989yk,Klumper:1992vt} for the homogeneous case. There, only scaling zeros appear (when they do) for sub-leading eigenvalues.

 It turns out that the right/left sectors of  the equations have exactly the same form as those of $\vd=0$, and are not affected by the changes in the impurity contribution to the driving term.  This means that,  when  $\vbu=0$ (i.e. the isotropic case), any $\vd$  such that  $|\Im \vd|<\gamma$ and $|\Re \vd|< \infty$, the critical exponents in the non-homogeneous case are exactly the same of the   homogeneous model. For any $\vbu$ such that $|\Im \vbu|<\gamma/2$, the same holds provided $|\Im(\vbu+\vd)|<\gamma$.  In more general circumstances, i.e. arbitrary $\vd$ and $J\geq1$, the impurity contribution to the driving-term also vanishes and  the sole difference,  compared to the homogeneous case, is  a possible recombination between right/left sectors due to possibly  different analyticity structures, here encoded by the behavior of zeros $\Theta_{\pm}$ \footnote{ See the case $J=1$ below, where these different behaviors occur in other domains of $\vd$, sometimes along with the presence of non-scaling zeros.}.
 
 Formally, (\ref{Eexponents}) holds for an arbitrary impurity of parameter $\vd$ and spin $J$.  Since the zeros satisfy the subsidiary conditions $\Y^{(J)}(\Theta^{(J)}\pm i \gamma/2)=0$, their precise positions do not matter:  due to expressions like $\int_{\Y^{(J)}({\cal C})}  \frac{{\rm d}w}{w}$ all that   matters is the homotopy class of the path ${\cal C}$ encircling  the points $\Theta  \pm i \gamma/2$ associated only to scaling zeros. More precisely, extra driving terms due to these zeros can be converted into  convolution terms by adding extra closed paths encircling them.

Finally, the exponents (\ref{Eexponents}) can be calculated using di-log trick and dilogarithms identities\cite{Bazhanov:1989yk, Klumper:1992vt, Kirillov95}.

To conclude this discussion, {\sl no explicit and continuous dependence on the parameter $\vd$ is found}. A thorough numerical study leads us to conclude that for $|\Im \vd|<\gamma$ and $|\Re \vd|< \infty$, the critical exponents are precisely the same as in the homogeneous case. In particular, the ground-state corrections, where no zeros are present, are consistent with $N_{\pm}=h_{\pm}=0$ and $c=1-6/p(p+1)$. In other words, introducing a column of modified tiles with $\tilde{u}=i\vd$ and $|\Im \vd|<\gamma$ and $|\Re \vd|< \infty$ gives rise to the same continuum limit as the homogeneous system.

Now let us consider what happens when $\vd$ does {\sl not} belong to this domain.

\subsubsection{(1,2) defect}

Sending $\vd \to -\infty $, we expect to recover the twisted partition function (\ref{eq:twist-part-func}) with $(r,s)=(1,2)$. For the $A_{3},~A_{4},~A_{5}$ models these are explicitly given by \cite{Petkova:2000ip}
\begin{equation}
Z_{{\cal D}_{(1,2)}}^{A3}(\tau, \bar{\tau}) = \chi_{(1,1)} \bar{\chi}_{(2,2)}+ \chi_{(2,2)} \bar{\chi}_{(1,3)} +\mbox{c.c.} \, , \nonumber
\end{equation}
\begin{equation}
Z_{{\cal D}_{(1,2)}}^{A4}(\tau, \bar{\tau}) = \chi_{(2,2)} \bar{\chi}_{(2,2)}+ 
\left(\chi_{(1,1)} \bar{\chi}_{(3,3)} +\chi_{(2,2)} \bar{\chi}_{(2,4)} +\chi_{(3,3)} \bar{\chi}_{(1,3)} +\chi_{(3,1)} \bar{\chi}_{(1,3)} +\mbox{c.c.}\right) \,  , \nonumber
\end{equation}
\begin{multline}\label{eq:A-5-twst-part-fnc}
Z_{{\cal D}_{(1,2)}}^{A5}(\tau, \bar{\tau}) = \chi_{(1,1)} \bar{\chi}_{(4,4)}+ \chi_{(2,2)} \bar{\chi}_{(3,3)} +\chi_{(3,3)} \bar{\chi}_{(2,4)} +\chi_{(4,4)} \bar{\chi}_{(1,3)} +\chi_{(1,3)} \bar{\chi}_{(4,2)} +  \\
\chi_{(2,4)}  \bar{\chi}_{(3,1)} +\chi_{(4,2)} \bar{\chi}_{(1,5)} +\mbox{c.c.} \,  ,
\end{multline}
where we have chosen to write characters in terms of Kac indices with the same parity, i.e. covering half of the exponents in Kac's table in a checkerboard manner. In general, left and right sectors now belong to different representations. For instance, the left $(1,1)$ chiral field appears combined with the right one $(p-1,p-1)$.

In terms of the NLIE's,  expressions for $\vd \to - \infty$ are not formally  different from those for  $\vd$ finite in the scaling limit (\ref{scaledNLIE},~\ref{EigImpurityScaled}). Nevertheless, if one follows  continuously  $\vd$ for the previous ground-state (or largest eigenvalue), ones  finds that a pair of zeros of the eigenvalue eventually enter the analytical strip, where they collide exactly on the real axis $\Im v =0$, and then depart from each other, with one tending to the origin, while the other is dragged along with the variation of $\vd$, i.e. $\Theta \to \infty$ when $\vd \to -\infty$.
It follows that the exponents are still given by (\ref{Eexponents}), but the number of zeros seems to be modified. 
It happens, however, that  none of these two zeros behave as $\Theta \simeq {\log 2 L \over p+1}$, for one is slow non-scaling, while the other is fast non-scaling. Therefore, the only effect of taking $\vd\to - \infty$ is the modification of the twist sector, say for the right movers. Explicitly,  we have the right moving  sector NLIE's:
\begin{multline}
    \log \frac{\y_{+}^{(J)}(v)}{\y^{(J)}(\infty)} =-2\delta_{J1} {\rm e}^{-(p+1)v}  +\pi i ~\delta_{J2}        +\sum_M  \frac{G_{J M}^{[p-2]}}{2 \pi i}\int_{-\infty}^{\infty} \log \left[\frac{{\rm e}^{(p+1) (v-s)} -i}{{\rm e}^{(p+1) (v-s)} +i }\right] {\log'} \Y_{+}^{(M)}(s) {\rm d}s\Bigg]. \ 
\end{multline}
In consequence, the homogeneous ground-state with $h_{\pm}=0$ becomes $h_-=0$, $h_+=h_{(1,2)}=h_{(p-1,p-1)}$.
However,  the ground-state for $\vd\to -\infty$ does not correspond to the ground-state for $\vd \to 0$ when one smoothly varies $\vd$. Generally, the previous exponents correspond to an excited state. For $p=4$, the ground-state is found in the  sector $l=2$. At large but finite $\vd$, we find two zeros associated to impurity eigenvalue $J=2$, which provides extra contributions to the driving-term of the equation with $J=1$. As $\vd \to -\infty$, one zero tends to the origin of the complex plane, while the other is dragged along to $\infty$. Like the excited stated explained above, these two zeros are slow non scaling and fast non scaling, respectively. This implies that the scaling limit of the NLIE acquires an additional phase of $2 \pi \im$ (one due to fast non-scaling zero and the other due to the $\vd$ dependent driving-term), which can therefore be dropped. The scaled equations are then equivalent to the ones leading to the $(h_{(2,2)},h_{(2,2)})$ field in the homogeneous case. More generally, for $p$ even the scaled equations for the right/left moving sectors are virtually equal, and correspond to the homogeneous model with a different sector, but no scaling zeros. In consequence, we have $(h_-,h_+)=(h_{(\frac{p}{2},\frac{p}{2})},h_{(\frac{p}{2},\frac{p}{2})})$. 

\begin{figure}[htb]
\begin{center}
\includegraphics[width=0.48\linewidth]{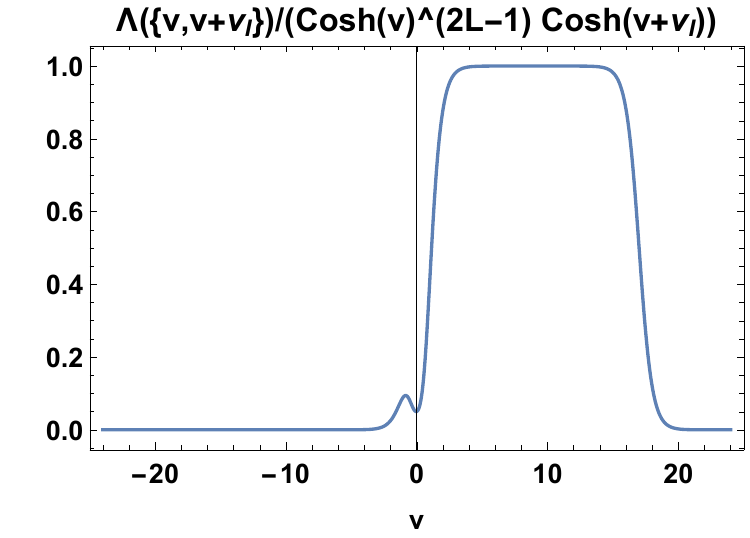}
\end{center}
\caption{ Normalized ground-state eigenvalue for $A_5$ with $(1,2)$ defect realized as $\vd \to -\infty$. Here we fixed $\vd=-17$ to highlight different asymptotics and, therefore, sectors for left/right movers.  At any finite $v_I$ left and right asymptotics are the same (finite plateau), thus the same value of $l$ is found for both type of movers. As $v_I \to -\infty$, the plateau completely emerges, effectively leading to different values of $l$. 
}
\label{DifferentAsymptotic}
\end{figure}

For $p=5$ the situation is different. We start by taking $\vd$ large but finite as in the previous case. The ground-state is found to be in sector $l=3$ and no zeros appear inside the analyticity strips. One would tend to think that the corresponding field is like the homogeneous one $(h_{(3,3)},h_{(3,3)})$. Nevertheless, the fundamental eigenvalue with $J=1$ is not a trigonometric polynomial of order $2 L$ because $[2]_{q^l}=0$, so that two zeros are at infinity. This makes the asymptotic limit $\Lambda_{ \infty}$ manifestly different for left/right movers. For instance, while for left movers it remains unchanged, one can see in figure \ref{DifferentAsymptotic} that an “intermediate” asymptotic limit exist for $0 \ll -v  \ll -\vd$ for right movers. This intermediate asymptotic corresponds to $l=2$, so when $\vd$ is sent to $-\infty$ we have the same equations as for the homogeneous model for left/right movers but with a different twist sector, say while the left sector maintains  the $h_{(3,3)}$ exponent, the right one corresponds to $h_{(2,2)}$. In general, for $p$ odd, left/right sectors result from different equations and give rise to exponents in the main diagonal of the Kac's table $(h_-,h_+)=(h_{(\frac{p+1}{2},\frac{p+1}{2})},h_{(\frac{p-1}{2},\frac{p-1}{2})})$ for the ground-state. 

These results are consistent with what one expects from the corresponding Verlinde line, i.e. the inserted defect acts by fusion on a chiral character to produce a new combination:
\begin{equation}
\phi_{(1,2)} \chi_{(r,s)}=\sum_{s'} G_{ss'}^{[p]} \chi_{(r,s')},
\end{equation}
 So, for the simplest case where $(r,s)=(1,1)$, a combination like $\chi_{(1,2)} \bar{\chi}_{(1,1)}=\chi_{(p-1,p-1)} \bar{\chi}_{(1,1)}$ is now possible. Likewise, while $\chi_{(r,r)} \bar{\chi}_{(r,r)}$ does not appear for $r<p/2$, $\chi_{(\frac{p}{2},\frac{p}{2})} \bar{\chi}_{(\frac{p}{2},\frac{p}{2})}$ does for $p$ even. Whereas for $p$ odd we find a new combination of sectors: $\chi_{(\frac{p-1}{2},\frac{p-1}{2})}\bar{\chi}_{(\frac{p+1}{2},\frac{p+1}{2})}$.

\subsubsection{(2,1) defect}
By setting the defect parameter $\vd = \pm \im \pi/2  $, we expect to recover the partition functions (\ref{eq:twist-part-func}) with $(r,s)=(2,1)$. For the $A_{3},~A_{4},~A_{5}$ models these are explicitly given by
\begin{equation}
Z_{{\cal D}_{(2,1)}}^{A3}(\tau, \bar{\tau}) = \chi_{(2,2)} \bar{\chi}_{(2,2)}+ \left(\chi_{(1,1)} \bar{\chi}_{(1,3)} +\mbox{c.c.}\right), \nonumber
\end{equation}
\begin{equation}
Z_{{\cal D}_{(2,1)}}^{A4}(\tau, \bar{\tau}) = \chi_{(1,1)} \bar{\chi}_{(2,4)} +\chi_{(2,2)} \bar{\chi}_{(3,3)} +\chi_{(2,2)} \bar{\chi}_{(1,3)} +\chi_{(2,4)} \bar{\chi}_{(3,1)} +\mbox{c.c.}, \nonumber
\end{equation}
\begin{multline}
Z_{{\cal D}_{(2,1)}}^{A5}(\tau, \bar{\tau}) = \chi_{(3,3)} \bar{\chi}_{(3,3)}+\Big(\chi_{(1,1)} \bar{\chi}_{(3,5)}+ \chi_{(2,2)} \bar{\chi}_{(4,4)} +\chi_{(2,2)} \bar{\chi}_{(2,4)} +\chi_{(3,3)} \bar{\chi}_{(1,3)} +\\\chi_{(3,5)} \bar{\chi}_{(3,1)} +  
\chi_{(2,4)}  \bar{\chi}_{(4,2)} +\chi_{(3,1)} \bar{\chi}_{(1,5)} +\mbox{c.c.}\Big). \label{twist21}
\end{multline}

Like the $(1,2)$ case, although finite $L$ equations are manifestly different, their scaling limit is formally identical  to the one of the  homogeneous problem, except  for possibly  different zero structures and twist sectors, which results in different combinations between left and right sectors. The parameter $\vd=i \pi/2$ replaces the driving-terms of (\ref{NLIEDirect}) by
\begin{equation}
\delta_{J1}  (2L-1) \log \left[\frac{{\rm e}^{(p+1)v}-1}{{\rm e}^{(p+1)v}+1}\right] +\delta_{J,p-2}~\log \left[\frac{{\rm e}^{(p+1)v}-1}{{\rm e}^{(p+1)v}+1}\right], \nonumber
\end{equation}
therefore, such a modification  is immaterial in terms of the scaling limit, since, while the term proportional to the system size gives exactly the same result as compared to $\vd=0$, the other contribution just vanishes. The important modification comes from possible additional zeros, which indeed appear.

For example, for $p=4$ we have a system of two NLIE's with an additional negative non-scaling zero for $J=2$ and a scaling positive zero for $J=1$. In consequence we get $(h_-,h_+)=(h_{(2,2)}, h_{(3,3)})$ for the ground-state. In general, for $p$ even one has $(h_-,h_+)=\left(h_{\left(\frac{p}{2},\frac{p}{2}\right)}, h_{\left(\frac{p}{2}+1,\frac{p}{2}+1\right)}\right)$. As for $p$ odd, the situation is more familiar since left/right combination for the ground-state equation already appear in the homogeneous case. For $p=5$ we find the ground-state at sector $l=3$ and a slow non scaling zero for $J=2$ eigenvalue. Therefore, the ground-state exponents are simply given by the diagonal combination $(h_{(3,3)},h_{(3,3)})$. More generally, one expects to find the ground-state at the twist sector $l = \frac{p+1}{2}$. Therefore, we get $(h_-,h_+)=\left(h_{\left(\frac{p+1}{2},\frac{p+1}{2}\right)},h_{\left(\frac{p+1}{2},\frac{p+1}{2}\right)}\right)$.  

One notices that the situation compared to the $(1,2)$ defect ground-states is reversed in terms of which values of $p$ allows for a diagonal combination of exponents. This is all consistent with the expected behavior of the corresponding Verlinde line, in this case,
\begin{equation}
\phi_{(2,1)} \chi_{(r,s)}=\sum_{r'} G_{rr'}^{[p-1]} \chi_{(r',s)}.
\end{equation}

A similar kind of analysis could be carried out  for the low-lying excited states. One must be careful, however, in identifying the proper analyticity structure like the twist sector, the zeros in the analyticity strip and their nature. The situation may become very subtle in cases where the structure is only apparent  when the system size is already large enough. This another reason to combine different approaches to identify the defects. 

 \subsubsection{From the transfer matrix to the Hamiltonian}
 
 We have so far discussed results in the vicinity of  the isotropic point $u={\gamma\over 2}$. We now need to address  what happens for the Hamiltonian. To start, we emphasize  that the results in this case do not simply follow from those  for the transfer matrix in the vicinity of $u=0$. This is because eigenenergies  of the Hamiltonian are obtained as  logarithmic {\sl derivatives} of eigenvalues of the transfer matrix, so that, in particular,  it is the ordering of these derivatives rather than  of the eigenvalues themselves that determines where levels "stand" in the spectrum. Thorough analysis shows that the strips for the different defects that we have identified in  the  case of the defect transfer matrix at $u={\gamma\over 2}$ (the bulk isotropic case) carry over to identical strips for the defect Hamiltonians. In order to limit the size of this paper,  we  postpone further discussion of this point to  appendix \ref{StripsTMH}, and  
 pass now to the crossed channel.

\subsection{The Crossed Channel}\label{sec:crossed-bethe}

In the crossed channel, the defect partition function obtained by acting repeatedly  with the fundamental transfer matrix (associated with the homogeneous Hamiltonian), except for one instance of a  different time-evolution line, which we call the defect transfer matrix - see figure \ref{fig:defect-line}. 
\begin{figure}[hbt]
     \begin{adjustbox}{max totalsize={1.2\textwidth}{1.2\textheight},center}
     \begin{tikzpicture}
           \fill[red!10] (0,2) -- (4,2) -- (4,3) -- (0,3) -- cycle ;
         \draw[black, thick] (0,0) -- (4,0) ;
         \draw[black, thick] (0,1) -- (4,1) ;
         \draw[black, thick] (0,2) -- (4,2) ;
         \draw[black, thick] (0,3) -- (4,3) ;
         \draw[black, thick] (0,4) -- (4,4) ;

         \draw[black, thick] (0,0) -- (0,4) ;
         \draw[black, thick] (1,0) -- (1, 4) ;
         \draw[black, thick] (2,0) -- (2,4) ;
         \draw[black, thick] (3,0) -- (3,4) ;
         \draw[black, thick] (4,0) -- (4,4) ;

        \node[] at (0.5,2.5) {$ {\gamma\over 2} + \tilde{u}$} ;
         \node[] at (1.5,2.5) {${\gamma\over 2} + \tilde{u}$} ;
         \node[] at (2.5,2.5) {${\gamma\over 2} + \tilde{u}$} ;
        \node[] at (3.5,2.5) {${\gamma\over 2} +  \tilde{u}$} ;

 \node[] at (0.5,0.5) {${\gamma\over 2}$} ;
         \node[] at (1.5,0.5) {${\gamma\over 2}$} ;
         \node[] at (2.5,0.5) {${\gamma\over 2}$} ;
        \node[] at (3.5,0.5) {${\gamma\over 2}$} ;

 \node[] at (0.5,1.5) {${\gamma\over 2}$} ;
         \node[] at (1.5,1.5) {${\gamma\over 2}$} ;
         \node[] at (2.5,1.5) {${\gamma\over 2}$} ;
        \node[] at (3.5,1.5) {${\gamma\over 2}$} ;

 \node[] at (0.5,3.5) {${\gamma\over 2}$} ;
         \node[] at (1.5,3.5) {${\gamma\over 2}$} ;
         \node[] at (2.5,3.5) {${\gamma\over 2}$} ;
        \node[] at (3.5,3.5) {${\gamma\over 2}$} ;

     \end{tikzpicture}

    \end{adjustbox}

    \caption{Defect line transfer matrix in the isotropic case.}
    \label{fig:defect-line}
\end{figure}

If we impose that this defect line is topological (transmissive), then it should commute with the usual time evolution. In the previous section, we presented a commutative family of defect lines which was obtained in the framework of integrable models thanks to the fusion hierarchy, Eq. (\ref{eq:fusion-Th}). 
As we have discussed, the fundamental transfer matrix, $J=1$, with a different spectral parameter ${\gamma\over 2}+\tilde{u}$ 
already introduces a defect line. We can derive the faces which comprise these transfer matrices and the fusion hierarchy among them by exploring special relations that result from the Yang-Baxter equation at singular points, see Appendix \ref{AnotherFusion}.  ~ Importantly, for the A$_p$ models the hierarchy truncates as $T^{(p)}(u)=0$ while $T^{\left(p-1\right)}(u) \propto {\cal R}$, the height reflection operator.

Due to commutativity, the functional relations among the transfer matrices (\ref{eq:fusion-Th}) imply equivalent relations among the eigenvalues, so one can derive higher-spin expressions from the fundamental one. 

In the context of building $l$TDLs, all we need in the crossed channel are the transfer matrices $T^{(J)}({\gamma\over 2}+\tilde{u})$, where we have already set $\vbu=0$. Like in the direct channel, it is convenient to use a modified parametrization setting 
\begin{equation}
\tilde{u}= {\rm i} \, \vd \, .
\end{equation}
Note the symmetry ${\gamma\over 2}+\tilde{u}\to \gamma-\left({\gamma\over 2}+\tilde{u}\right)$ becomes $v_C\to -v_C$. We now introduce the important notation 
\begin{equation}\label{signL}
\Lambda^{(J)}(\vc)=\text{eigenvalue of}\, {(-1)}^{  \R }~T^{(J)} \left({\gamma\over 2}+i\vc \right). 
\end{equation}
Similarly to (\ref{eq:fusion-T-M-def}), we introduce a short notation for the shifted function $f_{[k]}(\vc)=f(\vc+ {\rm i} \,k  \gamma)$. 

Recall that in Bethe-ansatz we parametrize the eigenvalue expression in terms of the  Bethe-roots $v_i$ 
\begin{equation}\label{eq:bethe_eig_tmat}
\Lambda^{(J)}(\vc)= \sum_{j=1}^{J+1} \lambda_j^{(J)}(\vc)  \, ,
\end{equation}
\begin{equation}
\lambda_j^{(J)}(\vc)= {\rm e}^{ \im \eta (J-2(j-1))}\Phi_{\left[{J\over 2}-(j-1)\right]}\left(\vc \right) \frac{q_{[- (J+1)/2 ]}(\vc  ) \, q_{[ (J+1)/2] }(\vc   )}{q_{\left[(J+1)/2 -(j-1)\right]}(\vc) \, q_{[(J+1)/2-j]}(\vc )}  \, ,
\end{equation}
where
\begin{equation}\label{eq:bethe-ansatz-qfunc}
q(\vc)= \prod_{i=1}^{\R } \sinh (\vc-v_i), \qquad \Lambda^{(0)}(\vc)=\Phi(\vc)= {\left(\sinh \vc\over \sin \gamma\right)}^{2 \R}.    
\end{equation}
The Bethe roots $v_n$ are fixed through the Bethe equations
\begin{equation}\label{Betheeqs}
{\rm e}^{2 \im \eta} \frac{\Phi_{[{1\over 2}]}(v_n) q_{[-1]}(v_n)}{\Phi_{[-{1 \over 2}]}(v_n) q_{[1]}(v_n)}=-1 \, ,\qquad n=1,\ldots,{\R } \, .
\end{equation}
Like in direct channel expressions, the parameter $\eta$ runs over the exponents of the corresponding diagram, Eq. (\ref{exponentsimp}), which separates the spectrum into sectors. In particular the ground-state corresponds  to $l=1$ or $l=p$. In general,  these sectors resolve the operator  content according to the algebra defined from the $l$TDL's \cite{Belletete:2018eua}.

\subsection{Generalized T-system}

It may be useful to start by a short review of the topic. In addition to the fusion hierarchy to be discussed next, other kinds of finite difference functional relations among the fused transfer matrices exist. In the context of quantum integrable spin chains they were first obtained in \cite{Klumper:1992vt} \footnote{Under the name  ``generalized inversion identities''.} as a means to explore analytical properties of the eigenvalues, an information which allows for the derivation of  non-linear integral equations(NLIEs). The similarity between these  equations and those obtained  by the Thermodynamic Bethe Ansatz (TBA) was soon noticed, even though their physical origin is quite different indeed: while in the former case one makes use of mathematical tools such as the analytical and algebraic structure of the T-system, in the latter, one seeks instead to minimize the free-energy functional in terms of n-strings distributions and the like \cite{TAKAHASHI}. On the other hand, with the advent of the Quantum Transfer Method \cite{TROTTER, MSUZUKI, KLUMPER93} it became clear  that the TBA equations could be obtained instead via the T-system for the fusion hierarchy of quantum transfer matrices\cite{Takahashi_2001}. This allowed complete  bypassing of the string hypothesis, see also \cite{DestriDeVega1992}.  Later on, the ubiquity of T-systems and Y-systems was recognized and derived for different models\cite{Kuniba_2011}. Additionally, in \cite{Krichever1997}, a connection  between quantum integrable systems and the classical theory of solitons was noted. In the latter case, the T-system features as the Hirota's bilinear relations and allows to re-derive the nested Bethe ansatz equations through the zero curvature condition and Bäcklund flow. By now there are different types of such relations for the different systems, see e.g. \cite{Tsuboi:1999ft,morin2021groundstate}.

For our purposes, we will need to derive bilinear relations in full generality so that the scaling limit of the fusion algebra can be completely analyzed.  The bilinear relations we obtain are, of course, a
generalization of the original T-system \cite{Klumper:1992vt, Kuniba:1998ygt}.

\subsubsection{Young-diagram presentation}
Following \cite{KUNIBA}, we introduce an Yangian analogue of the Young tableaux. In our case we thus  consider the fundamental representation eigenvalue as
\begin{equation}\label{fundreprep}
\Lambda^{(1)}(\vc)= \begin{ytableau}
~
\end{ytableau}_{\vc}= \begin{ytableau}
1
\end{ytableau}_{\vc}+\begin{ytableau}
2
\end{ytableau}_{\vc} \, ,
\end{equation}
with
\begin{equation}
\begin{ytableau}
1
\end{ytableau}_{\vc}
= {\rm e}^{\im \eta}\Phi\left(\vc+\frac{\im \gamma}{2}\right) \frac{q(\vc-\im \gamma)}{q(\vc)} \, , \qquad
\begin{ytableau}
2
\end{ytableau}_{\vc}
= {\rm e}^{-\im \eta}\Phi\left(\vc-\frac{\im \gamma}{2}\right) \frac{q(\vc+\im \gamma)}{q(\vc)} \, .
\end{equation}
The notation in (\ref{fundreprep}) implies that whenever one finds an empty tableau, one should sum over  all its possible fillings.
The Bethe equations (\ref{Betheeqs}) then imply that the poles associated with the Bethe roots in $q$ functions are removable, so the eigenvalue is an analytical function of the spectral parameter $\vc$. 

We can use this notation to represent higher-spin eigenvalues. For example, the spin-$2$ eigenvalue\footnote{Recall we work in units where spins $J$ are integers, so the fundamental has spin $J=1$ etc.}
\begin{equation}
\Lambda^{(2)}(\vc)= {\rm e}^{2 \im \eta}\Phi(\vc+\im \gamma) \frac{q(\vc-3\im \frac{\gamma}{2})}{q(\vc+\im \frac{\gamma}{2})}+\Phi(\vc) \frac{q(\vc-3\im \frac{\gamma}{2}) q(\vc+3\im
\frac{\gamma}{2})}{q(\vc-\im \frac{\gamma}{2})q(\vc+\im \frac{\gamma}{2})}+{\rm e}^{-2 \im \eta}\Phi(\vc-\im \gamma) \frac{q(\vc+3\im \frac{\gamma}{2})}{q(\vc-\im \frac{\gamma}{2})},
\end{equation}
is obtained by considering the product
\begin{equation}
\left(\begin{ytableau}
~
\end{ytableau}_{\vc-\im\frac{\gamma}{2}} \right)~\cdot~ \left(\begin{ytableau}
~
\end{ytableau}_{\vc+\im\frac{\gamma}{2}}\right)=
\begin{ytableau}
1 & 1
\end{ytableau}
+
\begin{ytableau}
1 & 2
\end{ytableau}
+
\begin{ytableau}
2 & 2
\end{ytableau}
+
\begin{ytableau}
2 & 1
\end{ytableau} \, ,\label{repbox}
\end{equation}
where in the double boxes, it is implied that the spectral parameter for the left box is shifted by
$-\im \frac{\gamma}{2}$, and by   $\im \frac{\gamma}{2}$ for the right box. Therefore, the first three terms on the right hand side  must make up the spin-2 transfer matrix
\begin{equation}
\Phi(\vc)\Lambda^{(2)}(\vc)=\begin{ytableau}
~ & ~
\end{ytableau} \, ,
\end{equation}
while the last diagram in (\ref{repbox}) does not satisfy the filling rules of a Young tableau. One can adopt the notation
\begin{equation}
\begin{ytableau}
2 & 1
\end{ytableau}
=\begin{ytableau}
1 \\ 2
\end{ytableau} \, , \label{quantumDeterminant}
\end{equation}
so that imaginary parts of shifts increase by $\gamma$ from left to right or from bottom to top (so here box 2 is at $\vc-{\rm i}{\gamma\over 2}$ and box 1 at $\vc+{\rm i}{\gamma\over 2}$)\footnote{Recall that the possible fillings in a $\mathsf{sl}_n$ Young tableau should satisfy  $j_1\leq j_2$ if the box filled with $j_2$ sits on the right of the box with $j_1$ and $j_1< j_2$ if the box with $j_2$ is below the box with $j_1$, thus the notation is consistent.}. The function (\ref{quantumDeterminant}) is proportional to the identity, i.e.,  independent of the Bethe roots, thus of the eigenvectors. In the QISM formalism this is a central element of the algebra, also called \textit{quantum determinant}. 
Therefore, to get fusion hierarchies or fusion transfer matrices out of the box equations, all we need is to normalize the  multi-boxes by identifying trivial multiplicative
factors. Hence (\ref{repbox}) becomes
\begin{equation}
\left(\begin{ytableau}
~
\end{ytableau}_{\vc-\im\frac{\gamma}{2}} \right)~\cdot~ \left(\begin{ytableau}
~
\end{ytableau}_{\vc+\im\frac{\gamma}{2}}\right)=
\begin{ytableau}
~ & ~
\end{ytableau}+
\begin{ytableau}
~ \\ ~
\end{ytableau} \, .
\end{equation}
More generally, equation (\ref{eq:fusion-Th}) 
 can be reformulated as  
\begin{equation}
\left(\begin{ytableau}
~
\end{ytableau}_{\vc-\im\frac{k \gamma}{2}}\right)
\cdot
\left(\underbrace{\begin{ytableau}
~ & ~ & \ldots & ~
\end{ytableau}_{\vc+\im\frac{\gamma}{2}}}_{k}\right)
=
\underbrace{\begin{ytableau}
~ & ~ & \ldots & ~
\end{ytableau}_{\vc}}_{k+1}
+
\left({
\begin{ytableau}
1
\\
2
\end{ytableau}}_{\vc-\im\frac{k \gamma}{2}} \right) \cdot
\left(\underbrace{\begin{ytableau}
~ & ~ & \ldots & ~
\end{ytableau}_{\vc+\im \gamma}}_{k-1}\right),
\end{equation}
Here  we  have defined
\begin{equation}\label{eq:gen-form}
\underbrace{\begin{ytableau}
~ & ~ & \ldots & ~
\end{ytableau}_{\vc}}_{k} = \sum_{\{j\}} \prod_{m=1}^k  {\begin{ytableau}
j_m
\end{ytableau}}_{\vc+\im \gamma \left((m-1)-\left(\frac{k-1}{2}\right)\right)},
\end{equation}
where the sum over $\{j\}$ corresponds to the admissible fillings of the $\mathsf{sl}_2$ tableau of size $k$ on the left hand side. Further, one can express the eigenvalues of fused transfer matrix, i.e. $T^{(J)}$ with $J >1$, in terms of the Young tableau in the following way 
\begin{equation}\label{eq:fused-eigval-YT}
  \left( \prod_{j = 0}^{J-2} \Phi\left(\vc + {\rm i} \gamma \left(j - \frac{J-2}{2} \right) \right) \right) \Lambda^{(J)}(\vc) = \underbrace{\begin{ytableau}
~ & ~ & \ldots & ~
\end{ytableau}_{\vc}}_{J} \, .
\end{equation}
One can use the three equations above to re-derive Eq. \eqref{eq:fusion-Th}.

\subsubsection{General bilinear relation}

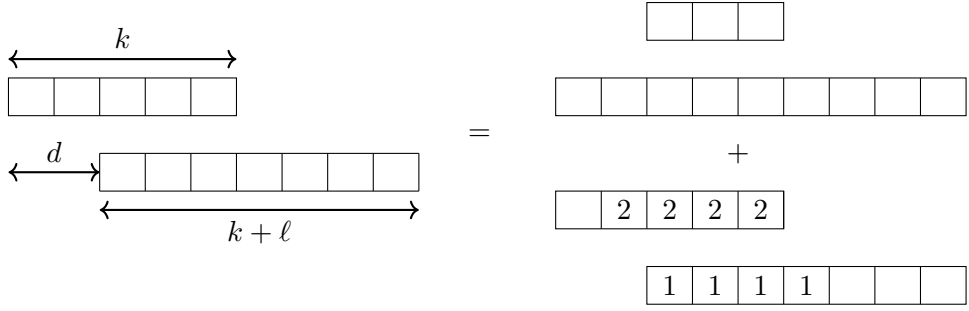
\begin{figure}[H]
\begin{center}
\begin{tikzpicture}

    \draw (0,2) rectangle (3,2.5);
    \draw (0.6,2) -- (0.6,2.5);
    \draw (1.2,2) -- (1.2,2.5);
    \draw (1.8,2) -- (1.8,2.5);
    \draw (2.4,2) -- (2.4,2.5);

    \draw[<->, thick] (-0.0,2.75) -- (3.0,2.75) node[midway, above] {$k$};

        \draw (1.2,1) -- (5.4,1);
    \draw (1.2,1.5) -- (5.4,1.5);
    \draw (1.2,1) -- (1.2,1.5);
 
    \draw (1.8,1) -- (1.8,1.5);
    \draw (2.4,1) -- (2.4,1.5);
    \draw (3,1) -- (3,1.5);
    \draw (3.6,1) -- (3.6,1.5);
    \draw (4.2,1) -- (4.2,1.5);
    \draw (4.8,1) -- (4.8,1.5);
    \draw (5.4,1) -- (5.4,1.5);

    \draw[<->, thick] (-0.,1.25) -- (1.2,1.25) node[midway, above] {$d$};
    \draw[<->, thick] (1.2,0.75) -- (5.4,0.75) node[midway, below] {$k+\ell$};
    \node at (6.2, 1.75) {$=$};

    \draw (8.4,3) rectangle (10.2,3.5);
 
    \draw (7.2,2) rectangle (12.6,2.5);
    \draw (9,3) -- (9,3.5);
    \draw (9.6,3) -- (9.6,3.5);

    \draw (7.8,2) -- (7.8,2.5);

    \draw (8.4,2) -- (8.4,2.5);
    \draw (9,2) -- (9,2.5);
    \draw (9.6,2) -- (9.6,2.5);
    \draw (10.2,2) -- (10.2,2.5);
    \draw (10.8,2) -- (10.8,2.5);
    \draw (11.4,2) -- (11.4,2.5);
    \draw (12,2) -- (12,2.5);

    \node at (9.6, 1.5) {$+$};

 \draw (7.2,1.) rectangle (10.2,0.5);
    \draw (7.8,1) -- (7.8,0.5);

    \draw (8.4,1) -- (8.4,0.5);
    \draw (9,1) -- (9,0.5);
    \draw (9.6,1) -- (9.6,0.5);
    \draw (10.2,1) -- (10.2,0.5);
   
 \draw (8.4,0.) rectangle (12.6,-0.5);

    \draw (9,0) -- (9,-0.5);
    \draw (9.6,0) -- (9.6,-0.5);
    \draw (10.2,0) -- (10.2,-0.5);
    \draw (10.8,0) -- (10.8,-0.5);
    \draw (11.4,0) -- (11.4,-0.5);
    \draw (12,0) -- (12,-0.5);

    \node at (8.1,0.75) {2};
    \node at (8.7,0.75) {2};
    \node at (9.3,0.75) {2};
    \node at (9.9,0.75) {2};
   
    \node at (8.7,-0.25) {1};
    \node at (9.3,-0.25) {1};
    \node at (9.9,-0.25) {1};
    \node at (10.5,-0.25) {1};

\end{tikzpicture}
  \caption{Multiplication of Tableaux. One may redistribute boxes as to form a bigger tableau whenever the shifts in spectral parameter allows the connection. In the picture
  one can either move down two leftmost boxes or move up four rightmost boxes. The remainder forms a number of quantum determinants.}\label{Bilinear}
\end{center}
\end{figure}

In the simpler case of $\mathsf{sl}_2$ fusion rules, we may 
consider a generic multiplication of Young diagrams such as on the left hand side of figure \ref{Bilinear}. There, we consider a smaller (leftmost) tableau and a possibly larger (rightmost) tableau. If the smaller one carries $k$ boxes, the larger one carries $k+\ell$ boxes, so the integer $\ell$ indicates by how many boxes the later exceeds the former. In this multiplication there is another parameter, $d$, which is the left displacement. We emphasize that, once a particular box carries  some value of spectral parameter,  all the others are determined by the relative positions respective to it. Moreover, we take advantage of the fact that the anti-symmetric fusion is trivial, as it provides the quantum determinant, so that only horizontal tableaux are non-trivial.  We can then use the horizontal position as a ``global coordinate'' for the shifts in $\im \gamma$. Here we mean that by fixing the spectral parameter of a box, say in the smaller tableau, this will fix all other boxes' values, be it in the same tableau or not. Boxes at the same horizontal position carry the same spectral parameter and the integer $d$ indicates by how many units of $\im \gamma$ the leftmost box of the possibly larger tableau exceeds the leftmost box of the smaller tableau. 

To re-express the product on the left hand side of figure \ref{Bilinear}, one should be able to move a certain quantity of boxes from one tableau in order to build  tableaux of different sizes. One has to guarantee the connectivity of boxes and that shifts/spectral parameters go to the correct positions. This is possible and non-trivial whenever $1 \leq d \leq k$. In this situation, there are  two possible moves: 1) we may try to move down the first left $d$ boxes, or 2) we may try to move up the last right $\ell+d$ boxes. One will notice that these two moves are complementary in that summing over all possible fillings in which these are allowed we find the uppermost term in the right hand side  of figure \ref{Bilinear}. 
The remaining diagrams that could not be arranged in that way will form a number of $(k-d+1)$ quantum determinants. To see this, let us denote the integers which fill boxes in the smaller tableau by $j_m$, $m=1,2,~\ldots,k$, all of them assuming values $1,2$ respecting the tableau rules for $\mathsf{sl}_2$. Also, let us denote  the associated integers for the larger tableau on the left hand side by  $n_m$, $m=1,2,~\ldots,k+\ell$. On the right hand side, denote the $p_m$ with $m=1,2,~\ldots,k-d$ the fillings for the smallest tableau on the upper side and by $a_m$, 
$m=1,2,~\ldots,d+k+l$ the ones for 
the largest  tableau. Let us try to form the product of tableaux on the upper part of the right hand side of figure \ref{Bilinear}. 

If $j_d=1$, then $j_d\leq n_1$, and we are free to perform the first move without any constraint left for the smallest tableau, i.e. $p_m=j_{m+d}$. However, for the new formed largest tableau we have $a_{m\leq d}=j_{m\leq d}=1$. Now suppose $j_d=2$, then we can only make the first move if $n_1=2$, which implies $n_m=2$ for all $m$, while we do also have $j_{m>d}=2$ which will form the smallest tableau, i.e. $p_m=j_{m+d}=2$, for $m=1,\ldots,k-d$. Furthermore, if $j_d=2$ and $n_1=1$ we cannot move boxes that way. Therefore, after the first move, we lack a product of tableaux on the right hand side with $p_1=1$,  $a_{m\geq d}=2$, while we still have non moved tableaux with $j_{m\geq d}=2$ and $n_1=1$. Now, to these remaining non moved tableaux we apply the second move. This is only possible if $n_{k-d+1}=2$ and completes what we lacked. However, if $n_{k-d+1}=1$, then $n_{m\leq k-d+1}=1$ and we are left with non-moved tableaux such $j_{m\geq d}=2$ and $n_{m\leq k-d+1}=1$. What is left is represented on the bottom part of  the right hand side of figure \ref{Bilinear}. The number of $1$'s and $2$'s are the same and shifted accordingly to form the $k-d+1$ quantum determinants.       

As a result, we find the following set of bilinear
relations among the transfer matrices eigenvalues:
\begin{equation}
\Lambda^{\left(k\right)}_{{\left[\frac{d}{2}+\frac{\ell}{4}\right]}}\left(\vc \right) \Lambda_{{\left[-\frac{d}{2}-\frac{\ell}{4}\right]}}^{\left(k+\ell\right)}\left(\vc\right)=\Lambda_{_{{\left[\frac{\ell}{4}\right]}}}^{\left(k-d\right)}\left(\vc \right) \Lambda_{\left[-\frac{\ell}{4}\right]}^{\left(k+d+\ell\right)}\left(\vc
\right)+\Lambda^{\left(d-1\right)}_{\left[\frac{k+1}{2}+\frac{\ell}{4}\right]}\left(\vc \right) \Lambda^{\left(d-1+\ell\right)}_{\left[-\frac{k+1}{2}-\frac{\ell}{4}\right]}\left(\vc\right). \label{Bilinearev0}
\end{equation}
where $k,\ell,d$ are positive integers with the restriction that $1 \leq d \leq k$.  Of course, another set of bilinear relations can be derived by exchanging the placement (shifts) of the smaller and bigger tableaux on the left hand side. Therefore,
\begin{equation}
\Lambda^{\left(k+\ell\right)}_{{\left[\frac{d}{2}+\frac{\ell}{4}\right]}}\left(\vc \right) \Lambda_{{\left[-\frac{d}{2}-\frac{\ell}{4}\right]}}^{\left(k\right)}\left(\vc\right)=\Lambda_{_{{\left[\frac{\ell}{4}\right]}}}^{\left(k+d+\ell\right)}\left(\vc \right) \Lambda_{\left[-\frac{\ell}{4}\right]}^{\left( k-d\right)}\left(\vc
\right)+\Lambda^{\left(d-1+\ell\right)}_{\left[\frac{k+1}{2}+\frac{\ell}{4}\right]}\left(\vc \right) \Lambda^{\left(d-1\right)}_{\left[-\frac{k+1}{2}-\frac{\ell}{4}\right]}\left(\vc\right). \label{Bilinearev01}
\end{equation}

Our generalized T-system encompasses bilinear relations previously obtained in the literature \cite{Klumper:1992vt,Kuniba:1998ygt}. The equations are generally asymmetric with respect to tableaux sizes, so that by varying the integer $\ell$ one can interpolate between fusion hierarchy functional relations and the usual T-system. Here we have used the eigenvalue expressions to infer general relations valid for the transfer matrices themselves. 
In order to promote our approach to a rigorous proof, it would be desirable to show completeness of the spectrum, or that, regardless of  a definite eigenvalue expression, they would still comply with the Yangian version of the Young tableau. It is not our objective to dig into these matters further. The transfer matrices can be built and the relations can be verified at this level.
Furthermore, by now, it is not hard to propose an induction hypothesis over the different integers appearing in our generalized version of the T-system. This would lead to a natural extension of the proof in \cite{Klumper:1992vt} for the standard T-system.

\subsection{Scaling limit and Defect entropies}

Recall that our goal is to obtain, for lattice realizations of  each module of the algebra $\hbox{Vir}\otimes \overline{\hbox{Vir}}$, $V_{(r's')}\otimes \overline{V}_{(r's')}$ with character $|\chi_{(r's')}|^2$,  lattice defect operators with  eigenvalues, in the scaling limit, given  by(\ref{eq:verlinde-line-op})
\begin{equation}
\left.\widehat{\cal{D}}_{(rs)} \right|_{V_{(r's')}\otimes \overline{V}_{(r's')}}=
{S_{(rs)(r's')}\over S_{(11)(r's')}}=
 {(-1)}^{(r+s)(r'+s')}
{\sin {\pi rr'\over p}\sin{\pi ss'\over p+1}\over \sin{\pi r'\over p}\sin{\pi s'\over p+1}} \, . \label{rsrs}
\end{equation}
The eigenvalues of these defect operators acting on the ground-state
\begin{equation}
\left.\widehat{\cal{D}}_{(rs)} \right|_{V_{(11)}\otimes \overline{V}_{(11)}}=
{\sin {\pi r\over p}\sin{\pi s\over p+1}\over \sin{\pi \over p}\sin{\pi \over p+1}} \, . \label{rsrs}
\end{equation}
are particularly useful as ``degeneracies'' and can be measured e.g. with entanglement entropies. 
 We have in particular
\begin{equation}
g_{{\cal D}_{(11)}}=1,~g_{{\cal D}_{(12)}}=2\cos{\pi\over p+1},~g_{{\cal D}_{(21)}}=2\cos{\pi\over p} \, .
\end{equation}
It turns out that the generalized T-system provides an efficient way to calculate the eigenvalues of the$\widehat{\cal{D}}_{(rs)}$, and justify the identifications made previously.  

So far however, the identifications of $\widehat{\cal{D}}^{(\rm latt)}$  in 
section \ref{sec:21-discretization} were made only  up to a normalization factor.

It is now time to tackle this point. First, since the lattice regularization of the TDL, $T(\tilde{u}={\gamma\over 2}+iv_C)$, acts on a periodic system, one may be tempted to think that, as is usually the case for transfer matrices of periodic, homogeneous models at their critical points, the logarithm of its eigenvalues  for low-lying energy states should have the form 
\begin{equation}\label{familiar}
\Lambda \approx e^{2\R e_0 +{\mathcal O}({1\over \R})} \,,
\end{equation}
where $e_0$ would be the non-universal ground state energy per side, while the ${1\over \R}$ terms would encode the conformal weights via the usual logarithmic mapping formula. In particular, in Eq. (\ref{familiar}) there is no term of ${\mathcal O}(1)$. On the other hand,  the g-factors we are interested in  {\sl are}  precisely  terms of this form!

This apparent paradox  (see also \cite{tavares2024}) occurs because while we are interested indeed in the low-lying energy states of the periodic model, the transfer matrix used to discretize the TDL is {\sl not} evaluated in the same   spectral parameter region as the corresponding physical transfer matrix, which is defined, instead,  in the vicinity of $\tilde{u}=0$. This means that the familiar form  (\ref{familiar}) in fact does not hold. We expect instead to have 
\begin{equation}
\Lambda_{n}^{(J)}(\vc) = {\rm e}^{2 \R e_0^{(J)}(\vc)+ f_n^{(J)}(\vc)+ {\mathcal O}({1\over \R})} \, \label{Lamscal},
\end{equation}
while as mentioned above $f_n^{(J)}$ would vanish in the ordinary case $\vc\sim 0$. The identification of the defects will only be complete when the ${\rm e}^{2 \R e_0^{(J)}(\vc)}$ term (``bulk term'') is properly factored out. Luckily we know these terms from the Bethe-ansatz.  There are  two different expressions  for each $J$.
For $|\Im \vc|< (J+1) \frac{\gamma}{2}$, $e_0^{(J)}(\vc)$  is simply given by (\ref{bulk}), whereas for $|\Im \vc -\pi/2|\leq (p-J)\frac{\gamma}{2}$ one has to use in Eq. (\ref{Lamscal}) the same expression (\ref{bulk}) but with $J\to p-1-J$ and the argument  $\vc-i \pi/2$ instead of $\vc$. In what follows we will set
\begin{equation}\label{eq:corr-norme}
\begin{split} 
    \tilde{e}_0^{(J)}(v) = &
\begin{cases}
e_{0}^{(J)}(v), & \text{for } |\operatorname{Im} v| < (J+1)\gamma/2 \, , \\
e_{0}^{(p-1-J)}(v \mp \frac{\pi {\rm i}}{2} ), & \text{for } (J+1)\gamma/2 < |\operatorname{Im} v| \leq \pi/2 \, ,
\end{cases}
\end{split}
\end{equation}
This is true for $J>0$. For $J=0$ we have  $ e_0^{(J)}(v_I)=\log {\sinh \vc \over \sin \gamma }$ for $|\Im v_I|< \gamma/2$, whereas for $\gamma/2<|\Im v_I|\leq \pi/2 $ equation (\ref{eq:corr-norme}) remains correct. 
As we shall see latter, the different bulk behaviors imply different defect realizations because the subtraction of these non-universal contributions should be done for all states, and because the additive terms in the discretized version of the defect algebra,  (\ref{Bilinearev0}), survive the scaling limit under the condition that they exhibit  the same common bulk behavior. In consequence, a single functional relation may produce different algebraic relations when the parameter  $\vc$ belongs to different regions of the complex plane, leading to different results for   $f_n^{(J)}$. For instance if $n=0$ corresponds to the identity field, one finds $f_0^{(1)}(\vc)=\log g_{{\cal D}_{(11)}}=0$ for $|\Im \vc|<  \gamma$, whereas $f_0^{(1)}(\vc)=\log \left[{(-1)}^{\R}g_{{\cal D}_{(21)}}\right]$ for $\gamma< |\Im \vc |\leq \frac{\pi}{2}$, and $f_0^{(1)}(\pm \infty)=\log g_{{\cal D}_{(12)}}$. Bethe-ansatz calculations allow for a reasonably simple derivation of these results as discussed below.

\subsubsection{The \texorpdfstring{$(1,s)$}{Lg} case}
First, let us consider the situation when $\vc\rightarrow \pm \infty$.
 This limit exists if one normalizes each transfer matrix by a proper ${\rm e}^{\mathcal{O}(\R)}$ factor, which we may take to be $\cosh^{2\R}(v)$. For example, if we take the Bethe ansatz expression, we obtain
\begin{equation}
\lim_{\vc \rightarrow \infty}\frac{\Lambda_n^{(J)}(\vc)}{\cosh^{2\R}(\vc)}  = \frac{\sin (J+1) \eta} {\sin(\eta)}\equiv {[J+1]}_{{\rm e}^{\im \eta}}\label{1sentropies} \, ,
\end{equation}
 We insist that  this is true for all states, and  
 that the state dependency enters solely through $\eta$. Due to the relations (\ref{eq:fusion-Th}) or (\ref{Bilinearev0}), the numbers in \eqref{1sentropies} satisfy a series of relations, namely
\begin{equation}
[J+1]_{{\rm e}^{\im \eta}} [J+\ell+1]_{{\rm e}^{\im \eta}}= [J-d+1]_{{\rm e}^{\im \eta}} [J+d+\ell+1]_{{\rm e}^{\im \eta}}+[d]_{{\rm e}^{\im \eta}} [d+\ell]_{{\rm e}^{\im \eta}},
\end{equation}
with the same restrictions as in (\ref{Bilinearev0}). Moreover, for the A$_p$ model we have the truncation $[p+1]_{\eta}=0$, which gives rise to the possible A$_p$ exponents (\ref{exponentsimp}).

Meanwhile we can use known results about the relationship between the various sectors of the minimal models CFT and the Bethe-ansatz equations.  Writing the conformal partition function as (\ref{lattZ}),
it turns out \cite{Belletete:2017gwt}  that the sector with fixed $s'$ is obtained by choosing $\eta={s'\pi\over p+1}$. For a fixed $s'$, the action of the $\uR^{-1}$ operator adds to (\ref{1sentropies}) a factor  $(-1)^{r'+s'+\R}$ factor from Eq. (\ref{rsrs}), producing the correct result after multiplication by $(-1)^\R$ - see appendix  \ref{sec:Yeigenvalues} for a thorough discussion, and section \ref{sec:examples} for examples.  

This confirms the identification with  the $(r,s)=(1,1+J)$ defects, generalizing earlier observations in the case $J=1$ and the defect $(1,2)$. 

Note that this result was obtained in \cite{Belletete:2018eua} by purely algebraic means, using in particular the fact that the sector of fixed $s'$ can be identified as an irreducible representation of the relevant quotient of the affine Temperley-Lieb algebra (although the result in this reference  must be slighlty corrected due to the sign issues mentioned above, see appendix  \ref{sec:Yeigenvalues}.

\subsubsection{Finite spectral parameter}
\label{gfactors}

As announced earlier, see e.g. figure \ref{fig:crossed-channel }, we expect defects of type $(r,s)=(1+J,1)$ to be obtained instead from a row of modified faces of spin $J$ with spectral parameter $\tilde{u}=\pm {\pi\over 2}$ that is $\vc=\pm i{\pi\over 2}$. In fact, we also mentioned 
that defects of type $(r,s)=(J,1)$ (note the appearance of $J$ instead of $1+J$) can be obtained via the same construction but with $|\Im \vc|< \frac{(J+1) \gamma}{2}$ and $|\Re \vc|<\infty$.

\begin{figure}[htb]
  \begin{center}
  \includegraphics[width=0.4 \linewidth]{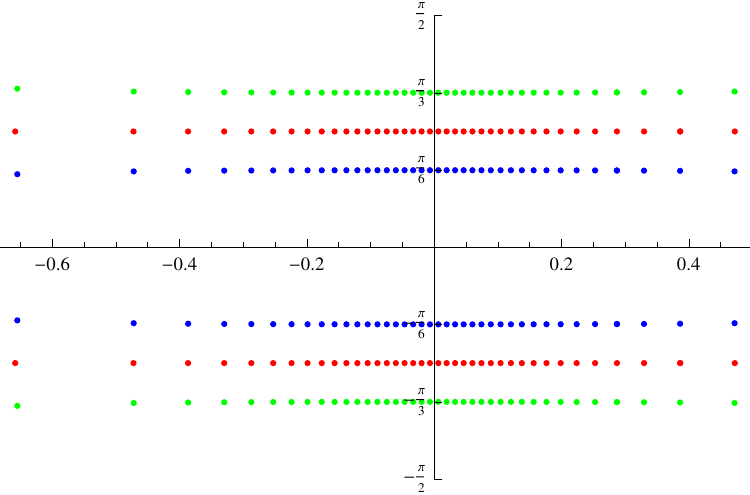}
  \caption{Eigenvalue's zeros of the $A_5$ model ($\gamma={\pi\over 6}$)associated to the largest eigenvalue of the fundamental transfer matrix ($J=1$). Blue corresponds to $J=1$; Red corresponds to $J=2$; Green corresponds to $J=3$. We chose the system size $2\R=80$.}
\label{ZerosEig}
\end{center}
\end{figure}

For the largest eigenvalue $\Lambda_{0}^{(1)}(\vc)$, Bethe roots lie along the real line and zeros (in the variable $\vc$)  of the eigenvalue are close to lines $\Im(\vc)=\pm \gamma$. This further implies that the  $\Lambda_{0}^{(J)}(\vc)$ have zeros at lines $\Im(\vc)=\pm ({J+1 \over 2})\gamma$. Exceptions to this are $J=0$ and $J=p-1$, for which we find $\Lambda_{0}^{(0)}(\vc) = \sinh^{2 \R}(\vc)$ and $\Lambda_{0}^{(p-1)}(\vc) = \pm \cosh^{2\R}(\vc) $. Therefore, if we consider that the $\vc$ complex plane has been compactified (using  $\Lambda_{n}^{(J)}(\vc)= \Lambda_{n}^{(J)}(\vc+\pi \im)$), we may claim that the above mentioned line of zeros divide this surface into two regions: $|\Im \vc|< ({J+1 \over 2})\gamma$ (centered at zero) and the complement $|\Im \vc-\frac{\pi}{2}|< \frac{(p-J) \gamma}{2}$, centered at $  {i \pi \over 2}$. In the thermodynamic limit such zeros are expected to  become dense and separate the two regions of the surface for each fused eigenvalue, see Fig. \ref{ZerosEig}.
For the  first excited states we will  still have such dense lines of zeros separating different regions of the complex plane, but a finite number of zeros  may also appear close to $\Im \vc=0,~\pi/2$. Nevertheless they generally behave as $\Theta \propto {\gamma \over \pi} \log 2\R$\footnote{Recall we denote generically by $\Theta$ a zero of any eigenvalue.}

Define
    \begin{equation}
\begin{split}
\Y^{(J)}(\vc)=\frac{\Lambda^{(J)}(\vc-\im \gamma/2) \Lambda^{(J)}(\vc+\im \gamma/2)}{\Lambda^{(0)}\left(\vc-\im \frac{(J+1) \gamma}{2}\right) \Lambda^{(0)}\left(\vc+\im \frac{(J+1) \gamma}{2}\right)}\, ,
\\
\y^{(J)}(\vc)=\frac{\Lambda^{(J-1)}(\vc) \Lambda^{(J+1)}(\vc)}{\Lambda^{(0)}\left(\vc-\im \frac{(J+1) \gamma}{2}\right) \Lambda^{(0)}\left(\vc+\im \frac{(J+1) \gamma}{2}\right)},
\end{split}
\end{equation}
then $\Y^{(J)}(\vc)=1+\y^{(J)}(\vc)$, thanks to the relations (\ref{Bilinearev0}), with $\ell=0$ and $d=1$. The Y-system is then the set of equations 
\begin{equation} \label{yfunctions}
\y^{(J)}\left(\vc-\im {\gamma \over 2}\right) \y^{(J)}\left(\vc+\im {\gamma \over 2}\right) = \Y^{(J-1)}(\vc) \Y^{(J+1)}(\vc).
\end{equation}
By exploring the analyticity properties we can derive a finite set of non linear integral equations\cite{Klumper:1992vt} from the $Y$-system,
\begin{multline}
    \log \frac{\y^{(J)}(\vc)}{\y^{(J)}(\infty)} =\delta_{J1}  2\R \log \left[\frac{{\rm e}^{(p+1)\vc}-1}{{\rm e}^{(p+1)\vc}+1}\right]  +  \\\sum_{M} G_{JM}^{[p-2]} \left[ \sum_{\Theta \in M}\log\left[\frac{{\rm e}^{(p+1) (\vc-\Theta)} -1}{{\rm e}^{(p+1) (\vc-\Theta)} +1 }\right]          + \frac{1}{2 \pi i}\int_{-\infty}^{\infty} \log\left[\frac{{\rm e}^{(p+1) (\vc-s)} -i}{{\rm e}^{(p+1) (\vc-s)} +i }\right] {\log'} \Y^{(M)}(s) {\rm d}s\right], \ \label{SNLIET}
\end{multline}
where 
$\Theta \in M$ refers to the eigenvalue zeros associated to a  fused eigenvalue $\Lambda_n^{(M)}(\vc)$ - in other words,   excitations over the ground-state are  solely parameterized in terms of the  number of   eigenvalue zeros in the strip $|\Im \Theta|<\frac{\gamma}{2}$.

Equations (\ref{SNLIET}) allow for self-consistent determination of $Y$-functions. 
Once they are solved one may compute the eigenvalue expressions via:
\begin{multline}
\log \frac{\Lambda^{(J)}(\vc)}{\Lambda_{\infty}^{(J)}}= 2\R e_0^{(J)}(\vc)+  \sum_{\Theta \in J}\log\left[\frac{{\rm e}^{(p+1) (\vc-\Theta)} -1}{{\rm e}^{(p+1) (\vc-\Theta)} +1 }\right]    \\    +\frac{1}{2 \pi i}\int_{-\infty}^{\infty} \log\left[\frac{{\rm e}^{(p+1) (\vc-s)} -i}{{\rm e}^{(p+1) (\vc-s)} +i }\right] {\log}' \Y^{(J)}(s) {\rm d}s, \label{RSOSEigenvalue}
\end{multline}
Using the analyticity structure so far assumed, it may be apparent that we do not obtain eigenvalues for $\vc$ in the whole complex plane. In particular one can shift spectral parameters so as to extend (\ref{RSOSEigenvalue}) to the region $|\Im \vc|<(J+1)\gamma/2$. However, this is, of course, all we need since the reflection relation allows for the evaluation of eigenvalues, say, in the complementary strip: $\Lambda^{(J)}(\vc)=\epsilon \Lambda^{(p-1-J)}(\vc+\frac{\pi\im }{2})$, where $\epsilon=\pm1$ is the corresponding eigenvalue of ${\cal R}$ \footnote{For simplicity we may assume $2\R=0~ \mbox{mod} ~4$.} 

To obtain g-factors, we write the eigenvalue expression in terms of the scaled functions, $Y_+^{(J)}$.
Observe, however, that we are interested in corrections of order ${\cal O}(1)$ for large $\R$. At this stage, we have
\begin{equation}
\log \Lambda^{(J)}(\vc) \approx  2\R e_0^{(J)}(\vc)+\pi \im \nu^{(J)}_+ + \log (\mbox{\text{sign}} \Lambda_{\infty}^{(J)})+\frac{1}{2} \log \Y_+^{(J)}(-\infty), \label{RSOSScaledEigenvalue}
\end{equation}
the cases $J=0$ and $J=p-1$ being trivial, since $\Lambda^{(0)}(\vc)= \sinh^{2\R}(\vc)$ and $\Lambda^{(\frac{p-1}{2})}(\vc)=  \epsilon \cosh^{2\R}(\vc)$.
We take the ``scaling limit'' in (\ref{SNLIET}), which gives the algebraic system
\begin{equation}
X_J^2=1+X_{J+1}X_{J-1}, \qquad J=1,\ldots,p-1,
\label{AlgebraicSystem}
\end{equation}
where $X_{J} = {(-1)}^{\nu_{+}^{(J)}} \mbox{\text{sign}} \Lambda_{\infty}^{(J)} \sqrt{\Y_+^{(J)}(-\infty)}$ for $J=2,\ldots,p-2$, $X_1={(-1)}^{\nu_{+}^{(1)}} \mbox{\text{sign}} \Lambda_{\infty}^{(1)}$, $X_{p-1} = \epsilon $, $X_0=X_p=0$. The distinguished form of $X_1$ is related to the fact that the scaling limit of (\ref{SNLIET}) produces a driving-term that goes to $-\infty$ when $\vc \to -\infty$. See also (\ref{scaledNLIE})

\begin{figure}[htb]
\begin{center}
\includegraphics[width=0.48\linewidth]{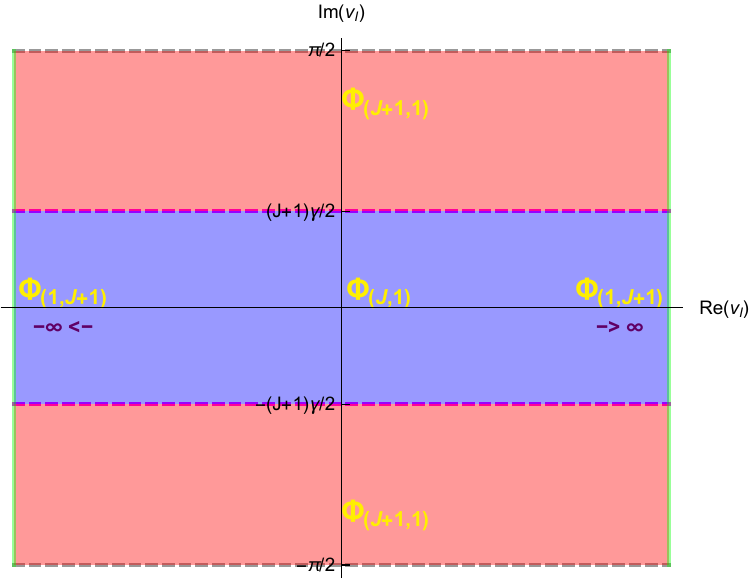}
\includegraphics[width=0.48\linewidth]{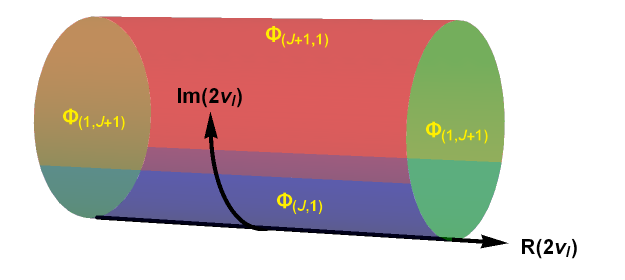}
\end{center}
\caption{ a) Different defect realizations in $\vc$ plane. b) Compactified version in variable $2 \vc$.  }
\label{DefectRegions}
\end{figure}

In general, the system (\ref{AlgebraicSystem}) for $p\geq 5$ provides spurious solutions on top of  the physical  ones, and  is by itself  not sufficient to determine the quantities $X_i$. A possible way to rule out the spurious solutions is to use the generalized T-system to collect other algebraic relations. In other words, we plug Eq. (\ref{RSOSScaledEigenvalue}) into (\ref{Bilinearev0}) to find these\footnote{In terms of NLIEs one may also write expressions of the new functions in terms of the usual $Y$-system.}. We will come back to this in section \ref{continuum}.  Another way to go is to impose that quantities $\sqrt{\Y_+^{(J)}(-\infty)}$ should be non-negative by definition, which further constrains the solutions.
The final possible set of values are 
\begin{equation}\label{Xqtilde}
X_J= {(-1)}^{({r'}+{s'})} [J]_{\tilde{q}^{r'}},
\end{equation}
with $\tilde{q}={\rm e}^{i \frac{\pi}{p}}$.
Our notation is suggestive: as in (\ref{rsrs}) Kac integers $r',~s'$ label the collection of fields in a given conformal tower. For any such a state, the  $O(1)$ expectation value of $T^{(J)}(\vc)$ is precisely $X_J$, for $|\Im \vc|<(J+1) \gamma/2$ and $|\Re \vc|<\infty$. To connect with Bethe data, one observes that sectors parameterized in terms of number of zeros and twist sector are such that
\begin{equation}
{(-1)}^{\nu_+^{(J)}} = {(-1)}^{({r'}+{s'})} \mbox{sign}\left([J+1]_{q^{s'}}[J]_{\tilde{q}^{ r'}}\right), \qquad \epsilon={(-1)}^{s'+1},
\label{sectors}
\end{equation}
at sufficiently large $\R$. While the sectors (\ref{sectors}) may look ambiguous for cases where quantum number vanishes, it is always possible to make sense of these limits as they merely reflect the fact that the sign of the eigenvalue at its asymptotic $\vc \to \infty$ must cross the real line a certain number times to arrive at $\vc \to 0$ with the expected sign given by $X_J$.
We find, after factoring out the $\tilde{e}_0^{(J)}$ terms, that 
\begin{equation} \label{expectedvalues}
\Lambda^{(J)}(\vc) \to 
\begin{cases}
{(-1)}^{ (r'+s')}[J]_{\tilde{q}^{ r'}}, \qquad & \text{for } |\Im \vc|<(J+1) \gamma/2,
\\
(-1)^{\R}[J+1]_{\tilde{q}^{ r'}}, \qquad & \text{for } (J+1) \gamma/2<|\Im \vc| \leq \pi/2.
\end{cases}
\end{equation}
This is represented in Fig \ref{DefectRegions}. 

We now go back to write results in terms of the actual transfer matrices. Recall that the eigenvalue of $T^{(J)}(\gamma/2+\im v_I)$ has been denoted ${(-1)}^{\R}\Lambda^{(J)}(v_I)$ and more generally $T^{(J)}(u+\tu)$ has eigenvalues ${(-1)}^{\R} \Lambda(\vbu+v_I)$, with $u=\gamma/2+\im v_B$ and $\tu=\im v_I$. The defect line realized by the transfer-matrix of spin $J$  carries, generally, a spectral parameter dependence on both $\vbu$  and $\vc$, so (\ref{expectedvalues}) translates to generic values of $u$ and $\tilde{u}$ as follows
\begin{equation} \label{expectedvalues2}
T^{(J)}(u+\tilde{u}) \to 
\begin{cases}
{(-1)}^{ (r'+s'+\R)}[J]_{\tilde{q}^{ r'}}, \qquad & \text{for } |\Re (u+\tilde{u})-\gamma/2|<(J+1) \gamma/2,
\\
[J+1]_{\tilde{q}^{ r'}}, \qquad & \text{for } (J+1) \gamma/2<|\Re (u+\tilde{u})-\gamma/2| \leq \pi/2,
\end{cases}
\end{equation}
for $0<\Re u< \gamma$ and $|\Im u|<\infty$, so that the bulk leads to the minimal model at hand.

To proceed, we obtain the eigenvalues of the defect operators by dividing by the $J^{th}$ power of the eigenvalue value of $\uR$ in the scaling limit. The latter is obtained by setting  $J=1$ in the first of eqs. (\ref{expectedvalues2}), leading to
\begin{equation} \label{expectedvalues1m}
\uR^{-J}T^{(J)}(u+\tu) \to 
\begin{cases}
{(-1)}^{(J+1) (r'+s'+R)}[J]_{\tilde{q}^{ r'}}, \qquad &  \text{for } |\Re (u+\tilde{u})-\gamma/2|<(J+1) \gamma/2,
\\
(-1)^{J(r'+s'+R)}[J+1]_{\tilde{q}^{ r'}} \qquad & \text{for } (J+1) \gamma/2<|\Re (u+\tilde{u})-\gamma/2| \leq \pi/2.
\end{cases}
\end{equation}
We stress that these expressions are valid even for $J=0$, where $T^{(0)}$ is defined in Eq. \ref{eq:fusion-T-M-def}.

Now recall that the first case corresponds to the defect line $\widehat{\cal D}_{(J~1)}$, the second to the defect line $\widehat{\cal D}_{(J+1~1)}$. We  should have therefore the values of  
\begin{eqnarray}
\left.\widehat{\cal{D}}_{(r1)} \right|_{V_{(r's')}\otimes \overline{V}_{(r's')}}=
{S_{(r1)(r's')}\over S_{(11)(r's')}}= (-1)^{(r+1)(r'+s')} {\sin {\pi rr'\over p}\over \sin{\pi r'\over p}}\label{gfactorr}
\end{eqnarray}
We thus see  that (again, after factoring out the $\tilde{e}_0^{(J)}$ terms)
\begin{equation} \label{expectedvalues1}
\uR^{-J}T^{(J)}(u+\tu) \to 
\begin{cases}
(-1)^{\R(J+1)}\left.\widehat{\cal{D}}_{(J1)} \right|_{V_{(r's')}\otimes \overline{V}_{(r's')}}, \qquad &  \text{for } |\Re (u+\tilde{u})-\gamma/2|<(J+1) \gamma/2,
\\
 (-1)^{\R J}\left.\widehat{\cal{D}}_{(J+1 ~1)} \right|_{V_{(r's')}\otimes \overline{V}_{(r's')}},\qquad & \text{for } (J+1) \gamma/2<|\Re (u+\tilde{u})-\gamma/2| \leq \pi/2.
\end{cases}
\end{equation}
For $\R$ even, all signs disappear. For $\R$ odd however, we see from this equation that it is not possible to normalize intelligently the object on the left so that it {\it always} converges to the conformal defect in the continuum limit, since the sign depends also on the domain to which the parameter $\tu$ belongs. Of course, one can decide to focus on a single type of domain, and then there exists a uniform ($J$ independent) normalization, that could be absorbed in a redefinition of $e_0$.

Before we close this section let us recall that the diagonalization of a  certain number of generalized symmetries allows the resolution of the spectrum into sectors. 
Nevertheless, such a resolution, in general, might not comply with the usual labeling in terms of Kac indices or another way to separate conformally convariant states belonging to the same representation. 
This is expected for our defects, in particular for the generators $\widehat{\cal D}_{(1,2)}$ and $\widehat{\cal D}_{(2,1)}$, presumably because they are topological. 
 To test this, one should show that the expected values of these symmetry operators are indeed the same for the different fields belonging to the same conformal tower, as implied by (\ref{rsrs}). Within Bethe ansatz calculations this can be done by taking one step further, i.e.  evaluating ${\cal O}(1/\R)$ corrections to $\Lambda^{(1)}(\vc)$ by use of di-logarithmic relations.
Formally, the scaled equations are not any different as compared to the direct channel ones (\ref{scaledNLIE}) and (\ref{Eexponents}). They are actually simpler in terms of analyticity hypothesis such as the structure of zeros.  Like in the direct channel, the input for the dilog functions are the homotopy class of paths defined by $\y^{(J)}(v)$ when $v$ runs over the real line in addition to some  encircled  zeros ($\Theta\pm  \im \gamma/2$) of $\Y^{(J)}(v)$ functions . Moreover, the asymptotics which define integration limits, say for right movers, are $\y_+^{(J)}(\infty)=\y^{(J)}(\infty)$ and $\y_+^{(J)}(-\infty)\neq \y^{(J)}(-\infty)$,  where the inequality results from the fact that the scaling limit has to be taken before the asymptotics is evaluated. Since this has been done in the literature \cite{Klumper:1992vt}, we shall not bother the reader with more technicalities. In any case, it is manifest that defect expected values enter the evaluation of critical exponents through $\y^{(J)}_+(\pm \infty)$. We repeat, while low-lying states belong to infinite towers, for rational models we have only a finite number of defect expected values, which hints to the separation according to Kac modules.

\newpage

\section{Explicit Examples}\label{sec:examples}
In this section, we first consider the Ising model, and present analytical results for the defect Hamiltonian as well as  the defect operators, showing how they exactly match the (many) known results. We then  move on to numerical study  of the  Ising, Tri-Critical Ising and Tetra-Critical Ising  models, and discuss many aspects and examples  of our general construction, including questions related with the fusion of defects and of defect operators that will be studied more systematically in the next section. Finally, to illustrate  the generality of our results, we will also discuss the A$_{10}$ RSOS model, with and without defects, and compare the results with expectations from the $\mathcal{M}(11,10)$ Minimal Model of CFT.

\subsection{A\texorpdfstring{$_3$}{Lg} RSOS model : Ising}\label{sec:Ising}
It can be shown that the $2L$ site A$_3$ RSOS model contains two copies of $L$ site Transverse Field Ising (TFI) model. The discussion is essentially similar to the discussion for $D_4$ RSOS model and three state Potts Model in \cite{Sinha:2023hum}. Let us take the example of 8 sites periodic $A_4$ RSOS model. The Hilbert space splits under the action of the Temperley-Lieb algebra into two subspaces generated by the states 
\begin{equation}\label{eq:RSOS-Hilbert-space}
    \ket{x_0, 2, x_1 , 2,x_2, 2, x_3 , 2 }\, , \ket{2,x_{\frac{1}{2}}, 2, x_{\frac{3}{2}},2,x_{\frac{5}{2}}, 2, x_{\frac{7}{2}}} \, , 
\end{equation}
(the 
 half integer labeling is special to this section, and makes mapping onto standard formulations of the Ising model easier)
where $x_i$'s are 1 or 3. We refer to these two subspaces as  the odd sector and the  even sector respectively. Forgetting the (implied) heights equal to 2, we can represent states in these sectors instead as 
\begin{equation}\label{eq:TFI-Dual-TFI-Hilbert-space}
    \ket{\bar{x}_0, \bar{x}_1,\bar{x}_2, \bar{x}_3} \, , \ket{\bar{x}_{\frac{1}{2}}, \bar{x}_{\frac{3}{2}}, \bar{x}_{\frac{5}{2}}, \bar{x}_{\frac{7}{2}}}
\end{equation}
where we can trade heights for spins via $\bar{x}_i = \, \uparrow$  or $\downarrow$  when  $x_i = 1$ or 3 to map to  the usual Ising model. The states of the  odd sector in the RSOS Hilbert space are mapped to states in the TFI chain, whereas the states  of the even sector are mapped to states in the ``dual'' TFI chain \cite{Li:2023ani}.  Hence, each sector of the $2L$ site RSOS chain is isomorphic to $L$ site TFI. 

Let us now consider the action of $e_1$ on the odd sector
\begin{equation}
    e_1 \ket{x_0, 2, x_1, 2,x_2, 2, x_3} = \delta_{x_0, x_1} \sqrt{2} \ket{x_0, 2, x_1, 2,x_2, 2, x_3} \, , 
\end{equation}
which is equivalent to 
\begin{equation}
    \frac{1}{\sqrt{2}}\left(  1 + \sigma_0^{z} \sigma_1^{z} \right) \ket{\bar{x}_0, \bar{x}_1, \bar{x}_2, \bar{x}_3} = \delta_{\bar{x}_0, \bar{x}_1} \sqrt{2} \ket{\bar{x}_0, \bar{x}_1, \bar{x}_2, \bar{x}_3} \, . 
\end{equation}
where the $\sigma$ are Pauli matrices. Now, if we act  with $e_2$ on an element of the odd sector, we get 
\begin{equation}
    e_2 \ket{x_0, 2, x_1, 2,x_2, 2, x_3} = \frac{1}{\sqrt{2}} \sum_{a \in \{ 1,3\}} \ket{x_0, 2, a, 2,x_2, 2, x_3} \, , 
\end{equation}
which is equivalent to 
\begin{equation}
    \frac{1}{\sqrt{2}} \left(1 + \sigma_1^{x}\right) \ket{\bar{x}_0, \bar{x}_1, \bar{x}_2, \bar{x}_3}= \frac{1}{\sqrt{2}}\sum_{x \in \{\uparrow, \downarrow \}} \ket{ \bar{x}_0, x, \bar{x}_2, \bar{x}_3 }
\end{equation}
In fact, for the odd sector in  general, we have  that 
\begin{equation}\label{eq:ei_op_pauli_odd}
        e_{2i} \equiv \frac{1}{\sqrt{2}} \left(1 + \sigma_{i}^{x}\right) ,  \, \  e_{2i + 1} \equiv \frac{1}{\sqrt{2}} \left(1 + \sigma_{i}^{z}\sigma_{i + 1}^{z} \right) \, .
\end{equation}
Similarly, for the even sector it turns out  that 
\begin{equation}\label{eq:ei_op_pauli_even}
        e_{2i} \equiv \frac{1}{\sqrt{2}}\left(1 + \sigma^{z}_{i - \frac{1}{2}}\sigma^{z}_{i + \frac{1}{2}}\right) , \, \ e_{2i + 1} \equiv \frac{1}{\sqrt{2}}\left(1 + \sigma^{x}_{i + \frac{1}{2}}\right) \, .
\end{equation}
Hence, we see that in each sector individually $-\sum e_i$ behaves like the TFI Hamiltonian  and therefore there is one groundstate in each sector, as we discussed before. We are now ready to compare results for the RSOS and TFI models. We shall first discuss the crossed channel, and then the direct channel. 

\bigskip 

It is known that the IR limit of either sector the A$_3$ RSOS model, or equivalently TFI model, is described by the $c = \frac{1}{2}$ Ising CFT \cite{DiFrancesco:1997nk} or the $\mathcal{M}(4,3)$ minimal model. This CFT has three primary fields : $(1,1) \equiv \mathbf{1} , \, (1,2) \equiv \sigma, \, {\rm and  } \, (1,3) \equiv \epsilon $ of conformal dimensions 0, $\frac{1}{16}$, and $\frac{1}{2}$. There are only three TDLs corresponding to these three fields ${\cal D}_{(1,1)} \equiv \mathbf{1} , \, {\cal D}_{(1,2)} \equiv N,$ and ${\cal D}_{(1,3)} \equiv \eta$ \cite{Petkova:2000ip}, and they obey the following fusion rules
\begin{equation}\label{eq:Z2-TY-fusion}
    \eta^2 = \mathbf{1} \, , \quad N^2 = \mathbf{1} + \eta \, , \quad \eta N = N \eta = N \, .  
\end{equation}
These three TDLs together form the $\mathbb{Z}_2$ Tambara-Yamagami category and the $N$ line is called the duality line, see \cite{Chang:2018iay} for more details. These TDLs give rise to line operators and defect Hilbert space (Hamiltonian), depending on the direction in which they run. We will now discuss their lattice realizations.

\subsubsection{Crossed Channel}
We will first describe the situation where the TDLs are running perpendicular to the axis of cylinder, under which condition they behave as  line operators. Their action  can be deduced from Eq. \eqref{eq:verlinde-line-op}:
\begin{table}[H]\renewcommand{\arraystretch}{1.2}
\centering
    \begin{tabular}{|c|c|c|c|} 
    \hline 
\centering         &  $\mathbf{1}$&  $\sigma$& $\epsilon$\\ \hline 
\centering        \quad $\widehat{\mathcal{D}}_{(1,1)} \equiv \widehat{\mathbf{1}}$ \quad &  $1$&  $1$& $1$\\ \hline 
  \centering      \quad  $\widehat{\mathcal{D}}_{(1,2)} \equiv \widehat{N}$&  $\sqrt{2}$\quad &  0& $-\sqrt{2}$\\ \hline 
 \centering       \quad  $\widehat{\mathcal{D}}_{(2,1)} \equiv \widehat{\eta}$\quad &  $1$&  $-1$& $1$\\ 
         \hline
    \end{tabular}
    \caption{Action of Verlinde line operators on states corresponding to primary fields in Ising CFT}
\end{table}

We will first describe the lattice realization of the $\widehat{N}$ line operator. As we discussed in \autoref{sec:line-op} before, the lattice $Y$ operator
    \begin{equation}
        Y = (-q)^{-1/2} \,   \, g^{-1}_{1} \ldots g_{2 \R - 1}^{-1}  \uR^{-1} 
        +  (-q)^{1/2} \, \uR \, g_{2 \R -1}  \ldots g_1    \,  ,
    \end{equation}
provides a realization of  the $\widehat{\cal D}_{(1,2)}$ line operator. Recall also that the $Y$  are, up to a factor, transfer matrices  at spectral parameter $\pm$i$\infty$. Therefore   they map states from the odd sector to the even sector and vice versa, or equivalently states of the TFI chain to the dual TFI chain and vice versa.

 Now, let us study the product $Y e_i$. We know that  
\begin{equation}
    Y e_i = e_i Y \, , 
\end{equation}
as $Y$ lies in the center of affine TL \cite{Belletete:2018eua}. Consider $Y e_i \ket{\psi}$, where first $\ket{\psi}$ lies in the odd RSOS sector.
\begin{equation}\label{eq:Y_move_1}
\begin{split}
 & Y e_{2i} \ket{\psi}  =  Y \left( \frac{1}{\sqrt{2}} (1 + \sigma_{i}^{x}) \right) \ket{\psi} \\ 
& =  e_{2i} Y \ket{\psi}    =   \frac{1}{\sqrt{2}} \left(1 + \sigma^{z}_{i - \frac{1}{2}}\sigma^{z}_{i + \frac{1}{2}}\right)   \ket{ Y \psi}
\end{split}
\end{equation}
where the equality in the first line follows from Eq. \eqref{eq:ei_op_pauli_odd}, whereas the equality in second line follows from Eq. \eqref{eq:ei_op_pauli_even} as $\ket{Y \psi}$ lies  in the  even sector since $Y$ maps RSOS states from even to odd sector and vice-versa. Similarly, we can show that for $e_{2i + 1}$
\begin{equation}\label{eq:Y_move_2}
\begin{split}
    & Y e_{2 i + 1 } \ket{\psi} =  Y \left( \frac{1}{\sqrt{2}} (1 + \sigma_{i}^{z} \sigma_{i + 1}^{z}) \right) \ket{\psi}\\
    & =  e_{2 i + 1 } Y \ket{\psi} =   \frac{1}{\sqrt{2}}(1 + \sigma^{x}_{i + \frac{1}{2}})  \ket{Y \psi} \, .
\end{split}
\end{equation}

Hence \eqref{eq:Y_move_1}, \eqref{eq:Y_move_2} imply  
\begin{equation}
    Y \sigma_i^x = \sigma_{i - \frac{1}{2}}^{z}\sigma_{i + \frac{1}{2}}^{z} Y ,  \   Y (\sigma_{i}^{z}\sigma_{i+1}^{z})  = \sigma_{i + \frac{1}{2}}^{x} Y  \ , \, i \in \mathbb{Z} \,  .
\end{equation}
Similarly, when  taking $\ket{\psi}$  in the even sector, instead of the odd sector which we did above, we would obtain 
\begin{equation}
    Y \sigma_{i - \frac{1}{2} }^z\sigma_{i + \frac{1}{2} }^z = \sigma_i^x Y ,  \ Y \sigma_{i + \frac{1}{2}}^{x} = \sigma_{i}^z \sigma_{i + 1}^z Y \ ,  \, i \in \mathbb{Z} \, .
\end{equation}
The above two equations imply that 
\begin{equation}
    Y \sigma_{i - \frac{1}{2} }^z\sigma_{i + \frac{1}{2} }^z = \sigma_i^x Y ,  \ Y \sigma_{i + \frac{1}{2}}^{x} = \sigma_{i}^z \sigma_{i + 1}^z Y \ ,  \, i \in \frac{1}{2} \mathbb{Z} \, ,
\end{equation}
which is exactly the Kramers-Wannier duality transformation for a spin-chain which contains two copies of Ising \cite{Li:2023ani}. In fact, it is not hard to check that the $Y$ operator acts, up to a constant phase, exactly like the $\mathcal{N}$ operator in \cite{Li:2023ani}.
Let us now study a slightly different operator, $\uR^{-1}Y $, which  does not mix the sectors. It is therefore   well-defined  on the TFI chain, and hence can be written in terms of Pauli operators. In fact, in \autoref{sec:Yop-shao-seib}, we show that 
\begin{equation}
   \uR^{-1} Y  = \text{e}^{2 \pi i \frac{\Ll}{8}} \frac{1+\varepsilon}{\sqrt{2}} \frac{1-i \sigma^x_\Ll}{\sqrt{2}} \frac{1-i \sigma^z_\Ll \sigma^z_{\Ll-1}}{\sqrt{2}} \cdots \frac{1-i \sigma^z_2 \sigma^z_1}{\sqrt{2}} \frac{1-i \sigma^x_1}{\sqrt{2}} \, ,  
\end{equation}
coincides with  the D operator of \cite{Seiberg:2024gek}, argued in that reference to  flow to the (1,2) Verlinde line operator in the Ising CFT. 

In \autoref{sec:anyon-chain-RSOS}, we also show that, up to a sign, the $Y$ operator is unitarily equivalent to the  ``topological symmetry operator'' of \cite{Gils_2009}.

\bigskip

We now discuss the (2,1) defect operator in the context of the  Ising model. As  discussed in \autoref{sec:line-op}, we expect $T\left( \frac{\pi}{2}\right)$ to provide a  lattice realization of the $\widehat{\cal{D}}_{(2,1)}$ line operator. 

Let us therefore see  how the transfer matrix behaves when the spectral parameter is $\frac{\pi}{2}$. Its action can be represented as in the diagram below:
\begin{figure}[H]
    \centering
            \begin{adjustbox}{max totalsize={0.9\textwidth}{0.9\textheight},center}
            \begin{tikzpicture}[thick, scale=0.9]
            \draw[black, thick] (0,0) -- (18,0);
            \draw[black, thick] (0,3) -- (18,3);
            \draw[black, thick] (0,0) -- (0,3);
            \draw[black, thick] (3,0) -- (3,3);
            \draw[black, thick] (6,0) -- (6,3);
            \draw[black, thick] (9,0) -- (9,3);
            \draw[black, thick] (12,0) -- (12,3);
            \draw[black, thick] (15,0) -- (15,3);
            \draw[black, thick] (18,0) -- (18,3);
            \node[] at (0,-0.3) {\Large{$x_0$}};
            \node[] at (3,-0.3) {\Large{$2$}};
            \node[] at (6,-0.3) {\Large{$x_1$}};
            \node[] at (9,-0.3) {\Large{$2$}};
            \node[] at (10.5,-0.3) {\Large{$\ldots$}};
            \node[] at (12,-0.3) {\Large{$x_{\Ll-1}$}};
            \node[] at (15,-0.3) {\Large{$2$}};
            \node[] at (18,-0.3) {\Large{$x_{\Ll} = x_0$}};
            \node[] at (0,3.4) {\Large{$2$}};
            \node[] at (3,3.4) {\Large{$x'_{\frac{1}{2}}$}};
            \node[] at (6,3.4) {\Large{$2$}};
            \node[] at (9,3.4) {\Large{$x'_{\frac{3}{2}}$}};
            \node[] at (12,3.4) {\Large{$2$}};
            \node[] at (10.5,3.4) {\Large{$\ldots$}};
            \node[] at (15,3.4) {\Large{$x'_{\Ll - \frac{1}{2}}$}};
            \node[] at (18,3.4) {\Large{$2$}};
            \node[] at (1.5,1.5) {\Large{$\frac{\pi}{2}$}};
            \node[] at (4.5,1.5) {\Large{$\frac{\pi}{2}$}};
            \node[] at (7.5,1.5) {\Large{$\frac{\pi}{2}$}};
            \node[] at (10.5,1.5) {\Large{$\ldots$}};
            \node[] at (13.5,1.5) {\Large{$\frac{\pi}{2}$}};
            \node[] at (16.5,1.5) {\Large{$\frac{\pi}{2}$}};
    \end{tikzpicture}
\end{adjustbox}
    \caption{ Transfer matrix element  $\langle 2,x'_{\frac{1}{2}},2,x'_{\frac{3}{2}},\ldots, x'_{\Ll-\frac{1}{2}} \left| T\left(\frac{\pi}{2}\right) \right| x_0, 2, x_1, 2, \ldots, 2, x_{\Ll-1}, 2 \rangle$.}
    \label{fig:(2,1)IsingDefectOperator}
\end{figure}
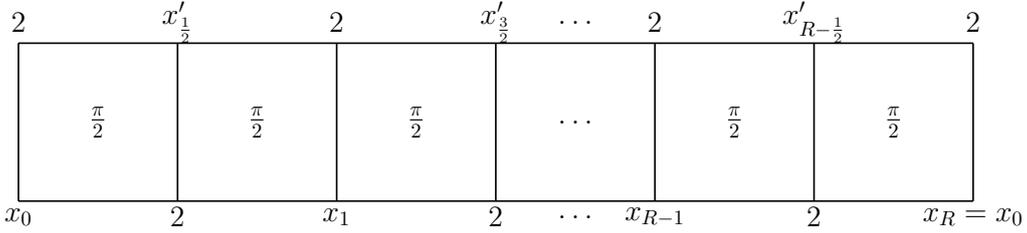

The first box can written using the face weight formula given in \eqref{eq:standard_face_weight}, 
\begin{equation}
\begin{split}
      W \!  \!  \left(\begin{array}{ll}
2 & x'_{\frac{1}{2}} \\
x_0 & 2
\end{array} \Bigg| \ \frac{\pi}{2}\right) &= \sqrt{\frac{\gf_{x'_{\frac{1}{2}}} \gf_{x_0}}{\gf_2 \gf_2}} \frac{\sin \frac{\pi}{2}}{\sin \gamma} + \delta_{x_0, x'_{\frac{1}{2}}} \frac{\sin (\gamma - \frac{\pi}{2})}{\sin \gamma} \\
& = 1 - \delta_{x_0, x'_{\frac{1}{2}}} 
\end{split}
\end{equation}
which is $0$, if $x_0 = x'_{\frac{1}{2}}$ and $1$ otherwise. Hence, to have a non-zero matrix element from  the figure above, we must have $x'_{\frac{1}{2}} =  x^*
_0$, where we define $1^* = 3$ and $3^* = 1$. Let us now examine the second box in figure \ref{fig:(2,1)IsingDefectOperator}. 
\begin{equation}
\begin{split}
    W \!  \! \left(\begin{array}{ll}
x'_{\frac{1}{2}} & 2 \\
2 & x_1
\end{array} \Bigg| \ \frac{\pi}{2} \right) &= \frac{\sin(\gamma - \frac{\pi}{2})}{\sin \gamma} + \frac{\sin \frac{\pi}{2}}{\sin \gamma}\sqrt{\frac{\gf_2 \gf_2}{\gf_{x'_{\frac{1}{2}}} \gf_{x_1}}} \delta_{x'_{\frac{1}{2}}, x_1} \\
 & = - 1 + 2 \, \delta_{x'_{\frac{1}{2}}, x_1} = -1 + 2 \, \delta _{x^*_0, x_1}
\end{split}
\end{equation}
which is $-1$ if $x_0 = x_1$ and $ 1$ if $x_0 \neq x_1$. So, we collect negative sign if two neighboring physical spins $(x_i \, 's)$ are the same. In a row, for instance in figure \ref{fig:(2,1)IsingDefectOperator}, there are $\Ll$ physical spins, therefore the number of times spin can possibly change is $\Ll $. Also, due to the PBC, the number of times the height changes is even - say $2k$, when we start from $x_0$ and end back at $x_0$. Therefore, as we start from $x_0$ and end at $x_0$, the number of times we get $-$ sign is $\Ll - 2k$, which is $-1$ if $\Ll$ is odd and $1$ if $\Ll$ is even. Hence
\begin{equation}
   T\left( \frac{\pi}{2} \right) \ket{x_0, 2, x_1, 2, \ldots, 2, x_{\Ll-1}, 2} = (-1)^{\Ll} \ket{2, x^*_0, 2, x^*_1, 2, \ldots, 2, x^*_{\Ll-1}} \, .
\end{equation}
One can check that $T\left(\frac{\pi}{2}\right)$ acts in the same way on states in even sector, i.e.
\begin{equation}
    T\left( \frac{\pi}{2} \right) \ket{2,x_{\frac{1}{2}}, 2,x'_{\frac{3}{2}}, \ldots, x_{\Ll-\frac{1}{2}} }= (-1)^{\Ll} \ket{x_{\Ll-\frac{1}{2}}^{*}, 2,x_{\frac{1}{2}}^{*}, 2,x_{\frac{3}{2}}^{*}, \ldots,x_{\Ll - \frac{3}{2}}^{*}, 2 }\, .
\end{equation}
Therefore we observe that 
\begin{equation}
  \uR^{-1}   T\left(\frac{\pi}{2}\right) = (-1)^\Ll \widehat{\eta} \, , 
\end{equation}
where $\widehat{\eta}$  keeps the height $2$ fixed, but interchanges $1$ and $3$, which corresponds to $\uparrow/\downarrow$ when we map sectors of RSOS model to TFI chain. If we just consider one sector of the RSOS Hilbert space,  $\widehat{\eta}$ then acts exactly like the spin flip operator, i.e. $\prod \sigma_i^{x}$, which we know flows to the (2,1) or the (1,3) line operator in the Ising CFT  \cite{Aasen:2016dop}, hence our general result  
\begin{equation}
   \widehat{{\cal D}}_{(2,1)}^{(\rm latt)} = (-1)^\R\uR^{-1} T\left(\frac{\pi}{2}\right) \, ,
\end{equation}
  is verified in  the case of TFI. As we expect $\uR \sim 1$ for states with low momentum in the   limit of large systems, we could have also worked with $T\left(\frac{\pi}{2}\right)$ as the lattice realization of the $\widehat{\cal{D}}_{(2,1)}$ line operator in the $A_3$  RSOS model. 

For any given state in A$_3$ RSOS Hilbert space, using the definition given in Eq. \eqref{def:heightreflection}, it is easy to see that $\mathcal{R}$ swaps $1$ and $3$, and keeps $2$ fixed. hence it also exactly acts like a spin flip operator. We expected the same as we had noted in section \ref{sec:fusion-Tmatrices-(1s)} that ${\cal R}$ is exactly $\widehat{{\cal D}}_{(1,p)}^{(\rm latt)}$.

\subsubsection{Direct Channel}
From the partition function of the cylinder with a line operator inserted, using a modular transformation, one can determine the partition function associated to  the defect Hilbert space.  In the case of the  Ising model, the defect Hilbert space can be figured out from Eq. \eqref{eq:twist-part-func}, and is the following 
\begin{equation}
\begin{split}    
    {\cal H}_{(1,2)} =& \left(\frac{1}{16},0 \right) \oplus  \left(0,\frac{1}{16} \right)  \oplus\left(\frac{1}{2},\frac{1}{16} \right) \oplus \left(\frac{1}{16} , \frac{1}{2}\right)  \, , \\
    {\cal H}_{(2,1)} =& \left(\frac{1}{2},0 \right) \oplus  \left(0,\frac{1}{2} \right) \oplus \left(\frac{1}{16},\frac{1}{16} \right) \, .  \\
\end{split}
\end{equation}
We now discuss the lattice Hamiltonians which realize these  defect Hilbert spaces in the IR.  We start with the  $N$ TDL.
\bigskip

As  discussed in \autoref{sec:def-ham}, when we set $\tilde{u} = \pm {\rm i} \infty$ in Eq. \eqref{eq:H12} and \eqref{eq:anti-chiral-(1,2)}, we get the following Hamiltonians: 
 \begin{equation}
H_{(1,2)} = -\frac{\gamma}{\pi\sin\gamma}\sum_{i = 1}^{2L}e_i + \frac{\gamma}{\pi\sin\gamma}\left(qe_{k}e_{k+1} + q^{-1}e_{k+1}e_{k}\right) \, .  
\end{equation}
and 
\begin{equation}
H_{\overline{(1,2)}} = -\frac{\gamma}{\pi\sin\gamma}\sum_{i = 1}^{2L}e_i + \frac{\gamma}{\pi\sin\gamma}\left(q^{-1}e_{k} e_{k+1} + qe_{k+1}e_{k}\right) \, . 
\end{equation}
Since the $e_i$ operators do not change the sector, we can write these Hamiltonians in terms of Pauli operators. For instance  in the odd sector, using \eqref{eq:ei_op_pauli_odd} and   setting $k = 2k_0$, we get 
\begin{equation}\label{eq:dual-def-Ham-Is}
    H_{(1,2)} =\sum_{i = 1}^L \left( -\frac{1}{4} \left(1 + \sigma_{i}^z\sigma_{i+1}^z \right) - \frac{1}{4} \left(1 + \sigma_i^x \right) \right) + \frac{1}{4} \left(1 + \sigma_{k_0}^x + \sigma_{k_0}^z\sigma_{k_0+1}^z + \sigma_{k_0}^y \sigma_{k_0+1}^z \right) \, , 
\end{equation}
\begin{equation}
    H_{\overline{(1,2)}} = \sum_{i = 1}^L \left( -\frac{1}{4} \left(1 + \sigma_{i}^z\sigma_{i+1}^z \right) - \frac{1}{4} \left(1 + \sigma_i^x \right) \right)  + \frac{1}{4} \left(1 + \sigma_{k_0}^x + \sigma_{k_0}^z\sigma_{k_0+1}^z - \sigma_{k_0}^y \sigma_{k_0+1}^z \right) \, , 
\end{equation}
which are related by the local unitary operator $\sigma_{k_0}^z$. Further, the Hamiltonian in Eq. \eqref{eq:dual-def-Ham-Is} is the ``duality defect Hamiltonian''~\cite{Oshikawa1997, Grimm1992, Grimm2001}, which is known to flow to the (1,2) Defect Ising CFT.

\bigskip

Now, we discuss the $(2,1)$ defect Hamiltonian for Ising CFT. When we set $\tilde{u} = \pm \frac{\pi}{2}$ and again substitute the form of TL generators from Eq. \eqref{eq:ei_op_pauli_odd}, we get 
\begin{equation}\label{eq:21-def-ising}
 H_{(2,1)} = H_{\overline{(2,1)}} =\sum_{i = 1}^L \left( -\frac{1}{4} \left(1 + \sigma_{i}^z\sigma_{i+1}^z \right) - \frac{1}{4} \left(1 + \sigma_i^x \right) \right) + \frac{1}{2} \left(1 + \sigma_{k_0}^z \sigma_{k_0 + 1}^z + \sigma_{k_0}^x \right) \, , 
\end{equation}
which can be shown to be unitarily equivalent to the anti-periodic boundary condition TFI model using the local unitary $\sigma_{k_0}^z$. It is known that the anti-periodic boundary condition Hamiltonian realizes the (2,1) defect in the Ising CFT \cite{Aasen2016}.

Note here the difference between crossed and direct channel. While in the crossed channel we have to take into account the existence of odd and even sectors and introduce $\uR$ operators in our definitions of the defects, in the direct channel, everything simply happens in one sector.  

\subsection{Even and odd  sectors and numerics}\label{subsec:tau_op_sign}

 For other models, we will need to carry out numerical calculations, and thus face  the complication due to the existence of even and odd sectors.  

 Although we discussed this in section \ref{specialsubsec}, it is probably worthwhile going over the point again, especially for a reader who may have skipped earlier sections.  

 Unlike the CFTs they flow to, the groundstates of the lattice RSOS models are  doubly degenerate. All the Hamiltonians we consider, for models with or without defects, are written using Temperley-Lieb generators on a periodic Hilbert space. They have two groundstates, one that lies in the even and the other in the odd sector. A state $\ket{a_0, a_1, \ldots, a_{2L-1}}$ is said to lie in the \textit{even (odd) }sector if $a_0$ is even (odd).  

 For the periodic Hamiltonian with no defect, the two groundstates, $\ket{\psi^{(0)}_{E}}$ and $ \ket{\psi^{(0)}_{O}}$, in the even and odd sectors respectively, can be mapped to each other using $\uR$, i.e. $\uR \ket{\psi^{(0)}_{O}} = \ket{\psi^{(0)}_{E}}$ and vice-versa. Using them, we can also can construct eigenstates of $\uR$ with eigenvalues $1$ and $-1$. These are the two ground states of the (off diagonal) transfer matrix see section \ref{specialsubsec}. In the continuum CFT, the vacuum for these A-type unitary minimal model CFTs is the state $\ket{0,0}$, which has 0 momentum. As the log of the shift operator eigenvalue is related to momentum, we call the state with $\uR$ eigenvalue $1$ the \textit{symmetric} state, and the state with $\uR$ eigenvalue $-1$ as the \textit{antisymmetric} state, and denote them by $\ket{\psi^{(0)}_{S}}$ and $\ket{\psi^{(0)}_{A}}$ respectively.
 
We can similarly define symmetric and antisymmetric states for excited energy eigenstates, again the momentum eigenvalue of the symmetric states agree with the continuum whereas the same for antisymetric ones does not. This definition also extends to defect Hamiltonian, where instead of $\uR$, we use the local translation operator.

Note that $\ket{\psi^{(0)}_{S}}$ and $\ket{\psi^{(0)}_{A}}$ are eigenstates of the transfer matrix for any value of the  spectral parameter, because of commutation with  the shift operator. Further, the corresponding eigenvalues only differ by a sign. We see then that, by adding $\uR^{-1}$ in Eq. \eqref{Dguess} together with the extra sign in Eq. (\ref{expectedvalues1}) for $\R$ odd (and raised to  the proper power for higher defects) we ensure that the expectation value of the defect operator is the same for both the $\ket{\psi^{(0)}_S}$ and $\ket{\psi^{(0)}_A}$. 

In practice now, when confronted with numerics for a lattice model, we can  proceed blindly and evaluate the expectation values of the defects with all their correcting factors, for all states.

However, since we will in what follows focus almost exclusively on the realizations of defects as in eqs. (\ref{eq:1s-discretization}) and (\ref{eq:r1-discretization}), we can simply do numerics on the symmetric eigenstates  - whose momentum has always vanishing finite part.  Note that for those states, the formula written earlier for lattice topological defects can be simplified: we can remove the powers of $\uR$, and we will always get the same result (in the continuum limit) as if we had acted with the full $\widehat{\cal D}_{(1,s)}^{(\rm latt)}$ operator. For the ${\cal D}_{(r,1)}$ defects, we will then still get a sign wrong when $\R$ is odd and $r$ even, which we can correct by adding a $(-1)^{\R(r-1)}$. This will then give correct results in all cases. 
To emphasize this point in this section, we refer to the eigenvalues of the defect operators thus obtained as $\lattDS_{(r,s)}$, for instance
\begin{equation}\label{eq:lattDSdef}
   \lattDS_{(1,2)} = Y \, , \quad \lattDS_{(2,1)} = T\left(\frac{\pi}{2}\right) \, .
\end{equation}
 Whether we use these or the $\lattD_{(r,s)}$ however, some finite size results will be affected by factors of the type
$\exp\left[{2i\pi\over 2\R}(h-\bar{h})\right]$.

 To drive these points home, we now show results for the action of the defect line operators  $\lattDS_{(r,s)}$ and $\widehat{\cal D}_{(r,s)}^{(\rm latt)}$ on symmetric and antisymmetric states  in the case of the $A_3$ RSOS model. 
\begin{table}[H]\renewcommand{\arraystretch}{1.2}
\centering   
\begin{tabular}{ |P{2.2cm} | P{2.8cm} |P{3.8cm}|P{1.0cm} |P{3.5cm}|P{1.0cm} | }
\hline
 \multicolumn{6}{|c|}{${\cal M}(4,3)$ States and expectation values - 16 RSOS sites } \\
 \hline
 State  &  $\tau$ & $\lattDS_{(2,1)}$ {\color{blue}($\widehat{{\cal D}}_{(2,1)}^{(\rm latt)}$)}   & $\widehat{{\cal D}}_{(2,1)}$   &  $\lattDS_{(1,2)}$ {\color{blue}($\widehat{{\cal D}}_{(1,2)}^{(\rm latt)}$)}  &  $\widehat{{\cal D}}_{(1,2)}$ \\ [2ex]
 \hline
  $\ket{0,0}_S  $  & $1$&  $1$  {\color{blue}$( 1)$}   & $1$  & $\sqrt{2}$ ${\color{blue}(\sqrt{2})}$ &  $\sqrt{2}$  \\
  $\ket{0,0}_A   $  & $-1$ & $-1$  {\color{blue}$( 1)$}   & $1$  &  $-\sqrt{2}$ ${\color{blue}(\sqrt{2}})$ &  $\sqrt{2}$  \\
 $\ket{\frac{1}{16},\frac{1}{16}}_S\ $ &  1  &  $-1$ {\color{blue}$( -1)$}  & $-1$ & 0 ${\color{blue}(0})$& $0$  \\
 $\ket{\frac{1}{16},\frac{1}{16}}_A $ &  $-1$  &  $1$ {\color{blue}$( -1)$}  & $-1$  & $0$ ${\color{blue}(0})$ & $0$  \\
 $\ket{\frac{1}{2},\frac{1}{2}}_S $  &  1   &  $1$  {\color{blue}$( 1)$} &   $1$  & $-\sqrt{2}$ ${\color{blue}(-\sqrt{2}})$ & $-\sqrt{2}$  \\
 $\ket{\frac{1}{2},\frac{1}{2}}_A$  &  $-1$   &   $-1$  {\color{blue}$( 1)$} &   $1$  & $\sqrt{2}$ ${\color{blue}(-\sqrt{2}})$& $-\sqrt{2}$ \\
  $L_{-1}\ket{\frac{1}{16},\frac{1}{16}}_S\ $ &  $\exp(\frac{{\rm i} \pi}{8} )$  &  $-\exp(\frac{{\rm i} \pi}{8} )$ {\color{blue}$( -1)$}  & $-1$ & 0 ${\color{blue}(0)}$ & $0$  \\
  $\bar{L}_{-1}\ket{\frac{1}{16},\frac{1}{16}}_S\ $ &   $\exp(-\frac{{\rm i} \pi}{8} )$ & $-\exp(-\frac{{\rm i} \pi}{8} )$ {\color{blue}$( -1)$} & $-1$ & 0 ${\color{blue}(0)}$ & $0$  \\
    $L_{-1}\ket{\frac{1}{16},\frac{1}{16}}_A \ $ &  $- \exp(\frac{{\rm i} \pi}{8} )$  &  $\exp(\frac{{\rm i} \pi}{8} )$ {\color{blue}$( -1)$}  & $-1$ & 0 ${\color{blue}(0)}$ & $0$  \\
 $\bar{L}_{-1}\ket{\frac{1}{16},\frac{1}{16}}_A $ &  $-\exp(-\frac{{\rm i} \pi}{8} )$  &  $\exp(-\frac{{\rm i} \pi}{8} )$ {\color{blue}$(-1)$}  & $-1$  & $0$ ${\color{blue}(0)}$ & $0$  \\
  $L_{-2}\ket{0,0}_S $  &  $\exp(\frac{{\rm i} \pi}{4} )$   &  $\exp(\frac{{\rm i} \pi}{4} )$  {\color{blue}$( 1)$} &   $1$  & $\sqrt{2} $  \, {\color{blue} ($\sqrt{2}\exp(-\frac{{\rm i} \pi}{4} ))$ } & $\sqrt{2}$  \\
   $\bar{L}_{-2}\ket{0,0}_S$  & $\exp(-\frac{{\rm i} \pi}{4} )$   &   $\exp(-\frac{{\rm i} \pi}{4} )$  {\color{blue}$( 1)$} &   $1$  & $\sqrt{2} $  \,  {\color{blue} $(\sqrt{2}\exp(\frac{{\rm i} \pi}{4} ))$ } & $\sqrt{2} $ \\
 $L_{-2}\ket{0,0}_A$  & $ -\exp(\frac{{\rm i} \pi}{4} )$   &   $ -\exp(\frac{{\rm i} \pi}{4} )$  \, \,  {\color{blue}$( 1)$} &   $1$  & $-\sqrt{2}$ {\color{blue} $(\sqrt{2}\exp(-\frac{{\rm i} \pi}{4} )) 
 $ } & $\sqrt{2}$ \\
 $\bar{L}_{-2}\ket{0,0}_A$  & $ -\exp(-\frac{{\rm i} \pi}{4} )$   &   $ -\exp(-\frac{{\rm i} \pi}{4} )$  {\color{blue}$( 1)$} &   $1$  & $-\sqrt{2}$ {\color{blue} $(\sqrt{2}\exp(\frac{{\rm i} \pi}{4} ))$ } & $\sqrt{2}$ \\
 \hline
\end{tabular}
\caption{In this table and the next, we list the action of the lattice discretization of the $(2,1)$ and the $(1,2)$ Verlinde lines, along with the results from the continuum. Observe how the $\widehat{{\cal D}}_{(r,s)}^{(latt),0}$ operators provide the most convenient discretization when restricted to symmetric states, while their signs are off for anti-symmetric ones. }
\label{tab:A3:16:phy-unphy}
\end{table}

\begin{table}[H]\renewcommand{\arraystretch}{1.2}
\centering   
\begin{tabular}{ |P{2.2cm} | P{2.6cm} |P{3.2cm}|P{1.0cm} |P{3.6cm}|P{1.0cm} | }
\hline
 \multicolumn{6}{|c|}{${\cal M}(4,3)$ States and expectation values - 18 RSOS sites } \\
 \hline
 State  &  $\tau$ & $\lattDS_{(2,1)}$ {\color{blue}($\widehat{{\cal D}}_{(2,1)}^{(\rm latt)}$)}   & $\widehat{{\cal D}}_{(2,1)}$   &  $\lattDS_{(1,2)}$ {\color{blue}($\widehat{{\cal D}}_{(1,2)}^{(\rm latt)}$)}  &  $\widehat{{\cal D}}_{(1,2)}$ \\ [2ex]
 \hline
  $\ket{0,0}_S  $  & $1$&  $1$  {\color{blue}$( 1)$}   & $1$  & $\sqrt{2}$ ${\color{blue}(\sqrt{2})}$ &  $\sqrt{2}$  \\
  $\ket{0,0}_A   $  & $-1$ & $-1$  {\color{blue}$( 1)$}   & $1$  &  $-\sqrt{2}$ ${\color{blue}(\sqrt{2}})$ &  $\sqrt{2}$  \\
 $\ket{\frac{1}{16},\frac{1}{16}}_S\ $ &  1  &  $-1$ {\color{blue}$( -1)$}  & $-1$ & 0 ${\color{blue}(0})$& $0$  \\
 $\ket{\frac{1}{16},\frac{1}{16}}_A $ &  $-1$  &  $1$ {\color{blue}$( -1)$}  & $-1$  & $0$ ${\color{blue}(0})$ & $0$  \\

 $\ket{\frac{1}{2},\frac{1}{2}}_S $  &  1   &  $1$  {\color{blue}$( 1)$} &   $1$  & $-\sqrt{2}$ ${\color{blue}(-\sqrt{2}})$ & $-\sqrt{2}$  \\
 $\ket{\frac{1}{2},\frac{1}{2}}_A$  &  $-1$   &   $-1$  {\color{blue}$( 1)$} &   $1$  & $\sqrt{2}$ ${\color{blue}(-\sqrt{2}})$& $-\sqrt{2}$ \\
 
  $L_{-1}\ket{\frac{1}{16},\frac{1}{16}}_S\ $ &  $\exp(\frac{{\rm i} \pi}{9} )$  &  $-\exp(\frac{{\rm i} \pi}{9} )$ {\color{blue}$( -1)$}  & $-1$ & 0 ${\color{blue}(0)}$ & $0$  \\
  $\bar{L}_{-1}\ket{\frac{1}{16},\frac{1}{16}}_S\ $ &   $\exp(-\frac{{\rm i} \pi}{9} )$ & $-\exp(-\frac{{\rm i} \pi}{9} )$ {\color{blue}$( -1)$} & $-1$ & 0 ${\color{blue}(0)}$ & $0$  \\
  
    $L_{-1}\ket{\frac{1}{16},\frac{1}{16}}_A \ $ &  $-\exp(\frac{{\rm i} \pi}{9} )$  &  $\exp(\frac{{\rm i} \pi}{9} )$ {\color{blue}$( -1)$}  & $-1$ & 0 ${\color{blue}(0)}$ & $0$  \\
 $\bar{L}_{-1}\ket{\frac{1}{16},\frac{1}{16}}_A $ &  $-\exp(-\frac{{\rm i} \pi}{9} )$  &   $\exp(-\frac{{\rm i} \pi}{9} )$ {\color{blue}$(-1)$}  & $-1$  & $0$ ${\color{blue}(0)}$ & $0$  \\
 
  $L_{-2}\ket{0,0}_S $  &    $\exp(\frac{{\rm i} 2\pi}{9} )$   &  $\exp(\frac{{\rm i} 2\pi}{9} )$  {\color{blue}$( 1)$} &   $1$  & $\sqrt{2} $  {\color{blue} $(\sqrt{2} \exp(-\frac{{\rm i}2 \pi}{9}) )$ } & $\sqrt{2}$  \\
  
   $\bar{L}_{-2}\ket{0,0}_S$  & $\exp(-\frac{{\rm i} 2\pi}{9} )$   &   $\exp(-\frac{{\rm i} 2\pi}{9} )$  {\color{blue}$( 1)$} &   $1$  & $\sqrt{2} $ {\color{blue} $(\sqrt{2} \exp(\frac{{\rm i}2 \pi}{9}) )$ } & $\sqrt{2}$ \\
   
 $L_{-2}\ket{0,0}_A$  & $ -\exp(\frac{{\rm i} 2\pi}{9} )$   &   $ -\exp(\frac{{\rm i} 2\pi}{9} ) $\, \,  {\color{blue}$( 1)$} &   $1$  & $-\sqrt{2}$ {\color{blue} $(\sqrt{2} \exp(-\frac{{\rm i}2 \pi}{9}) )$} & $\sqrt{2}$ \\
 $\bar{L}_{-2}\ket{0,0}_A$  & $-\exp(-\frac{{\rm i} 2\pi}{9} )$   &   $ -\exp(-\frac{{\rm i} 2\pi}{9} )$  {\color{blue}$( 1)$} &   $1$  & $-\sqrt{2}$  {\color{blue} $(\sqrt{2} \exp(\frac{{\rm i}2 \pi}{9}) )$ } & $\sqrt{2}$  \\
 \hline
\end{tabular}
\caption{See caption of Table \ref{tab:A3:16:phy-unphy}, the data here is for 18 sites.}
\label{tab:A3:phy-unphy}
\end{table}

\subsection{A\texorpdfstring{$_4$}{Lg} RSOS model : Tri-Critical Ising}
The A$_4$ RSOS model leads to the tri-critical Ising (TCI) CFT - $\mathcal{M}(5,4)$, which is a Virasoro minimal model with $c=7/10$ \footnote{We do not discuss here the fermionic $c = 7/10$  CFT with superconformal symmetry - SVIR$_3$, see \cite{Makabe:2017ygy}.}. In contrast with the Ising case, there is no simple map to a natural  spin chain model such as the Blume-Capel model. 

\bigskip 

The ${\cal M}(5,4) $ CFT has 6 primary fields, whose labels and conformal dimensions - and thus the corresponding Verlinde lines - are given in the table below
\begin{table}[H]\renewcommand{\arraystretch}{1.2}
        \centering
        \begin{tabular}{ |m{2cm}|m{1cm}|m{1cm}|m{1cm}|m{1cm}|m{1cm}|m{1cm}|}
    \hline
         &    $\mathbf{1}$  &  $\epsilon$  & $\epsilon'$  & $\epsilon''$  & $\sigma'$& $\sigma$  \\
    \hline
   Kac label       & (1,1)  & (1,2) & (1,3) &(1,4)  &  (2,1) & (2,2) \\
    \hline
   Conformal dimension        &   0 &  $\frac{1}{10}$ & $\frac{3}{5}$    &  $\frac{3}{2}$ &  $\frac{7}{16}$  & $\frac{3}{80}$  \\
   \hline
  Verlinde line        &   $\mathbf{1}$ &  $\eta W$ & $W$    &  $\eta$ &  $N$  & $WN$  \\
   \hline

            \end{tabular}
        \caption{Primary fields in TCI, their Kac labels, conformal dimensions and the associated Verlinde lines.}
        \label{tab:TCI-primary}
    \end{table}
Note we  will also sometimes denote these lines by their Kac labels, for instance $WN = {\cal D}_{(2,2)}$. The $W$ line is also known as the Fibonacci line since 
\begin{equation}
    W^2 = \mathbf{1} + W \, , 
\end{equation}
whereas the $\mathbf{1}$, $\eta$, and $N$ lines have the same fusion rules as in Eq. \eqref{eq:Z2-TY-fusion}. These TDLs again give rise to line operators and defect Hilbert space,  whose lattice realizations we discuss next.

\subsubsection{Crossed Channel}
We first describe the action of line operators on states of the TCI Hilbert space. 
\begin{table}[H]\renewcommand{\arraystretch}{1.2}
\centering
\begin{tabular}{|c|c|c|c|c|c|c|}
\hline
& $\mathbf{1}$ & $\epsilon$ & $\epsilon'$ & $\epsilon''$ & $\sigma'$ & $\sigma$ \\
\hline
$\widehat{\mathcal{D}}_{(1,3)}  = \widehat{W}$  & $\zeta$ & $-\zeta^{-1}$ & $-\zeta^{-1}$ & $\zeta$ & $\zeta$ & $-\zeta^{-1}$ \\
\hline
$ \widehat{\mathcal{D}}_{(1,4)}  = \widehat{\eta}$  & 1 & 1 & 1 & 1 & $-1$ & $-1$ \\
\hline
$\widehat{\mathcal{D}}_{(2,1)}  = \widehat{N}$ & $\sqrt{2} $& $-\sqrt{2}$ & $\sqrt{2}$ & $-\sqrt{2} $ & 0 & 0 \\
\hline
\end{tabular}
\caption{Action of Verlinde line operators on states corresponding to primary fields in the  TCI CFT.}
\end{table}
\noindent where $\zeta = \frac{1 + \sqrt{5}}{2}$ is the golden ratio. The action of other line operators, i.e. ${{\cal \widehat D}_{(1,2)}} = \widehat{\eta}\,\widehat{W},$ and ${\cal \widehat D}_{(2,2)} =  \widehat{W} \, \widehat{N} $, can be determined from the actions of $\widehat{\eta}, \, \widehat{W},  $ and $\widehat{N}$ given in the table above.
\par 
As  discussed in section \ref{sec:line-op}, and like in  the Ising case, the lattice realizations of $(1,2)$ and $(2,1)$ lines are given by the transfer matrix at spectral parameters $ \pm{\rm i} \infty$ and $\pm \frac{\pi}{2}$ respectively.  However, as we had noted in section \ref{sec:21-discretization}, such identification leaves open, in the $(2,1)$ case, the question of normalization.  For spectral parameter $\pi/2$, we will show later in this section using Bethe ansatz, that this factor is exactly $ \left( \sin (\gamma + \frac{\pi}{2}) / \sin \gamma \right)^{2 {\rm R}} =  \cot^{2 {\rm R}} \gamma $. Therefore, the data in the column of $\widehat{\cal D}^{\rm (latt)}_{(2,1)}$ corresponds to the transfer matrix in Eq. \eqref{eq:tmat_TLgen_KP} at spectral parameter $ \pi/2 $ divided by this factor, i.e. 
\begin{equation}\label{eq:21-norm-line-op}
    \widehat{\cal D}^{\rm (latt)}_{(2,1)} =(-1)^\R \uR^{-1}\tan^{2 \R} \left( \gamma \right) \, \,  T\left(\frac{\pi}{2}\right) \, .
\end{equation}
For A$_3$ RSOS model, the bulk factor was 1, as $\gamma = \pi/4$, so we did not encounter the issue of normalizing the $\widehat{\cal D}^{\rm (latt)}_{(2,1)}$ operator there. Even after removing this factor,  the expectation value for $T(\frac{\pi}{2})$ shows in general a non-trivial phase  for states with non-zero momentum, as shown in some cases  in the table below. There is strong numerical evidence however that these phases go to zero as R$\to\infty$.

\begin{table}[H]\renewcommand{\arraystretch}{1.2}
\centering   
\begin{tabular}{ |P{2cm}|P{4.0cm}|P{1.5cm} |P{4cm}|P{1.5cm} | }
\hline
 \multicolumn{5}{|c|}{${\cal M}(5,4)$ States and expectation values -  20 RSOS sites } \\
 \hline
 State  & $\lattDS_{(2,1)}$ {\color{blue} $(\widehat{{\cal D}}^{(\rm latt)}_{(2,1)})$}   & $\widehat{{\cal D}}_{(2,1)}$  &  $\lattDS_{(1,2)}$ {\color{blue} $(\widehat{{\cal D}}_{(1,2)}^{(\rm latt)})$}  &  $\widehat{{\cal D}}_{(1,2)}$ \\ [2ex]
 \hline
  $\ket{0,0} $  & 1.420720  & $\sqrt{2}$  & $\phi$ & $\phi$  \\

 $\ket{\frac{3}{80},\frac{3}{80}} $ &  0.073106  & 0 & $\phi^{-1}$ & $\phi^{-1}$ \\
 $\ket{\frac{1}{10},\frac{1}{10}} $  &     $-1.39876$ &   $-\sqrt{2}$  & $-\phi^{-1}$ & $-\phi^{-1}$ \\
 $\ket{\frac{7}{16},\frac{7}{16}} $  &    $-0.48105$ & 0  & $-\phi$& $-\phi$    \\
 $L_{-1}\ket{\frac{3}{80},\frac{3}{80}} $ & $0.5269 \, {\rm e}^{-{\rm i} \, 0.5430 }$
 {\color{blue} $(0.5269 \, {\rm e}^{-{\rm i} \, 0.2289 })$}   & 0 & $\phi^{-1}$ {\color{blue} \quad \quad  \quad \quad$(\phi^{-1} \, {\rm e}^{{\rm i}\,  0.3141})$
}& $\phi^{-1}$   \\
 $\Bar{L}_{-1}\ket{\frac{3}{80},\frac{3}{80}} $   & 
 $0.5269 \,  {\rm e}^{{\rm i} \, 0.5430 }$
 {\color{blue} $(0.5269 \, {\rm e}^{{\rm i} \, 0.2289 })$}  & 0 & $\phi^{-1}$ {\color{blue}\quad \quad \quad \quad $(\phi^{-1} \, {\rm e}^{-{\rm i}\,  0.3141})$
}& $\phi^{-1}$ \\
 $L_{-1}\ket{\frac{1}{10},\frac{1}{10}} $    &  $-1.30177 \, {\rm e}^{{\rm i}\, 0.1226}$ ${\color{blue} (-1.30177  {\rm e}^{{\rm i} \, 0.4368}) }
 $ & $-\sqrt{2}$ & $-\phi^{-1}$  {\color{blue} \quad \quad $(-\phi^{-1} \, {\rm e}^{{\rm i}\,  0.3141})$
}&$ -\phi^{-1} $ \\
 $\Bar{L}_{-1}\ket{\frac{1}{10},\frac{1}{10}} $    &   $-1.30177 \, {\rm e}^{-{\rm i}\, 0.1226}$ ${\color{blue} (-1.30177  {\rm e}^{-{\rm i} \, 0.4368}) }
 $ &  $-\sqrt{2}$ & $-\phi^{-1}$ {\color{blue} \quad \quad $(-\phi^{-1} \, {\rm e}^{-{\rm i}\,  0.3141})$
} & $-\phi^{-1}$ \\
 $\ket{\frac{3}{5},\frac{3}{5}} $      &  1.30321  & $\sqrt{2}$ & $-\phi^{-1}$ & $-\phi^{-1}$ \\
 \hline
\end{tabular}
\caption{In this table, we list the action of the lattice discretization of the $(2,1)$ and the $(1,2)$ Verlinde lines, along with the results from the continuum.  In black (blue) is data for defect line operator without (with) $(-1)^\R\uR^{-1}$ for $(2,1)$ line operator and $\uR^{-1}$ for $(1,2)$ line operator. $\phi = \frac{1 + \sqrt{5}}{2}$ is the golden ratio.   }
\label{tab:(12)-(21)-TCI}
\end{table}

The results in the table above confirm the fact  that, while  $\widehat{{\cal D}}_{(1,2)}^{(\rm latt)}$ behaves exactly like its continuum counterpart, the same is not true for $\widehat{{\cal D}}_{(2,1)}^{(\rm latt)}$. The former is topological on the lattice, while the latter is not. This is also clear from the fact that while $\widehat{{\cal D}}_{(1,2)}^{(\rm latt)}$ commutes with  the $e_i$ generators, the same is not true for $\widehat{{\cal D}}_{(2,1)}^{(\rm latt)}$. This explains at least in part  the slow convergence of the results.

\begin{figure}
    \centering
    \includegraphics[width=1\linewidth]{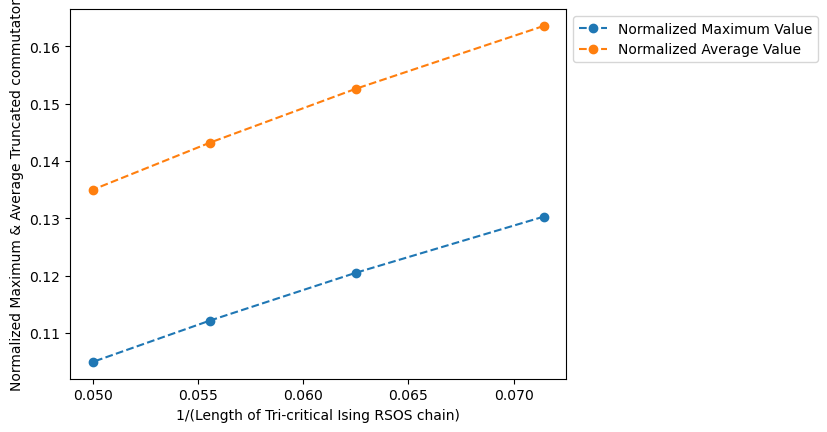}
    \caption{Maximum and Average absolute values of elements in the truncated commutator $[\widehat{{\cal D}}_{(2,1)}^{(\rm latt)}, e_i]$ as a function of inverse system-size, normalized using the operator $\widehat{{\cal D}}_{(2,1)}^{(\rm latt)}e_i$. The lines are constructed by fitting the data points with a polynomial fit.  Moreover, we only studied the lowest 8 energy eigensates. Normalized Maximum Value and Normalized Average Value are found to converge to 0.02088502 and 0.03777982 respectively. This confirms the  topological nature of the line operator in the scaling limit.}
    \label{fig:TCI-W-comm-E-zero}
\end{figure}

Nevertheless, one can see that the lattice results converge to the expected continuum ones as the size is increased. A related intriguing aspect concerns the case when $ \widehat{{\cal D}}_{(2,1)}$ is non-invertible in the CFT. Nonetheless, for any finite size, we find that the eigenvalues are small but non zero, and the lattice operators remain invertible, so that non-invertibility is an exclusive feature of the continuum limit.  In figure \ref{fig:TCI-W-comm-E-zero}, we explore the commutativity of $ \widehat{{\cal D}}_{(2,1)}$ with  TL generators and observe that the two operators appear to commute only in the continuum limit. 

To assess  the convergence of the results such as in Table (\ref{tab:(12)-(21)-TCI}) and go beyond the limitations of the numerics,  we can apply results from  the Bethe ansatz. First, we may use expression (\ref{bulk}) to remove extensive non-universal contributions. Then, using analytical calculations similar to section \ref{gfactors}, one can see that the next most important sub-leading finite-size effect to the expectation value of $T(\pm {\pi \over 2})$ is ${\cal{O}}(\R^{-{1 \over 2}})$ rather than the usual ${\cal{O}}(\R^{-{1}})$, beyond the universal value. 
 
\begin{figure}[htb]
\begin{center}
\includegraphics[width=0.48\linewidth]{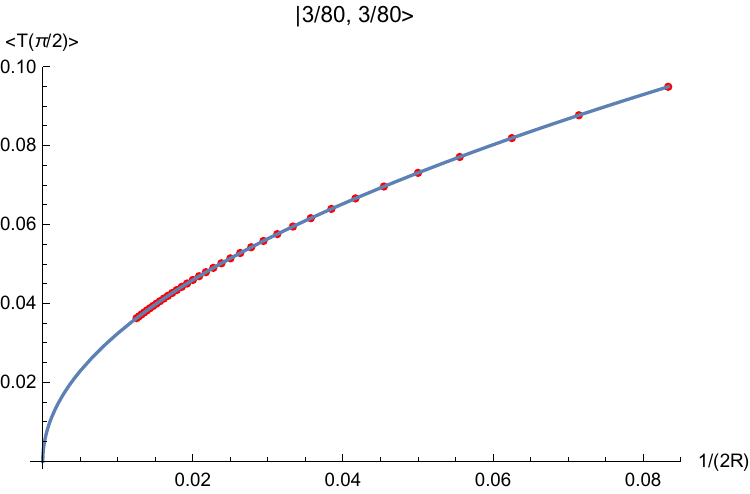}
\includegraphics[width=0.48\linewidth]{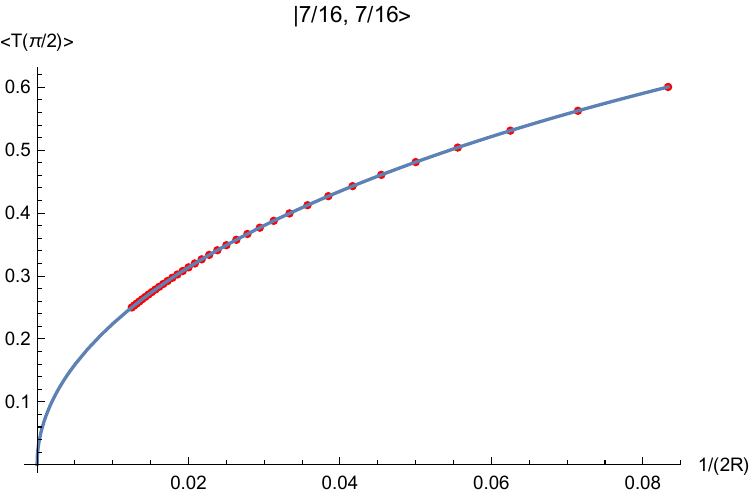}
\end{center}
\caption{Expectation values for $ \widehat{{\cal D}}^{(\rm latt)}_{(2,1)}$ for states $|\frac{3}{80},\frac{3}{80}\rangle$ a) and $|\frac{7}{16},\frac{7}{16}\rangle$ b). We have removed the bulk non-universal contributions  explicitly. The fitting function includes a few terms of a power series in $\R^{-\frac{1}{2}}$, with diverging derivatives in $\R \to \infty$.}
\label{CurvesDig}
\end{figure}

\begin{figure}[htb]
\begin{center}
\includegraphics[width=0.48\linewidth]{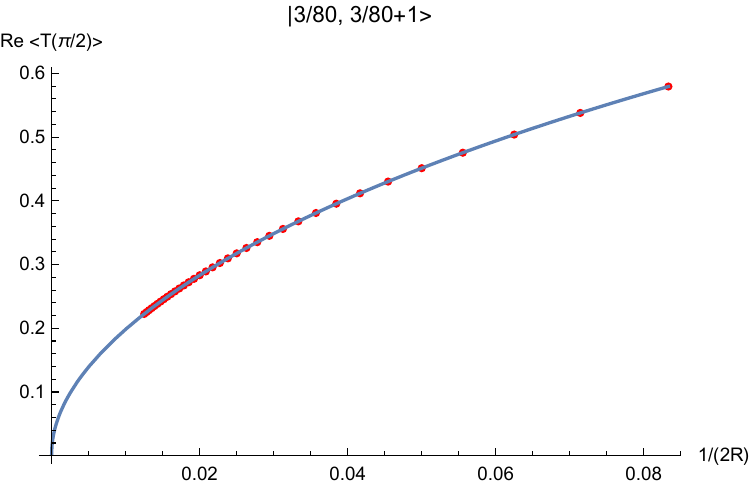}
\includegraphics[width=0.48\linewidth]{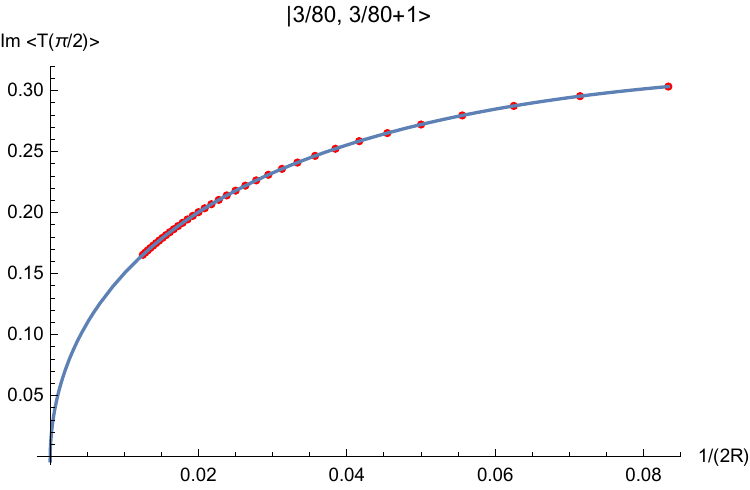}
\end{center}
\caption{Expectation values for $ \widehat{{\cal D}}^{(\rm latt)}_{(2,1)}$ at  descendant state  $\overline{L}_{-1}|\frac{3}{80},\frac{3}{80} \rangle$: real part a), imaginary part b).  }
\label{CurvesDesc}
\end{figure}

Therefore it is natural to propose the expansion\footnote{For $T(\pi/2)$ we have  $\vd=i(\gamma-\pi)/2$, so we should use $e_0^{(J=1)}(i\gamma/2-i\pi/2)$. Following the discussion after Eq.(\ref{Lamscal}) , we replace this by $e_0^{p-2}(i\gamma/2-i\pi)$ which is the same as $e_0^{(p-2)} (-i\gamma/2)$, using periodicity.} 
\begin{equation}\label{eq:bulk-contrib-scale-21}
\langle T\left( {\pi \over 2}\right)/ \exp\left(2 \R e_0^{(p-2)}\left(-i{\gamma \over2}\right)\right) \rangle \approx c_0 +c_1 {(2\R)}^{-{1 \over 2}}+c_2 {(2\R)}^{-1}+c_3 {(2\R)}^{-{3 \over 2}}+c_4 {(2\R)}^{-2},
\end{equation}
with $c_0$ the best estimate for $\langle\widehat{\cal{D}}_{(2,1)}^{(\rm latt)}\rangle$. The result of such a fitting  is shown on Figs. \ref{CurvesDig}-\ref{CurvesDesc}. For example, for system-sizes ranging from $2\R=12$ to $80$, one finds $c_0\approx-3.1\times 10^{-5} $ and  $c_1 \approx 0.32448$ for the $|\frac{3}{80},\frac{3}{80} \rangle$ state. The estimated errors for $c_0$ and $c_1$ are of order $10^{-6}$ and $10^{-5}$, respectively. Moreover, there are important (linear) covariance checks  among all fitting parameters, which make the value of  $c_0$  consistent with $0$ while $c_1$ definitely is not. Differences between data and the fit are of the order of  $10^{-8}$ - on the figure, it is impossible with the naked eye to see any difference between the two. One can also check that the factor which corresponds to bulk contribution in Eq. \eqref{eq:bulk-contrib-scale-21}, $\exp\left(2 \R e_0^{(p-2)}\left(-i{\gamma \over2}\right)\right)$, is the same as $\cot^{2 {\rm R}} \gamma$, using its expression given in Eq. \eqref{bulk}. Observe that, were analytical results from the Bethe-ansatz not available, it would be difficult, for sizes reachable using numerical methods, to confirm the eigenvalues of the 
$\widehat{{\cal D}}^{(\rm latt)}_{(2,1)}$ operators.

In contrast with the case of $\widehat{{\cal D}}_{(2,1)}$, we reiterate -  as  discussed in section \ref{sec:fusion-Tmatrices-(1s)} - that  the $Y_k$ operators, which are defined in Eq. \eqref{eq:prop-fac-high-fus}, are not only lattice topological defects, but also that  their expectation value when acting on primary states are exactly the same as what is expected from the continuum $\widehat{{\cal D}}_{(1,k)}$  defects, even for finite lattice sizes. 

At this stage, we have discussed the lattice realization of all Verlinde line operators in TCI except $\widehat{{\cal D}}_{(2,2)} $. Since in the continuum $\widehat{{\cal D}}_{(2,2)} = \widehat{{\cal D}}_{(1,3)}\widehat{{\cal D}}_{(2,1)}$, we finally define 
\begin{equation}
   \widehat{{\cal D}}_{(2,2)}^{(\rm latt)}  = \widehat{{\cal D}}_{(1,3)}^{(\rm latt)}\widehat{{\cal D}}_{(2,1)}^{(\rm latt)} = \left(\tan \gamma\right)^{2 \R} \uR^{-3} (-1)^\R\left( Y^2 - 1 \right) \,  T\left(\frac{\pi}{2}\right) \, .
\end{equation}
where we added the normalization  factor from Eq. (\ref{eq:21-norm-line-op}).
This operator is  not topological on the lattice (since $\widehat{{\cal D}}_{(2,1)}^{(\rm latt)}$ is not)  but again, in the $\R \to \infty $ limit, it behaves exactly like its continuum counterpart.

\subsubsection{Direct Channel }
We first give the defect Hilbert spaces for TCI CFT 
\begin{equation}
\begin{split}
    {\cal H}_{(1,2)}  & = \left(\frac{1}{10},0 \right) \oplus \left(0,\frac{1}{10} \right) \oplus \left(\frac{3}{5},\frac{1}{10} \right) \oplus\left(\frac{1}{10},\frac{3}{5} \right) \oplus \left(\frac{3}{5},\frac{3}{2} \right) \oplus \left(\frac{3}{2},\frac{3}{5} \right) \oplus \left(\frac{7}{16},\frac{3}{80} \right)   \\ & \oplus \left(\frac{3}{80},\frac{7}{16} \right) \oplus \left(\frac{3}{80},\frac{3}{80} \right)  \, ,   \\
    {\cal H}_{(1,3)}  & = \left(\frac{3}{5},0 \right) \oplus \left(\frac{1}{10},\frac{1}{10} \right) \oplus \left(\frac{3}{2},\frac{1}{10} \right)  \oplus \left(\frac{3}{5},\frac{3}{5} \right) \oplus \left(0,\frac{3}{5} \right) \oplus \left(\frac{1}{10},\frac{3}{2} \right) \oplus \left(\frac{7}{16},\frac{3}{80} \right)   \\ &  \oplus \left(\frac{3}{80},\frac{7}{16} \right) \oplus \left(\frac{3}{80},\frac{3}{80} \right)  \, ,    \\
    {\cal H}_{(1,4)}  & = \left(\frac{3}{2},0 \right) \oplus  \left(\frac{3}{5},\frac{1}{10} \right) \oplus   \left(\frac{1}{10},\frac{3}{5} \right) \oplus \left(0,\frac{3}{2} \right) \oplus  \left(\frac{7}{16},\frac{7}{16} \right) \oplus \left(\frac{3}{80},\frac{3}{80} \right)  \, ,    \\
    {\cal H}_{(2,1)}  & =\left(\frac{7}{16},0 \right) \oplus \left(\frac{3}{80},\frac{1}{10} \right) \oplus \left(\frac{3}{80},\frac{3}{5} \right) \oplus  \left(\frac{7}{16},\frac{3}{2} \right) \oplus \left(0,\frac{7}{16} \right) \oplus  \left(\frac{3}{2}, \frac{7}{16} \right) \oplus \left(\frac{1}{10} , \frac{3}{80}\right)  \\
     &  \oplus \left(\frac{3}{5},\frac{3}{80} \right)  \, ,  \\
    {\cal H}_{(2,2)}  & =  \left(\frac{3}{80},0 \right) \oplus \left(\frac{7}{16},\frac{1}{10} \right) \oplus  \left(\frac{3}{80},\frac{1}{10} \right) \oplus \left(\frac{7}{16},\frac{3}{5} \right) \oplus \left(\frac{3}{80},\frac{3}{5} \right) \oplus  \left(\frac{3}{80},\frac{3}{2} \right) \oplus \left(\frac{1}{10},\frac{7}{16} \right)  \\
    & \oplus \left(\frac{3}{5},\frac{7}{16} \right)  \oplus \left(0,\frac{3}{80} \right)  \oplus \left(\frac{1}{10},\frac{3}{80} \right) \oplus \left(\frac{3}{5},\frac{3}{80} \right) \oplus \left(\frac{3}{2},\frac{3}{80} \right)  \, . 
    \end{split}
\end{equation}

Let us go back to the lattice now. First, we inspect the one-impurity Hamiltonian, given in Eq. \eqref{eq:def_ham}. As we discussed in section \ref{sec:def-ham}, when we set the impurity parameter in this Hamiltonian $\pm {\rm i } \infty$ or $\pm \frac{\pi}{2}$, we should recover the (1,2) and (2,1) defect Hamiltonians. This can be checked by looking at the spectrum of conformal weights extracted from lattice data, and comparing with the expected  states of the ${\cal H}_{(1,2)}$ and ${\cal H}_{(2,1)}$ defect Hilbert spaces respectively.
Results are shown in tables \ref{tab:12-defect} and \ref{tab:21-defect}. 

\begin{table}[H]\renewcommand{\arraystretch}{1.2}
\centering
\begin{tabular}{ |m{3.5cm}| m{3.5cm}|m{3.5cm}|m{3.5cm}| }
\hline
 \multicolumn{4}{|c|}{States and Conformal dimensions - $(1,2)$ defect - 28 RSOS sites} \\
 \hline
 State(Descendant) & $h + \bar{h}$ & $h - \bar{h}$ & Theoretical ($h + \bar{h}$,$h - \bar{h}$) \\ [2ex]
 \hline
 
${\color{black}\ket{\frac{3}{80},\frac{3}{80}}}$        & 0.07436   & 0   & (0.075,0)     \\
${\color{black} \ket{\frac{1}{10},0}}$        & 0.10016   & 0.1   & (0.1,0.1)     \\
${\color{black} \ket{0,\frac{1}{10}}}$        & 0.10016   & $-0.1$   & (0.1,$-0.1$)     \\
${\color{black}\ket{\frac{7}{16},\frac{3}{80}}}$        & 0.47629   & 0.4   & $(0.475,0.4) $    \\
${\color{black}\ket{\frac{3}{80},\frac{7}{16}}}$        & 0.47629   & $-0.4$& $(0.475,-0.4) $    \\
${\color{black}\ket{\frac{3}{5},\frac{1}{10}}}$        & 0.5857   & $0.5$   & $(0.6,0.5) $    \\
${\color{black}\ket{\frac{1}{10},\frac{3}{5}}}$        & 0.5857   & $-0.5$   & $(0.6,-0.5) $    \\
$L_{-1}{\color{black}\ket{\frac{3}{80},\frac{3}{80}}}$        & 1.00485   & 1   & (1.075,1)     \\
$\bar{L}_{-1}{\color{black}\ket{\frac{3}{80},\frac{3}{80}}}$        & 1.00485   & $-1$   & (1.075,$-1$)     \\
 \hline
\end{tabular}
\caption{In this table, we show the low lying energy levels of the one impurity defect Hamiltonian with spectral parameter i$\infty$. The scaling is done by $L-1$, for which we obtain exact momentum values. }
\label{tab:12-defect}
\end{table}
Note that,  when the defect is topological on the lattice - like   the $(1,2)$ defect - the momenta values from the  continuum theory can already be obtained in finite size if we use as  scaling parameter $L$ with the appropriate shift. For $(1,s)$ defects, in any A type model, the correct scaling parameter is  to $L - s + 1$.
\bigskip 
\begin{table}[H]\renewcommand{\arraystretch}{1.2}
\centering
\begin{tabular}{ |m{3.5cm}| m{3.5cm}|m{3.5cm}|m{3.5cm}| }
\hline
 \multicolumn{4}{|c|}{States and Conformal dimensions - $(2,1)$ defect - 28 RSOS sites} \\
 \hline
 State(Descendant) & $h + \bar{h}$ & $h - \bar{h}$ & Theoretical ($h + \bar{h}$,$h - \bar{h}$) \\ [2ex]
 \hline
 
$\ket{\frac{1}{10},\frac{3}{80}} $       & 0.131418   & 0.06463945   & (0.1375,$0.0625$)     \\
 $\ket{\frac{3}{80},\frac{1}{10}} $       &0.131418 & $-0.06463945$ & (0.1375,$-0.0625$)     \\
 $\ket{\frac{7}{16},0} $            &0.39360  & 0.41604573   & $(0.4375,0.4375) $    \\
 $\ket{0,\frac{7}{16}} $               & 0.39360  & $-0.41604573$   & $(0.4375,-0.4375)$     \\
 $\ket{\frac{3}{5},\frac{3}{80}} $        & 0.673715   & $0.55547075$& $(0.6375
5,0.5625) $    \\
 $\ket{\frac{3}{80},\frac{3}{5}} $         &0.673715  & $-0.55547075$   & $(0.6375,-0.5625) $    \\
 \hline
\end{tabular}
\caption{In this table, we show the low lying energy level of the one impurity defect Hamiltonian with spectral parameter $\frac{\pi}{2}$. The scaling is done by $L$, like for A$_3$ RSOS for the $(2,1) $ case. }
\label{tab:21-defect}
\end{table}

We now discuss the situation  with  two impurities, which was considered  in section \ref{subsec:two-imp-ham}. Recall that there are several options, depending on whether or not we apply the JW projectors to define the Hamiltonians. For the case without projectors, setting the spectral parameter in Eq. \eqref{eq:two-imp-ham-no-JW} $\tu$ to i$\infty$ and $\frac{\pi}{2}$, we obtain the results  shown in tables \ref{tab:13-def-two-imp} and \ref{tab:31-def-two-imp} respectively.

\begin{table}[H]\renewcommand{\arraystretch}{1.2}
\centering
\begin{tabular}{ |p{3.5cm}| p{3.5cm}|m{3.5cm}|m{3.5cm}| }
\hline
 \multicolumn{4}{|c|}{States and Conformal dimensions - $(1,1)$ + ${\color{red} (1,3)}$ defect - up to 28 sites - $\tu = {\rm i } \infty$} \\
 \hline
 State(Descendant) & $h + \bar{h}$ & $h - \bar{h}$ & Theoretical ($h + \bar{h}$,$h - \bar{h}$) \\ [2ex]
 \hline
 
$\ket{0,0}$        & $-0.000876$   & 0  & 0     \\
${\color{black}\ket{\frac{3}{80},\frac{3}{80}}}$        & 0.07621   & 0   & (0.075,0)     \\
${\color{red}\ket{\frac{3}{80},\frac{3}{80}}}$        & 0.07621   & 0   & (0.075,0)     \\
$\ket{\frac{1}{10},\frac{1}{10}}$        & 0.20041   & 0   & (0.2,0)     \\
${\color{red} \ket{\frac{1}{10},\frac{1}{10}}}$        & 0.20041   & 0   & (0.2,0)     \\
${\color{red}\ket{\frac{7}{16},\frac{3}{80}}}$        & 0.4609   & 0.4   & $(0.475,0.4) $    \\
${\color{red}\ket{\frac{3}{80},\frac{7}{16}}}$        & 0.4609   & $-0.4$& $(0.475,-0.4) $    \\
${\color{red}\ket{\frac{3}{5},0}}$        & 0.5857   & 0.6   & $(0.6,0.6) $    \\
${\color{red}\ket{0,\frac{3}{5}}}$        & 0.5857   & $-0.6$   & $(0.6,-0.6) $    \\

 \hline
\end{tabular}
\caption{In this table, we show the low lying energy levels of the two impurity defect Hamiltonian. The scaling is done by $L-2$ to obtain accurate momentum values. Note, shifting the spectral parameters by $\pm \frac{\gamma}{2}$ does not make any difference when the imaginary part is large. In this table and the next, the results in red are those obtained when inserting the JW projector.}
\label{tab:13-def-two-imp}
\end{table}

\begin{table}[H]\renewcommand{\arraystretch}{1.2}
\centering
\begin{tabular}{ |p{3.5cm}| p{3.5cm}|m{3.5cm}|m{3.5cm}| }
\hline
 \multicolumn{4}{|c|}{States and Conformal dimensions - $(1,1)$ + ${\color{red} (3,1)}$ defect - up to 28 sites - $\tu =  \frac{\pi}{2} $} \\
 \hline
 State(Descendant) & $h + \bar{h}$ & $h - \bar{h}$ & Theoretical ($h + \bar{h}$,$h - \bar{h}$) \\ [2ex]
 \hline
$\ket{0,0}$        & $-0.00905$   & 0  & 0     \\
${\color{red}\ket{\frac{3}{80},\frac{3}{80}}}$        & 0.074036   & 0   & (0.075,0)     \\
$\ket{\frac{3}{80},\frac{3}{80}}$        & 0.078688   & 0  & (0.075,0)     \\
$\ket{\frac{1}{10},\frac{1}{10}}$        & 0.2200   & 0   & (0.2,0)     \\
${\color{red}\ket{\frac{1}{10},\frac{3}{5}}}$        & 0.68393   & $-0.5 $  & $(0.7,-0.5) $    \\
${\color{red}\ket{\frac{3}{5},\frac{1}{10}}}$        & 0.68393   & 0.5   & $(0.7,0.5)$   \\
${\color{red}\ket{\frac{7}{16},\frac{7}{16}}}$        & 0.80624   & 0   & $(0.875,0) $    \\
 \hline
\end{tabular}
\caption{In this table, we show the low lying energy level of the two impurity defect Hamiltonian. The scaling is done by $L-1$ to obtain accurate momentum values. We obtain exact momenta result as $(3,1) \equiv (1,4)$, which is topological on the lattice. }
\label{tab:31-def-two-imp}
\end{table}
For the Hamiltonian with the projectors \eqref{eq:two-imp-ham-JW} and the same choice of spectral  parameters $\tu$ in Eq.  to ${\rm i} \infty$ and $\frac{\pi}{2}$, we would have obtained only the data in red in the two tables above. In practice, the Jones-Wenzl projector simply removes the states corresponding to the $(1,1)$ defect. 

We also note that even though $(2,1)$ defect is not topological on the lattice in the case of the  TCI, upon its fusion, we should get $(1,1)$ and $(3,1) \equiv (1,4)$ defects, both of which admit realizations that are  topological on the lattice. While the result of the fusion does not satisfy this property, we nevertheless observe in table \ref{tab:31-def-two-imp} that the momentum values are exact  in finite size.

A very intriguing question is, what would  happen if we did not shift the spectral parameters by $\pm \frac{\gamma}{2}$ in the case  of the (2,1) defect, that is when we had set $\tu = \frac{\pi}{2}$ (the shift is moot for (1,2) defects since anyway the spectral parameters are sent to $i\infty$). This is after all what one would naively do to study fusion  without knowledge of the underlying integrable structure. In practice, this   corresponds to  studying the Hamiltonian in Eq. \eqref{eq:two-imp-ham-two-spec} with $\tu_k = \tu_{k+1} = \frac{\pi}{2}$. The results obtained are  summarized in table \ref{tab:two-imp-Ham-no-shift}.

\begin{table}[H]\renewcommand{\arraystretch}{1.2}
\centering
\begin{tabular}{ |p{3.5cm}| p{3.5cm}|m{3.5cm}|m{3.5cm}| }
\hline
 \multicolumn{4}{|c|}{States and Conformal dimensions - $(1,1)$ + ${\color{red} (3,1)}$ defect - up to 28 sites, $\tu_k =  \tu_{k+1} = \frac{\pi}{2}$.} \\
 \hline
 State(Descendant) & $h + \bar{h}$ & $h - \bar{h}$ & Theoretical ($h + \bar{h}$,$h - \bar{h}$) \\ [2ex]
 \hline
 
$\ket{0,0}$        & $-0.0033$ {\color{blue}($-0.00034$)}  & $0$   & (0,0)     \\
${\color{red}\ket{\frac{3}{80},\frac{3}{80}}}$        & 0.04843 {\color{blue}(0.05794)} & $0$   & (0.075,0)    \\
$\ket{\frac{3}{80},\frac{3}{80}}$        & 0.10435 {\color{blue}(0.09265)}   & 0  & (0.075,0)     \\
$\ket{\frac{1}{10},\frac{1}{10}}$        & 0.2053   & $0$   & (0.2,0)     \\
${\color{red}\ket{\frac{1}{10},\frac{3}{5}}}$        & 0.6494   & $-0.49194$    & $(0.7,-0.5)$    \\
${\color{red}\ket{\frac{3}{5},\frac{1}{10}}}$        & 0.6494   &$-0.49194$   & (0.7,0.5)     \\
${\color{red}\ket{\frac{7}{16},\frac{7}{16}}}$        & 0.68101   & $0$   & (0.875,0)     \\
 \hline
\end{tabular}
\caption{In this table, we show the low lying energy levels of the two-impurity defect Hamiltonian where we do not shift the spectral parameters by $\pm \frac{\gamma}{2}$. In blue, we have data using DMRG, where we have gone up to much larger system sizes (128 sites). Compare with the results in table \ref{tab:31-def-two-imp}. Again the scaling here is done by $L - 1$. }
\label{tab:two-imp-Ham-no-shift}
\end{table}
By comparing tables \ref{tab:31-def-two-imp} and \ref{tab:two-imp-Ham-no-shift}, we see that  in both cases we obtain states corresponding to the $(1,1)$ and $(3,1)$ defects in the $L \to \infty$ limit: the results are asymptotically the same, which is of course a good sanity check  on our construction. However,  the convergence is much faster when we shift the spectral parameters by $\pm \frac{\gamma}{2}$ (see  the discussion in section  \ref{sec:fusion-Tmatrices-(1s)}).

It is also possible  to try to fuse now the (1,2) and (2,1) defects. Conformal field theory predicts that this should give rise to the $(2,2)$ defect.  In the direct channel, this fusion corresponds, on the lattice, to setting the two impurities $\tu_{k}$ and $\tu_{k+1}$ in the defect Hamiltonian in Eq. \eqref{eq:two-imp-ham-two-spec}, to i$\infty$ and $\frac{\pi}{2}$ respectively. The results in table \ref{tab:22-defect} confirm that indeed what is observed  is the $(2,2)$ defect in the IR limit. 

\begin{table}[H]\renewcommand{\arraystretch}{1.2}
\centering
\begin{tabular}{ |m{3.5cm}| m{3.5cm}|m{3.5cm}|m{3.5cm}| }
\hline
 \multicolumn{4}{|c|}{States and Conformal dimensions - $(2,2)$ defect - up to 28 RSOS sites} \\
 \hline
 State(Descendant) & $h + \bar{h}$ & $h - \bar{h}$ & Theoretical ($h + \bar{h}$,$h - \bar{h}$) \\ [2ex]
 \hline
 
$\ket{\frac{3}{80},0} $       & 0.02977   & 0.034188  & (0.0375,$0.0375$)     \\
 $\ket{0,\frac{3}{80}} $       &0.02977 & $-0.034188$ & (0.0375,$-0.0375$)     \\
  $\ket{\frac{1}{10},\frac{3}{80}} $       &0.1419 &0.05899 & (0.1375,$0.0625$)     \\
 $\ket{\frac{3}{80}, \frac{1}{10}} $       &0.1419 & $-0.05899$ & (0.1375,$-0.0625$)     \\
  $\ket{\frac{7}{16},\frac{1}{10}} $       &0.44079 &0.313150 & (0.5375,$0.3375$)     \\
 $\ket{\frac{1}{10},\frac{7}{16}} $       &0.44079 & $-0.313150
 $ & (0.5375,$-0.3375$)     \\

 \hline
\end{tabular}
\caption{In this table, we show the low lying energy levels of the two-impurity defect Hamiltonian with neighboring spectral parameters ${\rm i } \infty$ and $\frac{\pi}{2}$. The scaling is done by $L-2$ to obtain more accurate results.}
\label{tab:22-defect}
\end{table}

Finally, we note that, instead of studying the two impurity Hamiltonian - where by construction the two impurities are placed next to each other - we could have also studied a  Hamiltonian with two impurities, separated by some distance, i.e.
\begin{equation}\label{eq:def_ham-two}
\begin{split}
 &   H^{k,k+1, l, l+1}(\tilde{u}_1,\tilde{u}_2) = - \sum_{_{\substack{i=0 \\ i \neq k, k + 1, l, l+1}}}^{2L - 1} \left( - \cot \gamma +  \frac{1}{ \sin \gamma} e_i\right)  -    \left( -\cot \gamma + \frac{1}{\sin \gamma} R_{k}(\tilde{u}_1)^{-1}e_{k + 1} R_{k}(\tilde{u}_1) \right)  \\
& - \left( -\cot \gamma + \frac{1}{\sin \gamma} R_{l}(\tilde{u}_2)^{-1}e_{l + 1} R_{l}(\tilde{u}_2) \right) -  \left( - \cot(\gamma - \tilde{u}_1 ) + \frac{\sin \gamma}{\sin (\gamma + \tilde{u}_1 ) \sin (\gamma - \tilde{u}) }e_k \right)  
\\
& -  \left( - \cot(\gamma - \tilde{u}_2 ) + \frac{\sin \gamma}{\sin (\gamma + \tilde{u}_2 ) \sin (\gamma - \tilde{u}_2) }e_l \right) \, . 
 \end{split}
\end{equation}
where $l> k+2$. Using the Yang-Baxter equation, the two defects could have been moved  around so they sit next to each other: indeed, one finds for this Hamiltonian  the same energies as before .

\subsection{A\texorpdfstring{$_5$}{Lg} RSOS model : Tetra-Critical Ising}
The A$_5$ RSOS model realizes the ${\cal M}(6,5)$ Virasoro minimal model or the Tetra-Critical Ising CFT on the lattice. This CFT has 10 primary fields and therefore 10 Verlinde lines. The central charge is   $c = {4\over 5}$. For compactness, we do not provide the general tables for the action of the Verlinde lines and the decomposition of the  defect Hilbert space, but they can be deduced from eqs. \eqref{eq:verlinde-line-op} and \eqref{eq:twist-part-func} respectively. 

\subsubsection{Crossed Channel}
\begin{figure}[H]
    \centering
    \includegraphics[width=1\linewidth]{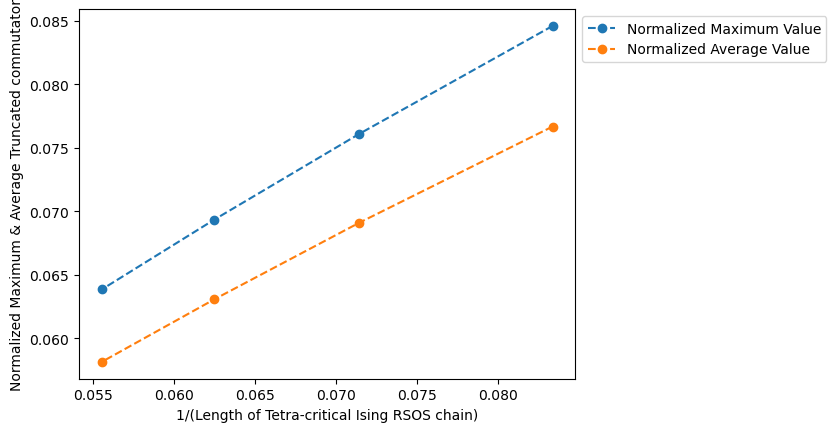}
    \caption{Maximum and Average absolute values of elements in the truncated commutator $[\widehat{{\cal D}}_{(2,1)}^{(\rm latt)}, e_i]$ as a function of inverse system-size, normalized using the operator $\widehat{{\cal D}}_{(2,1)}^{(\rm latt)}e_i$. The lines are constructed by fitting the data points with a polynomial fit, further we only study the lowest 8 energy eigenstates. Normalized Maximum Value and Normalized Average Value converge to 0.01243544 and 0.01180966 respectively. This confirms the  topological nature of the line operator in the scaling limit.}
    \label{fig:DiagPotts-W-comm-E-zero}
\end{figure}

Again we use the transfer matrix at spectral parameter $  {\rm i } \infty$ and $ \frac{\pi}{2}$ to define $\widehat{{\cal D}}^{(\rm latt)}_{(1,2)}$ and $\widehat{{\cal D}}^{(\rm latt)}_{(2,1)}$ respectively. 
\begin{table}[H]\renewcommand{\arraystretch}{1.2}
\centering   
\begin{tabular}{ |P{2cm}|P{4.0cm}|P{2cm} |P{2.5cm}|P{2cm} | }
\hline
 \multicolumn{5}{|c|}{${\cal M}(6,5)$ States and expectation values - up to 20 RSOS sites } \\
 \hline
 State  & $\widehat{{\cal D}}_{(2,1)}^{(\rm latt)}$   & $\widehat{{\cal D}}_{(2,1)}$  &  $\widehat{{\cal D}}_{(1,2)}^{(\rm latt)}$  &  $\widehat{{\cal D}}_{(1,2)}$ \\ [2ex]
 \hline
  $\ket{0,0} $  & 1.6216746  & $\phi$  & $\sqrt{3}$ & $\sqrt{3}$  \\

 $\ket{\frac{1}{40},\frac{1}{40}} $ &   0.6747658  & $\frac{1}{\phi}$ & 1 & 1 \\
$\ket{\frac{1}{15},\frac{1}{15}}$  &     $-0.540432$ &   $-\frac{1}{\phi}$  & $0$ & $0$ \\
 $\ket{\frac{1}{8},\frac{1}{8}}  $  &    $-1.606873$ & $-\phi$  & $-1$& $-1$    \\
 $\ket{\frac{2}{5},\frac{2}{5}} $ & $-1.0397844 $    & $-\frac{1}{\phi}$ & $-\sqrt{3}$& $-\sqrt{3}$   \\
 \hline
\end{tabular}
\caption{In this table, we list the action of the lattice discretization of the $(2,1)$ and the $(1,2)$ Verlinde lines, along with the results from the continuum. Scaling has been done with  $L$ for $\lattD_{(2,1)}$. $\phi = \frac{1 + \sqrt{5}}{2}$, the golden ratio here.}
\label{tab:(12)-(21)-M65}
\end{table}
Again, from the table above we observe that $\widehat{{\cal D}}_{(1,2)}$ is topological on the lattice while $\widehat{{\cal D}}_{(2,1)}$ is not, as we have stated before (we have again removed the bulk factor of $\cot^{2 {\rm R}} {\gamma}$ to get the correct normalization for $\widehat{{\cal D}}_{(2,1)}^{(\rm latt)}$).   In figure \ref{fig:DiagPotts-W-comm-E-zero}, we  study the commutation of $\widehat{{\cal D}}_{(2,1)}$ with the TL generators, and again note that commutators tend to zero in the scaling limit.

\subsubsection{Direct Channel}
We now discuss the direct channel for Tetra-critical Ising. Again, when we set the impurity parameter to $ {\rm i } \infty$ and $ \frac{\pi}{2}$ in Eq. \eqref{eq:def_ham}, we obtain the $(1,2)$ and $(2,1)$ defect Hamiltonians. The results are presented in table \ref{tab:12-def-Diag-Potts} and \ref{tab:21-def-Diag-Potts} below. 
\begin{table}[H]\renewcommand{\arraystretch}{1.2}
\centering
\begin{tabular}{ |m{3.5cm}| m{3.5cm}|m{3.5cm}|m{3.5cm}| }
\hline
 \multicolumn{4}{|c|}{States and Conformal dimensions - $(1,2)$ defect - up to 26 RSOS sites} \\
 \hline
 State(Descendant) & $h + \bar{h}$ & $h - \bar{h}$ & Theoretical $ (h + \bar{h}, h - \bar{h}) $ \\ [2ex]
 \hline
$\ket{\frac{1}{15},\frac{1}{40}}$        & 0.089946  & $0.041666$   &   $(0.09167,0.04167)$     \\
$\ket{\frac{1}{40},\frac{1}{15}}$        & 0.089946  & $-0.041666$   &   $(0.09167,-0.04167)$     \\
$\ket{\frac{1}{8},0}$                    & 0.12542    & $0.125$   &   (0.125,0.125)     \\
$\ket{0,\frac{1}{8}}$                    & 0.12542    & $-0.125$   &   $(0.125,-0.125)$     \\
$\ket{\frac{2}{5},\frac{1}{40}}$         & 0.43171 & $0.375$   &   (0.425,0.375)       \\
$\ket{\frac{1}{40},\frac{2}{5}}$         & 0.43171  & $-0.375$   &  $(0.425,-0.375)$       \\
$\ket{\frac{21}{40},\frac{1}{15}}$       & 0.58978  & $0.4583$   &  (0.59167,0.4583)     \\
$\ket{\frac{1}{15},\frac{21}{40}}$       & 0.58978  & $-0.4583$   &  $(0.59167,-0.4583) $     \\
$\ket{\frac{2}{3},\frac{1}{8}}$          & 0.76300  & $0.54167$   &  $(0.79167, 0.54167)$           \\
$\ket{\frac{1}{8}, \frac{2}{3}}$          & 0.76300  & $-0.54167$   & $(0.79167, -0.54167)$            \\
 \hline
\end{tabular}
\caption{In this table, we show the low lying energy level of the one impurity defect Hamiltonian with spectral parameter i$\infty$. The scaling is done by $L-1$, for which we obtain exact momentum values.}
\label{tab:12-def-Diag-Potts}
\end{table}

\begin{table}[H]\renewcommand{\arraystretch}{1.2}
    \centering
\begin{tabular}{ |m{3.5cm}| m{3.5cm}|m{3.5cm}|m{3.5cm}| }
\hline
 \multicolumn{4}{|c|}{States and Conformal dimensions in (2,1) - up to 26 RSOS sites} \\
 \hline
 State & $h + \bar{h}$ & $h -\bar{h}$ &Theoretical $(h + \bar{h}, h - \bar{h})$  \\ 
 \hline
 $\ket{\frac{1}{15},\frac{1}{15}} $       & 0.11859   & $0$   &   (0.1333,0) \\
  $\ket{\frac{1}{8},\frac{1}{40}} $         & 0.14483   &  $0.1027$   &  $(0.15,0.1)$ \\
 $\ket{\frac{1}{40},\frac{1}{8}} $        & 0.14483 &  $-0.1027$   &  $(0.15,-0.1)$\\
 $\ket{\frac{2}{5},0} $     & 0.35254  & $0.3590$  &  $(0.4,0.4)$ \\
 $\ket{0,\frac{2}{5}} $     & 0.35254   &  $-0.3590$   & $(0.4,-0.4)$  \\
  $\ket{\frac{21}{40},\frac{1}{40}} $        & 0.57212 &  $0.4737$   &  $(0.55,0.5)$\\
 $\ket{\frac{1}{40},\frac{21}{40}} $         & 0.57212   &  $-0.4737$   & $(0.55,-0.5)$   \\
  $\ket{\frac{2}{3},\frac{1}{15}} $        & 0.7909 &  $0.5839$   &  $(0.733,0.6
)$\\
 $\ket{\frac{1}{15},\frac{2}{3}} $         & 0.7909   &  $-0.5839$   &  $(0.733,-0.6
)$   \\

 \hline
\end{tabular}
    \caption{In this table, we show the low lying energy level of the one impurity defect Hamiltonian with spectral parameter $\frac{\pi}{2}$. The scaling is done by $L$, like for A$_3$ and A$_4$ RSOS. }
    \label{tab:21-def-Diag-Potts}
\end{table}
We can also study the two impurity Hamiltonian here, by setting the spectral parameters to i$\infty  \pm \frac{\gamma}{2}$ and $\frac{\pi}{2} \pm \frac{\gamma}{2}$, to obtain states corresponding to $(1,3)$ and $(3,1)$ defect respectively. Again, by adding the JW projector, we can get rid off the states corresponding to the $(1,1)$ defect. The results are presented in table \ref{tab:diag_potts_13} and \ref{tab:31-defect-diag-potts}.
\begin{table}[H]\renewcommand{\arraystretch}{1.2}
\centering
\begin{tabular}{ |m{3.5cm}| m{3.5cm}|m{3.5cm}|m{3.5cm}| }
\hline
 \multicolumn{4}{|c|}{States and Conformal dimensions - $(1,1)$ + ${\color{red} (1,3)}$ defect - up to 26 sites - $\tu = {\rm  i} \infty$} \\
 \hline
 State(Descendant) & $h + \bar{h}$ & $h - \bar{h}$ & Theoretical ($h + \bar{h}$,$h - \bar{h}$) \\ [2ex]
 \hline
 
$\ket{0,0}$        & $-0.0013$   & 0   & $(0,0)$    \\
${\color{red}\ket{\frac{1}{40},\frac{1}{40}}}$        & 0.05118   & 0   & $(0.05,0)$  \\
${\ket{\frac{1}{40},\frac{1}{40}}}$        & 0.05118  & 0  & (0.05,0)   \\

$\ket{\frac{1}{15},\frac{1}{15}}$        &  0.13639  & 0  &$(0.1333,0) $  \\

${\color{red} \ket{\frac{1}{15},\frac{1}{15}}}$        & 0.13639   & 0  &$(0.1333,0)$     \\
$\ket{\frac{1}{8},\frac{1}{8}}$        &  0.2494  &0 & $(0.25,0)  $   \\
${\color{red}\ket{\frac{1}{8},\frac{1}{8}}}$        &  0.2494  &0 & $(0.25,0)  $    \\

${\color{red}\ket{\frac{2}{5},\frac{1}{15}}}$        & 0.4443   & 0.3333  & $(0.4667,0.333)  $    \\
${\color{red}\ket{\frac{1}{15},\frac{2}{5}}}$        & 0.4443  & $-0.3333$  & $(0.4667,-0.333)  $   \\

${\color{red}\ket{\frac{21}{40},\frac{1}{40}}}$        & $0.5295$   & 0.5  & $(0.55,0.5)$     \\
${\color{red}\ket{\frac{1}{40},\frac{21}{40}}}$        & $0.5295$   & $-0.5$  & $(0.55,-0.5)$      \\

 \hline
\end{tabular}
\caption{In this table, we show the low lying energy level of the two impurity defect Hamiltonian. The scaling is done by $L-2$ to obtain exact momentum values. }
\label{tab:diag_potts_13}
\end{table}

\begin{table}[H]\renewcommand{\arraystretch}{1.2}
\centering
\begin{tabular}{ |m{3.5cm}| m{3.5cm}|m{3.5cm}|m{3.5cm}| }
\hline
 \multicolumn{4}{|c|}{States and Conformal dimensions - $(1,1)$ + ${\color{red} (3,1)}$ defect - up to 26 sites - $\tu =  \pi/2 $} \\
 \hline
 State(Descendant) & $h + \bar{h}$ & $h - \bar{h}$ & Theoretical ($h + \bar{h}$,$h - \bar{h}$) \\ [2ex]
 \hline
 
$\ket{0,0}$        & $-0.0013$   & 0   &  $(0,0)$      \\
${\color{red}\ket{\frac{1}{40},\frac{1}{40}}}$        & 0.03663   & 0   & $(0.05,0)$     \\
${\ket{\frac{1}{40},\frac{1}{40}}}$        & 0.05118   & 0  & $(0.05,0)$     \\

$\ket{\frac{1}{15},\frac{1}{15}}$        &  0.13639  & 0  & $(0.1333,0) $  \\
${\color{red} \ket{\frac{1}{15},\frac{1}{15}}}$        & 0.14642   & 0  &$(0.1333,0) $    \\
$\ket{\frac{1}{8},\frac{1}{8}}$        &  0.2494  &0 & $(0.25,0)  $    \\
${\color{red}\ket{\frac{21}{40},\frac{1}{8}}}$        & 0.51069   &  0.366115  & $(0.65,0.4)$     \\
${\color{red}\ket{\frac{1}{8},\frac{21}{40}}}$        & 0.51069   & $-0.366115$  & $(0.65,-0.4)$     \\
 \hline
\end{tabular}
\caption{In this table, we show the low lying energy level of the two impurity defect Hamiltonian. The scaling is done by $L-2$.}
\label{tab:31-defect-diag-potts}
\end{table}
Again, like in the previous subsection, we present in table \ref{tab:diag-potts-31-def-no-shift} the result for (3,1) defect without shifting the spectral parameter by $\frac{\gamma}{2}$.

\begin{table}[H]\renewcommand{\arraystretch}{1.2}
\centering
\begin{tabular}{ |m{3.5cm}| m{3.5cm}|m{3.5cm}|m{3.5cm}| }
\hline
 \multicolumn{4}{|c|}{States and Conformal dimensions - $(1,1)$ + ${\color{red} (3,1)}$ defect - up to 26 sites -  $\tu_k =  \tu_{k+1} = \frac{\pi}{2}$} \\
 \hline
 State(Descendant) & $h + \bar{h}$ & $h - \bar{h}$ & Theoretical ($h + \bar{h}$,$h - \bar{h}$) \\ [2ex]
 \hline
 
$\ket{0,0}$        & $-0.00839$ {\color{blue}($-0.0056$)}   & 0   & $(0 ,0)$    \\
${\color{red}\ket{\frac{1}{40},\frac{1}{40}}}$       & 0.02103   {\color{blue}($-0.02911$)} & 0   & $(0.05 ,0)$    \\
${\ket{\frac{1}{40},\frac{1}{40}}}$        & 0.06152  {\color{blue}$(0.05801)$}  &0   & $(0.05,0)$     \\

$\ket{\frac{1}{15},\frac{1}{15}}$        &  0.12624  & 0  &$( 0.1333,,0 )$    \\
${\color{red} \ket{\frac{1}{15},\frac{1}{15}}}$        & 0.17877   & 0 & $( 0.1333,,0 )$   \\
$\ket{\frac{1}{8},\frac{1}{8}}$        &  0.2670  & 0  & $(0.25,0 )$    \\
${\color{red}\ket{\frac{21}{40},\frac{1}{8}}}$        & 0.5677   & $0.35927$   & $(0.65,0.4)$     \\
${\color{red}\ket{\frac{1}{8},\frac{21}{40}}}$        & 0.5851   & $-0.35927$   & $(0.65,-0.4)$    \\
 \hline
\end{tabular}
\caption{In this table, we show the low lying energy level of two impurity defect Hamiltonian where we do not shift the spectral parameters by $\pm \frac{\gamma}{2}$. The scaling is done by $L-2$ In blue, we have data using DMRG, where we have gone up to much larger system sizes (128 sites).  }
\label{tab:diag-potts-31-def-no-shift}
\end{table}

\subsection{The A\texorpdfstring{$_{10}$}{Lg} RSOS model  }
We will now study the example of the A$_{10}$ RSOS model to illustrate the universality of our results. The CFT describing the IR states of this theory is the ${\cal M}(11,10)$ CFT, with 45 primary fields and central charge $c = \frac{52}{55}$. The conformal dimensions for the primary fields can be calculated using Eq. \eqref{eq:conf-dim-min} and the action of Verlinde line in the crossed and direct channel can be understood using Eq. \eqref{eq:verlinde-line-op} and \eqref{eq:twist-part-func}.

\subsubsection{Crossed Channel}

We will again show that the transfer matrix at spectral parameter i$\infty$ realizes the $\widehat{{\cal D}}^{}_{(1,2)}$ on the lattice, and is topological.
\begin{table}[H]\renewcommand{\arraystretch}{1.2}
\centering   
\begin{tabular}{ |P{2cm}|P{4.0cm}|P{2cm} |P{2.5cm}|P{2cm} | }
\hline
 \multicolumn{5}{|c|}{${\cal M}(11,10)$ States and expectation values - up to 12 RSOS sites } \\
 \hline
 State  & $\widehat{{\cal D}}_{(2,1)}^{(\rm latt)}$   & $\widehat{{\cal D}}_{(2,1)}$  &  $\widehat{{\cal D}}_{(1,2)}^{(\rm latt)}$  &  $\widehat{{\cal D}}_{(1,2)}$ \\ [2ex]
 \hline
  $\ket{0,0} $  &1.90824912  & 1.90211303  & 1.91898595 & 1.91898595  \\

 $\ket{\frac{3}{440},\frac{3}{440}} $ & 1.65063291  & 1.61803398 & 1.68250707 & 1.68250707 \\
 $\ket{\frac{1}{55},\frac{1}{55}} $ &   1.24601835  & 1.1755705 & 1.30972147 & 1.30972146 \\
 $\ket{\frac{3}{88},\frac{3}{88}} $ &   0.72929549 & 0.6180339 & 0.83083003 & 0.83083003 \\
 $\ket{\frac{3}{55},\frac{3}{55}} $ &   0.14504319  & 0 & 0.28462968 & 0.28462968 \\
 \hline
\end{tabular}
\caption{In this table, we list the action of the lattice discretization of the $(2,1)$ and the $(1,2)$ Verlinde lines, along with the results from the continuum. Scaling has been done with $L$ for $\lattD_{(2,1)}$. }
\label{tab:(12)-(21)-M1110}
\end{table}
Hence, we again observe that while $\widehat{{\cal D}}^{}_{(1,2)}$ is topological on the lattice,  $\widehat{{\cal D}}^{}_{(2,1)}$ is not. By using higher transfer matrices at spectral parameter i$\infty$, we can define $\widehat{{\cal D}}_{(1,s)}^{(\rm latt)}$, and they are topological on the lattice too. Further, we have normalized $\widehat{{\cal D}}^{(\rm latt)}_{(2,1)}$ correctly by removing the factor of $\cot^{2 {\rm R}}{\gamma}$.  

\subsubsection{Direct Channel}
We now discuss the direct channel for the A$_{10}$ RSOS model. Again, when we set the impurity parameter to $ {\rm i } \infty$ and $ \frac{\pi}{2}$ in Eq. \eqref{eq:def_ham}, we obtain the $(1,2)$ and $(2,1)$ defect Hamiltonians. The results are presented in tables \ref{tab:12-def-A10} and \ref{tab:21-def-A10} below. 
\begin{table}[H]\renewcommand{\arraystretch}{1.2}
\centering
\begin{tabular}{ |m{3.5cm}| m{3.5cm}|m{3.5cm}|m{3.5cm}| }
\hline
 \multicolumn{4}{|c|}{States and Conformal dimensions - $(1,2)$ defect - up to 20 RSOS sites} \\
 \hline
 State(Descendant) & $h + \bar{h}$ & $h - \bar{h}$ & Theoretical $ (h + \bar{h}, h - \bar{h}) $ \\ [2ex]
 \hline
$\ket{\frac{3}{55},\frac{3}{55}}$        & 0.102036 {\color{blue}(0.1042519)}  & $0$   &   $(0.109091,0)$     \\
$\ket{\frac{7}{88},\frac{3}{88}}$        & 0.107276 {\color{blue}(0.1092138)}  & $0.04545$   &   $(0.113636,0.04545)$     \\
$\ket{\frac{3}{88},\frac{7}{88}}$        & 0.107276 {\color{blue}(0.1092138)} & $-0.04545$   &   $(0.113636,-0.04545)$     \\
$\ket{\frac{6}{55},\frac{1}{55}}$        &0.122939 {\color{blue} (0.124097)} & $0.09091$   &   $(0.12727,0.09091)$     \\
$\ket{\frac{1}{55},\frac{6}{55}}$        & 0.122939 {\color{blue} (0.124097)} & $-0.09091$   &   $(0.12727,-0.09091)$     \\
 \hline
\end{tabular}
\caption{In this table, we show the low lying energy level of the one impurity defect Hamiltonian with spectral parameter i$\infty$. The scaling is done by $L-1$, for which we obtain exact momentum values. In blue, we have data using DMRG, where we have gone up to much larger system sizes (40 sites).}
\label{tab:12-def-A10}
\end{table}
\begin{table}[H]\renewcommand{\arraystretch}{1.2}
    \centering
\begin{tabular}{ |m{3.5cm}| m{3.5cm}|m{3.5cm}|m{3.5cm}| }
\hline
 \multicolumn{4}{|c|}{States and Conformal dimensions in (2,1) - up to 18 RSOS sites} \\
 \hline
 State & $h + \bar{h}$ & $h -\bar{h}$ &Theoretical $(h + \bar{h}, h - \bar{h})$  \\ 
 \hline
 $\ket{\frac{7}{88},\frac{3}{55}} $       & 0.11669 {\color{blue}(0.116746)}  & $0.0243412$   &   (0.134091,0.025) \\
 $\ket{\frac{3}{55},\frac{7}{88}} $       &  0.11669 {\color{blue}(0.116746)} & $-0.0243412$   &   $(0.134091,-0.025)$ \\

 $\ket{\frac{6}{55},\frac{3}{88}} $       & 0.130906  {\color{blue} (0.129128)}  & $0.0731811$   &   (0.143182,0.075) \\
 $\ket{\frac{3}{88},\frac{6}{55}} $       & 0.130906  {\color{blue} (0.129128)}  & $-0.0731811$   &   $(0.143182,-0.075)$ \\

 $\ket{\frac{63}{440},\frac{1}{55}} $       & 0.159673 {\color{blue} (0.154294)}  & $0.1226349$   &   $(0.161364,0.125)$ \\
 $\ket{\frac{1}{55},\frac{63}{440}} $       & 0.159673 {\color{blue} (0.154294)}  & $-0.1226349$   &   $(0.161364,-0.125)$ \\

 \hline
\end{tabular}
    \caption{In this table, we show the low lying energy level of the one impurity defect Hamiltonian with spectral parameter $\frac{\pi}{2}$. The scaling is done by $L$, like for A$_3$, A$_4$, and A$_5$ RSOS. In blue, we have data using DMRG, where we have gone up to much larger system sizes (32 sites).}
    \label{tab:21-def-A10}
\end{table}
We can also study the two impurity Hamiltonian here, by setting the spectral parameters to i$\infty  \pm \frac{\gamma}{2}$ and $\frac{\pi}{2} \pm \frac{\gamma}{2}$, to obtain states corresponding to $(1,3)$ and $(3,1)$ defect respectively.

\begin{table}[H]\renewcommand{\arraystretch}{1.2}
\centering
\begin{tabular}{ |m{3.5cm}| m{3.5cm}|m{3.5cm}|m{3.5cm}| }
\hline
 \multicolumn{4}{|c|}{States and Conformal dimensions - $(1,1)$ + ${\color{red} (1,3)}$ defect - up to 20 sites - $\tu = {\rm  i} \infty$} \\
 \hline
 State(Descendant) & $h + \bar{h}$ & $h - \bar{h}$ & Theoretical ($h + \bar{h}$,$h - \bar{h}$) \\ [2ex]
 \hline
 
$\ket{0,0}$        & $-0.0036878$   & 0   & $(0,0)$    \\
${\color{red}\ket{\frac{3}{440},\frac{3}{440}}}$        & 0.0119661   & 0   & $(0.013636,0)$  \\
${\ket{\frac{3}{440},\frac{3}{440}}}$        & 0.0119661 & 0  & (0.013636,0)   \\

${\color{red} \ket{\frac{1}{55},\frac{1}{55}}}$        & 0.037754  & 0  &$(0.0363636,0)$     \\
$\ket{\frac{1}{55},\frac{1}{55}}$        &  0.037754  & 0  &$(0.0363636,0) $  \\

${\color{red}\ket{\frac{3}{88},\frac{3}{88}}}$        &  0.073196  &0 & $(0.0681818,0)  $    \\
$\ket{\frac{3}{88},\frac{3}{88}}$        & 0.073196  &0 & $(0.0681818,0)  $   \\

${\color{red}\ket{\frac{3}{55},\frac{3}{55}}}$        &  0.11754  &0 & $(0.1090909,0)  $    \\
$\ket{\frac{3}{55},\frac{3}{55}}$        &  0.11754  &0 & $(0.1090909,0)  $   \\

 \hline
\end{tabular}
\caption{In this table, we show the low lying energy level of the two impurity defect Hamiltonian. The scaling is done by $L-2$ to obtain accurate momentum values. }
\label{tab:A10_13}
\end{table}

\begin{table}[H]\renewcommand{\arraystretch}{1.2}
\centering
\begin{tabular}{ |m{3.5cm}| m{3.5cm}|m{3.5cm}|m{3.5cm}| }
\hline
 \multicolumn{4}{|c|}{States and Conformal dimensions - $(1,1)$ + ${\color{red} (3,1)}$ defect - up to 18 sites - $\tu =  \pi/2 $} \\
 \hline
 State(Descendant) & $h + \bar{h}$ & $h - \bar{h}$ & Theoretical ($h + \bar{h}$,$h - \bar{h}$) \\ [2ex]
 \hline
 
$\ket{0,0}$        & $-0.004922$   & 0   & $(0,0)$    \\
${\color{red}\ket{\frac{3}{440},\frac{3}{440}}}$        & 0.001704   & 0   & $(0.013636,0)$  \\
${\ket{\frac{3}{440},\frac{3}{440}}}$        & 0.01111 & 0  & (0.013636,0)   \\

${\color{red} \ket{\frac{1}{55},\frac{1}{55}}}$        & 0.029868  & 0  &$(0.0363636,0)$     \\
$\ket{\frac{1}{55},\frac{1}{55}}$        &  0.037468  & 0  &$(0.0363636,0) $  \\

${\color{red}\ket{\frac{3}{88},\frac{3}{88}}}$        &  0.068572  &0 & $(0.0681818,0)  $    \\
$\ket{\frac{3}{88},\frac{3}{88}}$        & 0.0735518  &0 & $(0.0681818,0)  $   \\

${\color{red}\ket{\frac{3}{55},\frac{3}{55}}}$        &  0.11703  &0 & $(0.1090909,0)  $    \\
$\ket{\frac{3}{55},\frac{3}{55}}$        &  0.118462  &0 & $(0.1090909,0)  $   \\

 \hline
\end{tabular}
\caption{In this table, we show the low lying energy level of the two impurity defect Hamiltonian. The scaling is done by $L-2$.}
\label{tab:31-defect-A10}
\end{table}

\newpage

\section{Fusion of defects}
\label{continuum}

We have discussed in earlier sections how we could use the ``fusion technology'' from the quantum inverse scattering arsenal to obtain $l$TDL and dTDLs associated with  Kac labels $(r,1)$ and $(1,s)$. For the $(1,s)$ defects - which are topological on the lattice as well - the construction boils down to bringing several $(1,2)$ defect lines together, and projecting onto higher values of $s$. Without the projection, we would simply obtain the same direct sum as the one predicted from conformal field theory, as discussed in subsection \ref{sec:fusion-Tmatrices-(1s)}. Things are a bit different for the $(r,1)$ defects. In order to have a direct sum when bringing several $(2,1)$ defects together, we need to shift the corresponding spectral parameters by some very specific amounts multiple of $\gamma$. The reason why this works - and can still be interpreted as fusion of  topological defect lines -  is that dTDLs are obtained not only for specific values of the impurity spectral parameter, but in whole domains. As it turns out, these domains are broad enough and have enough overlap that the fusion equations used to obtain the higher-spin Boltzmann weights can in fact be used to obtain the lattice equivalent of the fusion of defects  in the continuum limit. 
We saw some examples of this in the previous section, and now tackle the point systematically.

\subsection{The defect algebra as the continuum limit of the generalized \texorpdfstring{$T$}{Lg}-system}
\label{ALGEBRA}

We  go back to equation (\ref{Bilinearev0}), and wish to study its scaling limit - where the size of the system is taken to infinity while we focus on low-energy excitations. In this limit, we should obtain something akin to fusion of topological defects in the CFT.
A crucial aspect of this limit comes from normalization issues. As discussed earlier, $\widehat{\cal{D}}^{(\rm latt)}$ is identified in our construction only up to a non-universal bulk term $\exp[-2 \R e_0(\tilde{u})]$. When considering Eq. (\ref{Bilinearev0}) in the limit of large $\R$, we thus have to compare the bulk behaviors of the different pieces on the left and right hand sides. Ultimately, we need to factor these out in order to obtain a relation for the topological defects themselves. Note that in this process some terms  in (\ref{Bilinearev0}) may just disappear  if they are suppressed by a sub-leading bulk behavior.

The usual $Y$-system (or the TBA equations) allows for the identification of the non-universal part of the eigenvalues (\ref{bulk}). If we recall the relations (\ref{eq:spec_case_fusion}),  for each defect transfer matrix of a given spin $J$ there are two possible bulk behaviors separated by strips as in (\ref{expectedvalues}).  For $J=0$, however, the expression (\ref{bulk}) is not valid and one finds the bulk behavior in terms of the normalization of the identity matrix: $\Lambda^{(0)}(v_I)= {\left(\frac{\sinh v_I}{\sin \gamma}\right)}^{2 \R} 1$.

Observe that while a generic transfer matrix of spin $J$ has strips defined by dense set of zeros that divide the complex plane, this is not true when  $J=0$ or ,$J=p-1$, as in this case the zeros are isolated ones with degree $2 \R$. What happens when these transfer matrices  are present in the fusion equations  can be inferred by letting the spectral parameter get  close enough to these zeros: the term then tend to vanish if the remaining bilinear terms do not possess the same isolated zeros of the same degree.

\begin{figure}[htb]
  \begin{center}
  \includegraphics[width=0.31 \linewidth]{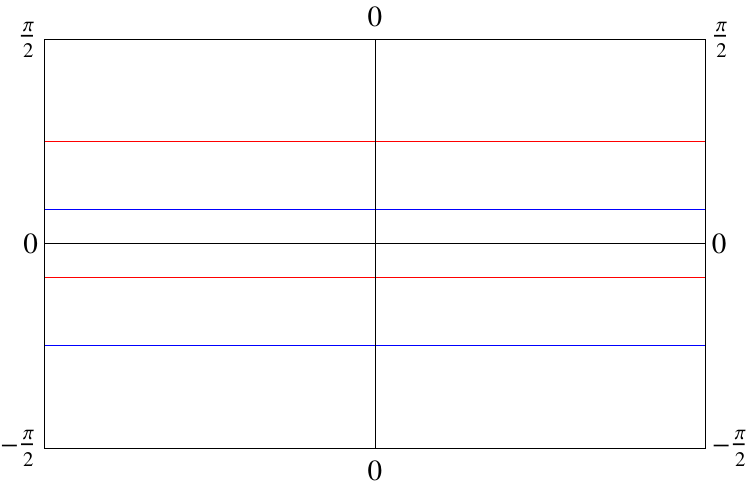}
  \includegraphics[width=0.31 \linewidth]{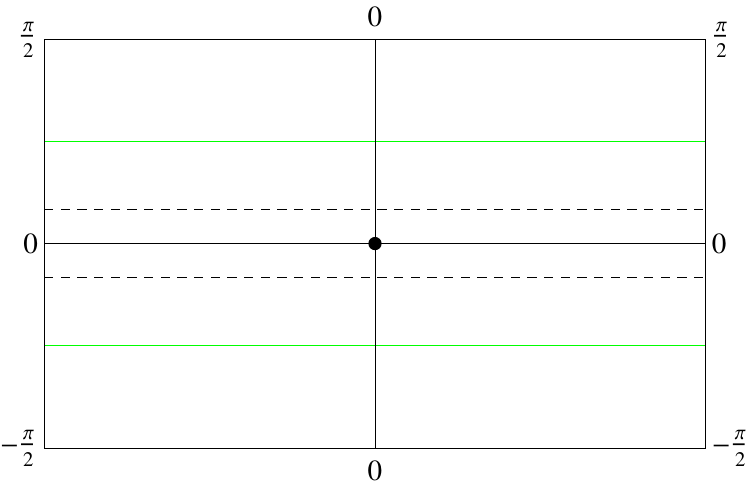}
  \includegraphics[width=0.31 \linewidth]{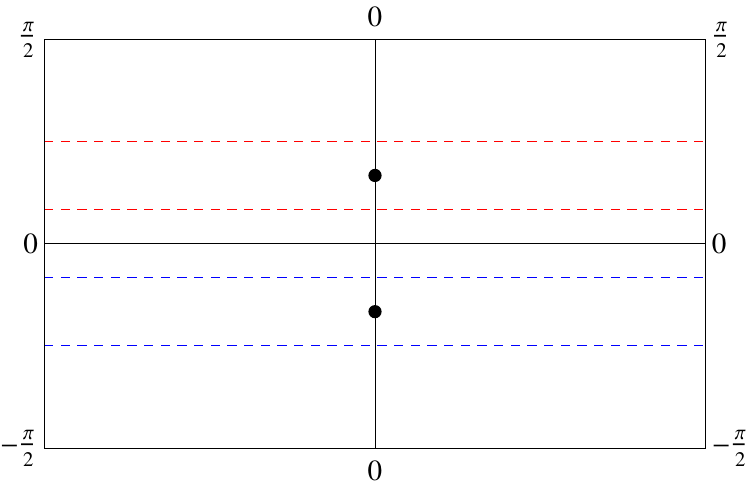}\caption{ figures represent  (dense) lines of zeros for the two eigenvalues of each bilinear term in Eq. (\ref{addedeq}). If the spectral parameter $v_I$  sits inside dashed lines, it eventually kills the bilinear term because of the difference bulk behaviors. }
\label{DefectAlgebraPic}
\end{center}
\end{figure}

To show how this works, let us consider the simplest non trivial relation of (\ref{Bilinearev0}) with $k=d=1$ and $\ell=0$:
\begin{equation}
\Lambda^{(1)} \left(v_I +\im \frac{\gamma}{2}\right) \Lambda^{(1)} \left(v_I -\im \frac{\gamma}{2}\right) = \Lambda^{(0)}(v_I) \Lambda^{(2)} (v_I) + \Lambda^{(0)}(v_I+\im \gamma) \Lambda^{(0)} (v_I -\im \gamma).\label{addedeq}
\end{equation}
In figure (\ref{DefectAlgebraPic}) we pictured lines of zeros corresponding to each bilinear term (the three terms in Eq. (\ref{addedeq}) are represented in the same order in the figure). Zeroes of the same eigenvalue  are identified by the same color. When the eigenvalue does not possess dense lines of zeros, we marked a black dot at the position of the isolated zeros of large degree and drew dashed lines which defines an extended region $|\Im (v_I-\Theta)|<\gamma/2$ around these zeros. By doing so, all the plots partition the complex plane in the same way: there are four regions if we consider that the plane is compactified. Moreover, the extended regions around isolated zeros follows the same structure of (\ref{expectedvalues}), so we are simply extending this result to $J=0$. Now let us see what it means concretely.

If we take any $v_I$ in the middle strip, i.e. $|\Im v_I|< \gamma/2$, the bulk contribution of the first term on the RHS is  smaller than the bulk contribution of LHS:
\begin{equation}
\Re( \log [\sinh (v_I+e_0^{(2)}(v_I))])<\Re (e_0^{(1)}(v_I+\im \gamma/2)+e_0^{(1)}(v_I-\im \gamma/2)) \, .
\end{equation}
Multiplying   the equation by the inverse of the bulk term of the LHS, the contribution from the  first term on RHS will  then  vanish at large $\R$. Of course, since the  identity is valid for any finite value of $\R$,  the second term on the RHS cannot possibly vanish in this limit. In fact, the   extensive non-universal contributions (the bulk terms)  must match, which we easily check since
\begin{equation}\label{goodId}
e_0^{(1)}(v_I+\im \gamma/2)+e_0^{(1)}(v_I-\im \gamma/2) = \log( \sinh(v_I+\im  \gamma) \sinh(v_I-\im  \gamma)),
\end{equation}
It follows that the product of the two fundamental transfer matrices on the LHS simply yields an identity operator on the RHS   at large $\R$. (This can be seen easily on figure \ref{DefectAlgebraPic} where for $|\Im v_I|< \gamma/2$,   $v_I$ sits between the dashed lines of the second term.)

Taking the strips so defined, i.e. with the appropriate extension to $J=0$ discussed above, let us denote the following scaling limits for $|\mbox{Re}(v_I)|<\infty$,
\begin{equation}
\uR^{-J} T^{(J)}\left(\frac{\gamma}{2}+\im v_I\right) \propto
\begin{cases}
 (-1)^{\R(J+1)} \widehat{\cal D}_{(J,1)}^{(\rm latt)} \mapsto  \widehat{\cal D}_{(J,1)},& \mbox{if } |\Im(v_I)|< \left(\frac{J+1}{2}\right) \gamma,  \\
 (-1)^{\R J} \widehat{\cal D}_{(J+1,1)}^{ (latt)}\mapsto \widehat{\cal D}_{(J+1,1)}, & \mbox{if } \left(\frac{J+1}{2}\right) \gamma <|\Im(v_I)|\leq \frac{\pi}{2},
\end{cases}\label{prr}
\end{equation}
 where as usual the symbol $\propto$ indicates equality up to a bulk term, and  we also introduced formally  $\widehat{\cal D}_{(0,1)}^{(\rm latt)}$ and $\widehat{\cal D}_{(p,1)}^{(\rm latt)}$ to indicate terms whose bulk terms are not compensated, and thus do not contribute to the identities to follow in the limit $\R\to\infty$ .

We see now that the simplest relation (\ref{Bilinearev0}) with $k=d=1$ and $\ell=0$ realizes an interesting algebraic relation on each one of the four strips:
\begin{align}
  \widehat{\cal D}^{(\rm latt)}_{(1,1)} \widehat{\cal D}^{(\rm latt)}_{(1,1)} & = \widehat{\cal D}^{(\rm latt)}_{(0,1)} \widehat{\cal D}^{(\rm latt)}_{(2,1)}+ \widehat{\cal D}^{(\rm latt)}_{(1,1)}\widehat{\cal D}^{(\rm latt)}_{(1,1)}\mapsto\widehat{\cal D}_{(1,1)},  \nonumber \\
  \widehat{\cal D}^{(\rm latt)}_{(1,1)} \widehat{\cal D}^{(\rm latt)}_{(2,1)} & = \widehat{\cal D}^{(\rm latt)}_{(1,1)} \widehat{\cal D}^{(\rm latt)}_{(2,1)}+ \widehat{\cal D}^{(\rm latt)}_{(0,1)} \widehat{\cal D}^{(\rm latt)}_{(1,1)}\mapsto  \widehat{\cal D}_{(2,1)}, \nonumber\\
  \widehat{\cal D}^{(\rm latt)}_{(2,1)} \widehat{\cal D}^{(\rm latt)}_{(1,1)} &  = \widehat{\cal D}^{(\rm latt)}_{(1,1)} \widehat{\cal D}^{(\rm latt)}_{(2,1)}+ \widehat{\cal D}^{(\rm latt)}_{(0,1)}\widehat{\cal D}^{(\rm latt)}_{(1,1)}\mapsto \widehat{\cal D}_{(2,1)}, \nonumber\\
  \widehat{\cal D}^{(\rm latt)}_{(2,1)} \widehat{\cal D}^{(\rm latt)}_{(2,1)} & = \widehat{\cal D}^{(\rm latt)}_{(1,1)} \widehat{\cal D}^{(\rm latt)}_{(3,1)}+ \widehat{\cal D}^{(\rm latt)}_{(1,1)}\widehat{\cal D}^{(\rm latt)}_{(1,1)}\mapsto
  \widehat{\cal D}_{(3,1)}+ \widehat{\cal D}_{(1,1)},
\end{align}
Note that in using (\ref{Bilinearev0}) to obtain identities for defect operators, we have had to insert also the proper powers of $\uR$ as well as $(-1)^\R$: it is easy to check that they match uniformly between the various terms of the equations. 

One can similarly consider relations with $k=2,~d=1$ and $\ell=0,~1,~2,\ldots,~p-1$ to find
\begin{equation}
\widehat{\cal D}_{(2,1)}^{(latt)} \widehat{\cal D}_{(k,1)}^{(latt)} \mapsto \begin{cases}
                    \widehat{\cal D}_{(2,1)}, & \mbox{if~} k=1, \\
                      \widehat{\cal D}_{(p-2,1)}, & \mbox{if~} k=p-1, \\
                    \widehat{\cal D}_{(k-1,1)}+\widehat{\cal D}_{(k+1,1)}, & \mbox{otherwise}.
                  \end{cases} \label{chiral fusion}
\end{equation}
We note that the possible   g-factors  may be computed directly  from this algebra. We could use the r-type defect operators themselves as a module for the action of $\widehat{\cal D}_{(2,1)}$, and find that, in this (faithful) representation,  the generator is given by the adjacency matrix of the $A_{p-1}$ Dynkin diagram. Therefore, the possible eigenvalues of $\widehat{\cal D}_{(2,1)}$ are ${(-1)}^{r'+s'}{[ 2 ]}_{\tilde{q}^{r'}}$, $r'=1,~2,\ldots,~p-1$ with $\tilde{q}={\rm e}^{i \frac{\pi}{p}}$, which of course is the  result well known from CFT. Clearly, there are underlying algebraic structures in the integrable  model that essentially bypass the continuum limit and give rise to similar results in finite size already - this is a well known fact in other contexts such as fusion of diagram algebras etc  - see \cite{Gainutdinov_2013} and references therein. 

Another important set of relations is obtained by considering (\ref{Bilinearev0}) with $k+d+\ell=p$ and $d=1$, so that the first bilinear term on the RHS vanishes due to the exact truncation $T^{(p)}(u)=0$, and we obtain for $v_I \to 0$:
\begin{align}
    \widehat{\cal D}^{(\rm latt)}_{(p-1,1)} \, \widehat{\cal D}^{(\rm latt)}_{(k,1)}&\mapsto 
   \widehat{\cal D}_{(p-k,1)}, 
     \qquad k=1,~2,\ldots,~p-1.  \label{contReflection}
\end{align}

It is also interesting  to notice that, since conformal weights can be represented using two different sets of Kac labels   $(r',s')$ (with $1\leq r'\leq p-1$ and $1\leq s'\leq p$) and $ (p-r',p+1-s')$ giving rise as well to  identical fusion rules, S-matrix elements etc in the CFT, there should be a similar identification for the lattice defect operators. Let us just give one example of this here. Introduce  
 \begin{equation}
 \Pi= \left(T^{(0)}_{[0]}(\gamma/2+iv_I)\right)^{-1} \cdot T^{(p-1)}_{[\frac{p+1}{2}]}(\gamma/2+iv_I) \, ={(-1)}^{p \R} {\cal R}, 
 \end{equation}
 which using analyticity property, can be shown to be defect parameter independent, hermitian and unitary, so $\Pi^2=1$\footnote{
To show this, we may use the following properties of the transfer matrices:
\begin{align}
{\left(T^{(J)}(\gamma/2+ iv)\right)}^t &=T^{(J)}(\gamma/2- iv), \nonumber\\
\bar{T}^{(J)}(\gamma/2+ iv)&=T^{(J)}(\gamma/2- i\bar{v}), \nonumber
\\
T^{(J)}(u+\pi)&=T^{(J)}(u), \nonumber
\end{align}
 where bar mark refers to complex conjugation. Consequently, all transfer matrices $T^{(J)}(\gamma/2+ iv_I)$ are hermitian at lines $\Im v_I =0 ~{\mbox{mod} ~\pi/2}$. Due to the commutativity of transfer matrices, so is $\Pi$.
Now, using the periodicity  and  the bilinear relation with $\ell=0$, $d=1$, $k=p-1$, we find that $\Pi(v_I+\im \gamma/2) \cdot \Pi(v_I-\im \gamma/2) =1 $. Since the eigenvalues of $\Pi(v_I)$, which we denote by $\Lambda_{\Pi}(v_I)$, are real at $\Im v_I =0 ~{\mbox{mod} ~\pi/2}$, they further satisfy $\bar{\Lambda}_{\Pi}(v_I)=\Lambda_{\Pi}(\bar{v}_I)$. This also results from the second and the third relations above. Therefore ${|\Lambda_{\Pi}(v_I+\im \gamma/2)|}^2=1$ all along these lines. Since the dependence of $\Lambda_{\Pi}$ on $v_I$ is meromorphic, either we have $\Lambda_{\Pi}(v_I+\im \gamma/2)=1$ or $\Lambda_{\Pi}(v_I+\im \gamma/2)=-1$, which is clearly constant. Hence, by analytic continuation theorem, $\Lambda_{\Pi}(v_I)$ is rather analytic and constant. Thus $\Pi(v_I)$ is also unitary and independent of $v_I$.  

This is, of course, all straightforward from the explicit representation we have at hand, in which the pole of order $2\R$ due to ${\left(T^{(0)}_{[0]}(\gamma/2+\im v_I)\right)}^{-1}$ is exactly canceled by the zero of the same order in ${T^{(p-1)}_{[\frac{p+1}{2}]}(\gamma/2+\im v_I)}$.}. It follows that  one may evaluate it in two different ways, which lead either to an r-type or an s-type defect,
\begin{align}
\Pi &\approx {(-1)}^{p\R} \tau^{(p-1) } \widehat{{\cal D}}_{(1,p)}^{(latt)},\\
&\approx {(-1)}^{p\R} \tau^{(p-1) } \widehat{{\cal D}}_{(p-1,1)}^{(latt)},
\end{align}
where the approximation is due to  sub-leading  $O(1/\R)$ corrections. It follows that
\begin{equation}
\widehat{{\cal D}}_{(p-1,1)} = \widehat{{\cal D}}_{(1,p)}, \qquad \widehat{{\cal D}}_{(p-1,1)}\cdot \widehat{{\cal D}}_{(1,p)} =1,
\end{equation}
in the scaling limit, as expected.

\subsection{Fusion of $(r,1)$ defects  without shifting the spectral parameters (cross-channel)}

While the result is convenient, it is of course unpleasant to have to shift the spectral parameters in the various dTDLs to perform  fusion. 

On the other hand, we saw in the numerical section several examples where we simply inserted several dTDLs and observed, in the scaling limit, the same results as those expected from the CFT. We now discuss why this is the case, and what more precisely can be expected.

A crucial point is that the $\widehat{\cal{D}}_{(2,1)}^{(\rm latt)}$ operators, while not to topological on the lattice {\sl do commute} with the transfer matrix due to the underlying integrability. This means in particular that they act as scalars on all the eigenstates of the transfer matrix, which, at low-energy, encode of course the ground and excited states of the CFT. It follows in turn that, since the eigenvalues of  $\widehat{\cal{D}}_{(2,1)}^{(\rm latt)}$ converge, for low-energy states, to the correct values in the  continuum limit, so does their product  - that is, the limit as $\R$ becomes large of the   numerical eigenvalues  of $\left(\widehat{\cal{D}}_{(2,1)}^{(\rm latt)}\right)^2$ is  the sum of the limits of the eigenvalues of $\widehat{\cal{D}}_{(3,1)}^{(\rm latt)}$ and $\widehat{\cal{D}}_{(1,1)}^{(\rm latt)}$ (and all of this is  obtained  {\sl without} shifting the spectral parameters). 

Of course, the convergence may not be uniform at all. To study this further, 
we consider  the quantity
\begin{equation}
\mbox{NF}(\R) = \lnorm{
\widehat{{\cal D}}^{(latt,\R)}_{(2,1)} \cdot \widehat{{\cal D}}^{(latt,\R) }_{(2,1)} -
\widehat{{\cal D}}^{(latt,\R) }_{(3,1)} -\widehat{{\cal D}}^{(latt,\R) }_{(1,1)}
},\label{normNFR}
\end{equation}
where, once again, all operators are evaluated at the same value of the spectral parameter. Moreover, we have introduced a suffix $\R$ to keep track explicitly of the  finite system-size $2\R$. Using the formula established previously  $\widehat{{\cal D}}^{(latt,\R)}_{(J+1,1)} =\uuR^{J} \exp({-2\R e_0^{(J)}})T^{(J)}((\pi-\gamma)/2)$ and  using the  spectral norm for matrices \footnote{Defined for a matrix $A$ as the square root of the largest eigenvalue of $A^\dagger A$.} we find however that   that $\mbox{NF}(\R)$ does not  converge to zero as $\R \to \infty$, see table \ref{FusionTest1} below.

\begin{table}[h]
  \centering
\begin{subtable}[t]{0.27\textwidth}\caption{}\label{taa}  
\begin{center}
\begin{tabular}{|c|c|}
\hline
 $2 \R$ & $\mbox{NF}(\R)$
 \\ \hline
 $4$ & $0.504532$
 \\ \hline
 $6$ & $0.934641$
 \\ \hline
 $8$ & $0.98402$
 \\ \hline
 $10$ &$1.00018$ 
 \\ 
 \hline
 \end{tabular}
 \end{center}
 \end{subtable} 
 \hfill
  \begin{subtable}[t]{0.27\textwidth }\caption{}\label{taab} 
  \begin{center}
 \begin{tabular}{|c|c|}
\hline
 $2 \R$ & $\mbox{NF}(\R)$
 \\ \hline
 $4$ & $0.0704755$
 \\ \hline
 $6$ & $0.0610188$
 \\ \hline
 $8$ & $0.0535826$
 \\ \hline
 $10$ & $0.0478148$
 \\ 
 \hline
 \end{tabular} 
 \end{center}
 \end{subtable}
\hfill
 \begin{subtable}[t]{0.27\textwidth}\caption{}\label{taac}
 \begin{center}
 \begin{tabular}{|c|c|}
\hline
 $2 \R$ & $\mbox{NF}(\R)$
 \\ \hline
 $4$ & $0.173609$
 \\ \hline
 $6$ & $0.150498$
 \\ \hline
 $8$ & $0.132619$
 \\ \hline
 $10$ & $0.118727$
 \\ 
 \hline
 \end{tabular}
 \end{center}
 \end{subtable}
 \caption{Assessment of fusion algebra for the $A_4$ model in  the large $\R$ limit. (\subref{taa}) By simply removing the non-universal extensive contribution from the transfer-matrix. (\subref{taab}) With further elimination of high-energy states, keeping only k=4 low-energy states. (\subref{taac}) k=6.
 }\label{FusionTest1}
\end{table}
This simply means that the $\ell_2$ norm is strongly affected (in fact, dominated) by highly excited states for which the eigenvalues of $l$TDLs converge to very different numbers (and in general are controlled by different bulk terms). 

In the spirit of other calculations like e.g. in \cite{Koo_1994}, we can instead truncate the Hilbert space by keeping only a definite number ($k$) of low-lying states for all transfer-matrices. The corresponding results are  now given in  (\ref{FusionTest1})-(b) with $k=4$ and (c) with $k=6$.
Obviously, the convergence is worse as $k$ increases (by definition of the norm itself). But we expect that  for any fixed $k$, the norm goes to zero as $\R\to\infty$. Then, in the familiar double limit process \cite{Koo_1994}, we will recover that the term in the norm on the rhs of Eq. (\ref{normNFR}) converges to zero. This is akin to getting the Virasoro algebra in the continuum limit of the Temperley-Lieb algebra in \cite{Koo_1994}.

\subsection{Fusion of $(r,1)$ defects in the direct channel}

We now consider fusion  of r-type defects in the direct channel. The crucial reason why we shifted the spectral parameters in many of our calculations is that, while the shift does not affect the IR limit of each individual defect, it gives rise, when the defects are combined, to an exact decomposition of the action of the combined impurity Hamiltonian into a sum of two independent channels. The mechanism behind this is discussed in detail in  appendix \ref{AnotherFusion}. We would like to re-express it here in slightly different terms, which will allow a better understanding of what happens when the spectral parameters are not shifted. 

To start, we introduce generically operators such as the one represented on figure \ref{doublecolumnn}. These operators act on  the ``internal'' labels $\alpha,\beta$, while depending on the ``external'' heights $x,\{y_i\}$ and $x,\{z_i\}$.   Explicitly, figure \ref{doublecolumnn} represents therefore the matrix element 
\begin{equation}
\langle\beta|M^{(\mathfrak{l})}(x,\{y_i\},\{z_i\})|\alpha\rangle.\label{Mdefined}
\end{equation}
\begin{figure}[b]
    \centering
\begin{tikzpicture}
    \draw (0,0) rectangle (2,4);
    \draw (1,0) -- (1,4);
    \foreach \y in {1,2,3} {
    \draw (0,\y) -- (2,\y);
}
\foreach \i in {0,...,3} {
    \node at (0.5,\i+0.5) {$u_1$};
    \node at (1.5,\i+0.5) {$u_2$};
}
\node[left] at (0,0) {$x$};
\node[left] at (0,1) {$y_1$};
\node[left] at (0,2) {$y_2$};
\node[left] at (0,3) {$y_3$};
\node[left] at (0,4) {$x$};

\node[right] at (2,0) {$x$};
\node[right] at (2,1) {$z_1$};
\node[right] at (2,2) {$z_2$};
\node[right] at (2,3) {$z_3$};
\node[right] at (2,4) {$x$};

\node[above] at (1,4.) {$\beta$};
\node[below] at (1,0) {$\alpha$};
\node[right] at (-2,2) {$M_{\alpha \beta} = $} ;

\end{tikzpicture}
    \caption{Double columns viewed as  operators with matrix elements indices  $\alpha$ and $\beta$  inside the defect Hilbert space. The  ``external'' heights comprise two ``propagation histories'' of length $\mathfrak{l}=4$ (number of faces): $x,~y_1,~y_2,~y_3,~x$ and $x,~z_1,~z_2,~z_3,~x$, which define the operator $M$. In general, two $M$'s with the same $x$ (hence acting on the same Hilbert space) but different propagation histories do not commute. However, they do if $u_2-u_1 = \gamma$.}
  \label{doublecolumnn}
\end{figure}
In what follows, we will often refer to a sequence of external heights as a ``propagation history''. We measure its  size   (and therefore of the operator $M$) by the number of faces it comprises along the time direction. Let us denote it $\mathfrak{l}$. Note that  such a double column of faces must be part of a non-contractible loop of faces on the cylinder, whose size is what we have denoted $2 \R$.

When the difference of parameters is $\pm \gamma$ not all linear combinations of internal heights $\alpha$  {\it connect} with  linear combinations of $\beta$, see the figure \ref{Disconnected2} in appendix \ref{AnotherFusion}. Regardless of the choices of propagation histories, there will be sets of $v^{+}_x$ vectors that cannot overlap with $v^{-}_x$ at the other extremity. This is the feature that produces well separated fusion channels in this case:   the $M$ operators then  have eigenvectors independent of the whole propagation histories, and commute.

Note that we have considered so far operators associated with the same heights at the four corners of the rectangle. Other choices would be possible too, although the adjacency rules of the $A_p$ diagram  render many of these operators trivial - i.e. proportional to the identity. In the case represented in figure \ref{doublecolumnn}, the  operators will be  $2\times 2$ matrices except for $x =1$ or $p$, where they'll be  $1\times 1$ matrices. Furthermore, the discussion could be extended to higher defects made out of several columns, and   leading to operators acting on larger spaces of intermediate/internal  (i.e., generalizing $\alpha,\beta$) heights. Here again, these operators would have eigenvectors independent of the whole propagation histories and commute whenever the spectral parameters of the different columns differ by some half-integer multiple of $\gamma$ \footnote{For instance,  when we fuse a defect $J=1$ with a $J=2$, the difference of parameters is $\pm 3 \gamma/2$. If we further recall that $J=2$ was formed by two fundamental faces $J=1$, then the three originating fundamental faces have spectral parameters that differ by units of $\gamma$.}.

In contrast, for general spectral parameters (especially when they are equal ($u_1=u_2$ in figure \ref{doublecolumnn})) and general propagation histories, the $M$ operators in (\ref{doublecolumnn})  and their generalizations do not allow for a separation of fusion channels, and do not commute at all. It is this commutation that we will now be interested in.

Note that  arbitrary propagation histories are not typical of the low-energy physics. To get closer to this, we  can  think of the problem in the cross-channel - that is,  with time running horizontally  on figure \ref{doublecolumnn}. A homogeneous transfer matrix or Hamiltonian propagating horizontally (thus acting on a system of length $2\R$ with periodic boundary-conditions) will have low-lying states, which we denote for convenience (as often in this paper) by the conformal weights they correspond to in the scaling limit, with a superscript $\R$ to keep this size dependency explicit. In what follows we will only consider $|h_{(1,1)},h_{(1,1)}\rangle^{(\R)}$ and 
$|h_{(2,2)},h_{(2,2)}\rangle^{(\R)}$. 

We now define a new version of the $M$ operators by sandwiching them between two such states. Explicitly this works as follows. We start by 
writing for instance
\begin{equation}
|h_{(1,1)},h_{(1,1)}\rangle^{({\mathfrak{\R}})}=\sum_{\{a_i\}} c_{\{a_i\}}|a_1,\ldots,a_{2 \R}\rangle,
\label{finitizedState}
\end{equation}
with $a_1=a_{2\R+1}$. Some of the heights will be at the positions of $x,\{y_i\},x$ on figure (\ref{Mdefined}): the others will be above and below, and we refer to those collectively as $a_{\text{out}}$. We then define a modified version of the operator $M$ by constructing the new matrix elements
\begin{equation}
\langle \beta |M_{(1,1)}^{(\mathfrak{l},\R)}(x)|\alpha\rangle\equiv \sum_{\{a_{\text{out}},\{y_i\},\{z_i\}\}}c^*_{\{a_{\text{out}},x,\{y_i\}\}}c_{\{a_{\text{out}},x,\{z_i\}\}}\langle \beta|M(x,\{y_i\},\{z_i\})|\alpha\rangle,
\end{equation}
where the sub-index $(1,1)$ labels the two states $|h_{(1,1)},h_{(1,1)}\rangle^{(\R)}$ that have been used on both sides.

Now these operators depend only on $x,\R,\mathfrak{l}$ - and the low-lying energy state we have used to project out high-energy propagation histories. We can finally calculate (numerically) the commutator of two such operators. Here is an example 
for the $A_4$ model, and  $x=2$ (resp. to Fig. \ref{doublecolumnn}), where we have measured
\begin{equation}
C = \lnorm{\Big[M^{(\mathfrak{l},\R)}_{(1,1)}/\lnorm{M^{(\mathfrak{l},\R)}_{(1,1)}},M^{(\mathfrak{l},\R)}_{(2,2)}/\lnorm{M^{(\mathfrak{l},\R)}_{(2,2)}}\Big]}, \label{norml2co}
\end{equation}
where as before  $\lnorm{\cdot}$ is the spectral norm. 
with the results given in table \ref{ACommutation}.
\begin{table}[ht]
  \centering
\begin{tabular}{|c| c c c c c c|}\hline
$2\R$\textbackslash $\mathfrak{l}$   & 2        & 4         & 6         & 8         & 10        & 12  \\ \hline
2 & 0.105204 & ~         & ~         & ~         & ~         & ~  \\
4 & 0.068526 &	0        & ~         & ~         & ~         & ~  \\
6 & 0.047401 &	0.043912 &	0        & ~         & ~         & ~  \\
8 & 0.029262 &	0.059179 & 	0.016328 &	0        & ~         & ~  \\
10& 0.019295 &	0.05027  &	0.044857 &	0.002206 &	0        & ~  \\
12& 0.013563 &	0.039653 &	0.049543 &	0.029219 &	0.005178 & 0  \\ \hline
\end{tabular}
\caption{The norm of commutators between  operators of increasing propagation history $\mathfrak{l}$ after projection onto the low-energy sector. Both defect parameters are set to $\tilde{u}=\pi/2$.}
\label{ACommutation}
\end{table}

What we would like to see is clear evidence that, when one takes the limit $\R$ large first and then $\mathfrak{l}$ large, the operators commute. While $\mathfrak{l}=2$ shows a nice behavior, the case $\mathfrak{l}=4$  exhibits non-monotonicity. Observe however that along each line ($2 \R$ fixed), the maximum value occurs at $\mathfrak{l}=\R$ (when possible). Now if we list these values as function of $\R$, we see that they decrease monotonically: for $2\R =4,~8,~12$ the maximum occurs at $\mathfrak{l}=2,~4,~6$ with values of $\approx ~0.069,~0.059,~0.050$. If this converges to zero, indeed, then for increasing values of $\R$, larger than any given $\mathfrak{l}$, it is also natural to expect that the values will again converge to zero,  which is in agreement with our claims.

\subsection{Fusion of $(r,1)$ and $(1,s)$ defects}
\label{9sub4}
We have so far seen that fusion of $(1,s)$ defects can be done exactly on the lattice (up to the slight complication of the $\uR$ factors) as well, where  these defects are already topological. Since they are  obtained in our construction for infinite values of the spectral parameter, the question of whether we should shift the spectral parameters to compose them does not arise.

For fusion of $(r,1)$ defects we have seen that a version of it  can be done exactly on the lattice - in the sense that we obtain well defined fusion channels - provided we shift spectral parameters appropriately. We have also seen that, if one doesn't do so but take all the impurity faces at their ``standard'' value of the impurity spectral parameter, fusion still works, although the decomposition into channels only holds in the continuum limit.

Now the question arises of how to fuse $(1,s)$ and $(r,1)$ defects. This doesn't seem to be possible exactly within the integrable formalism. That's because the  generalized bilinear relations (\ref{Bilinearev0}) do not relate r-type and s-type defects. Indeed, if we fix $|\Re v_I|<\infty$ all transfer matrices in these relations will result in $r$-type defects; similarly,  by sending $v_I \rightarrow \pm \infty$, one finds relations between chiral or anti-chiral defects, i.e. never mixing r-type to s-type or even s-type of chiral or anti-chiral nature. On the other hand, the fact that our construction is based on quantum integrability once again comes to the rescue. We can indeed form the product 
\begin{equation} \label{VerDefects}
\widehat{\cal D}_{(r,s)}^{(\rm latt)}\equiv \widehat{\cal D}_{(r,1)}^{(\rm latt)} \widehat{\cal D}_{(1,s)}^{(\rm latt)},
\end{equation}
for any $(r,s)$ in Kac's table. Since the two terms in the product satisfy the fusion algebra \cite{Belletete2020} in the continuum limit, and since they commute and act diagonally on eigenstates of the Hamiltonian, we are guaranteed that the product itself also satisfies the corresponding fusion relations.

For example, if we perform the same computations as in table \ref{ACommutation}, but now with defect parameters $\tu_1 = \pi/2$ and $\tu_2 \to\im \infty$ as to reproduce the composition of defects $(2,1)$ and $(1,2)$, we obtain results in table \ref{ACommutation2}.
\begin{table}[ht]
  \centering
\begin{tabular}{|c| c c c c c c|}\hline
$2\R$\textbackslash $\mathfrak{l}$   & 2        & 4         & 6         & 8         & 10        & 12  \\ \hline
2 & 0 & ~           & ~         & ~         & ~         & ~  \\
4 & 0 &	0        & ~         & ~         & ~         & ~  \\
6 & 0 &	0.011733 &	0        & ~         & ~         & ~  \\
8 & 0 &	0.022988 & 	0.012688 &	0        & ~         & ~  \\
10& 0 &	0.021762 &	0.030936 &	0.011192 &	0        & ~  \\
12& 0 &	0.017804 &	0.037218 &	0.030913 &	0.009461 & 0  \\ \hline
\end{tabular}
\caption{The norm of commutators between  operators of increasing propagation history $\mathfrak{l}$ after projection onto the low-energy sector. One defect parameter is set to $\tilde{u}=\pi/2$ while the other is set to $\tilde{u}\to\im \infty$ with the appropriate normalization.}
\label{ACommutation2}
\end{table}
While the numbers are of the same order as in table \ref{ACommutation}, and the first column is always zero , we see that contrary to the r-type fusion  the maximum values in each line does not seem to converge to zero: for $2\R = 4, 8, 12$, the maximum occurs at $\mathfrak{l} = 2, 4, 6$,
with values of $\approx 0,~ 0.023,~ 0.037$. This is an evidence that the fusion channels do not emerge, rather the product of defects $\widehat{{\cal D}}_{(2,1)}\cdot \widehat{{\cal D}}_{(1,2)}$ should be seen as a new single entity, which we denote $\widehat{{\cal D}}_{(2,2)}$.

Now, from the definition (\ref{eq:prop-fac-high-fus}) we see that
\begin{equation}
 \widehat{{\cal D}}_{(1,p)} \cdot \widehat{\cal D}_{(1,s)} = \widehat{\cal D}_{(1,p+1-s)}.
\end{equation}
Then, we can  verify that the defects (\ref{VerDefects}) satisfy the Kac's table symmetry:
\begin{multline}
\widehat{\cal D}_{(r,s)}=\widehat{\cal D}_{(r,1)} \cdot \widehat{\cal D}_{(1,s)} = \widehat{\cal D}_{(r,1)}\cdot(\widehat{\cal D}_{(p-1,1)}\cdot \widehat{\cal D}_{(1,p)})\cdot  \widehat{\cal D}_{(1,s)}=\\  \widehat{\cal D}_{(p-r,1)} \cdot \widehat{\cal D}_{(1,p+1-s)}= \widehat{\cal D}_{(p-r,p+1-s)},
\end{multline}
so the defect operators obey the expected  equivalence relations. 

Note once again that,  using the set of all independent defects as a module for the action of the defect operators themselves, we find that the eigenvalues of the generators $\{\widehat{{\cal D}}_{(2,1)},\widehat{{\cal D}}_{(1,2)}\}$ are conveniently parametrized  by $\{(-1)^{r'+s'}[2]_{\tilde{q}^{r'}},(-1)^{r'+s'}[2]_{q^{s'}} \}$, where $(r',s')\equiv (p-r',p+1-s')$ runs over the whole Kac's table and obeys the required equivalence relation.

\section{Conclusion}

To conclude this paper, we have seen that all TDL in diagonal Virasoro minimal models of CFT can be obtained as  continuum limit of well defined lattice objects using integrable RSOS realizations. While this may appear too technical for a conclusion, we feel it is  useful  to recall here the corresponding relevant formulas for the $\widehat{\cal {D}}$:
(see also figure  \ref{DefectRegions}): 
\begin{equation}
    \begin{split}
        & (-1)^{\R(J+1)}\uR^{-J} e^{-2R\tilde{e}_0^{(J)}(v_I)}T^{(J)}\left( \frac{\gamma}{2} + {\rm i} v_I \right) \mapsto \widehat{\mathcal{D}}^{}_{(J,1)}, 
 \text{for } |\mathrm{Im}\,v_I| < (J+1)\gamma/2,\; |\mathrm{Re}\,v_I| < \infty \, , \\
& (-1)^{\R J}\uR^{-J}e^{-2R\tilde{e}_0^{(J)}(v_I)}T^{(J)}\left( \frac{\gamma}{2} + {\rm i} v_I \right) \mapsto \widehat{\mathcal{D}}^{}_{(J+1,1)} \,  , 
\text{for } (J+1)\gamma/2 < |\mathrm{Im}\,v_I| \leq \pi/2,\; |\mathrm{Re}\,v_I| < \infty  \, , \\
&
\uR^{-J}\left((-q)^{J/2} {\sin \gamma\over \sin\left({J\over 2}\gamma-iv_I\right)}\right)^{2\R} T^{(J)}\left({\gamma\over 2}+iv_I\right)=\uR^{-J} Y_{J\over 2}\mapsto \widehat{\mathcal{D}}^{}_{(1,J+1)}\,
\,
\text{for } |\mathrm{Re}\,v_I| = \infty \, .
    \end{split}\label{summaryeq}
\end{equation}
where 
\begin{equation}\label{eq:corr-norm}
\begin{split} 
    \tilde{e}_0^{(J)}(v_I) = &
\begin{cases}
e_{0}^{(J)}(v_I), & \text{for } |\operatorname{Im} v_I| < (J+1)\gamma/2 \, , \\
e_{0}^{(p-1-J)}(v_I \mp \frac{\pi {\rm i}}{2} ), & \text{for } (J+1)\gamma/2 < |\operatorname{Im} v_I| \leq \pi/2 \, ,
\end{cases}
\end{split}
\end{equation}
and we defined $e_0^{(J)}$ in Eq. \eqref{bulk}\footnote{When we realize the $(2,1)$ line operator on the lattice using $T(\frac{\pi}{2} )$, i.e. $v_I = - {\rm i } \left( \frac{\pi}{2}  - \frac{\gamma}{2}\right)$,
  the factor in Eq. \eqref{eq:corr-norm}  is $\exp\left( - 2 \R e_0^{(p-2)}\left(-i{\gamma \over2}\right)\right) = \tan^{2 \R} \gamma $}. We also saw that, while the last type of defect in Eq. (\ref{summaryeq}) corresponds to a TDL that is topological on the lattice, this property is obeyed for the others only in the continuum limit. Similar results hold in the direct channel for the realization of the defect Hamiltonians or transfer matrices using the defect spectral parameters $\tu=iv_I$. 

 We also saw in this paper the importance of lattice fusion relations. Combined with the existence of domains for the continuum limit of our dTDLs, they form one more example of an algebraic structure of the CFT that is already present on the lattice, completing examples such as fusion of Virasoro and lattice algebras representations, structure of null vectors in Virasoro or lattice algebras, lattice modular invariance etc. This could provide a particularly useful tool to investigate defects in non-rational, non-unitary CFTs (such as loop models), where little is known so far. 
 
Generalizations to the case of non-diagonal minimal CFTs should be straightforward - see \cite{Sinha:2023hum} for the example of the three-state Potts model. It is also 
 natural to expect that similar properties will hold for other integrable models and associated CFTs - e.g. the RSOS models  based on the $SU(n)$ weight diagram and the corresponding minimal $SU(n)$ coset CFTs. This is clear for the equivalent of the $(1,s)$ defects, which can still be obtained in the limit $|\mathrm{Re}\,v| = \infty$, and will give rise to $l$TDLs as well. Finally, note that nothing in our construction depended on the unitarity of the underlying CFT:  generalizations to the  non-unitary case  should therefore be immediate.

 We emphasize that the construction of ${\cal D}_{(1,s)}$ TDLs extends to models which are based on the Temperley-Lieb algebra, even {\sl when non-integrable}, since it follows from  a simplified version of  Yang-Baxter moves (braid relations),  quite generally valid if the interaction depends only of the $e_i$ generators.  This applies  in particular to off-critical RSOS models perturbed by the $\Phi_{(21)}$ operator, or to  the massive Potts model. Remarkably, the construction in fact generalizes even to bond disordered Potts models. 

On the other hand, the construction of the other TDLs - in particular the ${\cal D}_{(2,1)}$ relies entirely on integrability\footnote{Recall that all these comments hold for the ``dense versions'' of the models. For their ``dilute versions'', the roles of ${\cal D}_{(1,s)}$ and ${\cal D}_{(r,1)}$ have to be switched.}. Even with this property, is not totally clear why inserting lines with modified spectral parameters in an integrable model should give rise to TDLs  in the continuum limit. As discussed in more detail in \cite{tavares2024}, one can in some cases consider that lines with modified spectral parameter describe {\sl perturbations} of topological defects, and thus can be expected to flow to conformal defects - in fact, topological defects if one can moreover argue that the perturbation is chiral \cite{Kormos:2009sk,tavares2024}. But for the case of spectral parameter $\tu=\pm {\pi\over2}$ (for instance) this is not so clear, even though the underlying integrability guarantees remarkable properties (like commutation with $L_0+\bar{L}_0$) from the onset.

The  lattice framework described in this paper is crucial for the quantitative investigation of several questions concerning TDLs that remain open. These include the computation of entanglement characteristics of subsystems containing TDLs in CFTs. This is particularly important when the  defect line coincides with the boundary of the subsystem. In this case, the field theory computations have been shown to be incompatible with ab-initio lattice computations~\cite{Roy2021a, Rogerson_2022, Roy2024}. Yet another open question concerns the fate of TDLs along renormalization group flows connecting two different CFTs, generalizing the framework of this paper and the results obtained in Ref.~\cite{tavares2024}. Finally, lattice incarnations enable realization of the TDLs in physical systems. Given the fine-tuned nature of the Hamiltonians required for TDLs, engineered quantum systems are natural candidates for their realization. In the age where noisy quantum devices are readily available and larger-scale quantum simulators are within reach, the proposed lattice framework is crucial for investigation of those questions involving TDLs where integrability or tensor network methods have limited success. These include transport and non-equilibrium characteristics as well as questions concerning thermalization. Generalizing the embedding of RSOS models with qubit registers~\cite{Roy:2024xdi}, the various TDLs discussed in this work and more broadly, a large family of low-dimensional QFTs, could be realized in near-term quantum simulators.

\bigskip

\noindent {\bf Acknowledgments:}
We thank Paul Fendley, Holger Frahm, Gleb Kotousov, John McGreevy, Abhinav Prem, Ingo Runkel, Sahand Seifnashri, and Bram Vancraeynest-De Cuiper for very helpful discussions. We also thank Jonathan Bellet\^ete, Azat Gainutdinov, Jesper Jacobsen, Linnea Grans-Samuelsson, and Fei Yan for discussions and related earlier collaborations. The work of H.S.  was supported by the French Agence Nationale de la Recherche (ANR) under grant ANR-21-CE40-0003 (project CONFICA).
\newpage
\begin{appendix}
\newpage
\section{Expression of the transfer matrix  in terms of affine TL generators}\label{sec:Tmatrix-appendix}
We derive in this section the analytical expression of the most  general transfer matrix when all spectral parameters   $\{u_0,u_1,\ldots u_{2\Ll-1} \}$ are a priori different. We first state the result
\begin{equation}\label{eq:Tmatrix-TL}
 T\left(  \{ u \} \right)   = \frac{\sin u_0 }{\sin^{2\Ll} \gamma } \left(\prod_{j = 1}^{2\Ll-1}\tilde{R}_j(u_j) \right) \,  \uR^{-1} +    \frac{\sin (\gamma - u_0)}{\sin^{2\Ll} \gamma} \uR \prod_{j = 1}^{2L-1} R_{2\Ll-j}(u_{2\Ll-j}) \, ,  
\end{equation}
\begin{equation}
    \text{ where } R_j(u_j) = \left( \sin(\gamma-u_j) \, \mathbb{1} + \sin(u_j) \,  e_j \right), \ \tilde{R}_j(u_j) =  \left( \sin(\gamma-u_j) \, e_j + \sin(u_j) \, \mathbb{1}\right) \, .  
\end{equation}

To make notations clear, here we have a TL chain where the first site is labeled by $0$ and the last site is labeled by $2L - 1$. Also note $2L \equiv 0$ and 
\begin{equation}\label{eq:order-R-mat}
\begin{split}
    \prod_{j = 1}^{2\Ll-1} R_{2\Ll-j}(u_{2\Ll-j}) & \equiv R_{2\Ll-1}(u_{2\Ll-1})\, R_{{2\Ll-2}}(u_{2\Ll-2})\dots R_{1}(u_1) \, , \\\
 \prod_{j = 1}^{2\Ll-1}\tilde{R}_j(u_j) & \equiv \tilde{R}_{1}(u_1)\tilde{R}_{2}(u_2) \dots \tilde{R}_{2\Ll-1}(u_{2\Ll-1}) \, .
\end{split}
\end{equation}
To prove the above identity, we first note that 
\begin{equation}
\begin{split}
    & \bra{b_0, b_1, \ldots, b_{2\Ll -1}} \tilde{R}_{2\Ll - 1}(u_{2\Ll - 1}) \uR^{-1} \ket{a_0, a_1, \ldots, a_{2\Ll -1}}
    \\ &= \left( \prod_{i = 0}^{2\Ll -2} \delta_{ a_{i + 1} ,b_i } \right) \left( \sin (\gamma - u_{2\Ll -1}) \, \frac{[a_{2\Ll}]^{1/2} [b_{2\Ll-1}]^{1/2}}{[a_{2\Ll - 1}]^{1/2} [b_{2\Ll}]^{1/2}} \delta_{a_{2\Ll - 1}, b_{2\Ll}}  + \sin (u_{2\Ll - 1}) \delta_{a_{2\Ll}, b_{2\Ll - 1}} \right) \, .
\end{split}
\end{equation}
Now, we use mathematical induction, so we first assume the expression 
\begin{equation}\label{eq:induc-tmat}
\begin{split}
    & \bra{b_0, b_1, \ldots, b_{2\Ll -1}} \tilde{R}_{2\Ll - k}(u_{2\Ll - k}) \ldots \tilde{R}_{2\Ll - 1}(u_{2\Ll - 1}) \,  \uR^{-1} \ket{a_0, a_1, \ldots, a_{2\Ll -1}}
    \\ &=\left( \prod_{i = 0}^{2\Ll - k -1} \delta_{a_{i + 1} , b_i }\right) \prod_{j = 1}^k   \Bigg( \sin (\gamma - u_{2\Ll -j})\, \frac{[a_{2\Ll - j + 1}]^{1/2} [b_{2\Ll-j}]^{1/2}}{[a_{2\Ll - j}]^{1/2} [b_{2\Ll- j + 1}]^{1/2}} \delta_{a_{2\Ll - j}, b_{2\Ll - j + 1}}  + \\
    & \quad  \sin (u_{2\Ll - j}) \, \delta_{a_{2\Ll - j + 1}, b_{2\Ll - j}} \Bigg)  \, , 
\end{split}
\end{equation}
where $k \leq 2\Ll - 2$, then by using the expression 
\begin{equation}
\begin{split}
    & \bra{b_0, b_1, \ldots, b_{2\Ll -1}} \tilde{R}_{2\Ll - k - 1}(u) \ket{a_0, a_1, \ldots, a_{2\Ll -1}}  \\
   &  =  \left(\prod_{i \neq 2\Ll - k - 1} \delta_{a_i, b_i}\right) \Bigg( \sin(  \gamma - u_{2\Ll - k - 1} ) \, \delta_{a_{2\Ll - k - 2} , b_{2\Ll - k}} \frac{[a_{2\Ll - k - 1}]^{1/2}[b_{2\Ll - k - 1}]^{1/2}}{[a_{2\Ll - k-2}]^{1/2} [b_{2\Ll - k} ]^{1/2} }  +  \\  
&  \quad   \sin(u_{2\Ll - k - 1}) \, \delta_{a_{2\Ll -k - 1 },b_{2\Ll -k - 1 } } \Bigg) \, , 
\end{split}
\end{equation}
we can show that
\begin{equation}
\begin{split}
    & \bra{b_0, b_1, \ldots, b_{2\Ll -1}} \tilde{R}_{2\Ll - k-1}(u_{2\Ll - k-1}) \tilde{R}_{2\Ll - k}(u_{2\Ll - k}) \ldots \tilde{R}_{2\Ll - 1}(u_{2\Ll - 1}) \uR^{-1} \ket{a_0, a_1, \ldots, a_{2\Ll -1}}
    \\ &=\left( \prod_{i = 0}^{2\Ll - k -2} \delta_{a_{i + 1} , b_i }\right) \prod_{j = 1}^{k+1}  \Bigg( \sin (\gamma - u_{2\Ll -j})\frac{[a_{2\Ll - j + 1}]^{1/2} [b_{2\Ll-j}]^{1/2}}{[a_{2\Ll - j}]^{1/2} [b_{2\Ll- j + 1}]^{1/2}} \delta_{a_{2\Ll - j}, b_{2\Ll - j + 1}}  + \\
 & \quad    \sin (u_{2\Ll - j}) \delta_{a_{2\Ll - j + 1}, b_{2\Ll - j}} \Bigg) \, . 
\end{split}
\end{equation}
Hence, let us take $k = 2\Ll -2$ in the above expression\footnote{Note, in our induction step we had taken $k \leq 2\Ll -2$, so we cannot take a value of $k$ greater than $2\Ll-2$.}, to get
\begin{equation}\label{eq:prod_tildeR}
\begin{split}
    & \bra{b_0, b_1, \ldots, b_{2\Ll -1}} \tilde{R}_{1}(u_{1}) \ldots \tilde{R}_{2\Ll - 1}(u_{2\Ll - 1}) \,  \uR^{-1} \ket{a_0, a_1, \ldots, a_{2\Ll -1}}
    \\ &=\delta_{a_{1} , b_0 } \prod_{j = 1}^{2\Ll-1}  \left( \sin (\gamma - u_{2\Ll -j}) \, \frac{[a_{2\Ll - j + 1}]^{1/2} [b_{2\Ll-j}]^{1/2}}{[a_{2\Ll - j}]^{1/2} [b_{2\Ll- j + 1}]^{1/2}} \delta_{a_{2\Ll - j}, b_{2\Ll - j + 1}}  + \sin (u_{2\Ll - j}) \, \delta_{a_{2\Ll - j + 1}, b_{2\Ll - j}} \right)  \, . 
\end{split}    
\end{equation}
Similarly, we can show 
\begin{equation}\label{eq:prod-R}
\begin{split}
   &  \bra{b_0, b_1, \ldots, b_{2\Ll - 1}} \uR \, R_{2\Ll - 1} (u_{2\Ll - 1}) \ldots R_{1}(u_1)\ket{a_0, a_1, \ldots b_{2\Ll - 1}}  \\
  & = \delta_{a_0, b_1} \prod_{j = 1 }^{2\Ll - 1} \left( \sin (\gamma - u_{2\Ll -j}) \, \delta_{a_{2\Ll - j}, b_{2\Ll - j + 1}}  +   \sin (u_{2\Ll - j}) \, \frac{[a_{2\Ll - j}]^{1/2} [b_{2\Ll- j + 1}]^{1/2}}{[a_{2\Ll - j + 1}]^{1/2} [b_{2\Ll-j}]^{1/2}} \delta_{a_{2\Ll - j + 1}, b_{2\Ll - j}}    \right) \, . 
\end{split}
\end{equation}
We can re-write Equation \eqref{eq:prod_tildeR} as follows
\begin{equation}\label{eq:prod-tildeR}
    \begin{split}
     & \bra{b_0, b_1, \ldots, b_{2\Ll -1}} \tilde{R}_{1}(u_{1}) \ldots \tilde{R}_{2\Ll - 1}(u_{2\Ll - 1}) \,  \uR^{-1} \ket{a_0, a_1, \ldots, a_{2\Ll -1}}\quad\quad\quad\quad \\ 
     & =  \delta_{a_1,b_0} \frac{[a_0]^{1/2} [b_1]^{1/2}}{ [a_1]^{1/2} [b_0]^{1/2}}
     \prod_{j = 1 }^{2\Ll - 1} \Bigg( \sin (\gamma - u_{2L -j})\, \delta_{a_{2\Ll - j}, b_{2\Ll - j + 1}}  +    \,  \\    
  & \sin (u_{2\Ll - j})  \frac{[a_{2\Ll - j}]^{1/2} [b_{2\Ll- j + 1}]^{1/2}}{[a_{2\Ll - j + 
     1}]^{1/2} [b_{2\Ll-j}]^{1/2}}   \delta_{a_{2\Ll - j + 1}, b_{2\Ll - j}} \Bigg) \, . 
    \end{split}
\end{equation}
Now, we can substitute Equations \eqref{eq:prod-R} and \eqref{eq:prod-tildeR} in Equation \eqref{eq:Tmatrix-TL} 
\begin{equation}
\begin{split}
   & \bra{b_0, b_1, \ldots b_{2\Ll -1} }T(\{ u \}) \ket{a_0, a_1, \ldots, a_{2\Ll -1}} \\ 
   & = \frac{1}{\sin^{2\Ll} \gamma}\left( \sin (\gamma - u_0) \, \delta_{a_0, b_1}  + \sin u_0  \,  \frac{[a_0]^{1/2} [b_1]^{1/2}}{ [a_1]^{1/2} [b_0]^{1/2}}  \delta_{a_1,b_0} \right) 
   \\
   & \, \, \prod_{j = 1 }^{2\Ll - 1} \left( \sin (\gamma - u_{2\Ll -j}) \, \delta_{a_{2\Ll - j}, b_{2\Ll - j + 1}}  +  \sin (u_{2\Ll - j}) \frac{[a_{2\Ll - j}]^{1/2} [b_{2\Ll- j + 1}]^{1/2}}{[a_{2\Ll - j + 1}]^{1/2} [b_{2\Ll-j}]^{1/2}}   \delta_{a_{2\Ll - j + 1}, b_{2\Ll - j}} \right) \, , \\
 & =   \prod_{j = 1 }^{2L } \left( \frac{\sin (\gamma - u_{2L -j})}{\sin \gamma} \, \delta_{a_{2L - j}, b_{2L - j + 1}}  +   \frac{\sin (u_{2L - j})}{\sin \gamma} \frac{[a_{2L - j}]^{1/2} [b_{2L- j + 1}]^{1/2}}{[a_{2L - j + 1}]^{1/2} [b_{2L-j}]^{1/2}}   \delta_{a_{2L - j + 1}, b_{2L - j}} \right) \, .
\end{split}
\end{equation}
Hence, we can now show

\bigskip

\begin{adjustbox}{max totalsize={\textwidth}{\textheight},center}

    \begin{tikzpicture}[thick, scale=1]
    \begin{scope}[very thick, every node/.style={sloped,allow upside down}]

            \node[] at (-2.5,1.5) {\LARGE{}{ $ \bra{b} T \left( \{ u \} \right) \ket{a } = $}};
            \draw[black, thick] (0,0) -- (18,0);
            \draw[black, thick] (0,3) -- (18,3);
            \draw[black, thick] (0,0) -- (0,3);
            \draw[black, thick] (3,0) -- (3,3);
            \draw[black, thick] (6,0) -- (6,3);
            \draw[black, thick] (9,0) -- (9,3);
            \draw[black, thick] (12,0) -- (12,3);
            \draw[black, thick] (15,0) -- (15,3);
            \draw[black, thick] (18,0) -- (18,3);
            \node[] at (0,-0.3) {\Large{$a_0$}};
            \node[] at (3,-0.3) {\Large{$a_1$}};
            \node[] at (6,-0.3) {\Large{$a_2$}};
            \node[] at (9,-0.3) {\Large{$a_3$}};
            \node[] at (12,-0.3) {\Large{$a_{2\Ll-2}$}};
            \node[] at (15,-0.3) {\Large{$a_{2\Ll-1}$}};
            \node[] at (18,-0.3) {\Large{$a_0$}};
            \node[] at (0,3.4) {\Large{$b_0$}};
            \node[] at (3,3.4) {\Large{$b_1$}};
            \node[] at (6,3.4) {\Large{$b_2$}};
            \node[] at (9,3.4) {\Large{$b_3$}};
            \node[] at (12,3.4) {\Large{$b_{2\Ll-2}$}};
            \node[] at (15,3.4) {\Large{$b_{2\Ll-1}$}};
            \node[] at (18,3.4) {\Large{$ b_0$}};
            \node[] at (1.5,1.5) {\Large{$u_0$}};
            \node[] at (4.5,1.5) {\Large{$u_1$}};
            \node[] at (7.5,1.5) {\Large{$u_2$}};
            \node[] at (10.5,1.5) {\Large{$\ldots$}};
            \node[] at (13.5,1.5) {\Large{$u_{2\Ll-2 }$}};
            \node[] at (16.5,1.5) {\Large{$u_{2\Ll-1}$}};
        \end{scope}
    \end{tikzpicture}
\end{adjustbox}
where the weight of a face is given by 
\begin{equation}
   W\left(\begin{array}{ll}
d & c \\
a & b
\end{array} \Bigg| \ u\right)= \left( \frac{\sin(\gamma - u)}{\sin \gamma} \, \delta_{a , c} + \delta_{b , d}  \sqrt{\frac{g_a g_c}{g_b g_d}} \frac{\sin u}{\sin \gamma}   \right) \, . 
\end{equation}
\newpage 
\section{Eigenvalue of the  \texorpdfstring{$Y$}{Lg} operator from modules of Affine Temperley-Lieb}
\label{sec:Yeigenvalues}
For A$_p$ RSOS model, using the modules of aTL$_{2\R}(q)$, we can predict how $Y$ operator will act on eigenstates of the lattice Hamiltonian. We first note that periodic RSOS model has the following decomposition (up to an isomorphism)\cite{Belletete:2017gwt}
\begin{equation}\label{decompH}
    \rho_{\rm per} \simeq \bigoplus_{s = 1}^{p}  \, \bchi_{0, q^{2p}} \, ,
\end{equation}
where $\rho_{\rm per}$ is the Hilbert space of the periodic model of length $2\R$ and $\bchi_{0,q^{2n}}$ are irreducible modules of aTL$_{2L}(q)$. In the thermodynamic limit, it is known that \cite{Pasquier:1989kd}
\begin{equation}\label{eq:TL-CFT-mod}
    {\rm Tr}_{\bchi_{0, q^{2s}}} q^{L_0 - c/24} \bar{q}^{\bar{L}_0 - c/24} = \sum_{r = 1}^{p-1} \chi_{r,s} \cdot \bar{\chi}_{r,s} \, , 
\end{equation}
where $\chi_{r,s}  \equiv \chi_{r,s}(q) $, is the character of the irreducible Virasoro representation corresponding to the primary field with conformal dimension $h_{r,s}$, and $\bar{\chi}_{r,s}$ is the anti-chiral part. $\bchi_{0, q^{2n}}$ is the unique irreducible quotient of the standard module $\mathcal{W}_{0,q^{2n}}$ - where  the standard modules of aTL$_{2\R}(q)$ are denoted $\mathcal{W}_{k, z^2}$,  $k$ being  the number of through lines ($2k$) and $z$ a phase see \cite{Belletete:2017gwt} for more details. From Eq. \eqref{eq:TL-CFT-mod}, one can postulate that 
\begin{equation}\label{eq:cont-lim-chi-Hilb}
    \bchi_{0, q^{2s}} \mapsto  \bigoplus_{r = 1}^{p-1} V_{(r,s)} \otimes \overline{V}_{\overline{(r,s)}} \, , 
\end{equation}
where we indicate taking the thermodynamic limit by $\mapsto$, see \cite{Gainutdinov:2012nq} for a more rigorous discussion.

In \cite{Graham2003}, it was shown that the $Y$ operator acts as a multiple of identity on the standard module - $\mathcal{W}_{k,z^{2}}$, with eigenvalue \cite{Belletete:2018eua}  $z(-q)^{k} + z^{-1}(-q)^{-k}$. In particular, we have that 
\begin{equation}\label{eq:Y-chi-action}
    Y\lvert_{\bchi_{0, q^{2s}}} = z + z^{-1} = q^{s} + q^{-s}. 
\end{equation}

Let us consider the A$_p$ RSOS model, where recall $q = {\rm e}^{{\rm i} \frac{\pi}{p+1}}$. In notations of the main text, the sectors in the $aTL$ algebra correspond to $\eta={s\pi\over p+1}$. The decomposition (\ref{decompH}) gives rise to the full partition function (\ref{lattZ}) with every primary occurring twice due to the symmetry $h_{(rs)}=h_{(p-r,p+1-s)}$. When $p$ is odd, no conformal weight appears twice in a given $\bchi_{0,q^{2s}}$. The signs of the eigenvalues (and correspondingly, the finite parts of the lattice momentum) alternate in the module: we have 
\begin{equation}
\tau=(-1)^{r+s}  \, , 
\end{equation}
for the lattice state corresponding to  highest-weight state of $V_{(r,s)}\otimes \overline{V}_{(r,s)}$, and the same sign for all the descendents (although with non-vanishing spin $h-\bar{h}$, the value of $\tau$ on these descendents is affected by conformal corrections which vanish in the  limit $\R\to\infty$). Note now that 
\begin{equation}
(-1)^{p-r+p+1-s}=-(-1)^{r+s} \, ,
\end{equation}
so each conformal weight appears once with a lattice state whose momentum has  finite part equal to $\pi$ and once with a lattice state whose momentum has no finite part. When $p$ is even, this is still true, although now the conformal weights for $s={p\over 2}$ appear twice in $\bchi_{0,q^{2s}}$, once with each possible finite value of the lattice momentum.

Note now that we have in general
\begin{equation}
{\cal D}_{(12)}|\phi_{(r,s)}\rangle=(-1)^{r+s} 2\cos{\pi s\over p+1}
\end{equation}
while it is known \cite{Gainutdinov:2012nq} that $Y=2\cos{\pi s\over p+1}$. We now check that 
\begin{equation}
\uR^{-1} Y |\phi_{(r,s)}\rangle= (-1)^{r+s}Y |\phi_{(r,s)}\rangle={\cal D}_{12}|\phi_{(r,s)}\rangle
\end{equation}
a result that extends, in the limit $\R\to\infty$, to all the Virasoro descendants.

\newpage
\section{Anyonic Chains and \texorpdfstring{$F$}{Lg}-symbols}\label{sec:anyon-chain-RSOS}
 Anyonic chains are 1 + 1 - dimensional quantum models defined using fusion categories \cite{Feiguin:2006ydp, Gils_2009, Gils_2013, Buican:2017rxc,Zini:2017bzi, Aasen:2020jwb, Lootens:2021tet, Inamura:2023qzl, Bhardwaj:2024kvy}. In this appendix, we will show the equivalence between anyonic chain constructed using two categories, $\mathcal{A}_{p}$ and ${\rm su}(2)_{p+1}$, and the A$_p$ RSOS models. Further, using $F-$symbols of the input category one can define symmetry operators of the anyonic chain. These have the same fusion algebra as the input category. In what follows,  we will also discuss  the relationship between these symmetry operators and  (fused) transfer matrices of the A$_p$ RSOS models.
\subsection{ $\mathcal{A}_{p}$ and ${\rm su}(2)_{p+1}$ categories} 
Let us first study the two closely related fusion categories, $\mathcal{A}_{p}$ and ${\rm su}(2)_{p+1}$. In both, the simple objects are given by $1, 2, \ldots, p-1, p$ and fusion rules are

\begin{equation}\label{eq:fus-rul-su-A}
    N^{j_1}_{j_2 j_3} = 
    \begin{cases} 
        1 \quad 
        \begin{aligned}
         & \text{if }    j_1 + j_2 \geq j_3 + 1 , \quad j_2 + j_3 \geq j_1 + 1 , \quad j_3 + j_1 \geq j_2 + 1 \, ,  \\ 
            & j_1 + j_2 + j_3 \in 2 \mathbb{Z} + 1  , \quad j_1 + j_2 + j_3 \leq 2p+1 \, , 
        \end{aligned}  \\ 
        0 \quad \text{otherwise.}
    \end{cases}
\end{equation}
These can also be written as 
\begin{equation}
    j_2 \otimes j_3 = \sum_{ \substack{
    \lvert j_2 - j_3  \rvert + 1 
    \\ \text{step} = 2 } }^{\substack{ {\rm min} \left(j_2 + j_3 - 1, \right. \\ \left.  2p + 1 - j_2 - j_3\right) }} j_1 \, .
\end{equation}
Note, in many references \cite{Gils_2009,Gils_2013, Aasen:2020jwb}, the simple objects for these categories are half integers and start from 0. Here  we follow a slightly different  convention where the simple objects are integers and start from 1, since  we want to connect the Anyonic chain with A-type RSOS model. A simple object $x$ in the conventions of \cite{Gils_2009,Gils_2013, Aasen:2020jwb} is $2 x + 1$ in ours.
For the two categories $\mathcal{A}_{p}$ and ${\rm su}(2)_{p+1}$, the quantum dimensions are the same too 
\begin{equation}
    d_h = \frac{  \sin\frac{ \pi h }{p + 1}}  { \sin  \frac{\pi}  {p + 1} }  = [h]\, , 
\end{equation}
where $  
[n]= \frac{q^{n} - q^{-n}}{q - q^{-1}}$ and $q = \text{e}^{\frac{ \text{i}  \pi }{p + 1}}$.
However, the $F$-symbol for the two categories are different. It follows that the Frobenius-Schur indicator for a simple object $a$ in both of these categories, which is given by 
\begin{equation}
    \chi_a = d_a \,  \left(F^{aaa}_{a}\right)^{*}_{0,0} \, , 
\end{equation}
 is always 1 in $\mathcal{A}_{p}$, but $(-1)^{a + 1}$ in ${\rm su}(2)_{p+1}$.

Let us first describe the $F$-symbol for the ${\rm su}(2)_{p+1}$ category, which are defined using the Racah-Wigner 6j symbol for the quantum group $ U_q (\mathfrak{sl}_2)$  \cite{Kirillov:1991ec}
\begin{equation}
\begin{split}
& \left\{\begin{array}{lll}
j_1 & j_2 & j_3 \\
j_4 & j_5 & j_6
\end{array}\right\}_q=\Delta\left(j_1, j_2, j_3 \right) \Delta\left(j_1, j_5, j_6\right) \Delta\left(j_5, j_3, j_4\right) \Delta\left(j_4, j_2, j_6\right) \\
& \times \sum_z  \frac{(-1)^z[z-1]!}{\left[z - \frac{j_1+j_2+j_3 + 1}{2}\right]!\left[z- \frac{j_1 + j_5 + j_6 + 1}{2} \right]!\left[z- \frac{j_2 + j_4 + j_6 + 1}{2}  \right]!\left[z- \frac{j_4+j_5+j_3 + 1}{2}    \right]!}
\\
& \times \quad \frac{1}{\left[ \frac{j_1 + j_2 + j_4 + j_5}{2} - z \right]!\left[ \frac{j_1+j_4+j_3+j_6}{2}-z\right]!\left[\frac{j_2+j_5+j_3+j_6}{2}-z\right]!}\, .
\end{split}
\end{equation} 
     Here, $z$ is an integer with the constraint that the terms in summation are finite. Further  
\begin{equation}
    \begin{split}
         [n]! & = \left\{\begin{array}{cl}\prod_{m=1}^n[m] & n>0, \\ 1 & n=0, \\ \infty & n<0,\end{array}\right. \\ \Delta\left(j_1, j_2, j_3\right) & = \left\{\begin{array}{cl}
    \sqrt{   \frac{\left[\frac{j_1+j_2-j_3 - 1}{2} \right]!\left[\frac{j_3+j_1-j_2 - 1}{2}\right]!\left[\frac{j_2+j_3-j_1-1}{2}\right]!}{\left[ \frac{j_1+j_2+j_3-1}{2}\right]!}} & N_{j_2 j_3}^{j_1}=1, \\ 0 & \text { otherwise },\end{array} \right.
    \end{split}
\end{equation}
The $F$-symbol in \cite{Kirillov:1991ec} can be written as  
\begin{equation}
\left(\wt{F}^{j_1 j_2 j_4}_{j_5}\right)_{j_3 ,  j_6} = 
        \left\{\begin{array}{ccc}j_1 & j_2 & j_3 \\ j_4 & j_5 & j_{6}\end{array}\right\}  = (d_{j_3}d_{j_6})^{\frac{1}{2}} (-1)^{ \frac{j_4 + j_5 + 2 j_3 - (j_1 + j_2 + 2) }{2}} \left\{\begin{array}{ccc}j_1 & j_2 & j_3 \\ j_4 & j_5 & j_{6}\end{array}\right\}_q \, , 
\end{equation}
where the symbols in curly brackets are 6j-symbols - and  come up when one studies the different bases in which a tensor product of three irreducible representations of $ U_q (\mathfrak{sl}_2)$ can be decomposed. \footnote{We again note that in \cite{Kirillov:1991ec}, the representations are labeled by half integers, and we alter the formulae above to match our conventions.} Now, it can also  be shown that 
\begin{equation}
   (-1)^{ \frac{j_4 + j_5 + 2 j_3 - (j_1 + j_2 + 2 )}{2} }  =   (-1)^{ \frac{j_1 + j_2 + j_4 + j_5}{2}} \, ,
\end{equation}
as the 6j symbols are non-zero only if $j_1 + j_2 - j_3 \in 2\mathbb{Z} + 1$. 
Hence, we have 
    \begin{equation}\label{eq:F_symb_su2}
\left(\wt{F}^{j_1 j_2 j_4}_{j_5}\right)_{j_3, \,  j_{6}} = 
        \left\{\begin{array}{ccc}j_1 & j_2 & j_3 \\ j_4 & j_5 & j_{6}\end{array}\right\}  = (d_{j_3}d_{j_6})^{\frac{1}{2}} (-1)^{\frac{j_1 + j_2 +j_4 + j_5}{2} } \left\{\begin{array}{ccc}j_1 & j_2 & j_3 \\ j_4 & j_5 & j_6\end{array}\right\}_q \, ,
\end{equation}

On the other hand, the $F$-symbol for $\mathcal{A}_{p}$, which we denote here by $\hat{F}$, is \cite{Aasen:2020jwb}
\begin{equation}\label{eq:F_symb_A}
\begin{split}
   & \quad \quad \quad \quad \left(\hat{F}^{a b d}_{e}\right)_{c,\,  f}   = (-1)^{s} \sqrt{d_c d_f} \left\{\begin{array}{ccc}a & b & c \\ d & e & f\end{array}\right\}_q \, ,  \\
    s   &= \frac{3\left(\frac{a+b+c+d+e+f}{2} - 3 \right)^2-\left(\frac{a+d}{2} -1\right)^2-\left(\frac{b+e}{2} -1\right)^2-\left( \frac{c+f}{2} - 1\right)^2}{2} \, .
    \end{split}
\end{equation}

The Hilbert space for the anyonic chain with $L$ sites is defined using the fusion tree.
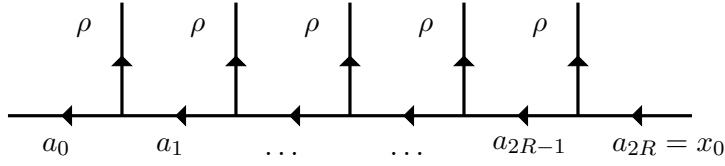
\begin{figure}[h]
\centering
\begin{tikzpicture}[scale = 1.5]
\begin{scope}[very thick, every node/.style={sloped,allow upside down}]
\draw (1,0)-- node {\midarrow} (0,0);
\draw (2,0)-- node {\midarrow} (1,0);
\draw (3,0)-- node {\midarrow} (2,0);
\draw (4,0)-- node {\midarrow} (3,0);
\draw (5,0)-- node {\midarrow} (4,0);
\draw (6,0)-- node {\midarrow} (5,0);
\draw (1,0)-- node {\midarrow} (1,1);
\draw (2,0)-- node {\midarrow} (2,1);
\draw (3,0)-- node {\midarrow} (3,1);
\draw (4,0)-- node {\midarrow} (4,1);
\draw (5,0)-- node {\midarrow} (5,1);
\node[above right] at (.5,.6) {$\rho$};
\node[above right] at (1.5,0.6) {$\rho$};
\node[above right] at (2.5,0.6) {$\rho$};
\node[above right] at (3.5,0.6) {$\rho$};
\node[above right] at (4.5,0.6) {$\rho$};
\node[above right] at (0.2,-0.45) {$a_0$};
\node[above right] at (1.2,-0.45) {$a_1$};
\node[above right] at (2.15,-0.45) {$\ldots$};
\node[above right] at (3.25,-0.45) {$\ldots$};
\node[above right] at (4.15,-0.45) {$a_{2\Ll -1}$};
\node[above right] at (5.2,-0.45) {$a_{2\Ll} = x_0 $};
\end{scope}
\end{tikzpicture}
\caption{The fusion tree that defines the anyonic chain. Here we set $\rho = 2$.}
\label{fig:fusiontree}
\end{figure}

As seen from above, the Hilbert space in these anyonic theories are dictated by the fusion rules, hence the Hilbert space for ${\rm su}(2)_{p+1}$ and $\mathcal{A}_{p }$ are the same. 
Further, it is easy to see that the Hilbert space of A$_{p}$ RSOS model is the same as that of  ${\rm su}(2)_{p+1}$ and $\mathcal{A}_{p}$, when we set $\rho = 2$.

Now, we will describe the Hamiltonians of the anyonic model, and relate them with those  of  the A$_p$ RSOS model. The Hamiltonian for  anyonic chains are written in terms of local projectors. Here, we discuss the case when we project into the trivial anyon, which we have labeled by 0. By doing $F-$ transformations, it can be checked that this local operator is \cite{Buican:2017rxc}
\begin{equation}\label{eq:loc_op_Fs}
\begin{split}
   & h_{j} \ket{\ldots a_{j-1}, a_j, a_{j + 1} \ldots }  = \\ & \delta_{a_{j-1}, a_{j+1}}  \sum_{ \tilde{a}_j \in \{ a_{j-1} - 1,a_{j-1} +  1\}  } \,    \left(F^{a_{j-1} \,  2 \,  2}_{a_{j-1}}\right)_{a_j, \,  1} \, \,  \left(F^{a_{j-1} \, 2 \, 2}_{a_{j-1}} \right)_{\tilde{a}_j, \, 1 }   \ket{\ldots a_{j-1}, \tilde{a}_j, a_{j+1} \ldots } \, , 
   \end{split}
\end{equation}
where we have used that the inverse of $F$-matrix is its transpose, which is true for both ${\rm su }(2)_{p}$ and $\mathcal{A}_{p}$ categories (this is true as both the $F$ matrices are unitary and their elements are real). The fusion rules dictates that both $a_j$ and $\tilde{a}_j$ be $a_{j-1} \pm 1$. The Hamiltonian is then written as
\begin{equation}
    H = - \sum_{j} h_j \, .
\end{equation}
Now, using the form of $F$ - symbols in Eq. \eqref{eq:F_symb_su2}, the following can be checked 
\begin{equation}\label{eq:id_Fs_su2}
    \left(\wt{F}^{a_{j-1} \, 2 \, 2}_{a_{j-1}} \right)_{a_{j-1} -1 \, , \, 1} = -\sqrt{ \frac{d_{a_{j-1} - 1} } {d_{2} d_{a_{j-1}}}  } \, , \quad  \left(\wt{F}^{a_{j-1} \, 2  \, 2}_{a_{j-1}} \right)_{a_{j-1} + 1 \, , \, 1} = \sqrt{ \frac{d_{a_{j-1} + 1 }  }{d_{2} d_{a_{j-1}}}} \, .  
\end{equation}
Using the form of $F$ - symbols for $\mathcal{A}_{p}$ category in Eq. \eqref{eq:F_symb_A}, the following was also seen to be true for all values of $p$ that we checked
\begin{equation}\label{eq:id_Fs_A}
    \left(\hat{F}^{a_{j-1} \, 2 \, 2}_{a_{j-1}} \right)_{a_{j-1} -1 \, , \, 1} = \sqrt{ \frac{d_{a_{j-1} - 1}}{d_{2} d_{a_{j-1}}}  }   \, , \quad   \left(\hat{F}^{a_{j-1} \, 2 \, 2}_{a_{j-1}} \right)_{a_{j-1} +1 \, , \, 0} = \sqrt{ \frac{d_{a_{j-1} + 1}}{d_{2} d_{a_{j-1}}}} \, . 
\end{equation}
Let $\wt{h}_i$ and $\hat{h}_i$ be the operators defined in Eq. \eqref{eq:loc_op_Fs} using the $F$-symbols for $\text{su}(2)_{p+1}$ and $\mathcal{A}_{p}$. As there is a minus sign in Eq. \eqref{eq:id_Fs_su2}, we can check that 
\begin{equation}\label{eq:T_h_op_su2k}
\begin{split}    
    \wt{h}_i & = T_h^{\dagger} \, \hat{h}_i   \, T_h \, , \\
    \text{where } \, \, T_h \ket{a_0, a_1, \ldots, a_{2\Ll - 1}}  & = (-1)^{\sum \frac{a_i}{2}}\ket{a_0, a_1, \ldots, a_{2\Ll - 1}} \, .
\end{split}
\end{equation}
As $T_h$ is a diagonal, unitary matrix, $\wt{h}_i$ and $\hat{h}_i$ are unitarily equivalent, and so are the Hamiltonians defined using them. Now, it can be checked that
\begin{equation}
e_i =  d_{2} \, \, \hat{h}_i \, , 
\end{equation}
where $e_i$ is the TL generators defined in Eq. \eqref{eq:ei-operator}. Hence, up to a scale, the A$_{p}$ RSOS Hamiltonian is equal (or unitarily equivalent) to the anyonic chain Hamiltonian corresponding to $\mathcal{A}_{p}$ 
(or ${\rm su}(2)_{p+1}$) category.

\subsection{\texorpdfstring{$F$}{Lg}-symbol for \texorpdfstring{$Y$}{Lg} operator}\label{sec:Yop-F-symb}
We now discuss a reformulation of the $Y$ operator for the A$_{p}$ RSOS model in terms of $F$ symbols. We start with the  Boltzmann weights for  faces 
\begin{equation}\label{eq:gauged_face_weight}
\begin{split}
 & \widetilde{W}_1\left(\begin{array}{ll}
d & c \\
a & b
\end{array} \Bigg| \ u\right)=  q^{\frac{1}{2}}\left(\delta_{b , d} (-1)^{\frac{a-c}{2}} \sqrt{\frac{\theta_{a} \theta_{c}}{\theta_{b} \theta_{d}}}   \frac{\sin u}{\sin (\gamma - u)} + \delta_{a , c} \right)\, ,  \\
    & \widetilde{W}_2\left(\begin{array}{ll}
d & c \\
a & b
\end{array} \Bigg| \ u\right)=  q^{-\frac{1}{2}}\left(\delta_{b , d} (-1)^{\frac{a-c}{2}} \sqrt{\frac{\theta_{a} \theta_{c}}{\theta_{b} \theta_{d}}}\frac{\sin u}{\sin (\gamma - u)} + \delta_{a , c}  \right)\, , 
\end{split}
\end{equation}
and
\begin{equation}\label{eq:bolz_wt_form}
 \theta_t=\frac{\sin (\gamma t)}{\sin \gamma}, \quad \gamma = \frac{\pi}{p+1} \, . 
\end{equation}
Note, if we set $S_a = (-1)^{\frac{a}{2}}$ in Eq. \ref{eq:standard_face_weight} and scale by $q^{\frac{1}{2}} \frac{\sin \gamma}{\sin(\gamma - u)}$ $\left(q^{-\frac{1}{2}} \frac{\sin \gamma}{\sin(\gamma - u)}\right)$  we get $\wt{W}_1$ ($\wt{W}_2$). For the case when spectral parameter is $ \rm{i} \infty$ or $- \rm{i} \infty$, the transfer matrix is 
\begin{equation}\label{eq:Y}
 \bra{b_0,  \ldots b_{2\Ll - 1}} \wt{T}_1(\text{i} \infty)\ket{a_0,  \ldots, a_{2\Ll-1}} = \prod_{i = 0}^{2\Ll-1} q^{-\frac{1}{2}} \left( (-1)^{\frac{a_i - b_{i + 1}}{2}}\delta_{a_{i+1}, b_{i }} \sqrt{\frac{[a_i ] [ b_{i+1} ]}{[a_{i+1} ][b_i ]}}  - q \delta_{a_i, b_{i + 1}} \ \right) \, , 
\end{equation}
\begin{equation}\label{eq:Ybar}
 \bra{b_0,  \ldots b_{2\Ll - 1}} \wt{T}_2( - \text{i} \infty) \ket{a_0,  \ldots, a_{2\Ll-1}} = \prod_{i = 0}^{2\Ll-1} q^{\frac{1}{2}} \left( (-1)^{\frac{a_i - b_{i + 1}}{2}} \delta_{a_{i+1}, b_{i }} \sqrt{\frac{[a_i] [ b_{i+1} ]}{[ a_{i+1}][ b_i]}} - q^{-1} \delta_{a_i, b_{i + 1}}\ \right) \, .  
\end{equation}
From the form of the 6j symbol, it can be checked that \cite{BERNARD1990721}
\begin{equation}\label{eq:BL_}
\begin{split}    
& \left\{\begin{array}{ccc} 2 & j+1 & j+2 \\ 2 & j+1 & j\end{array}\right\}=\left\{\begin{array}{ccc}2 & j+1 & j \\ 2 & j+1 & j+2\end{array}\right\}=\left(\frac{[ j] \, [ j+2]}{[ j+1]^2}\right)^{\frac{1}{2}}, \\ 
& \left\{\begin{array}{ccc}2 & j+1 & j \\ 2 & j+1 & j\end{array}\right\}=-\left\{\begin{array}{ccc}2 & j+1 & j+2 \\ 2 & j+1 & j+2\end{array}\right\}=-\frac{1}{[j+1]}, \\
& \left\{\begin{array}{ccc}2 & j & j+1 \\ 2 & j+2 & j+1\end{array}\right\}=\left\{\begin{array}{ccc}2 & j & j-1 \\ 2 & j-2 & j-1\end{array}\right\}=1 .
\end{split}
\end{equation}
The above identities are very useful in proving the identities in Eq. \eqref{eq:Y-op-6j} and \eqref{eq:Ybar-op-6j}.
By considering all possible cases of allowed height configurations and using identities in Appendix A of \cite{Bernard1991}, it can be shown that 
\renewcommand\arraystretch{1.2}

\begin{equation}\label{eq:Y-op-6j}
\begin{split}
        & q^{-\frac{1}{2}} \left( (-1)^{\frac{a_i - b_{i+1}}{2} }  \delta_{a_{i+1}, b_{i }} \sqrt{\frac{[a_i]  [b_{i+1}]}{[a_{i+1}][ b_i ]}} - q \delta_{a_i, b_{i + 1}} \ \right)  \\
      &  =  (-1)^{ \frac{(a_i - b_i )}{2} - \frac{(a_{i + 1}  - b_{i + 1} )}{2} + 1 }q^{ ( C_{a_{i + 1}} -  C_{b_{i + 1}}) -  (C_{a_{i }} - C_{b_{i }} )  } 
      \Biggl\{
\begin{matrix}
     & 2 & b_{i}    &  b_{i + 1}\\
     & 2 & a_{i + 1} & a_i
\end{matrix}
    \  \Biggr\} \, ,      
\end{split}
\end{equation}

\begin{equation}\label{eq:Ybar-op-6j}
    \begin{split}
        & q^{\frac{1}{2}} \left(  (-1)^{\frac{a_i - b_{i+1}}{2} } \delta_{a_{i+1}, b_{i }} \sqrt{\frac{[a_i ] [b_{i+1} ]}{[ a_{i+1} ][ b_i]}} - q^{-1} \delta_{a_i, b_{i + 1}} \ \right)  \\
      &  =(-1)^{ \frac{(a_i - b_i )}{2} - \frac{(a_{i + 1}  - b_{i + 1} )}{2} + 1 }q^{ (C_{a_{i }} - C_{b_{i }} ) - ( C_{a_{i + 1}} -  C_{b_{i + 1}}) } 
      \Biggl\{
\begin{matrix}
     & 2 & b_{i}    &  b_{i + 1}\\
     & 2 & a_{i + 1} & a_i
\end{matrix}
    \  \Biggr\}   \, ,   
    \end{split}
\end{equation}
where $C_a = (a-1)(a + 1)/4$ and 
\begin{equation}   \left\{\begin{array}{ccc}j_1 & j_2 & j_{3} \\ j_4 & j_5 & j_{6}\end{array}\right\}  = (-1)^{\frac{j_1 + j_2 + j_4 + j_5}{2}} \sqrt{d_{j_3} d_{j_6}} \left\{\begin{array}{ccc}j_1 & j_2 & j_{3} \\ j_4 & j_5 & j_{6}\end{array}\right\}_q
\equiv \,  \left(\wt{F}^{j_1 j_2 j_4}_{j_5}\right)_{j_{3} ,  j_{6}} \, , 
\end{equation}
where the 6j-symbol in the RHS is the Racah-Wigner 6j symbol in Equation A.7 of \cite{Aasen:2020jwb}. 
 If we substitute the above equations into Equations \eqref{eq:Y} and \eqref{eq:Ybar}, it is easy to see that the terms $(-1)^{\frac{(a_i - b_i )}{2} - \frac{(a_{i + 1}  - b_{i + 1}  )}{2} }$ and $q^{ ( C_{a_{i + 1}} -  C_{b_{i + 1}}) -  (C_{a_{i }} - C_{b_{i }} )  } $ cancel out because of PBC.  Substituting Eq. \eqref{eq:Y-op-6j}  and \eqref{eq:Ybar-op-6j} into Eq. \eqref{eq:Y} and \eqref{eq:Ybar}, we get 
\begin{equation}
 \bra{b_0,  \ldots b_{2\Ll - 1}} \wt{T}_1(\text{i} \infty)\ket{a_0,  \ldots, a_{2\Ll-1}} = \prod_{i = 0}^{2\Ll-1}  \,       \Biggl\{
\begin{matrix}
     & 2 & b_{i}    &  b_{i + 1}\\
     & 2 & a_{i + 1} & a_i
\end{matrix}
    \  \Biggr\}     = \prod_{i = 0}^{2\Ll-1} \left( \wt{F}^{2 \, b_i \, 2  }_{a_{i + 1}} \right)_{b_{i + 1}, a_i} \, , 
\end{equation}

\begin{equation}
\begin{split}
 \bra{b_0,  \ldots b_{2\Ll - 1}} \wt{T}_2(-\text{i} \infty)\ket{a_0,  \ldots, a_{2\Ll-1}} &= \prod_{i = 0}^{2\Ll-1}  \,       \Biggl\{
\begin{matrix}
     & 2 & b_{i}    &  b_{i + 1}\\
     & 2 & a_{i + 1} & a_i
\end{matrix}
    \  \Biggr\}  = \prod_{i = 0}^{2\Ll-1} \left( \wt{F}^{2 \, b_i \, 2 }_{a_{i + 1}} \right)_{b_{i + 1} , a_i} \,  .
\end{split}
\end{equation}
Hence, we obtain
\begin{equation}\label{eq:def-tild-Y}
 \wt{T}_1({\rm i } \infty) = \wt{T}_2( -{\rm i } \infty) \equiv \wt{Y}_{\frac{1}{2}}   \, ,
\end{equation}
i.e. $\wt{Y}$ is transfer matrix at spectral parameter i$\infty$, when the weights used are the ones in Eq. \eqref{eq:gauged_face_weight}. Recall, in \autoref{sec:fusion-Tmatrices-(1s)}  we studied transfer matrices at spectral parameter $\pm {\rm i} \infty$ 
\begin{equation}\label{eq:Y-Ybar-rel-tmat}
    \begin{split}
        Y_{\frac{k}{2}} &=  \lim_{u \to {\rm i} \infty} \left( (-q)^{\frac{k}{2}} \frac{\sin\gamma }{\sin \left( \frac{k+1}{2}\gamma - u \right)}\right)^{2\Ll}  T^{(k)}_{[0]} \, ,\\
        \overline{Y}_{\frac{k}{2}} &=  \lim_{u \to - {\rm i} \infty} \left( (-q)^{-\frac{k}{2}} \frac{\sin\gamma }{\sin \left(\frac{k+1}{2} \gamma - u\right)}\right)^{2\Ll}  T^{(k)}_{[0]} \, .
    \end{split}
\end{equation}
As we discussed before, $\wt{W}_1 \,  (\wt{W}_2)$ can be related with the weights in Eq. \eqref{eq:standard_face_weight}, with gauge factor $S_a  = 1$, by the following factor
\begin{equation}
\begin{split}
     \widetilde{W}_1\left(\begin{array}{ll}
d & c \\
a & b
\end{array} \Bigg| \ u \right) & =   q^{\frac{1}{2}} \, (-1)^{\frac{a-c}{2}} \, \frac{\sin \gamma}{\sin(\gamma - u)} \ W\left(\begin{array}{ll}
d & c \\
a & b
\end{array} \Bigg| \ u \right) \, , 
 \\
    \widetilde{W}_2\left(\begin{array}{ll}
d & c \\
a & b
\end{array} \Bigg| \ u \right) & = q^{-\frac{1}{2}} \,  (-1)^{\frac{a-c}{2}} \,  \frac{\sin \gamma}{\sin(\gamma - u)}W \left(\begin{array}{ll}
d & c \\
a & b
\end{array} \Bigg| \ u \right) \, . 
\end{split}
\end{equation}   
 For the  transfer matrices we get
\begin{equation}\label{eq:tmatrix-rel-two}
\begin{split}
  & \bra{b_0, \ldots, b_{2\Ll-1}} \wt{T}_1(u_1) \ket{a_0, \ldots, a_{2\Ll-1}}  \\
 &  = \left( q^{\frac{1}{2}} \,  \frac{\sin \gamma}{\sin(\gamma - u_1)} \right)^{2\Ll}
  \,  (-1)^{\sum \frac{a_i - b_i}{2}} \bra{b_0, \ldots, b_{2\Ll-1}} T^{(1)}_{[0]} \ket{a_0, \ldots, a_{2\Ll-1}} \, ,  \\
  & \bra{b_0, \ldots, b_{2\Ll-1}} \wt{T}_2(u_2) \ket{a_0, \ldots, a_{2\Ll-1}}   = \\
&  \left( q^{-\frac{1}{2}} \,  \frac{\sin \gamma}{\sin(\gamma - u_2)} \right)^{2\Ll}\,  (-1)^{\sum \frac{a_i -b_i}{2} } \bra{b_0, \ldots, b_{2\Ll-1}} T^{(1)}_{[0]} \ket{a_0, \ldots, a_{2\Ll-1}} \, .
\end{split}
\end{equation}
 If we set $u_1 = {\rm i} \infty$ and $u_2 = - {\rm i} \infty$ and use Eq. \eqref{eq:def-tild-Y} and \eqref{eq:Y-Ybar-rel-tmat}  with $k = 1$, we obtain 
\begin{equation}\label{eq:transfer-mat-iinfty}
\begin{split}
    \bra{b_0, \ldots, b_{2\Ll-1}}  \wt{Y}_{\frac{1}{2}} \ket{a_0, \ldots, a_{2\Ll-1}} & = (-1)^{\Ll}  (-1)^{\sum \frac{a_i - b_i}{2}}  \bra{b_0, \ldots, b_{2\Ll-1}}  Y_{\frac{1}{2}} \ket{a_0, \ldots, a_{2L-1}} 
   \\ &= (-1)^{\Ll}  (-1)^{\sum \frac{a_i - b_i}{2}}  \bra{b_0, \ldots, b_{2\Ll-1}}  \overline{Y}_{\frac{1}{2}} \ket{a_0, \ldots, a_{2\Ll-1}}  \\  \implies Y_{\frac{1}{2}} & = \overline{Y}_{\frac{1}{2}}
\end{split}
\end{equation}
(The above derivation only holds for A-type RSOS model. As the three state Potts model is of D-type, $Y$ and $\overline{Y}$ are different \cite{Sinha:2023hum}.)  Substituting  the  operator $T_h$ from Eq. \eqref{eq:T_h_op_su2k} in \eqref{eq:transfer-mat-iinfty}, it is not hard to see that
\begin{equation}\label{eq:tilde-Y-D1/2}
  \wt{Y}_{\frac{1}{2}} = \wt{T}_1({\rm i} \infty)  = (-1)^{\Ll} T_h^{\dagger} Y_{\frac{1}{2}}  T_h  =   T_h Y_{\frac{1}{2}} T_h \, ,  
\end{equation}
using that $(-1)^{\Ll} T_h = T_h^{-1} = T_h^{\dagger}$. Hence, $\wt{Y}_{\frac{1}{2}}$ is unitarily equivalent, up to a sign, to the $Y$ operator.

Note, that $Y_{\frac{1}{2}}$ is symmetric in the standard basis, i.e. $Y_{\frac{1}{2}} = Y_{\frac{1}{2}}^{T}$. Hence $\wt{Y}_{\frac{1}{2}}$ is also symmetric, as $T_h$ is also symmetric in this basis, since it is a diagonal operator.  Therefore 
\begin{equation}
     \bra{b_0,  \ldots b_{2\Ll - 1}} \wt{Y}_{\frac{1}{2}}\ket{a_0,  \ldots, a_{2\Ll-1}} =  \bra{a_0,  \ldots, a_{2\Ll-1}}  \wt{Y}_{\frac{1}{2}} \ket{b_0,  \ldots b_{2\Ll - 1}}    = \prod_{i = 0}^{2\Ll-1} \left( \wt{F}^{2 \, a_i \, 2  }_{b_{i + 1}} \right)_{a_{i + 1},\,  b_i} \, . 
\end{equation}
This $\wt{Y}_{\frac{1}{2}}$ operator has also been studied extensively  as a ``topological symmetry operator'' in \cite{Feiguin:2006ydp, Gils_2009, Gils_2013, Buican:2017rxc} for example. Following \cite{Buican:2017rxc}, we could define more generally  the $k$-dependent operators
\begin{equation}\label{eq:top-sym-op-su2}
\bra{b_0,  \ldots b_{2\Ll - 1}} \wt{Y}_{k} \ket{a_0,  \ldots, a_{2\Ll-1}} := \prod_{i = 0}^{2\Ll - 1} \left( \wt{F}^{2  \,  a_i \, 2k+1 }_{b_{i + 1}} \right)_{a_{i + 1} , \, b_i} \, , 
\end{equation}
which are shown in \cite{Buican:2017rxc} to have the same fusion rules as the input category for the anyonic chain, i.e.
\begin{equation}\label{eq:fusion-tild-Y-cat}
    \wt{Y}_{k_1} \circ \wt{Y}_{k_2} = \sum_{k_3} N^{k_3}_{k_1 k_2} \wt{Y}_{k_3} \, , 
\end{equation}
where $N$ is given in Eq. \eqref{eq:fus-rul-su-A}. Using these fusion rules we see that 
\begin{equation}\label{eq:fus-hier-tildY}
    \begin{split}
        \wt{Y}_{1} \, = \, & \wt{Y}_{\frac{1}{2}}^2 - \mathbb{1} \\
        \wt{Y}_{\frac{3}{2}} \, = \, &   \wt{Y}_{\frac{1}{2}}^{3} - 2 \wt{Y}_{\frac{1}{2}} \\
          \wt{Y}_2  \, = \, & \wt{Y}_{\frac{1}{2}}^4 - 3 \wt{Y}_{\frac{1}{2}}^2 + \mathbb{1} \, , \\
          \ldots \, = \, & \ldots \,
    \end{split}
\end{equation}
 Recall that $Y_{\frac{1}{2}}$ has a similar fusion hierarchy \eqref{eq:fusion-Y-TL}
\begin{equation}\label{eq:fus-hier-Y}
    \begin{split}
               Y_1 & = Y_{\frac{1}{2}}^2 - \mathbb{1} \, , \\
        Y_{\frac{3}{2}} & = Y_{\frac{1}{2}}^3  - 2 Y_{\frac{1}{2}} \, , \\
        Y_2 & = Y_{\frac{1}{2}}^4 - 3 Y_{\frac{1}{2}}^2 + \mathbb{1} \, . \\ 
       \ldots  & = \ldots 
    \end{split}
\end{equation}
Now, using Eq. \eqref{eq:tilde-Y-D1/2}, \eqref{eq:fus-hier-tildY},   and \eqref{eq:fus-hier-Y}, we see that
\begin{equation}\label{eq:unitary-equivelance-Y-Ytilde}
    \begin{split}
        \wt{Y}_{\frac{1}{2}} & = (-1)^{\Ll}T_h^{\dagger} Y T_h \, , \\
        \wt{Y}_{1} & = T_h^{\dagger} Y_{1} T_h \, , \\
        \wt{Y}_{\frac{3}{2}} & = (-1)^{\Ll}T_h^{\dagger} Y_{\frac{3}{2}} T_h \, , \\
        \wt{Y}_{2} & =T_h^{\dagger} Y_{2} T_h \,. \\
        \ldots  & = \ldots 
    \end{split}
\end{equation}
Hence, $Y_k$ and $\wt{Y}_k$ are unitarily equivalent, up to a sign, which shows that $\wt{Y}_k$, which are written in terms of $F$-symbols, can be written in terms of fused transfer matrices as well.

\newpage
\section{A (slightly) new calculation of fused weights}
\label{AnotherFusion}
\subsection{General construction}\label{AnotherFusion1}

As mentioned in the bulk of  this paper, because of the underlying Yang-Baxter equation, the spectrum of the transfer matrix (and thus the partition function) in the presence of columns with modified spectral parameters does not depend on the precise position of these columns.  One can thus bring them together (this  is of course not the same as the full lattice topological invariance),   and consider the result  to be a single embedded impurity. This means that, in picture \ref{Comp}-b, the two colored columns can be considered as  a single column of new, more complicated  faces, now parametrized by two  spectral parameters. 

\begin{figure}[htb]
    \centering
\begin{tikzpicture}
    \draw (0,0) rectangle (2,4);
    \draw (1,0) -- (1,4);
    \foreach \y in {1,2,3} {
    \draw (0,\y) -- (2,\y);
}
\foreach \i in {0,...,3} {
    \node at (0.5,\i+0.5) {$u_1$};
    \node at (1.5,\i+0.5) {$u_2$};
}
\node[left] at (0,0) {$y_0$};
\node[left] at (0,1) {$y_1$};
\node[left] at (0,2) {$y_2$};
\node[left] at (0,3) {$y_3$};
\node[left] at (0,4) {$y_4$};

\node[right] at (2,0) {$z_0$};
\node[right] at (2,1) {$z_1$};
\node[right] at (2,2) {$z_2$};
\node[right] at (2,3) {$z_3$};
\node[right] at (2,4) {$z_4$};

\node[above] at (1,4.) {$\beta$};
\node[below] at (1,0) {$\alpha$};

\end{tikzpicture}
    \caption{Fusion in the direct channel. Two column-to-column transfer matrices are brought together to form a new seam.  When $u_2-u_1=\gamma$ different fusion channels emerge exactly on the lattice, and we can compute the corresponding  ``fused matrices''.
     The  {\it external} heights are the $y_i,z_i$. The heights inside are called {\it internal}. Here, we consider that propagation has taken $\alpha$ to $\beta$.}
  \label{doublecolumnnbis}
\end{figure}
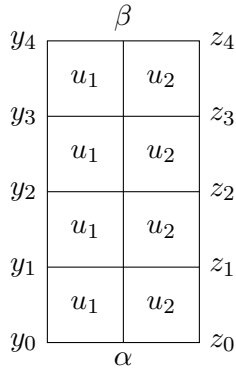

Our next  goal should be  to decompose the defect thus obtained into irreducible elements, if at all possible. 
For this, we observe that  the internal heights, i.e. those not connected to the exterior of the defect, may take any admissible value. As we build the partition function by joining row to row transfer matrices, it is convenient to think of the summation over these internal heights connecting impurity faces along the vertical seam  as a matrix product where each multiplying factor depends on the ``external heights'' - i.e. those  on the edge of the columns with modified spectral parameter. See figure \ref{doublecolumnnbis}  for an illustration.

\begin{figure}[hbt]
\begin{center}
  \includegraphics[width=0.6 \linewidth]{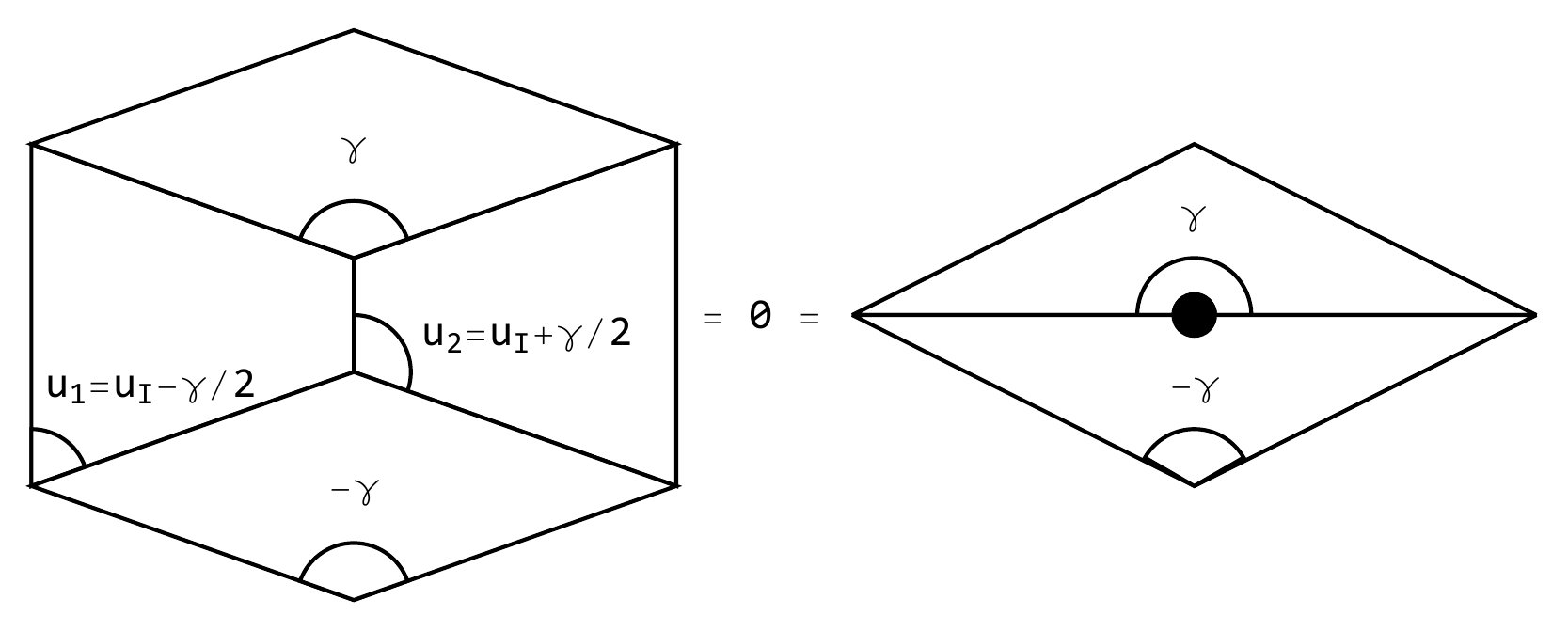}
  \\
  \caption{At the singular point $v=\gamma$ the face operator is proportional to a projector and is annihilated by the complementary operator at $v=-\gamma$.}\label{Disconnected}
  \end{center}
\end{figure}

Consider the propagation of internal heights, say from bottom to top. Given any vector associated to the internal height at the bottom of the column, the general action of the associated matrix mixes orthogonal states in way that it is impossible to disentangle different fusion channels. However, something special happens  when the difference of spectral  parameters is exactly $\gamma$. In this case,  the Yang-Baxter equation can be used to show the relation in figure \ref{Disconnected},
and thus  renders the transmitted  face along the column (with spectral parameter $\gamma$) ``singular'' - the determinant of the corresponding matrix of weights vanishes:
\begin{equation}\label{proj0}
\begin{bmatrix}
W \! \! \left(\begin{smallmatrix}a\pm2 & a \pm1 \\ a\pm1  & a \end{smallmatrix} \Big| \gamma \right)
\end{bmatrix}
= 0 \, ,
\end{equation}
while 
\begin{equation}
\begin{bmatrix}
W \! \! \left(\begin{smallmatrix}a & a-1 \\ a-1  & a \end{smallmatrix}\Big| \gamma \right)&W \! \! \left(\begin{smallmatrix}a & a+1 \\ a-1  & a \end{smallmatrix}\Big| \gamma
\right)\\
W \! \! \left(\begin{smallmatrix}a & a-1 \\ a+1  & a \end{smallmatrix}\Big| \gamma \right)&W \! \! \left(\begin{smallmatrix}a & a+1 \\ a+1  & a \end{smallmatrix}\Big| \gamma \right)
\end{bmatrix}
=\theta_2
\begin{bmatrix} \sqrt{\frac{\theta_{a-1}}{\theta_2 {\theta_a}}} \\ \sqrt{\frac{\theta_{a+1}}{\theta_2{\theta_a}}} \end{bmatrix} \begin{bmatrix}\sqrt{\frac{\theta_{a-1}}{\theta_2{\theta_a}}}& \sqrt{\frac{\theta_{a+1}}{\theta_2{\theta_a}}} \end{bmatrix} \, , 
\label{proj}
\end{equation}
which obviously has determinant zero.

In fact, one can normalize the face operators at $\pm \gamma$ in order to obtain complementary projectors. It follows, as expressed in (\ref{proj}), that one can use a different local basis in terms of vectors being projected in/out. Let 
\begin{equation}
v_{a}^+ = \begin{bmatrix}\sqrt{\frac{\theta_{a-1}}{\theta_2{\theta_a}}}, \sqrt{\frac{\theta_{a+1}}{\theta_2{\theta_a}}} \end{bmatrix}, \qquad
v_{a}^- =\begin{bmatrix}\sqrt{\frac{\theta_{a+1}}{\theta_2{\theta_a}}}, -\sqrt{\frac{\theta_{a-1}}{\theta_2{\theta_a}}} \end{bmatrix}, \label{orthogonalvectors}
\end{equation}
 be orthogonal vectors made from superpositions of canonical heights $a-1,~a+1$, in this order. Whenever a symbol $v_a^{\pm}$ is encountered, one has an expansion in terms of the usual heights for the A$_p$ model (we refer to those as  canonical heights) - see figure \ref{linearcomb}. 
\begin{figure}[H]
\begin{tikzpicture}[scale=1.2, every node/.style={scale=0.8}]

\draw[thick] (0,0) rectangle (1,1);
\draw[thick] (1,0) rectangle (2,1);
\draw[thick] (1,0) -- (1,1);
\node at (0.5,0.5) {$ u_1$};
\node at (1.5,0.5) {$u_2$};
\node at (0,-0.2) {$a$};
\node at (0,1.2) {$a+1$};
\node at (1,-0.2) {$v_a^{-}$};
\node at (1,1.2) {$v_{a+1}^{+}$};
\node at (2,-0.2) {$a$};
\node at (2,1.2) {$a+1$};

\node at (2.4,0.5) {$=$};

\begin{scope}[xshift=3cm]
\draw[thick] (0,0) rectangle (1,1);
\draw[thick] (1,0) rectangle (2,1);
\node at (0.5,0.5) {$ u_1$};
\node at (1.5,0.5) {$ u_2$};
\node at (0,-0.2) {$a$};
\node at (0,1.2) {$a+1$};
\node at (1,-0.2) {$a-1$};
\node at (1,1.2) {$a$};
\node at (2,-0.2) {$a$};
\node at (2,1.2) {$a+1$};
\node at (-0.2,0.5) {$\frac{1}{\theta_2}$};
\end{scope}


\begin{scope}[xshift=6.3cm]
\draw[thick] (0,0) rectangle (1,1);
\draw[thick] (1,0) rectangle (2,1);
\node at (0.5,0.5) {$ u_1$};
\node at (1.5,0.5) {$ u_2$};
\node at (0,-0.2) {$a$};
\node at (0,1.2) {$a+1$};
\node at (1,-0.2) {$a+1$};
\node at (1,1.2) {$a$};
\node at (2,-0.2) {$a$};
\node at (2,1.2) {$a+1$};
\node at (-0.675,0.5) {$ - \frac{1}{\theta_2} \sqrt{\frac{\theta_{a-1}}{\theta_{a+1}}}$};
\end{scope}


\begin{scope}[xshift=9.8cm]
\draw[thick] (0.2,0) rectangle (1.2,1);
\draw[thick] (1.2,0) rectangle (2.2,1);
\node at (0.7,0.5) {$ u_1$};
\node at (1.7,0.5) {$ u_2$};
\node at (0.2,-0.2) {$a$};
\node at (0.2,1.2) {$a+1$};
\node at (1.2,-0.2) {$a+1$};
\node at (1.2,1.2) {$a+2$};
\node at (2.2,-0.2) {$a$};
\node at (2.2,1.2) {$a+1$};
\node at (-0.675,0.5) {$- \frac{1}{\theta_2} \sqrt{\frac{\theta_{a-1}\theta_{a+2}}{\theta_a \theta_{a+1}}}$};
\end{scope}

\end{tikzpicture}
  \caption{Expansion of double faces in terms of canonical faces. }\label{linearcomb}

\end{figure}

In terms of $v^{\pm}$, when the impurity parameters differ by $\gamma$, different groups of double faces cannot connect along the seam, as illustrated in figure \ref{Disconnected2}.  We will study $v_{\pm}$ in context of crossed channel as well, see figure \ref{fig:def:vhp-vhm} to see the definition of $v_{\pm}$, which is the same for direct channel. 
\begin{figure}
\begin{tikzpicture}[scale=1.2, every node/.style={font=\footnotesize}]
  \foreach \x in {0,2} {
    \draw (\x,0) rectangle ++(2,2);
  }

  \node at (1,1) {$u_{I} - \gamma/2$};
  \node at (3,1) {$u_{I} + \gamma/2$};

  \node at (0,2.2) {$a$};
  \node at (4,2.2) {$a$};
  \node at (2,2.2) {$v_a^{+}$};

  \node at (0,-0.2) {$a \pm 1$};
  \node at (4,-0.2) {$a \pm 1$};
  \node at (2,-0.2) {$v_{a\pm1}^{-}$};

  \node at (4.755,1) {$=$};

  \foreach \x in {5.5,7.5} {
    \draw (\x,0) rectangle ++(2,2);
  }

  \node at (6.5,1) {$u_{I} - \gamma/2$};
  \node at (8.5,1) {$u_{I} + \gamma/2$};

  \node at (5.5,2.2) {$a$};
  \node at (9.5,2.2) {$a$};
  \node at (7.5,2.2) {$v_a^{+}$};

  \node at (5.5,-0.2) {$a \mp 1$};
  \node at (9.5,-0.2) {$a \pm 1$};
  \node at (7.5,-0.2) {$a$};

  \node at (10.2,1) {$= 0$};

\end{tikzpicture}
\centering
  \caption{Double-faces of vanishing weight. The symbols $v_a^+$, $v_a^-$ indicate a linear combination with coefficients given by the respective basis vector. }\label{Disconnected2}
\end{figure}
In consequence, the total Hilbert space is broken into two parts. More precisely, consider the vector basis where we take heights to be the  canonical ones, i.e. taking values from $1$ to $p$, except at middle of the defect double face, where we use $v_a^{\pm}$ if neighboring heights are both of the canonical type $a$ or just fix it to be $(a +b)/2$ if neighboring heights are different: $b =a \pm 2$. Since time evolution over a full period will not mix $v^+$ type with either a $v^-$ type or a canonical type, there is a decoupling between the two set of modules: basis elements with a $v^+$ and basis elements without $v^+$. Consequently, the original partition function is broken into a sum of two smaller ones, each with a reduced Hilbert space\footnote{Evaluating the dimension of each reduced Hilbert space is a very simple problem, for they can be viewed as partition functions of one dimensional statistical models. The local weights are defined by the adjacency matrices and completeness of the original (non-reduced) space results from the fusion rule: $\mbox{Tr}~G^{2 L} = \mbox{Tr}~G^{2L-2} (G^{(0)}+G^{(2)})$.}.

Likewise, these special values of spectral parameter shifts  allow us to consider different families of faces which do not mix since the internal heights of type $v^+$ cannot be connected to heights of type $v^-$. Hence the original impurity transfer-matrix is found to have a triangular structure: from top to bottom, $v^+$ cannot connect with $v^-$ but $v^-$ can connect with $v^+$.
On the other hand, once a particular fusion channel has been chosen, internal heights do not correspond to  new   degrees of freedom in the A$_p$ models\footnote{This is not the case e.g. in the $D_4$ model.}, for their values  are fixed by the  adjacent external heights. We may thus simply ignore them and treat each impurity transfer matrix as acting on a properly reduced Hilbert space, which only considers the external heights. Hence, the possible fusion channels are in one to one correspondence with  similar   channels for the adjacency matrix:
\begin{equation}
{ G} \cdot { G} = { G}^{(2)}+1, 
\end{equation}
where ${ G}^{(2)}$ gives the adjacency for the weights we defined via $v^-$, while ${ G}^{(0)}=1$ gives the adjacency for those defined via $v^+$. By direct inspection one can see that the $v^+$ channel gives an identity fused face: its insertion in the partition function can be regarded as a mere reduction of the system-size from $2L$ to $2(L-1)$. In contrast, the  $v^-$ fusion channel is non-trivial. It turns out to be related to a higher irreducible defect, which we call  spin 2\footnote{It is unfortunate that the "symmetric sector" (spin 2) appears with the minus label and minus sign in the corresponding linear combination of basis vectors, but this follows from standard convention choices for the face weights.}. 

\begin{figure}[hbt]
\begin{center}
  \includegraphics[width=0.6 \linewidth]{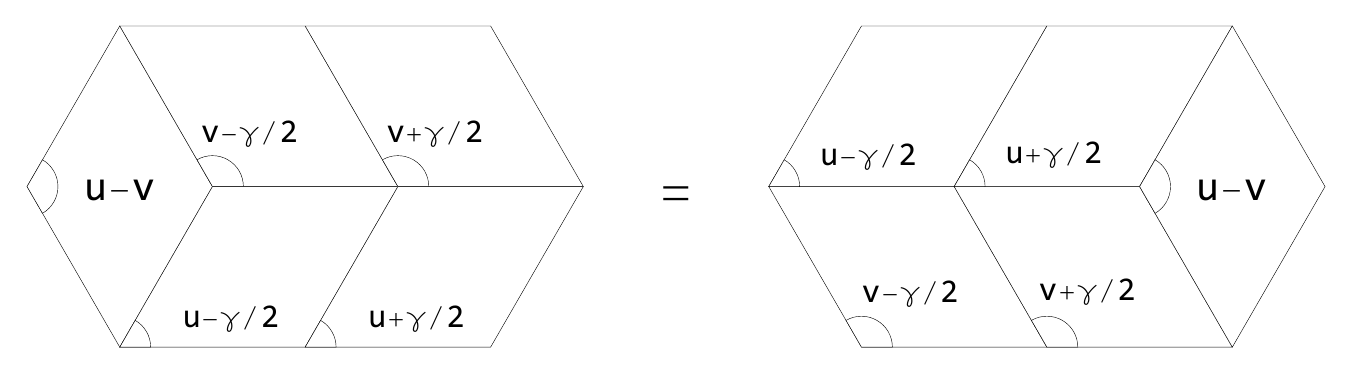}
  \\
  \caption{The Yang-Baxter equation for the composed defect translates into an equation for the fused faces when the impurity parameters differ by the crossing parameter $\gamma$.   }\label{doubleYB}
  \end{center}
\end{figure}

The integrability of the system with such  spin 2 fused faces is guaranteed by  the Yang-Baxter equation, as pictured in figure \ref{doubleYB}. More concretely, the usual argument for the commutation of transfer-matrices with generic column spectral parameters applies to this case with an important observation: if instead of the usual heights we insert $v^-$ in the intermediate external heights, equations in figure \ref{Disconnected2} will force the internal degree of freedom to be of the same kind, so the Yang-Baxter equation for the composition of the two impurity faces (with parameters differing by $\gamma$) factorizes, and gives rise indeed to  the Yang-Baxter for the spin-2 fused face. 

Of course one can continue the fusion procedure by looking at the singular points of  the faces obtained so far to obtain ``higher spin'' fused faces. 
Within our conventions,  the Boltzmann weights of fused faces are given by the formulae 
\begin{equation}
~^{(1J)}W \! \! \left(\begin{smallmatrix} 
d & c \\  
a & b \end{smallmatrix} \Big| u \right) = 
(-1)^{(-(1+J)/2+(d-b+c-a) (a-c)/4)}\sqrt{\frac{\gf_{\frac{c+a-1-J}{2}} \gf_{\frac{c+a+1+J}{2}}}{\gf_b \gf_d}} \frac{\sin(u+ (b d-a c-1) \gamma/2)}{\sin \gamma},  \nonumber
\end{equation}
for $b-a=c-d$,
\begin{equation}
~^{(1J)}W \! \! \left(\begin{smallmatrix} 
d & c \\  
a & b \end{smallmatrix} \Big| u \right) =-(-1)^{J ( (a+c-b-d)/4+1/2)}\sqrt{\frac{\gf_{\frac{c-a+1+J}{2}} \gf_{\frac{a-c+1+J}{2}}}{\gf_b \gf_d}} \frac{\sin( u+ (a c-b d-1) \gamma/2)}{\sin \gamma}, \label{IRFweightsSd}
\end{equation}otherwise. These expressions must be supplemented by the adjacency rules for the heights. These are  conveniently expressed using   new adjacency matrices. To each defect of spin $J$, one can associate the fused adjacency matrix $G^{(J)}$ obtained recursively from 
\begin{equation}
{ G} \cdot { G}^{(J-1)} = { G}^{(J)}+{ G}^{(J-2)}, 
\end{equation}
with the initial condition $G^{(1)}=G$. As usual, these matrices encode incidence rules - heights take values that can be labeled by their rows or columns, and heights  on neighboring sites along a lattice column (row) must correspond to a non-zero matrix element in the adjacency matrix $G^{(1)}$ (resp. $G^{(J)}$).  This  explains our notation in (\ref{IRFweightsSd}). For example, for the $A_4$ model we have 
\begin{equation}
\underbrace{\left(
\begin{array}{cccc}
 1 & 0 & 0 & 0 \\
 0 & 1 & 0 & 0 \\
 0 & 0 & 1 & 0 \\
 0 & 0 & 0 & 1 \\
\end{array}
\right)}_{G^{(0)}},~~
\underbrace{\left(
\begin{array}{cccc}
 0 & 1 & 0 & 0 \\
 1 & 0 & 1 & 0 \\
 0 & 1 & 0 & 1 \\
 0 & 0 & 1 & 0 \\
\end{array}
\right)}_{G^{(1)}},~~
\underbrace{\left(
\begin{array}{cccc}
 0 & 0 & 1 & 0 \\
 0 & 1 & 0 & 1 \\
 1 & 0 & 1 & 0 \\
 0 & 1 & 0 & 0 \\
\end{array}
\right)}_{G^{(2)}},~~
\underbrace{\left(
\begin{array}{cccc}
 0 & 0 & 0 & 1 \\
 0 & 0 & 1 & 0 \\
 0 & 1 & 0 & 0 \\
 1 & 0 & 0 & 0 \\
\end{array}
\right)}_{G^{(3)}}.
\end{equation}

When  $J=p-1$, the weights $~^{(1J)}W$ in Eq. (\ref{IRFweightsSd})   weights become proportional to the height-reflection (on the horizontal direction), i.e it connects reflected heights: $h \leftrightarrow p+1-h$. It follows that  no new defect is obtained at $J=p$, since
\begin{equation}
{ G} \cdot { G}^{(p-1)} = {G}^{(p-2)} \, . 
\end{equation}

To recap the total fusion procedure, to produce the defect of spin $J$, we have assembled $J$ faces of spin $1$, with spectral parameters differing by $\gamma$ from left to right. After this, we have projected out non interesting states thanks to the singular faces, hence  reducing the original Hilbert space. 

\subsection{The projectors}
\label{projectorsdef}

In this section we give more details about   the projectors we have used to produce the  fused faces in terms of the fundamental face. Consider the normalized version of the $R$ operator in (\ref{eq:R&Rtop}) (so that the coefficient in front of the identity operator is one) $\Rn_j(u) =  R_j(u) \times \sin \gamma/\sin(\gamma-u)
$. To obtain the fused face with $J=2$, we have noted that 
\begin{equation}
R_k(\gamma) \cdot \Rn_k(-\gamma) =0 ,
\label{unin}
\end{equation}
where $\Rn_k(-\gamma)=P_k^{(2)}$ is the Jones-Wenzl projector\footnote{ In principle, the normalized $\Rn_k(-\gamma)$  need not to be a projector in the usual sense, for the sole condition in (\ref{unin}) does not guarantee the idempotent property. Nevertheless, this is also unnecessary. Let us consider the situation where $\Rn_k(-\gamma)$ is diagonalizable. Because this matrix in not invertible, the minimal polynomial $P(X)$ contains a factor $X = \Rn_k(-\gamma)$. Furthermore, using Lagrange interpolation formula it is possible to build idempotent operators for both spaces: one for vectors with zero eigenvalues and one for vectors with nonzero eigenvalues. In the former case, the idempotent also has a factor $X$. This is actually what one needs to apply the arguments of the previous section and to construct orthogonal states which form the projector in this section. The factor $X$ can pass through double-faces using Yang-Baxter and be projected out thanks to (\ref{unin}) or $P(X)=p(X) \cdot X =0$. } acting on two strands and starting at site $k$, i.e. one strand lies between $k,~k+1$ and the other between  $k+1,~k+2$.

For $J=3$, besides the singular point of the fused face with $J=2$ at $u= -3 \gamma/2$, one also has to consider the required projection to build this latter fused face by means of two $J=1$ singular faces at $u=-\gamma$.  
In total, we have four fundamental faces that can be further simplified  
\begin{equation}
P_k^{(3)}=\Rn_{k+1}(-\gamma) (\Rn_k(-2\gamma) \Rn_{k+1}(-\gamma))\Rn_{k}(-\gamma)= \Rn_{k+1}(-\gamma) \Rn_k(-2\gamma) \Rn_{k+1}(-\gamma) \, ,
\end{equation}
where the brackets in the first equation indicate that the corresponding term  comes from the two faces, fused in the previous step, while the external $\Rn$ operators are the required projections to fuse it. The RHS gives the Jones-Wenzl projector $P_k^{(3)}$ acting on the three strands at positions between  $k$ to $k+3$.

\begin{figure}[hbt]
\begin{center}
  \includegraphics[width= 0.2\linewidth]{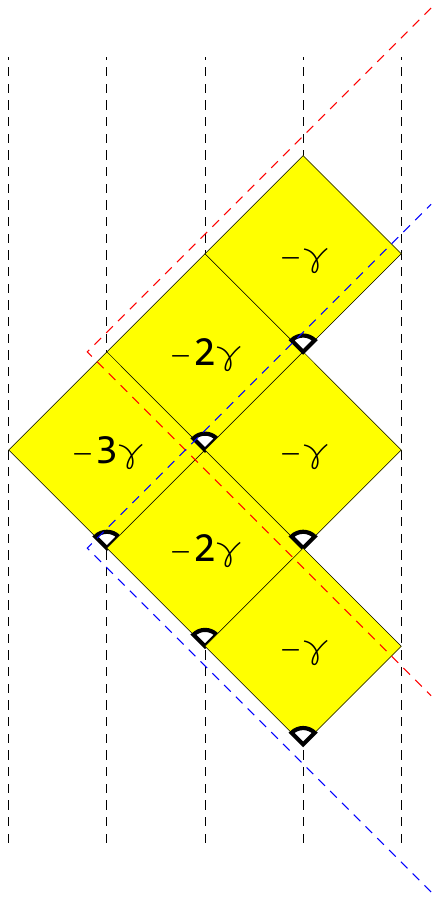}
  \includegraphics[width= 0.3\linewidth]{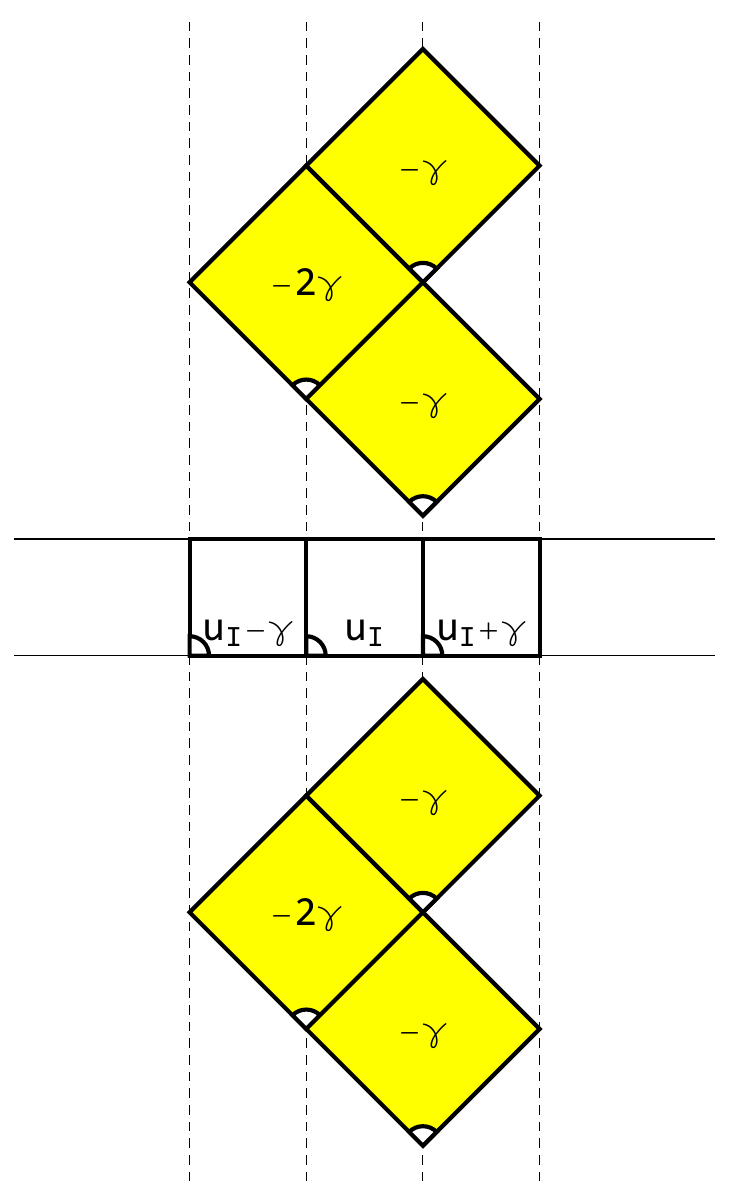}
\\
  \caption{On the left, the $P^{(4)}_k$ Jones-Wenzl projector has been presented in terms of normalized $R$ operators (yellow face). Dashed red and blue lines encompass faces which form two copies of $P^{(3)}_{k+1}$ projector, thus leading to  the recursion relation. Dashed black lines represent local sites where the heights are supported. On the right, three fundamental faces are used to produce an impurity of spin 3. }\label{Projector}
  \end{center}
\end{figure}

It is not hard to show\footnote{The singular points for fused faces of spin $J$ are  $\pm \frac{J+1}{2} \gamma$ when fused faces are balanced with respect to the shifts in $\gamma$ } that generally we have
\begin{equation}\label{eq:proj-R-mat}
P_k^{(J)}=\prod_{i=1}^{J-1} \prod_{j=1}^i \Rn_{k+J-1-(i-j+1)}  (-(i-j+1) \gamma) \, ,
\end{equation}
see figure \ref{Projector}. In the picture it is also evident that these projectors satisfy the familiar recursion relation \cite{KauffmanLins+1994+13+21}:
\begin{equation}
P_k^{(J)}
=
P_{k+1}^{(J-1)} \Rn_k(-(J-1) \gamma) P_{k+1}^{(J-1)}
=
P_{k+1}^{(J-1)}-\frac{\theta_{J-1}}{\theta_{J} }P_{k+1}^{(J-1)} \cdot e_k \cdot P_{k+1}^{(J-1)}. 
\end{equation}
These expressions for the Jones-Wenzl projectors allow us to obtain formulas for the impurity transfer-matrix and Hamiltonians in terms of  Temperley-Lieb generators, with the proviso that the resulting expression should be projected in the end. 

\subsection{Crossed-channel}
In the crossed-channel the fused faces define  functional relations among the  transfer matrices. Consider the product of transfer matrices in figure \ref{YBRotated}.
\begin{figure}[H]
    \centering
            \begin{adjustbox}{max totalsize={0.6\textwidth}{0.8\textheight},center}
            \begin{tikzpicture}[thick, scale= 0.7]
            \draw[black, thick] (0,0) -- (18,0);
            \draw[black, thick] (0,3) -- (18,3);
            \draw[black, thick] (0,6) -- (18,6);

            \draw[black, thick] (0,0) -- (0,6);
            \draw[black, thick] (3,0) -- (3,6);
            \draw[black, thick] (6,0) -- (6,6);
            \draw[black, thick] (9,0) -- (9,6);
            \draw[black, thick] (12,0) -- (12,6);
            \draw[black, thick] (15,0) -- (15,6);
            \draw[black, thick] (18,0) -- (18,6);
            \node[] at (1.5,1.5) {\Large{$v$}};
            \node[] at (4.5,1.5) {\Large{$v$}};
            \node[] at (7.5,1.5) {\Large{$v$}};
            \node[] at (10.5,1.5) {\Large{$v$}};
            \node[] at (13.5,1.5) {\Large{$v$}};
            \node[] at (16.5,1.5) {\Large{$v$}};
            \node[] at (1.5,4.5) {\Large{$u$}};
            \node[] at (4.5,4.5) {\Large{$u$}};
            \node[] at (7.5,4.5) {\Large{$u$}};
            \node[] at (10.5,4.5) {\Large{$u$}};
            \node[] at (13.5,4.5) {\Large{$u$}};
            \node[] at (16.5,4.5) {\Large{$u$}};

    \end{tikzpicture}
\end{adjustbox}
  \caption{ Product of two row-to-row transfer matrices}\label{YBRotated}
\end{figure}
The heights on the boundary  are regarded as fixed for a particular (matrix) element of the resulting product, whereas the summation over internal heights is understood as a trace of an ordered product of matrices propagating horizontally on figure \ref{YBRotated}, not all equal since each individual matrix depends also on the external heights. 

One may ask what happens to such a
product when the intertwiner \footnote{Recall that the commutativity of the two row-to-row transfer matrices is guaranteed thanks to the local condition ascribed by the Yang-Baxter equation. The additional (individual) face $W(u-v)$ which articulates the interchange of positions of  faces $W(u)$ and $W(v)$ comprising the product of transfer matrices is what we refer to as the ``intertwiner''.}  becomes singular.  
Similarly to what happens in the direct channel, at $u-v = \gamma$, we obtain
\begin{equation}\label{eq:fused_wt_hpm2}
\begin{bmatrix}
W \! \! \left(\begin{smallmatrix}a\pm2 & a \pm1 \\ a\pm1  & a \end{smallmatrix} \Big| \gamma \right)
\end{bmatrix}
= 0,
\end{equation}
and
\begin{equation}\label{singde}
\begin{bmatrix}
W \! \! \left(\begin{smallmatrix}a & a-1 \\ a-1  & a \end{smallmatrix}\Big| \gamma \right)&W \! \! \left(\begin{smallmatrix}a & a+1 \\ a-1  & a \end{smallmatrix}\Big| \gamma
\right)\\
W \! \! \left(\begin{smallmatrix}a & a-1 \\ a+1  & a \end{smallmatrix}\Big| \gamma \right)&W \! \! \left(\begin{smallmatrix}a & a+1 \\ a+1  & a \end{smallmatrix}\Big| \gamma \right)
\end{bmatrix}
=\theta_2
\begin{bmatrix} \sqrt{\frac{\theta_{a-1}}{\theta_2 {\theta_a}}} \\ \sqrt{\frac{\theta_{a+1}}{\theta_2{\theta_a}}} \end{bmatrix} \begin{bmatrix}\sqrt{\frac{\theta_{a-1}}{\theta_2{\theta_a}}}& \sqrt{\frac{\theta_{a+1}}{\theta_2{\theta_a}}} \end{bmatrix} \, , 
\end{equation}
Notice that the propagation direction for the intertwiner is along the horizontal seam between the two transfer matrices. The Yang-Baxter equation imposes again some constraints to the double faces. 
\begin{figure}[H]
    \centering
        \begin{adjustbox}{max totalsize={0.4\textwidth}{0.4\textheight},center}
    \begin{tikzpicture}
    \draw[black, thick] (0,0) -- (2,0);
    \draw[black, thick] (0,2) -- (2,2);
    \draw[black, thick] (0,4) -- (2,4);
    \draw[black, thick] (2,0) -- (2,2);
    \draw[black, thick] (2,0) -- (2,4);
    \draw[black, thick] (0,4) -- (0,0);
    \draw[black, thick] (0,0) -- (2,0);
    \draw[black, thick] (0,2) -- (2,2);
    \draw[black, thick] (0,4) -- (2,4);
    \draw[black, thick] (2,0) -- (2,2);
    \draw[black, thick] (2,0) -- (2,4);
    \draw[black, thick] (0,4) -- (0,0);
    \node[] at (1.,1) {\large{$u$}};
    \node[] at (1.,3) {\large{$u + \gamma$}};
     \node[] at (0,-0.3) {\large{$a \mp 1$}};
     \node[] at (2,-0.3) {\large{$a$}};
     \node[] at (2,4.3) {\large{$a$}};
     \node[] at (-0.3,2) {\large{$a$}};
     \node[] at (2.3,2) {\large{$v_a^{+}$}};

     \node[] at (2,4.3) {\large{$a$}};

     \node[] at (0,4.3) {\large{$a \pm 1$}};
 \node[] at (2.9,2) {\large{$=$}};

     \draw[black, thick] (4,0) -- (6,0);
    \draw[black, thick] (4,2) -- (6,2);
    \draw[black, thick] (4,4) -- (6,4);
    \draw[black, thick] (6,0) -- (6,2);
    \draw[black, thick] (6,0) -- (6,4);
    \draw[black, thick] (4,4) -- (4,0);
    \draw[black, thick] (4,0) -- (6,0);
    \draw[black, thick] (4,2) -- (6,2);
    \draw[black, thick] (4,4) -- (6,4);
    \draw[black, thick] (6,0) -- (6,2);
    \draw[black, thick] (6,0) -- (6,4);
    \node[] at (5.,1) {\large{$u$}};
    \node[] at (5.,3) {\large{$u + \gamma$}};
     \node[] at (4,-0.3) {\large{$a \mp 1$}};
     \node[] at (6,-0.3) {\large{$a$}};
     \node[] at (6,4.3) {\large{$a$}};
     \node[] at (3.6,2) {\large{$v_{a \mp 1}^{-}$}};
     \node[] at (6.3,2) {\large{$v_a^{+}$}};

     \node[] at (6,4.3) {\large{$a$}};

     \node[] at (4,4.3) {\large{$a \mp 1$}};
 \node[] at (7,2) {\large{$= 0. $}};

    \end{tikzpicture}
    \end{adjustbox}
    \caption{Zero double-faces. The symbol $v_a^+, v_a^-$ indicates a linear combination with coefficients given by the respective basis
  vector.}
\label{Zerodouble}\end{figure}

In particular, we make use of conditions pictured in Fig.\ref{Zerodouble} to study product of transfer matrices, like in figure \ref{YBRotated}. 
We take advantage of the fact that we can make a change of basis in the computation of  the matrix products, i.e. we may insert different resolutions
of the identity while computing the trace that produces the  matrix element in Fig.\ref{YBRotated}. 
Again, $v_{a}^{\pm}$ stands for orthogonal vector basis element like in (\ref{orthogonalvectors}). When these symbols appear as the internal degrees of freedom one expands the double faces in a sum of the canonical
double faces (i.e. with the usual  heights as the internal degrees of freedom) multiplied by the vector basis coefficients. We explain this with an example in the two figures below. 
\begin{figure}[H]
    \centering
            \begin{adjustbox}{max totalsize={0.9\textwidth}{0.9\textheight},center}
           \begin{tikzpicture}[thick, scale=1]
           \draw[black, thick] (0,0) -- (6,0);
           \draw[black, thick] (0,3) -- (6,3);
           \draw[black, thick] (0,6) -- (6,6);

            \draw[black, thick] (0,0) -- (0,6);
            \draw[black, thick] (3,0) -- (3,6);
            \draw[black, thick] (6,0) -- (6,6);
           \node[] at (1.5,1.5) {\Large{$u$}};
           \node[] at (4.5,1.5) {\Large{$u$}};
            \node[] at (1.5,4.5) {\Large{$u + \gamma$}};
         \node[] at (4.5,4.5) {\Large{$u+ \gamma$}};
            \node[] at (0,6.2) {\large{$a+1$}};
            \node[] at (0,-0.2) {\large{$a-1$}};
            \node[] at (6,6.2) {\large{$a-1$}};
            \node[] at (6,-0.2) {\large{$a+1$}};
            \node[] at (6.5,3) {\large{$=$}} ;

           \draw[black, thick] (9,0) -- (12,0);
           \draw[black, thick] (9,3) -- (12,3);
            \draw[black, thick] (9,6) -- (12,6);

           \draw[black, thick] (9,0) -- (9,6);
          \draw[black, thick] (12,0) -- (12,6);
            \node[] at (10.5,1.5) {\Large{$u$}};
            \node[] at (10.5,4.5) {\Large{$u + \gamma$}};
           \node[] at (9,6.2) {\large{$a+1$}};
            \node[] at (7.7,2.7) {\Large{$  \sum\limits_{b \in \{ v_a^+,v_a^- \} }  $}};
            \node[] at (8.8,3) {\large{$ a$}};
            \node[] at (12.3,3) {\large{$b$}};
 
            \node[] at (9,-0.2) {\large{$a-1$}};
            \node[] at (12,6.2) {\large{$a$}};
            \node[] at (12,-0.2) {\large{$a$}};
             \node[] at (13,3) {\LARGE{$\times$}};
            
           \draw[black, thick] (15,0) -- (18,0);
            \draw[black, thick] (15,3) -- (18,3);
            \draw[black, thick] (15,6) -- (18,6);

            \draw[black, thick] (15,0) -- (15,6);
           \draw[black, thick] (18,0) -- (18,6);
            \node[] at (16.5,1.5) {\Large{$u$}};
            \node[] at (16.5,4.5) {\Large{$u + \gamma$}};
            \node[] at (15,6.2) {\large{$a$}};
            \node[] at (13.7,2.7) {\Large{$  \sum\limits_{b \in \{ v_a^+,v_a^- \} }  $}};
            \node[] at (14.8,3) {\large{$ b$}};
            \node[] at (18.3,3) {\large{$a$}};
 
            \node[] at (15,-0.2) {\large{$a$}};
            \node[] at (18,6.2) {\large{$a-1$}};
            \node[] at (18,-0.2) {\large{$a+1$}};

    \end{tikzpicture}
\end{adjustbox}
\caption{A two row transfer matrix can be resolved in the above way.}
   \label{fig:two-row-resol}
\end{figure}
\begin{figure}[H]
    \centering
            \begin{adjustbox}{max totalsize={0.8\textwidth}{0.8\textheight},center}
            \begin{tikzpicture}[thick, scale=1]
            \draw[black, thick] (0,0) -- (3,0);
            \draw[black, thick] (0,3) -- (3,3);
            \draw[black, thick] (0,6) -- (3,6);

            \draw[black, thick] (0,0) -- (0,6);
            \draw[black, thick] (3,0) -- (3,6);
            \node[] at (1.5,1.5) {\Large{$u$}};
            \node[] at (1.5,4.5) {\Large{$u + \gamma$}};
            \node[] at (0,6.2) {\large{$a+1$}};
            \node[] at (-0.3,3) {\large{$a$}};
            \node[] at (3.3,3) {\large{$v_a^+$}};
 
            \node[] at (0,-0.2) {\large{$a-1$}};
            \node[] at (3,6.2) {\large{$a$}};
            \node[] at (3,-0.2) {\large{$a$}};

            \node[] at (3.8,3) {\large{$=$}} ; 

            \draw[black, thick] (6,0) -- (9,0);
            \draw[black, thick] (6,3) -- (9,3);
            \draw[black, thick] (6,6) -- (9,6);
            \draw[black, thick] (6,0) -- (6,6);
            \draw[black, thick] (9,0) -- (9,6);
            \node[] at (7.5,1.5) {\Large{$u$}};
            \node[] at (7.5,4.5) {\Large{$u + \gamma$}};
            \node[] at (6,6.2) {\large{$a+1$}};
            \node[] at (5.8,3) {\large{$a$}};
            \node[] at (10,3) {\large{$a-1 \quad  + $ } };
            \node[] at (6,-0.2) {\large{$a-1$}};
            \node[] at (9,6.2) {\large{$a$}};
            \node[] at (9,-0.2) {\large{$a$}};
            \node[] at (4.8,3) {\LARGE{$\sqrt{\frac{\theta_{a-1}}{\theta_{2}\theta_a}}$}};
            \node[] at (11.6,3) {\LARGE{$\sqrt{\frac{\theta_{a+1}}{\theta_{2}\theta_a}}$}} ; 

            \draw[black, thick] (13,0) -- (16,0);
            \draw[black, thick] (13,3) -- (16,3);
          \draw[black, thick] (13,6) -- (16,6);

            \draw[black, thick] (13,0) -- (13,6);
            \draw[black, thick] (16,0) -- (16,6);
            \node[] at (14.5,1.5) {\Large{$u$}};
            \node[] at (14.5,4.5) {\Large{$u + \gamma$}};
            \node[] at (13,6.2) {\large{$a+1$}};
            \node[] at (12.8,3) {\large{$a$}};
            \node[] at (17,3) {\large{$a+1  $ } };
 
            \node[] at (13,-0.2) {\large{$a-1$}};
            \node[] at (16,6.2) {\large{$a$}};
           \node[] at (16,-0.2) {\large{$a$}};
            
    \end{tikzpicture} 
   \end{adjustbox}
       \begin{adjustbox}{max totalsize={0.8\textwidth}{0.8\textheight},center}
             \begin{tikzpicture}[thick, scale=1]
            \draw[black, thick] (0,0) -- (3,0);
            \draw[black, thick] (0,3) -- (3,3);
            \draw[black, thick] (0,6) -- (3,6);

            \draw[black, thick] (0,0) -- (0,6);
            \draw[black, thick] (3,0) -- (3,6);
            \node[] at (1.5,1.5) {\Large{$u$}};
            \node[] at (1.5,4.5) {\Large{$u + \gamma$}};
            \node[] at (0,6.2) {\large{$a+1$}};
            \node[] at (-0.3,3) {\large{$a$}};
            \node[] at (3.3,3) {\large{$v_a^-$}};
 
            \node[] at (0,-0.2) {\large{$a-1$}};
            \node[] at (3,6.2) {\large{$a$}};
            \node[] at (3,-0.2) {\large{$a$}};

            \node[] at (3.8,3) {\large{$=$}} ; 

          \draw[black, thick] (6,0) -- (9,0);
          \draw[black, thick] (6,3) -- (9,3);
            \draw[black, thick] (6,6) -- (9,6);

            \draw[black, thick] (6,0) -- (6,6);
            \draw[black, thick] (9,0) -- (9,6);
            \node[] at (7.5,1.5) {\Large{$u$}};
            \node[] at (7.5,4.5) {\Large{$u + \gamma$}};
            \node[] at (6,6.2) {\large{$a+1$}};
            \node[] at (5.8,3) {\large{$a$}};
            \node[] at (10,3) {\large{$a-1 \quad  - $ } };
 
            \node[] at (6,-0.2) {\large{$a-1$}};
            \node[] at (9,6.2) {\large{$a$}};
            \node[] at (9,-0.2) {\large{$a$}};
            \node[] at (4.8,3) {\LARGE{$\sqrt{\frac{\theta_{a+1}}{\theta_{2}\theta_a}}$}};
          \node[] at (11.6,3) {\LARGE{$\sqrt{\frac{\theta_{a-1}}{\theta_2\theta_a}}$}} ; 

            \draw[black, thick] (13,0) -- (16,0);
            \draw[black, thick] (13,3) -- (16,3);
           \draw[black, thick] (13,6) -- (16,6);

            \draw[black, thick] (13,0) -- (13,6);
            \draw[black, thick] (16,0) -- (16,6);
            \node[] at (14.5,1.5) {\Large{$u$}};
            \node[] at (14.5,4.5) {\Large{$u + \gamma$}};
            \node[] at (13,6.2) {\large{$a+1$}};
            \node[] at (12.8,3) {\large{$a$}};
            \node[] at (17,3) {\large{$a+1  $ } };
 
            \node[] at (13,-0.2) {\large{$a-1$}};
            \node[] at (16,6.2) {\large{$a$}};
           \node[] at (16,-0.2) {\large{$a$}};
            
    \end{tikzpicture}
\end{adjustbox}
\caption{Here we define what we mean by a site labeled by $v_a^+ / v_a^-$.} 
   \label{fig:def:vhp-vhm}
\end{figure}

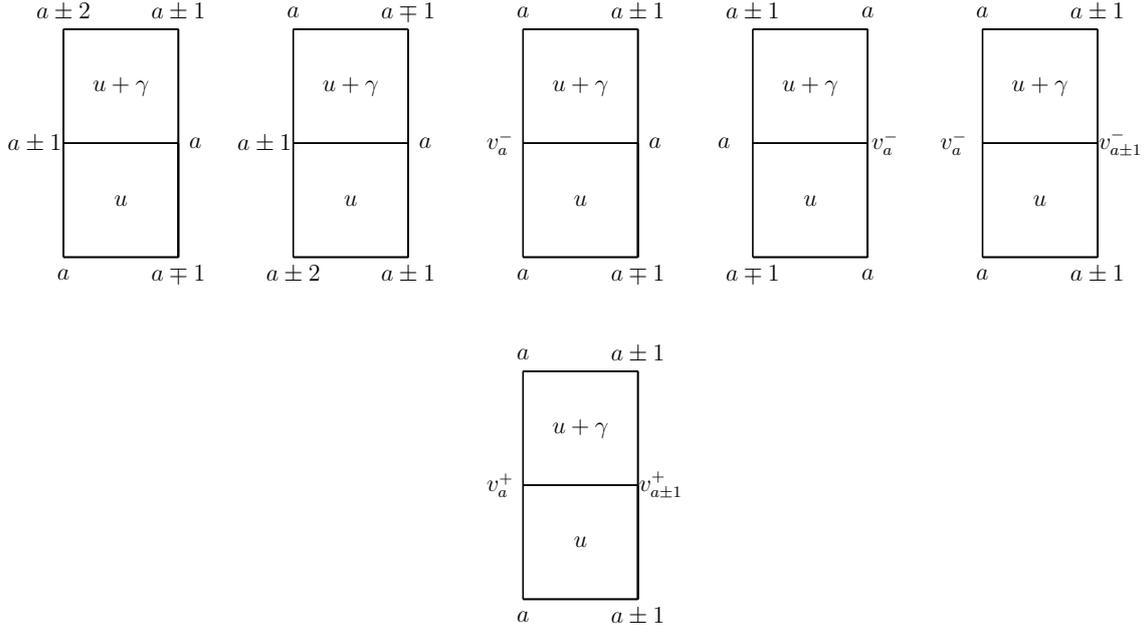
\begin{figure}[H]
    \centering
    \begin{adjustbox}{max totalsize={1\textwidth}{1\textheight},center}
    \begin{tikzpicture}
    \draw[black, thick] (0,0) -- (2,0);
    \draw[black, thick] (0,2) -- (2,2);
    \draw[black, thick] (0,4) -- (2,4);
    \draw[black, thick] (2,0) -- (2,2);
    \draw[black, thick] (2,0) -- (2,4);
    \draw[black, thick] (0,4) -- (0,0);
    \draw[black, thick] (0,0) -- (2,0);
    \draw[black, thick] (0,2) -- (2,2);
    \draw[black, thick] (0,4) -- (2,4);
    \draw[black, thick] (2,0) -- (2,2);
    \draw[black, thick] (2,0) -- (2,4);
    \draw[black, thick] (0,4) -- (0,0);
    \node[] at (1.,1) {\large{$u$}};
    \node[] at (1.,3) {\large{$u + \gamma$}};
     \node[] at (0,-0.3) {\large{$a $}};
     \node[] at (2,-0.3) {\large{$a \mp 1$}};
     \node[] at (2,4.3) {\large{$a \pm 1$}};
     \node[] at (-0.5,2) {\large{$a \pm 1$}};
     \node[] at (2.3,2) {\large{$a$}};
    \node[] at (0,4.3) {\large{$a \pm 2$}};



     \draw[black, thick] (4,0) -- (6,0);
    \draw[black, thick] (4,2) -- (6,2);
    \draw[black, thick] (4,4) -- (6,4);
    \draw[black, thick] (6,0) -- (6,2);
    \draw[black, thick] (6,0) -- (6,4);
    \draw[black, thick] (4,4) -- (4,0);
    \draw[black, thick] (4,0) -- (6,0);
    \draw[black, thick] (4,2) -- (6,2);
    \draw[black, thick] (4,4) -- (6,4);
    \draw[black, thick] (6,0) -- (6,2);
    \draw[black, thick] (6,0) -- (6,4);
    \node[] at (5.,1) {\large{$u$}};
    \node[] at (5.,3) {\large{$u + \gamma$}};
     \node[] at (4,-0.3) {\large{$a \pm 2$}};
     \node[] at (6,-0.3) {\large{$a \pm 1$}};
     \node[] at (3.5,2) {\large{$a \pm 1$}};
     \node[] at (6.3,2) {\large{$a$}};

     \node[] at (6,4.3) {\large{$a \mp 1$}};

     \node[] at (4,4.3) {\large{$a$}};

     \draw[black, thick] (8,0) -- (10,0);
    \draw[black, thick] (8,2) -- (10,2);
    \draw[black, thick] (8,4) -- (10,4);
    \draw[black, thick] (10,0) -- (10,2);
    \draw[black, thick] (10,0) -- (10,4);
    \draw[black, thick] (8,4) -- (8,0);
    \draw[black, thick] (8,0) -- (10,0);
    \draw[black, thick] (8,2) -- (10,2);
    \draw[black, thick] (8,4) -- (10,4);
    \draw[black, thick] (10,0) -- (10,2);
    \draw[black, thick] (10,0) -- (10,4);
    \node[] at (9.,1) {\large{$u$}};
    \node[] at (9.,3) {\large{$u + \gamma$}};
     \node[] at (8,-0.3) {\large{$a $}};
     \node[] at (10,-0.3) {\large{$a \mp 1$}};
     \node[] at (7.6,2) {\large{$v_{a}^{-}$}};
     \node[] at (10.3,2) {\large{$a$}};
     \node[] at (10,4.3) {\large{$a \pm 1$}};
     \node[] at (8,4.3) {\large{$a$}};
     \draw[black, thick] (12,0) -- (14,0);
    \draw[black, thick] (12,2) -- (14,2);
    \draw[black, thick] (12,4) -- (14,4);
    \draw[black, thick] (14,0) -- (14,2);
    \draw[black, thick] (14,0) -- (14,4);
    \draw[black, thick] (12,4) -- (12,0);
    \draw[black, thick] (12,0) -- (14,0);
    \draw[black, thick] (12,2) -- (14,2);
    \draw[black, thick] (12,4) -- (14,4);
    \draw[black, thick] (14,0) -- (14,2);
    \draw[black, thick] (14,0) -- (14,4);
    \node[] at (13.,1) {\large{$u$}};
    \node[] at (13.,3) {\large{$u + \gamma$}};
     \node[] at (12,-0.3) {\large{$a \mp 1$}};
     \node[] at (14,-0.3) {\large{$a$}};
     \node[] at (11.5,2) {\large{$a$}};
     \node[] at (14.3,2) {\large{$v_a^-$}};
     \node[] at (14,4.3) {\large{$a $}};
     \node[] at (12,4.3) {\large{$a \pm 1$}};
     \draw[black, thick] (16,0) -- (18,0);
    \draw[black, thick] (16,2) -- (18,2);
    \draw[black, thick] (16,4) -- (18,4);
    \draw[black, thick] (18,0) -- (18,2);
    \draw[black, thick] (18,0) -- (18,4);
    \draw[black, thick] (16,4) -- (16,0);
    \draw[black, thick] (16,0) -- (18,0);
    \draw[black, thick] (16,2) -- (18,2);
    \draw[black, thick] (16,4) -- (18,4);
    \draw[black, thick] (18,0) -- (18,2);
    \draw[black, thick] (18,0) -- (18,4);
    \node[] at (17.,1) {\large{$u$}};
    \node[] at (17.,3) {\large{$u + \gamma$}};
     \node[] at (16,-0.3) {\large{$a $}};
     \node[] at (18,-0.3) {\large{$a \pm 1$}};
     \node[] at (15.5,2) {\large{$v_a^-$}};
     \node[] at (18.4,2) {\large{$v^-_{a \pm 1}$}};

     \node[] at (18,4.3) {\large{$a \pm 1$}};
     \node[] at (16,4.3) {\large{$a$}};
     \draw[black, thick] (8,-6) -- (10,-6);
    \draw[black, thick] (8,-4) -- (10,-4);
    \draw[black, thick] (8,-2) -- (10,-2);
    \draw[black, thick] (10,-6) -- (10,-4);
    \draw[black, thick] (10,-6) -- (10,-2);
    \draw[black, thick] (8,-2) -- (8,-6);
    \draw[black, thick] (8,-6) -- (10,-6);
    \draw[black, thick] (8,-4) -- (10,-4);
    \draw[black, thick] (8,-2) -- (10,-2);
    \draw[black, thick] (10,-6) -- (10,-4);
    \draw[black, thick] (10,-6) -- (10,-2);
    \node[] at (9.,-5) {\large{$u$}};
    \node[] at (9.,-3) {\large{$u + \gamma$}};
     \node[] at (8,-6.3) {\large{$a $}};
     \node[] at (10,-6.3) {\large{$a \pm 1$}};
     \node[] at (7.6,-4) {\large{$v_{a}^{+}$}};
     \node[] at (10.4,-4) {\large{$v_{a \pm 1}^{+}$}};
     \node[] at (10,-1.7) {\large{$a \pm 1$}};
     \node[] at (8,-1.7) {\large{$a$}};
    \end{tikzpicture}
    \end{adjustbox}
    \caption{Two sets of double faces, with 5 and 1 elements, that do not mix. }
        \label{DoubleWeights}
  \end{figure}

The conditions implied by figure \ref{Zerodouble} tell us that one may separate the double faces into two groups that do not mix when computing the resulting transfer matrix of
figure \ref{YBRotated}. 
Like in the direct channel, this leads  to a decoupling of the internal(intermediate) heights.
Therefore, effectively, one should consider two sets of weights, like in figure \ref{DoubleWeights}.
These are the only weights that appear when we resolve the two row transfer matrix in figure \ref{YBRotated}. For instance, consider the weight in  figure (\ref{fig:anonal-two-face}) below:
\begin{figure}[H]
    \centering
    \begin{adjustbox}{max totalsize={0.2\textwidth}{0.2\textheight},center}
    \begin{tikzpicture}
    \draw[black, thick] (0,0) -- (2,0);
    \draw[black, thick] (0,2) -- (2,2);
    \draw[black, thick] (0,4) -- (2,4);
    \draw[black, thick] (2,0) -- (2,4);
    \draw[black, thick] (0,4) -- (0,0);
    \draw[black, thick] (0,0) -- (2,0);
    \draw[black, thick] (0,2) -- (2,2);
    \draw[black, thick] (0,4) -- (0,0);
    \node[] at (1.,1) {\large{$u$}};
    \node[] at (1.,3) {\large{$u + \gamma$}};
     \node[] at (0,-0.3) {\large{$a $}};
     \node[] at (2,-0.3) {\large{$a \pm 1$}};
     \node[] at (2,4.3) {\large{$a \pm 1$}};
     \node[] at (-0.3,2) {\large{$v_a^{+}$}};
     \node[] at (2.4,2) {\large{$v_{a \pm 1}^{-}$}};

     \node[] at (0,4.3) {\large{$a$}};
    \end{tikzpicture}        
    \end{adjustbox}
    \caption{A configuration which is not in either of two sets above.  }
    \label{fig:anonal-two-face}
\end{figure}
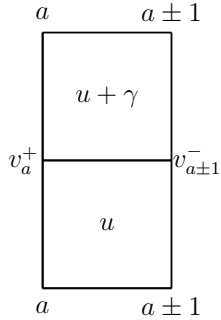
While its explicit value is non-zero - and  therefore it is not precluded from the construction in figure \ref{Zerodouble} -  when one considers the periodic boundary condition in the horizontal direction, one cannot connect it to the other possible faces depicted in figure \ref{DoubleWeights}. For example, to its left one may join double faces from the bottom set of figure \ref{DoubleWeights}, while on the right one may join double faces from the top set. Closing the set  of double faces under the periodic boundary condition would require an additional double face of type listed in figure \ref{Zerodouble}, which is  impossible.

Explicit computation of the weights shows  that the set on the top row of figure \ref{DoubleWeights} provides a higher spin transfer matrix with vertical neighboring heights constrained by the adjacency matrix $G^{(2)}$. 
Meanwhile, the group on the bottom row of figure \ref{DoubleWeights} is an operator proportional to the identity, to which we associate the adjacency matrix $G^{(0)}=1$. Now, if one looks back at the product of transfer matrices, this leads to the first instance of the fusion hierarchy
\begin{equation}
 T^{(1)}_{[-\frac{J}{2}]}T^{(J)}_{[\frac{1}{2}]} = T^{(0)}_{[-\frac{J-1}{2}]} T^{(J+1)}_{[0]}+ T^{(0)}_{[-\frac{J+1}{2}]} T^{(J-1)}_{[1]}, \label{prefusion}
\end{equation}
with $J=1$ and where we have denoted $T^{(J)}_{[m]}=T^{(J)}(u_k+ i m \gamma )$, with
\begin{equation}\label{fusedTmat}
\left(T^{(J)}(u_k)\right)_{\mathbf{x}}^{\mathbf{x}'}= \prod_{m=1}^{2 \R} ~^{(J1)}
W \! \! \left(
\begin{smallmatrix} 
x'_{m} & x'_{m+1} \\  
x_{m} & x_{m+1} 
\end{smallmatrix} 
\Big| u_k  
\right) \, .
\end{equation}
The weights for the fused faces are given by the formula
\begin{equation}
~^{(J1)}W \! \! \left(\begin{smallmatrix} 
d & c \\  
a & b \end{smallmatrix} \Big| u \right) = 
(-1)^{((1+J)/2+(b-d+c-a) (a-c)/4)}\sqrt{\frac{\gf_{\frac{c+a-1-J}{2}} \gf_{\frac{c+a+1+J}{2}}}{\gf_b \gf_d}} \frac{\sin(u+ (b d-a c-1) \gamma/2)}{\sin \gamma},  \nonumber
\end{equation}
for $c-b=d-a$,
\begin{equation}
~^{(J1)}W \! \! \left(\begin{smallmatrix} 
d & c \\  
a & b \end{smallmatrix} \Big| u \right) =(-1)^{J ( (a+c-b-d)/4-1/2)}\sqrt{\frac{\gf_{\frac{c-a+1+J}{2}} \gf_{\frac{a-c+1+J}{2}}}{\gf_b \gf_d}} \frac{\sin( u+ (a c-b d-1) \gamma/2)}{\sin \gamma}, 
\end{equation}otherwise. Now heights on sites  linked by vertical edges are constrained by  the adjacency matrix $G^{(J)}$, while the ones linked by horizontal edges are constrained by  $G^{(1)}$. In particular, $J=0$ corresponds to a transfer matrix proportional to identity, $J=1$ gives the fundamental transfer matrix, and $J=p-1$ is proportional to the height reflection operator.

We notice that we are free to perform gauge transformations of the type
\begin{equation}
~^{(J1)}W \! \! \left(\begin{smallmatrix}d & c \\ a & b \end{smallmatrix}\Big| u \right) \rightarrow ~^{(J1)}W \! \! \left(\begin{smallmatrix}d & c \\
a & b \end{smallmatrix}\Big| u \right) \frac{f(a, d)}{f(b,c)} \, ,
\end{equation}
without modifying the transfer matrix. The function $f(a,d)$ is an arbitrary function of the vertical edge heights. It obviously cannot change the transfer matrix
on account of the horizontal periodic boundary conditions.  It is also
possible to include such a transformation for the horizontal edges, say with another function $f_2(a, b)$, whose contributions  will cancel out due to  the vertical periodic boundary
conditions. This corresponds to a similarity transformation between the transfer matrices.

Like in the direct channel, the fusion procedure readily generalizes to higher levels of the fusion hierarchy - that is, involving more than two rows.  This follows from the fact that each set of fused faces does also satisfy the Yang Baxter equation for higher representations  and, as such, the newly defined Boltzmann weights can be analyzed at their  singular points to further decompose higher products of transfer matrices. Therefore, the induction step from one fusion level to the next  boils down to the proof of  the Yang-Baxter equations for different higher representations, which can be carried out along the same lines.

\newpage

\section{Relationship with other fusion constructions }\label{sec:fus:KP}

The fused weights in section \ref{sec:fused-Tmat-Th}  are different from those  of \cite{Klumper:1992vt} and \cite{Bazhanov:1989yk}. We discuss in this  appendix how these different conventions/constructions are related. 

We  first review the construction of A$_{p}$ RSOS ($p_1,p_2$) face weights based on \cite{Klumper:1992vt} and \cite{Bazhanov:1989yk}. For an integer $p_1 \geq 1$, we define the fused face $\overline{W}^{p_1,1}$ as in figure \ref{fig:fused-p,1}. 
\begin{figure}[H]
    \centering
    \begin{tikzpicture}

\coordinate (A1) at (0,0);
\coordinate (A2) at (2,0);
\coordinate (AP) at (8,0);
\coordinate (AP1) at (10,0);

\coordinate (B1) at (0,2);
\coordinate (B2) at (2,2);
\coordinate (BP) at (8,2);
\coordinate (BP1) at (10,2);

\draw (A1) rectangle (B1);
\draw (A2) rectangle (B2);
\draw (AP) rectangle (BP);
\draw (AP1) rectangle (BP1);


\node at (-0.2, -0.3) {$a_1$};
\node at (2, -0.3) {$a_2$};
\node at (8, -0.3) {$a_{p_1}$};
\node at (10.2, -0.3) {$a_{p_1+1}$};

\node at (-0.2, 2.3) {$b_1$};
\node at (2, 2.3) {$b_2$};
\node at (8, 2.3) {$b_{p_1}$};
\node at (10.2, 2.3) {$b_{p_1+1}$};

\node at (1, 1) {\scriptsize{$u - (p_1 - 1) \gamma$}};
\node at (5, 1) {$\cdots$};
\node at (4.5, 1) {$\cdots$};
\node at (4, 1) {$\cdots$};
\node at (3.5, 1) {$\cdots$};
\node at (3, 1) {$\cdots$};

\node at (9, 1) {$u$};
\node at (7, 1) {$u - \gamma$};
\draw[black] (6,0) -- (6,2);
\draw[black] (0,2) -- (10,2); 
\draw[black] (0,0) -- (10,0); 

\fill (2, 0) circle (2pt);
\fill (6, 0) circle (2pt);

\fill (8, 0) circle (2pt);

\end{tikzpicture}
    \caption{Fused $(p_1,1)$ face. All sites covered with solid circle are summed over. Labels on neighboring sites still have to respect the condition that they must be neighbors on A$_{p}$ Dynkin diagram.}
    \label{fig:fused-p,1}
\end{figure}
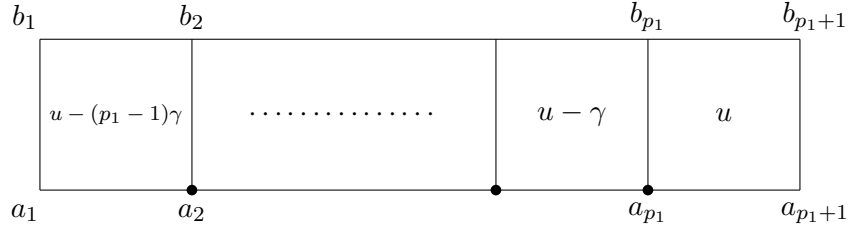
The weight for such a face is given by 
\begin{equation}\label{eq:gauge_face_wt_KP_fusion}
\overline{W}^{p_1, 1}\left(\begin{array}{ll}
b_1 & b_{p_1+1} \\
a_1 & a_{p_1+1}
\end{array} \Bigg| \ u\right)=\sum_{a_2, \ldots a_{p_1}} \prod_{l=1}^{p_1} \, \overline{W}\left(\begin{array}{ll}
b_l & b_{l+1} \\
a_l & a_{l+1}
\end{array} \Bigg| \  u+(l-p_1)\,  \gamma\right) . 
\end{equation}
where $b_2, \ldots, b_{p_1}$ are arbitrary (but satisfy the   condition that $\lvert b_l - b_{l + 1} \rvert$ =1): the weight of the row does not depend on their exact values because of the patterns of spectral parameters. Finally   $\overline{W}$ denote weights with gauge factor $S_a = (-1)^{a/2}/\sqrt{\theta_a}$ in Eq. \eqref{eq:standard_face_weight} \footnote{In \cite{Klumper:1992vt}, the gauge factor chosen for doing fusion was just $S_a = (-1)^{a/2}$, however with that gauge factor the fused weights, such as $\overline{W}^{p_1,1}$, will depend on what $b_i$'s one chooses in Eq. \eqref{eq:gauge_face_wt_KP_fusion}}.

For integers $p_1,p_2$ greater than or equal to 1, we define 
\begin{equation}
    \overline{W}^{p_1,p_2}\left(\begin{array}{ll}
a_{p_2+1} & b_{p_2+1} \\
a_1 & b_1
\end{array} \Bigg| \ u\right)= \prod_{m = 0}^{p_2-2}  s^{p_1}_m(u)^{-1} \sum_{a_2,\ldots,a_{p_2}}\prod_{l = 1}^{p_2} \overline{W}^{p_1, 1}\left(\begin{array}{ll}
a_{l + 1} & b_{l+1} \\
a_l & b_{l}
\end{array} \Bigg| \ u + (l-1) \, \gamma \right)\, , 
\end{equation}
where 
\begin{equation}
    s_m^{q}(u) = \prod_{j = 0}^{q-1} \frac{\sin( u + (m - j)\,  \gamma)}{\sin \gamma} \, , \quad \gamma = \frac{\pi}{p + 1} \,.
\end{equation}
Note, we do not require that neighboring sites of fused faces be neighbors on Dynkin diagram, instead we demand 
\begin{equation}\label{eq:cond-neighb-wts}
   0 \leq \frac{a_i - a_j + m}{2} \leq m \, , \quad \quad m < a_i + a_j <  2 p - m + 4 \, ,
\end{equation}
where $m$ is $p_1$ or $p_2$ if $a_i$ and $a_j$ are horizontal or vertical neighbors respectively.

These weights satisfy a \textit{generalized} Yang-Baxter equation
\begin{equation}
\begin{split}
\sum_g    \overline{W}^{q,s}\left(\begin{array}{ll}
e & g \\
f & a
\end{array} \Bigg| \ v\right)
  \overline{W}^{m,s}\left(\begin{array}{ll}
g & c \\
a & b
\end{array} \Bigg| \ u+v\right)
    \overline{W}^{m,q}\left(\begin{array}{ll}
e & d \\
g & c
\end{array} \Bigg| \ u\right) \,   =    \\
\sum_g    \overline{W}^{m,q}\left(\begin{array}{ll}
f & g \\
a & b
\end{array} \Bigg| \ u\right)
  \overline{W}^{m,s}\left(\begin{array}{ll}
e & d \\
f & g
\end{array} \Bigg| \ u+v\right)
    \overline{W}^{q,s}\left(\begin{array}{ll}
d & c \\
g & b
\end{array} \Bigg| \ v\right) \, , 
\end{split}
\end{equation}
which is a consequence of usual Yang-Baxter. Let us set $m = 1$ and $q,s = r$, then we can show the $(1,r)$ face with spectral parameter $u = {\rm i} \infty$ satisfies the defect YB in figure \ref{fig:first_cond_top_inv}.
\begin{figure}[H]
    \centering
\begin{tikzpicture}

    \fill[red!10] (6,0) -- (8,0) -- (9,1.5) -- (7,1.5) -- cycle;
\draw[black, thick] (6,0) -- (8,0) -- (9,1.5) -- (7,1.5) -- cycle;
    \fill[red!10] (6,0) -- (8,0) -- (9,-1.5) -- (7,-1.5) -- cycle;
\draw[black, thick] (6,0) -- (8,0) -- (9,-1.5) -- (7,-1.5) -- cycle;
\draw[black, thick] (8,0) -- (9,1.5) -- (10,0) -- (9,-1.5) -- cycle;

\node at (5.8,0) {$f$} ; 
\node at (6.8,1.5) {$e$} ; 
\node at (9.25,1.5) {$d$} ; 
\node at (6.8,-1.5) {$a$} ; 
\node at (9.25,-1.5) {$b$} ; 
\node at (10.2,0) {$c$} ; 
\node at (7.8,-0.3) {$g$} ; 
\node at (5.8,0) {$f$} ; 
\draw[red, thick] (6.3,0) arc (0 : 60 : 0.3) ;
\draw[red, thick] (7.2,-1.5) arc (0 : 120 : 0.2) ;
\draw[red, thick] (8.15,-0.15) arc (-60 : 60 : 0.2) ;
\node at (9,0) {$v$} ; 
\node at (1,0) {$v$} ; 

\draw[black, thick] (0,0) -- (1,1.5) -- (2,0) -- (1,-1.5) -- cycle;
\fill[red!10] (3,1.5) -- (1,1.5) -- (2,0) -- (4,0) -- cycle;
\fill[red!10] (3,-1.5) -- (1,-1.5) -- (2,0) -- (4,0) -- cycle;
\draw[black, thick] (3,1.5) -- (1,1.5) -- (2,0) -- (4,0) -- cycle;
\draw[black, thick] (3,-1.5) -- (1,-1.5) -- (2,0) -- (4,0) -- cycle;

\node at (-.2,0) {$f$} ; 
\node at (0.8,1.5) {$e$} ; 
\node at (3.25,1.5) {$d$} ; 
\node at (0.8,-1.5) {$a$} ; 
\node at (3.25,-1.5) {$b$} ; 
\node at (4.2,0) {$c$} ; 
\node at (2.2,-0.3) {$g$} ; 
\node at (5,0) {   $= \sum_{g}$} ;
\node at (-1,0) {   $\sum_{g}$} ;

\draw[red, thick] (0.15,-0.15) arc (-60 : 60 : 0.2) ;
\draw[red, thick] (2.2,0) arc (0 : 120 : 0.2) ;
\draw[red, thick] (1.2,-1.5) arc (0 : 60 : 0.2) ;

\node at (2.5,0.75) { \small $u = \im \infty$ }  ; 
\node at (7.5,-0.75) { \small $u = \im \infty$ }  ; 

\node at (2.5,-0.75) { \small $u + v = \im \infty$ }  ; 
\node at (7.5,0.75) { \small $u + v = \im \infty$ }  ;

\end{tikzpicture}
    \caption{Yang-Baxter equation for face model. We can set $ q,s = r$, and the defect face corresponds to defect of type $(1,r+1)$ in the continuum CFT.}
    \label{fig:YB-fig-fus}

\end{figure}
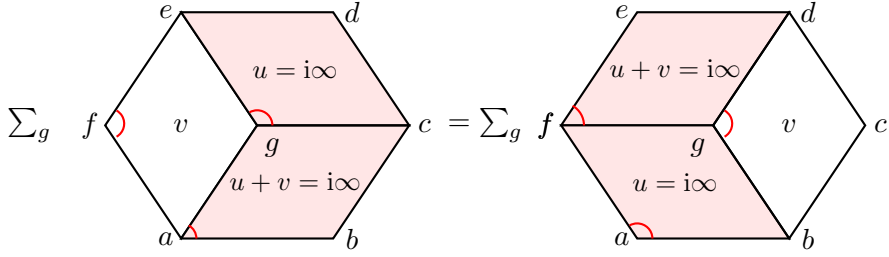

We now wish to show the unitarity of $\overline{W}^{1,r}$ face weights. For simplicity   we will restrict the discussion to  $\overline{W}^{1,2}$ face weights, the generalization to $\overline{W}^{1,r}$ face weight following easily. 
\begin{figure}[H]
    \begin{tikzpicture}[scale = 1.5]

    \draw (0,0) rectangle (1,1); 
    \draw (1.25,1.25) rectangle (2.25,2.25); 
    \node at (0.5,0.5) {\large $-u$};
    \node at (1.75,1.75) {\large $u$};

    \node at (0, 1.15) {\large $a_1$};
    \node at (0, -0.15) {\large $a_2$};
    \node at (1, 1.15) {\large $c$};
    \node at (1, -0.15) {\large $a_3$};

    \node at (1.25, 2.4) {\large $a_1$};
    \node at (1.25, 1.1) {\large $c$};
    \node at (2.25, 2.4) {\large $d$};
    \node at (2.25, 1.1) {\large $a_3$};


    \node at (5.5,1) {\large $ = \sum_c  \overline{W}^{1,2} \left( \begin{matrix} a_1 & c \\ a_2 & a_3 \end{matrix} \middle\vert -u \right)  \overline{W}^{2,1} \left( \begin{matrix} a_1 & d \\ c & a_3 \end{matrix} \middle\vert u \right)  \propto \, \delta_{d,a_2}$};
    \filldraw [black] (1,1) circle (1pt);
    \filldraw [black] (1.25,1.25) circle (1pt);

\end{tikzpicture}

   \caption{Unitarity of (1,2) fused weight, $c$ is summed over }
    \label{fig:unitarity_fused}
\end{figure}
Now, we know from definition of fused weights that 
\begin{equation}\label{eq:fused_wts}
\begin{split}
      \overline{W}^{2, 1}\left(\begin{array}{ll}
a_1 & d \\
c & a_3
\end{array} \Bigg| \ u\right) &=\sum_{t} \overline{W}\left(\begin{array}{ll}
a_1 & \alpha \\
c & t
\end{array} \Bigg| \  u - \gamma\right)  \, \overline{W}\left(\begin{array}{ll}
\alpha & d \\
t & a_3
\end{array} \Bigg| \  u  \right) \, , \\
     \overline{W}^{1,2}\left(\begin{array}{ll}
a_1 & c \\
a_2 & a_3
\end{array} \Bigg| \ -u\right) & = s^{1}_{0}(-u)^{-1}\sum_{\tilde{t}} \overline{W}\left(\begin{array}{ll}
\tilde{t} & \beta \\
a_2 & a_3
\end{array} \Bigg| \  - u \right)  \, \overline{W}\left(\begin{array}{ll}
a_1 & c \\
\tilde{t} & \beta
\end{array} \Bigg| \  -u + \gamma  \right) \, , 
\end{split}
\end{equation}    
where $\alpha$ and $\beta$ are any heights. Now, note $t$ has to satisfy the constraint that it is within distance 1 of $c$, $a_3$, and $\alpha$. When we choose $\beta$, we must ensure that it is distance 1 of $c$ and $a_3$,  hence we can always choose $\beta$ such that it equals $t$. This is important as when we multiply the two terms in Equation \eqref{eq:fused_wts}, we can take summation $t$ outside and take $\beta = t$, to get 

\begin{equation}
\begin{split}
 & \sum_c \overline{W}^{2, 1}\left(\begin{array}{ll}
a_1 & d \\
c & a_3
\end{array} \Bigg| \ u\right)   \overline{W}^{1,2}\left(\begin{array}{ll}
a_1 & c \\
a_2 & a_3
\end{array} \Bigg| \ -u\right)\\
&  = \sum_{t,c} \overline{W}\left(\begin{array}{ll}
a_1 & \alpha \\
c & t
\end{array} \Bigg| \  u - \gamma\right)  \, \overline{W}\left(\begin{array}{ll}
\alpha & d \\
t & a_3
\end{array} \Bigg| \  u  \right) \,  \overline{W}^{1,2}\left(\begin{array}{ll}
a_1 & c \\
a_2 & a_3
\end{array} \Bigg| \ -u\right)  \\ &  =  s^{1}_{0}(-u)^{-1}
    \sum_{t, \tilde{t}, c}  \overline{W}\left(\begin{array}{ll}
a_1 & \alpha \\
c & t
\end{array} \Bigg| \  u - \gamma\right)  \, \overline{W}\left(\begin{array}{ll}
\alpha & d \\
t & a_3
\end{array} \Bigg| \  u  \right) \overline{W}\left(\begin{array}{ll}
\tilde{t} & t \\
a_2 & a_3
\end{array} \Bigg| \  - u \right)  \, \overline{W}\left(\begin{array}{ll}
a_1 & c \\
\tilde{t} & t
\end{array} \Bigg| \  -u + \gamma  \right) \\
& \propto  \sum_{t}  \overline{W}\left(\begin{array}{ll}
\alpha & d \\
t & a_3
\end{array} \Bigg| \  u  \right) \overline{W}\left(\begin{array}{ll}
\alpha & t \\
a_2 & a_3
\end{array} \Bigg| \  - u \right) \propto \delta_{d,a_2}
\end{split}
\end{equation}
where we used unitarity for $\overline{W}^{1,1}$  in the 3$^{\rm rd}$ line to replace the two weights with spectral parameter $u- \gamma$ and $-u + \gamma$ with a proportionality factor. We again used unitarity in 4$^{\rm th}$ line to finally get the result. The proportionality factor can be exactly calculated using Eq. \eqref{eq:YBRSOS} and it is 
\begin{equation}
    s_0^1(-u)^{-1}  \frac{\sin (\gamma - u) \sin(\gamma + u)}{\sin^2 \gamma} \frac{\sin(2 \gamma - u) \sin u}{\sin^2 \gamma}    = - \frac{\sin(2 \gamma - u) \sin (\gamma + u) \sin (\gamma - u)}{\sin^{3} \gamma} \, .
\end{equation}
Using these weights, we can construct generalized transfer matrices, 
\begin{equation}
   \bra{a}\overline{T}^{p_1,p_2}(u)\ket{b} = \prod_{j = 1}^{2L} \overline{W}^{p_1,p_2}\left(\begin{array}{ll}
b_{j} & b_{j+1} \\
a_j & a_{j+1}
\end{array} \Bigg| \ u\right) \, ,
\end{equation}
which satisfy the following relations 
\begin{equation}\label{eq:KP-fusion-tmatrices}
    \overline{T}^{p_1,p_2}_0\overline{T}^{p_1,1}_{p_2} = f^{p_1}_{p_2} \overline{T}^{p_1,p_2-1}_0 + f^{p_1}_{p_2-1}\overline{T}^{p_1,p_2+1}_0 \, , 
\end{equation}
where 
\begin{equation}
    \overline{T}^{p_1,p_2}_k = \overline{T}^{p_1,p_2}(u  + k \gamma) \, , \quad f^{p_1}_{p_2} = [s^{p_1}_{p_2}]^{2L}   \, . 
\end{equation}
We can finally discuss how to relate the fused transfer matrices of this section, with the ones used in \autoref{sec:fused-Tmat-Th}. Starting from  the form of Boltzmann weights for the two set of weights for the simple transfer matrix, and following the constructions, it is not hard to see that 
\begin{equation}
\begin{split}
   T^{(1)}_{[0]} =  T(u) & = U_{h}^{-1} \overline{T}^{1,1}(u) U_{h} \, , \\
    U_h \ket{a_0, \ldots a_{2L - 1}} & =  \left(\prod_{i =0 }^{2L - 1}  \frac{(-1)^{\frac{a_i}{2}}}{\sqrt{\gf_{a_i}}}\right)\ket{a_0, \ldots a_{2L - 1}} \, .
\end{split}
\end{equation}
For the higher fusion transfer matrices, using the two fusions in Eq. \eqref{eq:fusion-Th} and \eqref{eq:KP-fusion-tmatrices}, we finally find
\begin{equation}\label{eq:rel-Th-KP}
    T^{(k)}_{[\frac{k-1}{2}]} =  U_{h}^{-1} \overline{T}^{1,k}(u) U_{h} \, . 
\end{equation}

\newpage

\section{Technical issues related with the spectral parameters}\label{StripsTMH}

\subsection{Defect identifications for the defect Hamiltonian and transfer matrices in the direct channel}\label{app:def_Ham-tmat}

We discuss in this appendix the potential difference between the low-energy limits of the defect Hamiltonian and the defect transfer matrix as the bulk spectral parameter $u$ is varied  for a fixed $\tu$ (or $\vbu$ is varied for fixed $v_I$ in other notations).

Let us consider the largest eigenvalue of the row to row transfer matrix in the direct channel as function of the bulk parameter $\vbu$ and the defect parameter $\vd$. The Hamiltonian derived from this transfer matrix depends only on $v_I$, and the ordering of eigen-energies is fixed once this parameter is set. Meanwhile, one can easily see  that, as $\vbu$ is varied,  the ordering of the transfer matrix eigenvalues (recall that, by integrability, the eigenvectors do not depend on $\vbu$) may change, so that the correspondence between largest eigenvalues and low-lying energies is affected. This is clear in the context of bulk transitions, e.g. ferromagnetic vs. antiferromagnetic transition, where by varying $\vbu$ we may find that the dominant eigenvalue of the transfer matrix is associated with an eigenstate which is the  ground-state of $-H$ instead of $H$. In this paper, however, we keep $|\Im \vbu|< \gamma/2$ (the Hamiltonian is obtained by sending $i\vbu$ to $-{\gamma\over 2}^+$) so that the underlying bulk theory is still given by the minimal model of A type.  

\begin{figure}[htb]
\begin{center}
\begin{subfigure}{.48\textwidth}
\includegraphics[width= \linewidth]{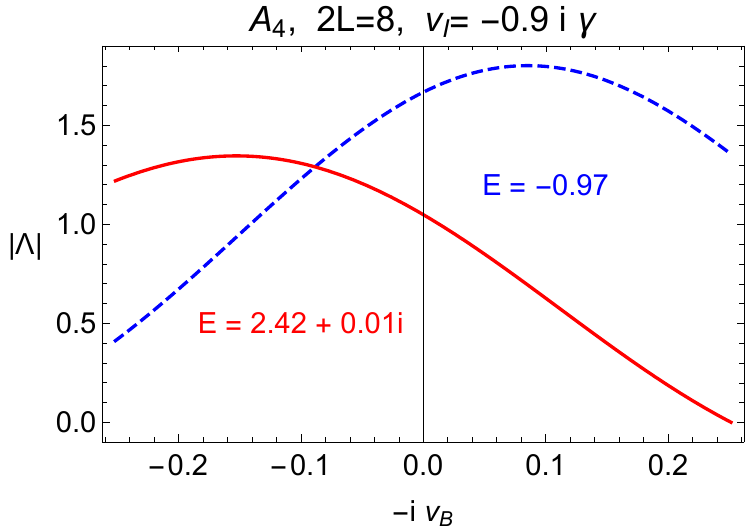}
\caption{}
\end{subfigure} \begin{subfigure}{.48\textwidth}
\includegraphics[width=\linewidth]{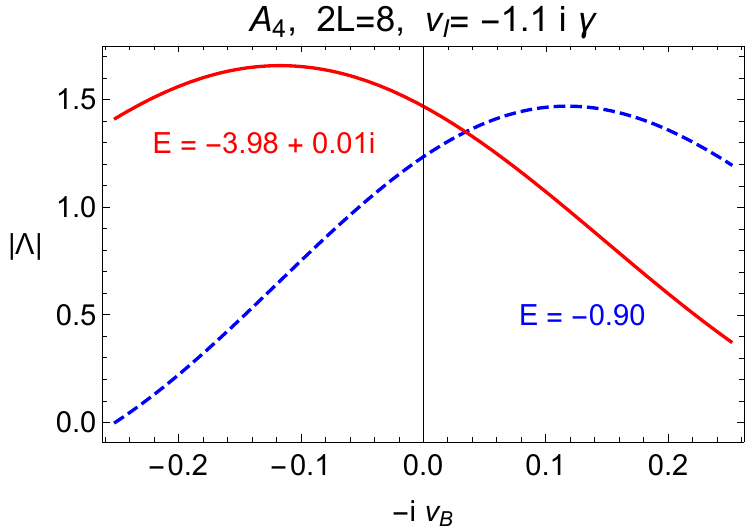}
\caption{}
\end{subfigure}\\
\begin{subfigure}{.48\textwidth}
\includegraphics[width= \linewidth]{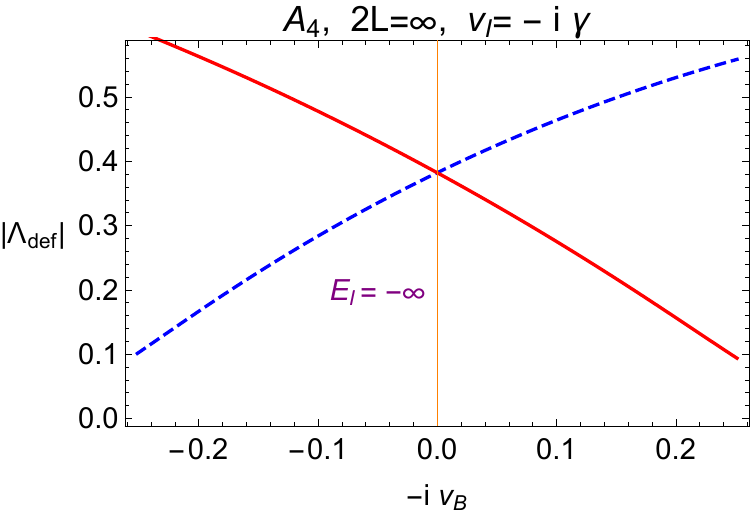}
\caption{}
\end{subfigure}
\caption{ Distinction between phase transitions for the Hamiltonian and transfer matrix. For fixed values of the defect parameter, the effective low-energy Hamiltonian description does not always correspond to the dominant eigenvalues of the transfer matrix, as the latter depend also  on the bulk parameter. But it does at the isotropic point $\vbu=0$ for sufficiently large system-size. Here the data is for the $A_4$ model so $\gamma={\pi\over 5}$.}
\label{TransitionImp}
\end{center}
\end{figure}

We start with a case discussed in  figure \ref{TransitionImp}-a), where  we took a finite system of size $2L=8$ for the $A_4$ model, and an  impurity parameter  $\vd=-0.9 ~i \gamma$.

We consider first the  transfer-matrix eigenvalues. The dashed blue line is obtained by taking the largest eigenvalue at the isotropic point $\vbu=0$  and then continuing analytically to other values of $\vbu$ in the ferromagnetic regime. Instead,  the red line is obtained by taking the largest eigenvalue at  $-i\vbu$ to $-{\gamma\over 2}^+$, and performing a similar continuation. Observe the presence of a crossing. By studying the same phenomenon in higher sizes (note that the position of the crossing depends on $L$), we can ascertain that, to the right of the crossing, the state corresponding to the blue curve does not belong to the scaling limit defined by low-energy excitations over the state from the red curve, and conversely to the left of the crossing. 

As for  the  Hamiltonian, it  is not hermitian (in finite system size), and eigen-energies may have imaginary parts, that tend to vanish as the size is increased.   On the figure, we wrote down  in blue and red  the eigen-energies  obtained by taking the logarithmic derivatives (at  $i\vbu$ to $-{\gamma\over 2}^+$ ) of the transfer matrix eigenvalues represented on the blue and red curves. 

We see on this figure \ref{TransitionImp}-a) that the dominant eigenvalue corresponds to the ground-state of the Hamiltonian (the lowest eigenenergy) only in the right most portion. When $-i\vbu$ is  sufficiently negative however, this is not true any longer.

Now, extending the result for the Hamiltonian, we checked that the blue line corresponds, for the transfer matrix also,  to the ground-state of the system with identity defect in the whole region to the right of the crossing. The red line, which, recall, does not belong to the scaling limit to the right of the crossing, corresponds, on the other hand, to the ground-state for $(2,1)$ defect on the left of the crossing. 

Hence, for the chosen  value of $\vd$, we see that,  depending on the bulk parameter $\vbu$, the transfer matrix  provides  a realization  of either of these two defects.  In  the vicinity of $\vbu=0$ - that is, the isotropic point -  we get  the identity defect, just like we did for the Hamiltonian. This observation extends to all other cases considered in the bulk of this paper. 
In figure \ref{TransitionImp}-b), we set $\vd=-1.1~i \gamma$ instead. In this case, the blue and red lines play reversed roles, but once again the identification of the defect carried out for the Hamiltonian applies  in the vicinity of the isotropic point $\vbu=0$  (in this case, we get the $(2,1)$ defect).

Results in  figures a) and b) are general: Hamiltonian and transfer matrix in the vicinity of the isotropic point $\vbu=0$ give rise, in the scaling limit,  to the same defect for the same $\vd$.  The level crossing between transfer-matrix eigenvalues may occur at either smaller or larger values of $-i \vbu$, depending of the value of $\vd$ itself. At exact point $\vd = \pm i \gamma $, the Hamiltonian becomes singular (see equations (\ref{eq:def_Ham_simp} - \ref{eq:f-def-ham})), which marks its phase transition. This is also a phase transition point for the transfer-matrix at the isotropic point, figure \ref{TransitionImp}-c). However, to see this unambiguously one has to take the large system-size limit to clean eigenvalues from finite-size contributions beyond $O(1)$.

It may sound a bit strange that while we discuss  defect phase transitions, it is the bulk parameter that is being varied. We repeat that, as far as the Hamiltonian is concerned,  $\vbu$ plays no role as long as it leads to the correct bulk theory. As for the transfer matrices, equations (\ref{bulk}) and (\ref{EigImpurityScaled}) for the identity defect show that the $O(1)$ correction  contains a non-universal term $e^{(1)}(\vbu+\vd)$ that depends only on $\vbu+\vd$, and thus allows the tradeoff between the two parameters to assess the phase transition. The competing states also have a non-universal contribution of the same type, say $e^{(p-2)}(\vbu+\vd\pm \pi \im /2)$, where the correct sign is chosen so that $|\Im (\vbu+\vd)\pm \pi /2| < p \gamma/2$. It is the crossover between these $O(1)$ corrections that triggers the transitions. In fact, the transition extends from points $\vbu+\vd=\pm \im \gamma$ to lines in the complex plane defined by the condition 
\begin{equation}
\Re ( ~e^{(p-2)}(\vbu+\vd\pm \pi \im /2)-e^{(1)}(\vbu+\vd)~)=0,
\end{equation}
where in figure \ref{TransitionImp}-c) we denote $\Lambda_{\text{def}}= \exp (e^{(1)}(\vbu+\vd)),~\exp (e^{(p-2)}(\vbu+\vd\pm \pi \im /2)) $, corresponding respectively to  the dashed blue and red lines. 

While transition lines for the transfer matrices are $v_B$ dependent, only at $v_B=0$ do they match the corresponding transitions of the Hamiltonian, leading to parallel lines to the real $\vd$-axis: $|\Im \vd|=\gamma$. This is reminiscent of the distortion occurring in  going from the lattice to the continuum, for the direction of propagation. Only at $\vbu =0$ the statistical model reproduces the partition function with a time-evolution as given by $H^{k,k+1}(\tu)$.

To conclude, Hamiltonian  and transfer-matrix results for the defects coincide when  the latter is at the isotropic point. For other values of $\vbu$, the correspondence of defects  - albeit similar in nature - is more complicated, and doesn't seem worth pursuing here. 

\subsection{On the Hermiticity of the defect Hamiltonians}

\par
In this subsection, we will first discuss what happens for  values of the defect spectral parameter other than $0, \pm \frac{\pi}{2}$, and $\pm {\rm i} \infty$ -  where, as mentioned, lattice defect Hamiltonians are not hermitian, but, if $\tu$ belongs to the proper strip, are still expected to  realize the $(1,1)$, $(2,1)$, and $(1,2)$ defect Hamiltonians in the continuum limit. 

To be more specific, let us consider the Hamiltonian in Eq. \eqref{eq:def_Ham_simp}
\begin{equation}
    H^{k,k+1}(\tu) = H + \frac{1}{\sin \gamma} \left( f(\tu)e_ke_{k+1}  + f(-\tu)e_{k+1}e_k \right) \, . 
\end{equation}
When $\tu$ is purely imaginary, $f(\tu)^{\star} = f(-\tu)$, and $H^{k,k+1}(\tu)$  is Hermitian (this situation was studied in  \cite{tavares2024} in the context of the flow between $(1,2)$ and $(1,1)$ defects). The same holds when  $\Re(\tu) = \pm \frac{\pi}{2}$ (this situation was studied in \cite{tavares2024} in the context of the flow between $(1,2)$ and $(2,1)$ defects).
But when $\tu$ is real and generic (that is, except for  $\tu = 0, \pm  \frac{\pi}{2}$, where $f(\tu) = f(-\tu) = 0$) $f(\tu)^{\star} \neq f(- \tu)$, and $H^{k,k+1}(\tu)$ is not hermitian.
\par
\begin{figure}[H]
    \centering    \includegraphics[width=0.75\linewidth]{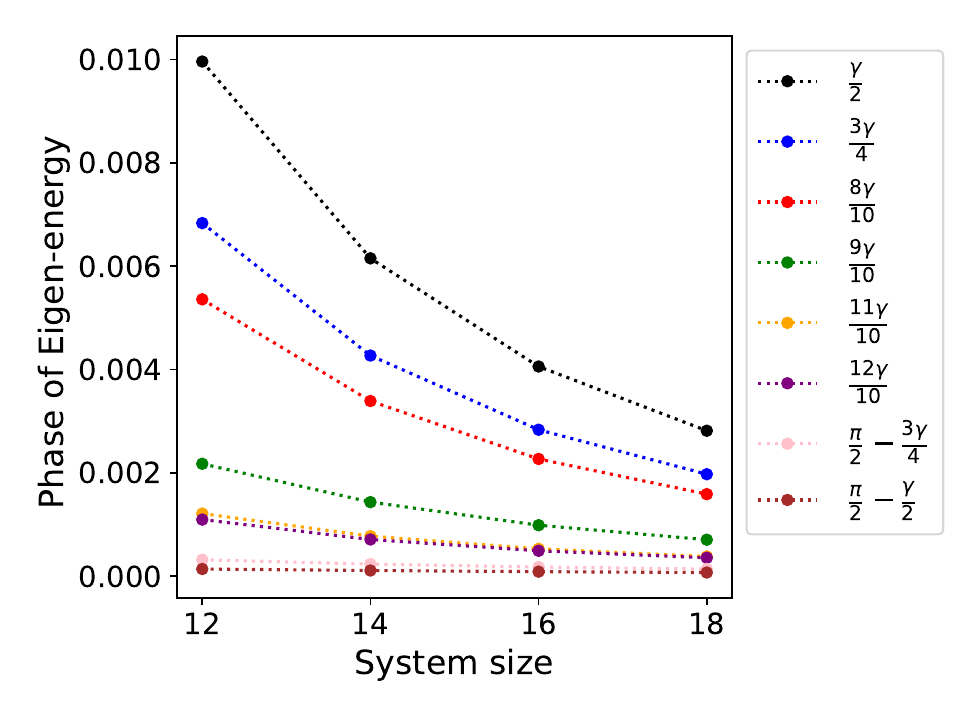}
    \caption{For each impurity parameter, we find a low-lying level for which the eigen-energy is complex. We extract its phase  and plot it with system size for A$_4$ RSOS.  }
    \label{fig:phase-A4-complex-eigval}
\end{figure}
\begin{figure}[H]
    \centering
    \includegraphics[width=0.75\linewidth]{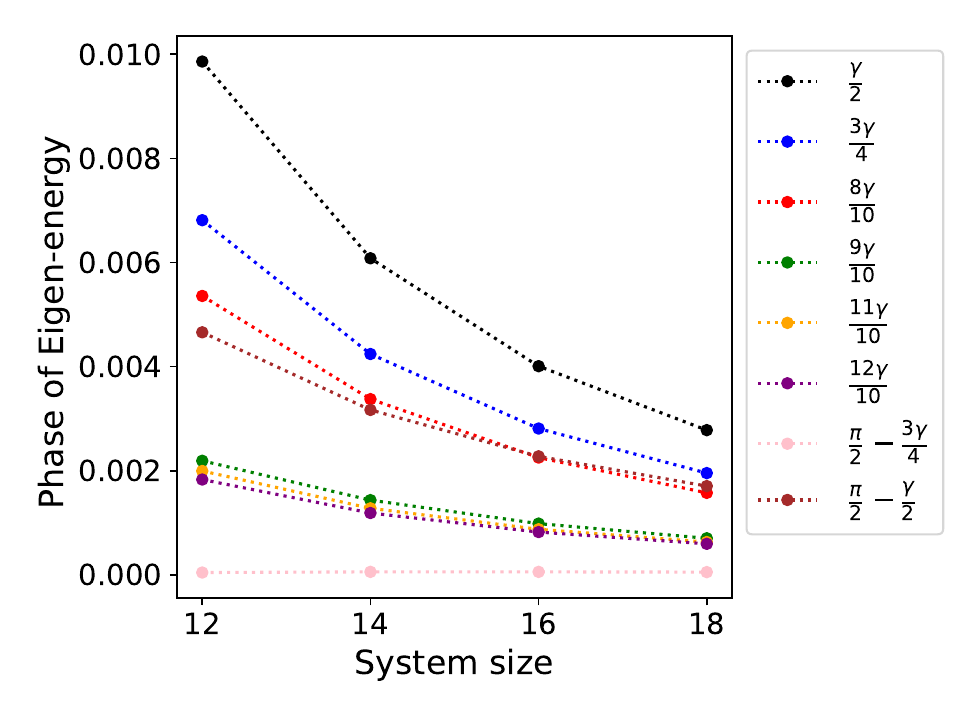}
    \caption{This figure is same as figure \ref{fig:phase-A4-complex-eigval} but for A$_5$ RSOS.}
    \label{fig:phase-A5-complex-eigval}
\end{figure}

In these non-hermitian cases, it turns out that  the eigen-energies are generically complex. We  can then arrange these eigen-energies in ascending order of their real part. We  observe in our problem that, either the low-lying eigen-energies are real, or, if they are complex their phases are small and tend to zero  with increasing system size. This is illustrated with some   examples in figure \ref{fig:phase-A4-complex-eigval} and \ref{fig:phase-A5-complex-eigval} for the  A$_4$ and A$_5$ RSOS models. 


\begin{figure}[H]
    \centering
    \includegraphics[width=0.7\linewidth]{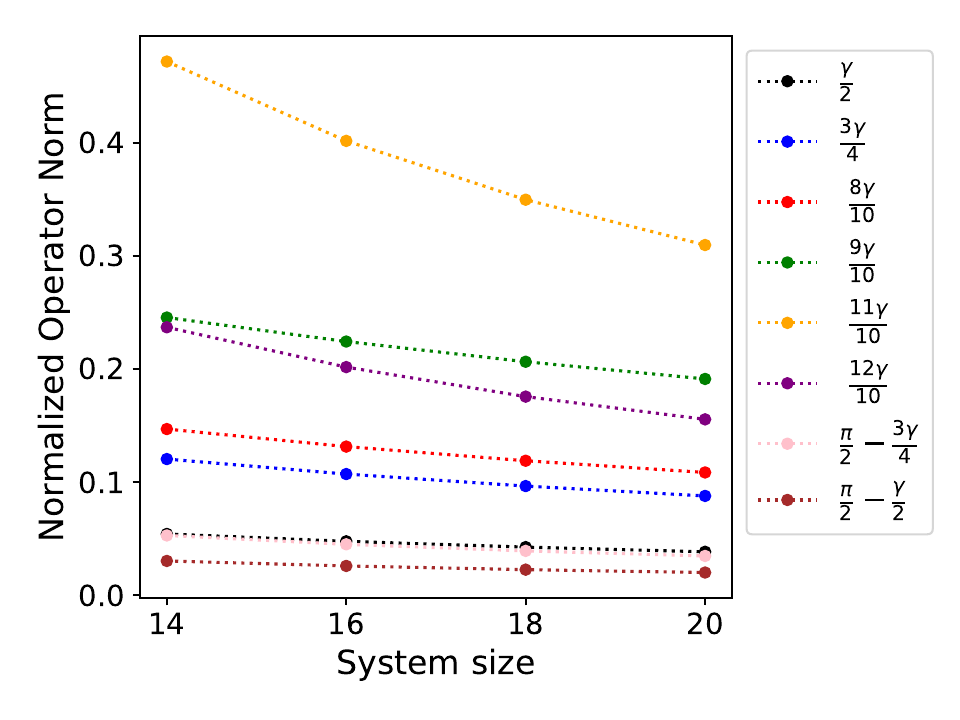}
    \caption{In this figure we show how with increasing system size the normalized operator: $\frac{\lvert H^{k,k+1}(\tu) - H^{k,k+1}(\tu)^{\dagger} \rvert }{\lvert H^{k,k+1}(\tu) + H^{k,k+1}(\tu)^{\dagger}\rvert}$  decreases for different values of spectral parameter between $0$ and $\frac{\pi}{2}$ for A$_4$ RSOS model.}
    \label{fig:non-herm-A4}
\end{figure}

\begin{figure}[H]
    \centering
    \includegraphics[width=0.7\linewidth]{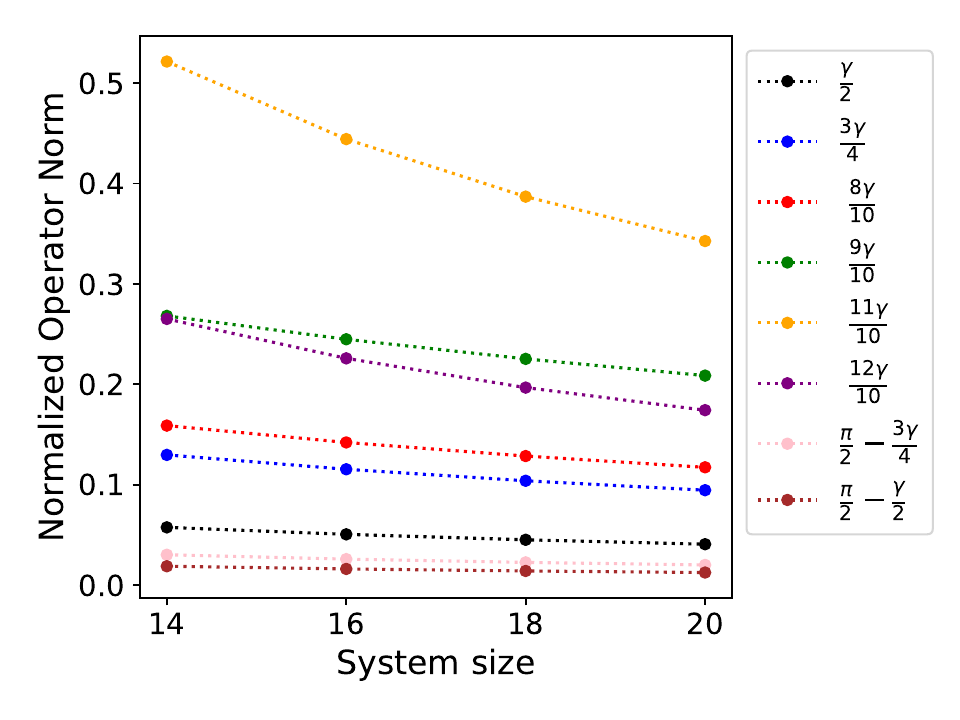}
    \caption{This figure is same as figure \ref{fig:non-herm-A4} but for A$_5$ RSOS.}
    \label{fig:non-herm-A5}
\end{figure}

We can more generally   form some simple quantities to evaluate  the ``non-hermiticity'' of the Hamiltonians, and see that it decreases with  increasing system size. See the figures \ref{fig:non-herm-A4} and \ref{fig:non-herm-A5} for some evidences for this

Next, it is interesting to see what happens for the identification of the defect as we move $v_I$ through the strips we identified with the Bethe-ansatz. 
\begin{figure}[H]
    \centering
    \includegraphics[width=0.7\linewidth]{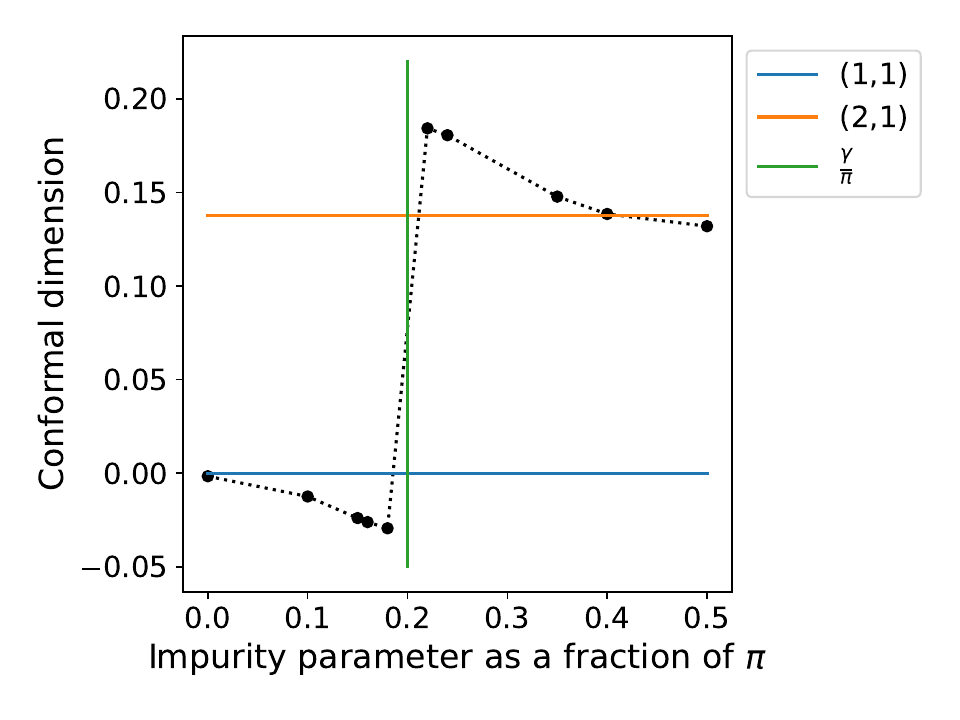}
    \caption{ Here we study the ground state of Hamiltonian as the  impurity parameter $v_I$ is varied in $[0,\frac{\pi}{2}]$ and calculate the corresponding conformal dimension from the real part of the eigen-energy. In this figure we study the A$_4$ model and  systems of size 14 to 20.}
    \label{fig:conf-dim-11-21-A4}
\end{figure}

\begin{figure}[H]
    \centering
    \includegraphics[width=0.7\linewidth]{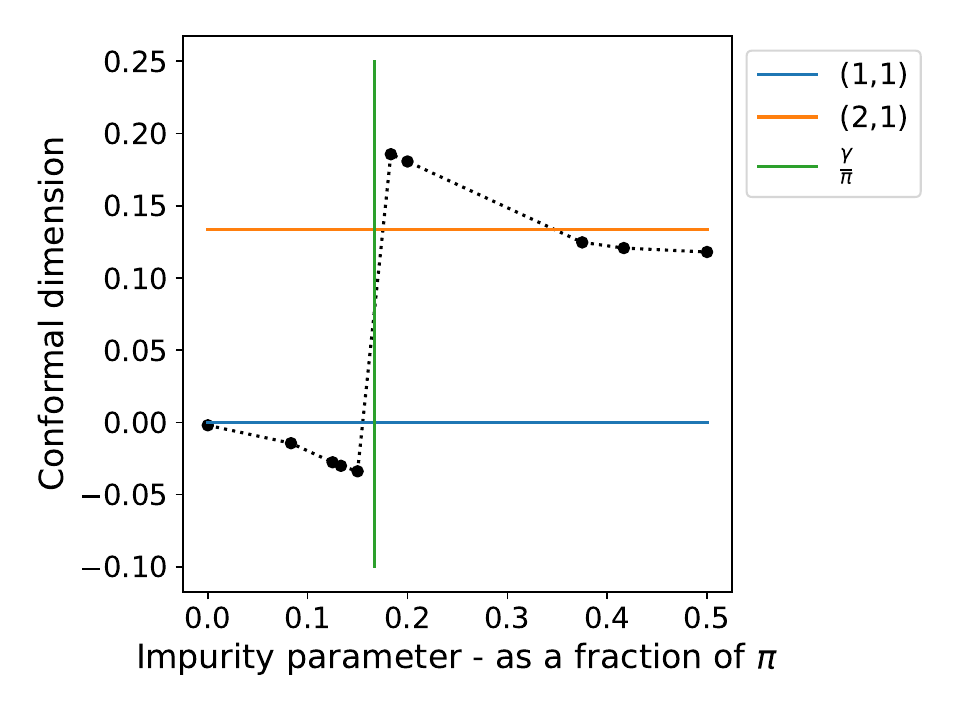}
    \caption{This figure is same as figure \ref{fig:conf-dim-11-21-A4} but for A$_5$ RSOS. }
    \label{fig:conf-dim-11-21-A5}
\end{figure}

\newpage
\section{\texorpdfstring{$Y$}{Lg} operator and D operators in the Ising model }\label{sec:Yop-shao-seib}
 In this appendix, we  show the equivalence between our $Y$ operator for A$_3$ RSOS model, the D operator of \cite{Seiberg:2024gek} (which implements KW duality in the Ising model) and the  symmetry operator of  \cite{Gils_2009, Buican:2017rxc} (which was defined using $F-$ symbol of the su$(2)_3$ category).

The D operator was defined in \cite{Seiberg:2024gek} as 
\begin{equation}
  \text{D} =  \text{e}^{2 \pi i \frac{\Ll}{8}} \frac{1+\varepsilon}{\sqrt{2}} \frac{1-i \sigma^x_\Ll}{\sqrt{2}} \frac{1-i \sigma^z_\Ll \sigma^z_{\Ll-1}}{\sqrt{2}} \cdots \frac{1-i \sigma^z_2 \sigma^z_1}{\sqrt{2}} \frac{1-i \sigma^x_1}{\sqrt{2}} \, ,  
\end{equation}
which by using $\varepsilon = \prod_{i = 1}^{\Ll} \sigma^x_i$ can  also be written as 
\begin{equation}
    \text{D} = \text{D}_1 + \text{D}_2 \, , 
\end{equation}
where 
\begin{equation}
    \text{D}_1 =  \frac{\text{e}^{2 \pi i \frac{\Ll}{8}}}{\sqrt{2}} \frac{1-i \sigma^x_\Ll}{\sqrt{2}} \frac{1-i \sigma^z_\Ll \sigma^z_{\Ll-1}}{\sqrt{2}} \cdots \frac{1-i \sigma^z_2 \sigma^z_1}{\sqrt{2}} \frac{1-i \sigma^x_1}{\sqrt{2}} \, ,
\end{equation}
and 
\begin{equation}
    \text{D}_2 = \frac{\text{e}^{2 \pi i \frac{\Ll}{8}}}{\sqrt{2}} \frac{\sigma^x_\Ll - i  }{\sqrt{2}} \frac{1 + i \sigma^z_\Ll \sigma^z_{\Ll-1}}{\sqrt{2}} \cdots \frac{1 + i \sigma^z_2 \sigma^z_1}{\sqrt{2}} \frac{\sigma^x_1 - i }{\sqrt{2}} \, . 
\end{equation}
Now using the form of TL generators 
\begin{equation}
 e_{2i-1} = \frac{1}{\sqrt{2}} \left( 1 + \sigma_i^x  \right) \,  ,   \quad  e_{2i} = \frac{1}{\sqrt{2}} \left( 1 + \sigma_i^z \sigma_i^{z+1}\right) \,  ,
\end{equation}
we can write 
\begin{equation}
    \begin{split}
        g_{2j - 1} = \text{e}^{ \frac{ 3 \text{i} \pi}{8}} \frac{1}{\sqrt{2}} \left(\sigma^x_j - i \right)  \, , \quad g_{2j} = \text{e}^{ \frac{ - \text{i} \pi}{8}} \frac{1}{\sqrt{2}} \left(1 + i \sigma^z_j \sigma^z_{j+1} \right) \, , \\
 g_{2j - 1}^{-1} = \text{e}^{ \frac{\text{i} \pi}{8}} \frac{1}{\sqrt{2}} \left(1 - i \sigma^x_j  \right)  \, , \quad g_{2j }^{-1} = \text{e}^{ \frac{ \text{i} \pi}{8}} \frac{1}{\sqrt{2}} \left(1 - i \sigma^z_j \sigma^z_{j+1} \right) \, .
    \end{split}
\end{equation}
\footnote{One has to be careful about the branch cuts here, must get $(-q)^{1/2} = \text{e}^{ \frac{ - 3\pi i}{8}}$ and $(-q)^{-1/2} = \text{e}^{ \frac{  3\pi i}{8}}$}Hence, we have 
\begin{equation}
    \text{D}_1 =  \frac{\text{e}^{ \frac{- \pi i}{8}}}{\sqrt{2}} g_{2\Ll -1}g_{2\Ll -2} \ldots g_{1} \,  ,
\end{equation}
and 
\begin{equation}
    \text{D}_2 = \frac{\text{e}^{ \frac{ \pi i}{8}}}{\sqrt{2}}  g_{2\Ll-1}^{-1}g_{2\Ll-2}^{-1} \ldots g_{1}^{-1} \,  . 
\end{equation}
Note that $\uR^{2} = T$, where $T$ is the lattice translation operator for Ising. Further, we can write 
\begin{equation}
\begin{split}
        Y & = (-q)^{-1/2} \,   \, g^{-1}_{1} g^{-1}_{2} \ldots g_{2\Ll - 1}^{-1}  \uR^{-1} 
        +  (-q)^{1/2} \, \uR \, g_{2\Ll -1}   g_{2\Ll -2}  \ldots g_1    \,  , \\ 
 & = \text{e}^{ \frac{ 3\pi i}{8}} g^{-1}_{2\Ll} g^{-1}_{1} \ldots g_{2\Ll - 2}^{-1}  \uR^{-1}  +  \text{e}^{ \frac{ - 3\pi i}{8}}\uR\, g_{2\Ll-2}   g_{2\Ll -3}  \ldots g_1 g_{2\Ll}  \, , \\ 
 & = \text{e}^{ \frac{ 3\pi i}{8}} g^{-1}_{2L} g^{-1}_{1} \ldots g_{2L - 2}^{-1}  \uR^{-1}  +  \text{e}^{ \frac{ - 3\pi i}{8}} g_{2L-1}   g_{2L -2}  \ldots  g_{1} \uR \, .
\end{split}
    \end{equation}
Similarly, we can write $\overline{Y}$ as 
\begin{equation}
    \overline{Y} =  \text{e}^{ \frac{ 3\pi i}{8}} g_{2L-1}^{-1} g_{2L-2}^{-1} \ldots g_{1}^{-1} \uR +  \text{e}^{ \frac{  - 3\pi i}{8}} g_{2L} g_1 \ldots g_{2L-2} \uR^{-1} \, . 
\end{equation}
For diagonal models we know that $Y = \overline{Y}$, hence we can write \begin{equation} \label{eq:YtoD}
\begin{split}
       & Y \uR^{-1} = \frac{\text{e}^{i \pi /4} }{\sqrt{2}} Y \uR^{-1} + \frac{\text{e}^{-i \pi /4} }{\sqrt{2}} \overline{Y} \uR^{-1}  \\ 
       &  \implies Y \uR^{-1} = \frac{\text{e}^{\frac{5 \pi i}  {8}}}{\sqrt{2}}g_{2L}^{-1}g_1^{-1} \ldots g_{2L-2}^{-1} T^{-1} + \frac{\text{e}^{\frac{- \pi i  }{8}}}{\sqrt{2}} g_{2L - 1}g_{2L-2} \ldots g_1  \\ 
      & + \frac{\text{e}^{\frac{ \pi i  }{8}}}{\sqrt{2}} g_{2L-1}^{-1}g_{2L-2}^{-1} \ldots g_{1}^{-1} + \frac{\text{e}^{\frac{ - 5 \pi i }{8}}}{\sqrt{2}} g_{2L} g_1 \ldots g_{2L-2} T^{-1} \, .  
\end{split}
\end{equation}
Now, as 
\begin{equation}\label{eq:prod-of-braid}
    q = \left( g_{2L} g_1 \ldots g_{2L-2} \right) \left( g_{2L-2} g_{2L-3} \ldots g_1 g_{2L} \right) \, ,  
\end{equation}
the first and fourth term in Equation \eqref{eq:YtoD} cancel out and we have 
\begin{equation}
    Y \uR^{-1} = \frac{\text{e}^{\frac{- \pi i  }{8}}}{\sqrt{2}} g_{2L - 1}g_{2L-2} \ldots g_1  +  \frac{\text{e}^{\frac{ \pi i  }{8}}}{\sqrt{2}} g_{2L-1}^{-1}g_{2L-2}^{-1} \ldots g_{1}^{-1}  = \text{D}_1 + \text{D}_2 = \text{D} \, . 
\end{equation}
To prove the relation in Equation \eqref{eq:prod-of-braid} above, one just has to write the braid operators in terms of Pauli operators. Note, this relation is only valid for TFI, and not a general RSOS model.
\newpage

\section{Topological defect conditions in Aasen-Mong-Fendley}\label{sec:equiv_def_com_F_app}
In this appendix, we will show the equivalence of the conditions for topological invariance discussed in the present  paper with the \textit{defect commutation relation} introduced in the work of Aasen, Fendley, and Mong (AFM) \cite{Aasen:2020jwb}.

In AFM, the two defect commutation relations are given below.

\begin{figure}[H]
    \centering
    \begin{tikzpicture}
        
        \fill[green!10] (0,0) -- (1,-1) -- (1,-1.5) -- (0,-0.5) --cycle;
        \fill[green!10] (1,-1) -- (2,0) -- (2,-0.5) -- (1,-1.5) -- cycle;

        \fill[green!10] (5,0.5) -- (6,-0.5) -- (6,0) -- (5,1) -- cycle ;
        \fill[green!10] (4,-0.5) -- (5,0.5) -- (5,1) -- (4,0) -- cycle ;

        \draw[black, thick] (0,0) -- (1,1) -- (2,0) -- (1,-1) -- cycle;
        \draw[black, thick] (0,0) -- (1,-1) -- (1,-1.5) -- (0,-0.5) --cycle;
        \draw[black, thick] (1,-1) -- (2,0) -- (2,-0.5) -- (1,-1.5) -- cycle;

        \draw[black, thick] (4,-0.5) -- (5,0.5) -- (6,-0.5) -- (5,-1.5) -- cycle;
        \draw[black, thick] (5,0.5) -- (6,-0.5) -- (6,0) -- (5,1) -- cycle ;
        \draw[black, thick] (4,-0.5) -- (5,0.5) -- (5,1) -- (4,0) -- cycle ;
    \node at (1,-0.6) {$g$};
    \node at (1,-1.8) {$f$};
    \node at (-0.2,0) {$d$};
    \node at (-0.2,-0.5) {$e$};
    \node at (2.2,0) {$b$};
    \node at (2.2,-0.5) {$a$};
    \node at (1,1.2) {$c$} ;
    \node at (1,0) {$\bar{u}$} ; 
    \node at (5,0.2) {$g$};
    \node at (5,-1.8) {$f$};
    \node at (3.8,0) {$d$};
    \node at (3.8,-0.5) {$e$};
    \node at (6.2,0) {$b$};
    \node at (6.2,-0.5) {$a$};
    \node at (5,1.2) {$c$} ;
    \node at (5,-0.5) {$\bar{u}$} ; 
    \node at (-1.0,-0.25) {$
    \sum_g$} ;
    \node at (3.0,-0.25) {$= \sum_g$} ;
    \filldraw[black] (1,-1) circle (2pt) ; 
    \filldraw[black] (5,0.5) circle (2pt) ; 

\draw[red, thick]  (1.1,-0.9) arc (0 : 180 : 0.1)  ; 
\draw[red, thick]  (0.85,-1.4) arc (180 : 90 : 0.15)  ; 
\draw[red, thick]  (1.9,-0.64) arc (225 : 90 : 0.15)  ; 
\draw[red, thick]  (5.1,-1.4) arc (0 : 180 : 0.1)  ; 
\draw[red, thick]  (5,0.65) arc (90 : 225 : 0.15)  ; 
\draw[red, thick]  (6,-0.3) arc (90 : 135 : 0.2)  ; 

    \end{tikzpicture}
    \caption{The first defect commutation relation in AFM. Black dots on a site $a$ indicates a multiplicative factor of $g_a$.}
    \label{fig:def-com-rel-1}

\end{figure}
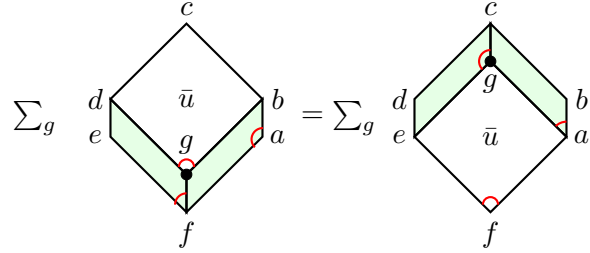

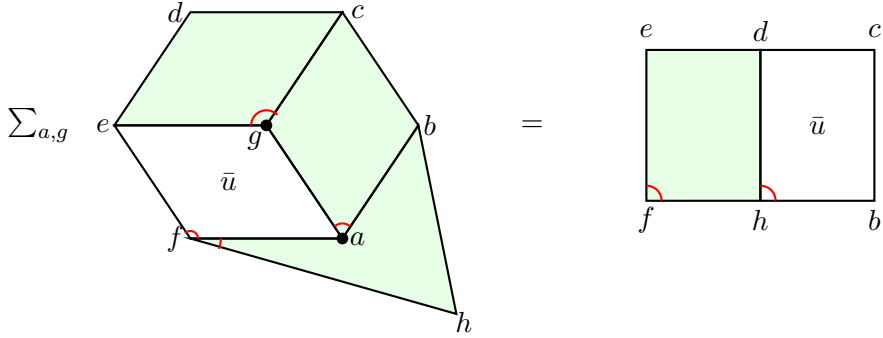
\begin{figure}[H]
    \centering
\begin{tikzpicture}

    \fill[green!10] (6,0) -- (8,0) -- (9,1.5) -- (7,1.5) -- cycle;
     \fill[green!10] (8,0) -- (9,1.5) -- (10,0) -- (9,-1.5) -- cycle;
     \fill[green!10] (7,-1.5) -- (9,-1.5) -- (10,0) -- (10.5, -2.5) -- cycle ; 

\draw[black, thick] (6,0) -- (8,0) -- (9,1.5) -- (7,1.5) -- cycle;
\draw[black, thick] (6,0) -- (8,0) -- (9,-1.5) -- (7,-1.5) -- cycle;
\draw[black, thick] (8,0) -- (9,1.5) -- (10,0) -- (9,-1.5) -- cycle;
\draw[black, thick] (7,-1.5) -- (9,-1.5) -- (10,0) -- (10.5, -2.5) -- cycle ; 
    
\node[] at (5,0) {$ \sum_{a,g} $} ;
\node[] at (5.85,0) {$e$};
\node[] at (7.85,-0.2) {$g$};

\node[] at (10.15,0) {$b$};

\node[] at (6.8,1.5) {$d$};
\node[] at (6.8,-1.5) {$f$};
\node[] at (9.2,1.5) {$c$};
\node[] at (9.2,-1.5) {$a$};
\node[] at (10.6,-2.6) {$h$};

\node[] at (11.5,0) {$ = $} ;
    \fill[green!10] (13, - 1) -- (13,1) -- (14.5,1) -- (14.5,-1 ) -- cycle;
    \draw[black, thick]  (13, - 1) -- (13,1) -- (14.5,1) -- (14.5,-1 ) -- cycle;
    \draw[black, thick]  (14.5, - 1) -- (14.5,1) -- (16,1) -- (16,-1 ) -- cycle;

\node[] at (13,-1.25) {$f$} ; 
\node[] at (14.5,-1.25) {$h$} ; 
\node[] at (16,-1.25) {$b$} ; 
\node[] at (13,1.25) {$e$} ; 
\node[] at (14.5,1.25) {$d$} ; 
\node[] at (16,1.25) {$c$} ; 
\node at (7.5,-0.75) {$\bar{u}$} ;

\node at (15.25,0) {$\bar{u}$} ;

\draw[red, thick]  (13,-0.8) arc (90 : 0 : 0.2)  ; 
\draw[red, thick]  (14.5,-0.8) arc (90 : 0 : 0.2)  ; 
\draw[red, thick]  (7.1,-1.5) arc (0 : 135 : 0.1)  ; 
\draw[red, thick]  (7.4,-1.5) arc (0 : -20 : 0.4)  ; 
\draw[red, thick]  (7.8,0) arc (180 : 44 : 0.2)  ; 
\draw[red, thick]  (9,-1.3) arc (90 : 135 : 0.15)  ; 
\draw[red, thick]  (9,-1.3) arc (90 : 45 : 0.15)  ; 

    \filldraw[black] (9,-1.5) circle (2pt) ; 
    \filldraw[black] (8,0) circle (2pt) ;

\end{tikzpicture}  
\caption{The second defect commutation relation in AFM.}
    \label{fig:def-com-rel-2}
\end{figure}

We have introduced  the notation $\bar{u}$ for the spectral parameter to distinguish  conventions  in our work from those in AFM. In the figure below, we show how the weights in the two works are related. 
\begin{figure}[H]
    \centering
\begin{tikzpicture}
    \draw[black, thick] (0,0) -- (0,2) -- (2,2) -- (2,0) -- cycle ;
    \draw[black, thick] (4,0) -- (4,2) -- (6,2) -- (6,0) -- cycle ;
    \node at (3,1) {$=$} ; 
    \node[] at (1,1) {$\bar{u}$} ; 
    \node[] at (5,1) {$u$} ; 
    \node[] at (-1,1) {$\sqrt{ \gf_a \gf_c}$} ; 
    \node [] at (0,-0.2) {$a$} ;
    \node [] at (2,-0.2) {$b$} ;
    \node [] at (4,-0.2) {$a$} ;
    \node [] at (6,-0.2) {$b$} ;
    \node [] at (0,2.2) {$d$} ;
    \node [] at (2,2.2) {$c$} ;
    \node [] at (4,2.2) {$d$} ;
    \node [] at (6,2.2) {$c$} ;

\draw[red, thick]  (0.2,0) arc (0 : 90 : 0.2)  ; 
\draw[red, thick]  (4.2,0) arc (0 : 90 : 0.2)  ;

\end{tikzpicture}  
\caption{When a face has spectral parameter $\bar{u}$, the face's weight is according to the convention in AFM. Above is how the weight compares with the weight convention that we use.}
    \label{fig:AFM-our-convention}
\end{figure}
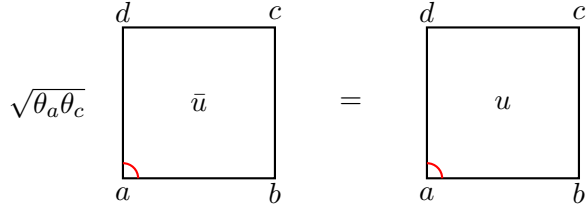

In what follows in this section, we will show that the two properties we listed above, our condition for topological invariance and defect commutation relation, are equivalent in the following sense. 
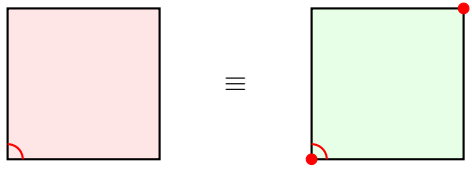
\begin{figure}[H]
    \centering
\begin{tikzpicture}
    \fill[red!10] (0,0) -- (2,0)-- (2,2) -- (0,2) -- cycle ; 
    \fill[green!10] (4,0) -- (6,0)-- (6,2) -- (4,2) -- cycle ; 
    \draw[black, thick] (0,0) -- (2,0)-- (2,2) -- (0,2) -- cycle ; 
    \draw[black, thick] (4,0) -- (6,0)-- (6,2) -- (4,2) -- cycle ; 
    \draw[red, thick]  (0.2,0) arc (0 : 90 : 0.2)  ; 
    \draw[red, thick]  (4.2,0) arc (0 : 90 : 0.2)  ; 
    \filldraw[red] (4,0) circle (2pt) ; 
    \filldraw[red] (6,2) circle (2pt) ; 
    \node at (3,1) {$\equiv$} ; 
\end{tikzpicture}    
    \caption{The way defect faces in two different conventions are related.}
    \label{fig:eq-face-cond}
\end{figure}
If a green face satisfies the defect commutation relations in figures \ref{fig:def-com-rel-1} and \ref{fig:def-com-rel-2}, then the red face which is defined as in figure \ref{fig:eq-face-cond} satisfies the conditions for topological invariance in figures \ref{fig:first_cond_top_inv} and \ref{fig:top-def-rel-2}. Further, if a red face satisfies the conditions in figures  \ref{fig:first_cond_top_inv} and \ref{fig:top-def-rel-2}, then a green face which satisfies the condition in figure \ref{fig:eq-face-cond} satisfies the relations in figures \ref{fig:def-com-rel-1} and \ref{fig:def-com-rel-2}. A necessary relation that we require to prove the equivalence is the crossing relation for defect face, by which we mean the following two configurations have the same Boltzmann weights. 
\begin{figure}[H]
    \centering
\begin{tikzpicture}
    \fill[blue!10] (0,0) -- (0,2) -- (2,2) -- (2,0) -- cycle ; 
    \fill[red!10] (6,0) -- (6,2) -- (8,2) -- (8,0) -- cycle ; 

    \draw[black, thick] (0,0) -- (0,2) -- (2,2) -- (2,0) -- cycle ; 
    \node at (4,1) {\Large $= \sqrt{ \frac{\gf_e \gf_d}{\gf_a \gf_c}}$ } ; 
    \draw[black, thick] (6,0) -- (6,2) -- (8,2) -- (8,0) -- cycle ; 
    \draw[red, thick]  (0.2,0) arc (0 : 90 : 0.2)  ; 
    \draw[red, thick]  (6.2,0) arc (0 : 90 : 0.2)  ; 

    \node [] at (0,-0.2) {$d$} ;
    \node [] at (2,-0.2) {$c$} ;
    \node [] at (0,2.2) {$a$} ;
    \node [] at (2,2.2) {$e$} ;

    \node [] at (6,-0.2) {$a$} ;
    \node [] at (8,-0.2) {$d$} ;
    \node [] at (6,2.2) {$e$} ;
    \node [] at (8,2.2) {$c$} ;

\end{tikzpicture}  
\\
\vspace{0.1cm}
\begin{tikzpicture}

    \fill[blue!10] (0,0) -- (0,2) -- (2,2) -- (2,0) -- cycle ; 
    \fill[red!10] (4,0) -- (4,2) -- (6,2) -- (6,0) -- cycle ; 
        \draw[black, thick]  (0,0) -- (0,2) -- (2,2) -- (2,0) -- cycle ;
        \draw[black, thick]  (4,0) -- (4,2) -- (6,2) -- (6,0) -- cycle ; 
            \draw[red, thick]  (0.2,0) arc (0 : 90 : 0.2)  ; 
    \draw[red, thick]  (4.2,0) arc (0 : 90 : 0.2)  ; 

    \node [] at (0,-0.2) {$d$} ;
    \node [] at (2,-0.2) {$c$} ;
    \node [] at (0,2.2) {$a$} ;
    \node [] at (2,2.2) {$e$} ;
    \node [] at (4,-0.2) {$a$} ;
    \node [] at (6,-0.2) {$d$} ;
    \node [] at (4,2.2) {$e$} ;
    \node [] at (6,2.2) {$c$} ;
\node [] at (3,1) { $=$ } ; 
        \filldraw[red] (2,0) circle (1.5pt) ; 
    \filldraw[red] (0,2) circle (1.5pt) ; 
    \filldraw[red] (4,2) circle (1.5pt) ; 
    \filldraw[red] (6,0) circle (1.5pt) ;

\end{tikzpicture}
    \caption{Crossing symmetry for defect face. Both equalities are equivalent.}
    \label{fig:cross-sym-def}
\end{figure}
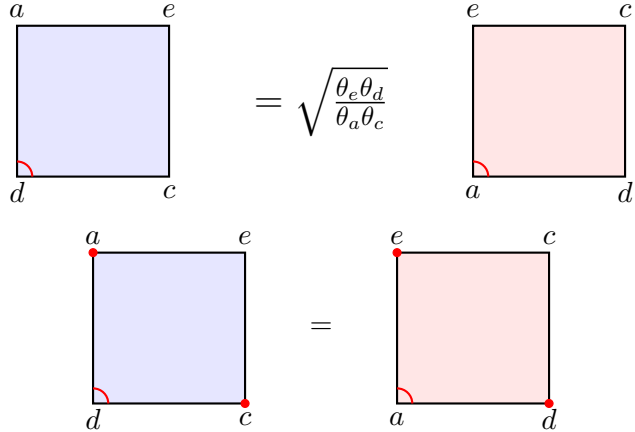
If we take the red defect face to be a face with spectral parameter $\im \infty$, and the blue face has spectral parameter $- \im \infty$, then it satisfies the conditions in figures \ref{fig:top-def-rel-2} and \ref{fig:cross-sym-def} due to unitarity (Eq. \eqref{eq:YB-Eq-face}) and crossing relation - Eq. \eqref{eq:crossing_sym}.

\bigskip

First, we will show that our first condition for topological invariance is equivalent to first defect commutation relation. Note that the following diagrams have the same Boltzmann weights, as each image is the rotated version of the next. 
\begin{figure}[H]
    \centering
\begin{tikzpicture}[scale = 0.8]
    \fill[red!10] (6,0) -- (8,0) -- (9,1.5) -- (7,1.5) -- cycle;
      \fill[red!10] (6,0) -- (8,0) -- (9,-1.5) -- (7,-1.5) -- cycle;
    \draw[black, thick] (6,0) -- (8,0) -- (9,1.5) -- (7,1.5) -- cycle;
    \draw[black, thick] (6,0) -- (8,0) -- (9,-1.5) -- (7,-1.5) -- cycle;
    \draw[black, thick] (8,0) -- (9,1.5) -- (10,0) -- (9,-1.5) -- cycle;
    \node at (5,0) {   $\sum_{g}$} ;

\node at (5.8,0) {$f$} ; 
\node at (6.8,1.5) {$e$} ; 
\node at (9.25,1.5) {$d$} ; 
\node at (6.8,-1.5) {$a$} ; 
\node at (9.25,-1.5) {$b$} ; 
\node at (10.2,0) {$c$} ; 
\node at (7.9,-0.2) {$g$} ; 
\draw[red, thick] (6.3,0) arc (0 : 60 : 0.3) ;
\draw[red, thick] (7.2,-1.5) arc (0 : 120 : 0.2) ;
\draw[red, thick] (8.3,0) arc (0: 60 : 0.3) ;
\draw[red, thick] (8.3,0) arc (0: -60 : 0.3) ;

\node at (9,0) {$u$} ; 
\node at (14.5,-0.75) {$u$} ; 

\node at (11,0) {   $= \sum_{g}$} ;

\fill[red!10] (12,0) -- (13,1.5) -- (14,0) -- (13,-1.5) -- cycle;

\draw[black, thick] (12,0) -- (13,1.5) -- (14,0) -- (13,-1.5) -- cycle;
\fill[red!10] (15,1.5) -- (13,1.5) -- (14,0) -- (16,0) -- cycle;
\draw[black, thick] (15,1.5) -- (13,1.5) -- (14,0) -- (16,0) -- cycle;
\draw[black, thick] (15,-1.5) -- (13,-1.5) -- (14,0) -- (16,0) -- cycle;

\node at (11.8,0) {$a$} ; 
\node at (12.8,1.5) {$f$} ; 
\node at (15.25,1.5) {$e$} ; 
\node at (12.8,-1.5) {$b$} ; 
\node at (15.25,-1.5) {$c$} ; 
\node at (16.2,0) {$d$} ; 
\node at (13.6,-0.2) {$g$} ; 

\draw[red, thick] (14.3,0) arc (0 : -120 : 0.3) ;
\draw[red, thick] (13.3,1.5) arc (0: -60 : 0.3) ;

\draw[red, thick] (12.3,0) arc (0 : 60 : 0.3) ;
\draw[red, thick] (12.3,0) arc (0 : -60 : 0.3) ;

    \fill[red!10] (20,0) -- (21,1.5) -- (22,0) -- (21,-1.5) -- cycle;
    \fill[red!10] (18,0) -- (20,0) -- (21,1.5) -- (19,1.5) -- cycle;
    \draw[black, thick] (18,0) -- (20,0) -- (21,1.5) -- (19,1.5) -- cycle;

\draw[black, thick] (18,0) -- (20,0) -- (21,-1.5) -- (19,-1.5) -- cycle;
\draw[black, thick] (20,0) -- (21,1.5) -- (22,0) -- (21,-1.5) -- cycle;

\node at (17.8,0) {$b$} ; 
\node at (18.8,1.5) {$a$} ; 
\node at (21.25,1.5) {$f$} ; 
\node at (18.8,-1.5) {$c$} ;
\node at (21.25,-1.5) {$d$} ; 
\node at (22.2,0) {$e$} ; 
\node at (19.7,0.2) {$g$} ; 

\draw[red, thick] (19.3,1.5) arc (0 : -120 : 0.3) ;
\draw[red, thick] (19.8,0) arc (180 : 300 : 0.2) ;
\draw[red, thick] (21,1.2) arc (-90 : -120 : 0.3) ;
\draw[red, thick] (21,1.2) arc (-90 : -60 : 0.3) ;

\node at (19.5,-0.75) {$u$} ; 
\node at (17,0) {   $= \sum_{g}$} ;

\end{tikzpicture}
\begin{tikzpicture}
    
\end{tikzpicture}
\begin{tikzpicture}[scale = 0.8]
\draw[black, thick] (6,0) -- (8,0) -- (9,1.5) -- (7,1.5) -- cycle;
    \fill[red!10] (6,0) -- (8,0) -- (9,-1.5) -- (7,-1.5) -- cycle;
    \fill[red!10] (8,0) -- (9,1.5) -- (10,0) -- (9,-1.5) -- cycle;

\draw[black, thick] (6,0) -- (8,0) -- (9,-1.5) -- (7,-1.5) -- cycle;
\draw[black, thick] (8,0) -- (9,1.5) -- (10,0) -- (9,-1.5) -- cycle;

\node at (5.8,0) {$d$} ; 
\node at (6.8,1.5) {$c$} ; 
\node at (9.25,1.5) {$b$} ; 
\node at (6.8,-1.5) {$e$} ; 
\node at (9.25,-1.5) {$f$} ; 
\node at (10.2,0) {$a$} ; 
\node at (7.8,-0.3) {$g$} ; 

\draw[red, thick] (8.7,-1.5) arc (180 : 120 : 0.3) ;
\draw[red, thick] (7.7,0) arc (180 : 60 : 0.3) ;
\draw[red, thick] (9.7,0) arc (180 : 120 : 0.3) ;
\draw[red, thick] (9.7,0) arc (180 : 240 : 0.3) ;

\node at (7.5,0.75) {$u$} ; 
\node at (2.5,0.75) {$u$} ; 
\node at (13,0) {$u$} ;

\fill[red!10] (0,0) -- (1,1.5) -- (2,0) -- (1,-1.5) -- cycle;
\draw[black, thick] (0,0) -- (1,1.5) -- (2,0) -- (1,-1.5) -- cycle;
\fill[red!10] (3,-1.5) -- (1,-1.5) -- (2,0) -- (4,0) -- cycle;
\draw[black, thick] (3,1.5) -- (1,1.5) -- (2,0) -- (4,0) -- cycle;
\draw[black, thick] (3,-1.5) -- (1,-1.5) -- (2,0) -- (4,0) -- cycle;

\node at (-.2,0) {$e$} ; 
\node at (0.8,1.5) {$d$} ; 
\node at (3.25,1.5) {$c$} ; 
\node at (0.8,-1.5) {$f$} ; 
\node at (3.25,-1.5) {$a$} ; 
\node at (4.2,0) {$b$} ; 
\node at (2.2,-0.3) {$g$} ; 
\node at (5,0) {   $= \sum_{g}$} ;
\node at (-1,0) {   $\sum_{g}$} ;

\draw[red, thick] (1,-1.2) arc (90 : 120: 0.3) ;
\draw[red, thick] (1,-1.2) arc (90 : 60: 0.3) ;
\draw[red, thick] (2.2,0) arc (0 : 120 : 0.2) ;
\draw[red, thick] (2.7,-1.5) arc (180 : 60 : 0.3) ;

\node at (11,0) {   $= \sum_{g}$} ;

\fill[red!10] (15,1.5) -- (13,1.5) -- (14,0) -- (16,0) -- cycle;
\fill[red!10] (15,-1.5) -- (13,-1.5) -- (14,0) -- (16,0) -- cycle;
\draw[black, thick] (12,0) -- (13,1.5) -- (14,0) -- (13,-1.5) -- cycle;
\draw[black, thick] (15,1.5) -- (13,1.5) -- (14,0) -- (16,0) -- cycle;
\draw[black, thick] (15,-1.5) -- (13,-1.5) -- (14,0) -- (16,0) -- cycle;

\node at (11.8,0) {$c$} ; 
\node at (12.8,1.5) {$b$} ; 
\node at (15.25,1.5) {$a$} ; 
\node at (12.8,-1.5) {$d$} ; 
\node at (15.25,-1.5) {$e$} ; 
\node at (16.2,0) {$f$} ; 
\node at (14.2,-0.3) {$g$} ;

\draw[red, thick] (13.7,0) arc (180 : 120 : 0.3) ;
\draw[red, thick] (13.7,0) arc (180 : 240 : 0.3) ;
\draw[red, thick] (14.7,1.5) arc (180 : 300: 0.3) ;
\draw[red, thick] (15.7,0) arc (180 : 240: 0.3) ;

\node at (14,2) {$=$};
\end{tikzpicture}

    \caption{The red face here is the defect face, whereas the white face carries spectral parameter $u$.}
    \label{fig:Top-Def-face-1}
\end{figure}
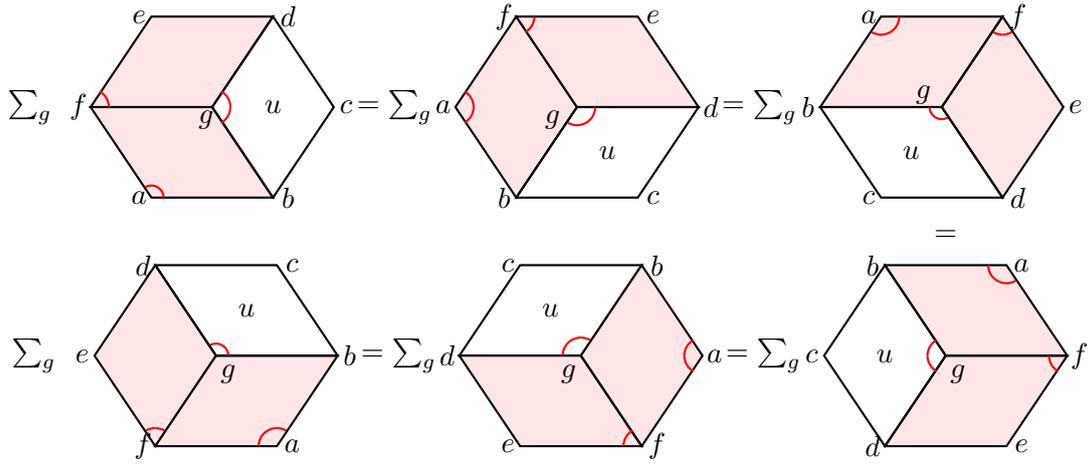

Similarly, the following configurations also have the same Boltzmann weight. 

\begin{figure}[H]
    \centering
\begin{tikzpicture}[scale = 0.8]
\draw[black, thick] (6,0) -- (8,0) -- (9,1.5) -- (7,1.5) -- cycle;
    \fill[red!10] (6,0) -- (8,0) -- (9,-1.5) -- (7,-1.5) -- cycle;
    \fill[red!10] (8,0) -- (9,1.5) -- (10,0) -- (9,-1.5) -- cycle;

\draw[black, thick] (6,0) -- (8,0) -- (9,-1.5) -- (7,-1.5) -- cycle;
\draw[black, thick] (8,0) -- (9,1.5) -- (10,0) -- (9,-1.5) -- cycle;

\node at (5.8,0) {$a$} ; 
\node at (6.8,1.5) {$f$} ; 
\node at (9.25,1.5) {$e$} ; 
\node at (6.8,-1.5) {$b$} ; 
\node at (9.25,-1.5) {$c$} ; 
\node at (10.2,0) {$d$} ; 
\node at (7.8,-0.3) {$g$} ; 
\draw[red, thick] (7.2,1.5) arc (0 : -120 : 0.2) ;
\draw[red, thick] (6.3,0) arc (0 : -60 : 0.3) ;
\draw[red, thick] (8.15,-0.15) arc (-60 : 60 : 0.2) ;
\node at (7.5,0.75) {$u$} ; 
\node at (1,0) {$u$} ; 
\node at (14.5,0.75) {$u$} ;

\draw[black, thick] (0,0) -- (1,1.5) -- (2,0) -- (1,-1.5) -- cycle;
\fill[red!10] (3,1.5) -- (1,1.5) -- (2,0) -- (4,0) -- cycle;
\fill[red!10] (3,-1.5) -- (1,-1.5) -- (2,0) -- (4,0) -- cycle;
\draw[black, thick] (3,1.5) -- (1,1.5) -- (2,0) -- (4,0) -- cycle;
\draw[black, thick] (3,-1.5) -- (1,-1.5) -- (2,0) -- (4,0) -- cycle;

\node at (-.2,0) {$f$} ; 
\node at (0.8,1.5) {$e$} ; 
\node at (3.25,1.5) {$d$} ; 
\node at (0.8,-1.5) {$a$} ; 
\node at (3.25,-1.5) {$b$} ; 
\node at (4.2,0) {$c$} ; 
\node at (2.2,-0.3) {$g$} ; 
\node at (5,0) {   $= \sum_{g}$} ;
\node at (-1,0) {   $\sum_{g}$} ;

\draw[red, thick] (0.15,-0.15) arc (-60 : 60 : 0.2) ;
\draw[red, thick] (2.2,0) arc (0 : 120 : 0.2) ;
\draw[red, thick] (1.2,-1.5) arc (0 : 60 : 0.2) ;

\node at (11,0) {   $= \sum_{g}$} ;

\fill[red!10] (12,0) -- (13,1.5) -- (14,0) -- (13,-1.5) -- cycle;
\fill[red!10] (15,-1.5) -- (13,-1.5) -- (14,0) -- (16,0) -- cycle;
\draw[black, thick] (12,0) -- (13,1.5) -- (14,0) -- (13,-1.5) -- cycle;
\draw[black, thick] (15,1.5) -- (13,1.5) -- (14,0) -- (16,0) -- cycle;
\draw[black, thick] (15,-1.5) -- (13,-1.5) -- (14,0) -- (16,0) -- cycle;

\node at (11.8,0) {$b$} ; 
\node at (12.8,1.5) {$a$} ; 
\node at (15.25,1.5) {$f$} ; 
\node at (12.8,-1.5) {$c$} ; 
\node at (15.25,-1.5) {$d$} ; 
\node at (16.2,0) {$e$} ; 
\node at (14.2,-0.3) {$g$} ; 

\draw[red, thick] (13,1.3) arc (270 : 230 : 0.2) ;
\draw[red, thick] (13,1.3) arc (270 : 310 : 0.2) ;

\draw[red, thick] (14.15,0) arc (0 : -120 : 0.15) ;
\draw[red, thick] (14.8,1.5) arc (180 : 300 : 0.2) ;

\end{tikzpicture}

\begin{tikzpicture}[scale = 0.8]
\node at (20,2) {$=$} ;
    \fill[red!10] (8,0) -- (9,1.5) -- (10,0) -- (9,-1.5) -- cycle;
    \fill[red!10] (6,0) -- (8,0) -- (9,1.5) -- (7,1.5) -- cycle;
    \draw[black, thick] (6,0) -- (8,0) -- (9,1.5) -- (7,1.5) -- cycle;
    \draw[black, thick] (6,0) -- (8,0) -- (9,-1.5) -- (7,-1.5) -- cycle;
    \draw[black, thick] (8,0) -- (9,1.5) -- (10,0) -- (9,-1.5) -- cycle;
    \node at (5,0) {   $\sum_{g}$} ;

\node at (5.8,0) {$e$} ; 
\node at (6.8,1.5) {$d$} ; 
\node at (9.25,1.5) {$c$} ; 
\node at (6.8,-1.5) {$f$} ; 
\node at (9.25,-1.5) {$a$} ; 
\node at (10.2,0) {$b$} ; 
\node at (7.9,-0.2) {$g$} ; 
\draw[red, thick] (7.8,0) arc (180 : 60 : 0.2) ;
\draw[red, thick] (7.2,-1.5) arc (0 : 120 : 0.2) ;
\draw[red, thick] (9,-1.2) arc (90: 120 : 0.3) ;
\draw[red, thick] (9,-1.2) arc (90: 60 : 0.3) ;

\node at (7.5,-0.75) {$u$} ; 
\node at (14.5,-0.75) {$u$} ; 

\node at (11,0) {   $= \sum_{g}$} ;

\fill[red!10] (12,0) -- (13,1.5) -- (14,0) -- (13,-1.5) -- cycle;

\draw[black, thick] (12,0) -- (13,1.5) -- (14,0) -- (13,-1.5) -- cycle;
\fill[red!10] (15,1.5) -- (13,1.5) -- (14,0) -- (16,0) -- cycle;
\draw[black, thick] (15,1.5) -- (13,1.5) -- (14,0) -- (16,0) -- cycle;
\draw[black, thick] (15,-1.5) -- (13,-1.5) -- (14,0) -- (16,0) -- cycle;

\node at (11.8,0) {$d$} ; 
\node at (12.8,1.5) {$c$} ; 
\node at (15.25,1.5) {$b$} ; 
\node at (12.8,-1.5) {$e$} ; 
\node at (15.25,-1.5) {$f$} ; 
\node at (16.2,0) {$a$} ; 
\node at (14.1,-0.2) {$g$} ; 

\draw[red, thick] (13.8,0) arc (180 : 240 : 0.2) ;
\draw[red, thick] (13.8,0) arc (180 : 120 : 0.2) ;

\draw[red, thick] (14.8,-1.5) arc (180 : 60 : 0.2) ;

\draw[red, thick] (15.7,0) arc (180 : 120 : 0.3) ;

    \fill[red!10] (18,0) -- (20,0) -- (21,-1.5) -- (19,-1.5) -- cycle;
    \fill[red!10] (18,0) -- (20,0) -- (21,1.5) -- (19,1.5) -- cycle;
    \draw[black, thick] (18,0) -- (20,0) -- (21,1.5) -- (19,1.5) -- cycle;

\draw[black, thick] (18,0) -- (20,0) -- (21,-1.5) -- (19,-1.5) -- cycle;
\draw[black, thick] (20,0) -- (21,1.5) -- (22,0) -- (21,-1.5) -- cycle;

\node at (17.8,0) {$c$} ; 
\node at (18.8,1.5) {$b$} ; 
\node at (21.25,1.5) {$a$} ; 
\node at (18.8,-1.5) {$d$} ; 
\node at (21.25,-1.5) {$e$} ; 
\node at (22.2,0) {$f$} ; 
\node at (19.7,-0.4) {$g$} ; 
\draw[red, thick] (20.7,1.5) arc (180 : 240 : 0.3) ;

\draw[red, thick] (19.8,0) arc (180 : 300 : 0.2) ;
\draw[red, thick] (21.8,0) arc (180 : 240 : 0.2) ;
\draw[red, thick] (21.8,0) arc (180 : 120 : 0.2) ;

\node at (21,0) {$u$} ; 
\node at (17,0) {   $= \sum_{g}$} ;

\end{tikzpicture}

    \caption{Again the red face is the defect face and the white face carries spectral parameter $u$.}
        \label{fig:Top-Def-face-2}

\end{figure}
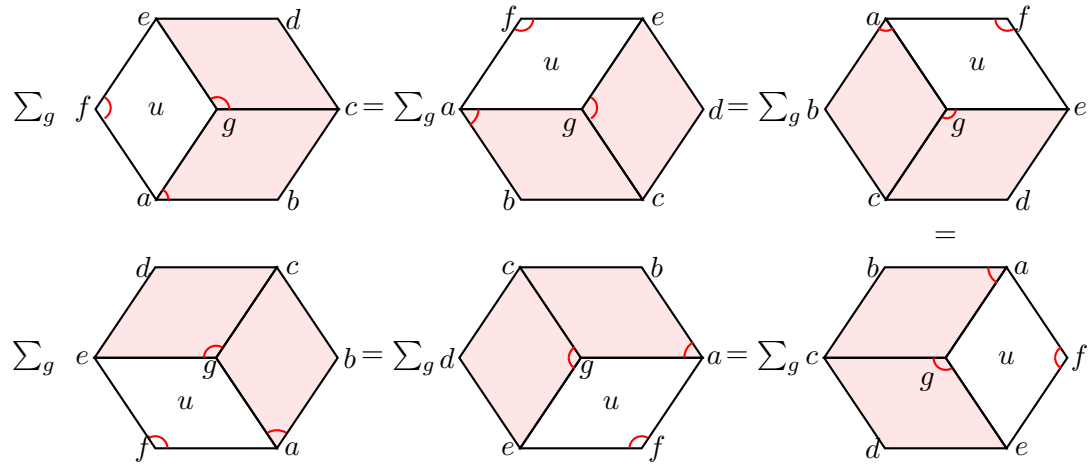

Now, the first condition for topological invariance in figure \ref{fig:first_cond_top_inv} says that the first configuration in figure \ref{fig:Top-Def-face-1} and \ref{fig:Top-Def-face-2} have equal weight. Hence, all the configurations in figures \ref{fig:Top-Def-face-1} and \ref{fig:Top-Def-face-2} have the same weight if the red defect face satisfies the first condition for topological invariance.

\bigskip 

In AFM, the first condition for topological defect is given in figure \ref{fig:def-com-rel-1}. If we convert everything with the convention followed here, we get the following equality. 

\begin{figure}[H]
    \centering
    \begin{tikzpicture}
        
        \fill[green!10] (0,0) -- (1,-1) -- (1,-1.5) -- (0,-0.5) --cycle;
        \fill[green!10] (1,-1) -- (2,0) -- (2,-0.5) -- (1,-1.5) -- cycle;

        \fill[green!10] (6,0.5) -- (7,-0.5) -- (7,0) -- (6,1) -- cycle ;
        \fill[green!10] (5,-0.5) -- (6,0.5) -- (6,1) -- (5,0) -- cycle ;

        \draw[black, thick] (0,0) -- (1,1) -- (2,0) -- (1,-1) -- cycle;
        \draw[black, thick] (0,0) -- (1,-1) -- (1,-1.5) -- (0,-0.5) --cycle;
        \draw[black, thick] (1,-1) -- (2,0) -- (2,-0.5) -- (1,-1.5) -- cycle;

        \draw[black, thick] (5,-0.5) -- (6,0.5) -- (7,-0.5) -- (6,-1.5) -- cycle;
        \draw[black, thick] (6,0.5) -- (7,-0.5) -- (7,0) -- (6,1) -- cycle ;
        \draw[black, thick] (5,-0.5) -- (6,0.5) -- (6,1) -- (5,0) -- cycle ;
    \node at (1,-0.6) {$g$};
    \node at (1,-1.8) {$f$};
    \node at (-0.2,0) {$d$};
    \node at (-0.2,-0.5) {$e$};
    \node at (2.2,0) {$b$};
    \node at (2.2,-0.5) {$a$};
    \node at (1,1.2) {$c$} ;
    \node at (1,0) {$u$} ; 
    \node at (6,0.2) {$g$};
    \node at (6,-1.8) {$f$};
    \node at (4.8,0) {$d$};
    \node at (4.8,-0.5) {$e$};
    \node at (7.2,0) {$b$};
    \node at (7.2,-0.5) {$a$};
    \node at (6,1.2) {$c$} ;
    \node at (6,-0.5) {$u$} ; 
    \node at (-1.5,-0.25) {\large $
    \sum_g \sqrt{ \frac{\theta_g}{\theta_c}}$} ;
    \node at (3.5,-0.25) {\large $= \sum_g \sqrt{\frac{\theta_g}{\theta_f}} $ } ;


\draw[red, thick]  (1.1,-0.9) arc (0 : 180 : 0.1)  ; 
\draw[red, thick]  (0.85,-1.4) arc (180 : 90 : 0.15)  ; 
\draw[red, thick]  (1.9,-0.64) arc (225 : 90 : 0.15)  ; 
\draw[red, thick]  (6.1,-1.4) arc (0 : 180 : 0.1)  ; 
\draw[red, thick]  (6,0.65) arc (90 : 225 : 0.15)  ; 
\draw[red, thick]  (7,-0.3) arc (90 : 135 : 0.2)  ; 

    \end{tikzpicture}
    \caption{The first condition for defect to be topological in AFM. Note, the Boltzmann weight for face with spectral parameter $u$ in our work and in AFM differ by Quantum dimensions, hence there are multiplicative factors above.}
    \label{fig:Fendley-top-cond-1}
\end{figure}

To remove these multiplicative factor, we add red dots to vertices as in the figure below. 
\begin{figure}[H]
    \centering
\begin{tikzpicture}
    
        \fill[green!10] (0,0) -- (1,-1) -- (1,-1.5) -- (0,-0.5) --cycle;
        \fill[green!10] (1,-1) -- (2,0) -- (2,-0.5) -- (1,-1.5) -- cycle;

        \fill[green!10] (6,0.5) -- (7,-0.5) -- (7,0) -- (6,1) -- cycle ;
        \fill[green!10] (5,-0.5) -- (6,0.5) -- (6,1) -- (5,0) -- cycle ;

        \draw[black, thick] (0,0) -- (1,1) -- (2,0) -- (1,-1) -- cycle;
        \draw[black, thick] (0,0) -- (1,-1) -- (1,-1.5) -- (0,-0.5) --cycle;
        \draw[black, thick] (1,-1) -- (2,0) -- (2,-0.5) -- (1,-1.5) -- cycle;

        \draw[black, thick] (5,-0.5) -- (6,0.5) -- (7,-0.5) -- (6,-1.5) -- cycle;
        \draw[black, thick] (6,0.5) -- (7,-0.5) -- (7,0) -- (6,1) -- cycle ;
        \draw[black, thick] (5,-0.5) -- (6,0.5) -- (6,1) -- (5,0) -- cycle ;
    \node at (1,-0.6) {$g$};
    \node at (1,-1.8) {$f$};
    \node at (-0.2,0) {$d$};
    \node at (-0.2,-0.5) {$e$};
    \node at (2.2,0) {$b$};
    \node at (2.2,-0.5) {$a$};
    \node at (1,1.2) {$c$} ;
    \node at (1,0) {$u$} ; 
    \node at (6,0.2) {$g$};
    \node at (6,-1.8) {$f$};
    \node at (4.8,0) {$d$};
    \node at (4.8,-0.5) {$e$};
    \node at (7.2,0) {$b$};
    \node at (7.2,-0.5) {$a$};
    \node at (6,1.2) {$c$} ;
    \node at (6,-0.5) {$u$} ; 
    \node at (-1.5,-0.25) {\large $
    \sum_g   $} ;
    \node at (3.5,-0.25) {\large $= \sum_g  $ } ;


\draw[red, thick]  (1.1,-0.9) arc (0 : 180 : 0.1)  ; 
\draw[red, thick]  (0.85,-1.4) arc (180 : 90 : 0.15)  ; 
\draw[red, thick]  (1.9,-0.64) arc (225 : 90 : 0.15)  ; 
\draw[red, thick]  (6.1,-1.4) arc (0 : 180 : 0.1)  ; 
\draw[red, thick]  (6,0.65) arc (90 : 225 : 0.15)  ; 
\draw[red, thick]  (7,-0.3) arc (90 : 135 : 0.2)  ; 
    \filldraw[red] (6,0.5) circle (1.5pt) ; 
    \filldraw[red] (6,1) circle (1.5pt) ; 
    \filldraw[red] (1,-1.5) circle (1.5pt) ; 
    \filldraw[red] (1,-1) circle (1.5pt) ; 
    \filldraw[red] (0,0) circle (1.5pt) ; 
    \filldraw[red] (5,0) circle (1.5pt) ; 

    \filldraw[red] (7,-0.5) circle (1.5pt) ; 
    \filldraw[red] (2,-0.5) circle (1.5pt) ;

\end{tikzpicture}
    \caption{The equality above is equivalent to the equality in figure \ref{fig:Fendley-top-cond-1}. Red dots on a site $a$ indicates a multiplicative factor of $\sqrt{\gf_a}$. }
    \label{fig:Fendley-top-cond-02}
\end{figure}
By rotating the figure above, we get
\begin{figure}[H]
    \centering
\begin{tikzpicture}
        \fill[green!10] (6,0) -- (8,0) -- (9,1.5) -- (7,1.5) -- cycle;
\draw[black, thick] (6,0) -- (8,0) -- (9,1.5) -- (7,1.5) -- cycle;
    \fill[green!10] (8,0) -- (9,1.5) -- (10,0) -- (9,-1.5) -- cycle;
\draw[black, thick] (6,0) -- (8,0) -- (9,-1.5) -- (7,-1.5) -- cycle;
\draw[black, thick] (8,0) -- (9,1.5) -- (10,0) -- (9,-1.5) -- cycle;

\node at (5.8,0) {$e$} ; 
\node at (6.8,1.5) {$d$} ; 
\node at (9.25,1.5) {$c$} ; 
\node at (6.8,-1.5) {$f$} ; 
\node at (9.25,-1.5) {$a$} ; 
\node at (10.2,0) {$b$} ; 
\node at (7.8,-0.2) {$g$} ; 

\draw[red, thick] (9,-1.3) arc (90 : 60 : 0.3) ;
\draw[red, thick] (9,-1.3) arc (90 : 120 : 0.3) ;
\draw[red, thick] (7.2,-1.5) arc (0 : 120 : 0.2) ;
\draw[red, thick] (7.8,0) arc (180: 60 : 0.2) ;
\node at (7.5,-0.75) {$u$} ; 
\node at (2.5,0.75) {$u$} ; 

\fill [green!10] (0,0) -- (1,1.5) -- (2,0) -- (1,-1.5) -- cycle;
\draw[black, thick] (0,0) -- (1,1.5) -- (2,0) -- (1,-1.5) -- cycle;
\fill[green!10] (3,-1.5) -- (1,-1.5) -- (2,0) -- (4,0) -- cycle;
\draw[black, thick] (3,1.5) -- (1,1.5) -- (2,0) -- (4,0) -- cycle;
\draw[black, thick] (3,-1.5) -- (1,-1.5) -- (2,0) -- (4,0) -- cycle;

\node at (-.2,0) {$e$} ; 
\node at (0.8,1.5) {$d$} ; 
\node at (3.25,1.5) {$c$} ; 
\node at (0.8,-1.5) {$f$} ; 
\node at (3.25,-1.5) {$a$} ; 
\node at (4.2,0) {$b$} ; 
\node at (2.2,-0.3) {$g$} ; 
\node at (5,0) {   $= \sum_{g}$} ;
\node at (-1,0) {   $\sum_{g}$} ;

\draw[red, thick] (2.2,0) arc (0 : 120 : 0.2) ;

\draw[red, thick] (1,-1.3) arc (90 : 120 : 0.2) ;
\draw[red, thick] (1,-1.3) arc (90 : 60 : 0.2) ;

\draw[red, thick] (2.8,-1.5) arc (180 : 60 : 0.2) ;

    \filldraw[red] (2,0) circle (1.5pt) ; 
    \filldraw[red] (8,0) circle (1.5pt) ; 
    \filldraw[red] (1,1.5) circle (1.5pt) ; 
    \filldraw[red] (7,1.5) circle (1.5pt) ; 
    \filldraw[red] (1,-1.5) circle (1.5pt) ; 
    \filldraw[red] (3,-1.5) circle (1.5pt) ; 

    \filldraw[red] (9,-1.5) circle (1.5pt) ; 
    \filldraw[red] (9,1.5) circle (1.5pt) ; 

\end{tikzpicture}    
    \caption{The configurations above is exactly the same as in figure \ref{fig:Fendley-top-cond-02}.}
    \label{fig:Fendley-top-cond-03}
    \end{figure}

However, the above two configurations also occur in figures \ref{fig:Top-Def-face-1} and \ref{fig:Top-Def-face-2}. Hence, the two conditions are equivalent. If a green defect face satisfies first defect commutation relation in figure \ref{fig:def-com-rel-1}, then the face in RHS of figure \ref{fig:eq-face-cond} satisfies our first condition in figure \ref{fig:first_cond_top_inv}. Further, if a red face satisfies our first condition in figure \ref{fig:first_cond_top_inv}, then a green face which obeys the equality in figure \ref{fig:eq-face-cond} satisfies the relation in figure \ref{fig:def-com-rel-1}.

\bigskip 

Now, we will show that if a defect face satisfies our condition for topological invariance and crossing symmetry, then is satisfies the second defect commutation relation of AFM. 
Using the relations in figures \ref{fig:first_cond_top_inv}, \ref{fig:top-def-rel-2}, and \ref{fig:cross-sym-def} all the following configurations have the same Boltzmann weight. 
\begin{figure}[H]
    \centering
\begin{tikzpicture}
     \fill[blue!10] (7,-1.5) -- (9,-1.5) -- (10,0) -- (10.5, -2.5) -- cycle ; 
    \fill[red!10] (8,0) --(7,-1.5) --  (9,-1.5) -- (10,0) -- cycle ; 
    \fill[red!10] (6,0) -- (7,1.5) -- (8,0) -- (7,-1.5) -- cycle ;

\draw[black, thick] (8,0) --(7,-1.5) --  (9,-1.5) -- (10,0) -- cycle ; 
\draw[black, thick] (8,0) --(10,0) --  (9,1.5) -- (7,1.5) -- cycle ; 

\draw[black, thick] (7,-1.5) -- (9,-1.5) -- (10,0) -- (10.5, -2.5) -- cycle ; 
\draw[black, thick] (6,0) -- (7,1.5) -- (8,0) -- (7,-1.5) -- cycle ;

\node[] at (5,0) {$ \sum_{a,g} $} ;
\node[] at (5.85,0) {$e$};
\node[] at (7.5,-0.2) {$g$};

\node[] at (10.15,0) {$b$};

\node[] at (6.8,1.5) {$d$};
\node[] at (6.8,-1.5) {$f$};
\node[] at (9.2,1.5) {$c$};
\node[] at (9.2,-1.5) {$a$};
\node[] at (10.6,-2.6) {$h$};

\node[] at (11.5,0) {$ = $} ;
\node[] at (11.5,-0.2) { \tiny $ 1 $} ;
    \fill[red!10] (13, - 1) -- (13,1) -- (14.5,1) -- (14.5,-1 ) -- cycle;
    \draw[black, thick]  (13, - 1) -- (13,1) -- (14.5,1) -- (14.5,-1 ) -- cycle;
    \draw[black, thick]  (14.5, - 1) -- (14.5,1) -- (16,1) -- (16,-1 ) -- cycle;

\node[] at (13,-1.25) {$f$} ; 
\node[] at (14.5,-1.25) {$h$} ; 
\node[] at (16,-1.25) {$b$} ; 
\node[] at (13,1.25) {$e$} ; 
\node[] at (14.5,1.25) {$d$} ; 
\node[] at (16,1.25) {$c$} ; 
\node at (8.5,0.75) {$u$} ;

\node at (15.25,0) {$u$} ;

\draw[red, thick]  (13,-0.8) arc (90 : 0 : 0.2)  ; 
\draw[red, thick]  (14.5,-0.8) arc (90 : 0 : 0.2)  ; 
\draw[red, thick]  (7,-1.3) arc (90: 120 : 0.2)  ; 

\draw[red, thick]  (8.2,0) arc (0 : 120 : 0.2)  ; 
\draw[red, thick]  (8.8,-1.5) arc (180 : 45 : 0.2)  ; 
\draw[red, thick]  (7,-1.3) arc (90: 120 : 0.2)  ; 
\draw[red, thick]  (7,-1.3) arc (90: 60 : 0.2)  ; 
\draw[red, thick] (10.45,-2.3) arc (120: 145 : 0.5) ;

    \filldraw[red] (13,-1.) circle (1.5pt) ; 
    \filldraw[red] (16,-1) circle (1.5pt) ; 

    \filldraw[red] (7,-1.5) circle (1.5pt) ; 
    \filldraw[red] (10,-0) circle (1.5pt) ; 

\end{tikzpicture}
    
\begin{tikzpicture}
\node[] at (4.55,-0.2) { \tiny $ 2 $} ;
\node[] at (11.5,0) {$ = $} ;
\node[] at (11.5, -0.2) {\tiny $ 3 $} ;
\node[] at (5,0) {$ = \sum_{a,g} $} ;
\node[] at (12,0) {$\sum_{a,g}$} ; 

\fill[red!10] (6,0) -- (8,0) -- (9,1.5) -- (7,1.5) -- cycle;
\fill[red!10] (8,0) -- (9,1.5) -- (10,0) -- (9,-1.5) -- cycle;
\fill[blue!10] (7,-1.5) -- (9,-1.5) -- (10,0) -- (10.5, -2.5) -- cycle ; 

\draw[black, thick] (6,0) -- (8,0) -- (9,1.5) -- (7,1.5) -- cycle;
\draw[black, thick] (6,0) -- (8,0) -- (9,-1.5) -- (7,-1.5) -- cycle;
\draw[black, thick] (8,0) -- (9,1.5) -- (10,0) -- (9,-1.5) -- cycle;
\draw[black, thick] (7,-1.5) -- (9,-1.5) -- (10,0) -- (10.5, -2.5) -- cycle ; 

\node[] at (5.85,0) {$e$};
\node[] at (7.85,-0.2) {$g$};
\node[] at (10.15,0) {$b$};
\node[] at (6.8,1.5) {$d$};
\node[] at (6.8,-1.5) {$f$};
\node[] at (9.2,1.5) {$c$};
\node[] at (9.2,-1.5) {$a$};
\node[] at (10.6,-2.6) {$h$};
\node at (7.5,-0.75) {$u$} ;

\draw[red, thick]  (7.2,-1.5) arc (0 : 135 : 0.2)  ; 
\draw[red, thick]  (7.8,0) arc (180 : 45 : 0.2)  ; 
\draw[red, thick]  (9,-1.3) arc (90 : 135 : 0.15)  ; 
\draw[red, thick]  (9,-1.3) arc (90 : 45 : 0.15)  ; 
\filldraw[red] (7,-1.5) circle (1.5pt) ; 
\filldraw[red] (10,0) circle (1.5pt) ; 
\draw[red, thick] (10.45,-2.3) arc (120: 145 : 0.5) ;


\fill[red!10] (13,0) -- (15,0) -- (16,1.5) -- (14,1.5) -- cycle;
\fill[red!10] (15,0) -- (16,1.5) -- (17,0) -- (16,-1.5) -- cycle;
\fill[red!10] (14,-1.5) -- (16,-1.5) -- (17,0) -- (17.5, -2.5) -- cycle ; 

\draw[black, thick] (13,0) -- (15,0) -- (16,1.5) -- (14,1.5) -- cycle;
\draw[black, thick] (13,0) -- (15,0) -- (16,-1.5) -- (14,-1.5) -- cycle;
\draw[black, thick] (15,0) -- (16,1.5) -- (17,0) -- (16,-1.5) -- cycle;
\draw[black, thick] (14,-1.5) -- (16,-1.5) -- (17,0) -- (17.5, -2.5) -- cycle ; 

\node[] at (12.85,0) {$e$};
\node[] at (14.85,-0.2) {$g$};
\node[] at (17.15,0) {$b$};
\node[] at (13.8,1.5) {$d$};
\node[] at (13.8,-1.5) {$f$};
\node[] at (16.2,1.5) {$c$};
\node[] at (16.2,-1.5) {$a$};
\node[] at (17.6,-2.6) {$h$};
\node at (14.5,-0.75) {$u$} ;

\draw[red, thick]  (14.2,-1.5) arc (0 : 135 : 0.2)  ; 
\draw[red, thick]  (14.8,0) arc (180 : 45 : 0.2)  ; 
\draw[red, thick]  (14.7,-1.5) arc (0 : -45 : 0.3)  ;

\draw[red, thick]  (16,-1.3) arc (90 : 135 : 0.15)  ; 
\draw[red, thick]  (16,-1.3) arc (90 : 45 : 0.15)  ; 
\filldraw[red] (16,-1.5) circle (1.5pt) ; 
\filldraw[red] (17.5,-2.5) circle (1.5pt) ; 

\end{tikzpicture}   
  \caption{The first equality is a consequence of second condition for topological invariance. The second equality follows from the first condition for topological invariance and the third follows from crossing symmetry.}
    \label{fig:def-com-rel-top-inv}
    \end{figure}
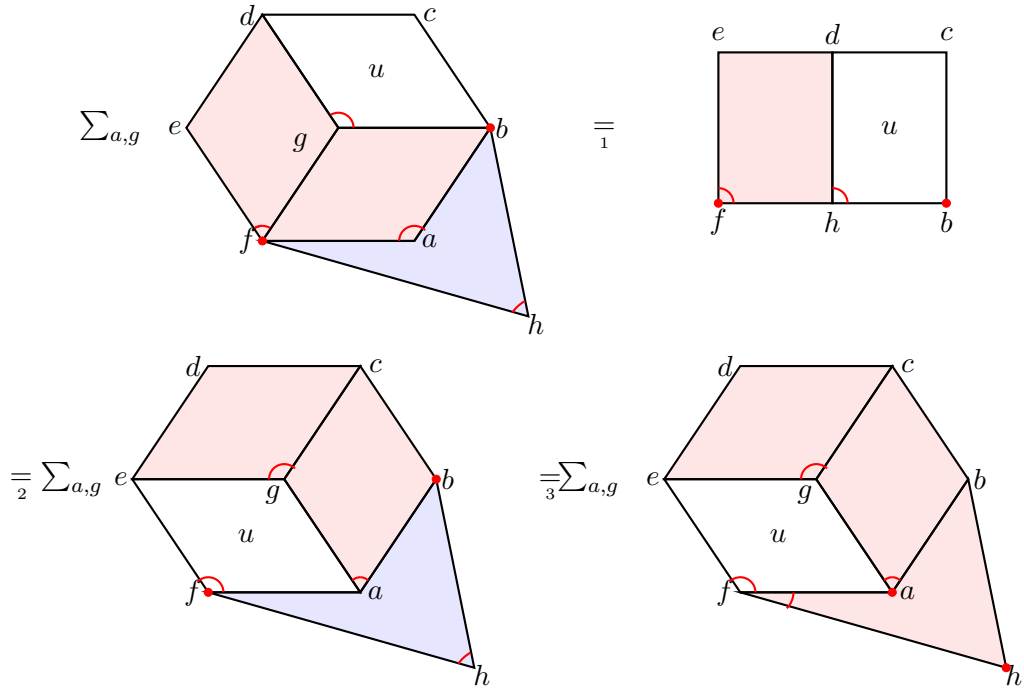

If we substitute the relations in figures \ref{fig:AFM-our-convention} and \ref{fig:eq-face-cond} in the equality in figure \ref{fig:def-com-rel-top-inv}, we see the following. 
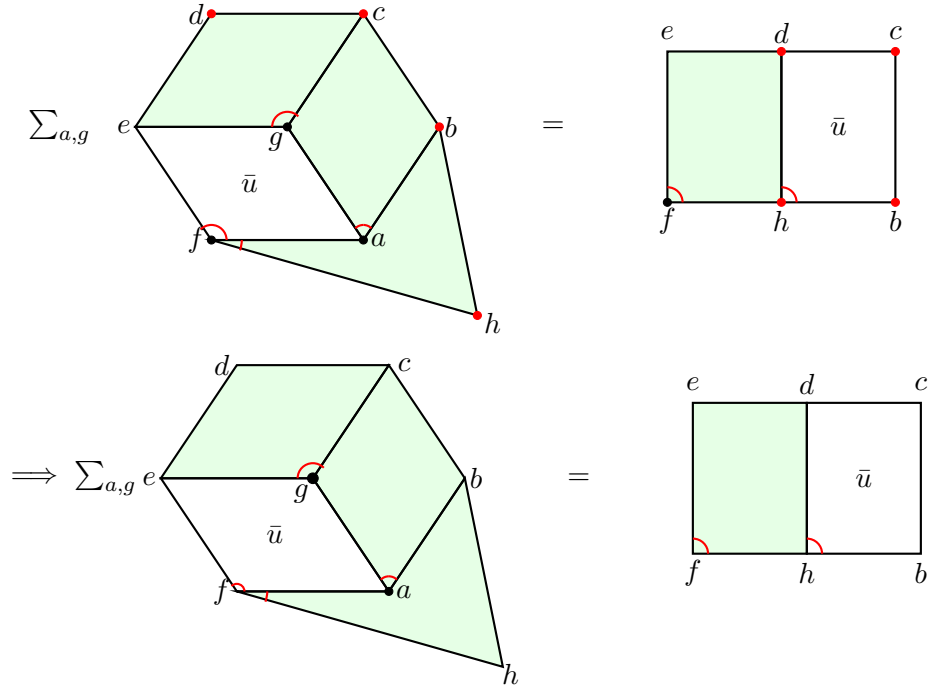
\begin{figure}[H]
    \centering
\begin{tikzpicture}
        \fill[green!10] (6,0) -- (8,0) -- (9,1.5) -- (7,1.5) -- cycle;
     \fill[green!10] (8,0) -- (9,1.5) -- (10,0) -- (9,-1.5) -- cycle;
     \fill[green!10] (7,-1.5) -- (9,-1.5) -- (10,0) -- (10.5, -2.5) -- cycle ; 

\draw[black, thick] (6,0) -- (8,0) -- (9,1.5) -- (7,1.5) -- cycle;
\draw[black, thick] (6,0) -- (8,0) -- (9,-1.5) -- (7,-1.5) -- cycle;
\draw[black, thick] (8,0) -- (9,1.5) -- (10,0) -- (9,-1.5) -- cycle;
\draw[black, thick] (7,-1.5) -- (9,-1.5) -- (10,0) -- (10.5, -2.5) -- cycle ; 
    
\node[] at (5,0) {$ \sum_{a,g} $} ;
\node[] at (5.85,0) {$e$};
\node[] at (7.85,-0.2) {$g$};

\node[] at (10.15,0) {$b$};

\node[] at (6.8,1.5) {$d$};
\node[] at (6.8,-1.5) {$f$};
\node[] at (9.2,1.5) {$c$};
\node[] at (9.2,-1.5) {$a$};
\node[] at (10.7,-2.6) {$h$};

\node[] at (11.5,0) {$ = $} ;
    \fill[green!10] (13, - 1) -- (13,1) -- (14.5,1) -- (14.5,-1 ) -- cycle;
    \draw[black, thick]  (13, - 1) -- (13,1) -- (14.5,1) -- (14.5,-1 ) -- cycle;
    \draw[black, thick]  (14.5, - 1) -- (14.5,1) -- (16,1) -- (16,-1 ) -- cycle;

\node[] at (13,-1.25) {$f$} ; 
\node[] at (14.5,-1.25) {$h$} ; 
\node[] at (16,-1.25) {$b$} ; 
\node[] at (13,1.25) {$e$} ; 
\node[] at (14.5,1.25) {$d$} ; 
\node[] at (16,1.25) {$c$} ; 
\node at (7.5,-0.75) {$\bar{u}$} ;

\node at (15.25,0) {$\bar{u}$} ;

\draw[red, thick]  (13,-0.8) arc (90 : 0 : 0.2)  ; 
\draw[red, thick]  (14.5,-0.8) arc (90 : 0 : 0.2)  ; 
\draw[red, thick]  (7.2,-1.5) arc (0 : 135 : 0.2)  ; 
\draw[red, thick]  (7.4,-1.5) arc (0 : -20 : 0.4)  ; 
\draw[red, thick]  (7.8,0) arc (180 : 44 : 0.2)  ; 
\draw[red, thick]  (9,-1.3) arc (90 : 135 : 0.15)  ; 
\draw[red, thick]  (9,-1.3) arc (90 : 45 : 0.15)  ; 

    \filldraw[black] (9,-1.5) circle (1.5pt) ; 
    \filldraw[black] (8,0) circle (1.5pt) ; 
    \filldraw[red] (10.5,-2.5) circle (1.5pt) ; 

        \filldraw[black] (13,-1) circle (1.5pt) ; 
        \filldraw[red] (14.5,-1) circle (1.5pt) ; 
        \filldraw[red] (16,-1) circle (1.5pt) ; 
        \filldraw[red] (10,0) circle (1.5pt) ; 
        \filldraw[red] (9,1.5) circle (1.5pt) ; 
\filldraw[black] (7 , -1.5) circle (1.5pt) ; 
        \filldraw[red] (16,1) circle (1.5pt) ; 
        \filldraw[red] (14.5,1) circle (1.5pt) ; 
        \filldraw[red] (7,1.5) circle (1.5pt) ;

\end{tikzpicture} 

\begin{tikzpicture}
        \fill[green!10] (6,0) -- (8,0) -- (9,1.5) -- (7,1.5) -- cycle;
     \fill[green!10] (8,0) -- (9,1.5) -- (10,0) -- (9,-1.5) -- cycle;
     \fill[green!10] (7,-1.5) -- (9,-1.5) -- (10,0) -- (10.5, -2.5) -- cycle ; 

\draw[black, thick] (6,0) -- (8,0) -- (9,1.5) -- (7,1.5) -- cycle;
\draw[black, thick] (6,0) -- (8,0) -- (9,-1.5) -- (7,-1.5) -- cycle;
\draw[black, thick] (8,0) -- (9,1.5) -- (10,0) -- (9,-1.5) -- cycle;
\draw[black, thick] (7,-1.5) -- (9,-1.5) -- (10,0) -- (10.5, -2.5) -- cycle ; 
    
\node[] at (4.8,0) {$ \implies \sum_{a,g} $} ;
\node[] at (5.85,0) {$e$};
\node[] at (7.85,-0.2) {$g$};

\node[] at (10.15,0) {$b$};

\node[] at (6.8,1.5) {$d$};
\node[] at (6.8,-1.5) {$f$};
\node[] at (9.2,1.5) {$c$};
\node[] at (9.2,-1.5) {$a$};
\node[] at (10.6,-2.6) {$h$};

\node[] at (11.5,0) {$ = $} ;
    \fill[green!10] (13, - 1) -- (13,1) -- (14.5,1) -- (14.5,-1 ) -- cycle;
    \draw[black, thick]  (13, - 1) -- (13,1) -- (14.5,1) -- (14.5,-1 ) -- cycle;
    \draw[black, thick]  (14.5, - 1) -- (14.5,1) -- (16,1) -- (16,-1 ) -- cycle;

\node[] at (13,-1.25) {$f$} ; 
\node[] at (14.5,-1.25) {$h$} ; 
\node[] at (16,-1.25) {$b$} ; 
\node[] at (13,1.25) {$e$} ; 
\node[] at (14.5,1.25) {$d$} ; 
\node[] at (16,1.25) {$c$} ; 
\node at (7.5,-0.75) {$\bar{u}$} ;

\node at (15.25,0) {$\bar{u}$} ;

\draw[red, thick]  (13,-0.8) arc (90 : 0 : 0.2)  ; 
\draw[red, thick]  (14.5,-0.8) arc (90 : 0 : 0.2)  ; 
\draw[red, thick]  (7.1,-1.5) arc (0 : 135 : 0.1)  ; 
\draw[red, thick]  (7.4,-1.5) arc (0 : -20 : 0.4)  ; 
\draw[red, thick]  (7.8,0) arc (180 : 44 : 0.2)  ; 
\draw[red, thick]  (9,-1.3) arc (90 : 135 : 0.15)  ; 
\draw[red, thick]  (9,-1.3) arc (90 : 45 : 0.15)  ; 

    \filldraw[black] (9,-1.5) circle (1.5pt) ; 
    \filldraw[black] (8,0) circle (2pt) ;

\end{tikzpicture} 
\caption{The second relation is exactly the second defect commutation relation in AFM. }
    \label{fig:2-def-com-rel}
\end{figure}

We will now show that the red defect face satisfying the second condition for topological invariance (cf. figure \ref{fig:top-def-rel-2}) is a consequence of the first and second defect commutation relation. 
We start with the first defect commutation relation in figure \ref{fig:def-com-rel-1}. 
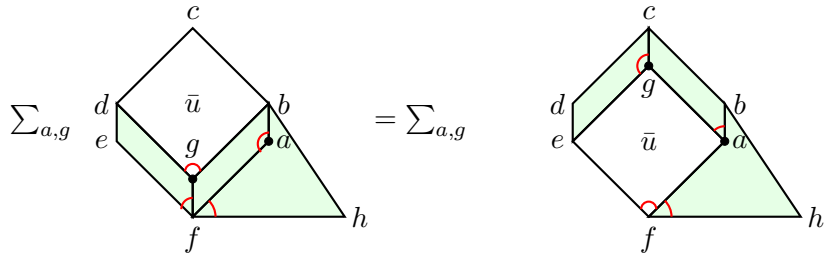
\begin{figure}[H]
    \centering
    \begin{tikzpicture}
        
        \fill[green!10] (0,0) -- (1,-1) -- (1,-1.5) -- (0,-0.5) --cycle;
        \fill[green!10] (1,-1) -- (2,0) -- (2,-0.5) -- (1,-1.5) -- cycle;

       \fill[green!10] (2,0) -- (2,-0.5) -- (1,-1.5) -- (3,-1.5) -- cycle;
        \draw[black, thick] (2,0) -- (2,-0.5) -- (1,-1.5) -- (3,-1.5) -- cycle;
        \draw[black, thick] (0,0) -- (1,1) -- (2,0) -- (1,-1) -- cycle;
        \draw[black, thick] (0,0) -- (1,-1) -- (1,-1.5) -- (0,-0.5) --cycle;
        \draw[black, thick] (1,-1) -- (2,0) -- (2,-0.5) -- (1,-1.5) -- cycle;

       \fill[green!10] (8,0) -- (8,-0.5) -- (7,-1.5) -- (9,-1.5) -- cycle;
        \draw[black, thick] (8,0) -- (8,-0.5) -- (7,-1.5) -- (9,-1.5) -- cycle;
        \fill[green!10] (7,0.5) -- (8,-0.5) -- (8,0) -- (7,1) -- cycle ;
        \fill[green!10] (6,-0.5) -- (7,0.5) -- (7,1) -- (6,0) -- cycle ;
        \draw[black, thick] (6,-0.5) -- (7,0.5) -- (8,-0.5) -- (7,-1.5) -- cycle;
        \draw[black, thick] (7,0.5) -- (8,-0.5) -- (8,0) -- (7,1) -- cycle ;
        \draw[black, thick] (6,-0.5) -- (7,0.5) -- (7,1) -- (6,0) -- cycle ;

    \node at (1,-0.6) {$g$};
    \node at (1,-1.8) {$f$};
    \node at (-0.2,0) {$d$};
    \node at (-0.2,-0.5) {$e$};
    \node at (2.2,0) {$b$};
    \node at (2.2,-0.5) {$a$};
    \node at (1,1.2) {$c$} ;
    \node at (1,0) {$\bar{u}$} ; 
    \node at (3.2,-1.5) {$h$};

    \draw[red, thick]  (1.3,-1.5) arc (0: 45 : 0.3)  ; 
\draw[red, thick]  (1.1,-0.9) arc (0 : 180 : 0.1)  ; 
\draw[red, thick]  (0.85,-1.4) arc (180 : 90 : 0.15)  ; 
\draw[red, thick]  (1.9,-0.64) arc (225 : 90 : 0.15)  ; 
    \filldraw[black] (2,-0.5) circle (1.5pt) ; 
    \filldraw[black] (1,-1) circle (1.5pt) ;

    \node at (7,1.2) {$c$} ;
    \node at (7,-0.5) {$\bar{u}$} ; 
    \node at (-1.0,-0.25) {$ \sum_{a,g}$} ;
    \node at (4.0,-0.25) {$= \sum_{a,g}$} ;
    \filldraw[black] (7,0.5) circle (1.5pt) ; 
    \filldraw[black] (8,-0.5) circle (1.5pt) ;    

\draw[red, thick]  (7.1,-1.4) arc (0 : 180 : 0.1)  ; 
\draw[red, thick]  (7,0.65) arc (90 : 225 : 0.15)  ; 
\draw[red, thick]  (8,-0.3) arc (90 : 135 : 0.2)  ; 
\draw[red, thick]  (7.3,-1.5) arc (0: 45 : 0.3)  ; 
    \node at (7,0.2) {$g$};
    \node at (7,-1.8) {$f$};
    \node at (5.8,0) {$d$};
    \node at (5.8,-0.5) {$e$};
    \node at (8.2,0) {$b$};
    \node at (8.2,-0.5) {$a$};
    \node at (9.2,-1.5) {$h$};

    \end{tikzpicture}
    \caption{The above two configurations are the configurations in figure \ref{fig:def-com-rel-1} with a face with heights $a,f,h,$ and $b$ multiplied on both sides.}
    \label{fig:second_def_rel_1}

\end{figure}

Now, on the configuration in RHS in the figure above, we use the second defect commutation relation in figure \ref{fig:def-com-rel-2}. 
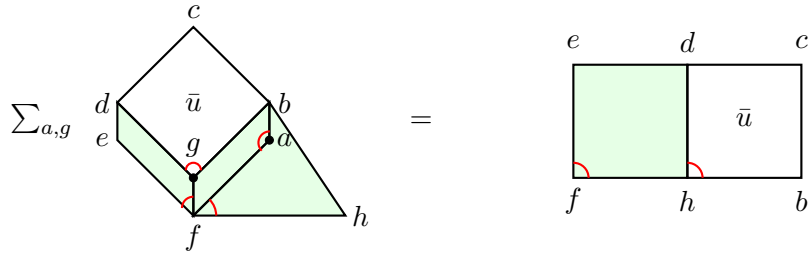
\begin{figure}[H]
    \centering
\begin{tikzpicture}
        \fill[green!10] (0,0) -- (1,-1) -- (1,-1.5) -- (0,-0.5) --cycle;
        \fill[green!10] (1,-1) -- (2,0) -- (2,-0.5) -- (1,-1.5) -- cycle;

       \fill[green!10] (2,0) -- (2,-0.5) -- (1,-1.5) -- (3,-1.5) -- cycle;
        \draw[black, thick] (2,0) -- (2,-0.5) -- (1,-1.5) -- (3,-1.5) -- cycle;
        \draw[black, thick] (0,0) -- (1,1) -- (2,0) -- (1,-1) -- cycle;
        \draw[black, thick] (0,0) -- (1,-1) -- (1,-1.5) -- (0,-0.5) --cycle;
        \draw[black, thick] (1,-1) -- (2,0) -- (2,-0.5) -- (1,-1.5) -- cycle;
    \node at (-1.0,-0.25) {$ \sum_{a,g}$} ;
    \node at (4,-0.25) {$=$} ;

    \draw[red, thick]  (1.3,-1.5) arc (0: 45 : 0.3)  ; 
\draw[red, thick]  (1.1,-0.9) arc (0 : 180 : 0.1)  ; 
\draw[red, thick]  (0.85,-1.4) arc (180 : 90 : 0.15)  ; 
\draw[red, thick]  (1.9,-0.64) arc (225 : 90 : 0.15)  ; 
    \filldraw[black] (2,-0.5) circle (1.5pt) ; 
    \filldraw[black] (1,-1) circle (1.5pt) ;

    \node at (1,-0.6) {$g$};
    \node at (1,-1.8) {$f$};
    \node at (-0.2,0) {$d$};
    \node at (-0.2,-0.5) {$e$};
    \node at (2.2,0) {$b$};
    \node at (2.2,-0.5) {$a$};
    \node at (1,1.2) {$c$} ;
    \node at (1,0) {$\bar{u}$} ; 
    \node at (3.2,-1.5) {$h$};
    
         \fill[green!10] (6,0.5) -- (7.5,0.5) -- (7.5,-1) -- (6,-1) -- cycle;

        \draw[black, thick] (6,0.5) -- (7.5,0.5) -- (7.5,-1) -- (6,-1) -- cycle;
        \draw[black, thick] (7.5,0.5) -- (9,0.5) -- (9,-1) -- (7.5,-1) -- cycle;
            \node at (6,-1.3) {$f$};
            \node at (7.5,-1.3) {$h$};
            \node at (9,-1.3) {$b$};
            \node at (6,0.8) {$e$};
            \node at (7.5,0.8) {$d$};
            \node at (9,0.8) {$c$};
    \node at (8.25,-0.2) {$\bar{u}$} ; 
    \draw[red, thick]  (6.2,-1) arc (0: 90 : 0.2)  ; 
    \draw[red, thick]  (7.7,-1) arc (0: 90 : 0.2)  ;

\end{tikzpicture}
        \caption{The above follows from substituting the relation in figure \ref{fig:def-com-rel-2} in figure \ref{fig:second_def_rel_1}.}
    \label{fig:second_def_rel_2}
\end{figure}
We can set $\bar{u} = 0$ in the figure above and then simplify to get the following.
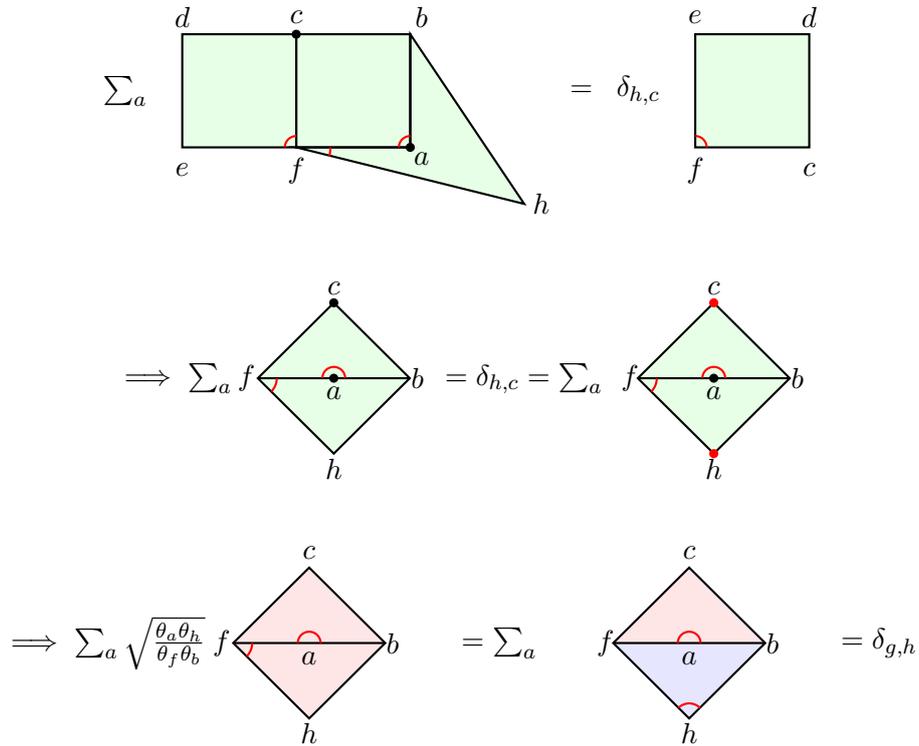
\begin{figure}[H]
    \centering
\begin{tikzpicture}[scale = 1.5]
    \node at (0,0) {$\sum_a$} ; 
    \node at (4,0) {$=$} ; 

    \fill[green!10] (0.5,-0.5) -- (1.5,-0.5) -- (1.5,0.5) -- (0.5,0.5) -- cycle; 
    \fill[green!10](1.5,-0.5) -- (2.5,-0.5) -- (2.5,0.5) -- (1.5,0.5) -- cycle; 
    \fill[green!10] (1.5,-0.5) -- (2.5,-0.5) -- (2.5,0.5) -- (3.5,-1) -- cycle; 
    \fill[green!10] (5,-0.5) -- (6,-0.5) -- (6,0.5) -- (5,0.5) -- cycle; 

    \draw[black, thick] (0.5,-0.5) -- (1.5,-0.5) -- (1.5,0.5) -- (0.5,0.5) -- cycle; 
    \draw[black, thick] (1.5,-0.5) -- (2.5,-0.5) -- (2.5,0.5) -- (1.5,0.5) -- cycle; 
    \draw[black, thick] (1.5,-0.5) -- (2.5,-0.5) -- (2.5,0.5) -- (3.5,-1) -- cycle; 
    \draw[black, thick] (5,-0.5) -- (6,-0.5) -- (6,0.5) -- (5,0.5) -- cycle;
    \node at (0.5,-0.7) {$e$} ; 
    \node at (1.5,-0.7) {$f$} ; 
    \node at (2.6,-0.6) {$a$} ; 
     \node at (0.5,0.65) {$d$} ; 
    \node at (1.5,0.65) {$c$} ; 
    \node at (2.6,0.65) {$b$} ; 
    \node at (5,-0.7) {$f$} ; 
    \node at (6,-0.7) {$c$} ; 
    \node at (5,0.65) {$e$} ; 
    \node at (6,0.65) {$d$} ; 
    \node at (3.65,-1) {$h$} ; 
\draw[red, thick]  (1.4,-0.5) arc (180: 90 : 0.1)  ; 
    \draw[red, thick]  (2.4,-0.5) arc (180: 90 : 0.1)  ; 
    \draw[red, thick]  (1.8,-0.5) arc (0: -15 : 0.3)  ; 
    \draw[red, thick]  (5.1,-0.5) arc (0: 90 : 0.1)  ; 
    \node at (4.5,0) {$\delta_{h,c}$};
    \filldraw[black] (1.5,0.5) circle (1pt) ; 
    \filldraw[black] (2.5,-0.5) circle (1pt) ; 
\end{tikzpicture}
\\
\begin{tikzpicture}
    \node at (-0.6,0) {$\implies \sum_a$} ;
    \draw[white, thick] (0,0) -- (0,2) ; 
        \fill[green!10] (0.5,0) -- (1.5,1) -- (2.5,0) -- (1.5,-1) -- cycle ;
    \draw[black, thick] (0.5,0) -- (1.5,1) -- (2.5,0) -- (1.5,-1) -- cycle ;
    \draw [black, thick] (0.5, 0) -- (2.5,0) ; 
    \node at (4,0) {$ = \delta_{h,c} = \sum_{a}$} ;
            \fill[green!10] (5.5,0) -- (6.5,1) -- (7.5,0) -- (6.5,-1) -- cycle ;
    \draw[black, thick] (5.5,0) -- (6.5,1) -- (7.5,0) -- (6.5,-1) -- cycle ;
    \draw [black, thick] (5.5, 0) -- (7.5,0) ; 
    \draw[red, thick]  (1.35,0) arc (180: 0 : 0.15)  ; 
        \draw[red, thick]  (6.35,0) arc (180: 0 : 0.15)  ; 

    \draw[red, thick]  (0.75,0) arc (0: -45 : 0.25)  ; 
        \draw[red, thick]  (5.75,0) arc (0: -45 : 0.25)  ; 

    \node at (0.35,0) {$f$} ; 
        \node at (2.6,0) {$b$} ; 
    \node at (1.5,-1.2) {$h$} ; 
    \node at (6.5,-1.2) {$h$} ; 
    \node at (1.5,1.2) {$c$} ; 
    \node at (6.5,1.2) {$c$} ; 
    \node at (1.5,-.2) {$a$} ; 
    \node at (6.5,-.2) {$a$} ; 

    \node at (5.4,0) {$f$} ; 
        \node at (7.6,0) {$b$} ; 
    \filldraw[black] (1.5,0) circle (1.5pt) ; 
    \filldraw[black] (6.5,0) circle (1.5pt) ; 
    \filldraw[black] (1.5,1) circle (1.5pt) ; 
    \filldraw[red] (6.5,1) circle (1.5pt) ; 
    \filldraw[red] (6.5,-1) circle (1.5pt) ; 
\end{tikzpicture}
\\
\begin{tikzpicture}
        \node at (-1.2,0) {$\implies \sum_a \sqrt{\frac{\gf_a \gf_h}{\gf_f \gf_b}}$} ;
    \draw[white, thick] (0,1) -- (0,2) ; 
        \fill[red!10] (0.5,0) -- (1.5,1) -- (2.5,0) -- (1.5,-1) -- cycle ;
    \draw[black, thick] (0.5,0) -- (1.5,1) -- (2.5,0) -- (1.5,-1) -- cycle ;
    \draw [black, thick] (0.5, 0) -- (2.5,0) ; 
    \node at (4,0) {$  = \sum_{a}$} ;
            \fill[red!10] (5.5,0) -- (6.5,1) -- (7.5,0) --   cycle ;
            \fill[blue!10] (5.5,0) -- (7.5,0) -- (6.5,-1) -- cycle ;

    \draw[black, thick] (5.5,0) -- (6.5,1) -- (7.5,0) -- (6.5,-1) -- cycle ;
    \draw [black, thick] (5.5, 0) -- (7.5,0) ; 
    \draw[red, thick]  (1.35,0) arc (180: 0 : 0.15)  ; 
        \draw[red, thick]  (6.35,0) arc (180: 0 : 0.15)  ; 

    \draw[red, thick]  (0.75,0) arc (0: -45 : 0.25)  ; 
        \draw[red, thick]  (6.5,-0.8) arc (90: 135 : 0.2)  ; 
        \draw[red, thick]  (6.5,-0.8) arc (90: 45 : 0.2)  ;

    \node at (0.35,0) {$f$} ; 
        \node at (2.6,0) {$b$} ; 
    \node at (1.5,-1.2) {$h$} ; 
    \node at (6.5,-1.2) {$h$} ; 
    \node at (1.5,1.2) {$c$} ; 
    \node at (6.5,1.2) {$c$} ; 
    \node at (1.5,-.2) {$a$} ; 
    \node at (6.5,-.2) {$a$} ; 

    \node at (5.4,0) {$f$} ; 
        \node at (7.6,0) {$b$} ; 
        \node at (9,0) {$ = \delta_{g,h}$} ; 

\end{tikzpicture}
        \caption{The second implication follows by substituting figure \ref{fig:eq-face-cond} into the line above. The last line is exactly second condition for topological invariance.}
    \label{fig:def_rel_to_unit}
\end{figure}

\newpage 
\section{Correspondence between the translation and line operator}\label{AppendixCorrespondence}
Let $H^{k}_{{\cal D}}$ be a defect Hamiltonian and $T^{k}_{{\cal D}}$ be a (unitary) local translation operator for this defect Hamiltonian, i.e. 
\begin{equation}
  T^{k}_{{\cal D}}  H^{(k,k+1)}_{{\cal D}} \left( T^{k}_{{\cal D}} \right)^{-1} = H^{(k-1,k)}_{{\cal D}} \, ,  
\end{equation}
and therefore $\uR \,  T^{k}_{{\cal D}}$ commutes with the defect Hamiltonian and can be used to calculate the momentum eigenvalue of eigenstates. Although there is no general prescription, in certain cases using the translation operator one can construct the line operator in the crossed channel corresponding to ${\cal D}$, i.e. $\widehat{{\cal D}}^{(\rm latt)}$, for example see \cite{Seifnashri:2023dpa, Sinha:2023hum} for invertible defects  and \cite{Seiberg:2024gek} for non-invertible defects. 

Let us consider first the case  of one impurity Hamiltonians ( Eq. \eqref{eq:def_ham}), for which the momentum and local translation operators are given in Eq. \eqref{eq:transl_op} and \eqref{TvsU} respectively. Setting $k = 1$, Eq. \eqref{eq:action-shift} implies 
\begin{equation}\label{eq:TmatA-shift}
R_{1}(\tilde{u}) \,  H^{(1,2)}(\tilde{u}) \,  R_{1}(\tilde{u})^{-1} =  H^{(0,1)}(\tilde{u}) \, .
\end{equation}
Successive application of the translation operator gives us 
\begin{equation}
\begin{split}
 R_{1}(\tilde{u})R_{2}(\tilde{u}) & \ldots R_{2\R -1}(\tilde{u}) \,  H^{(2\R - 1,2\R)}(\tilde{u}) \,  R_{2\R - 1}(\tilde{u})^{-1} \ldots R_{2}(\tilde{u})^{-1}R_{1}(\tilde{u})^{-1} =  H^{(2\R-1,0)}(\tilde{u}) \, , \\
\implies & \left[H^{(0,1)}(\tu),  R_{1}(\tilde{u})R_{2}(\tilde{u})\ldots R_{2\R -1}(\tilde{u}) \uR^{-1} \right] = 0 \, , \\
\implies & \left[ H^{(0,1)}(\tu) , T_A(\gamma - \tu)\right] = 0 \, , 
\end{split}
\end{equation}
where the first implication follows from the fact that $\uR^{-1} H^{(0,1)}(\tu) \uR = H^{(2\R - 1, 2\R)}$, while the second implication follows as $\tilde{R}_j(\gamma - \tu) = R_j(\tu)$. Recall, $T_A$ was defined in Eq. \eqref{eq:tmat_TLgen_KP}. Further, from the first line in Eq. \eqref{eq:TmatA-shift} we also obtain 
\begin{equation}
    \begin{split}
     &   \left[H^{(0,1)}(\tu), \uR  R_{2\R -1}(\tilde{u})^{-1} \ldots R_{2}(\tilde{u})^{-1} R_{1}(\tilde{u})^{-1}  \right] = 0 \, , \\
    \implies &  \left[H^{(0,1)}(\tu), \uR  R_{2\R -1}(-\tilde{u}) \ldots R_{2}(-\tilde{u}) R_{1}(-\tilde{u})  \right] = 0  \, , \\ 
    \implies & \left[H^{(0,1)}(\tu),T_{B}(-\tu) \right]  = 0  \, , 
    \end{split}
\end{equation}
where the first implication above follows from unitarity of $R$ matrix and $T_B$ was also defined in Eq. \eqref{eq:tmat_TLgen_KP}. 

Further, note that the defect shift operator for the defect Hamiltonian is related to these two line operators, $T_A(\gamma - \tu)$ and $T_B(-\tu)$ as follows
\begin{equation}\label{eq:rel-def-tmat}
\begin{split}
&     \left( \uR R_0(\tu) \right)^{2\R - 1} = R_{1}(\tilde{u})R_{2}(\tilde{u})\ldots R_{2\R -1}(\tilde{u}) \uR^{-1}  \propto T_{A}(\gamma - \tu)\, , \\
  &  \left(R_0(\tu)^{-1} \uR^{-1}  \right)^{2\R - 1} = \uR R_{2\R - 1}(\tilde{u})^{-1} \ldots R_{1}(\tilde{u})^{-1}  \propto T_B(-\tu)\, .
\end{split}
\end{equation}
Note, $R_0(\tu)^{-1} \uR^{-1} $ is the inverse of the usual defect shift operator, $\uR R_0(\tu)$. Hence if we knew the eigenvalues of $T_A(\gamma - \tu)$ or $T_B(- \tu)$, using the above equation we could determine the eigenvalues of $\uR R_{0}(\tu)$ and hence the momentum eigenvalues. The eigenvalues of $T_A(\gamma - \tu)$ and $T_B( - \tu)$ are not known generally, but when  $\tu = 0 $  
or $\pm {\rm i } \infty$, there are certain simplifications. When $\tu = 0$, i.e. the (1,1) defect case, we get from Eq. \eqref{eq:rel-def-tmat} that $\uR^{2\R} = \mathbb{1}$, which implies that the eigenvalues of the unitary operator $\uR$ are 
\begin{equation}\label{eq:eigval-mom-11}
    {\rm e}^{{\rm i } \frac{ 2\pi n}{2\R} } \, , \quad {\rm where} \, \, n \in \mathbb{Z} \, . 
\end{equation}
As $\uR$ is the shift operator for the (1,1) defect Hamiltonian, by taking log of eigenvalues in Eq. \eqref{eq:eigval-mom-12}, we get ${\rm i}\frac{2\pi (h - \bar{h})}{2\R }$, where $h$ and $\bar{h}$ are the conformal dimensions that appear in the partition function. Note, the fact that all states in A-type minimal model CFTs have $h - \bar{h} \in \mathbb{Z}$, lines up with the observation that $n \in \mathbb{Z}$ in Eq. \eqref{eq:eigval-mom-11}.

We will next discuss the case $\tu  = {\rm i } \infty$, the $\tu = -{\rm i} \infty$ being very similar. 
\par
When $\tu ={\rm i } \infty $, the defect Hamiltonian corresponds to the $(1,2)$ defect. The defect shift operator in that case is $\uR g_0$, which is not exactly $\uR \lim_{\tu \to {\rm i} \infty}R_{0}(\tu)$ but proportional to it.\footnote{$\uR \lim_{\tu \to {\rm i} \infty}R_{0}(\tu)$ diverges, therefore one has to rescale it to get the correct momentum eigenvalues. } The line operator corresponding to it are 
\begin{equation}
    \begin{split}
        (\uR g_0)^{2 \R - 1} & = \quad    g_1 \ldots g_{2\R -1}  \, \uR^{-1} \propto \lim_{\tu \to {\rm i}\infty} T_A(\gamma - \tu) \, ,  \\ 
        (g_0^{-1} \uR^{-1} )^{2 \R - 1} & = \quad \uR g_{2\R - 1} \ldots g_{1}^{-1}   \propto \lim_{\tu \to {\rm i}\infty} T_B( - \tu) \, . \\
    \end{split}
\end{equation}
Recall that in the limit of spectral parameter going to $-{\rm i } \infty$, the transfer matrix becomes the $\overline{Y}$ operator, which we now write as 
$  \overline{Y}   = \overline{Y}_A + \overline{Y}_B$, where
\begin{equation}
    \overline{Y}_A   = (-q)^{-\frac{1}{2}} \,  \uR  \, g^{-1}_{2\R - 1} \ldots g_{1}^{-1} \, , \quad     \overline{Y}_B   =  (-q)^{\frac{1}{2}} \, g_1 \ldots g_{2\R -1}  \, \uR^{-1} \, .
\end{equation}
The line operator $\overline{Y}$ is the same as $Y$. Now, one can see easily that $\overline{Y}_A^{-1} = \overline{Y}_B$, further as $g_i^{\dagger} = g_i^{-1}$, $\overline{Y}_A^{\dagger} = \overline{Y}_B$. Hence, $\overline{Y}_A$ and $\overline{Y}_B$ are unitary operators, which are inverses of each other, therefore their eigenvalues are phases, say ${\rm e}^{-{\rm i} \theta}$ and ${\rm e}^{{\rm i} \theta}$. Further, as they are inverses of each other, their eigenvectors are the same and therefore the eigenvalue of $\overline{Y}$ is $2 \cos \theta$. Now, we know the eigenvalues of $\overline{Y}$ in A$_p$ RSOS model are given by 
\begin{equation}
\begin{split}
   &  q^{s} + q^{-s} = 2 \cos \left( \frac{s \pi}{p+1}\right) \, , \quad {\rm where} \, \, 1 \leq s \leq p  \\
 \implies    &  \theta = 2 n \pi \pm \frac{s \pi}{p+1} \, , \quad {\rm where} \, \, n \in \mathbb{Z} \, .
\end{split}
\end{equation}

As $ (-q)^{\frac{1}{2}}(\uR g_0)^{2 \R - 1} = \overline{Y}_B$, the eigenvalue of $\uR g_0 $ is 
\begin{equation}\label{eq:eigval-mom-12}
\begin{split}    
  \left(  \exp(\rm i \theta)  (-q)^{-\frac{1}{2}} \right)^{\frac{1}{2\R - 1}} & =  \left( \exp   \left({\rm i} \theta + {\rm i} \frac{\gamma - \pi}{2}\right)  \right)^{\frac{1}{2\R-1}} \\
 & =   \exp\left( {\rm i} \frac{2 \pi}{2\R - 1} \left( \left( n - \frac{1}{4}\right)  \pm \frac{\left( s \pm \frac{1}{2} \right)}{2(p+1)}\right)\right) \\ 
\end{split}
\end{equation}
 $\uR g_0 $ is the defect shift operator for the $(1,2)$ defect Hamiltonian in Eq. \eqref{eq:H12},  using the log of eigenvalues in Eq. \eqref{eq:eigval-mom-12}, we get value of $h - \bar{h}$, where $h$ and $\bar{h}$ are the conformal dimension appearing in the twisted partition function. Note, from Eq. \eqref{eq:eigval-mom-12} we see that the spin chain must be treated to be of size $2\R - 1$, instead of $2\R$ like in the (1,1) defect case, to get accurate momentum scaling, as we have discussed in section \ref{sec:examples}. It can be checked by appropriately selecting $n$ and $s$, that we can recover the states which appear in partition function twisted by the $(1,2)$ field.

\par

We note that the (2,1) defect case is more complex: we can obtain it by setting $\tu = {\pm \pi/2}$, but we do not  have any analytical way to obtain the eigenvalue of $T_A(\gamma - \tu)$ or $T_B(-\tu)$ for these values of $\tu$.
\par
Now, let us consider the $(1,3)$ defect Hamiltonian $H^{0,1, 2}_{{\cal D}_{(1,2)}{\cal D}_{(1,2)}}$, the defect shift operator for this Hamiltonian is given by $\uR g_1g_0$. Let us study 
\begin{equation}
   \left( \uR g_1g_0 \right)^{2 \R - 2}  = \left(g_2 g_3 \ldots g_{2\R - 1} \right) \left(g_1 g_2 \ldots g_{2\R - 2} \right) \uR^{-2} \, . 
\end{equation}
Recall, to obtain the states of $(1,3)$ we have to use the Jones-Wenzl projector,  $P_1^{(1)}$. $P_1^{(1)}$ acts as identity on low-lying eigenstates of this Hamiltonian, which flow to states of $(1,3)$ defect Hilbert space, therefore $e_1$ must act as 0. Hence, if $\ket{\psi}$ lies in the projected Hilbert space then 
\begin{equation}
\begin{split}
   &  \bra{\psi} g_1\ket{\psi} = (-q)^{\frac{1}{2}} \\ 
   \implies  & \bra{\psi} \left( \uR g_1 g_0 \right)^{2\R - 2} \ket{\psi} = (-q)^{-1} \bra{\psi} \left(g_1 g_2 g_3 \ldots g_{2\R - 2} \right) \left(g_1 g_2 \ldots g_{2\R - 1}  g_{2\R - 1} \right) \uR^{-2}   \ket{\psi} \\ 
   & = (-q)^{-2}\bra{\psi} \overline{Y}_A \uR \overline{Y}_A \uR^{-1} \ket{\psi}\, . 
\end{split} 
\end{equation}
The eigenvalue of $\overline{Y}_A$ is the same as $\uR \overline{Y}_A \uR^{-1}$,\footnote{To see this note as $\uR$ and $\overline{Y}$ commute with each other, we can write $\overline{Y} = \uR \overline{Y}_A \uR^{-1} + \uR \overline{Y}_B \uR^{-1}$, and the eigenvalue of each component can be again found to be the same as before.} therefore the eigenvalue of $\left(\uR g_1 g_0\right)$ is given by 
\begin{equation}\label{eq:eigval-shift-13}
   \left( (-q)^{-2} \exp\left( {\rm i} 2\theta\right) \right)^{\frac{1}{2\R - 2}}  =  \exp\left( 2 {\rm i}\left(\theta - \gamma \right) \right)^{\frac{1}{2 \R - 2}} =  
  \exp \left(  \frac{\im \, 2  \pi}{2 \R - 2} \left(2 n  \pm \frac{s \mp 1}{p+1} \right) \right)
   \, . 
\end{equation}
Again, like in the case of $(1,2)$ defect, the log of the eigenvalues obtained in Eq. \eqref{eq:eigval-shift-13} gives us $h - \bar{h}$, where $h$ and $\bar{h}$ are the conformal dimension appearing in the $(1,3)$ twisted partition function. But in this case, from Eq. \eqref{eq:eigval-shift-13} we see that the length of the spin chain must be taken to be   $2 \R - 2$ in the finite size scaling analysis of the  momentum to get exact values, as we had noted in section \ref{sec:examples}.

\newpage
\end{appendix}

\newpage
\bibliography{main}

\end{document}